\documentclass[11pt,openright,final,smallheadings]{scrbook}

\usepackage[estonian, english]{babel}
\usepackage[utf8]{inputenc}
\usepackage[toc,title,page]{appendix}
\usepackage[T1]{fontenc} 
\usepackage{times}

\usepackage{longtable} 
\usepackage{aas_macros} 
\usepackage{verbatim} 
\usepackage{amsmath} 
\usepackage{amsfonts} 
\usepackage{amssymb}  
\usepackage{amscd} 
\usepackage{graphicx,epsfig} 
\usepackage{subfig} 
\usepackage{float} 
\usepackage{lscape} 
\usepackage{url}
\usepackage{sidecap}
\sidecaptionvpos{figure}{t}

\usepackage[english,estonian]{babel}

\usepackage{accents}

\usepackage{natbib}

\newcommand{\bn}{\begin{enumerate}}
\newcommand{\en}{\end{enumerate}}

\newcommand{\bed}{\begin{displaymath}}
\newcommand{\eed}{\end{displaymath}}

\newcommand{\ie}{{\it i.e. }}

\def\Mp{\,{\rm M_{\odot} /pc^2}}

\def\Mp3{\,{\rm M_{\odot} /pc^3}}
\def\sr{\sigma_R^2}
\def\st{\sigma_\theta^2}
\def\sz{\sigma_z^2}

\newcommand{\be}{\begin{equation}}
\newcommand{\ee}{\end{equation}}
\newcommand{\dd}[1]{{\rm d}#1\,}
\newcommand{\mm}[1]{\mathfrak{#1}\,}
\newcommand{\mc}[1]{\mathcal{#1}\,}
\newcommand{\ba}{\begin{array}} 
\newcommand{\ea}{\end{array}}
\newcommand{\pdif}[2]{\frac{\partial #1}{\partial #2}}
\newcommand{\ppdif}[2]{\frac{\partial^2 #1}{\partial #2^2}}
\newcommand{\psdif}[3]{\frac{\partial^2 #1}{\partial #2 \partial #3}}

\def\aj{AJ}%
\def\actaa{Acta Astron.}%
\def\araa{ARA\&A}%
\def\apj{ApJ}%
\def\apjl{ApJ}%
\def\apjs{ApJS}%
\def\ao{Appl.~Opt.}%
\def\apss{Ap\&SS}%
\def\aap{A\&A}%
\def\azh{AZh}%
\def\jrasc{JRASC}%
\def\mnras{MNRAS}%
\def\pasp{PASP}%
\def\pasj{PASJ}%
\def\sovast{Soviet~Ast.}%
\def\zap{ZAp}%
\def\nat{Nature}%
\def\aplett{Astrophys.~Lett.}%
\def\bain{Bull.~Astron.~Inst.~Netherlands}%
\def\aplett{Astrophys. Letters}

\bibpunct{(}{)}{;}{a}{}{,} 

\begin{document}

\frontmatter

\begin{sffamily}
\begin{center}
{DISSERTATIONES ASTRONOMIAE UNIVERSITATIS TARTUENSIS}\\
\textbf{}
\vspace{40mm}

{\LARGE{\textbf{JAAN EINASTO}}}\\
\vspace{20mm}

{\LARGE{Structure and Evolution of Regular Galaxies}}\\[10pt]


\end{center}
\end{sffamily}

\vfill

{\begin{figure*}[h] 
\centering 
\resizebox{0.55\textwidth}{!}{\includegraphics*{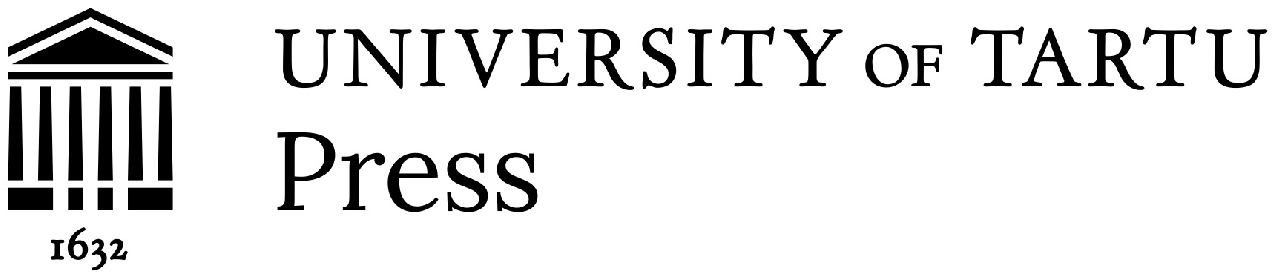}}
\end{figure*} 
}

\newpage
\thispagestyle{empty}
\noindent
This study was carried out at the Institute of Physics and Astronomy,
Estonian Academy of Sciences.

\vspace{5mm}

\noindent
The Dissertation was admitted on February 14, 1972, in partial fulfilment of
the requirements for the degree of Doctor of Science in astronomy and
celestial mechanics,  and allowed for defence by the Council of the
Institute of Physics, University of Tartu.

\vspace{5mm}

\begin{tabbing}

  Opponents: \hspace{0.7cm}

               \= Prof. E. K. Kharadze\\
              \> Abastumani Observatory \\
              \> Georgia \\
\\
	       \> Prof. T. A. Agekian\\
              \> Leningrad University \\
              \> Russia \\
\\
	      \> Prof. G. M. Idlis\\
              \> Alma-Ata University \\
              \> Kazakstan \\

\vspace{5mm}\\
            
Leading institute: Lebedev Physical Institue of the Soviet Academy of Sciences. \\             
\vspace{5mm}\\
Defence:      \> March 17, 1972, University of Tartu, Estonia \\

\end{tabbing}

\vfill
\begin{tabbing}
  Original in Russian --- I Volume: Chapters 1 -- 7, 195 pages\\
  \hskip 35mm  II Volume: Chapters 8 -- 23 with appendix, 330 pages
  \\
  \\
ISBN 978-9949-03-771-1
\\
\\
Copyright: Jaan Einasto, 2021 \\
\\
University of Tartu Press 2021\\
www.tyk.ee\\
\end{tabbing}

\setcounter{tocdepth}{2}
\tableofcontents

\chapter{Preface to the English edition}\label{ch000}

The Thesis was an attempt to combine data from three previously
independent areas: the structure and kinematics of stellar populations
of the Galaxy, models of galaxies, and models of the evolution of
galaxies. This synthesis was made with the goal to understand better
the structure and evolution of galaxies. When the work was finished it
was clear that there are difficulties and open problems in the
classical picture. Thus, immediately after the Thesis was completed, I
started together with my Tartu collaborators searching for solutions
to open problems. This search led to accepting the presence of dark
matter in galaxies. To understand the properties of dark matter in
galaxies, it was needed to study the environment of galaxies and the
distribution of galaxies in space, which culminated with the discovery
of the cosmic web. A short overview of the development of ideas
directly connected with the topic of the Thesis is given in the
Epilogue.

In some sense the Thesis is a time-capsule of the state of
affairs just at the verge of the paradigm shift in cosmology. The
Thesis was written in Russian and its most important parts were never
published. Thus, it would be useful to make this study available for
the astronomical community by translating the Thesis into English.

According to Soviet rules, doctoral theses must be written in Russian
and typed with a typewriter. It was common to base the thesis on
previously published papers, thus these papers must be retyped to form
a collection needed for the thesis. In early 1970's, I had finished
several cycles of papers on stellar kinematics and galactic models,
suited as the basis of the Thesis. Also, I had unpublished results on
the study of the dynamical and physical evolution of stellar systems.
I prepared my Thesis on the basis of this work. About half of it was
based on my papers published in Tartu Observatory Publications in
Russian, and a few papers in English in conference proceedings. New
results were written in 1971 as additional chapters of the Thesis.
The
text was typed only once with five carbon copies, all equations
hand-written. Copies of better quality were given to thesis reviewers
and to the Moscow office, where all theses completed in USSR were
collected and revised for acceptance.  The original copy of the Thesis
in Russian is scanned and can be accessed in Dspace link:
http://hdl.handle.net/10062/6113.  The translated  English version is
available in  link http://hdl.handle.net/10062/76090.

The Thesis consists of four parts, and is divided to 23 chapters. 
Chapters 4, 7, 20, 21, 22, 23 were unpublished and the present
translation is their first publication. These chapters were translated
in full. Chapters 7, 11, 17, 19 were published in Tartu Observatory
Publications or conference proceedings, but form the methodical and
data basis to understand the main topic of the Thesis, thus these
chapters were also translated in full. The rest of chapters, published in
Tartu Observatory Publications, describe the general background of the
topic, and are written in this English version as short summaries of
respective papers in Russian.

No original figures are available, only copies of very different
quality from paper copies and microfilms to figures on typed pages of
the Thesis. Copies were scanned and used to prepare figure files
suitable for publication. Tables were partly retyped and partly
scanned from the Thesis copy available.

The translation from Russian into English was made by myself, while my
colleague Peeter Tenjes translated chapter 11. My grandson Peeter
 corrected figure files. My colleagues in Tartu Observatory
helped to polish the text and to fix errors. The remaining errors are
my own responsibility.

\vskip 5mm
\hfill November 2021

\chapter{Preface}\label{ch001}

It is customary to divide the stellar astronomy into the theory of
stellar systems and observational astronomy. The classical theory of
stellar systems covers stellar dynamics and statistics, while
observational astronomy covers direct information about the structure
and composition of stellar systems and various astronomical and
astrophysical methods of obtaining it.

\citet{Parenago:1948aa} introduced the concept of practical stellar
dynamics to denote the study of the structure of stellar systems based
on observational data with the application of theoretical relations,
derived in the dynamics of stellar systems. Currently, the addition of
stellar dynamics and other theoretical disciplines — theories of
stellar evolution and chemical nucleosynthesis, gas dynamics,
relativistic astrophysics, etc. — are also applied to the study of the
structure of stellar systems. Retaining the convenient term of
practical stellar dynamics, it is reasonable to accept for its goal
the application of the results of the theory in the study of the
structure and evolution of particular stellar systems.

The basic method of practical stellar dynamics is the construction of
models of the objects under study. When considering theoretical
problems, usually only a certain aspect of the model is essential, and
there is no need to achieve representativeness of the model in other
details, secondary to the problem. The goal of practical stellar
dynamics is the developing of models of stellar systems, as
representative as possible, in which the synthesis of heterogeneous
observational information is based on the results of the theory of
stellar systems.

The main tasks of practical stellar dynamics include the study of the
evolution of stellar systems. The problem of evolution is considered
primarily as an observational one, \ie evolutionary conclusions are
drawn on the basis of a theoretical interpretation of suitable
observational data.

The body of works, which served as the basis for the present
dissertation, is devoted to the development of methods of practical
stellar dynamics and their application to investigate the structure
and evolution of regular galaxies like our Galaxy. We did not set as a
goal the further development of theory and obtaining new observational
data, since the already available theoretical and observational
information is much more than could be processed and combined in one
cycle of studies.

The author's interest in this subject arose already in the first half
of the 1940s, when the author, being a young amateur astronomer, read
with enthusiasm the articles by Ernst \"Opik, Taavet Rootsm\"ae, Aksel
Kipper and Grigori Kuzmin on the structure and evolution of stars and
stellar systems in the pages of Calendars of Tartu Observatory. Here I
would like to mention the pioneering works of \citet{Opik:1938aa} and
\citet{Rootsmae:1961}, which laid the foundation for revealing the
relation between ages and kinematical and spatial properties of
stellar populations in the Galaxy. The immediate impetus for beginning
the research on practical stellar dynamics was received in 1951, when
the author discussed with Pavel Parenago and Alla Massevich the
possible topic for my diploma thesis. They suggested a detailed study
of the kinematics of stars of the main sequence.
\citet{Parenago:1951aa} had just discovered that the main sequence is
kinematically inhomogeneous and wanted to have more detailed
information on this effect. This problem was very close to my own
interests as well as to the topic of the research of Prof. Rootsm\"ae,
so I agreed. This resulted in my diploma thesis
\citep{Einasto:1952tx}, as well as in my PhD thesis
\citep{Einasto:1954ts}. From this work grew a series of studies on
stellar kinematics, which served as the basis for the first section of
the first part of the present Thesis.

In 1952 and 1955—1956 author performed calculations for models of the
Galaxy by \citet{Kuzmin:1952aa, Kuzmin:1956ca}. In the course of this
work, I discovered that models can be refined by using some additional
data that were not taken into account at that time. The idea of
integrated use of observational information and theoretical results
was later applied in developing the concept of a consistent system of
local Galactic parameters and in constructing new empirical models of
the Galaxy. The corresponding series of studies is included in the
second section of the first part of the Thesis.

The construction of the Galactic model was hampered by two
difficulties. First, the result strongly depends on the method of
model building, in particular, on the choice of the initial
description function. Second, due to our position inside the Galaxy,
it is difficult to get a picture of its structure as a whole. To
enrich our understanding of the global structure of the Galaxy, the
study of other similar galaxies, among which Andromeda galaxy M31 is
the most suitable, plays an essential role. Thus, two cycles of works
arose, on the methods of building models of galaxies and on the study
of the structure of the M31 galaxy, which form the content of the
second and third parts of the dissertation.

\citet{Eggen:1962} showed that we can draw certain conclusions about
the evolution of the Galaxy from observational data on the spatial
kinematical structure of subsystems of stars of different ages. The
success of these authors prompted us to use the collected material to
identify the possible evolutionary path of the Galaxy. In addition to
the dynamical evolution, we also investigated the physical evolution,
following the example of \citet{Tinsley:1968}.
The addition of the
evolution issues allowed us to give the Thesis a more contemporary
character.

Constructed models of our Galaxy and the Andromeda Galaxy are more
detailed and representative than models known from the literature.
However, the available observational possibilities to improve the
models are far from being exhausted. On the other hand, the
methodology developed can also be applied to study the structure and
evolution of other galaxies.

Most of the results presented in the Thesis have been published. The
relevant papers have been reproduced partly unchanged, partly in
abridged or revised form.
Some results obtained recently have not yet
been published, so that the Thesis is of independent value.
The arrangement of the material is generally chronological with some
exceptions.

 \vskip 5mm
\hfill November 1971      
\mainmatter

\part{Spatial and kinematical structure of the Galaxy}

\chapter{Kinematical structure of the main sequence}\label{ch01}

It is well known that the velocity distribution of stars has
approximately the Schwarzschild character. The analysis of tangential
velocities of main sequence stars has shown that stars of the early
spectral type (hot giants) can be indeed presented by the Schwarzschild
law \citep{Einasto:1952tx,Einasto:1954ts}.  Stars of spectral classes
F, G, K, M have velocity distributions, which can be presented as a sum
of two Schwarzschild distributions with different velocity
dispersions. A similar picture is observed in populations of giant
stars.  Non-homogeneity  of kinematical characteristics of stellar
populations is evidently caused by the large dispersion of population
ages.  Hot giant stars of main sequence are relatively young. In
contrast, populations of stars of later spectral   types of the main
sequence as well as ordinary giant star populations are mixtures of
stars of rather various ages.

In calculations of kinematical characteristics of stars, the selection
of observational data is taken into account as well as the influence
of random observational errors \citep{Einasto:1955ty}. The Chapter is
published by \citet{Einasto:1954ts}, and is applied by
\citet{Einasto:1955ty,Einasto:1955tz}, and \citet{Tiit:1964mw}. 

Main results of the study can be summarised as follows.

1. A method is elaborated for treating the distribution of tangential
velocities under the assumption that the sample consists of two
groups of stars, with Schwarzschild velocity distributions with
different dispersions. By comparing the observed and theoretical 
distributions of tangential velocities, the method makes it possible 
to determine these dispersions as well as the fractions of stars
belonging to both groups.  The position of the centroids and the ratio
of the axes of the velocity ellipsoid were taken as given, since the
distribution of tangential velocities was only weakly dependent on
these parameters. In the method, the distorting effects are taken into
account: errors in selection of proper motions, tangential velocity
errors, and irregularities in the distribution of stars across the
sky. The method also makes it possible to determine the average errors
of the unknown quantities and to take into account the influence of
possible errors in the given parameters. In addition, it is possible
to calculate the mean variance of velocities corresponding to both
groups of stars taken together.

2. The analysis of  tangential velocities of main sequence stars in the
spectral range from A5 to M leads to the following results. The
distribution of velocities of  A stars is well represented by a
single Schwarzschild distribution. Starting from F stars, the observed
velocity distribution can be represented as the sum of two
Schwarzschild distributions with different dispersions: the stars are
separated here statistically into two kinematical groups. The
dispersions of velocities of both kinematical groups practically do not
depend on the spectral type, but the fraction of stars with low velocities
varies. It is minimal in spectral class G and increases towards the
earlier and later spectral classes.  In this connection the mean
velocity dispersion (both kinematical groups taken together) is maximal
at spectral class G and noticeably decreases in the transition to
spectral F and A classes, as well as to K and M. Stars of the middle
part of the main sequence can also be divided into two groups,
however, the spectral division does not coincide with the kinematical
one.  The spectral separation is observed, firstly, only in the
spectral range from F to G5.  Secondly, the velocity dispersion of one
spectrally separated group of stars coincides with the velocity
dispersion of the first kinematical group (small velocities), while the
velocity dispersion of the other group is much smaller than the
dispersion of the second kinematical group (high velocities), and is
close to the average dispersion in the second part of the main
sequence.

3. The discrepancy between the spectral and kinematical separation is
apparently due to the fact that only the spectral separation
corresponds to the partitioning of the main sequence into two
genetically unrelated parts, whereas the kinematical separation does not
correspond to this division. The first part of the main sequence has a
homogeneous kinematical structure and probably ends at the G5 spectral
class. The second part of the main sequence starts at spectral class F
and is characterised by a significant heterogeneity in the kinematical 
structure, with a possible continuous transition from stars with small
velocity dispersion to stars with large velocity dispersion. The
reason for the kinematical heterogeneity of the second part of the main
sequence is most likely due to the difference in ages of the
constituent stars. The youngest stars in the second part of the main
sequence are red dwarfs with emission lines in their spectra and the
lowest velocity dispersion. This allows us to conclude that velocities
of stars of the second part of the main sequence increase with time.

\vskip 5mm
\hfill 1954
\chapter{Velocity dispersions from 
  their observed velocities}\label{ch02}

In our recently published paper \citep{Einasto:1954ts}, we proposed a
simple method of determining the velocity dispersion from the total
tangential velocities of stars. The method is based on the fact that
the mean value of the squared tangential velocity is proportional to
the square of the velocity dispersion, if other parameters are
identical. The formulas are given to calculate the coefficient of
proportionality and to take into account the errors in parallax and
the mean error of the dispersion.  In a quite similar way, the mean
velocity dispersion can be found from radial and total spatial
velocities. For the proportionality coefficient velocities, a different
expression is obtained than for the tangential velocities.  The aim of
the present work \citep{Einasto:1955ty} is to describe the method in
more detail and to extend it to the case of radial and spatial
velocities.

The advantage of this method is, first of all, that it is very easy to
find the mean dispersion.  The result is only very slightly dependent
on the values of other parameters of velocity distributions. Another
advantage of the method is that the calculation of the dispersion
separately from radial and tangential velocities allows one to detect
possible systematic errors in the material, for example, in stellar
parallaxes.  Finally, in this way of calculating the dispersion, it is
very easy to account for observational errors in radial velocities,
proper motions and parallaxes. Here we give a short summary of the
paper.

The mean velocity dispersion of stellar populations can be calculated as follows:
\be
\sigma^2 = 1/3 (\sigma_R^2 + \sigma_\theta^2 + \sigma_z^2),
\label{2.1}
\ee
where $\sigma_R^2$, $\sigma_\theta^2$, $\sigma_z^2$ are velocity
dispersions in cylindrical coordinates. The mean velocity dispersion
is related to the mean observed velocity of stars, $V$: 
\be
\overline{V^2} = \beta \sigma^2,
\label{2.2}
\ee
where $V$ is the stellar velocity from observations, and 
$\beta $ is a dimensionless coefficient, depending on the nature of
$V$. As $V$ we can use the full spatial velocity of stars, $V_s$,
tangential velocity, $V_t$, or radial velocity, $V_r$.  Values of the
coefficient $\beta $ are calculated for all these cases, using the
generalised Kleiber theorem.  The kinematics of Me
dwarfs is studied using this method \citep{Einasto:1955tz}.

The results obtained in this paper can be seen as a generalisation of the
Kleiber theorem for moments of any order, including mixed moments,
and for velocity distributions, which are not spherical.  It is shown
that the respective coefficients have in the case of arbitrary
velocity distribution, including ellipsoidal, the same numerical
values, as in the case of spherical velocity distribution, if stars are
distributed uniformly over the whole sky, and if the velocity
distribution function is the same in all regions of the sky. Since
these conditions are in many cases well satisfied, one cannot agree
with the occurrence in the literature the statement, that in the light
of the modern understanding of velocity distributions, the
Kleiber theorem has lost his meaning.

  \vskip 5mm
\hfill   1955
\chapter{On the asymmetric shift of stellar velocity
  centroids}\label{ch03}

In this paper \citep{Einasto:1961aa}, we shall discuss one aspect of
the velocity distribution -- the asymmetric shift of the centroid of
the velocity ellipsoid.  A critical analysis of kinematical data,
collected in the next Chapter, shows that only part of the available
data can be used to study the asymmetric shift.  Here we give a short
summary of the paper.

For all populations we calculated  the following data: the mean
velocity dispersion,
\be
\sigma =
\sqrt{\frac{1}{3}( \sigma_R^2+\sigma_\theta^2+\sigma_x^2)},
\label{eq3.1}
\ee
and the mean heliocentric centroid velocity in rotational direction,
$\overline{V_\theta}$, where $\sigma_R$, $\sigma_\theta$, and
$\sigma_z$ are velocity dispersions in galactic cylindrical
coordinates.  The velocity dispersions were corrected for
observational errors using a method, proposed by us
\citep{Einasto:1955ty}.  The age of populations was determined from
Iben's evolutionary tracks, see Chapters 4 and 22. For halo
populations, the individual age determinations coincide within possible
errors. The relative age of these populations was estimated
theoretically, adopting for oldest halo populations the age of the
Galaxy, $10^{10}$ yr, and for other halo populations an age needed
for the population considered to collapse with free fall acceleration
to its observed dimensions.

{\begin{figure*}[h] 
\centering 
\hspace{2mm}
\resizebox{0.90\textwidth}{!}{\includegraphics*{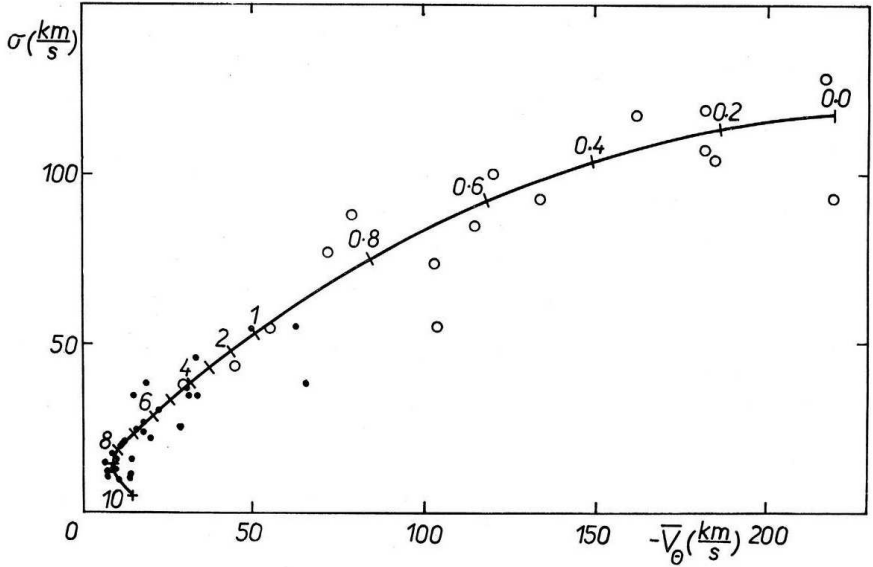}}
\caption{The Str\"omberg diagram for populations. In the horizontal
  axis, we show the heliocentric centroid velocity of the population in
  the direction of the Galactic rotation; in the vertical axis, we plot
  the mean velocity dispersion $\sigma = \sqrt{\frac{1}{3}(\sigma^2_R +
    \sigma^2_{\theta} + \sigma^2_z)}$. Open circles are for metal-poor
  populations, dots for populations with normal metal abundance.  The
  numbers give the birth-dates in $10^9$ years starting from the
  formation of the oldest populations, assuming for the age of the
  Galaxy $10^{10}$ years \citep{Einasto:1974ac}.
} 
  \label{Fig3.1}
\end{figure*} 
}

The Str\"omberg diagram for populations studied is given in Figure
\ref{Fig3.1}. Populations with metal deficit are represented by open
circles, populations with normal metal content by points, the
interstellar gas by a cross. The smooth curve shows the mean
dependence between $\sigma$ and $\overline{V_\theta}$ of populations
of different ages; the latter is indicated in $10^9$ yr, starting from
the formation of oldest galactic populations known.

The main results of this study may be formulated as follows.

  
1.  If we attribute all metal deficient subpopulations to the
  halo, then it appears that the halo is rather heterogeneous in its
  kinematical properties; it contains all subpopulations with velocity
  dispersion $\sigma \ge 50$~km/s. The corresponding axial ratio
  $\epsilon$ of equidensity ellipsoids, calculated from our recent
  model of the Galaxy \citep{Einasto:1970ad}, is equal to or larger
  than 0.10. Studying the structure of the Andromeda galaxy M31, we
  also came to the conclusion that its halo consists of a mixture of
  subpopulations with $\epsilon \ge 0.10$
  \citep{Einasto:1974ac}. These results show that intermediate
  subsystems of the Galaxy according to \citet{Kukarkin:1949aa} also
  belong to the halo.
  
2.  Direct age determinations of stellar populations are too
    inaccurate to estimate the duration of the initial galactic
    collapse. There exists, however, indirect observational
    \citep{Sandage:1969ab} and theoretical \citep{Eggen:1962} evidence
    that the collapse  proceeded in a short time scale compared with
    the age of the Galaxy.
    
3.    The populations of the galactic disc have mean velocity
    dispersions, $15 \le \sigma \le 50$~km/s, and respectively axial
    rations, $0.02 \le \epsilon \le 0.10$ \citep{Einasto:1970ad}. The
    age dependence of spatial and kinematical properties of these
    populations may be caused by the action of irregular gravitational
    forces \citep{Spitzer:1953aa, Kuzmin:1961aa}.
  
4.    The subsystem of interstellar gas and young stars rotate with
  a velocity smaller than the circulation one. Therefore,  young
  stellar subsystems are non-steady, and time is needed for them to
  obtain a steady structure. This result supports the recent discovery
  of the non-stationary state of young populations by
  \citet{Dixon:1967aa,Dixon:1967ab,Dixon:1968aa} and
  \citet{Joeveer:1968b}. 

  \vskip 5mm
\hfill August  1961

\chapter{Kinematical characteristics and ages of Galactic
  populations}\label{ch04}

The spatial and kinematical properties of galactic populations evolve
very slowly. Therefore, the study of these properties gives us certain
information on the past dynamical evolution of the Galaxy, in
particular on the evolution of star generating medium (interstellar
gas, as generally accepted). The detailed study of spatial structure
of stellar populations in our Galaxy is possible in most cases only in
the Solar neighbourhood. But the study of kinematical properties is
possible practically for all populations, which makes these studies
very useful for cosmogonic purpose.

In order to obtain adequate quantitative information for the study of
dynamical history of the Galaxy, the statistical data on stellar
velocities must satisfy the following requirements: populations
under study must be physically homogeneous; statistical samples of
stars must be free from selection effects, especially from velocity
selection; in order to correct the results for accidental observational
errors, the information on rms errors of observed quantities must be
known; the data to determine of the age of the sample must
be available.

We collected published data on stellar velocities and
determinations of kinematical parameters back to the fundamental work
by \citet{Parenago:1951aa}, which is the topic of this  Chapter.

\section{Introduction}

Characteristics of the spatial and kinematical parameters of star
samples are principal indicators in deciding to which Galactic
populations they belong.  Parameters of the spatial structure can be
found from observational data as a rule only for a restricted volume
of space near to the observer.  In contrast, kinematical parameters,
found in nearby volume, characterise the structure of the whole
population.  This allows to use kinematical parameters to investigate
the evolution of the Galaxy.

The first large surveys of kinematics of stellar populations were made
using radial velocities. Presently these studies have only historical
interest, since the samples were selected using parameters which are
not sufficient to select homogeneous populations. The first modern
compilation of stellar spatial velocities was compiled by
\citet{Parenago:1951aa}, a more recent one by \citet{Delhaye:1965wj}.

To apply kinematical characteristics of star samples to the study of
the structure and evolution of the Galaxy, the samples must be
representative.  This means that they should be free from observational
selection effects, and the influence of random and systematic errors
must be known. Samples collected by \citet{Parenago:1951aa} and
\citet{Delhaye:1965wj} do not satisfy these conditions accurately
enough.

In this paper, we collect published kinematical data of stellar
populations with the aim of finding representative samples. Also, we
shall try to find the ages of populations.

{
\begin{table*}[ht] 
  \caption{Kinematical data  on  Galaxy populations}
\label{Table4.1a}                         
{\footnotesize
  \centering
\begin{tabular}{llcrllcccl}
\hline  \hline
ID&  Sample   &Method&$n$&$\sigma$ & $-V_\theta$&$k_\theta$&$k_z$&$t$&Ref.\\
  \hline
(1)&(2)&(3)&(4)&(5)&(6)&(7)&(8)&(9)&(10)\\ 
\hline  
0  & Interst. H &21-cm&   &$5.5 \pm 0.4$ & $14.5 \pm 2.6$          &   &   & & 3 -- 6 \\
    & Interst. Ca&          &   &${\bf  6.0 \pm 1.0}$ &${\bf  15.5 \pm 0.8}$ &  &   & & 7 -- 9\\
     &                &          &    &$5.6 \pm 0.4$         &$15.4 \pm 0.8$    &  &        &0.00& \\
1  &$\delta$ Cep& $V_i$& 20&$8.4 \pm0.8$  &$16.4 \pm 1.6$      &0.53&0.20& & 10\\
  &                      &$V_r$&100&${\bf 10.0 \pm 0.5}$&${\bf 13.8 \pm 1.2}$& &      & &11,~12\\
  &                      &         &      &$9.5 \pm 0.4$          &$14.4 \pm 1.0$&    &       & & \\                                                  
  2&Supergiants&$V_i$&213&$9.8\pm 0.3$& $11.7 \pm 0.6$& 0.66& 0.36&      & 1\\
  3&B                &$V_r$&560&$10.3\pm 0.4$&$14.2 \pm 0.8$&       &        &0.01&13,~14\\
  4&Open clusters&$V_r$&    &$11.9 \pm 1.4$&                        &       &        &0.03&15\\
  5&Ap                 &$V_i$&62&$11.2 \pm 0.7$&$ 7.6\pm 0.8$&         &         & 0.3   &16\\
  &                &$V_r$&147&${\bf 10.0\pm 0.7}$&${\bf 15.0\pm 2.5}$&&     &    &17\\
  &                &         &      &$10.6\pm 0.6$&$8.0\pm 0.8$&                     &     &     &   \\
  6&B7-A8      & $V_i$&114&$12.4\pm 0.5$&$9.4\pm 1.0$&0.31&0.14   &0.18& 1\\
  7&A5-A9     &$V_t$&150&$11.1\pm 0.6$&                       &       &          &0.53&18\\
  8&B9-F0,~Ap&$V_i$&111&$11.9\pm 0.6$&$7.9\pm 1.0$&0.31&0.18&          &19\\
  9&A0-F3      &$V_i$& 89& ${\bf 14.0\pm 0.7}$&${\bf 7.7\pm 1.6}$&0.21&0.20&0.40&20\\
  10&A0          &$V_i$&475&$15.2\pm 0.8$& $14.2\pm 0.8$&              0.44&0.25&0.18&21\\
  11&F0-F4    &$V_t$&264&${\bf 12.9\pm 0.6}$&                &                     &       &0.95&18\\
  &A9-F4&     $V_i$&290&${\bf 18.5\pm 0.6}$&${\bf 10.4\pm 0.9}$&0.36&0.18&   & 1\\
  &          &              &     &$12.9\pm 0.6$&$10.4\pm 0.9$&                          &       &    &   \\
  12a&F5-F7& $V_t$&230&${\bf 15.6\pm 0.8}$&              &                           &       &1.54&18\\
  12b&F5-F7&$V_i$&177&$24.1\pm 0.8$&$17.7\pm 1.6$& 0.46  &  0.28&    &1\\
  12c&F4-F8&$V_i$& 88&${\bf 20.8\pm 1.0}$&${\bf 12.2\pm 2.4}$&0.42&0.41&   & 20\\
  13&F8-G2&$V_t$&261& ${\bf 20.4\pm 1.0}$&                                &       &       &2.6&18\\
  14&G3-G9&$V_t$&175&${\bf 22.9\pm 1.3}$&                                &       &        &3.1&18\\
  15&E0-E7&$V_t$&123&${\bf 18.1\pm 1.1}$&                                 &       &        &      &18\\
  16&F8-K6& $V_i$&522&$35.7\pm 0.7$&$31.4\pm 1.3$                 &0.35&0.23&     &1\\
  &F9-K6&$V_i$&228&${\bf 22.2\pm 0.7}$&${\bf 20.2\pm 1.6}$&0.34&0.26&    &20\\
  &          &        &      &$22.2\pm 0.7$&$20.2\pm 1.6$                 &0.34&0.26&      & \\
 \hline
\end{tabular}
\\
}
\end{table*}
}

{
\begin{table*}[ht] 
  \caption{Kinematical data  on  Galaxy populations}
\label{Table4.1b}                         
{\footnotesize
  \centering
\begin{tabular}{llcrllcccl}
\hline  \hline
ID&  Sample   &Method&$n$&$\sigma$ & $-V_\theta$&$k_\theta$&$k_z$&$t$&Ref.\\
  \hline
(1)&(2)&(3)&(4)&(5)&(6)&(7)&(8)&(9)&(10)\\ 
  \hline
  17&M     &$V_t$&347&${\bf 15.6\pm 1.0}$&              &                 &        &         &18\\
  &M     &$V_i$&170&32.9                          &22.5            &0.31        & 0.24&    & 1\\
  &K8-M67&$V_i$&112&${\bf 24.5\pm 1.0}$&$16.2\pm 2.5$&0.46&0.31&    &20\\
  &M          &$V_i,~V_t$&305&${\bf 26.3\pm 0.5}$&${\bf 17.1\pm 1.4}$&0.62&0.34& &22,~23\\
  &             &                 &     &$24.5\pm 1.0$&$16.2\pm 2.5$                &0.46&0.31& &\\
  18&dMe    &$V_i$&106& ${\bf 15.2\pm 0.9}$&$10.2\pm 1.4$&0.37&0.18& &24,~25\\
  19&Strong-line&$V_t$&898&$16.8\pm 0.9$&$9.9\pm0.6$&          &       &  &26\\
  &  ~~~-''-     &$V_s$&258&$17.3\pm 1.0$&            &          &        &  &27,~28\\
  &                 &         &       &$17.0\pm 0.7$&$9.9\pm 0.6$&        &        &   & \\
  20a&Weak-line&$V_t$&300&${\bf 27.2\pm2.1}$&              &        &        &4 &18\\
  & ~~~~-''-        &$V_t$&581&$25.9\pm 0.8$&$18.1\pm 1.0$&    &       &   &26\\
  &~~~~-''-         &$V_s$&267&$27.4\pm 0.8$&                       &    &        &   &27,~28\\
  &                       &         &      &$26.8\pm 0.6$&$18.1\pm 1.0$&   &        &    & \\
  20b& HV dwarfs&$V_r$&91& $25.7\pm 2.3$&$29\pm5$         &    &        &     &29\\
  21&gA-gG8        &$V_i$&404&$19.6\pm 0.4$&$11.5\pm 0.8$&0.46&0.29&     & 1\\
  22&gG9-gM       &$V_i$&921&$23.7\pm 0.4$&$17.5\pm 0.7$&0.47&0.29&     & 1\\
  &MIII              &$V_i$&226&$24.0\pm 0.7$&$21.0\pm 1.7$&0.85&0.34&     &19\\
  &                    &        &       &$23.8\pm 0.4$&$18.7\pm 0.7$&0.55&0.30&    &   \\
  23&Red var.     &$V_i$&130&$30.8\pm 1.2$&$22.6\pm 2.3$&0.55&0.44&     &  1\\
  24a&HV giants&$V_r$&308&$38.5\pm 2.0$&$66\pm 4$&             &       &     &29\\
  24b&~~~-''-    &$V_i$&20& $74\pm 8$&  $103\pm 16$& 0.36&0.25      &   &1\\
  25a&gM&$V_i$&6&$62\pm 12$&$74\pm 21$&    &             &    &42\\
  25b&gM&$V_i$&22&$40\pm 4$&$35\pm 7$& &       &    &42\\
  25c&gM&$V_i$&73&$26\pm 1.4$&$18\pm 2.5$& &  &   &42\\
  25d&gM&$V_i$&67&$22\pm 1.3$&$15\pm2.3$&   &  &    &42\\
  25e& gM&$V_i$&18&$17\pm 1.9$&$8\pm 3.4$&    &  &    &42\\
  26a&Subgiants&$V_i$&112&$31.7\pm 1.3$&$27.0\pm 2.5$&0.41&0.31&5&1\\
  26b&~~~-''-    &$V_i$&51 & $40.4\pm2.7$&$41\pm 5$     &0.42&0.31&   &30\\
        &                &        &     &$34\pm 1$      &$32\pm 2$     &       &       &    &  \\
 \hline
\end{tabular}
\\
}
\end{table*}
}

{
\begin{table*}[h] 
  \caption{Kinematical data  on  Galaxy populations}
\label{Table4.1c}                         
{\footnotesize
  \centering
\begin{tabular}{llcrllcccl}
\hline  \hline
ID&  Sample   &Method&$n$&$\sigma$ & $-V_\theta$&$k_\theta$&$k_z$&$t$&Ref.\\
  \hline
(1)&(2)&(3)&(4)&(5)&(6)&(7)&(8)&(9)&(10)\\ 
  \hline
  27&Plan.nebul.&$V_r$&96&$35\pm 3$&$29\pm 3$          & 0.60  &0.20&   &31\\           
       &W-dwarf&$V_t$&50&$33\pm 3$&$40\pm 3$       &0.38&0.20&      &32\\ 
  &~~~-''-        &$V_t$&27&$37\pm 5$&$37\pm 5$       &0.43&0.25&      &33\\
  &                   &         &     &$35\pm 2$&$34\pm 2$&      0.50&0.21& 5&    \\
  28&Subdwarf &$V_i$&141&$98\pm 5$&$136\pm 9$&   0.59&0.26&    &34\\
  &~~~-''-     &$V_r$&46&$75\pm 10$&$127\pm 19$&        &       &     &29\\
  &                 &         &    &$93\pm 4$&$134\pm 8$&    0.59&0.26&9.5&   \\
  29a&LPer var.&$V_r$&37&{\bf 54 }     & 50           &0.42&0.55&         &35,~36\\
  29b& ~~~-''- &-''-&76& {\bf 88}      &79           &0.94&0.42&9.2&-''-\\
  29c&~~~-''-  &-''-&129&{\bf 55}     &63            &1.10&1.47&    &-''-\\
  29d&~~~-''-  &-''-&129&{\bf 46}&     34            &0.59&0.19&    &-''-\\
  29e&~~~-''-   &-''-&134&{\bf 37}&    31            &0.49&0.76&     &-''-\\
  29f&~~~-''-    &-''-&83&{\bf 38} &    19             &1.59&0.36&     &-''-\\
  29g&~~~-''-   &-''-&51&{\bf 35}&     15.6          &        &      &2   &-''-\\
  30a&RR Lyr var&$V_i,V_r$&34&$77\pm 11$&$72\pm 21$&&&    &37\\
  30b&~~~-''-   &$V_i,V_r$&98&$118\pm 10$&$162\pm 24$&0.77&0.25& &37\\
  30c&~~~-''-   &$V_r$&  38&$38\pm 5$&$30\pm 11$&0.83&0.25&   &38\\
  30d&~~~-''-  &$V_r$& 21&$129\pm 16$&$217\pm 31$&0.41&0.59&10.0&38\\
  30e&~~~-''-  &$V_r$&16&$55\pm 11$&$55\pm 11$&            &       &9.0&39\\
  30f&~~~-''- &$V_r$&10&$85\pm 22$&$115\pm 22$&           &        &     &39\\
  30g&~~~-''-&$V_r$&27&$105\pm 17$&$185\pm 17$&         &        &      &39\\
  30h&~~~-''-&$V_r$&14&$55\pm 12$&$104\pm 29$&          &        &       &40\\
  30i&~~~-''-&$V_r$&11&$44\pm 11$&$45\pm 27$&             &        & 9.0&40\\
  30j&~~~-''-&$V_r$&37&$100\pm 14$&$120\pm 29$&         &        &      &40\\
  30k&~~~-''-&$V_r$&46&$108\pm 13$&$182\pm 30$&        &        &       &40\\
  30l&~~~-''-&$V_r$&21&$93\pm 17$&$220\pm 45$&           &        &10.0&40\\
  31&Glob.cl.&$V_r$&70&$120\pm 12$&$182\pm 30$&         &         &9.7&40\\
  32&HB stars&$V_r$&12&$92\pm 12$&                      &          &         &     &41\\
 \hline
\end{tabular}
\\
}
\end{table*}
}

\section{Kinematical characteristics of Galaxy populations}

Our compilation of kinematical data on samples of stars is given in
Tables~\ref{Table4.1a}, \ref{Table4.1b} and \ref{Table4.1c}.  In the
compilation, we used samples from the compilation by
\citet{Parenago:1951aa} which satisfied our criteria of
representativeness. A special attention was given to samples for which
it was possible find ages, and which in this way contributed to the
understanding of the evolution of the Galaxy.

Designations in the Table are as follows. In the first column, we give
the number of the sample, used also in Figures. The second column
gives the type of samples, the third column the method to find
kinematical characteristics: 21-cm -- using radio-line of neutral
hydrogen; $V_i$ -- using components of spatial velocities; $V_r$ --
using radial velocities; $V_t$ -- using tangential velocities; $V_s$
-- using full spatial velocities.
In the fourth column, we give the number of objects in samples $n$.
The fifth column is the mean velocity dispersion
\be
\sigma= \sqrt{\frac{1}{3}(\sigma_R^2+\sigma_\theta^2+\sigma_z^2)}.
\label{eq4.1}
\ee
In the sixth column $\overline{V}_\theta$ -- the heliocentric centroid
velocity in the direction  
of the Galaxy rotation. In two following columns -- ratios of velocity
dispersions
\be
k_\theta = \frac{\sigma_\theta^2}{\sigma_R^2}, \hspace{1cm}
k_z={\sigma_z^2 \over \sigma_R^2}.
\label{eq4.2}
\ee
In the  column (9)  -- the age of the sample in billions of years; in
the last column (10) the reference.

We note that in the Table~\ref{Table4.1b} samples No.~25 CN limits of M giants
are as follows: 25a: $CN \le -0.17$,~25b: $-0.17<CN\le -0.09$,~25c:
$-0.09<CN\le -0.01$, ~25d: $-0.01<CN\le 0.07$,~25e: $0.07<CN$.
In the Table~\ref{Table4.1c} in samples No.~29 periods of long-period
variables are: 29a: $P <150$,~29b: $150 \le P < 200$, 29c: $200
\le P< 250$, 29d: $250\le P <300$, 29e: $300\le P < 350$, 29f:
$350\le P < 410$, 29g: $410 \le P$, periods are in days.  In the
same Table in samples 30 RR Lyrae variables are the following: 30a: type
I, 30b: type II, 30c: type I, 30d: type II, 30e: the
$\Delta\,s$ parameter by \citet{Preston:1959tr} is the following: 30e:
 $0 \le \Delta\,S \le 2$, 30f: $3 \le \Delta\,S \le 4$, 30g: $5
\le \Delta\,S \le 10$, 30h: type c, 30i: type ab with period $P
<0.4$~days, 30j: type ab with $0.4\le P <0.4$, 30k: period $0.4 \le
P<0.6$, 30l: period $0.6 \le P$.

Reference numbers in Tables are the following: 1 --
\citet{Parenago:1951aa}, 3 -- \citet{Kwee:1954wu}, 4 --
\citet{Westerhout:1957aa}, 5 -- \citet{Schmidt:1957aa}, 6 --
\citet{Venugopal:1967vl} , 7 -- \citet{Plaskett:1931tz}, 8 --
\citet{Melnikov:1947aa}, 9 -- \citet{Blaauw:1952ws}, 10 --
\citet{Parenago:1947aa}, 11 -- \citet{Takase:1963wh}, 12 --
\citet{Kraft:1963to}, 13 -- \citet{Feast:1965tk}, 14 --
\citet{Rubin:1964vk}, 15 -- \citet{Johnson:1961wb}, 16 --
\citet{Eggen:1959ui}, 17 -- \citet{Day:1969vc}, 18 --
\citet{Einasto:1954ts}, 19 -- \citet{Eggen:1960wm}, 20 --
\citet{Wehlau:1957td}, 21 -- \citet{Alexander:1958wm}, 22 --
\citet{Dyer:1956aa}, 23 -- \citet{Mumford:1956vk}, 24 --
\citet{Einasto:1955tz}, 25 -- \citet{Gliese:1958tb}, 26 --
\citet{Vyssotsky:1953ub}, 27 -- \citet{Roman:1950wo}, 28 --
\citet{Roman:1952uj}, 29 -- \citet{Michaowska:1960uk}, 30 --
\citet{Eggen:1960wt}, 31 -- \citet{Wirtz:1922wj}, 32 --
\citet{Parenago:1947aa}, 33 -- \citet{Pavlovskaya:1956aa}, 34 --
\citet{Parenago:1949wo}, 35 -- \citet{Feast:1963wx}, 36 --
\citet{Smak:1965wb}, 37 -- \citet{Pavlovskaya:1953aa}, 38 --
\citet{Notni:1956aa}, 39 -- \citet{Preston:1959tr}, 40 --
\citet{Kinman:1959wa}, 41 -- \citet{Philip:1969tt}, 42 --
\citet{Yoss:1962uk}. 

The dispersion $\sigma$ was calculated using published values of
$\sigma_r,~\sigma_\theta,~\sigma_z$, or found using velocity components
$V_i$ or $V_r$ of individual stars, applying methods described in
Chapter 2.

Errors of $\sigma$ and $\overline{V}_\theta$ were taken from
published data or calculated using methods described in Chapter 2. The
determination of age estimates is described below.

In Fig~\ref{Fig4.1} the solid line shows the mean relation between
$\overline{V}_\theta$ and $\sigma$. We did not try to find a
mathematical expression for the relationship, since in this case some
important details needed to understand the evolution of the Galaxy 
would be lost, see Chapter 21.

{\begin{figure*}[h] 
\centering 
\hspace{2mm}
\resizebox{0.90\textwidth}{!}{\includegraphics*{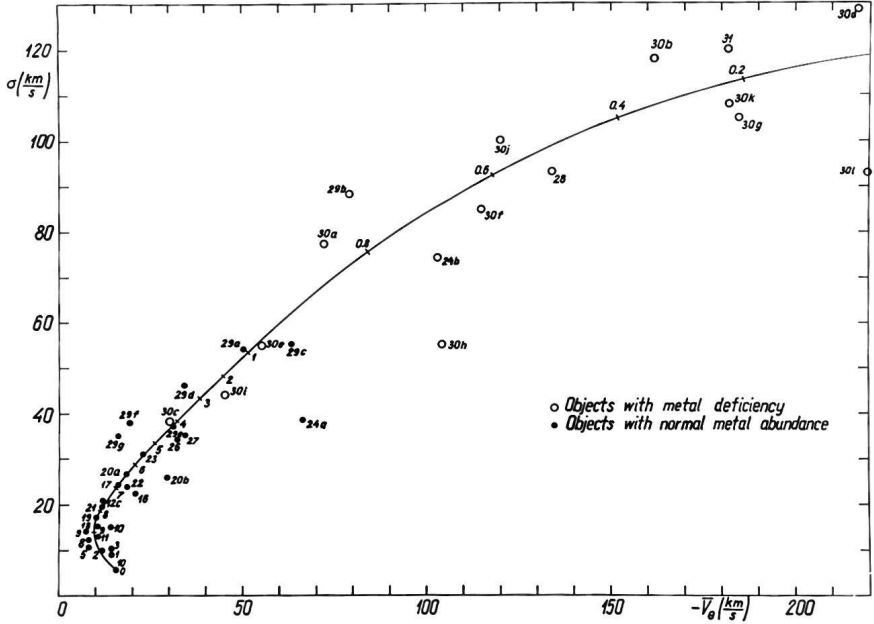}}
\caption{The relationship between the mean velocity dispersion $\sigma$
  and the heliocentric rotation velocity $V_\theta$ for various
  subsystems of the Galaxy. Subsystems of different chemical
  composition are marked with different symbols, numbers are according
to Tables \ref{Table4.1a}, \ref{Table4.1b} and  \ref{Table4.1c}. The
line shows the mean relationship of 
subsystems of various ages in billions of years, starting from the
formation of the oldest populations.} 
  \label{Fig4.1}
\end{figure*} 
}

\section{The influence of selection and observational errors}

In the study of the kinematical structure of main sequence star
samples, we paid essential attention to the influence of selection and
random errors \citep{Einasto:1954ts, Einasto:1955ty}.  Both effects
were taken into account by \citet{Wehlau:1957td}.  In other studies
these factors were ignored, or discussed using methods which did not
guarantee sufficient accuracy of results.

{\begin{figure*}[h] 
\centering 
\hspace{2mm}
\resizebox{0.50\textwidth}{!}{\includegraphics*{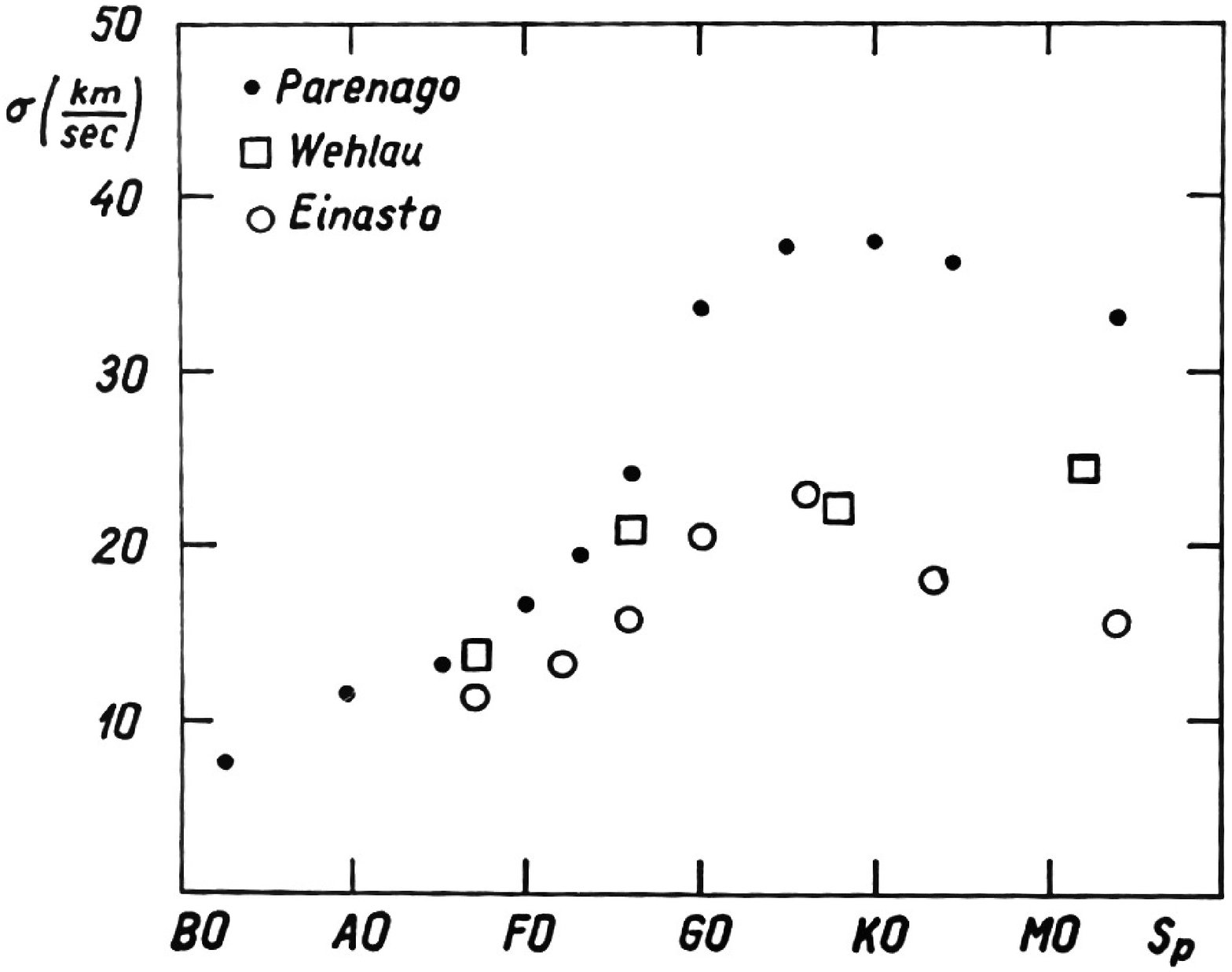}}
\caption{The mean velocity  dispersion of main sequence stellar
  populations according to three authors. Due to ignoring
  observational errors, dispersions found by Parenago are considerably
  higher, and dispersions found by Wehlau slightly higher. } 
  \label{Fig4.2}
\end{figure*} 
}

To illustrate the effect of these factors, we show in Fig.~\ref{Fig4.2}
kinematical characteristics of the stars of the main sequence according to
\citet{Parenago:1951aa}, \citet{Wehlau:1957td} and
\citet{Einasto:1954ts}.  We see that velocity dispersions $\sigma$
according to Parenago are much higher than dispersions obtained by other
investigators. Data by \citet{Wehlau:1957td} and
\citet{Einasto:1954ts} are generally in good mutual agreement, there
are only minor differences. We do not have original data by
\citet{Wehlau:1957td}, thus we cannot estimate the cause of
remaining differences. It is possible that the correction for observational
errors by Wehlau was not correct. On the other hand, it is not
excluded that we have overcorrected our samples for errors. For now, we
accepted data from our determinations for main sequence stars.

Kinematical characteristics, corrected for selection and observational
error effects, are printed in Tables \ref{Table4.1a}, \ref{Table4.1b} and \ref{Table4.1c} in boldface.

The need to take into account these factors was known long ago,
however, in most studies it was ignored. An exception is the work by
\citet{Michaowska:1960uk}, where the absence of stars with low
velocities was taken into account. In similar studies by
\citet{Yasuda:1961aa} and \citet{Eggen:1964ua}, the selection effect
was not taken into account, and we could not use these
samples. Moreover, \citet{Eggen:1969wd} divides stars into flat and disc
populations using components of spatial velocities ($V_R$,
$V_\theta$): stars within a defined region belong to the flat component,
and stars outside this region to the disc component.  This division
ignores the presence of tails in velocity distribution of the flat
component, and the presence of stars with low velocities in the disc
component.

Also, we could not use the large catalogue of spatial velocity
components by \citet{Eggen:1962wv}, since the system of photometric
parallaxes is voluntary, as shown by \citet{Weller:1968to}.

\section{Determination of  population ages} 

One of the main goals of the present paper is the establishment of a
relationship between kinematical characteristics and ages of
populations.

The ages of samples of stars from the upper part of the main sequence
were found either from the zero-age curve, shown in
Fig.~\ref{Fig22.1}, or from the mean spectral type. The maximal age of
main sequence stars can be found from Iben evolutionary tracks, see
Chapter 22.  Fast movement away from the main sequence starts at point
Nr. 3 of \citet{Iben:1967aa} track. We found that the respective age
is related to the visual luminosity with a simple equation ($M_V
\le5$): 
\be
\log\,t_3=8.19 + 0.353\,M_V.\nonumber
\ee
This equation gives the maximal age of stars. Early stars of the main
sequence have the mean age equal to half of the maximal age.

Stars of late spectral type of main sequence have large velocity
dispersions and form a thick subsystem. For this reason, the spatial
density of stars of this type is relatively small.  Samples of stars
 are collected near the Solar vicinity, thus old stars are
represented with a smaller frequency.  For this reason, for F8 to G9 type
stars of the main sequence we accepted mean ages slightly less than half
of the maximal age. For stars of later spectral type, it is difficult
to estimate this effect quantitatively, thus we did not try to find
their mean age.

We attribute to sub-giants, white dwarfs and planetary nebulae an
age equal to a half of the age of the whole Galaxy.  These objects have
large spread of ages, however, older stars dominate, which confirms our
estimate.  

The youngest Mira type variables with initial masses about 
$2.5\,M_\odot$ have ages  slightly less than a billion
years. However, the spread of periods of Miras of identical ages is
large (\citet{Smak:1966tk}, \citet{Feast:1965tk}).  For this reason, we
accepted for Miras of the type 29g a larger age -- 2 billion years.

To star samples with the largest velocity dispersion and centroid
velocity (short period cepheids with $P \ge 0.6$ days), we attribute
ages equal to the age of the whole Galaxy. As explained 
in Chapter 23, we accepted for the age of the Galaxy the round value 10
billion years.

Figure~\ref{Fig4.1} shows that approximately at the point with
coordinates $-\overline{V}_\theta = \sigma=50$~km/sec, a transition
takes place from stars with metal deficit to stars of normal metal
content.  We attribute stars with metal deficit to the halo, and
stars with normal metal content to the disc and the core. Calculations of the
physical evolution of the Galaxy show (\citet{Sandage:1969ac},
\citet{Cameron:1971aa}) that the metal enrichment proceeds in the early
phase of Galaxy evolution very rapidly, and that the mean chemical
composition changes little later.  For this reason, the second phase of
the chemical evolution of the Galaxy has a long duration about 9
billion years, see Chapter 23.  We conclude that at the point with
$-\overline{V}_\theta = \sigma=50$~km/sec, the halo formation was
finished, and the formation of disc started.

{\begin{figure*}[h] 
\centering 
\hspace{2mm}
\resizebox{0.50\textwidth}{!}{\includegraphics*{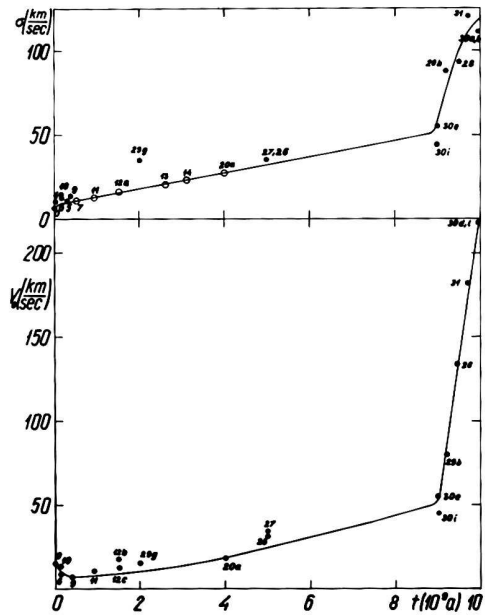}}
\caption{The dependence of the mean velocity dispersion $\sigma$ and
  heliocentric rotation velocity $V_\theta$ on the age $t$ of
  populations, according to Table \ref{Table4.1a}. Open circles in the
    top panel show populations where it was possible to correct data
    for observational errors.  } 
  \label{Fig4.3}
\end{figure*} 
}

The problem of the duration of the formation of the halo, and the
possibility that some halo and disc stars formed at the same time,
is widely discussed. \citet{Eggen:1962}, and \citet{Sandage:1969ab,
  Sandage:1970ab} argued that the halo formed quickly during several
hundred million years. On the other hand, \citet{Rood:1968aa} support
a more extended period of halo formation, and that some halo stars
formed after the start of disc star formation.  Our data, shown in
Fig.~\ref{Fig4.1}, suggest that some overlap in the formation of halo and
disc stars is possible. Our calculations of the Galaxy evolution
suggest that the formation of the halo could be slower than accepted by
\citet{Eggen:1962}.  However, the argument by Eggen et al.  concerning
the short formation scale of the halo must be correct.

Using these arguments, we accepted the ages of halo objects.  Globular
clusters have the largest axial ratios of equidensity ellipsoids
(Chapter 20) and the smallest heavy element abundance. For globular
clusters, we accepted ages longer than the mean age of the whole halo. To Mira
variables with periods 150 -- 200 days, we accepted an age
$9.2\times10^9$~yr.  Miras of this type can be located in 
relatively rich globular clusters (\citet{Arp:1963ur},
\citet{Sandage:1966ul}, \citet{Rosino:1966uf}), which according to
other data are younger than normal globular clusters.

We show in Fig.~\ref{Fig4.3} the dependence of $\sigma$ and
$\overline{V}_\theta$ on the age of subsystem $t_S$.  For disc objects,
we used data by \citet{Einasto:1954ts}, where observational selection
and error effects were studied in great detail. In the determination
of the function $\overline{V}_\theta(t)$, we notice that selection and
error effects shift values of $\sigma$ and   $\overline{V}_\theta$ in a
similar way, thus data points do not exit from the mean relationship
curve.  We can consider the Fig~\ref{Fig4.3} as a parametric
presentation of the mean relationship in Fig.~\ref{Fig4.1}.

{\scriptsize
\begin{table*}[h] 
  \caption{Kinematical data  on  Galaxy populations}
\label{Table4.2}                         
\centering
\begin{tabular}{cccccccc}
\hline  \hline
ID&$k_\theta$&$-\Delta\,e$ &$-\Delta\,h$&$(k_\theta)_0$&$k_z$&$-\Delta\,e$&$(k_z)_0$\\
\hline
  10&0.44&0.01&0.00&0.43&0.25&0.005&0.245\\
  12&0.44&0.01&0.00&0.43&0.32&0.005&0.315\\
  16&0.34&0.01&0.01&0.32&0.26&0.005&0.255\\
  17&0.46&0.01&0.03&0.42&0.31&0.005&0.305\\
  18&0.37&0.01&0.00&0.36&0.18&0.005&0.175\\
  21&0.46&0.01&0.00&0.45&0.29&0.005&0.285\\
  22&0.55&0.01&0.10&0.44&0.30&0.005&0.295\\
  23&0.55&0.01&0.10&0.44&0.44&0.005&0.435\\
  26&0.41&0.01&0.02&0.38&0.31&0.005&0.305\\
  27&0.50&0.01&0.06&0.43&0.21&0.005&0.205\\
 \hline
\end{tabular}
\end{table*}
}

\section{Ratios of semiaxes of velocity ellipsoids}

Ratios of semiaxes of velocity ellipsoids are important parameters, 
characterising the local structure of the Galaxy. Our collection of
kinematical data allows to get new estimates of these parameters.

As we see later in Chapter 21, very young stellar populations are not
in a stationary state. On the other hand, very old populations have
ratios of velocity ellipsoid semiaxes, different from ratios found
for young flat populations.  For this reason, we shall use in the
determination of mean values of ratios $k_\theta$ and $k_z$ only
subsystems in the velocity dispersion interval from 15 to 50
km/sec. Mean values of $k_\theta$ and $k_z$ in this $\sigma$ interval,
collected from data given in Table~\ref{Table4.1a} - \ref{Table4.1c}, are given in
Table~\ref{Table4.2}. 

To find mean values of $k_\theta$ and $k_z$, the influence of
observational errors must be taken into account. This error makes the
velocity ellipsoid rounder. Furthermore, it is needed to take into
account the kinematical heterogeneity of observation data. The
asymmetric shift of the velocity ellipsoid increases the dispersion ratio
$k_\theta$. A theory of this factor was developed by
\citet{Eelsalu:1958aa}.  We estimated corrections $\Delta\,e$ and
$\Delta\,h$ using data on the inhomogeneity of star samples, and mean
errors of stellar parallaxes. Results are given in Table~\ref{Table4.2}
as well as the corrected values of  $(k_\theta)_0$ and $(k_z)_0$. The
overall mean values and their
estimated errors are: $(k_\theta)_0=0.410\pm
0.015$ and $(k_z)_0= 0.278\pm 0.010$.

\section{Circular velocity in the Solar vicinity}

Our collection of data allows to calculate the circular velocity near
the Sun, using the theoretical Str\"omberg  asymmetry equation
\citep{Einasto:1964wx}
\be
V_i^2 -(G_i +m_0)\,\sigma_{Ri}^2 = V^2,
\label{eq4.3}
\ee
where $V_i$ is the galactocentric rotation velocity of subpopulation
$i$, $\sigma_{Ri}$ is the velocity dispersion in radial direction of
this subsystem, $V$ is the circular velocity, 
\be
G_i=G\{\rho(R)\}_i = \frac{\partial\ln\rho_i}{\partial\ln\,R}
\label{eq4.4}
\ee
is the logarithmic gradient of the density, and
\be
m_0=(1-k_\theta)+n_R(1-k_z)
\label{eq4.5}
\ee
is a quantity, equal for all Galaxy subsystems. Here we used designations
identical to Eq.~(\ref{eq11.2.10}).

Galactocentric centroid velocity $V_i$ can be expressed through the
$\theta$-component of the heliocentric centroid velocity
$\overline{V}_\odot$ using the equation
\be
V_i= V+(\overline{V}_\odot - V_\odot),
\label{eq4.6}
\ee
where $V_\odot$ is the $\theta$ component of the Solar velocity in
respect to the circular velocity.

Using our collected data as well as data by \citet{Blaauw:1965aa} on
the density gradient, we found for flat populations
\be
V_\odot = -9.0 \pm 0.2~ {\rm km/sec},
\label{eq4.7}
\ee
which yields, using data of intermediate and spherical populations,
\be
V=226 \pm 21~ {\rm km/sec}.
\label{e14.8}
\ee
Error of this determination of the circular velocity is fairly large,
it is determined by errors in the density gradient and velocity
dispersion.

We note that this method to determine the circular velocity in the Solar
neighbourhood was applied earlier by \citet{Parenago:1951aa}.

\vskip 5mm
\hfill September 1971

\chapter{ Model of the Galaxy and the system of Galactic parameters:
  Preliminary version}\label{ch05} 

This Chapter presents our first attempt to bring together the
available data on the structure of the Galaxy in a model, and to find
the system of Galactic parameters. The model was presented in the talk
by \citet{Einasto:1965aa} in the conference ``Kinematics and dynamics
of stellar systems and physics of the interstellar medium'' in
Alma-Ata in summer 1963.

\section{Introduction}

In order to bring together the available data on the structure of
stellar systems and to determine their gravitational field,
appropriate models are used. Naturally, particular attention is
paid to the building of the model of our Galaxy. Work in this
direction has been going on in Tartu for more than ten years. At first
we were interested mainly only in the radial distribution of masses in
the Galaxy \citep{Kuzmin:1952aa, Kuzmin:1956aa}. As for the spatial
distribution, the models were not specified, or special models were
used by \citet{Kuzmin:1956ca,Kuzmin:1962bb}. At the present time we
set the task of building a more detailed model of the Galaxy without
trying to base it on any special assumptions. 

The problem of constructing a model of the Galaxy is closely related
to the problem of constructing a system of Galactic parameters, and
the application of the equations of stellar systems hydrodynamics. The
hydrodynamics of stellar systems is considered in the paper by
\citet{Kuzmin:1965cc}, and the problem of determining the Galactic
parameters is discussed in the paper by \citet{Kutuzov:1965bb}. The
present paper considers the problem of constructing a model of the
Galaxy and determining the system of Galactic parameters from a
practical point of view, and provides some preliminary results of
calculations. 

The main task in the construction of the Galactic model is the
determination of the mass distribution function in it. In the first
rough approximation we can assume that surfaces of equal densities in
the Galaxy are similar ellipsoids of rotation, having a common axis
and a symmetry plane. In such an assumption, the radial mass
distribution is found from the circular velocity by a solution of the
integral equation: 
\be
V^2(x)=G\,\int_0^x{\mu(a)\,\dd{a} \over \sqrt{x^2-e^2a^2}},
\label{eq5.1}
\ee
where $V$ is the circular velocity in the symmetry plane of the
system; $G$ is the gravitational constant; $x$ is the distance from
the symmetry axis; $a$ is the semi-major axis of an ellipsoid of equal
density; $e =\sqrt{1 - \epsilon^2}$, with $\epsilon$ being the ratio
of the ellipsoid minor to major axes and, finally, $\mu(a)\,\dd{a}$ is the
mass contained between ellipsoids with semi-major axes $a$ and
$a+\dd{a}$. It is reasonable to express the values of $a$ and $x$ in
units of the Sun's distance from the center of the system $R_0$. In
this case, for the mass function $\mu(a)$ we have the expression: 
\be
\mu(a)=4\pi\,R_0^2\,\epsilon\,\rho_0\,a^2\,\rho^\star(a).
\label{eq5.2}
\ee
where $\rho^\star(a)$ is the volume mass density on the surface of an
ellipsoid with semi-major axis $a$ in units of the circumsolar density
$\rho_0$. 

Several variants of equation (\ref{eq5.1}) and methods of its solution
have been proposed by different authors. For example,
\citet{Wyse:1942wd} and  \citet{Schwarzschild:1954to}  consider a flat
model of the stellar system. In this case $\epsilon \rightarrow 0$ and
$\rho \rightarrow \infty$,    but $\mu$ remains finite. Instead of
$\mu$ they use the surface density $P$, making an integral equation
for it. The surface density is related to the mass function by a
simple integral relation (see \citet{Kuzmin:1956aa}). 

\citet{Kuzmin:1952aa, Kuzmin:1956aa},
\citet{Perek:1951aa,Perek:1954aa}  and \citet{Takase:1955tp} already
represented the Galaxy as an inhomogeneous ellipsoid. In this case we
have a spatial model of the system. However, the surfaces of equal 
density are not in fact similar ellipsoids. To eliminate this drawback,
\citet{Kuzmin:1956aa} proposed a generalised spheroidal model
consisting of a large number of individual spheroids. The presence of
spheroids was taken into account by introducing some mean values of
$\epsilon$ and $e^2$ as functions of $a$. 

In Kuzmin's generalised model we already have three unknown functions
--- $\rho(a)$, $\epsilon(a)$, and $e^2(a)$. It is clear that it is
impossible to solve one integral equation with three unknown
functions. If we attract additional observational material on the
density distribution, and the ratio of semi-major axes of individual
subsystems, and are interested not only in determining the mass or
surface density function but also the spatial density of the Galaxy,
it is more natural and simple to consider subsystems in explicit form,
without resorting to the average $\epsilon$ and $e^2$. This is the
path followed by most of the authors who have recently studied the
mass distribution in the Galaxy (\citet{Schmidt:1956},
\citet{Perek:1954aa},  \citet{Idlis:1961}). 

With respect to individual subsystems of the Galaxy, if they are
physically homogeneous groups of stars, we can assume with a much
better approximation than for the Galaxy as a whole that surfaces of
equal density are similar ellipsoids of rotation. The mass function
$\mu(a)$ and the ratio of ellipsoid semiaxes $\epsilon_i$ are, of
course, different for different subsystems. Summing up the
contributions of individual subsystems to $V^2$, we obtain: 
\be
V^2(x)=4\pi\,G\,R_0^2\,\sum_i\,\epsilon_i{\rho_0}_i\,\int_0^x{a^2\rho_i^\star(a)\dd{a} \over \sqrt{x^2-e_i^2a^2}}.
\label{eq5.3}
\ee
In equation (\ref{eq5.3}), both the circular velocity functions $V$
and the subsystem density distribution functions $\rho_i^\star$ are
known with some accuracy. The parameters of this formula, $R_0$,
$\epsilon_i$ and ${\rho_0}_i$, are also known
approximately. Therefore, the expression (\ref{eq5.3}) should be
considered not as an integral equation for determining the mass
distribution in the Galaxy but as an equation for mutual agreement
and specification of the functions and parameters appearing in it. 

\section{Observational data}

The following observational data are available to build a model of the
Galaxy:

(a) for the nearest neighbourhood of the Sun, there are kinematical  data
with respect to all subsystems of the Galaxy, and data on the spatial
distribution of most subsystems (excluding subsystems of absolutely
faint stars);

(b) for wide regions of the Galaxy, comparable to the size of the
whole stellar system, the spatial distribution is known only for a few
subsystems, among which, fortunately, representatives of all main
components of the Galaxy appear; in addition, the rotation of some
flat component subsystems is known;

(c) data on rotation and surface density distribution of other
galaxies, which to some extent supplement the information about our
Galaxy, especially for central and peripheral regions. \\

As we can see, the observational data are very limited, so the full 
description of the Galaxy in the six-dimensional phase space is out 
of the question. However, they are sufficient to determine the overall 
trends of the mass density distribution for the main components of the
Galaxy. It is possible to determine a number of other functions
characterising the structure of the Galaxy. 

As a function approximating the density distribution of Galactic
components, we have chosen a generalised exponential law 
\be
\rho^\star (a) = e^{-{m_0 \over \nu}\,(a^\nu -1)},
\label{eq5.4}
\ee
where $\nu$ is some positive number, characterising the degree of mass
concentration to the center of the system, and $m_0$ is the density
logarithm gradient near the Sun: 
\be
m = -{\partial\,\ln\,\rho(a) \over \partial a}= - {R \over Mod}{\partial\log \rho(R) \over \partial R }.
\label{eq5.5}
\ee

In particular cases $\nu = 2$ and $\nu =1$, we have Gaussian and
ordinary exponential distributions, which have already been repeatedly
applied to describe the  spatial density of galaxy subsystems
(\citet{Perek:1951aa},  \citet{Takase:1955tp},
\citet{Perek:1958wz}). On the other hand,
\citet{de-Vaucouleurs:1948aa, de-Vaucouleurs:1953tq}  showed that
surface brightness of elliptical galaxies and spherical components
(core and halo) of spiral galaxies can be represented using function
(\ref{eq5.4}) with $\nu =0.25$. If we assume that the mass-to-light
ratio for the spherical component of a given galaxy does not change
with the distance from the system's centre, the surface brightness is
proportional to the surface mass density. By solving the corresponding
integral equation, we found the spatial density distribution. It turned
out that the same function, (\ref{eq5.4}) is obtained with sufficient
accuracy but with $\nu =0.18$. Thus, the spherical components of
galaxies can be described by the formula (\ref{eq5.4})  with a small
value of $\nu$. 

The generalised exponential distribution (\ref{eq5.4}) is convenient,
because it is defined in an infinite interval, and therefore takes
into account the presence in the stellar system of stars with very
elongated orbits, whose velocities are close to the parabolic. On the
other hand, the density decreases quickly enough at large distances,
so that the mass of the model is finite, and the model does not have
such an extensive envelope as the Kuzmin model, derived from the third
integral theory \citep{Kuzmin:1956ca}. Furthermore, the distribution at
different $\nu$ gives a very different course of the density logarithm
gradient, decreasing ($\nu < 1$) or increasing ($\nu > 1$) with
increasing $a$. 

It is sufficient to represent the Galaxy as a composite model for
three components: planar (Flat), intermediate (Disc), and spherical
(Sph). The parameters characterising the structure of the components
are given in Table \ref{Table5.1}. The values obtained on the basis of
observational material are given in the column 0, and the values
obtained by equating the observed values are given in the column 1.

{
\begin{table*}[ht] 
  \caption{Parameters of Galaxy populations}
\label{Table5.1}                         
{\footnotesize
  \centering
\begin{tabular}{lccrllllcccl}
\hline  \hline
  Pop.   & $i$&$\nu_i$&$\overline{|z|_i}$ & $\epsilon_{i0}$&$\epsilon_{i1}$&$m_{i0}$&$m_{i1}$& $\rho_{0i0}$&$\rho_{0i1}$&$\frac{M_i}{M}$&Ref.\\
  \hline
  Flat              &1&2&     145&0.022&0.022&2.35&2.35&$30\pm 5$&25.0&0.041& 17 - 20\\
  Disc&2&1&     400&  0.09& 0.13 &4.00&3.30&$55\pm 10$&53.3&0.692&17, 21, 22 \\
  Sph.    &3&1/3&2300& 0.60& 0.60&3.10&3.91&$2\pm 2$&1.89&0.267& \\
\hline 
\end{tabular}
\\
}
{Notes: $\overline{|z|_i}$ is given in parsecs; densities $\rho_{0i0}$
  in $M_\odot$ per  kiloparsec$^3$.\\
References are: 17 --
\citet{Oort:1958aa}, 18 -- \citet{Westerhout:1957aa}, 19 --
\citet{Schmidt:1957aa}, 20 -- \citet{Kopylov:1955ab}, 21 --
\citet{Kopylov:1955aa}, 22 -- \citet{Kukarkin:1949aa}; references  
for spherical components
are: \citet{Kukarkin:1949aa}, \citet{de-Vaucouleurs:1953tq},
\citet{Perek:1954aa}, \citet{Schmidt:1956}, \citet{Schmidt:1957aa}, 
\citet{Notni:1956aa}, \citet{Oort:1958aa},
\citet{Baade:1958aa},
\citet{Oort:1960aa}, \citet{Johnson:1961wb},
\citet{Wallerstein:1962uf}, \citet{Lozinskaya:1963ta}.
}
\end{table*}
}

To calculate the flattenings of the components, we determined the
flattenings of the various subsystems of the Galaxy, using the formula  
\be
\epsilon = {\sqrt{m}\zeta \over  s R_0},
\label{eq5.6}
\ee
where $\zeta$ is a parameter characterising the distribution of stars
in the direction perpendicular to the Galactic plane, and $s$ is a 
dimensionless coefficient of the order of unity. The formula
(\ref{eq5.6}) is derived under the assumption that within subsystems the surfaces of
equal density are similar ellipsoids. The parameter
$\zeta$ can take, for example, one of the following values: 
\be
\ba{ll}
\zeta_l={R_0 \over l}=\left(-{\partial^2\,\ln \rho(z) \over \partial z^2}\right)_{z=0}^{-1/2}, & \zeta_0=z_0={1 \over \rho(0)}\int_0^\infty{\rho(z)\dd{z}}, \\
\\
\zeta_1=\overline{|z|}={\int_0^\infty{z\rho(z)\dd{z}} \over \int_0^\infty{\rho(z)\dd{z}}},&
  \zeta_2^2={\int_0^\infty{z^2\rho(z)\dd{z}} \over \int_0^\infty{\rho(z)\dd{z}}}.
\ea
\label{eq5.7}
\ee
Here $l$ is the parameter introduced by \citet{Kutuzov:1965bb}, $z_0$
is the equivalent half-thickness of the Galaxy \citep{Kuzmin:1952aa}. The numerical
value of the coefficient $\zeta$ depends on the particular kind of
density distribution. If we accept the law (\ref{eq5.4}) for the
density, then for $\nu = 2$ and $\nu = 1$ we have the values given in
Table \ref{Table5.2}. It should be said that $\zeta_0=1$, regardless
of the particular kind of density distribution.

{\begin{table*}[h] 
\caption{}
\centering 
\hspace{2mm}
\resizebox{0.80\textwidth}{!}{\includegraphics*{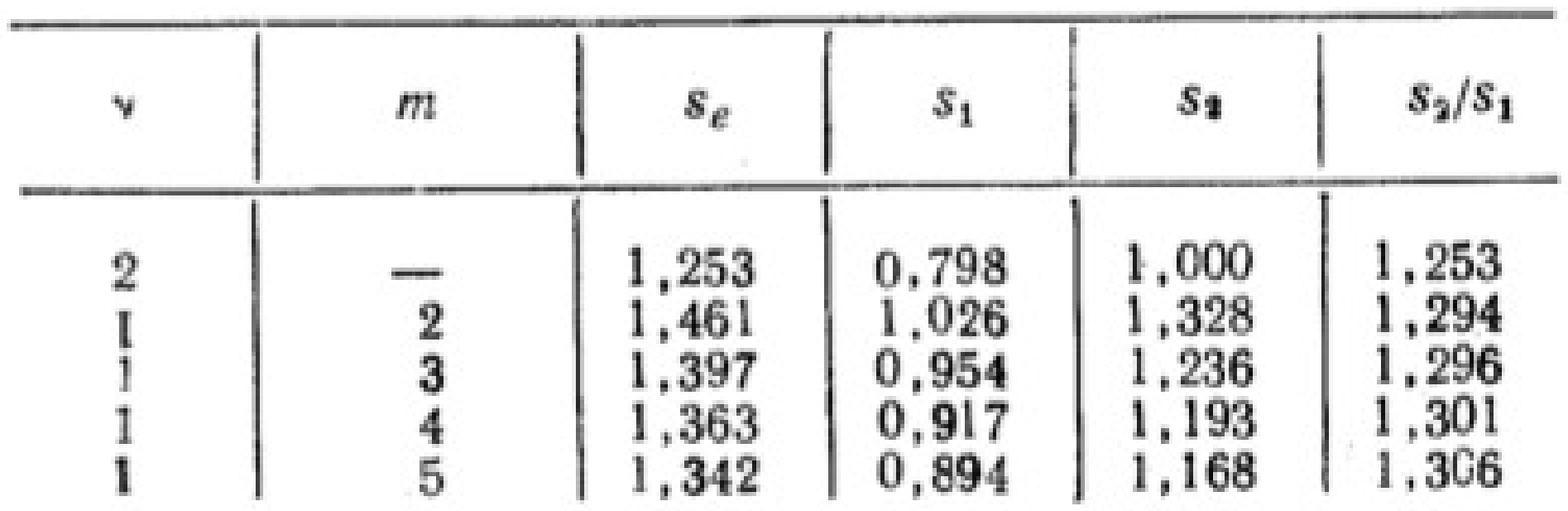}}
\label{Table5.2}
\end{table*} 
}

Comparing our derived values $\epsilon$ with the results of
\citet{Schmidt:1956} and \citet{Idlis:1961aa}, we can say the
following. The ratio of semiaxes for the planar component agrees well
with the results of other authors. Only for the central regions of the
Galaxy Idlis took $\epsilon = 0.25$ in order to have a smaller spatial
density for a given surface density. It is difficult to agree with
this, however, as direct estimates (\citet{Westerhout:1957aa},
\citet{Lozinskaya:1963ta}) indicate that the thickness of the
interstellar hydrogen layer, the main subsystem of the flat component
of the Galaxy, does not increase as we approach the centre of the
system, but, on the contrary, decreases. The data on the ratio of
half-axes for the intermediate component agree with Schmidt's data
(Idlis does not consider this component in his model). The data for
the spherical component differ strongly. Schmidt took $\epsilon =
0.16$ in this case, which is clearly insufficient. Idlis took for
the peripheral regions of the Galaxy an average of $\epsilon = 0.5$,
which is quite acceptable. However, in the centre of the system he
took an underestimated value of $\epsilon = 0.25$. Photographs of
spiral galaxies, visible from the edge, show that the nuclei of these
systems have an $\epsilon$ of the order of $0.5-0.7$
(\citet{Johnson:1961aa}, \citet{de-Vaucouleurs:1959vq}). The apparent 
decrease of $\epsilon$ for the inner regions in the subsystems of
globular clusters and short-period cepheids, noted by Idlis, is caused
by the fact that these subsystems are not homogeneous but consist of
a mixture of objects of intermediate and spherical components
(\citet{Notni:1956aa}, \citet{Baade:1958aa}). 

The $\nu$ parameter was chosen so that the law (\ref{eq5.4})
satisfactorily represented the available data on the density
distribution of the planar (\citet{Westerhout:1957aa},
\citet{Schmidt:1957aa}), intermediate  (\citet{Kukarkin:1949aa},
\citet{de-Vaucouleurs:1959vq}), and spherical
(\citet{Kukarkin:1949aa}, \citet{de-Vaucouleurs:1948aa,
  de-Vaucouleurs:1953tq}, \citet{de-Vaucouleurs:1959vq},
\citet{Oort:1960aa}) subsystems. 

The gradient $m$ for the intermediate component was taken from
\citet{Idlis:1961aa} summary,  while for the planar and spherical ones
it was calculated anew. It turned out that the earlier determinations
of this gradient for the spherical subsystems were exaggerated. 

In addition to data on the structure of the individual components, in
order to derive a system of Galactic parameters and build a model of
the Galaxy, we also need knowledge of the course of the circular
velocity. The observational material needed for this purpose is
available in the form of radio astronomical determinations of the
differential rotation of interstellar hydrogen. We used the data of
Dutch scientists in the treatment by \citet{Kwee:1954wu} and
\citet{Agekyan:1962}. Based on this material, we compiled for eight
values of $x$ the normal points of the Galactic differential rotation
function $U$. This function is related to the circular velocity $V$ by
the formula 
\be
V(x)=U(x)+x\,V_0,
\label{eq5.8}
\ee
where $V_0$ is the circular velocity in the vicinity of the Sun. The
results, together with estimates of their average errors, are given in
Table \ref{Table5.3}. The values obtained on the basis of observational
material are given in the column 0, and the values obtained by
smoothing  the observed values are given in the column 1.

{\begin{table*}[h] 
\caption{Differential rotation function $U(x)$ of the Galaxy}
\centering 
\hspace{2mm}
\resizebox{0.80\textwidth}{!}{\includegraphics*{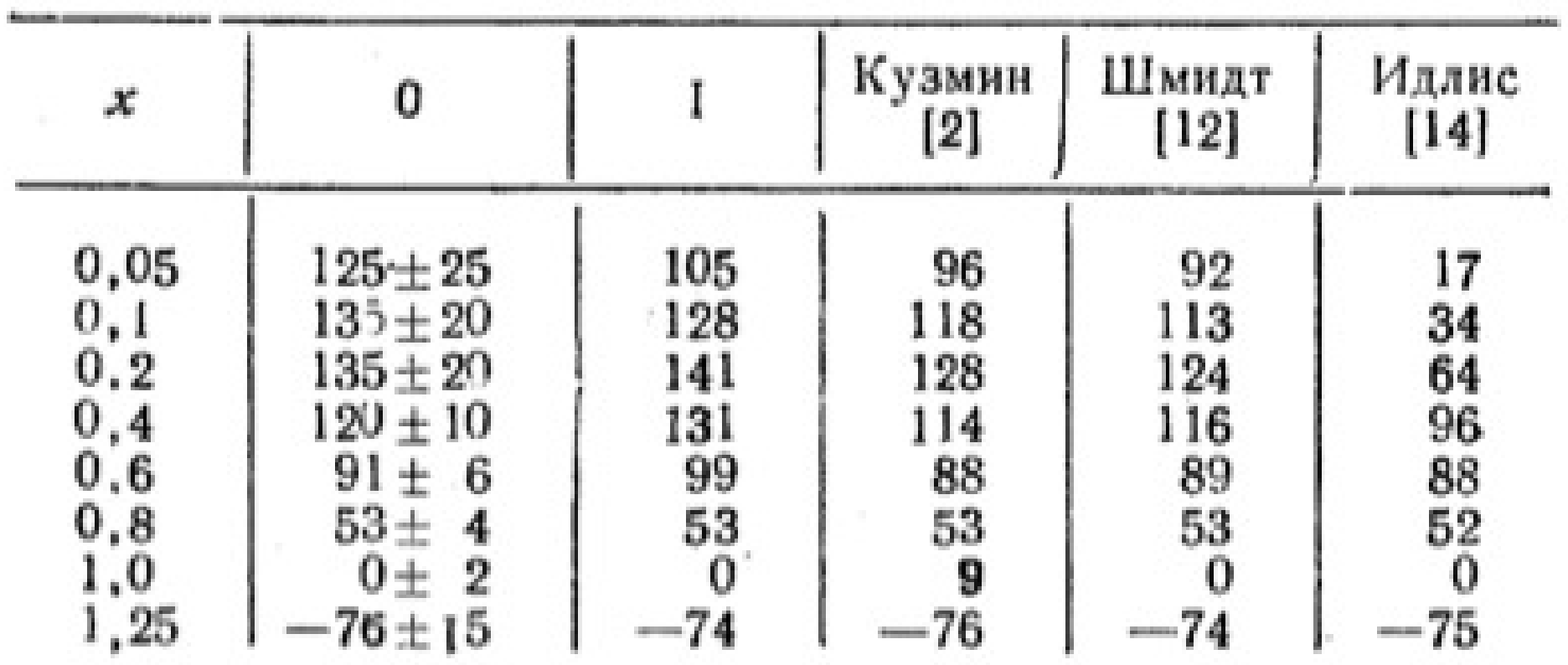}}
\label{Table5.3}
\end{table*} 
}

Finally, we used a number of other parameters, the circumsolar values
of which, together with their errors, are given in Table
\ref{Table5.4}.

{\begin{table*}[h] 
\caption{Near-solar values of galactic parameters}
{\centering 
\hspace{2mm}
\resizebox{0.95\textwidth}{!}{\includegraphics*{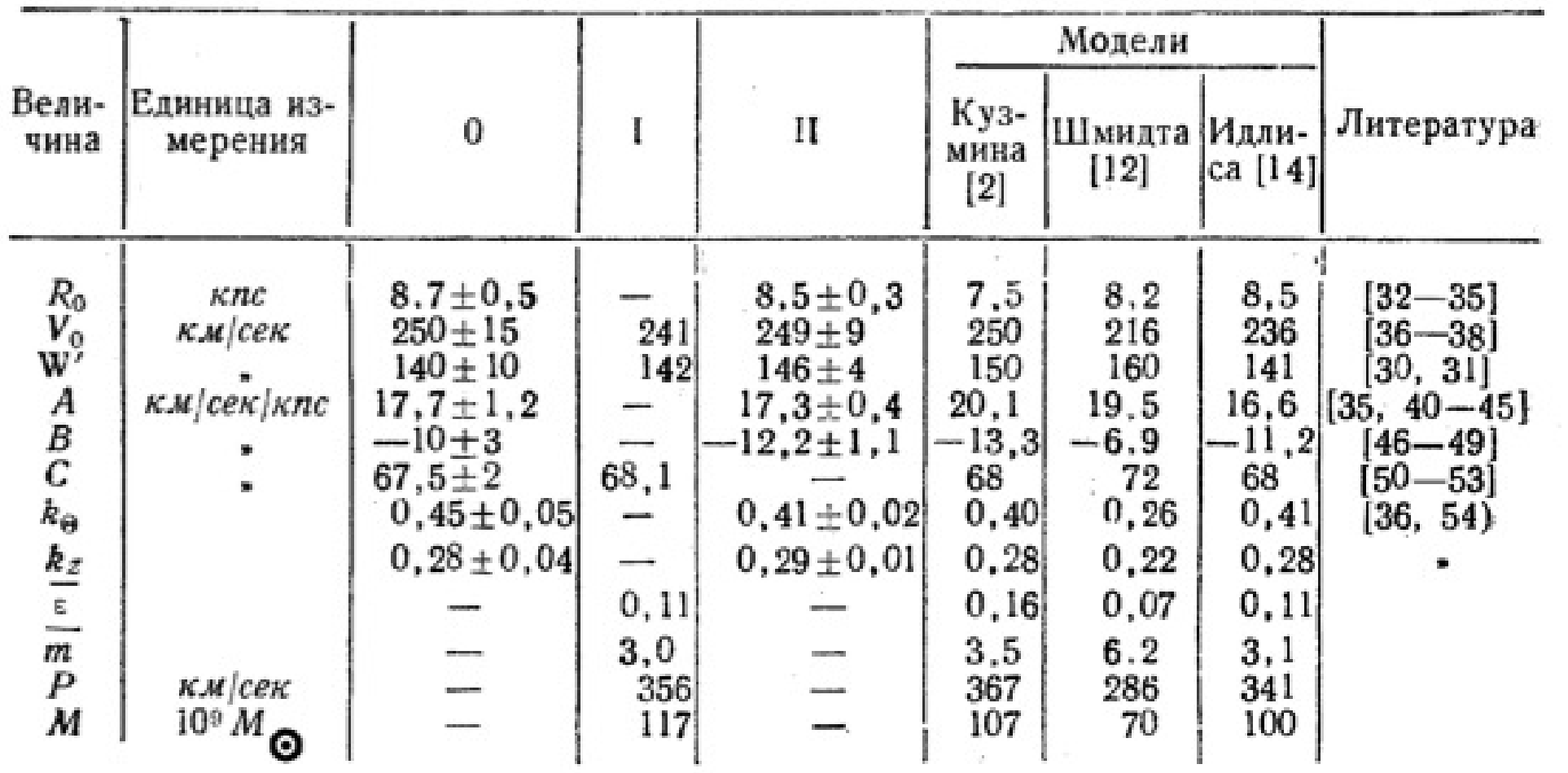}}
\label{Table5.4}\\
}
{ References: (30) \citet{Kwee:1954wu}, (31) \citet{Agekyan:1962},
  (32) \citet{Baade:1951cc}, (33) \citet{Whitford:1961wp}, (34)
  \citet{Weaver:1954uf}, (35) \citet{Feast:1958ub}, (36)
  \citet{Parenago:1951aa}, (37) \citet{Fricke:1949tl}, (38)
  \citet{Fricke:1949vp}, (40) \citet{Stibbs:1956un}, (41)
  \citet{Petrie:1956va}, (42) \citet{Gascoigne:1957wx}, (43)
  \citet{Walraven:1958tx}, (44) \citet{Janak:1958tz}, (45)
  \citet{Pskovskii:1959ua}, (46) \citet{van-de-Kamp:1937ux}, (47)
  \citet{Raymond:1938us}, (48) \citet{Vyssotsky:1948wm}, (49)
  \citet{Morgan:1951vk}, (50) \citet{Oort:1932tb}, (51)
  \citet{Kuzmin:1952ab}, (52) \citet{Kuzmin:1955aa}, (53)
  \citet{Eelsalu:1961aa}, (54) \citet{Dyer:1956aa}.  }
\end{table*} 
}

In deriving the distance to the centre of the Galaxy, only independent
estimates were taken into account, the dynamical determinations were not
used.  The latter depend on other Galactic parameters; in
the subsequent least-squares processing, however, it was assumed that
all the estimates of the parameters to be equated must be independent.  

The circular velocity is determined mainly dynamically by the
asymmetric shift of the velocity centroids of the stars
\citep{Parenago:1951aa}. These data as well as the corresponding
formula do not appear anywhere else in the construction of the system
of Galactic parameters, so that the velocity definition can be
considered independent. In addition, general dynamical considerations
\citep{Fricke:1949tl, Fricke:1949vp} were taken into account,
according to which $V_0$ cannot be much less than 275 km/s. 

The kinematical parameter,
$W= -1/2\,(\dd{U}/\dd{x})_{x=1}$,
characterises the circumsolar value of the Galactic differential
rotation function, based on the behaviour of the $U$ function (see Table
\ref{Table5.3}).  This value is somewhat smaller than that usually
accepted (\citet{Schmidt:1956}, \citet{Lozinskaya:1963ta}) and
nearly coincides with the value obtained by \citet{Idlis:1961aa}
(see Table \ref{Table5.4}). 

We consider the Oort-Kuzmin parameters $A, B$, and $C$ in the
dynamical sense, \ie corresponding to the gravitational acceleration
along $R$ and $z$.

When calculating the Oort parameter $A$, only values giving $A>15$
km/sec/kpc were taken into account. The significantly lower values obtained
by some authors are distorted, apparently, by local features of some
subsystems of stars, or by drawbacks in the methodology of material
processing. In calculating the average error of $A$, we took into account
the fact that many authors used practically the same observational material. 

The parameter $B$ is known to be different in the GC and FK3
system. Most astronomers prefer the FK3 system, in which $B =-7$ is
obtained. This value leads, however, to various dynamical  difficulties
\citep{Kuzmin:1956aa}. Therefore, we took the average of the $B$
values in the GC and FK3 systems, increasing the mean error, which, in
addition to the random error, also takes into account the unknown
systematic error. 

The parameter $C$ is taken from the determinations by 
\citet{Kuzmin:1952ab} and \citet{Eelsalu:1958aa}, for definition see
Chapter 6. The markedly larger 
values, obtained by some authors, are distorted, as pointed out by
\citet{Eelsalu:1958aa,Eelsalu:1961aa}, by the imperfection of the
applied methodology. 

The ratios of velocity dispersions of stars:
\be
k_\theta={\sigma_\theta^2 \over \sigma_R^2}, \hspace{1cm} k_z={\sigma_z^2 \over \sigma_R^2},
\label{eq5.9}
\ee
were taken from \citet{Parenago:1951aa} and \citet{Dyer:1956aa}. In
these works, the components of the spatial velocities of stars are
used, and the material is divided by physical features into separate
subsystems. The definitions of the dispersion relations, obtained from
the proper motions \citep{Hins:1948aa}, were not taken into
account. Apparently, they are distorted by systematic errors, which
was pointed out by \citet{Trumpler:1953tn}.

\section{Construction of the Galactic model and determination of Galactic
  parameters}

When building a model of the Galaxy, it is assumed that there are a
number of theoretical relations linking the parameters determined from
observations. Equations (\ref{eq5.1}) or (\ref{eq5.3}) and the Poisson's
formula are usually used as such relations. We use the Poisson
equation in the form, suggested by \citet{Kuzmin:1952ab} 
\be
4\pi\,G\rho_t=C^2 - 2\,(A^2 - B^2),
\label{eq5.10}
\ee
where $\rho_t$ is the total spatial density of mass. Also we use the Lindblad equation:
\be
k_\theta ={-B \over A-B},
\label{eq5.11}
\ee
and  the following expressions  derived from the definition of $A$, $B$, and $W$:
\be
A\,R_0 - W =0,
\label{eq5.12}
\ee
\be
R_0(A-B) - V_0 = 0.
\label{eq5.13}\ee
To these we can add another differential consequence of Eqn. (\ref{eq5.3})
\be
{\dd{V^2}(x) \over \dd{x}} = 4\pi\,GR_0^2\,\sum_i\epsilon_i{\rho_0}_i {\dd{\int_0^x{{a^2\rho_i^2(a)\dd{a} \over \sqrt{x^2-e_i^2a^2}}}} \over \dd{x}}
\label{eq5.14}
\ee
and \citet{Kuzmin:1961aa} equation:
\be
k_z^{-1} = 1 + k_\theta^{-1}.
\label{eq5.15}
\ee

Due to random and unaccounted for systematic errors in the definition
of parameters or description functions as well as the inaccuracy of
the applied theory, these equations are not fulfilled quite
accurately. In order for the equations to be satisfied, one has to
change the parameters and description functions somewhat. So far this
has been done by trial-and-error procedure,  and the result has depended heavily on the
taste of the author. Moreover, a certain system of rounded values of
the main parameters was often taken (for example, by
\citet{Schmidt:1956} for $A$, $B$, $C$, $W$), while all other
parameters were not taken from observations but were calculated by
formulas (\ref{eq5.10}) - (\ref{eq5.13}). 

An objective way to derive the best system of Galactic parameters and
to construct the corresponding model of the Galaxy is the application
of the least-squares method. In this case equations (\ref{eq5.3}) and
(\ref{eq5.10}) - (\ref{eq5.15}) are considered not as expressions for
determination of this or that quantity but as fundamental equations
to the equalisation of the system of Galactic parameters \citep{Kutuzov:1965bb}.   

We performed the parameter equalisation twice, with and without the use of
the model of mass distribution in the Galaxy. Such a way of solving
the problem was chosen in order to find out the suitability of the
model (\ref{eq5.4}) to describe the structure of the Galactic
components. 

In the first case, the following quantities were equated by the
least-squares method: the circumsolar densities of the intermediate
and spherical components $\rho_{02}$ and $\rho_{03}$; the circumsolar
value of the circular velocity $V_0$, kinematical parameter $W$; the
Kuzmin parameter $C$; values of the function $U$ for eight points (see
Table \ref{Table5.3}). Equations (\ref{eq5.3}), (\ref{eq5.10}), and (\ref{eq5.14})
were used as fundamental equations, and expression (\ref{eq5.8}) was
substituted for $V(x)$ in (\ref{eq5.3}) and (\ref{eq5.14}), and
expression (\ref{eq5.4}) was substituted for
$\rho_i^\ast(a)$. Calculation results are given in Tables
\ref{Table5.1}, \ref{Table5.3}, and \ref{Table5.4} (option 1). 

In the second case, the circumsolar values of the following quantities
were fixed: the Oort parameters $A$, $B$, the circular velocity
$V_0$, parameter $W$, the distance from the centre of the Galaxy
$R_0$, and the ratios of velocity dispersions, $k_\theta$ and
$k_z$. As fundamental equations we used formulas (\ref{eq5.11}) -
(\ref{eq5.13}) and (\ref{eq5.15}). The results are given in Table
\ref{Table5.4}  (option II).

{\begin{figure*}[h] 
\centering 
\hspace{2mm}
\resizebox{0.40\textwidth}{!}{\includegraphics*{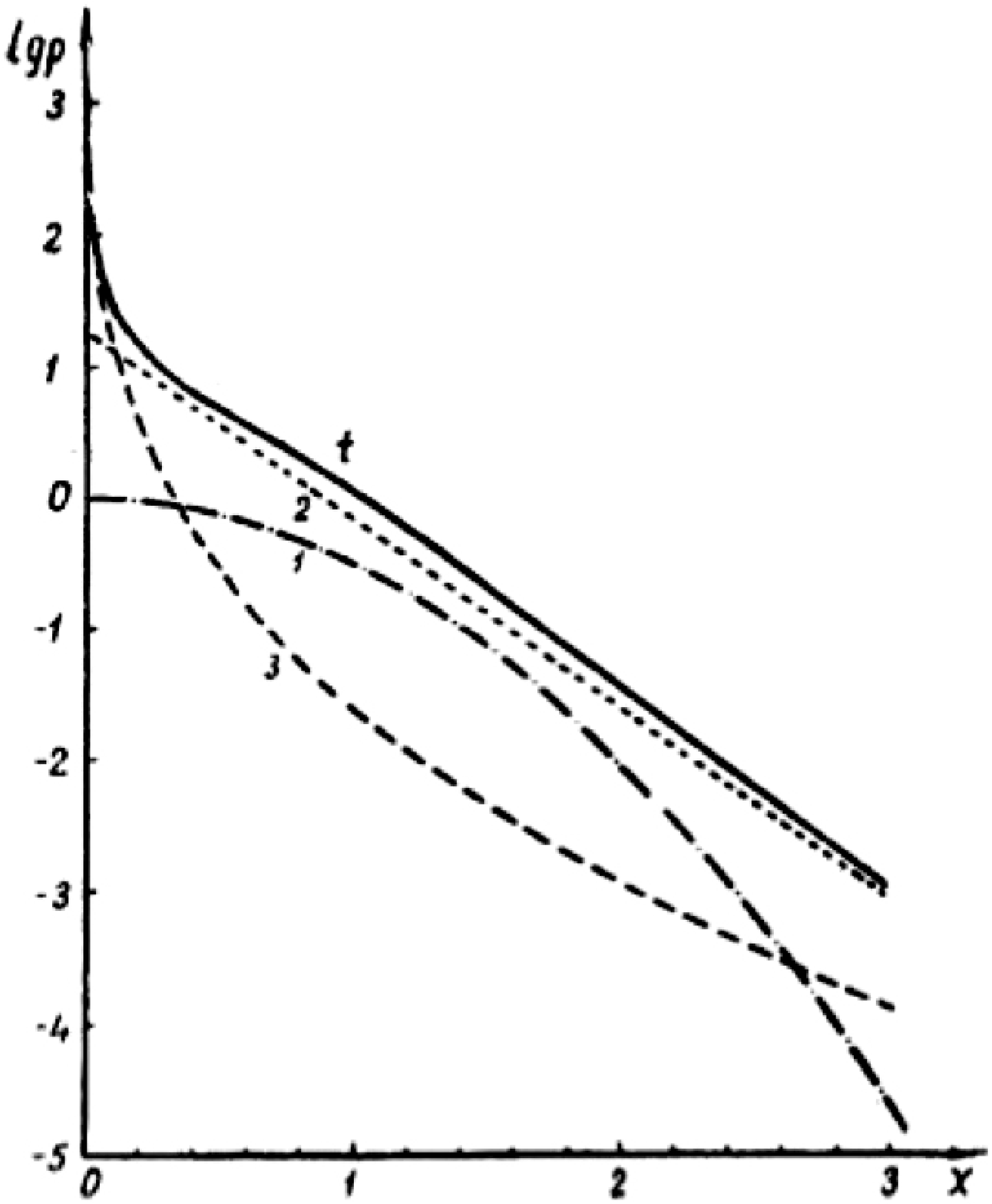}}
\hspace{2mm}
\resizebox{0.40\textwidth}{!}{\includegraphics*{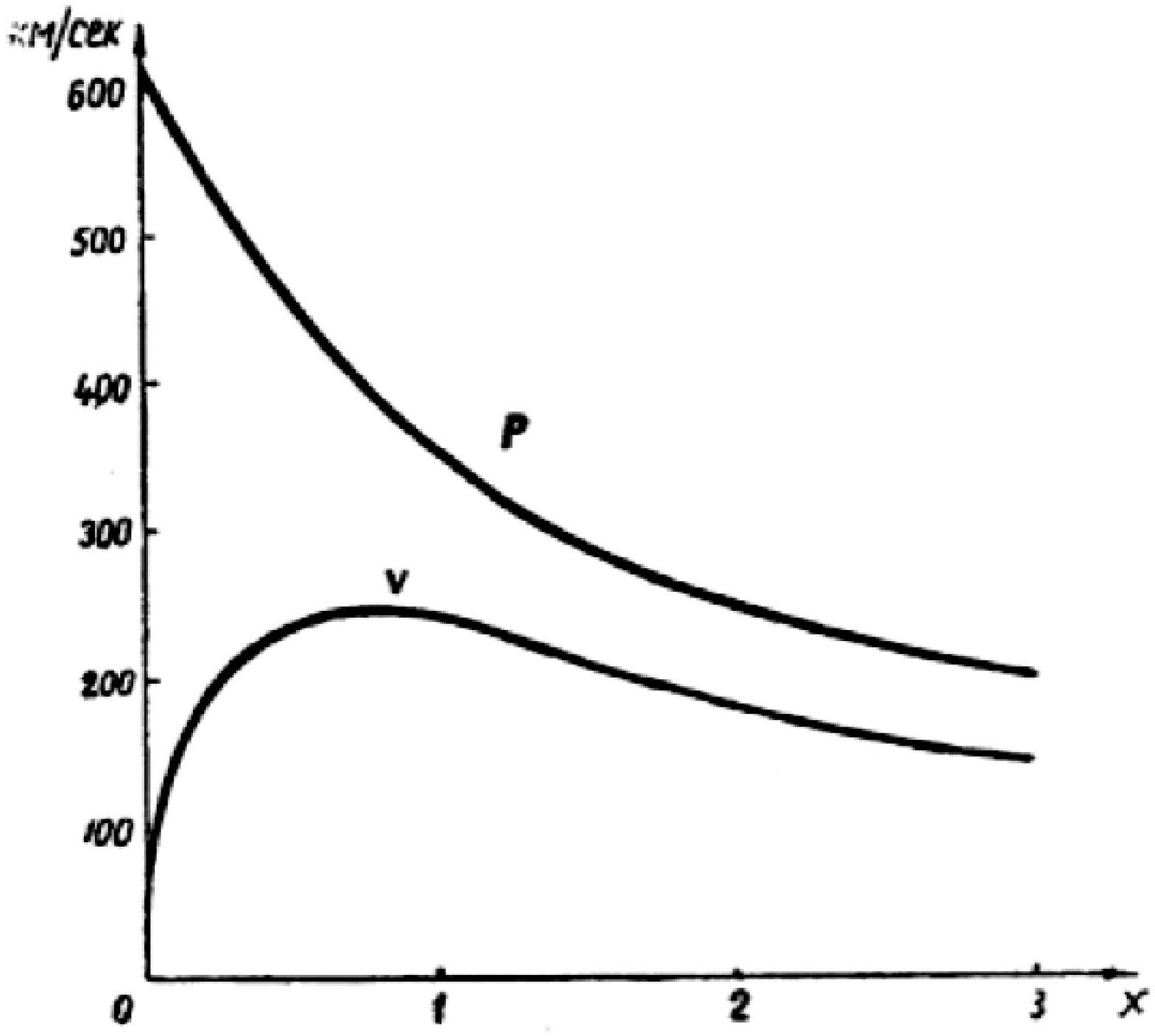}}\\
\caption{{\em Left:} Distribution of the logarithm of the density in units of
  the circumsolar total density, $\rho$ is the total density; 1, 2,
  and 3 are the densities of the planar, intermediate, and spherical
  components. {\em Right:} Dependence of the circular and parabolic velocity
  on the distance to the centre of the Galaxy. 
} 
  \label{Fig5.1}
\end{figure*} 
}

{\begin{figure*}[h] 
\centering 
\hspace{2mm}
\resizebox{0.40\textwidth}{!}{\includegraphics*{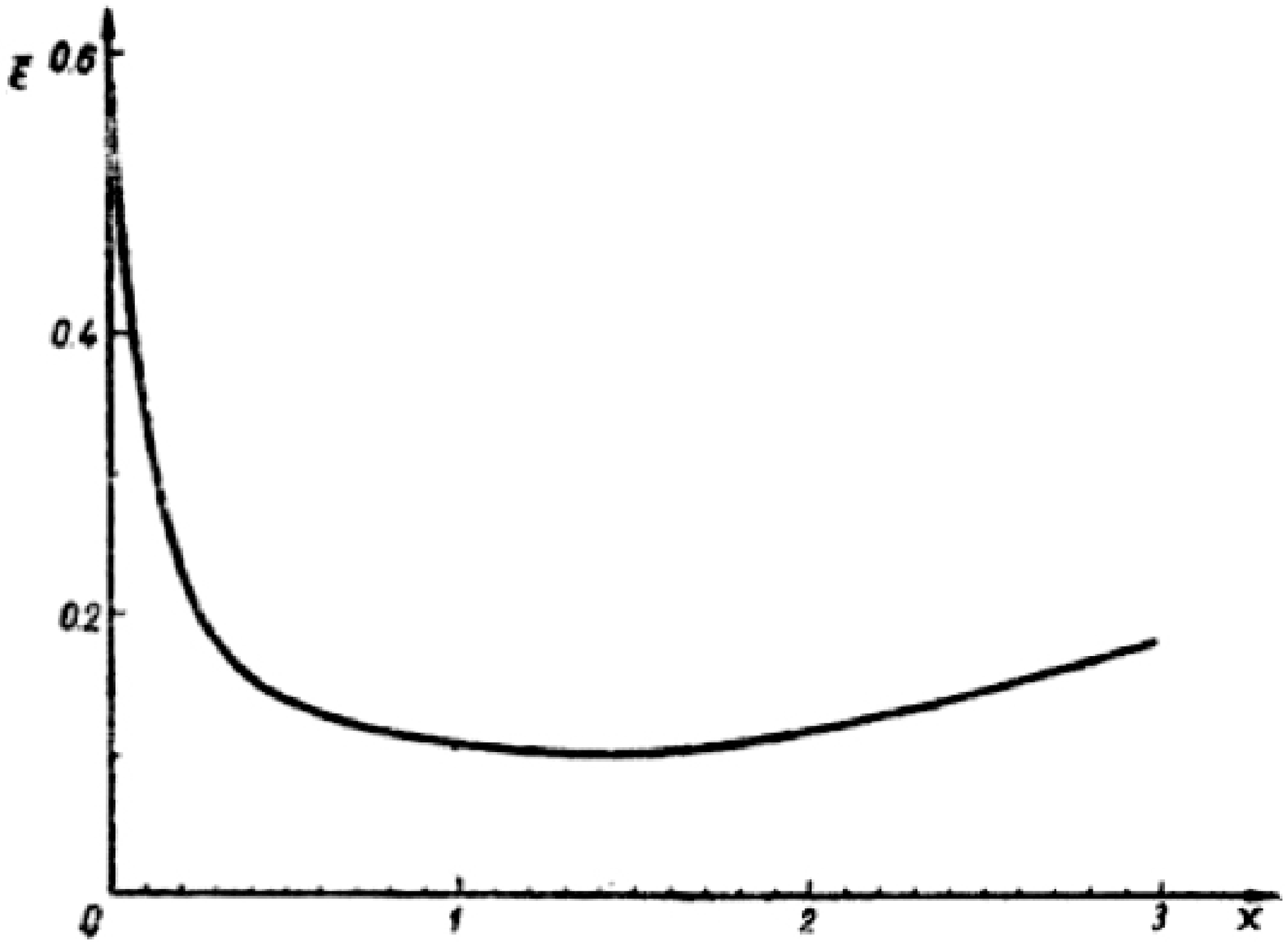}}
\hspace{2mm}
\resizebox{0.40\textwidth}{!}{\includegraphics*{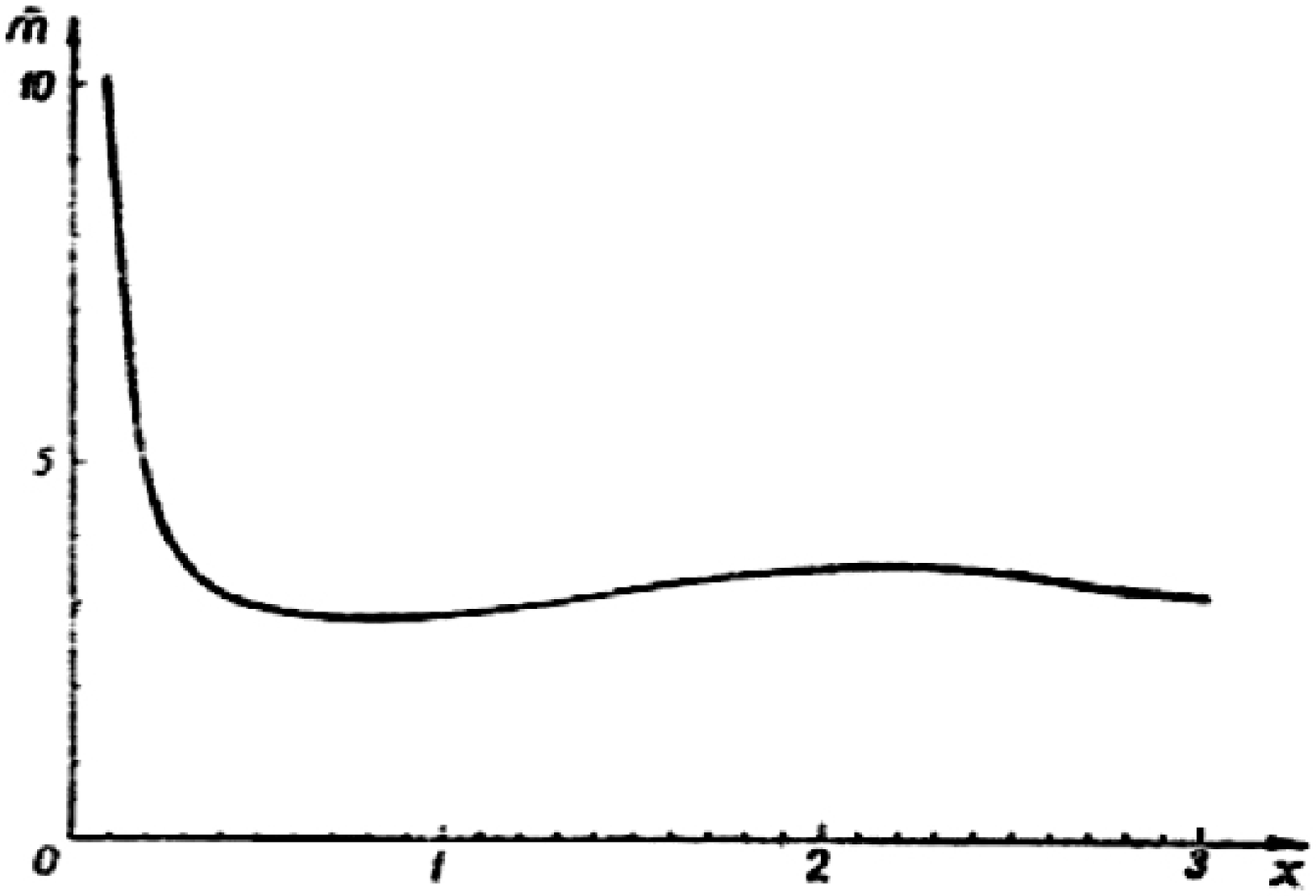}}\\
\caption{{\em Left:} The dependence of the average ratio of the semi-axes
  on the distance to the centre of the Galaxy. {\em Right:} Dependence of
  the average gradient of the logarithm of the density $m$ on the
  distance to the centre of the Galaxy.  }  
  \label{Fig5.2}
\end{figure*} 
}

In addition, we calculated the mean values of the ratio of the
semi-axes $\overline{\epsilon}$ and the gradient $\overline{m}$ (with
the weight $\rho_t$), the total mass of the Galaxy, 
\be
M=4\pi\,R_0^3\sum_i\epsilon_i{\rho_0}_i\,M_{2j}^\star,
\label{eq5.16}
\ee
and the parabolic velocity $P=\sqrt{2\Phi}$ (where $\Phi$ is the gravitational potential):
\be
P^2(x)= 2\int_x^\infty{V^2(x)\dd{x} \over x} = 8\pi\,G~R_0^2 \sum_i\epsilon_i{\rho_0}_i\,\Phi_i^\star(x).
\label{eq5.17}
\ee
In formula (\ref{eq5.17}) $\Phi_i^\star(x)$ is the mass of the
subsystem in units $4\pi\,R_0^3\epsilon{\rho_0}_i$. The coefficient
$M_{2j}^\star$ is a special case, $j=2$, of the moments of $M_{ji}^\star$ of
the function $\rho_j^\star (a)$. If we take expression (\ref{eq5.4}) for
$\rho_j^\star (a)$, in the general case we have: 
\be
M_{ji}^\star = \int_0^\infty{a^j\rho_i^\star(a)\dd{a}} = {1 \over \nu}\exp\left({m_0 \over \nu}\right)\,\left({m_0 \over \nu}\right)^{-{1+j \over \nu}}\,\Gamma\left({1+j \over \nu}\right),
\label{eq5.18}
\ee
where $\Gamma$ is gamma-function.

The functions $\Phi_i^\star(x)$ in formula (\ref{eq5.17}) are
gravitational potentials of subsystems in plane $z=0$ in units
$4\pi\,GR_0^2\epsilon_i\rho_{0i}$. They are calculated by the formula 
\be
\Phi_i^\star(x) = \int_0^\infty{\rho_i^\star(a) \chi_i\left({a \over x}\right)\,a\dd{a}},
\label{eq5.19}
\ee
where
\be
\chi_i\left(\frac{a}{x}\right)=
\left\{
  \ba{ll}
  \frac{1}{e_i}\arcsin(\frac{e_ja}{a}), & a\le x,\\
  \frac{1}{e_i}\arcsin(e_i), & a\ge x.
  \ea
\right.
\label{eq5.20}
\ee
For $x=0$ (Galactic center) we have as a special case
\be
\Phi_i^\star(0) = {\arcsin(e_i) \over e_i}\,M_{1i}^\star,
\label{eq5.21}
\ee
and for large $x$ we obtain the following asymptotic decomposition:
\be
\Phi^\star(x) \approx \frac{M_2^\star}{a}\left[1 + \frac{1}{2}\frac{1}{3}\left(\frac{e}{x}\right)^2\frac{M_4^\star}{M_2^\star} + \frac{1}{2}\frac{3}{4}\frac{1}{5}\left(\frac{e}{x}\right)^4\frac{M_6^\star}{M_2^\star} + \dots \right].
  \label{eq5.22}
  \ee

  \section{Conclusions}
  
The equalisation in the first variant with the original initial data
(column 0 in Tables \ref{Table5.1}, \ref{Table5.3}, and
\ref{Table5.4}) did not yield satisfactory results. The obtained
system of Galactic parameters differed markedly from that in the
second variant, the calculated circular velocity curve had an
unacceptable discrepancy with the observed curve. However, it turned
out that the obtained circular velocity curve was very sensitive to
changes in the parameters of formulas (\ref{eq5.3}) and (\ref{eq5.4})
for the intermediate and spherical components. The planar component
makes such a small contribution to the expression for $V^2$ (less than
5\%) that the $V^2$ curve almost does not depend on its
parameters. After slightly changing $\epsilon$ and $m$ (Table
\ref{Table5.1}, variant I) of the intermediate and spherical
component, we carried out the equation anew; the results were
satisfactory this time. Their agreement both with the original data
and with the results of equalisation without the model (variant II) is
good. Errors of parameters after adjustment in variant II turned out
to be notably smaller. 
  
The logarithms of densities of components and total densities, the mean values
of the ratio of half axes and the gradient of the density logarithm,
and the circular velocity are given in Figures
\ref{Fig5.1} and \ref{Fig5.2}. In the calculations, we took the
Galactic parameters according to variant I. The values of density
$\rho_3$ and $\rho_t$ at $x=0$ correspond to average density in
ellipsoid with semi-major axis $a = 0.0016$, gradient $m$ is
calculated for distance $x = 0.001$ (at $\nu<1$ $m \rightarrow \infty$
if $x \rightarrow 0$). 
  
  Let us now compare the obtained preliminary systems of Galactic
  parameters with those derived from the models by
  \citet{Kuzmin:1955aa}, \citet{Schmidt:1956} and \citet{Idlis:1961aa}
  (see Table \ref{Table5.4}). 
  
  Kuzmin's system differs from the others in the "classical" values
  of the parameters $A_0$, $B_0$ and $R_0$. The mean $\epsilon$ of the
  model is somewhat exaggerated, since the values of $A_0$ and $B_0$
  were taken too large, and, besides, a very rough model of the Galaxy
  was taken when calculating $\epsilon$ (formula (56) in
  \citet{Kuzmin:1952ab}). 
  
  Schmidt's system is markedly different from ours. It assumes $B$ in
  the FK3 system, which leads to underestimated values of the circular
  and parabolic velocity as well as the ratio $k_\theta$ but, on the
  other hand, to an unacceptably high value of the gradient
  $m_0$. Recently, \citet{Schmidt:1961uj, Schmidt:1962aa}  has become
  convinced of the necessity to reduce the density gradient. 
  
  Schmidt's model has too small dimensions and sharp boundaries due to
  large density gradient. This leads to an excessively fast decrease
  of the circular velocity with increasing distance, and a small value
  of the escape velocity. 
  
Our system of parameters agrees well with Idlis's system. But Idlis's
model has a significant drawback — by adopting Parenago's law for the
circular velocity, Idlis was not able to take into account the
presence in the Galaxy of a very dense and quite massive nucleus and
an extensive halo.

\vskip 0.5cm
\hfill October 1963 
\chapter{System of Galactic parameters\label{ch06}}

In this Chapter, we present a method of the determination of the system
of Galactic parameters, using equations, connecting parameters as
fundamental equations to the equalisation of the system of Galactic
parameter, using observational estimates of all parameters. The idea of
the method was described by \citet{Kutuzov:1964aa}.  The new system
was presented by \citet{Einasto:1964aa, Einasto:1964wx} to the
Commission 33 Meeting, IAU XII General Assembly, 
Hamburg, 1964.

In the following years, I continued to collect observational data on all
Galactic parameters which enter in the system of parameters. Results
of this search were collected in Chapter 6 of the original Thesis on
56 pages of typed text with 14 tables and a reference list with 159
entries. However, these data were not used to construct a new model,
because in 1971 I was busy calculating the physical evolution of
galaxies, and there was no time to find a new model of the
Galaxy. Soon it was evident that the model should include a massive,
extensive and almost spherical component — corona. This changes the
gravitational potential field of the Galaxy, which influences Galactic
parameters, thus the collected data on parameters became obsolete. For
this reason, this collection is not reproduced in the current English version
of the Thesis. I present here only the main idea of the method to  determine
 the system of Galactic parameters, and  the 1964 version of
parameters. The
preliminary 1963 version of the Galactic model with parameters was
published by \citet{Einasto:1965aa} and is presented in Chapter 5, the
1970 version of Galactic parameters was described by
\citet{Einasto:1970ad} and is presented in Chapter 7.  A preliminary
model of the Galaxy with massive corona and respective system of
parameters was presented by \citet{Einasto:1979ux} and is described
in the Epilogue. Our final model of the Galaxy with an improved system of
Galactic parameters was published by \citet{Haud:1989fo}.

\vskip 0.5cm

As Galactic parameters, the following parameters or constants are
usually considered:

\hspace{1cm}$R_0$ —  the distance of the Sun from the Galactic
centre,

\hspace{1cm}$A$, $B$ — Oort’s rotational parameters, referring to the
circular motion,

\hspace{1cm}$\rho_0$  — the Galactic mass density in the solar neighbourhood.

Given the first three of these parameters, the circular velocity
\be
V = R_0(A-B)
\label{eq6.1}
\ee
can be found.

To obtain the numerical values of these parameters, their direct
estimates can be used. On the other hand, the estimates of the
following quantities can be applied for this purpose: the Kuzmin parameter
\be
C= \frac{\sigma_z}{\zeta},
\label{eq6.2}
\ee
the gradient of the function of  differential rotation velocity 
\be
W = - \frac{1}{2}\,\left(\frac{\dd{U}}{\dd\xi}\right)_{\xi=1},
\label{eq6.3}
\ee
and the ratio of velocity dispersions
\be
k_\theta = {\sigma_\theta^2 \over \sigma_z^2}.
\label{eq6.4}
\ee
In these formulae $\sigma_R^2$, $\sigma_\theta^2$ and $\sigma_z^2$,
are the velocity dispersions in Galactic cylindrical coordinate
system,  $\zeta$ is the $z$-coordinate dispersion, $U$ is the function of
differential Galactic rotation, obtained from radio observations \citep{Kuzmin:1956ca},
 and $\xi = R/R_0$. Three parameters, $C,~ W$ and $k_\theta$, are connected
 with the former ones by equations (see Chapters 5 and 7):
 \be
 4\pi\,G\,\rho_0 = C^2 -2\,(A^2 - B^2),
 \label{eq6.5}
 \ee
 \be
 W = A\,R_0,
 \label{eq6.6}
 \ee
 \be
 k_\theta = \frac{-B}{A-B}.
 \label{eq6.7}
 \ee

In the old system of Galactic parameters \citep{Schmidt:1956}, the
parameters $A$ and $B$ were taken from proper motion studies ($A =
19.5$  and
$B = - 6.9$ km/sec/kpc), and $R_0$ was calculated by means of formula (\ref{eq6.6})
from $W = 160$ km/sec, which yielded $R_0 = 8.2$ kpc, in good agreement
with directly obtained value. For  the circular
velocity the value $V = 216$ km/sec was obtained.

In the new system of Galactic parameters, accepted at the Australia
symposium on Galactic structure \citep{Oort:1965aa}, the following rounded
values were proposed:

\hspace{1cm}$R_0=10$ kpc,

\hspace{1cm}$V = 250$ km/sec,

\hspace{1cm}$A=15$ km/sec/kpc,

\hspace{1cm}$B= -10$ km/sec/kpc,

\hspace{1cm}$\rho_0= 0.145~M_\odot$/pc$^3$.

This yields:

\hspace{1cm}$C= 90$ km/sec/kpc,

\hspace{1cm}$W=150$ km/sec,

\hspace{1cm}$k_\theta = 0.40$.

It should he emphasised, however, that it is possible to widen the
amount of observational data used, introducing new equations and new
parameters, for  which observed estimates are present. Furthermore,
by applying the least-squares method, it is possible to improve the
procedure of obtaining the system of parameters,

Besides the parameters characterising the Galactic structure in
general, the population parameters may be used. Practically it is
possible to use the heliocentric centroid velocities $V_i$, the
dispersions $\sigma_{Ri}^2$, and the radial logarithmic density
gradients $m_i= -R_0\,\dd{\ln\,\rho_i}/\dd{R}$.
 From these parameters, the circular velocity $V$ can be
 calculated by means of theoretical Str\"omberg’s asymmetry equation:
 \be
 V^2 = V_i^2 +(m_i-m_0)\,\sigma_{Ri}^2,
 \label{eq6.8}
 \ee
where $V_i$ is the galactocentric rotation velocity of the population,
and \citep{Kuzmin:1962ac}
\be
m_0 = 1-k_\theta + \frac{1}{4}\,m\,(1-k_z),
\label{eq6.9}
  \ee
 $m$ being the logarithmic gradient of the total mass density. The
 velocity $V_i$ is connected with the observed centroid velocity $V_i$
 by the obvious formula:
 \be
 V_i = V + (V_i - V),
 \label{eq6.10}
 \ee
 where $-V$ is the $\theta$-component of basic solar motion. This
 method for determination of the circular velocity was used earlier by
 \citet{Parenago:1951aa}.

In addition to the relation (\ref{eq6.7}), the general Galactic
parameter $k_\theta$  is
connected with another general parameter $k_z= \sigma_z^2/\sigma_R^2$
by means of formula
\be
k_z = \frac{k_\theta}{1-k_\theta},
\label{eq6.11}
\ee
found by \citet{Kuzmin:1961aa} from the theory of irregular gravitation
forces. Finally, we can use the angular velocity of the circular
motion
\be
\omega = \frac{V}{R}.
\label{eq6.12}
\ee

To summarise, we conclude that the system of Galactic parameters can
now be considered as consisting of $n = 10$ parameters

$R_0$,~$A$,~$B$,~$C$,~$\omega$,~$V$,~$W$,~$k_\theta$,~$k_z$, and
$\rho_0$. 

These ten parameters are connected by $l = 6$ equations (\ref{eq6.1}),
(\ref{eq6.5}) -- (\ref{eq6.7}),
(\ref{eq6.11}), and (\ref{eq6.12}), called in the theory of least squares as
fundamental. In future, with the development of the theory and the
improvement of observational data, the number of parameters in the
system as well as the number of fundamental equations may
increase. If $n - l = 4$ parameters are known, then the remaining $l$ ones
can be calculated from $l$ fundamental equations. It is natural to call
$n - l$ parameters as {\em principal Galactic parameters}. The principal
parameters can be chosen arbitrarily provided there are no functional
relationships between them, but from the practical point of view it is
advisable to call so the most frequently used ones, namely

$R_0,~A,~B,~C$.

All the Galactic parameters considered have observational estimates,
independent of other Galactic parameters. In determining the system of Galactic 
parameters with the method of least squares, one should
use all these independent estimates.

\begin{table*}[ht] 
\caption{Galactic parameters} 
\label{Tab6.1}
\centering 
\begin{tabular}{lccc}
\hline  \hline
  Quantity&	Unit	&Observed&		Calculated\\
  \hline
$R_0$&	kpc             &	$9.4	\pm	0.8$        &$9.05	\pm 0.4$\\
$A$ &	km/sec/kpc&	$15.2	\pm	1.4$&$15.7\pm 0.4$\\
$B$ &		``        &		                       &  $-10.3	\pm 0.4$\\
$C$ &	``                &	$70	\pm	5$            & $71	\pm 2$\\
$\omega$ &       ``      & $26.0 \pm 0.7$       & $26.0\pm  0.6$\\
$V$ &	km/sec      &	$226 \pm	21$	& $235\pm 10$\\
$W$ &	     ``          &	$142 \pm	6$&	$142\pm 6$\\
$k_\theta$ &                &	$0.408 \pm 0.013$&$0.396\pm 0.011$\\
$k_z$ &	                 &	$0.270 \pm 0.009$&$0.284\pm 0.006$\\
$\rho_0$ &$M_\odot$/pc$^3$&$0.091\pm 0.010$&	$0.088\pm 0.006$\\
\hline 
\end{tabular}
\end{table*}

In Table \ref{Tab6.1} the observed estimates of Galactic parameters together
with their standard external errors are given. The Table is based on a
critical analysis of modern observational data which will be published
in detail elsewhere.

In the Table, the value of angular velocity of the circular motion, $\omega$ ,
is used as the observed one instead of Oort’s parameter $B$. After our
critical survey of published data, the value of $B$ is in fundamental
systems GC, FK3 and N30 $- 11.7,~ -8.4$, and $-5.4$ km/sec/kpc
respectively. On the other hand, $\omega$ , derived as the difference between
$A$ and $B$ from the same proper motion data, equals in these fundamental
systems to $26.4, 25.5$ and $27.0$ km/sec/kpc. The scatter is much
smaller than in the case of $B$, and the mean value is more reliable.

The results of least-square solution are also represented in Table~\ref{Tab6.1}. 
The values obtained differ not very much from those adopted at the
Australia symposium. Therefore, it seems to us that it would too early
to accept now a new system of Galactic parameters. It would be very
useful to concentrate the attention to the practical as well as to the
theoretical aspects of the problem in order to increase the amount and the
quality of the observational data and their treatment. Only on such
basis, a new revision of the Galactic parameter system can be made.

 \vskip 3mm
\hfill July 1964
\vskip 2mm
\hfill Revised July 1971
\vskip 2 mm
\hfill Adapted  September 2021

\chapter{Galactic model\label{ch07}}

This Chapter presents our first model of the Galaxy, where both 
spatial and hydrodynamical descriptive functions were found.  In
calculations of hydrodynamical functions, we used methods developed by
\citet{Einasto:1970aa}, which form Chapter 11 of the Thesis.
Preliminary results of this model were reported at the IAU Commission
33 Meeting at the XIV General Assembly in Brighton by
\citet{Einasto:1970ad}. Here we give full data on the model.

\section{Introduction}

Galactic models have been constructed for various goals. Most detailed
models can be considered as a compact form of representing large
quantities of observational data for further theoretical analysis.
One possibility to use models is the study of evolution of galaxies.
For this task, models must represent the structure of actual galaxies
rather accurately and must be physically correct. 

The evolution of galaxies shall be discussed in the last Chapters of the
Thesis.  The goal of this paper is to analyse existing models of the
Galaxy and to find parameters for a new one. Next we use the
gravitational field of the model to calculate the spatial and
kinematical structure of test populations of various age.

\section{Analysis of existing Galactic models}

Recently a review of existing models of the Galaxy was published by
\citet{Perek:1962aa}.  The modern era of Galactic modelling was
introduced  by \citet{Schmidt:1956} and \citet{Kuzmin:1956aa}, and we
start our analysis with a description of these models.

The \citet{Schmidt:1956} model is the first one  where detailed
information on Galactic populations was used.  Model components
represent real populations of various flattening and isodensity
surfaces.  For some populations, kinematical characteristics were
calculated (velocity dispersions perpendicular to the plane of
Galaxy).  

\citet{Kuzmin:1956aa} used in the calculation of his model several
additional conditions which allowed to find model parameters with a
greater confidence, see below.

{\begin{table*}[ht] 
\caption{Galactic models} 
\centering 
\resizebox{0.98\textwidth}{!}{\includegraphics*{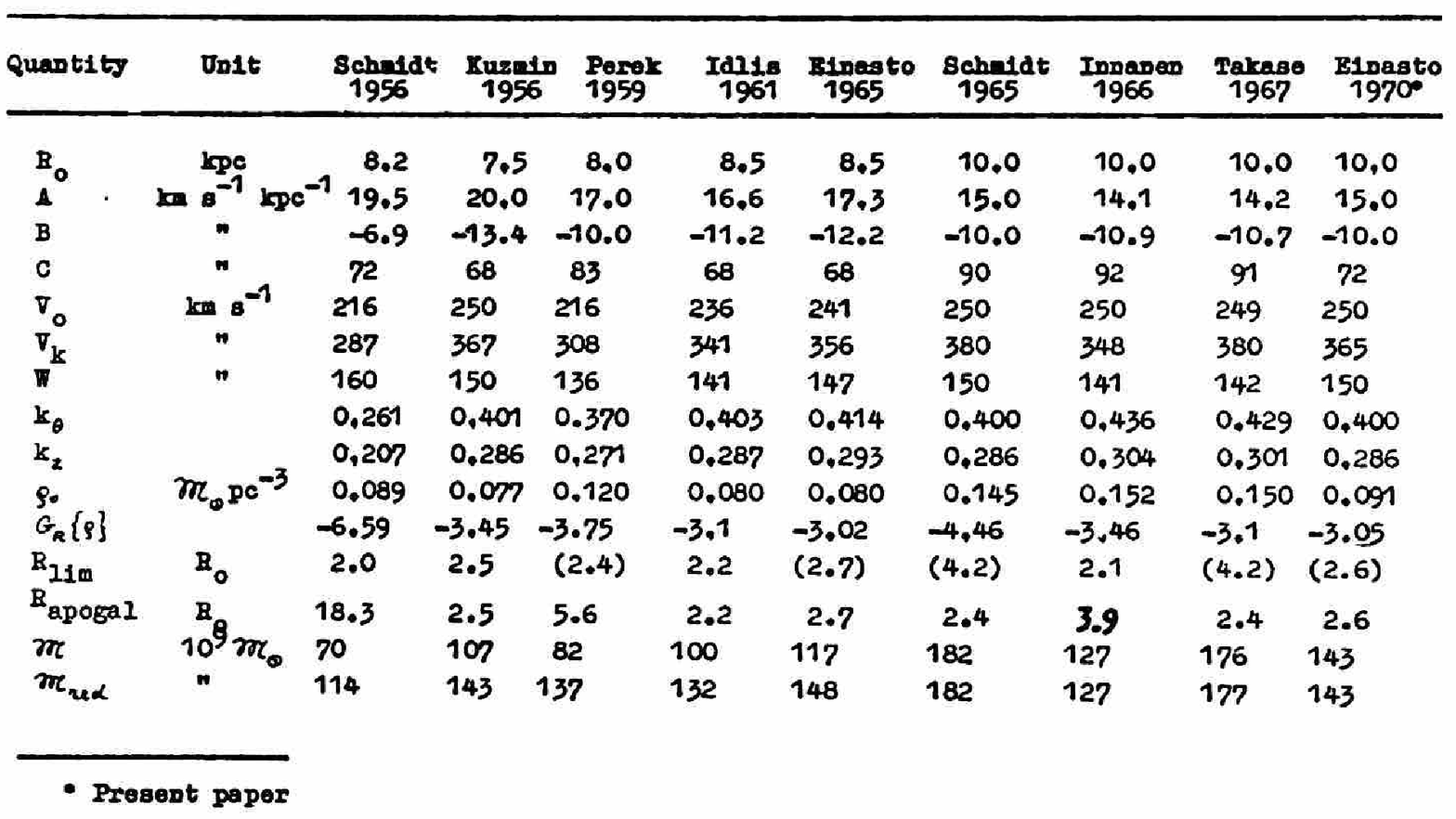}}
\label{Tab7.1}
\end{table*} 
}

In Table~\ref{Tab7.1} we give the main data on models of the Galaxy,
constructed since 1956: \citet{Schmidt:1956}, \citet{Kuzmin:1956aa},
\citet{Perek:1959aa}, \citet{Idlis:1961}, \citet{Einasto:1965aa},
\citet{Schmidt:1965aa}, \citet{Innanen:1966aa}, \citet{Takase:1967aa}
and the present model \citet{Einasto:1970ad}.  We use the following designations: \\
\hspace{1cm}$R_0$ — solar distance from the centre of the Galaxy;\\
\hspace{1cm}$A,~B,~C$ — Oort-Kuzmin dynamical parameters;\\
\hspace{1cm}$V_0$ — circular velocity at $R=R_0$;\\
\hspace{1cm}$V_k$ — escape velocity at $R=R_0$;\\
\hspace{1cm}$W$ — radial gradient of the function of differential Galactic
rotation (see below);\\
\hspace{1cm}$k_\theta,~k_z$ — ratios of velocity dispersions (see below);\\
\hspace{1cm}$\rho_0$ — total density of matter near the Sun;\\
\hspace{1cm}$G\{\rho(R)\}_0 = \left({\partial \ln{\rho} \over \partial \ln{R}}\right)_0$ —
logarithmic density gradient at $R=R_0$;\\
\hspace{1cm}$R_{lim}$ — radius of the outer boundary of the model (for models
with infinite boundary radius, an effective radius is given in
parenthesis, which corresponds to the distance in the symmetry plane
where the spatial density is $10^{2.5}$ times smaller than near $R=R_0$);\\
\hspace{1cm}$R_{apogal}$ — apogalactic distance of stars moving near the Sun with
Oort's limiting velocity;\\
\hspace{1cm}$\mm{M}$ — mass of the model;\\
\hspace{1cm}$\mm{M}_{red}$ — mass of the model, reduced to $R_0=10$~kpc and
$V_0=250$~km/s, supposing $\mm{M} \propto R_0\,V_0^2$.\\

Following \citet{Kuzmin:1952ab, Kuzmin:1954}, we use Oort-Kuzmin
parameters $A,~B,~C$ in dynamical sense. The dynamical parameter $C$
was introduced by \citet{Kuzmin:1952ab}, it is related to the
gravitational potential $\Phi$ of the Galaxy as follows
\begin{equation}
C^2 = -\left( \frac{\partial^2\Phi}{\partial z^2} \right)_{z=0},
\label{eq7.1}
\end{equation} 
where $z$ is the distance from the Galactic plane. The parameter $C$
determines the dependence of $\Phi$ on $z$ near the Galactic plane.
It is a supplement to the Oort parameters
$A$ and $B$ for the motion with circular velocity. Oort parameters $A$
and $B$ are related to the potential as follows:
\begin{equation}
\begin{array}{ll}
(A-B)^2 &= \frac{1}{R} \left( \frac{\partial\Phi}{\partial R}
\right)_{z=0}, \\
(A-B)(3A+B) &= \left( \frac{\partial^2\Phi}{\partial R^2} \right)_{z=0} , 
 \label{eq7.2}
 \end{array}
 \end{equation}
where $R$ is the distance from the Galactic axis.

In all models, it is assumed that equidensity surfaces are
concentric ellipsoids of rotational symmetry with constant or changing
flattening (ratio of vertical to radial axes) \citep{Perek:1962aa}.
Models differ from each other in three important ways: (i) in the
choice of principal Galactic parameters, (ii) in the extrapolation of
mass function to large galactocentric distances, and (iii) in the
choice of the principal descriptive function. These aspects are
closely related.

In early models presented until mid-1960's, Galactic parameters were
accepted in the old distance scale.  However, there exist other more
principal differences. 

We use as the principal function in mass modelling of galaxies the mass
 function
\be
\mu(a)= 4\pi\epsilon\rho(a)\,a^2,
\label{eq7.3}
\ee
where $\rho(a)$ is the spatial density, $\epsilon$ is the flattening
parameter (the ratio of minor to major semiaxis of the isodensity
surface). The function $\mu (a)$ is the mass of a spheroidal 
sheet per unit interval of $a$.  
Knowing the mass function of the spheroidal model, we may calculate the
circular velocity \citep[see][]{Kuzmin:1952ab}
\begin{equation}
  V^2 = G \int_0^R \frac{\mu (a)~da }{ \sqrt{R^2 - a^2e^2}} ,
  \label{eq7.4}
\end{equation}
where $G$ is the gravitational constant, and $e^2 = 1 -
\epsilon^2$. Here we can identify the circular velocity $V$ with the
rotation speed $V_\theta$  of flat population objects. The most
reliable data on the rotation velocity are provided by the 21-cm radio
data. Radio observations give the differential rotation velocity
function, $U(x)$, which is connected with the rotation velocity,
$V_\theta$, by formula \citep{Kuzmin:1956aa}:
\be
V_\theta(x) = U(x) + V_0\,x,
\label{eq7.5}
\ee
where $x=R/R_0$, and $V_0$ is circular velocity near the Sun. Radio
data enable us to determine $U(x)$ only for $x \le 1$, \ie inside the
Sun orbit in Galaxy, for $x >1$ the mass distribution function is to
be extrapolated.

As we see from Eq.~(\ref{eq7.5}), the model depends critically on the
adopted value of the circular velocity near the Sun, $V_0$. This
velocity can be determined on the basis of the adopted values for the
Sun's distance $R_0$, and Oort parameters, $A$ and $B$, using the
definition formula
\be
V_0 = (A-B)\,R_0.
\label{eq7.6}
\ee
All three parameters, $A$, $B$ and $R_0$,  may be influenced by
systematic and random errors, thus, an independent check of $A$, $B$
and $R_0$ or $V_0$ is necessary.

Parameters $A$, $B$ and $R_0$ can be checked using kinematical data
and applying the Lindblad formula
\be
k_\theta = {\sigma_\theta^2 \over \sigma_R^2} = {-B \over A-B},
\label{eq7.7}
\ee
and the radio data on the radial gradient of the function $U(x)$.
This gradient provides us with a new local Galactic constant, $W$,
which is connected with other constants by the following equation
\citep{Einasto:1964aa}:
\be
W = - {1 \over 2}{\dd{U} \over \dd{x}}|_{x=1} = A\,R_0.
\label{eq7.8}
\ee

{\begin{table*}[h] 
\caption{Components of the Galaxy model} 
\centering 
\hspace{2mm}
\resizebox{0.80\textwidth}{!}{\includegraphics*{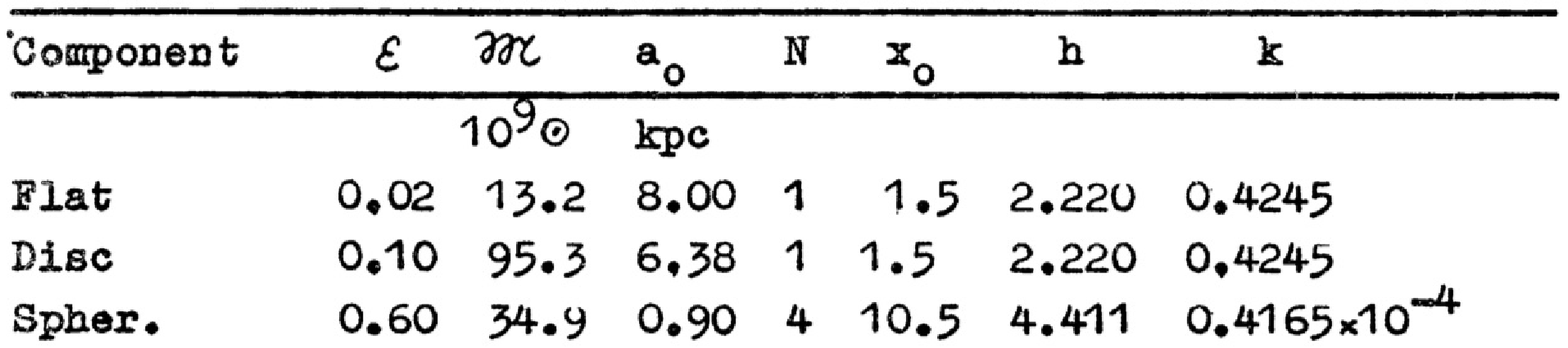}}
\label{Tab7.2}
\end{table*} 
}

{\begin{table*}[h] 
\caption{Descriptive functions of the Galaxy model} 
\centering 
\hspace{2mm}
\resizebox{0.60\textwidth}{!}{\includegraphics*{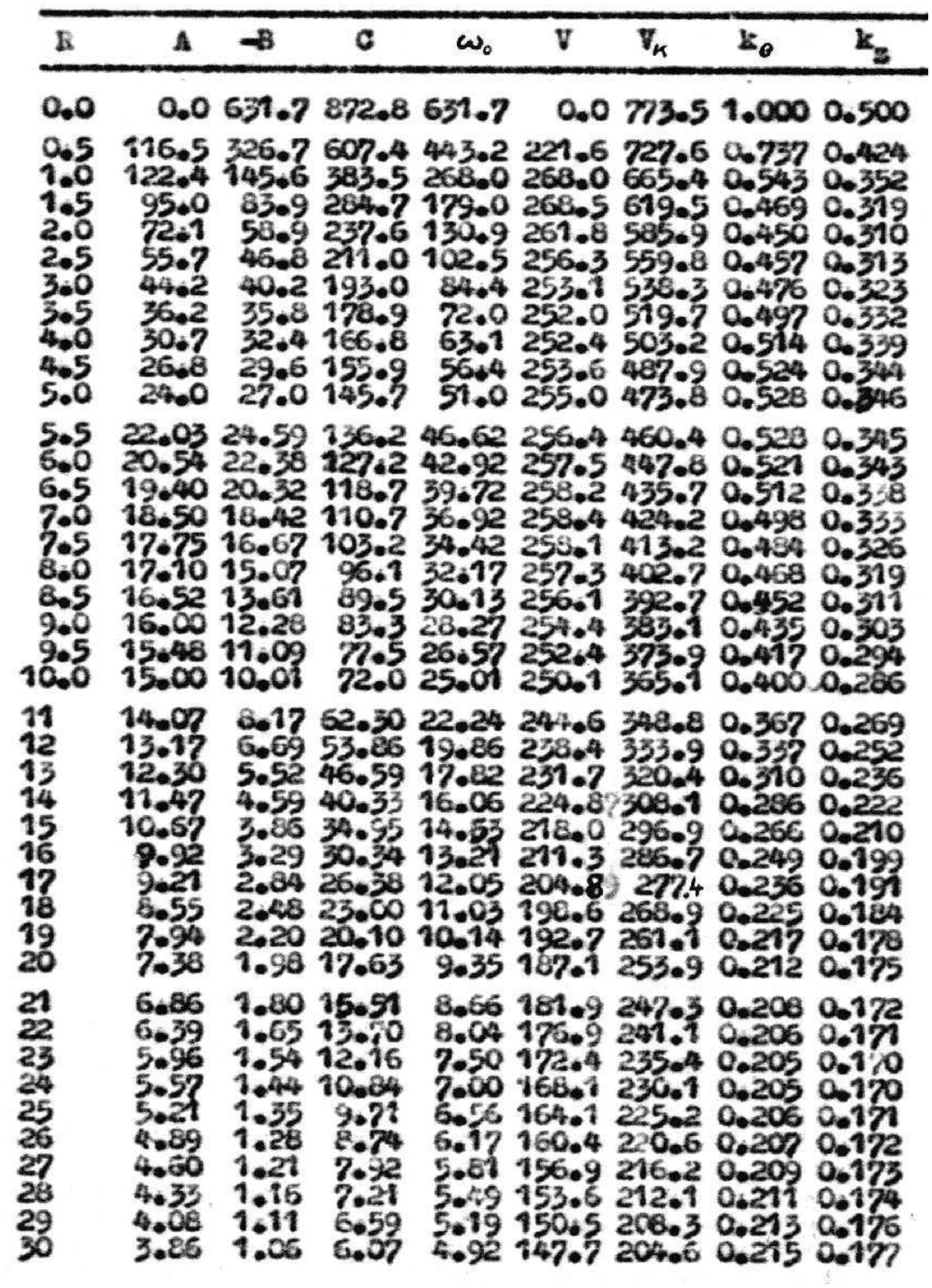}}
\label{Tab7.3}
\end{table*} 
}

\citet{Schmidt:1956} accepted in his model the local parameter system,
based on Oort $B$, found from proper motions of stars in FK3 and N30
systems, and used for  $k_\theta$ data on proper motions of faint
stars by \citet{Hins:1948aa}. In \citet{Kuzmin:1952aa}, model parameter
$B$ was taken in GC system, and $k_\theta$ was accepted using
\citet{Parenago:1951aa} analysis.  These two systems differ from each
other, thus, it is not sufficient to use only Eq.~(\ref{eq7.7}) and
(\ref{eq7.8}) to check parameters.

The value of the circular velocity near the Sun, $V_0$, can be checked
independently of local Galactic parameters in two ways
\citep{Kuzmin:1956aa}.  Firstly, the radial gradient of the mass
distribution function must be in accordance with the mean observed
radial gradient of the spatial density, $G_R(\rho)_0$. Secondly, the
limiting radius of the model, $R_{lim}$, must be equal to
$R_{apogal}$, the apogalactic distance of stars moving near the Sun
with  maximal galactocentric velocities, given by Oort's limiting
velocity.

{\begin{figure*}[h] 
\centering 
\resizebox{0.47\textwidth}{!}{\includegraphics*{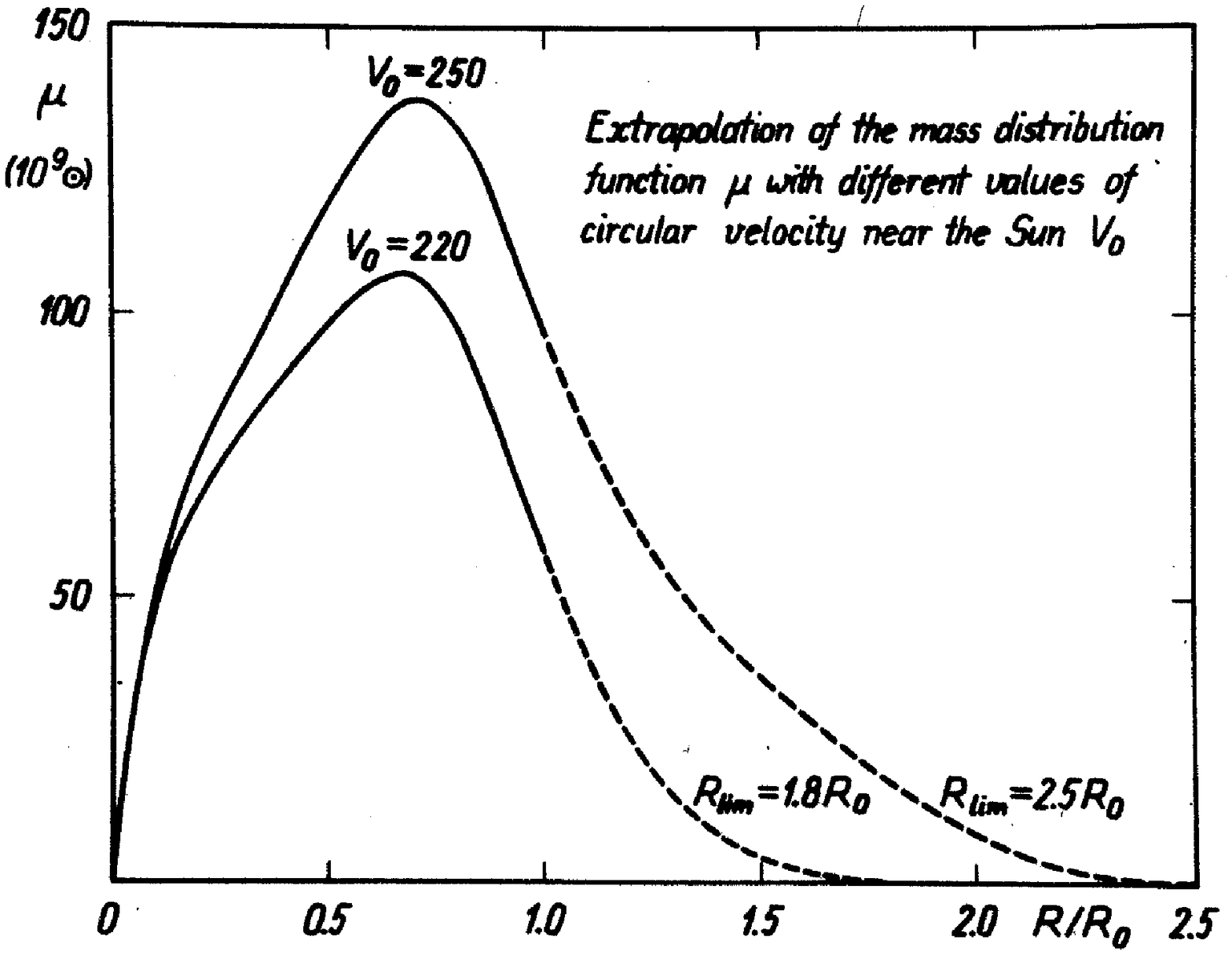}}
\resizebox{0.49\textwidth}{!}{\includegraphics*{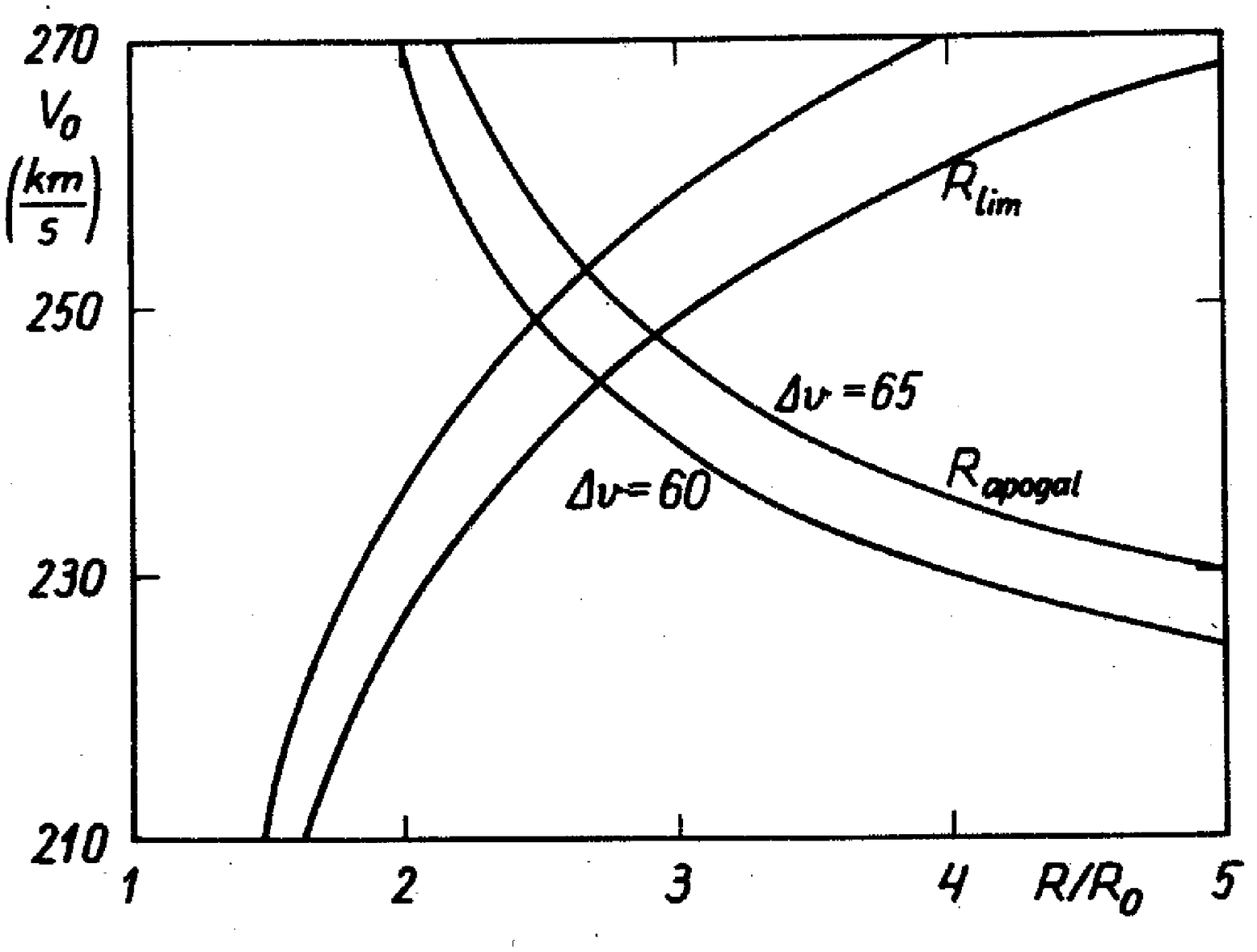}}
\caption{{\em Left:} The extrapolation of the mass distribution
  function beyond the Sun's distance $R>R_0$ (dashed lines) with
  different values for the circular velocity near the Sun,
  $V_0$. Limiting radii of models, $R_{lim}$, are indicated. {\em
    Right:} The dependence of limiting radii, $R_{lim}$, and
  apogalactic distances, $R_{apogal}$, on circular velocity near the
  Sun, $V_0$. Two cases of smooth extrapolation of the mass
  distribution function with different $R_{lim}$ are
  indicated. Apogalactic distances are given for two values of Oort's
  limiting velocity, $\Delta\,v$. }
  \label{Fig7.1}
\end{figure*} 
}

As recognised by \citet{Stromberg:1924aa}  and \citet{Oort:1928aa}, no
stars with apices within Galactic longitudes $l = 20^\circ - 85^\circ$
and heliocentric velocities exceeding 65 km/s are found. The limiting
velocity may correspond to the velocity required to reach the boundary of
the Galaxy \citep{Bottlinger:1933aa}, or to the velocity of escape
\citep{Oort:1928aa}.  As shown by \citet{Kuzmin:1956aa}, only the
first alternative, $R_{apogal} = R_{lim}$, can be correct. Adopting
the second alternative, as done by \citet{Schmidt:1956}, we get
$R_{lim} \ll R_{apogal} \approx \infty$, but this situation is impossible
since stars with velocities smaller than the escape velocity belong to
the system and must be included into the mass distribution
function. If these stars are included to the model, we get
\be
R_{lim} = R_{apogal}.
\label{eq7.9}
\ee

{\begin{figure*}[h] 
\centering 
\hspace{2mm}
\resizebox{0.50\textwidth}{!}{\includegraphics*{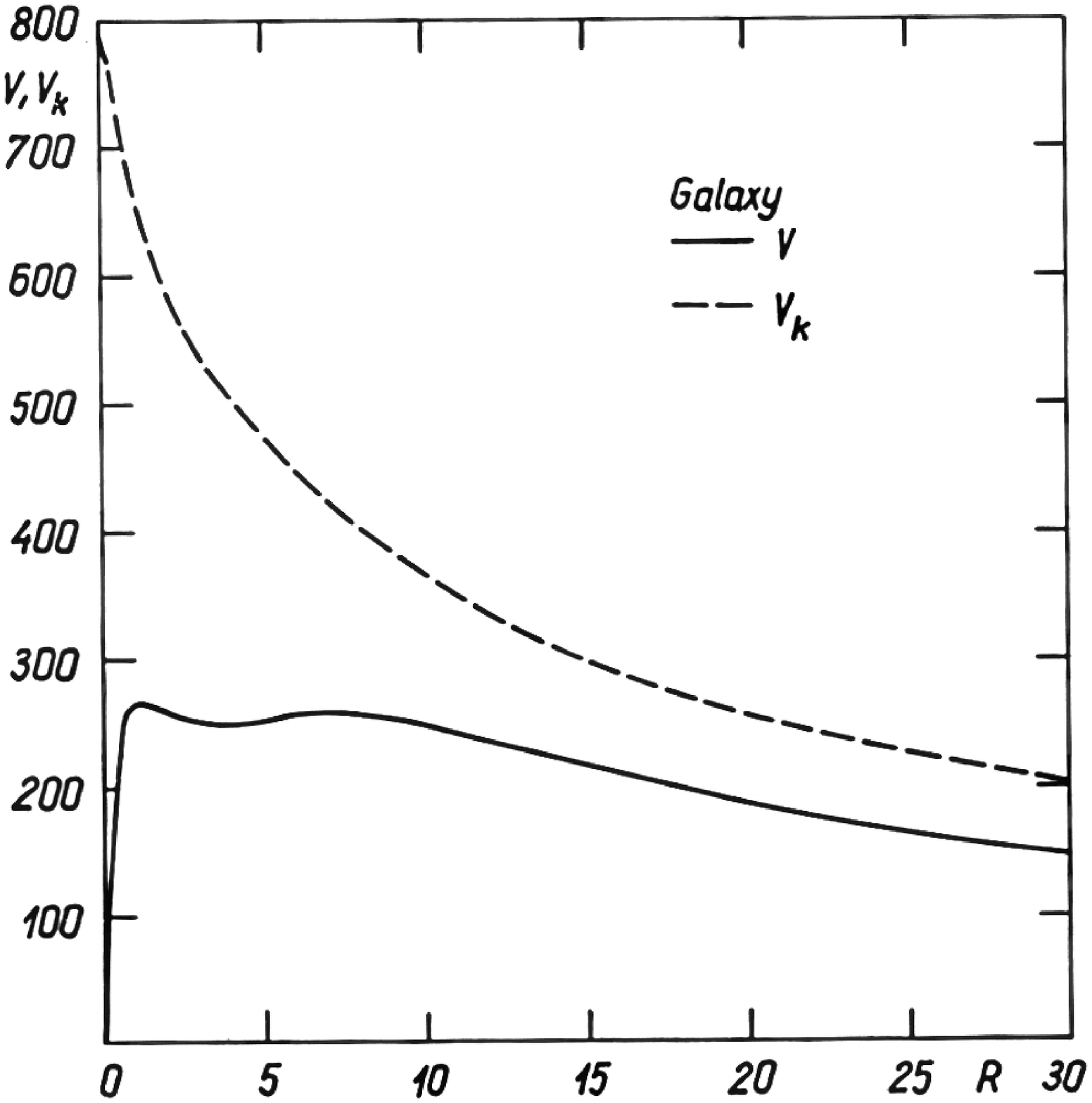}}
\caption{ Circular and escape velocities as functions of distance from
  Galactic centre.    } 
  \label{Fig7.3}
\end{figure*} 
}

{\begin{figure*}[h] 
\centering 
\hspace{2mm}
\resizebox{0.50\textwidth}{!}{\includegraphics*{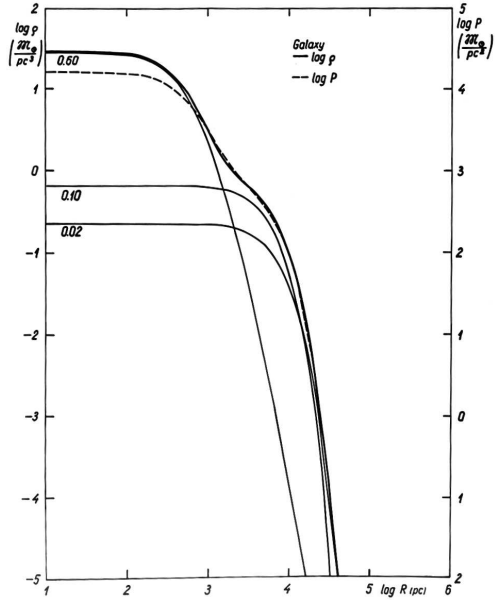}}\\
\caption{ Spatial densities $\rho$, and projected
densities $P$ of components
  as functions of distance from
  Galactic centre.     } 
  \label{Fig7.4}
\end{figure*} 
}

{\begin{figure*}[h] 
\centering 
\resizebox{0.48\textwidth}{!}{\includegraphics*{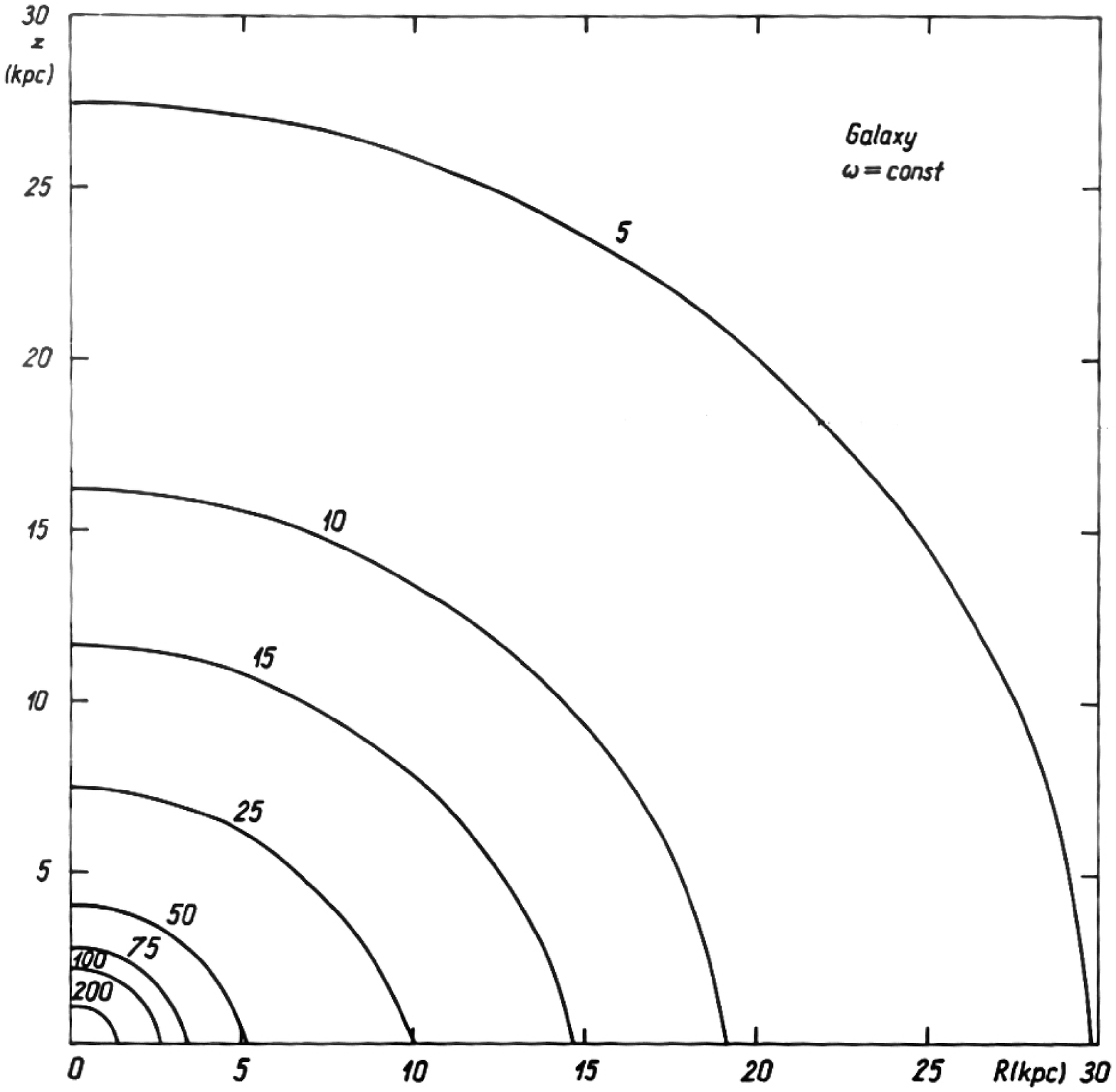}}
\resizebox{0.48\textwidth}{!}{\includegraphics*{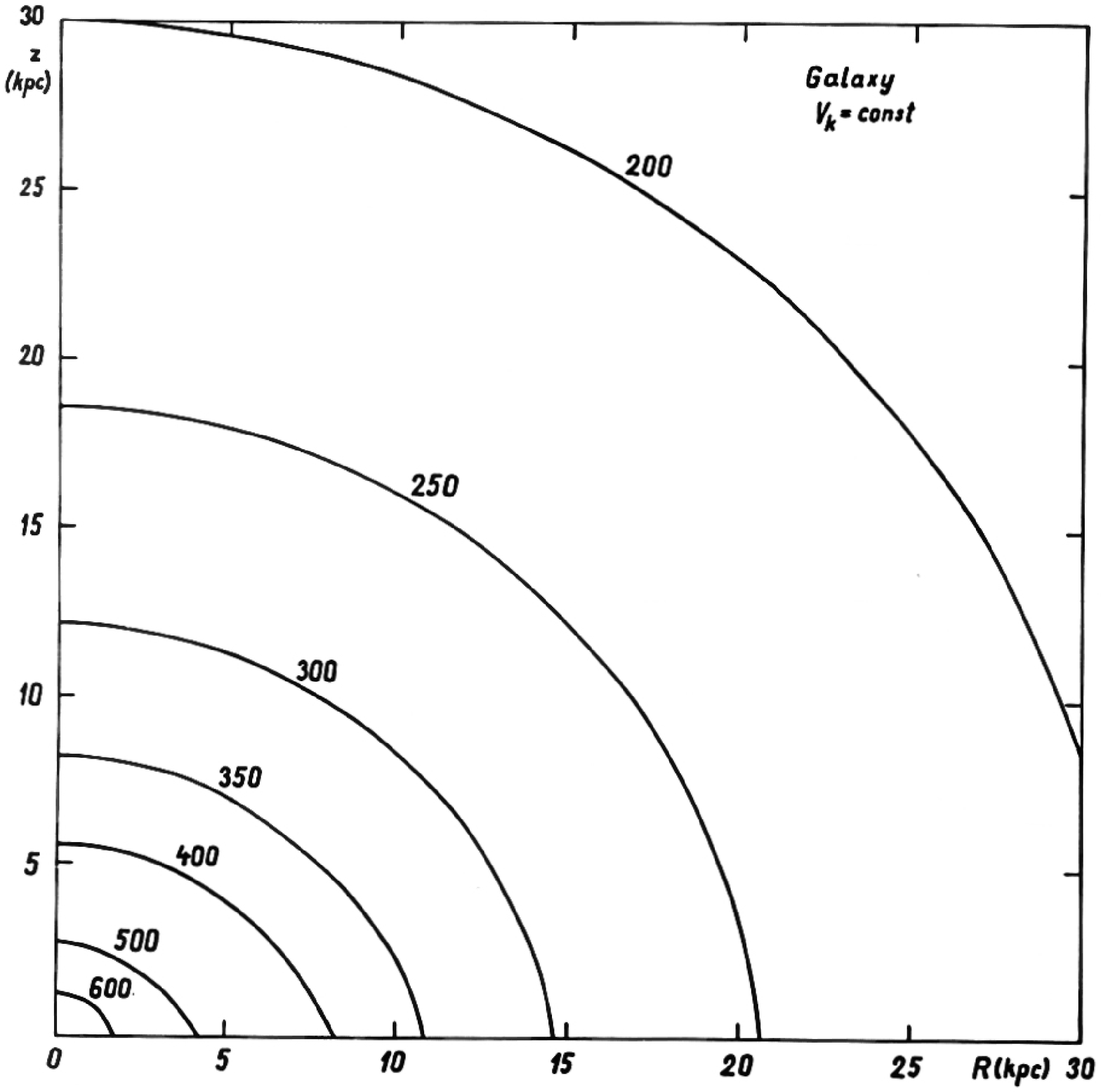}}
\caption{Isolines of angular velocity $\omega_0$ and  escape velocity
  $V_k$ are shown in left and right panels, respectively  } 
  \label{Fig7.5}
\end{figure*} 
}

%

\begin{table*}[h]
\centering    
\caption{Parameters of test  components of the Galaxy} 
\begin{tabular}{lcccrrrrrr}
  \hline  \hline
 Popul.&$\epsilon$&$a_0$&$N$&$x_0$&
                                    $(\sigma_z)_0$&$\sigma_z$&$\sigma_R$&$V_\theta$&
                                                                                     $t$\\
  &&kpc&&&&km/s&&& $10^9$yr\\
  \hline
Flat     &  0.02& 8.0& 0.5& 0.0&  8.8& 8.8& 16.4& 250& 0.9\\
Disc 1 &  0.05& 7.4& 1.0& 1.5&20.4&19.9&37.3& 239&3.9\\
Disc 2&  0.10& 6.4&  1.5& 3.0&34.7&34.4&64.5& 216&7.6\\
Halo 1& 0.20& 4.5&  2.0& 4.5&52.8&51.7&84.3& 185&9.1\\
Halo 2& 0.40& 1.9&  3.0& 7.5&75.2&71.4&92.8&142&9.4\\
Halo 3& 0.60& 0.9&  4.0&10.5&92.6&85.9&100.9& 96&9.7\\
Halo 4& 0.80& 0.6&  5.0&13.5&108.3&98.5&109.8&46&10.0\\
  \hline
\label{Tab7.4}   
\end{tabular}
\end{table*}

{\begin{figure*}[h] 
\centering 
\resizebox{0.75\textwidth}{!}{\includegraphics*{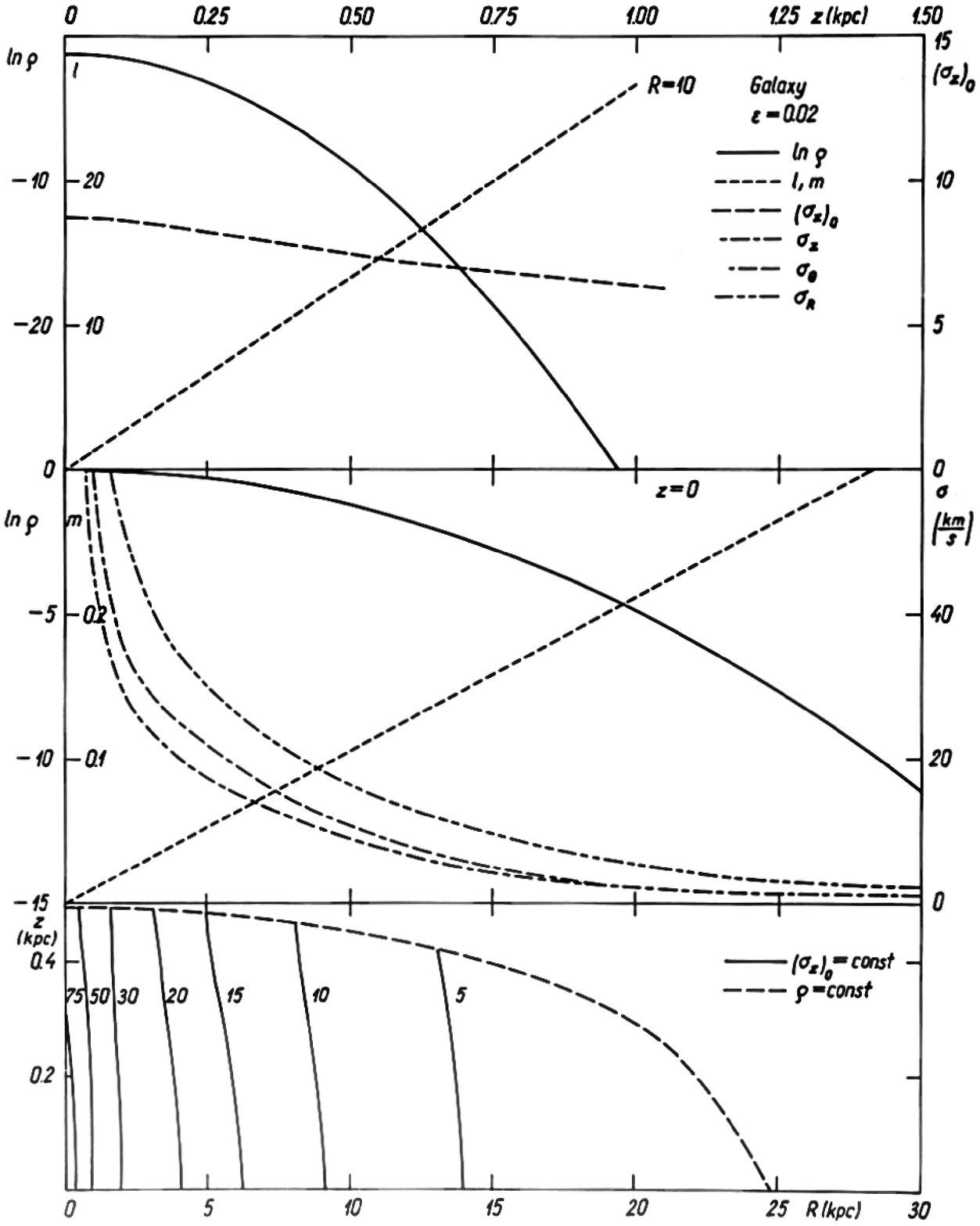}}
\caption{Descriptive functions for the test population of flattening
  $\epsilon=0.02$.} 
  \label{Fig7.7}
\end{figure*} 
}

{\begin{figure*}[h] 
\centering 
\resizebox{0.75\textwidth}{!}{\includegraphics*{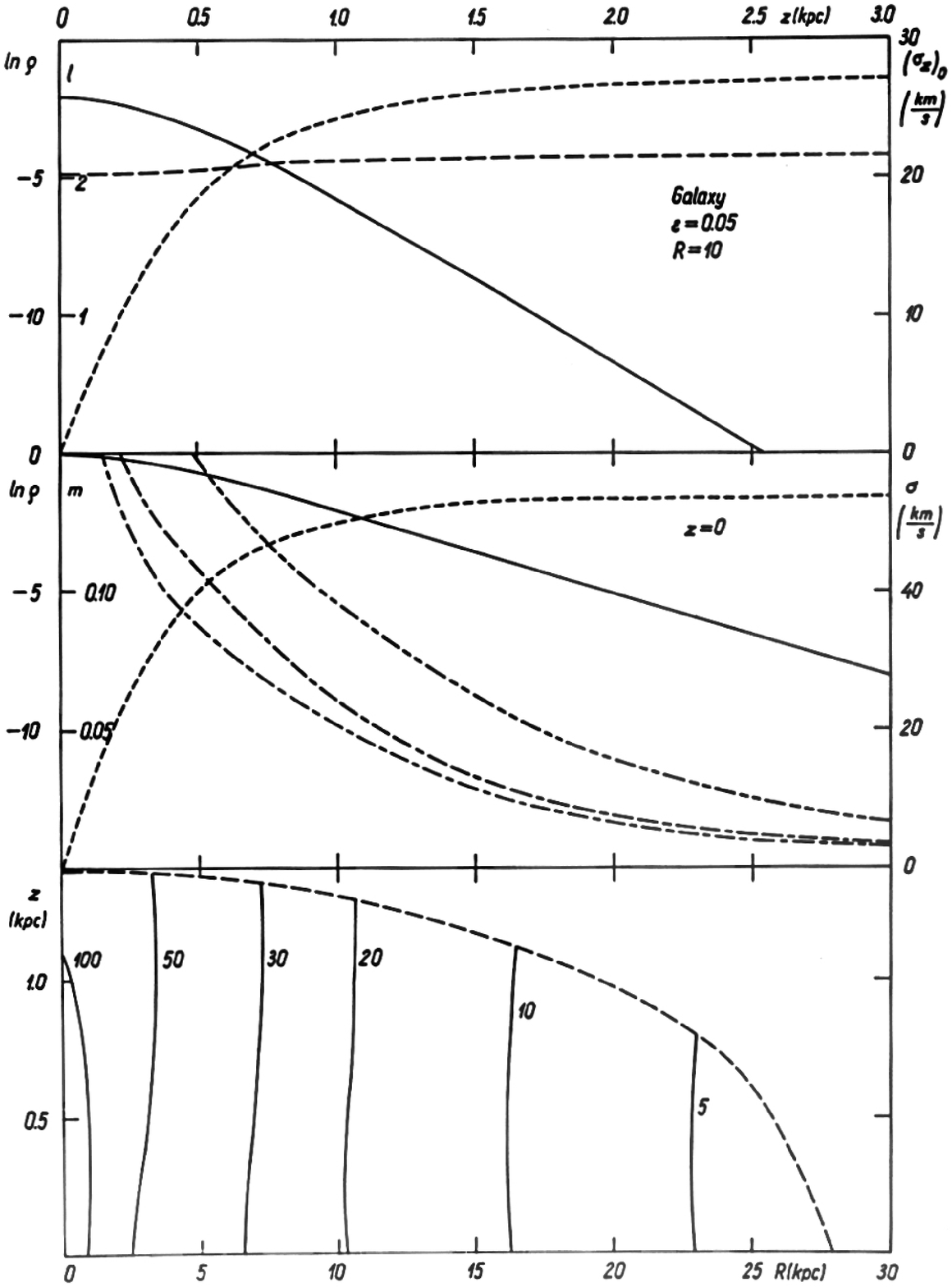}}\\
\caption{Descriptive functions for the test population of flattening
$\epsilon=0.05$.} 
  \label{Fig7.8}
\end{figure*} 
}

{\begin{figure*}[h] 
\centering 
\resizebox{0.75\textwidth}{!}{\includegraphics*{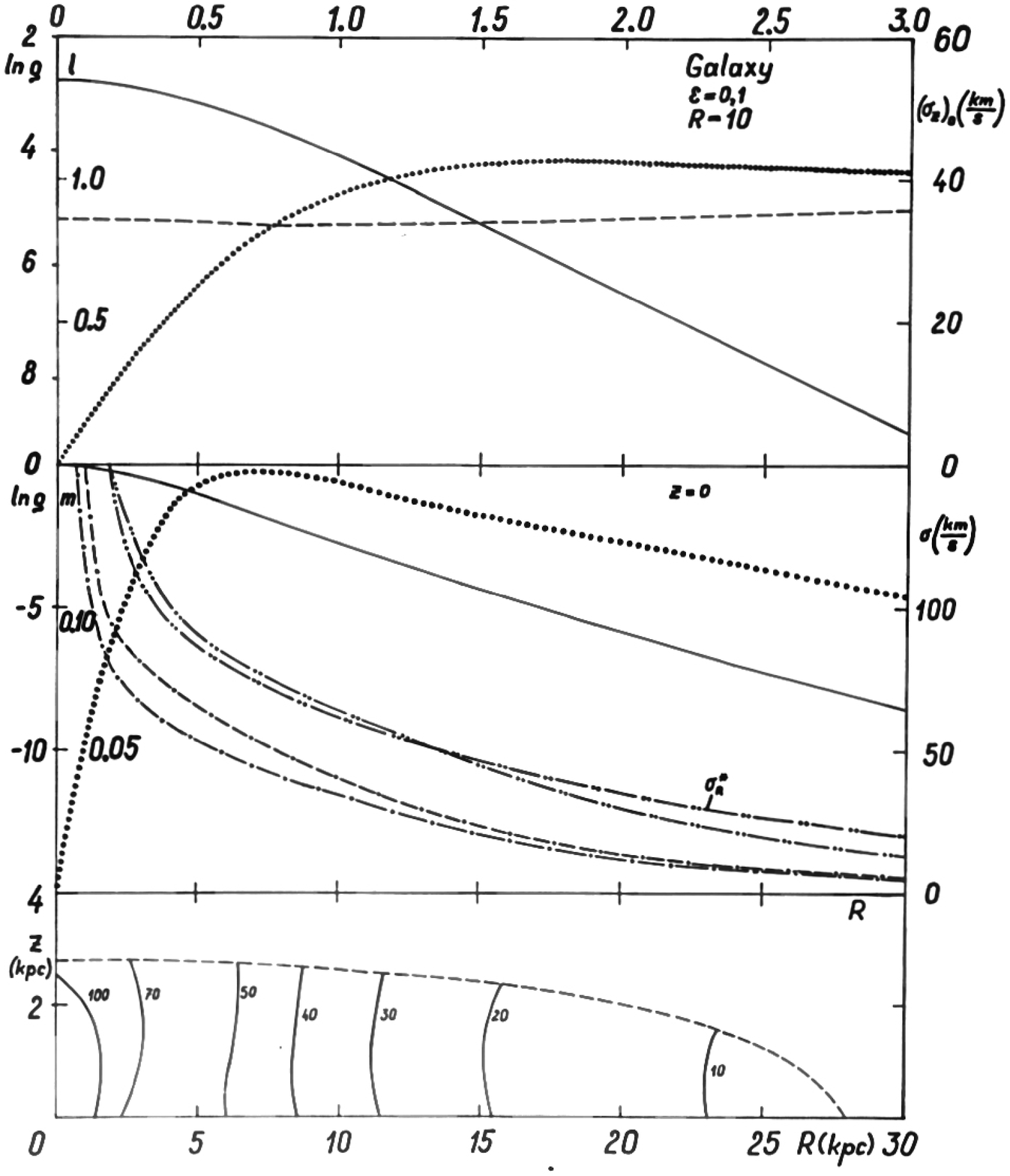}}
 \caption{Descriptive functions for the test population of flattening
   $\epsilon=0.10$.} 
  \label{Fig7.9}
\end{figure*} 
}

{\begin{figure*}[h] 
\centering 
\resizebox{0.75\textwidth}{!}{\includegraphics*{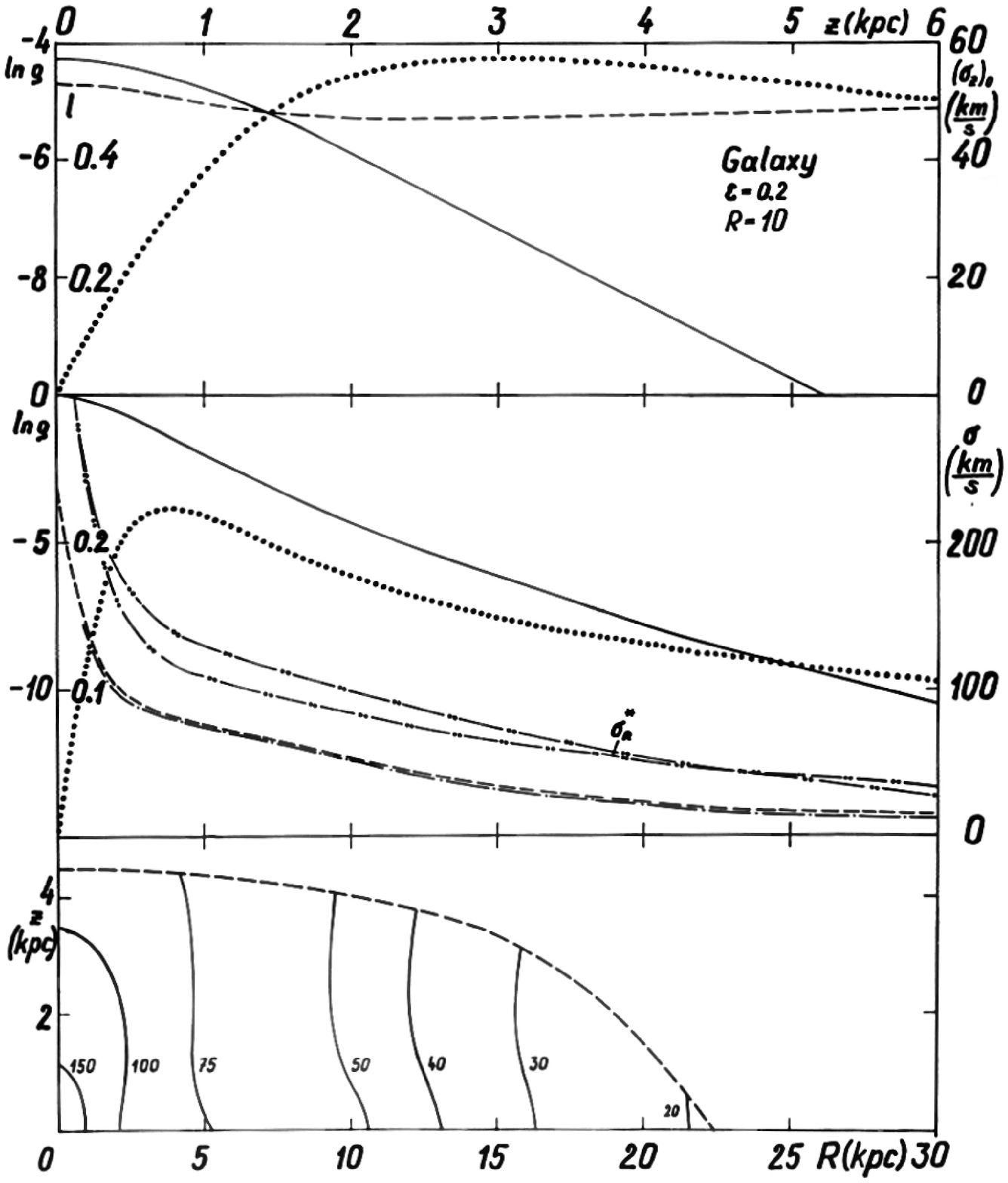}}
 \caption{Descriptive functions for the test population of flattening
 $\epsilon=0.20$.} 
  \label{Fig7.10}
\end{figure*} 
}

{\begin{figure*}[h] 
\centering 
\resizebox{0.75\textwidth}{!}{\includegraphics*{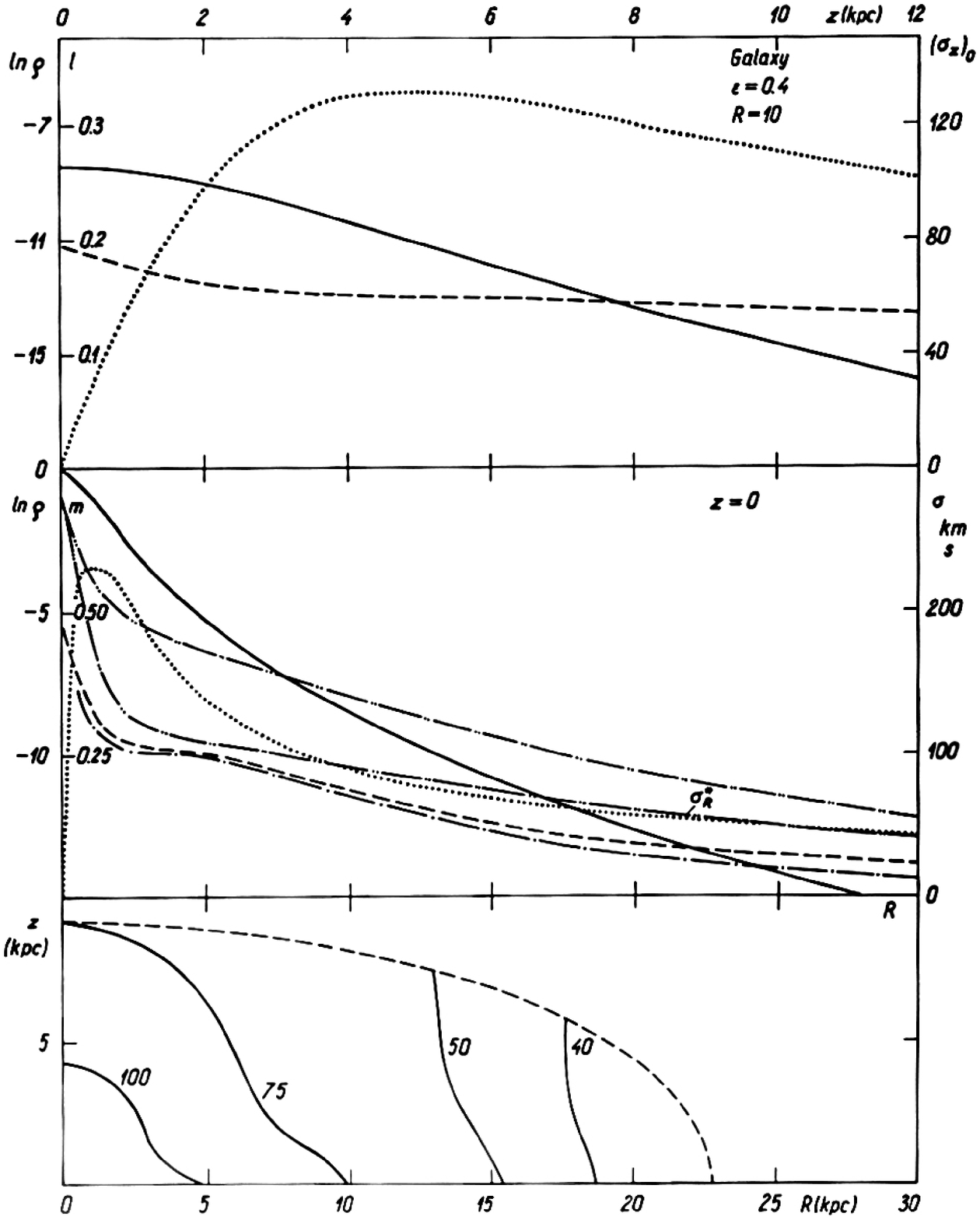}}
\caption{Descriptive functions for the test population of flattening
   $\epsilon=0.40$..} 
  \label{Fig7.11}
\end{figure*} 
}

{\begin{figure*}[h] 
\centering 
\hspace{2mm}
\resizebox{0.75\textwidth}{!}{\includegraphics*{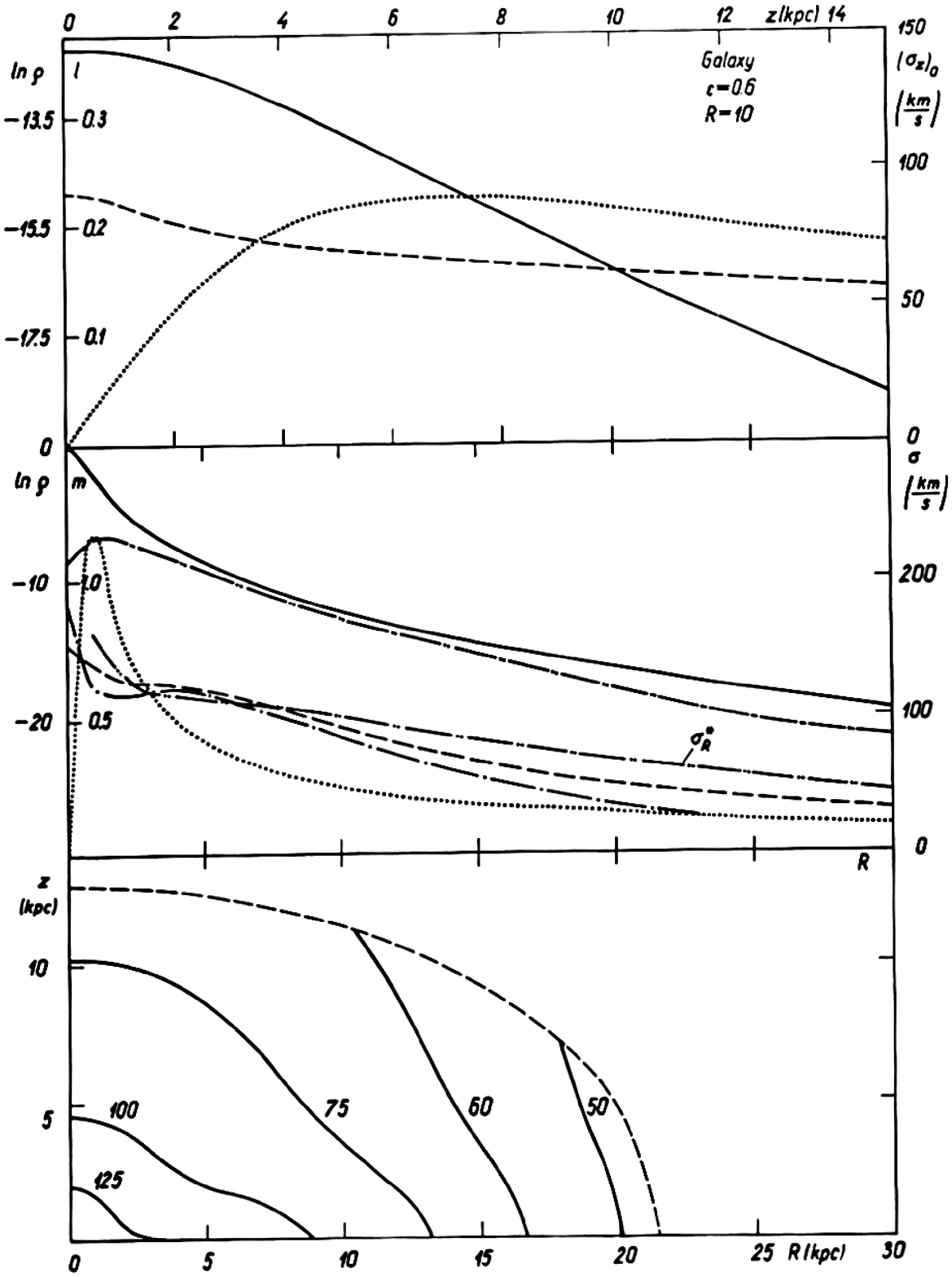}}\\
 \caption{Descriptive functions for the test population of flattening
   $\epsilon=0.60$.} 
  \label{Fig7.12}
\end{figure*} 
}

{\begin{figure*}[h] 
\centering 
\resizebox{0.75\textwidth}{!}{\includegraphics*{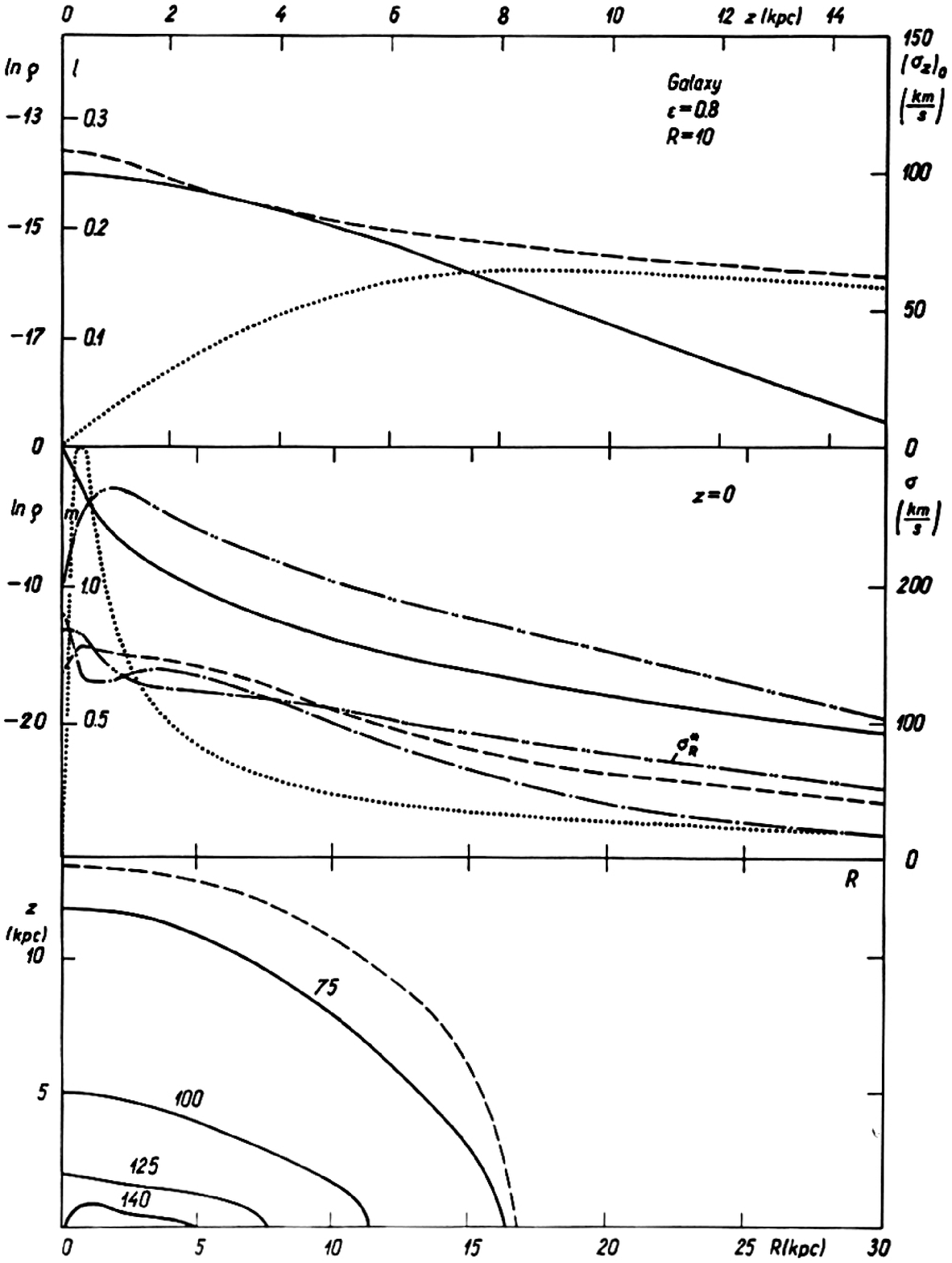}}
\caption{Descriptive functions for the test population of flattening $\epsilon=0.8$.} 
  \label{Fig7.13}
\end{figure*} 
}

{\begin{figure*}[h] 
\centering 
\hspace{2mm}
\resizebox{0.40\textwidth}{!}{\includegraphics*{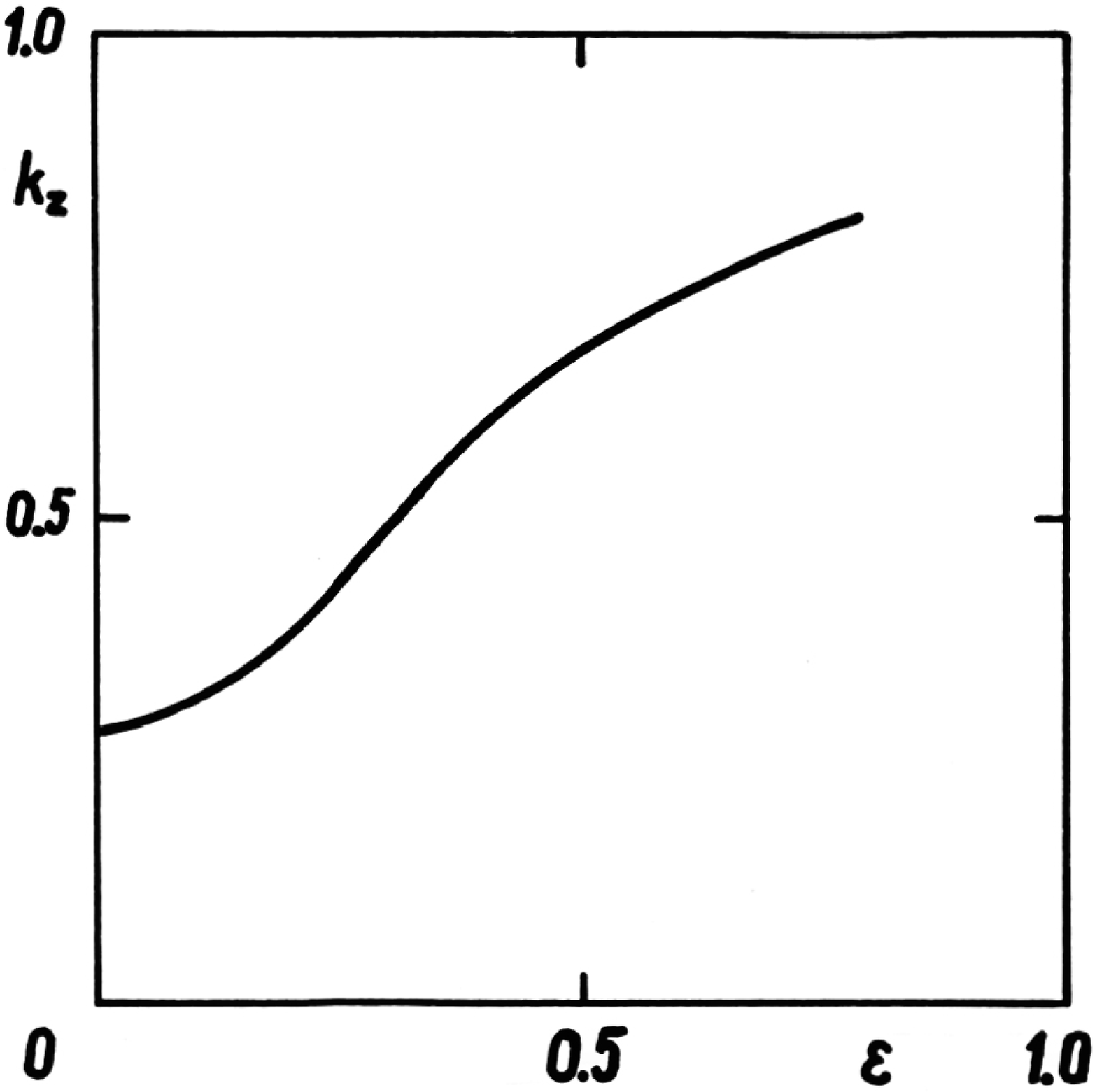}}
\caption{The dependence of $k_z=\sigma_z^2/\sigma_R^2$ on $\epsilon$
  at $R=10$~kpc and $z=0$. } 
  \label{Fig7.14}
\end{figure*} 
}

In left panel of Fig.~\ref{Fig7.1} the mass distribution function, $\mu(a)$, is
given for two values of the circular velocity, $V_0$. In both cases,
the function of differential rotation, $U(x)$, is the same, and the
extrapolation of $\mu(x)$ beyond the Sun's distance, $R > R_0$, is
smooth. In right panel of Fig.~\ref{Fig7.1} the dependence of $R_{lim}$ and
$R_{apogal}$ on $V_0$ is shown\footnote{The possibility of determining
  the circular velocity, $V_0$, with rather great accuracy on the
  basis of Oort's limiting velocity was discovered by the author, who
  made most calculations for the \citet{Kuzmin:1956aa} model. Details
  of the method were elaborated together with
  Kuzmin.   Fig.~\ref{Fig7.1}  was prepared by the
  author in February 1956 but was not published in the final version
  of \citet{Kuzmin:1956aa} paper.}.

Both methods, the use of the gradient of spatial density and Oort's
limiting velocity, give the circular velocity values  $V_0 \approx 
250$~km/s \citep{Kuzmin:1956aa}.  These possibilities for checking the
circular velocity were mentioned by \citet{Schmidt:1965aa}. But in
respect of extrapolation of the mass distribution function, his new
model is overcorrected, as $R_{lim} > R_{apogal}$.  The cause is a too
low decrease of the mass density in outer regions of the model
according to power law $\rho \propto a^{-4}$.  The same behaviour has
the model constructed by \citet{Takase:1967aa}.

For our Galaxy, we have no direct argument against the power law
$\rho \propto a^{-4}$ and $R_{lim} > R_{apogal}$.  Photometric
observations of other galaxies show, however, that galaxies have
well-defined outer boundaries \citep{Arp:1971aa} with exponential
density law \citep{de-Vaucouleurs:1969aa}.  If we accept for density
the power law $\rho \propto a^{-4}$, and for the circular velocity the
\citet{Bottlinger:1933aa} profile as in the model by
\citet{Takase:1967aa}, then with increasing $a$ the mass-to-light
ratio becomes very great, which is difficult to accept.  On the other
hand, if we accept $R_{lim} > R_{apogal}$, then in peripheral regions
of the Galaxy, all stars have small velocity dispersion, and no such
stars can reach in their orbits the Solar region. Data on other galaxies
suggest that peripheral regions of galaxies belong to halo population
with large velocity dispersion. For these reasons, it is not likely
that among these halo stars there are none with $R_{perigal} \le
R_0$.

The structure of the inner parts of models was discussed in detail by
\citet{Einasto:1965aa}.  Here we shortly discuss new models, suggested
after the publication of the model by \citet{Einasto:1965aa}.

One of aspects of the structure of galaxies is the presence of a dense
nucleus.  To describe this aspect of the structure of the Galaxy,
\citet{Schmidt:1965aa} and \citet{Innanen:1966aa} used density laws
with infinite density at centre.  It is clear that in this region
their models have only the approximate meaning.  Such models have the
peculiarity having zero velocity dispersion at the very center. This
aspect was ignored by \citet{Innanen:1968aa}, where velocity
dispersions were estimated for the \citet{Innanen:1966aa} model.
Presently we have little information on the density of matter near the
centre of the Galaxy.  Available data are probably underestimates.

\section{A new model of Galaxy}

We suppose that the mass distribution of the Galaxy can be represented
by a sum of three ellipsoidal components with various axial
ratios $\epsilon$. One ellipsoid represents spherical populations
(halo and bulge), the second ellipsoid represents the disc, and the
third one the flat component.

For mass density distribution of components of the Galaxy, we use the
modified exponential profile:
\be
\rho(a) =\rho_0\,\exp[x_0
-(x_0^{2N}+\xi^2)^{1/(2N)}],
\label{eq7.3.1}
\ee
where
\be
\rho_0 =  {\mm{M} \over a_0^3}{h \over 4\pi\epsilon}
\label{eq7.3.1A}
\ee
is the central density, and 
\be
\xi = a/(k\,a_0),\hspace{1cm} a^2 = R^2 + z^2/\epsilon^2.
\label{eq7.3.2}
\ee
Here $\mm{M}$ is the mass of the components, $a_0$ is its harmonic
mean radius, defined by the formula
\be
a_0^{-1} = \mm{M}^{-1}\int_0^\infty\mu(a)\,a^{-1}\dd{a},
\label{eq7.3.3}
\ee
and  $N$ and $x_0$  are  structural parameters.  $h$, $k$ are
dimensionless normalising parameters needed to get for the mass and
harmonic mean radius definition formulae Eq.~(\ref{eq7.3.3}) and
\be
\mm{M}= \int_0^\infty\,\mu(a)\dd{a},
\label{eq7.3.4}
\ee
as suggested by \citet{Einasto:1968ad}.  We use the harmonic mean
radius $a_0$ is the effective radius of the model.
The parameter $x_0$ is introduced
to avoid a too sharp density peak and the resulting minimum in velocity
dispersion near the centre of the model. If we take $x_0=0$, we get
the exponential density profile as suggested by
\citet{Einasto:1965aa}:
\be
\rho(a) = \rho_0\,\exp[-\xi^{1/N}].
\label{eq7.3.5}
\ee

We adopted structural parameters of components on the basis of analogy
with the Andromeda galaxy \citep{Einasto:1969aa, Einasto:1970aa}. The
spherical component combines halo and bulge populations, which for M31
had $\epsilon=0.30$ and $\epsilon=0.80$, respectively. For Galaxy we
used an intermedium value $\epsilon=0.6$. For disc and flat components
we used $\epsilon=0.10$ and $\epsilon=0.02$. These components
represent actual populations in the flatness range  $0.05 \le
\epsilon \le 0.20$ and $\epsilon  < 0.05$.

The mean radius of spherical component was determined using
photometric data by \citet{Arp:1971aa} on the bulge of Galaxy,  and data
by \citet{Perek:1962aa} on the distribution of RR Lyrae variables. 
The mass of this component was determined using the rotation
velocity in inner parts of the Galaxy \citep{Schmidt:1965aa}. Mean radii
and masses of other two components were estimated on the basis of
available data on the distribution of disc population and young
stars. The final values were found by a trial-and-error procedure to have at
$R=R_0$ the adopted values of Oort-Kuzmin parameters. Parameter $x_0$
was found by a similar procedure to have a monotonous increase of the
velocity dispersion by decreasing $a_0$. The adopted parameters are given
in Table~\ref{Tab7.2}.

We calculated for our model all principal descriptive functions. For
the Galactic plane $z=0$ we found: Oort-Kuzmin parameters $A,~B,~C$,
angular velocity $\omega_0$, circular velocity $V_c$, escape velocity
$V_k$, and ratios of velocity dispersions $k_\theta$ and $k_z$. These
data are given in Table~\ref{Tab7.3}.  Circular velocity and escape
velocity as functions of distance from the Galactic centre are shown in
Fig.~\ref{Fig7.3}.   Spatial densities $\rho$ and projected
densities $P$ of components are shown in  Fig.~\ref{Fig7.4}. Isolines
of angular velocity $\omega_0$  and escape velocity $V_k$ are given in
Fig.~\ref{Fig7.5}.

A few comments on the results obtained.  Of special interest is the
angular velocity $\omega_0$. It was calculated using the first
hydrodynamical equation by \citet{Einasto:1970ac}, see Chapter 11, which for small
velocity dispersion, $\sigma_R \rightarrow 0$,  has the form
\be
V_{\theta}^2 = R\,K_R=V_c^2,
\label{eq7.3.6}
\ee
where $K_R= -\partial \Phi/\partial R$, and $\Phi$ is the
gravitational potential. On the $z=0$ surface $\omega_0 = V_c/R$, where
$V_c$ is circular velocity. Fig.~\ref{Fig7.5} shows that in the $R,z$
surface isolines of $\omega_0$ are slightly flattened ellipsis. From
the theory of stationary galaxies it follows that isolines of
$\omega_0$ should be spheres  \citep{Kuzmin:1956aa}. Assumptions of
stationary galaxies are not exact, thus, small deviations from spheres
are likely.

When we compare description functions of our Galaxy with similar
functions for the Andromeda galaxy \citep{Einasto:1970ac}, we see that
they are very similar. The only large difference is in the structure
of central regions. Presently we have little data on the structure of
the nucleus of the Galaxy. For this reason, the nucleus was not
included as a component.  Available data suggest that the nucleus of
the Galaxy is similar to the nucleus of M31. If this is correct, then
the central density of the Galaxy would be of the order
$10^6\,M_\odot pc^{-3}$, and the velocity of escape about 1500
km/s, \ie\ two times higher than adopted in the present model. To
understand the structure of the central regions of the Galaxy, new
observational data are needed.

\section{Kinematics of population}

We used the model to find various characteristics of the spatial and
kinematical structure of test populations. These test populations are
listed in Table~\ref{Tab7.4}, they represent various actual Galactic
populations. The following data are listed: flattening $\epsilon$,
effective radius $a_0$, structural parameters $N$ and $x_0$,
normalising parameter $k$, vertical velocity dispersion $(\sigma_z)_0$
at the Sun distance in Jeans approximation, vertical velocity
dispersion $\sigma_z$ at the Sun distance, radial velocity dispersion
$\sigma_R$ at Sun distance, rotational velocity $V_\theta$ at Sun
distance, and estimated age $t$.

For test populations, we calculated various descriptive functions,
shown in Figs.~\ref{Fig7.7} to \ref{Fig7.13}.  These Figures have
three panels.

In the upper panel, we show with solid lines the density logarithm,
$\ln \rho$, with dotted lines the vertical density gradient,
$l = - \partial \log \rho/\partial z$, and with dashed lines the
velocity dispersion $(\sigma_z)_0$ in Jeans approximation, see Chapter
11.  Velocity dispersion $(\sigma_z)_0$ was calculated within the
range $0 \le z \le z_u$, where $z_u$ is the outer vertical limit of
the population. All functions are plotted as functions of $z$ at
$R=10$~kpc, the adopted distance of the Sun from Galactic centre. The
vertical scale $z$ is shown at the top of the panel, the density scale
is on the left border, and the velocity dispersion scale on the right
border.

Central panels of Figures show with solid lines the logarithm of the
density, $\rho$, with dotted lines radial density gradient,
$m= -\partial \log \rho/\partial R$, and with various dot-dashed lines
velocity dispersions $\sigma_R,~\sigma_\theta, \sigma_z$. These data
are given as functions of $R$ on the plane of the Galaxy, $z=0$. The
radial distance scale $R$ is shown at the lower border of the Figure,
the density scale is on the left vertical border, and the velocity
dispersion scale on the right border.  Densities are given in units of
the central density, gradients $l$ and $m$ in kpc$^{-1}$, dispersions
in km/s.
Vertical velocity dispersions $\sigma_z$ were calculated from the
second hydrodynamical equation (\ref{eq11.1.5}), using for $k_z$
Kuzmin equation (\ref{eq16.4}) from the theory of irregular forces.
Radial velocity dispersions $\sigma_R$ were found from definition
equation (\ref{eq11.1.8}) from $\sigma_z$ and $k_z$, taking into
account the Kuzmin equation (\ref{eq16.4}).
For the test populations with $\epsilon \ge 0.10$ radial velocity
dispersions were also calculated from the first hydrodynamical
equation (\ref{eq11.1.4}) from circular velocity $V_c$ and rotational
velocity $V_\theta$, this dispersion is marked as $\sigma_R^\ast$.
Figs.~\ref{Fig7.9} -- \ref{Fig7.13} show that in disc populations both
versions of the radial velocity dispersion are very similar, but in
halo populations $\sigma_R$ is larger than $\sigma_R^\ast$.

In bottom panels  we show with dashed lines isolines, $\rho=const$ and
with solid lines isolines 
$(\sigma_z)_0=const$, in the plane of $R,~z$-coordinates,  shown at
the bottom and the left border of the panel. Velocity dispersion
$\sigma_z$ was calculated within ranges $0 \le R \le 30$~kpc and
$0 \le z \le z_u$, where $z_u$ is the outer vertical limit of the
population.  We see that in disc populations lines
$(\sigma_z)_0=const$ are vertical, \ie the velocity dispersion at
given radial distance is constant.  In halo populations lines
$(\sigma_z)_0=const$ are similar to lines of constant density. A
similar picture is observed in the M31 galaxy, see
Fig.~\ref{Fig20.11}.

Now we discuss our results in some detail.

Velocity dispersions of disc populations have maxima at center and
decrease with the distance from the Galactic center.  In halo
populations velocity dispersions near the center have a moderate
minimum, a maximum not far from the center, and decrease at larger
distance.  The radial density gradient is zero at $R=0$, and increases
continuously with increasing distance for disc populations,
$\epsilon=0.02,~0.05$.  For halo populations the gradient has a
maximum at certain distance from the center, and slowly decreases at
larger distance.

A remarkable property is the form of isolines of vertical velocity dispersion
$\sigma_z=const$.  Outside central regions of the Galaxy, isolines of velocity
dispersion of flat and intermediate populations are almost
vertical, \ie\ velocity dispersions depend very weakly on $z$.
Available observational data support this picture. A different picture
was obtained by \citet{Innanen:1968aa}.  
Velocity dispersion isolines for his model 
\citep{Innanen:1966aa} were almost parallel to density isolines
$\rho=const$, \ie\ velocity dispersion $\sigma_z$ rapidly decrease
with the increase of $z$.  The reason for this discrepancy between
models is the Schmidt density law with a sharp boundary, accepted by
Innanen.  

On the $z=0$ surface, we can take into account corrections to
$\sigma_z$, using the gradient of the tilt of the velocity
ellipsoid. Applying equations from Chapter 11, we have in the Jeans
approximation
\be
Q^\star(\sigma_z^2)_0 = R^2\,C^2,
\label{eq7.4.1}
\ee
where $C$ is the dynamical Kuzmin constant, $(\sigma_z)_0$ — the
vertical velocity dispersion in Jeans approximation, and
\be
-Q^\star = R^2\left({\partial^2 \ln \sigma_z^2 \over \partial z^2} +
{\partial^2 \ln \rho \over \partial z^2}\right).
\label{eq7.4.2}
\ee
Taking into account the gradient of the tilt of the velocity ellipsoid,
we get
\be
Q\sigma_z^2 = R^2\,C^2,
\label{eq7.4.3}
\ee
where
\be
Q = Q^\star - q^\prime,
\label{eq7.4.4}
\ee
and $q^\prime$ is given by the Eq.~(\ref{eq11.3.3}). In calculations, we
used $Q^\star$ not from Eq.~(\ref{eq7.4.2}) but from
Eq.~(\ref{eq7.4.1}), since $(\sigma_z)_0$ is already known. Parameter
$q^\prime$ was calculated using Eq.~(\ref{eq11.3.3}). Results of
calculations are given in upper panels of Figs. \ref{Fig7.7} to
\ref{Fig7.13}.  We see that the tilt of velocity ellipsoid decreases
velocity dispersions slightly. Near the axis of the system $R=0$, this
factor increases the velocity dispersion. In this way, according to
the virial theorem, the mean value of the dispersion does not change.

The first hydrodynamical Eq.~(\ref{eq11.1.1}) is usually applied to
calculate the centroid velocity, $V_\theta$, from the velocity
dispersion,  $\sigma_R$.  The last quantity is calculated using the
definition equation
\be
k_z = {\sigma_z^2 \over \sigma_R^2},
\label{eq7.4.5}
\ee
from $\sigma_z$ and $k_z$.  The vertical dispersion can be found from
the second hydrodynamical Eq.~(\ref{eq11.1.2}), and $k_z$ from
Eq.~(\ref{eq11.2.17}) and Eq.~(\ref{eq11.2.2}).

This method of finding $\sigma_R$ gave satisfactory results only for
flat populations. The higher the velocity dispersion, the more the
calculated centroid velocity differs from the observed one. Starting
from some $\sigma_R$, the pressure component of the hydrodynamical 
equation becomes larger than the right part of the equation, and the
centroid velocity becomes imaginary.  This absurd result shows that
we have made an error in our calculations.

This hydrodynamical equation can be written in the form
Eq.~(\ref{eq11.1.4})
\be
V_\theta^2 - p\,\sigma_R^2 = V_c^2,
\label{eq7.4.6}
\ee
where the dimensionless coefficient $p$ has on the symmetry plane of
the galaxy the form Eq~(\ref{eq11.2.10}). Initially, we suspected that
the non-correct result was caused by the wrong calculation of the
coefficient $p$. However, a careful analysis brought us to the
conclusion that non-accuracy in the calculation of $p$ cannot be so
large to bring us to such an absurd result. It remains to check the
correctness of the determination of the dispersion $\sigma_R$. 

As written above, we calculated the dispersion $\sigma_R$ using
Eq.~(\ref{eq7.4.5}).  The vertical dispersion $\sigma_z$ is determined
accurately.  There remains to search the error in the coefficient
$k_z$.  It was calculated using the Kuzmin Eq.~(\ref{eq11.2.17}),
which was defined only for flat populations.

In the Solar neighbourhood, the dispersion $\sigma_R$ can be calculated
directly using Eq.~(\ref{eq7.4.6}), since $V_\theta$ can be found from
observations, and the coefficient $p$ calculated either from
observational data or from model. In the latter case, the
Eq.~(\ref{eq11.2.10}) is applied. The obtained dispersion $\sigma_R$ was
used to adjust $k_z$ using Eq.~(\ref{eq7.4.5}).

The relationship between the velocity dispersion and the centroid
velocity was determined using the mean velocity dispersion
$\sigma$. Thus, we need a relationship between the mean dispersion
$\sigma$ and the dispersion in the radial direction $\sigma_R$. It has the
form
\be
\sigma_z^2 = {3k_z \over 1+k_\theta + k_z}\,\sigma^2.
\label{eq7.4.7}
\ee
The coefficient $k_\theta$ was calculated from the relation
\be
\left({1-k_\theta \over 1-k_z}\right) = \left({1-k_\theta \over
    1-k_z}\right)_0,
\label{eq7.4.8}
\ee
where index 0 is for $k_\theta$ and $k_z$ on the plane of the
population. 

To determine the dispersion $\sigma_R$, another method can be used,
which applies the first hydrodynamical equation as a differential
equation for $\sigma_R$.  To solve this task, we need to express
$V_\theta$ as a function of $R$.  We are interested in solving this
equation for spherical populations, which have low rotational
velocities, $V_\theta \ll V$. Since $V_\theta$ is low, small
non-accuracies in this quantity do not influence our result. Thus we
accepted
\be
V_\theta(R) = \beta\,V_c(R),
\label{eq7.4.9}
\ee
where $\beta$ is a parameter, constant for a given population. Its
value can be estimated from kinematical data of populations near the
Sun.  The solution of the first hydrodynamical equation can be written
in the form
\be
\sigma_R^2(R) = \alpha \int_R^\infty K_R(R^\prime){\rho(R^\prime) \over
  \rho(R)} {L(R^\prime) \over L(R)} \dd{R^\prime},
\label{eq7.4.10}
\ee
where
\be
\alpha = 1 - \beta^2
\label{eq7.4.11}
\ee
and
\be
L(R) = \exp\left[-\int_R^{R1} p(R^\star) {\dd{R^\star} \over R^\star}\right].
\label{eq7.4.12}
\ee
Here $R1$ is a certain distance which later disappears, and
\be
p(R) = (1-k_\theta) + n_R(1-k_z),
\label{eq7.4.13}
\ee
where $n_R$ is expressed by Eq.~(\ref{eq11.2.12}). If we take $p=0$,
then $L=1$ and we get the velocity dispersion in Jeans
approximation. We calculated $\sigma_R$ for $p=0$ and for $p$
according to Eq~(\ref{eq7.4.13}).

Results of these calculations are shown in Figs.~\ref{Fig7.7} to 
\ref{Fig7.13}.  These results show, first, that the radial velocity
dispersion of populations is not constant as expected from the theory,
using the assumption that the ellipsoidal distribution of velocities
is fulfilled accurately. Furthermore, our results show that ratios of
velocity dispersions depend on dispersion. If the dispersion is
small, then $k_z$ approximates to the value given by Kuzmin
equation. When the dispersion increases, then $k_z$ also increases, see
Fig.~\ref{Fig7.14}.  Such a result is expected, since iso-surfaces of
densities approach  spheres, which in the limit 
 approach spheres,  determined by the escape velocity.

\section{Density distribution of populations}

For all test populations, we calculated spatial densities in meridional
surfaces $R, z$, projected densities $P$, and density gradients in
radial and vertical directions.

Consider first the density distribution in radial directions.  Density
gradients near the Sun,
\be
m= -{\partial \log \rho \over \partial R},
\label{eq7.5.1}
\ee
are close to values expected from observations
\citep{Kukarkin:1949aa,Parenago:1954aa, Blaauw:1965aa}. In central
regions of the Galaxy, the density gradient $m$ of the halo populations is much
larger than near the Sun.  This is in good agreement with results by
\citet{Baade:1958aa} for the Galaxy and
\citet{Sharov:1968aa,Sharov:1968ab} for M31.  This agreement
demonstrates that our test population model and its scale parameters
were chosen properly. 

The vertical density distribution is often accepted according to an
exponential law \citep{Kukarkin:1949aa,Parenago:1954aa}
\be
\rho(z)=\rho_0\,\exp^{-|z|/\beta},
\label{eq7.5.2}
\ee
where $\beta$ is a parameter, inversely proportional to the gradient
of density logarithm,
\be
\beta^{-1} Mod = l = - {\partial \log \rho \over \partial z}.
\label{eq7.5.3}
\ee
If the density behaves according to Eq.~(\ref{eq7.5.2}), then the
gradient $l$ should be constant.

Our calculations show that the gradient is not constant. This shows
that  Eq.~(\ref{eq7.5.2}) can be used only as the first
approximation. Density gradients found in this paper describe well
the observed distributions of stars.  We cannot expect a full
coincidence, since the observed populations of stars are not completely
homogeneous but consist of several close populations with slightly
variable properties, for instance, sums of several our close test populations.

\vskip 2mm
\hfill September 1971


\part{Methods to calculate spatial and hydrodynamical models of 
  regular stellar systems} 

\chapter{Classification of models. Conditions of physical
  correctness}\label{ch08}

The usual method to describe the structure of galaxies quantitatively
is to construct respective mathematical models. A review of models of
galaxies is given by \citet{Perek:1962aa}. A more detailed discussion
of modelling methods is made by \citet{Kutuzov:1968aa}, which forms
Chapter 8 of the original Thesis. In this English version, we describe
one aspect of the calculation of models — conditions of physical
correctness.

In general, models of galaxies can be characterised as
spatial/hydrodynamical, theoretical/empirical, general/detailed.
Theoretical models are devoted to explaining particular theoretical
properties, such as the model by \citet{Kuzmin:1952ac,Kuzmin:1954} to
explain the third integral of motions. Examples of empirical spatial
models are models of the Galaxy by \citet{Idlis:1956aa},
\citet{Schmidt:1956} and \citet{Einasto:1965aa}. An example of a
hydrodynamical model is the model of M31 by \citet{Einasto:1970ac}.

The conditions of physical correctness can be expressed in the
following way  \citep{Einasto:1969ab}:\\
(a) the spatial density $\rho(a)$ must be non-negative
and finite,
\be
0 \le \rho(a) <\infty;
\label{8.1}
\ee
\\
(b) the density should decrease with growing distance from the centre
of the system:
\be
G\{\rho(R)\} = \frac{\partial \ln\rho}{\partial\ln R} \le 0;
  \label{8.2}
\ee
\\
(c) the descriptive functions should not have breaks;\\
(d) some moments
of the mass-function should be finite:
\be
\mm{M}_i=\int_0^\infty{\mu(a)\,a^i \dd{a}} <\infty,
  \label{8.3}
  \ee
  where $\mu(a)=4\pi\epsilon\rho(a)\,a^2$ is the mass function;\\
  (e) the model should allow stable circular motions. In that case
  \be
  G\{F_R^0(R)\} =\frac{\partial \ln F^0_R}{\partial\ln R}> -1.
  \label{eq8.4}
  \ee
Here $F_R^0(R)=F_R(R)/\mm{M}$ is the normalised acceleration function
-- the ratio of radial acceleration of the model to the radial
acceleration of a mass-point model with the same mass $\mm{M}$. For a mass
point $F_R^0(R) \equiv 1$.

  Real stellar systems have finite dimensions and finite
  densities. Therefore, all moments of the mass function, $\mm{M}_i$,
  $i \ge -2$, are finite. But the requirement of the finiteness of all
  moments is too strict. Therefore, we suppose that only moments of the
  order, $-2 \le i \le 2$, must be finite. Moments $\mm{M}_{-1}$ and
  $\mm{M}_0$ determine the central potential and the mass of the
  model, respectively.  Moment $\mm{M}_1$ defines the effective
  radius of the model,  see Eq.~(\ref{eq7.3.3}).
  
\vskip 5mm
\hfill July 1969 
\chapter{Description functions and parameters}\label{ch09}

The problem of constructing an empirical model of a particular stellar
system consists of fixing description functions and determination of
their parameters. The aim of this paper is to discuss description
functions and parameters of stellar systems and their models in
general terms.  The full text of this Chapter is published by
\citet{Einasto:1968aa}. We here give a short summary of the main points.

We assume in galactic modelling that galaxies consist of a sum of
ellipsoidal components of axial symmetry around the $z$-axis, and that
all components have the identical symmetry plane $z=0$.  Components have
the vertical to radial axis ratio: $\epsilon = c/a$, where $c$ is the
minor semiaxis of the density ellipsoid, and $a$ is the major semiaxis
of the density ellipsoid: $a^2 = x^2+y^2+(z/\epsilon)^2$.

The main description functions for Galactic modelling are the following:\\
\hspace{1cm}$\rho(a)$ — spatial density of mass;\\
\hspace{1cm}$\mu(a)$ — mass function;\\
\hspace{1cm}$P(a)$ — projected mass density (definition below);\\
\hspace{1cm}$l(a)$ — spatial luminosity density;\\
\hspace{1cm}$L(R)$ — projected luminosity density;\\
\hspace{1cm}$V(R)$ — circular velocity.\\
In these equations, $a$ is the major semiaxis of the equal density
ellipsoid, and $R$ is the radial distance from the centre of the
galaxy.

Description functions are connected by the following equations: the mass function,
\be
\mu(a)= 4\pi\epsilon\rho(a)\,a^2,
\label{eq9.1}
\ee
the projected density function,
\be
P(A)= {1 \over 2\pi\,E}\int_A^{A^\circ} {\mu(a)\,\dd{a} \over a\,
    \sqrt{a^2-A^2}},
  \label{eq9.2}
  \ee
  here $A$ is the major semiaxis of the projected density,
  \be
  A^2 =X^2 + E^{-2}\,Y^2,
  \label{eq9.3}
  \ee
  where $E=\epsilon/A$ is the axial ratio of the projected density
  ellipsoid.  The projected and spatial density ellipsoid axial ratios are
  related as;
  \be
  E^2 = cos^2i + \epsilon^2\,sin^2i,
  \label{eq9.4}
  \ee
  where $i$ is the inclination angle of the symmetry axis of the
  galaxy to the line of sight.
  
Spatial mass and luminosity densities are related as:
  \be
  \rho(a)= f\,l(a),
  \label{eq9.5}
  \ee
  where $f$ is the mass-to-light ratio of a given component.

  The circular velocity is related with
  the mass function as follows:
  \be
   V(R)^2 = G \int_0^R \frac{\mu (a)~da }{ \sqrt{R^2 - a^2e^2}},
   \label{eq9.6}
   \ee
   where $G$ is the gravitational constant, and $e^2 = 1 -
\epsilon^2$. In these equations, we assume that the stellar system can
be divided into a number of homogeneous components, each with its axial
ratio $\epsilon$ and mass-to-light ratio
$f=\rho(a)/l(a)=P(A)/L(A)$.

The descriptive parameters of stellar systems and their models can be
divided into three groups:

a) the model-parameters —  parameters in the analytical
expressions of descriptive functions in the special models
of stellar systems;

b) the gross-parameters — integral quantities, characterising the
structure of a stellar system as a whole; 

c) the local galactic parameters — values of the galactic
descriptive functions or their combinations for the
surrounding of the Sun.

The model- and gross-parameters can be divided into three
kinds of parameters: scale-, concentration-, and flattening
parameters, the latter two kinds of parameters can be called
together as the structural ones. There are two scale parameters, one
of them determines the scale of the model in the 
space, the other the scale of the density (or the mass). The
concentration and flattening model-parameters determine the
concentration of mass to the centre of the model and the form
of the equidensity surfaces.

The gross-parameters are defined by means of the moments of the
mass-function, $\mu(a)$. It is proposed to use the effective radius, $R_e$, 
and the mass, $\mm{M}$, of the system as scale gross-parameters. 
A dimensionless concentration gross-parameter is the
index $\nu$ in the generalised exponential model, Eq.~(\ref{eq5.4}).
Dimensionless flattening parameter is the axial ratio of equidensity
ellipsoid $\epsilon$.

In practical use, it is common to accept mass functions of components
using suitable approximation density profiles, and to find the
circular velocity from Eq.~(\ref{eq9.6}) by summing the
contribution of all components.

  \vskip 5mm
\hfill April 1967
\chapter{Calculation  of models of spatial structure}\label{ch10}

In previous papers in this series, I discussed the classification of
stellar system models, the description functions and parameters of
models, and the type of relations between the description
functions. This article is devoted to the methodology of model
building. In this case, the model of a stellar system is defined as a
collection of mass or luminosity distribution functions, constructed
for a particular stellar system and given in a numerical form. We assume
that the stellar system has axial and planar symmetry, and that it
consists of a finite number of physically homogeneous 
spheroidal components.
The full text of the Chapter is published
by  \citet{Einasto:1968ad}. Here we give a short summary.

The basic input data to find a composite model of a galaxy are luminosity
distributions of components, and in some cases rotation velocities of
components.  The total luminosity distribution of the galaxy is the
sum of distributions of its components:
\be
L(X,Y) = \sum_{k=1}^n{L_k(A_k)},
\label{eq10.1}
\ee
where $n$ is the number of components,
\be
A_k^2 = X^2 + E_k^{-2}\,Y^2,
\label{eq10.2}
\ee
is the major semiaxis of the projected equidensity ellipse, $X$ and
$Y$ are rectangular projected coordinates, and
\be
E_k^2 = cos^2\,i + \epsilon_k^2\,sin^2\,
\label{eq10.3}
\ee
is the axial ratio of the projected density ellipsoid of the
component, and $i$ is the inclination angle of the symmetry axis of the
galaxy to the line of sight. The projected luminosity distribution is
related to the spatial mass function as follows:
\be
L_k(A_k)= {1 \over 2\pi\,E_k}\int_{A_k}^{A^\circ} {\mu_k(a)\,\dd{a} \over f_k\, a\,
  \sqrt{a^2-A_k^2}},
\label{eq10.4}
\ee
where
\be
\mu_k(a)= 4\pi\epsilon_k\rho_k(a)\,a^2,
\label{eq10.5}
\ee
is the mass function of the component $k$, $f_k$ is the mass-to-light
ratio of the components, and $\epsilon_k$ is the
axial ratio of its spatial density ellipsoid.

Similarly, the total velocity function is a sum of circular velocity functions
of components:
\be
V(R)^2 = \sum_{k=1}^n{V_k(R)^2},
\label{eq10.6}
\ee
where the circular velocity function of components is:
\be
V_k(R)^2 = G \int_0^R \frac{\mu_k (a)~da }{ \sqrt{R^2 - a^2e_k^2}}.
\label{eq10.7}
\ee

The task of modelling a stellar system is to calculate from the
available observational data all necessary functions and parameters to
describe a stellar system.  The modelling procedures vary widely
depending on the availability of observational material, on the way of
solving the main integral equations (\ref{eq10.4}) and (\ref{eq10.7}),
and the resulting form of model representation.  The procedures of
extrapolating and interpolating the observed description functions are
quite different.  All these aspects allow further methodological
refinement and specification in the classification of the models,
proposed by \citet{Einasto:1968aa}.

Let us now consider in a little more detail some 
aspects of stellar system modelling.

{\bf A.} It is well known that in the case of stellar systems visible from
the outside, observations allow to determine:

1) the projected luminosity distribution function
$L_S(X, Y)$ in photometric system $S$ (it is assumed that
$L_S(X, Y)$ is corrected for the effects of light absorption);

2) the rotational velocity of selected subsystems $V_\theta$, from which,
under known assumptions, it is possible to calculate the circular 
velocity  $V(R)$;

3) the dispersion of velocities of stars in the system;

4) the integral spectrum of the system, which allows to find the
stellar composition of the system for the given
 velocity function and the ratio of mass to light $f$.

 The most complete model can be constructed if all these data are
 known.  In this case one can speak about a synthetic model of the
 stellar system. If only the projected density is known 
 from observations, one can construct a photometric
 model of the system.  In case only the spectral data are given (points
 2-4), then a dynamical  model can be constructed.

 A special case in the stellar systems modelling is the
 construction of a model of our Galaxy, because neither the $L(A)$
 function nor the $V(R)$ function can be found directly from
 observations.  The velocity function can be calculated indirectly
 from the differential rotational velocity of the subsystems $U(x)$,
 using the formula
\be
V(x)=U(x)+x\,V_0,
\label{eq10.9}
\ee
where $V(x)$ is circular velocity, $x = R/R_0$, $V_0$ is circular
velocity in the vicinity of the Sun, and $R_0$ is the distance of the
Sun from the centre of the Galaxy.  The behaviour of the function
$L(A)$ can only be judged from the analogy with other galaxies. On the
other hand, it is possible to determine circumsolar values for a
number of other description functions.

The coupling formulas (\ref{eq10.4}) to (\ref{eq10.7}) are integral
equations in respect to the basic
description function $\mu_k(a)$.  The integral equations can be solved
numerically, in this case we obtain the description functions also in
a numerical form, and the result is a numerical model.
The integral equations can also be solved parametrically.
By setting the analytical form of one description function,
we can calculate all other description functions.
The model parameters can be found by approximating the observed
functions by the corresponding analytic description functions.

{\bf B.} The construction of the model has other important aspects. It
is well known that the functions $L(X,Y)$ and $V(R)$ cannot be deduced
from observations up to the outer limit of the system.  The peripheral
regions of the system are so weak that they are lost against the
general background of the sky. When constructing a model of the whole
stellar system, the description functions are therefore
extrapolated. The extrapolation can be done numerically, as done in
the model of the Galaxy by \citet{Kuzmin:1952aa}, following the rules of
physical correctness, see Chapter 8. The other possibility is to use
analytical expression for the main description function the spatial
density $\rho(a)$ or 
some other description function, again following rules of physical
correctness.

In most models of galaxies, the mass density function of components,
$\rho_k(a)$, is given by a suitable analytical expression, and the
construction of the model reduces to the determination of parameters
of the density function for all components.

\vskip 5mm
\hfill September 1967 
\chapter{Calculation of  hydrodynamical models\label{ch11}}

The theoretical foundation of the hydrodynamics of stellar systems was
discussed among others by
\citet{Kuzmin:1952aa,Kuzmin:1952ac,Kuzmin:1954, Kuzmin:1956aa,
  Kuzmin:1956ca, Kuzmin:1962ac,
  Kuzmin:1963aa, Kuzmin:1963ca, Kuzmin:1965cc}.  In this Chapter, we
analyse the application of the hydrodynamical theory to find practical
solutions to calculating models of real stellar systems.  This analysis
was made with the goal to calculate hydrodynamical models of the
Andromeda galaxy M31  \citep{Einasto:1970ac} and our own Galaxy
(Chapter 7).  The Chapter was published by \citet{Einasto:1970aa}.

\section{Equations to calculate kinematical functions}
\bigskip
As known, gravitational acceleration can be calculated from the mass
distribution \citep{Schmidt:1956, Perek:1962aa}. Components of
gravitational acceleration in radial and vertical directions $K_R = -
\partial\Phi/\partial R $ and $K_z = -\partial\Phi / \partial z$
($\Phi$ --- gravitational potential) are related to kinematic
functions according to hydrodynamical equations \citep{Kuzmin:1965cc,
  Einasto:1969ab}:
\be
  \frac{1}{R} \left( \sr -\st \right) + \frac{1}{\rho} \pdif{}{R} \left( \rho \sr \right) + \frac{1}{\rho} \pdif{}{z} \left[ \rho\gamma \left( \sr - \sz \right)\right] - \frac{V^2_\theta}{R} = -K_R , 
  \label{eq11.1.1}
\ee
\be
  	\frac{1}{R} \gamma \left( \sr - \sz \right) + \frac{1}{\rho} \pdif{}{R} \left[ \rho\gamma \left( \sr - \sz \right)\right] + \frac{1}{\rho} \pdif{}{z}\left( \rho\sz\right) = -K_z .
  	\label{eq11.1.2}
\ee
  In these equations $\sigma_R$, $\sigma_{\theta}$, $\sigma_z$ are velocity dispersions in cylindrical galactocentric coordinates $R, \theta , z$; $V_{\theta}$ is the centroid velocity; $\rho$ is the matter density and 
\be 
  	\gamma = \frac{1}{2} \tan 2\alpha , \label{eq11.1.3}
\ee
  where $\alpha$ is the inclination angle of the major axis of the velocity ellipsoid with respect to the galactic symmetry plane. It is assumed that the major axis of the velocity ellipsoid lies in the meridional plane of the galaxy, and the remaining components of the centroid velocity are $V_R = V_z = 0$.

Calculating the necessary derivatives, Eqs.~(\ref{eq11.1.1}) and (\ref{eq11.1.2}) can be written as
\be 
V_{\theta}^2 - p\sr = R K_R =V_c^2, \label{eq11.1.4}
\ee 
and
\be 
\frac{1}{\rho} \pdif{(\rho\sz )}{z} + q \frac{\sz}{R} = -K_z , \label{eq11.1.5}
\ee 
where
\be 
p = \left( 1-k_\theta \right) + G_R\{\rho\} + G_R\{\sr\} + \frac{R}{z}\gamma\left( 1-k_z \right) \left[ G_z\{\rho\} + G_z\{\gamma\} + G_z\{ 1-k_z\} \right] , \label{eq11.1.6}
\ee
and
\be 
q = \gamma \left( \frac{1}{k_z} -1 \right) \left[ 1+ G_R\{\rho\} + G_R\{\gamma\} + G_R\{ \sr - \sz\}\right] . \label{eq11.1.7}
\ee 
In these equations
\be 
k_\theta = \frac{\st}{\sr} , ~~~~~~ k_z = \frac{\sz}{\sr}, \label{eq11.1.8}
\ee 
and $G\{ ~\}$ is the logarithmic derivative, e.g.
\be 
G_R\{\rho\} = G\{\rho(R)\}=\frac{\partial \ln\rho}{\partial\ln R} . \label{eq11.1.9}
\ee 

Equations (\ref{eq11.1.4}) and (\ref{eq11.1.5}) include five unknown
kinematical functions: $\sigma_z$, $V_\theta$, $k_\theta$, $k_z$,
$\gamma$. To calculate these functions, we have only two equations at
present, thus the system of hydrodynamical equations is not closed. To
solve the problem, one needs to have three additional independent
relations between these unknown functions. It is convenient to give
these additional relations for $k_\theta$, $k_z$, and $\gamma$, which
determine the
shape and the orientation of the velocity ellipsoid. In this case
Eq.~(\ref{eq11.1.5}) allows to calculate the dispersion $\sigma_z$, 
giving the scale of the velocity dispersion, and Eq.~(\ref{eq11.1.4})
allows to calculate the centroid velocity $V_\theta$, giving the shift of the
velocity ellipsoid with respect to the local standard of rest.

The calculation of the hydrodynamical model of a galaxy reduces thus to
the problem of finding equations for auxiliary kinematical functions
$k_\theta$, $k_z$, and $\gamma$. In papers about stellar dynamics, this
problem has not yet been discussed with sufficient thoroughness. For this
reason, we first give a review of various  solutions of the problem.

\section{Methods to close the system of hydrodynamical equations}

\subsection{The method by Jeans-Oort}

From the classical stellar dynamics, the following relations result for ratios $k_z$ and $k_\theta$:
\be 
k_z=1, \label{eq11.2.1}
\ee 
\be 
k_\theta = \frac{-B}{A-B}, \label{eq11.2.2}
\ee 
where $A$ and $B$ are Oort kinematical parameters. Expressing $A$ and
$B$ in terms of the centroid velocity $V_\theta$ and its radial
gradient, we have 
\be 
k_\theta = \frac{1}{2} \left[ 1+ G_R\{ V_\theta \} \right] .  \label{eq11.2.3}
\ee 

In this way Eqs.~(\ref{eq11.1.6}) and (\ref{eq11.1.7}) will have the form
\be 
p= \left( 1-k_\theta \right) + G_R\{ \rho \} + G_R \{ \sr\} , \label{eq11.2.4}
\ee 
\be 
q=0, \label{eq11.2.5}
\ee 
and the quantity $\gamma$ vanishes altogether  from the system of
corresponding equations. The first hydrodynamical equation can be
solved by successive approximations ($V_\theta$ contains also in
$p$) by taking $p=0$ as a zero-approximation. The second equation has
a simple solution 
\be 
\sz = \rho^{-1} \int_z^{\infty} K_z\, \rho\, \dd{z} , \label{eq11.2.6}
\ee
being appropriate to be called as the Jeans solution. This method has been
used  by \citet{Oort:1940aa}  and \citet{Innanen:1967aa}. 

\subsection{The method by Innanen and Kellett}

In spherical stellar systems (e.g. in globular clusters), the major
axis of the velocity ellipsoid is directed toward the centre of the
system \citep{Michie:1961aa,Agekyan:1969aa}, and in this case  
\be 
\tan\alpha = \frac{z}{R} . \label{eq11.2.7}
\ee
In elliptical and spiral galaxies, the velocity ellipsoid is also
tilted with respect to the symmetry plane, but the corresponding
inclination angle is different. Following \citet{Oort:1965aa} and 
\citet{Innanen:1968aa} we get 
\be 
\gamma = \frac{z}{R} . \label{eq11.2.8}
\ee 

From observations it is known that $\sigma_z \ne \sigma_R$,  and thus
Eq.~(\ref{eq11.2.1}) is not valid. For this reason, Innanen and
Kennett used the relation 
\be 
k_z = \mu^{-2} , \label{eq11.2.9}
\ee
where $\mu$ is a constant. Ascribing different values to $\mu$ with
the help of successive approximations, authors calculated $\sigma_z$ by
integrating Eq.~(\ref{eq11.1.5}). As a zero approximation the velocity
dispersion was calculated from Eq. (\ref{eq11.2.6}). Applying the method
to a three-component model of the Galaxy, authors found that in case of
 flat and intermediate subsystems the method of successive
approximations converges for $\mu < 3$. In the case of the halo the
process converges only for $\mu < 1.2$. 

As the selection of the parameter $\mu$ remained open, and the
centroid velocity was not calculated, Innanen and Kennett did not
finalise the problem of solving the hydrodynamical equations. In
addition, Eq.~(\ref{eq11.2.8}) can be used only as a rather rough first
approximation (see below).  

\subsection{The method based on the Kuzmin theory}

The problem of solving the hydrodynamical equations was studied also
by G. G. Kuzmin. However, he studied the problem for a flat component
only. As we see below, the method can be generalised for a
spatial mass distribution.

In case of $z=0$, the velocity ellipsoid is not tilted with respect to
the galactic plane, and $\gamma = 0$, but
$\partial\gamma /\partial z \ne 0$. Thus, the Eq.~(\ref{eq11.1.6}) for
the parameter $p$ will have the form
\be
p=\left( 1-k_\theta \right) +
n_R \left(1-k_z\right) + G_R\{ \rho\} + G_R \{\sr\}
, \label{eq11.2.10} \ee where \be n_R=R\left(
  \pdif{\gamma}{z}\right)_{z=0} . \label{eq11.2.11}
\ee

On the basis of his theory of the third integral of motion of stars as
a quasi-integral, \citet{Kuzmin:1962ac}  demonstrated that 
\be 
n_R= - \frac{1}{4} G_R\{ C_c^2\} \left[ 1+ \frac{B_c (A_c -B_c)}{C_c^2} \right]^{-1} , \label{eq11.2.12}
\ee 
where $A_c$, $B_c$, $C_c$ are the Oort-Kuzmin dynamical parameters, \ie 
\be 
\left( \ppdif{\Phi}{z} \right)_{z=0} = -C_c^2 , \label{eq11.2.13}
\ee 
and $A_c$ and $B_c$ correspond to the circular velocity. They are
related with the gravitational potential according to equations 
\be 
\ba{lll} 
\left(\ppdif{\Phi}{R} \right)_{z=0} & = & (A_c - B_c ) (3A_c -B_c), \\
 
 \left( \frac{1}{R} \pdif{\Phi}{R}\right)_{z=0} & = &  -(A_c - B_c)^2.  
 \label{eq11.2.14} 
 \ea 
 \ee

The Poisson equation has in  $A_c$, $B_c$, $C_c$ terms for $z=0$ the form
\be 
4\pi\, G \rho_t = C_c^2 - 2(A_c^2 - B_c^2 ),
\label{eq11.2.15}
\ee 
where $\rho_t$ is the total matter density. Using the Poisson
equation in this form, and taking into account that $A_c, B_c \ll
C_c$, \citet{Kuzmin:1962ac} derived an approximate formula 
\be 
n_R = -\frac{1}{4} G_R\{ \rho_t \}. \label{eq11.2.16}
\ee 

It is not possible to calculate the dispersion ratio $k_z$ on the
basis of classical stellar dynamics, since the form of the velocity
ellipsoid in $z$-direction is determined by irregular forces, which
were not taken into account in the classical theory. The theory of
irregular forces gives us the following relation between the velocity
dispersion ratios \citep{Kuzmin:1961aa,Kuzmin:1963ca} 
\be 
k_z^{-1} = 1 + k_\theta^{-1} . \label{eq11.2.17}
\ee 
This equation was derived for the case of flat subsystems, and its validity
in the general case is not clear yet. It allows to calculate
$k_z$ when $k_\theta$ is known. The quotient $k_\theta$ can be
calculated in the case of flat components from the Lindblad's formula
(\ref{eq11.2.3}). 

\section{Solution of  hydrodynamical equation for $z=0$}

Let us discuss now in a somewhat more detail the questions related
with solving the hydrodynamical equations in the galactic plane. 

All three auxiliary kinematical functions $k_\theta$, $k_z$ and $n_R$
can be calculated from the mass distribution model. To calculate
$k_\theta$ we may take $V_\theta \sim V_c$, as a first approximation. 

The first hydrodynamical equation allows to calculate quite easily
the centroid velocities $V_\theta$ of galactic components. The density
gradient $G_R\{\rho\}$ is given by the mass distribution model;
velocity dispersions and their gradients can be calculated by solving
the second hydrodynamical equation with respect to $\sigma_z$, assuming
that auxiliary functions $k_\theta$, $k_z$, $n_R$ are known. Thus, the
problem reduces to the solution of the second hydrodynamical 
equation. 

In the galactic plane, the second hydrodynamical equation (\ref{eq11.1.2})
reduces to identity $0\equiv 0$. Hence, let us take a derivative from
the equation with respect to $z$ at $z=0$. This gives us 
\be 
Q\,\sz = R^2 C_c^2 , \label{eq11.3.1}
\ee 
where
\be 
Q = - \left( q' + R^2 \ppdif{\ln\rho}{z} + R^2 \ppdif{\ln\sz}{z} \right), \label{eq11.3.2}
\ee 
while
\be 
q' = n_R \left( \frac{1}{k_z} -1\right) \left[ G_R\{\rho\} + G_R\{ n_R\} + G_R\{ 1-k_z\} \right] . \label{eq11.3.3}
\ee 
Equations similar to (\ref{eq11.3.1})--(\ref{eq11.3.3}) were derived
by \citet{Kutuzov:1964aa}. However, they were not used, as it was not
known how to calculate $\partial^2\ln\sz / \partial z^2$. 

In the case of ellipsoidal density distribution, there exists a relation
\be 
R^2 \ppdif{\ln\rho}{z} = \epsilon^{-2}_\rho G_R\{\rho\} , \label{eq11.3.4}
\ee 
where $\epsilon_\rho$ is the ratio of the semiaxis of isodensity surfaces. 
By making a similar assumption in case of $\sz$, we find
\be 
R^2 \ppdif{\ln\sz}{z} = \epsilon^{-2}_\sigma G_R\{\sz\} , \label{eq11.3.5}
\ee
where $\epsilon_\sigma$ is the ratio of semiaxis of isosurfaces of
dispersions $\sz$. The gradient $G_R \{ \sz \}$ can be calculated from
Eq.~(\ref{eq11.3.1}). Ratio $\epsilon_\sigma$ can be calculated, if we
know also the distribution of the dispersions $\sz$ in the system axis
$R=0$. 

In the case of spherical subsystems, all three terms in the formula for $Q$
have a similar order of magnitude. In the case of intermediate and flat
subsystems, the second term dominates, and in the last case all 
remaining terms are even negligible
\citep[see][]{Kuzmin:1952ab}. Thus, for flat subsystems we have 
\be 
C=C_c , \label{eq11.3.6}
\ee 
where
\be 
C=\sigma_z /\zeta \label{eq11.3.7}
\ee 
is the Kuzmin's kinematical parameter, and
\be 
\zeta^{-2} = -\partial^2\ln\rho /\partial z^2 . \label{eq11.3.8}
\ee 

The second hydrodynamical equation allows also to calculate the mean
velocity dispersion in the symmetry plane of the galaxy. 

According to the definition
\be 
K_z = -\int _0^z \ppdif{\Phi}{z} \dd z. \label{eq11.3.9}
\ee 
Using the Poisson equation and neglecting dependences of $\partial\Phi
/\partial R$ and $\partial^2\Phi / \partial R^2$ on $z$ we have 
\be 
K_z = 4\pi G \int_0^z \rho_t\, \dd z + 2(A_c^2 - B_c^2) \, z. \label{eq11.3.10}
\ee 
We calculate the velocity dispersion within the Jeans approximation, \ie by
neglecting the tilt of the velocity ellipsoid outside the plane of
the galaxy. Substituting (\ref{eq11.3.10}) into (\ref{eq11.2.6}) we
derive 
\be 
\left( \rho_t \overline{\sz}\, \right)_0 = \frac{\pi GP^2}{2} \left( 1+ \frac{A_c^2 - B_c^2}{\pi G\rho_t}\frac{\overline{z}}{z_e} \right) ,\label{eq11.3.11} 
\ee 
where $P$ is the total projected matter density, $z_e = P/2\rho_t$ is
the effective half-thickness of the galaxy, and 
\be 
\overline{z} = \frac{\int_0^{\infty} \rho_t z\, \dd z}{\int_0^{\infty} \rho_t\,\dd z} . \label{eq11.3.12}
\ee 
Equation (\ref{eq11.3.11}) cannot be used in the central regions of the 
galaxy, where it is not justified to neglect dependences of
$\partial\Phi /\partial R$ and $\partial^2\Phi / \partial R^2$ on
$z$. Equation (\ref{eq11.3.11}) was derived by \citet{Kuzmin:1956ca},
and the corresponding correction term was calculated for a particular
galaxy model. 

The first hydrodynamical equation had been used by us earlier to estimate
the differences between the centroid velocity and the circular
velocity for various galactic subsystems \citep{Einasto:1961aa}. The
second hydrodynamical equation was used only in its simplest form
(\ref{eq11.3.6}) to estimate the value of the dynamical parameter
$C_c$ \citep{Kuzmin:1952ab,Kuzmin:1955aa}, but also in the form 
(\ref{eq11.3.11}) to calculate the mean velocity dispersion of stars
\citep{Kuzmin:1956ca}. In its complete form, the method of solving the
hydrodynamical equation is used here for the first time.

\section{Solution of hydrodynamical equations for $z\ne 0$}

 We saw above that there is no  satisfactory method to solve the
 hydrodynamical equations in case of $z\ne 0$ yet. In the present Section,
 we propose possible solutions to the problem by using the theory of
 the quadratic third integral of motion of stars, and ellipsoidal
 distribution of their velocities.
 
 At any point of the space $(R, z)$, the direction of the axis of the
 velocity ellipsoid defines an orthogonal system of coordinates. Moving
 along the axis of the velocity ellipsoid in case of axial and plane
 symmetry, we have a family of three orthogonal surfaces called
 according to \citet{Eddington:1915aa} principal velocity surfaces. If
 we accept the existence of the quadratic third integral of motion of
 stars, then these surfaces are  meridional planes and confocal
 ellipsoids and hyperboloids. As \citet{Eddington:1915aa} limited his
 analysis with a Schwarzschild velocity distribution, this was first 
 derived in general form by \citet{Kuzmin:1952ac}. Designating
 corresponding curvilinear coordinates as $x_i$, we have the following
 relations between $x_i$ and  cylindrical coordinates $R$,
 $\theta$, $z$ \citep{Kuzmin:1952ac}
 \be
 \frac{R^2}{x^2-z_0^2} +
 \frac{z^2}{x^2} = 1, ~~~~~ x_3 = \theta,
 \label{eq11.4.1}
 \ee
 where
 \be
 x^2 = \left\{ \ba{l}
   x_1^2 \ge \, z_0^2\\
   x_2^2 \le \, z_0^2.
   \ea \right.
 \label{eq11.4.2}
 \ee
 In these
 formulae $z_0$ is a constant, corresponding to common foci of
 ellipsoids and hyperboloids. They lie on the galactic axis at distances
 $z=\pm z_0$ from the centre.
 
 From the results presented above, it follows that one axis of the
 velocity ellipsoid coincides with the axis $V_\theta$ in cylindrical
 coordinates, and the inclination angle of the other axis with respect to the
 plane can be given as \citep{Kuzmin:1952ac} 
 \be 
 \gamma = \frac{Rz}{R^2 + z_0^2 -z^2} . \label{eq11.4.3}
 \ee
 
 Let us now look what expressions can be derived for velocity
 dispersions from the theory of the quadratic third integral. 
 
 It follows from the Jeans theorem that the phase spatial density
 depends on velocities and coordinates only through integrals of
 the motion of stars \citep{Kuzmin:1952ac}: 
 \be 
 \ba{lll}
 I_1 & = & v_1^2 + v_2^2 + v_\theta^2 - 2\Phi,\\
 I_2 & = & R v_\theta , \\
 I_3 & = & \left( \frac{x_2}{z_0}\right)^2 v_1^2 + \left( \frac{x_1}{z_0} \right)^2 v_2^2 + \left( \frac{x_1 x_2}{z_0^2} \right)^2 v_\theta^2 - 2\Phi^{\ast} ,
 \ea
 \label{eq11.4.4}
 \ee 
 where $v_i$ are velocity components along the main axis of the
 velocity ellipsoid, and $\Phi^{\ast}$ is a function, related to the
 gravitational potential $\Phi$. We assume that the velocity distribution
 is ellipsoidal with respect to all $v_i$. In this case, it is
 necessary that the phase density is a linear function of $I_1$ and
 $I_3$, and a quadratic function of $I_2$ 
 \be 
 s = a_1 I_1 + a_2 I_3 - 2 \frac{b_1}{z_0} I_2 +
 \frac{b_2}{z_0^2}I_2^2 .
 \label{eq11.4.5}
 \ee 
 In this case, the phase density is a function of velocity components
 according to a quadratic expression $(x_1 x_2 = \pm z\, z_0)$ 
 \be
 \ba{ll}
 s =&  \left[ a_1 + a_2 \left(\frac{x_2}{z_0}\right)^2 \right] v_1^2 +
 \left[ a_1 + a_2 \left( \frac{x_1}{z_0}\right)^2 \right] v_2^2 +\\
& \left[ a_1 + a_2 \left( \frac{z}{z_0}\right)^2 + b_2 \left(
     \frac{R}{z_0}\right)^2 \right] \left[ v_\theta -
   \overline{v_\theta} \right]^2 ,
 \ea
 \label{eq11.4.6}
 \ee
 where
 \be 
 \overline{v_\theta} = V_\theta = \frac{b_1 z_0 R}{a_1z_0^2 + a_2 z^2
   + b_2 R^2 } .
 \label{eq11.4.7}
 \ee 
 In this case, for the axial ratios of the velocity ellipsoid we find
 \be 
 k_{12} = \frac{\sigma_2^2}{\sigma_1^2} = \frac{a_1 z_0^2 + a_2
   x_2^2}{a_1 z_0^2 + b_2 x_1^2} ,
 \label{eq11.4.8}
 \ee 
 \be 
 k_{13} = \frac{\sigma_3^2}{\sigma_1^2} = \frac{a_1 z_0^2 + a_1
   x_2^2}{a_1 z_0^2 + a_2 z^2 + b_2R^2} .
 \label{eq11.4.9}
 \ee 
 Velocity dispersions in cylindrical coordinates are
 \be 
 \ba{lll}
 \sr & = & \sigma_1^2 \cos^2\alpha + \sigma_2^2 \sin^2\alpha ,  \\
 \sz & = & \sigma_1^2 \sin^2\alpha + \sigma^2_2 \cos^2\alpha ,  \\
 \st & = & \sigma_3^2 ,
 \ea
 \label{eq11.4.10}
 \ee 
 where $\alpha$ is the inclination angle of the major axis of
 the velocity ellipsoid with respect to the plane $z=0$. From
 (\ref{eq11.4.10}) we have 
 \be 
 k_z = \frac{\sin^2\alpha + k_{12} \cos^2\alpha}{\cos^2\alpha + k_{12},
   \sin^2\alpha}
 \label{eq11.4.11}
 \ee 
 and
 \be 
 k_\theta = \frac{k_{13}}{\cos^2\alpha + k_{12} \sin^2\alpha} .
 \label{eq11.4.12}
 \ee 
 
 In a special case of the galactic plane
 \be 
 \ba{llllll}
 x_1^2 & = & R^2+z_0^2, ~~~~~ & k_{12} & = & k_z \\
 x_2^2 & = & 0,               & k_{13} & = & k_\theta,
 \ea\label{eq11.4.13}
 \ee 
 giving with the help of (\ref{eq11.4.8}) and (\ref{eq11.4.9})
 \be 
 k_z (R,0) = \frac{a_1z_0^2}{(a_1 + a_2) z_0^2 + a_2 R^2}, \label{eq11.4.14}
 \ee 
 \be 
 k_\theta (R, 0) = \frac{a_1z_0^2}{a_1z_0^2 + b_2 R^2} . \label{eq11.4.15}
 \ee 
 Using the Kuzmin formula (\ref{eq11.2.17}) we see that $a_1 = a_2 = b_2 = a$. Therefore ($b = b_1/a$),
 \be 
 V_\theta = \frac{b z_0 R}{z_0^2 + z^2 + R^2} , \label{eq11.4.16}
 \ee 
 \be 
 k_{12} = \frac{z_0^2 + x_2^2}{z_0^2 + x_1^2}, \label{eq11.4.17}
 \ee 
 \be 
 k_{13} = \frac{z_0^2 + x_2^2}{z_0^2 + z^2 + R^2}. \label{eq11.4.18}
 \ee 
 
 Equations (\ref{eq11.4.7})--(\ref{eq11.4.9}) were derived already by
 \citet{Eddington:1915aa}, and
 Eqs.~(\ref{eq11.4.14})--(\ref{eq11.4.18}) by \citet{Idlis:1969aa},
 although in a somewhat different form and assuming a Schwarzschild
 velocity distribution.

 \section{Solution of  hydrodynamical equations for  $z\ne
   0$; a general case} 
 
 It was demonstrated already by \citet[][p.~47]{Eddington:1915aa} that
 the assumption of the Schwarzschild velocity distribution leads to an
 internal contradiction ---  it is not possible to find a mass
 distribution satisfying simultaneously the Poisson equation and an
 equation, resulting from the calculations of the phase density. It is
 not difficult to see that there will be a similar contradiction when
 assuming the existence of a precise quadratic third integral of
 the motion of stars, and an ellipsoidal velocity distribution. In the present
 Section, we study the possibility to calculate auxiliary kinematic
 functions, starting from the quadratic third integral as a
 quasi-integral, and an approximately ellipsoidal velocity
 distribution. 
 
The assumption about the symmetrical velocity distribution in $v_\theta$
 direction is obviously in contradiction with observations. Thus, when
 we calculate now the ratio of dispersions $k_\theta$, we do not use
 the equations resulting from the ellipsoidal velocity
 distribution. Instead, we start from the equation of micromotions (a
 term introduced by \citet{Kuzmin:1965cc}), giving us 
 \be
 \ba{ll}
 2k_\theta =& 1+ G_R\{ V_\theta\} + \gamma (1-k_z)\frac{R}{z} G_z\{
 V_\theta\} + \frac{R}{\rho\sr V_\theta} \pdif{(\rho\overline{v_R^2
     v_\theta})}{R} +\\
& \frac{2\overline{v_R^2v_\theta} - \overline{v_\theta^3}}{\sr
  V_\theta} + \frac{R}{\rho\sr V_\theta} \pdif{(\rho\overline{v_R v_z
    v_\theta})}{z} .
\label{eq11.5.1}
\ea
 \ee 
Near the plane $z=0$, the first two terms on the right side of
(\ref{eq11.5.1}) are dominating, and we have the usual Lindblad's
formula (\ref{eq11.2.3}). In order to use this equation in a general case, it is
necessary, first, to analyse the vertical gradient of $v_\theta$, and
also third moments of velocity components, being beyond the scope of
the present paper.  
 
With respect to $v_1$ and $v_2$ axis, as the first approximation,
velocity distribution can be assumed to be ellipsoidal. When we
derived the expression for the ratio of dispersions $k_z$ in Section~4, probably a
weak point was the assumption that the third integral of
motion is a precise integral. If we admit that the integral is a
quasi-integral only, it is necessary in its expansion to take into
account also higher-order terms. In this case, coefficients of terms
$v_1^2$ and $v_2^2$ in Eq.~(\ref{eq11.4.6}) are no longer linear
functions of $x_2^2$ and $x_1^2$, respectively, but have a more
general form. We expect that the first coefficient still is a function
of $x_2^2$ only, but the second coefficient will be a function of
$x_1^2$ in a more general form of $f(x^2)$. The ratio of dispersions
$k_{12}$ in this case will have the form 
 \be 
 k_{12}=\frac{f(x_2^2)}{f(x_1^2)} .\label{eq11.5.2}
 \ee 
 
The theory of the integrals of motion of stars does not allow to fix
uniquely the form of the function $f(x^2)$, since  the form of the
velocity ellipsoid is determined by irregular forces, which was not
taken into account in the theory of integrals of motions. The action of
irregular forces has been studied so far only in the case of very flat
subsystems, giving us the Kuzmin's formula (\ref{eq11.2.17}), and 
we shall use it to calculate $f(x^2)$.

In the case of $z=0$, the ratio $k_z$ is finite and nonzero. Thus,
without limiting generality we may take $f(0)=1$ and,  taking into
account Eq. (\ref{eq11.4.13}), we have 
\be 
k_z (R,0) = 1/ f(R^2+z_0^2). \label{eq11.5.3}
\ee 
This equation allows to calculate $f(x^2)$ if $k_z$ and $k_\theta$ are
known, but $k_\theta$ can be found from the Lindblad's formula
(\ref{eq11.2.3}).  

Equation (\ref{eq11.5.3}) defines the function $f(x^2)$ only for $x^2
\ge z_0^2$. In the region $0< x^2 < z_0^2$ the function should be
interpolated taking into account that according to definition
$f(0)=1$.  

It is necessary to point out a shortcoming in the calculations of $f(x^2)$
from (\ref{eq11.5.3}) with the help of (\ref{eq11.2.3}) and
(\ref{eq11.2.17}). According to the latter equations $k_z(0,0)=0.5$, and
hence $f(z_0^2)=2$,  independent of $z_0$. On the other hand,
velocity distribution in the centre of a spherical system should have
also a spherical symmetry, and therefore, in these systems $z_0=0$ and
$f(z_0^2)=1$, as it follows from the definition of the function. But
in this case, there should be a discontinuity at the centre of the
system. It seems to us that when looking for more and more spherical
systems, the function $f(x^2)$ approaches unity not with a jump but
smoothly. In other words, the Kuzmin equation (\ref{eq11.2.17}) in
central parts of stellar systems is not valid in the general case. 

Finally, let us discuss the generalisation of the equation to
calculate the quantity $\gamma$.

In the theory of the quadratic third integral, the parameter $\gamma$
is found using Eq~(\ref{eq11.4.3}), numerical values of $\gamma$ at $(R,z)$ are
determined by the parameter $z_0$. But the orientation of the velocity
ellipsoid is determined by the gravitational potential of the whole
system, and the value of  $\gamma$ can be calculated also directly from
the potential. For this we use the differential equation, derived from
the theory of the third integral of motion of stars
\citep{Eddington:1915aa,Kuzmin:1952ac} 
\be 
3\left( \frac{1}{R} \pdif{\Phi}{R} - \frac{1}{z} \pdif{\Phi}{z} \right) - \frac{1}{\gamma} \psdif{\Phi}{R}{z} + \ppdif{\Phi}{R} - \ppdif{\Phi}{z} = 0. \label{eq11.5.4}
\ee

Strictly speaking, this equation can be used only in the case of the
quadratic third integral. However, differentiating (\ref{eq11.5.4})
with respect to $z$,  expressing potential derivatives via
Oort-Kuzmin parameters (\ref{eq11.2.13}), (\ref{eq11.2.14}),  and
calculating $n_R$, we shall have Eq.~(\ref{eq11.2.12}), which was  initially
derived from the theory of the third integral as a quasi-integral. In this
way, when $z\rightarrow 0$, Eq.~(\ref{eq11.5.4}) remains valid also
when we have the third integral as a quasi-integral. Therefore, we may
assume that we shall not make a significant error by using
(\ref{eq11.5.4}) for arbitrary $z$. 

In order to use (\ref{eq11.5.4}) for the calculation of the kinematical function
$\gamma$ it is necessary to calculate first the potential derivatives at
all points $(R,z)$ interesting us. However, these calculations can be
largely simplified when using the model potential as a sum of Kuzmin's
flat model potentials. 

\citet{Kuzmin:1956ca} demonstrated that the existence of the quadratic
third integral together with the natural assumptions about the
finiteness of the mass and non-negative density significantly
constrains the number of possible expressions for the galactic
potential. In the limiting case of $\epsilon_\rho \rightarrow 0$
the expression for the potential has a specific form 
\be 
\Phi (R,z) = \frac{GM}{r},
\label{eq11.5.5}
\ee 
where $M$ is the mass of the galaxy, and
\be 
r^2 = R^2 + (z\pm z_0)^2, \label{eq11.5.6}
\ee 
while $\mathrm{sign}~z_0 = \mathrm{sign}~z$. The matter density is in this case
\be 
\rho (R,z) = \rho_0 \left( 1+\frac{R^2}{z_0^2} \right)^{-2} . \label{eq11.5.7}
\ee 

Starting from the formulae above for the density and potential, Kuzmin
constructed a corresponding model of the Galaxy
\citep{Kuzmin:1956ca}. Comparison of the model with the empirical one
indicates significant deviations. In particular, in the Kuzmin model
there is nearly no nucleus and,  on the other hand, the decrease of the density
in outer parts is too slow. 

The existence of nuclei and clear outer limits
seem to be general properties of nearly all galaxies. For this reason,
the Kuzmin model can be used as the first, quite rough approximation
only. On the other hand, a composite Kuzmin model gives us quite
satisfactory results. In this case 
\be 
\Phi (R,z) = G \sum_{i=1}^n \frac{M_i}{r_i} , \label{eq11.5.8}
\ee 
where $M_i$ are masses of components, $r_i^2 = R^2 + (z\pm z_{0i})^2$
and 
$z_{0i}$ are scale parameters of components,  and $n$ is the number of
components. In the present case, the separation of the  galaxy into components is
purely mathematical only, and components do not necessarily correspond
to real galactic subsystems. 

Substituting (\ref{eq11.5.8}) into (\ref{eq11.5.4}) and taking into
account (\ref{eq11.4.3}), we derive the following rule to calculate
the mean value of the parameter $z_0^2$: 
\be 
\overline{z_0^2} = \frac{\sum_{i=1}^n f_i z_{0i}^2}{\sum_{i=1}^n f_i} , \label{eq11.5.9}
\ee 
where
\be 
f_i = \frac{M_i (z+z_{0i})}{r_i^5} . \label{eq11.5.10}
\ee 
The parameter $\overline{z_0^2}$ is defined in a way that it allows to
calculate $\gamma$ from (\ref{eq11.4.3}) by substituting $z_0^2$ with
$\overline{z_0^2}$ there. Hence, the expression for $\gamma$ remained
in its previous form. 

{\begin{figure*}[h] 
\centering 
\hspace{2mm}
\resizebox{0.80\textwidth}{!}{\includegraphics*{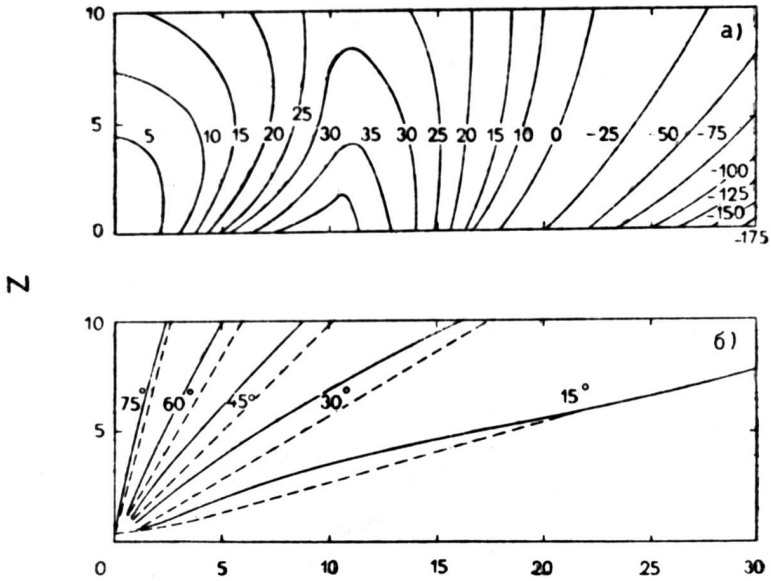}}
\caption{ {\em Top:}  Isolines of $\overline{z_0^2}$. 
  {\em Bottom:}  Isolines of the inclination angle, $\alpha$, of the major axis
  of the velocity ellipsoid with respect to the galactic 
symmetry plane for $\overline{z_0^2}=
  \overline{z_0^2(R,z)}$  (solid lines) and for $z_0=0.5$~kpc (dashed
  lines). 
} 
  \label{Fig18.1}
\end{figure*} 
}

The averaged parameter $\overline{z_0^2}$ is not a constant, but 
a function of coordinates $\overline{z_0^2} (R,z)$, see
Fig.~\ref{Fig18.1} for the M31 model, described in Chapter 18. 
Mean values of $\overline{z_0^2} (R,z)$ were calculated using
Eq.~(\ref{eq11.5.9}) for the \citet{Kuzmin:1956ca} model with
parameters, given in Table~\ref{Tab18.1}.

We saw above that
$z_0$ gives us the foci of confocal ellipsoids and hyperboloids; in
addition, at foci the velocity ellipsoids reduce to spheres. In the 
present case, the principal velocity-surfaces are no longer ellipsoids
and hyperboloids but have a more complicated form. Velocity ellipsoids
reduce to spheres at system axis at points $z=\pm z_{0e}$ ($z_{0e}$
are points where $\overline{z_0^2} (0,z)=z^2$). These points can be
called effective foci of the composite model.

Formulae derived here solve the problem, established at the beginning
of the paper. To our understanding, they allow to calculate a more
realistic hydrodynamical model of a galaxy, compared to previous
ones. But of course, our method is also only a preliminary one.
Further development of the construction of hydrodynamical models of
stellar systems is difficult without further development of the theory of
the third integral as a quasi-integral, together with the theory of
irregular forces in elliptical galaxies. And at last, we would like to
call attention to the paramount importance to solve these problems.

\medskip

\hfill March 1969

\chapter{Virial theorem and its application to the determination of
  masses of stellar systems}\label{ch12}

In the determination of masses of stellar systems and their subsystems,
the virial theorem can be used if the system or subsystem is
sufficiently isolated.  Galaxies consist of a number of populations
with mutual influence. An exception  is the nucleus of a galaxy which
is relatively isolated. Chapter 12 of the Thesis discussed general
properties of the virial theorem.  Here we consider one special case
--  the application of the virial theorem to find the mass of the
nucleus of M31, as done by \citet{Einasto:1970vz}.

The nucleus of a galaxy can be considered in a good approximation to
be an isolated dynamical system. In this case we may apply the tensor
virial theorem \citep{Kuzmin:1963aa}.  Assuming a rigid body rotation
and ellipsoidal shape for the nucleus we have

\be
\overline{\sigma_R^2} + \frac{1}{3}\omega^2\,\overline{a^2}= \frac{1}{2}\beta_R\,G\,\mm{M}a^{-1},
\label{eq12.1}
\ee
\be
\overline{\sigma_z^2}= \frac{1}{2}\beta_z\,G\mm{M}a^{-1}.
\label{eq12.2}
\ee
In these formulae $\omega$ is the constant angular velocity, $G$ the
gravitational constant, $\mm{M}$ the mass of the nucleus, and

\be
\overline{a^2}= \frac{1}{\mm{M}}\int_0^\infty\mu(a)a^2\dd{a},
\label{eq12.3}\ee
\be
a^{-1}=\frac{2}{\mm{M}^2}\int_0^{\mm{M}}\frac{M(a)\dd{M(a)}}{a},
\label{eq12.4}
\ee
where
\be
\mu(a)=4\pi\epsilon\rho(a)\,a^2
\label{eq12.5}
\ee
is the mass distribution function, and
\be
M(a)=\int_0^a\mu(a)\dd{a}
\label{eq12.6}
\ee
is the integral mass distribution function.

The constants $\beta_R$ and $\beta_z$ depend on the shape of the
system. Denoting $e^2=1-\epsilon^2$ we have

\be
\beta_R = \frac{1}{2e^2}\left[\frac{\arcsin e}{e}-\epsilon\right],
\label{eq12.7}
\ee
\be
\beta_z=\frac{\epsilon^2}{e^2}\left[\frac{1}{\epsilon}-\frac{\arcsin e}{e}\right].
\label{eq12.8}
\ee

From (\ref{eq12.1}) - (\ref{eq12.2}) we obtain
\be
k_z=\frac{\sigma_z^2}{\sigma_R^2}=\frac{\beta_z}{\beta_R}\left(1+\frac{\omega^2\overline{a^2}}{3\overline{\sigma_R^2}}\right).
\label{eq12.9}
\ee

As in the nucleus of the Andromeda galaxy $\omega^2a^2\ll\sigma_R^2$,
the mean axial ratio of the velocity ellipsoid depends sufficiently
only on the axial ratio of the system itself. The value of $k_z$,
found for the nucleus of the galaxy, can be adopted for $k_z(0,0)$.

\vskip 5mm
\hfill May 1971
\chapter{Some families of models of stellar systems}\label{ch13}

Methods of determining the mass-distribution in oblate stellar systems
were summarized by \citet{Perek:1962aa} who presented a classification of models
of stellar systems. A reclassification of models has been undertaken
by \citet{Kutuzov:1968aa}. The present paper deals with the methods
for determining the mass-distribution, the point of view being
different from that of Perek.

As we will see later, almost all expressions proposed earlier for
model construction can be interpreted as particular cases of one
general law. This enables us to study various models from a single
viewpoint.

To select suitable expressions for constructing the models, some
conditions are imposed, restricting the choice of
model-parameters. Great attention is given to the behaviour of models
in their outer region. These aspects were discussed in Chapter 8.

{\begin{table*}[h]
    {\small
      \centering    
\caption{} 
\begin{tabular}{cccccl}
\hline  \hline
$\alpha$ & $n$ & $\epsilon$& Case & Remarks & Model\\
  \hline
  0;1 & 1 & $\epsilon$& $A$ &$a,c,d,e$&\citet{Schmidt:1956,Schmidt:1965aa}\\
   2  & 1 & 0                & $B$  &$c,e$     &\citet{Schwarzschild:1954to}\\
    2  & 2 & $\epsilon$& $A$&$c,e$      &\citet{Perek:1962aa}\\
   2  & $n$&0                &$B$& $+$   &\citet{Wyse:1942wd}\\
  2  & $n$& $\epsilon$&$A$& $+$  &\citet{Burbidge:1959wd}\\
  \hline
\label{Tab13.1}   
\end{tabular}
\\
}
\end{table*} 
}

{\begin{table*}[h]
    {\small
    \centering    
\caption{} 
\begin{tabular}{cccccl}
\hline  \hline
$\beta$ & $\nu$ & $\epsilon$& Case & Remarks & Model\\
  \hline
  $>0$&2&1&  $A$&$\beta>2.5$&\citet{Lohmann:1964uw},\citet{Veltmann:1965aa}\\
   1&2& $\epsilon$&$B$&$+$&\citet{King:1962aa}\\
   2&1& $\epsilon$&$B$&$d$&\citet{Hubble:1930vx}\\
   2&2& $\epsilon$&$A$&$d$&\citet{Kuzmin:1956ca}\\
   2&3&    1&           $A$&$d$&\citet{Bottlinger:1933aa}\\
      2.5&2&1&         $A$&$d$&\citet{Lohmann:1964uw}\\
$\infty$&$\nu$& $\epsilon$& $A$&$+$&\citet{Einasto:1965aa}\\
$\infty$&2& $\epsilon$& $A$&$+$&Gauss\\
$\infty$&1& $\epsilon$& $A$&$+$&Expon.\\
$\infty$&1/4& $\epsilon$&$B$&$+$&\citet{de-Vaucouleurs:1948aa}\\
$<0$&1&1&$A,~B$&$\beta<-1$&\citet{Wallenquist:1959vc}\\
$<0$&2& $\epsilon$&$A$&$\beta<-1$&\citet{Perek:1962aa}\\
$-1$&$\nu$&1&$A$&$c,e$&\citet{van-Wijk:1949tv}\\
$-0$&$\infty$& $\epsilon$& $A$&  $c,e$&   Homogen.
\label{Tab13.2}   
\end{tabular}
\\
}
\end{table*} 
}

{\begin{table*}[h]
    {\small
      \centering    
\caption{} 
\begin{tabular}{ccccl}
\hline  \hline
$\alpha$ & $\beta$ & $\nu$ & Remarks & Model\\
  \hline
  3&$3/\nu$& $\nu$& $d$   &\citet{Brandt:1965ty}\\
  3& 2          &2      &$a,d$    &\citet{Parenago:1950aa}\\
 3&$3/2$    &2      &$d$      &\citet{Kuzmin:1956aa}\\
  3&1           & 3     &$d$     &\citet{Bottlinger:1933aa}\\
     3&$-1$  &2     & $c,e$   &\citet{Perek:1962aa}\\
    2&$-1$  & 1     &$a,c,d,e$&\citet{Schmidt:1956}\\
     0& $-1$  &3    &$a,c,d,e$&\citet{Oort:1927wy}
\label{Tab13.3}   
\end{tabular}
\\
}
\end{table*} 
}

Usually the distribution of mass only in very oblate systems is
considered. In such a case, the mass distribution can be evaluated
directly from the rotation data, since the rotation velocity equals
the circular one. In the ellipsoidal case, the pressure term (velocity
dispersion) cannot be neglected, as stated by \citet{Opik:1922}
 and later by \citet{Oort:1965aa}. In the central parts of oblate stellar
systems, the ellipsoidal component, the bulge, is often prevailing,
therefore, even in the oblate systems the exact mass-distribution can
be found only from a hydrodynamical model, using the data on both
rotation and dispersion. Furthermore, independent data on the distribution
of luminosity and mass-to-light ratio can also improve the model.

Basic results of the description of some families of models of stellar
systems were published by \citet{Einasto:1969ab}. Here we present in a
condensed way the main results of this study.

Some quite general families of the descriptive functions can be
constructed by means of the function 
\be
g(a)= g_0\,g^\ast(\xi),
\label{eq13.3.1}
\ee
where $\xi=a/a_0$ is dimensionless distance, $a_0$ and $g_0$ are scale
parameters, and
\be
g^\ast(\xi)=\left\{
  \ba{ll}
\xi^\alpha\,\prod^n_{i=0}(1+\chi_i/\beta_i\,\xi^{\nu_i})^{-\beta_i}
&\xi\le\xi_0,\\
0,&\xi\ge \xi_0.
\ea \right.
\label{eq13.3.2}
\ee
In the last formula, $\xi_0$ is the smallest positive root of the equation
\be
g^\ast(\xi)=0,
\label{eq13.3.3}
\ee
and $\alpha,~\chi_i,~\beta_i,~\nu_i$ are structural parameters.

The expression $g(a)$ can be identified with various descriptive
functions:\\
$A$) \hspace{3cm} $g(a) \equiv \mu(a)$;\\
$B$) \hspace{3cm} $g(A) \equiv L(A)$;\\
$C$) \hspace{3cm} $g(R) \equiv F_R(R)$;\\
$D$) \hspace{3cm} $g(R) \equiv V_\theta(R)$;\\
$E$) \hspace{3cm} $g(R) \equiv \sigma_R^2(R)$.

{\begin{figure*}[h] 
\centering 
\hspace{2mm}
\resizebox{0.45\textwidth}{!}{\includegraphics*{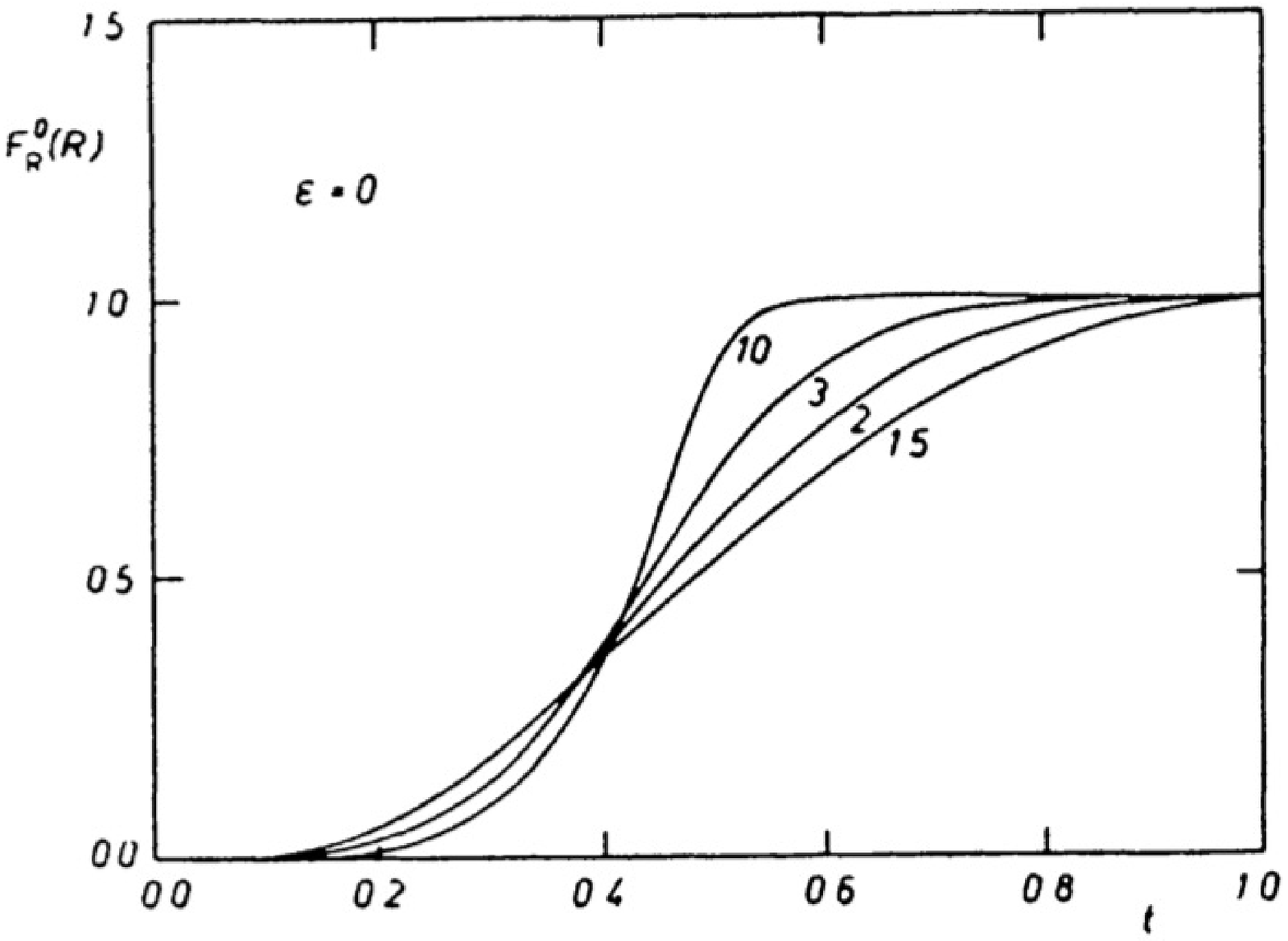}}
\hspace{2mm}
\resizebox{0.48\textwidth}{!}{\includegraphics*{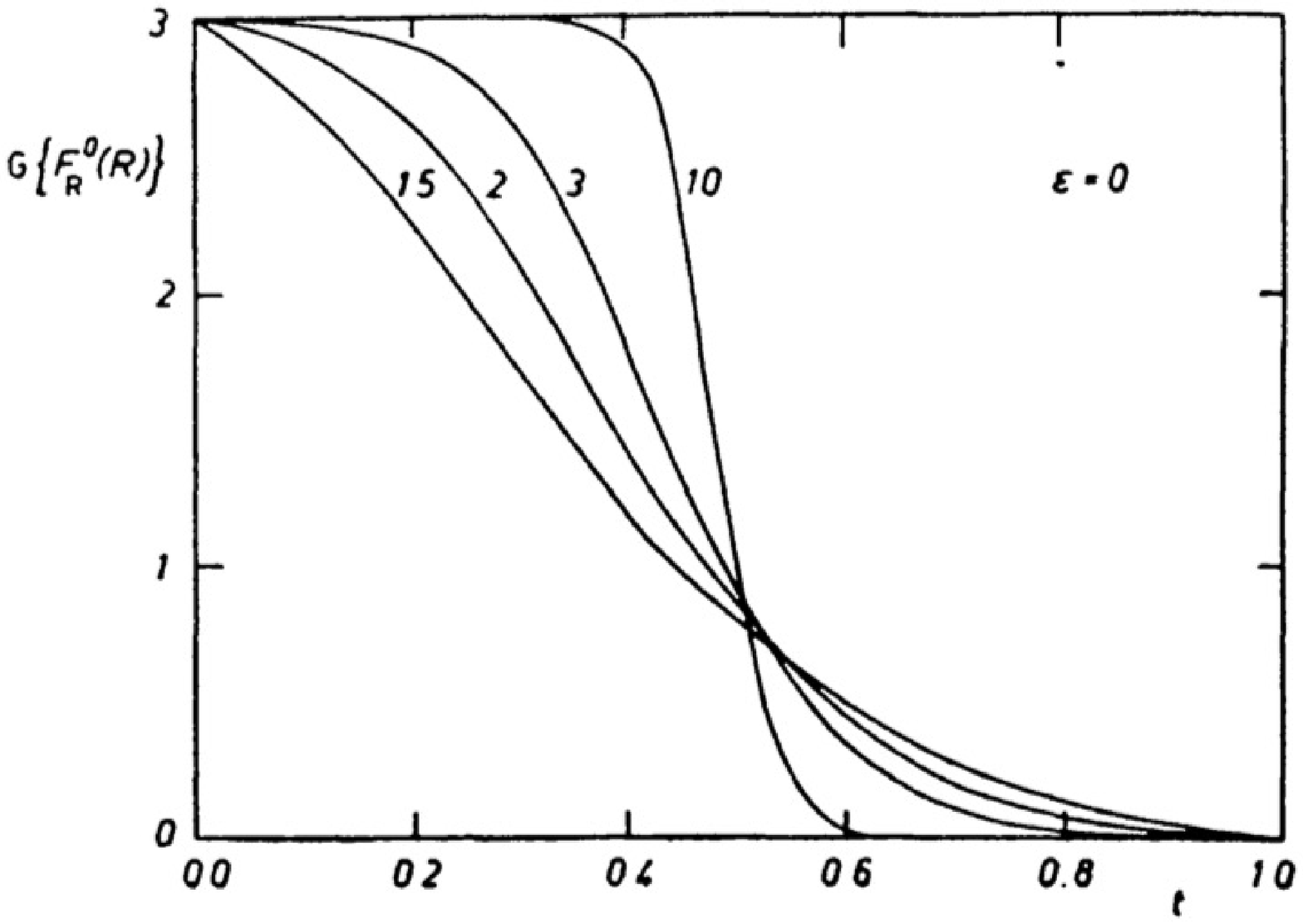}}
\caption{ The function $F_R^0(R)$ and its logarithmic gradient as
  functions of $t=a/(1+a)$ for the generalised Bottlinger model with
  parameters $\epsilon=0$ and various $\nu$, shown as index of curves.}
  \label{Fig13.1}
\end{figure*} 
}

The formula (\ref{eq13.3.2}) is too general and contains too many
parameters for practical use. We will consider three families of
special models based on the law (\ref{eq13.3.2}): \\
a) the polynomial model (cases $A$ and $B$)
\be
g^\ast(\xi) = \xi^\alpha\,\prod^n_{i=0}\,(1-\chi_i\,\xi),
\label{eq13.3.4}
\ee
b) the binomial model (cases $A$ and $B$)
\be
g^\ast(\xi) = \xi^\alpha\,\left(1+1/\beta\,\xi^\nu\right)^{-\beta},
\label{eq13.3.5}
\ee
c) the generalised Bottlinger model (case $C$)
\be
F_R^\ast(\xi) = \xi^\alpha\,\left(1+1/\beta\,\xi^\nu\right)^{-\beta}.
\label{eq13.3.6}
\ee
In cases $D$ and $E$, the binomial law (\ref{eq13.3.5}) can also be used.

Most models of galaxies and star clusters proposed earlier are
particular cases of models  (\ref{eq13.3.4}) $\dots$
(\ref{eq13.3.6}). Their review is given in Tables \ref{Tab13.1}
$\dots$    \ref{Tab13.3}. In Tables {\em Remarks} letters $a,b,c,d,e$
indicate that the model {\em does not} 
agree with condition $a,b,c,d,e$, discussed  in Chapter 8, $+$ indicates that the
model agrees with all these conditions.  Binomial models with $\beta = \infty$
are generalised exponential models \citep{Einasto:1965aa}, having for
spatial density (case A) the form
\be
\rho(a) = \rho_0\,\exp[-\xi^{1/N}].
\label{eq13.4}
\ee

The radial attraction function $F_R^0(R)$ and its logarithmic
gradient for the generalised Bottlinger model are shown in
Fig.~\ref{Fig13.1}, using as argument $t=a/(1+a)$. For the generalised
exponential model several descriptive functions are shown in Chapter 15.

 \vskip 5mm
 \hfill May 1968
 \vskip 2mm
\hfill Revised May 1971 
\chapter{Polynomial models}\label{ch14}

The analysis of galactic models has shown that most models were
constructed using analytical expressions for descriptive functions of
two families, polynomial or binomial models. In this Chapter, I
discuss concisely the polynomial models.  A more detailed discussion was
published by \citet{Einasto:1968ab, Einasto:1969ab}.

General families of the description functions can be constructed by
means of the function
\be
g(a)=g_0\,g^\ast(\xi),
\label{eq14.1}
\ee
where $\xi=a/a_0$ is the dimensionless distance, $a_0$ and $g_0$ are
scaling parameters, and
\be
g^\ast(\xi)=\left\{\ba{rc}
\xi^{\alpha}\,\Pi_{i=0}^n\,\left(1+{\chi_i \over
    \beta_i}\,\xi^{\nu_i}\right)^{-\beta_i}, & \xi \le \xi^0,\\
    0, & \xi\ge \xi^0.
    \ea
  \right.
  \label{eq14.2}
  \ee
In the last formula, $\xi^0$ is the smallest positive root of the
equation $g^\ast(\xi)=0$, and  $\alpha, ~\chi_i, ~\beta_i, ~\nu_i$  are
structural parameters. 

Eq.~(\ref{eq14.2}) is too general and contains too many parameters for
practical use. The polynomial model is defined as
\be
g^\ast(\xi) = \xi^{\alpha}\,\Pi_{i=0}^n\,(1-\chi_i\,\xi),
\label{14.3}
\ee
The polynomial model with parameters $\alpha = 0,~1$ and $n=1$ was used in
models of the Galaxy by \citet{Schmidt:1956,Schmidt:1965aa}. These
models are in conflict with our conditions of physical correctness a),
c) and d), discussed in Chapter 8. \citet{Wyse:1942wd} used this
profile to describe the projected density distribution of M31. The
polynomial profile was also used by \citet{Burbidge:1959wd,
  Burbidge:1960vy} in their series of studies of the rotation and mass
distribution of galaxies. In the latter cases, most conditions of physical
correctness were fulfilled, however, the density function has a break at
$\xi^0$. 

\vskip 5mm
\hfill April 1967
\chapter{Binomial models}\label{ch15}

Binomial models were discussed in detail by
\citet{Einasto:1968ac,Einasto:1969ab}, and formed the basis of Chapter
15 of the original Thesis.  Here I give a short summary, concentrating
on the case of the generalised exponential function.

{\begin{figure*}[h] 
\centering 
\hspace{2mm}
\resizebox{0.50\textwidth}{!}{\includegraphics*{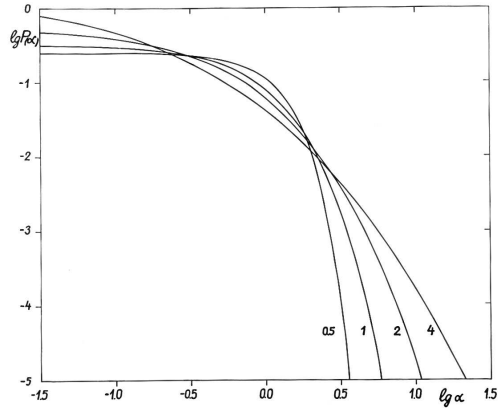}}\\
\caption{ Distribution of the projected density
  $P(\alpha)$ for the generalised exponential model. Parameter $N$ is
  shown as an index of curves. } 
  \label{Fig15.1}
\end{figure*} 
}

{\begin{figure*}[ht] 
\centering 
\hspace{2mm}
\resizebox{0.40\textwidth}{!}{\includegraphics*{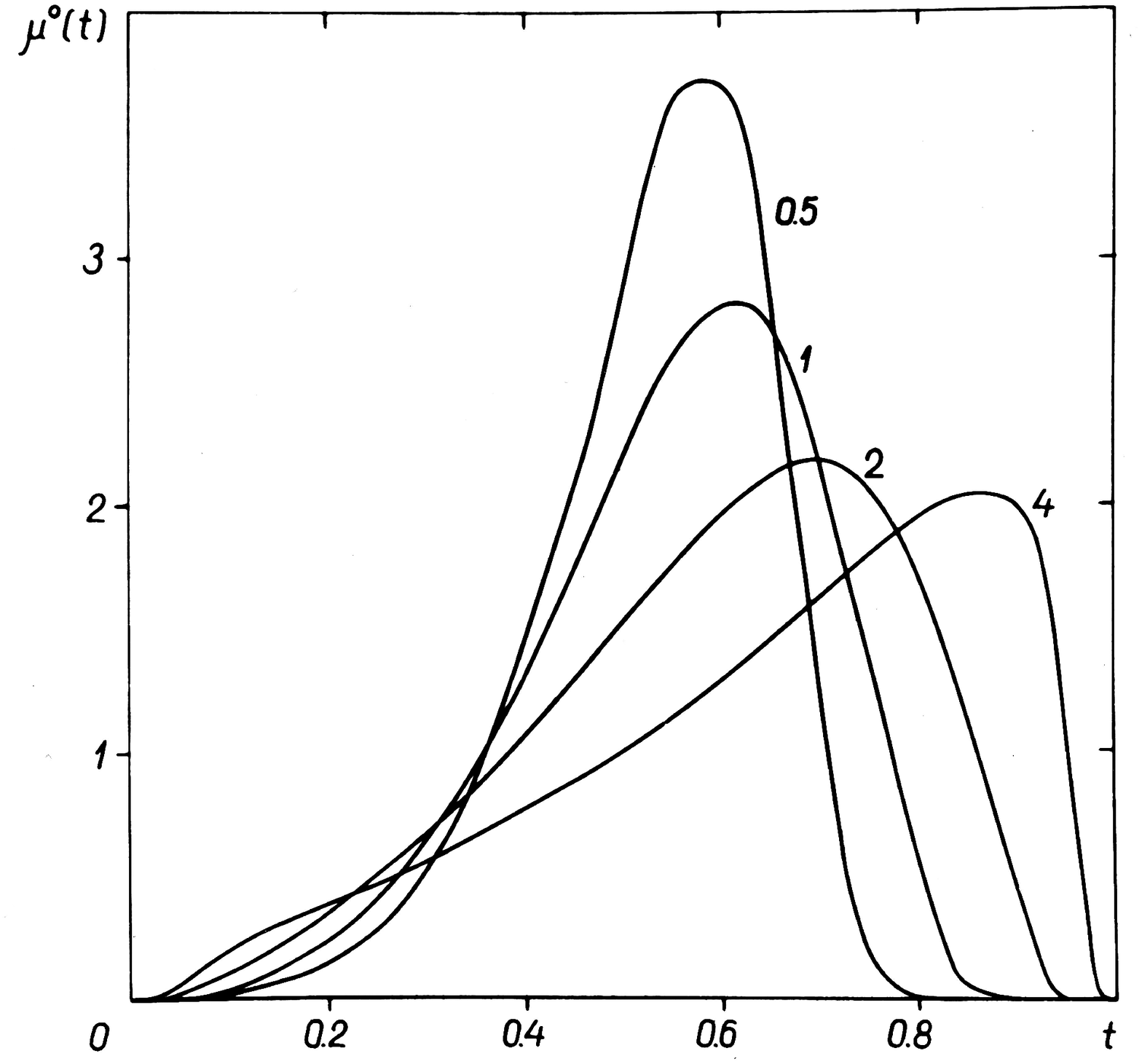}}
\hspace{2mm}
\resizebox{0.50\textwidth}{!}{\includegraphics*{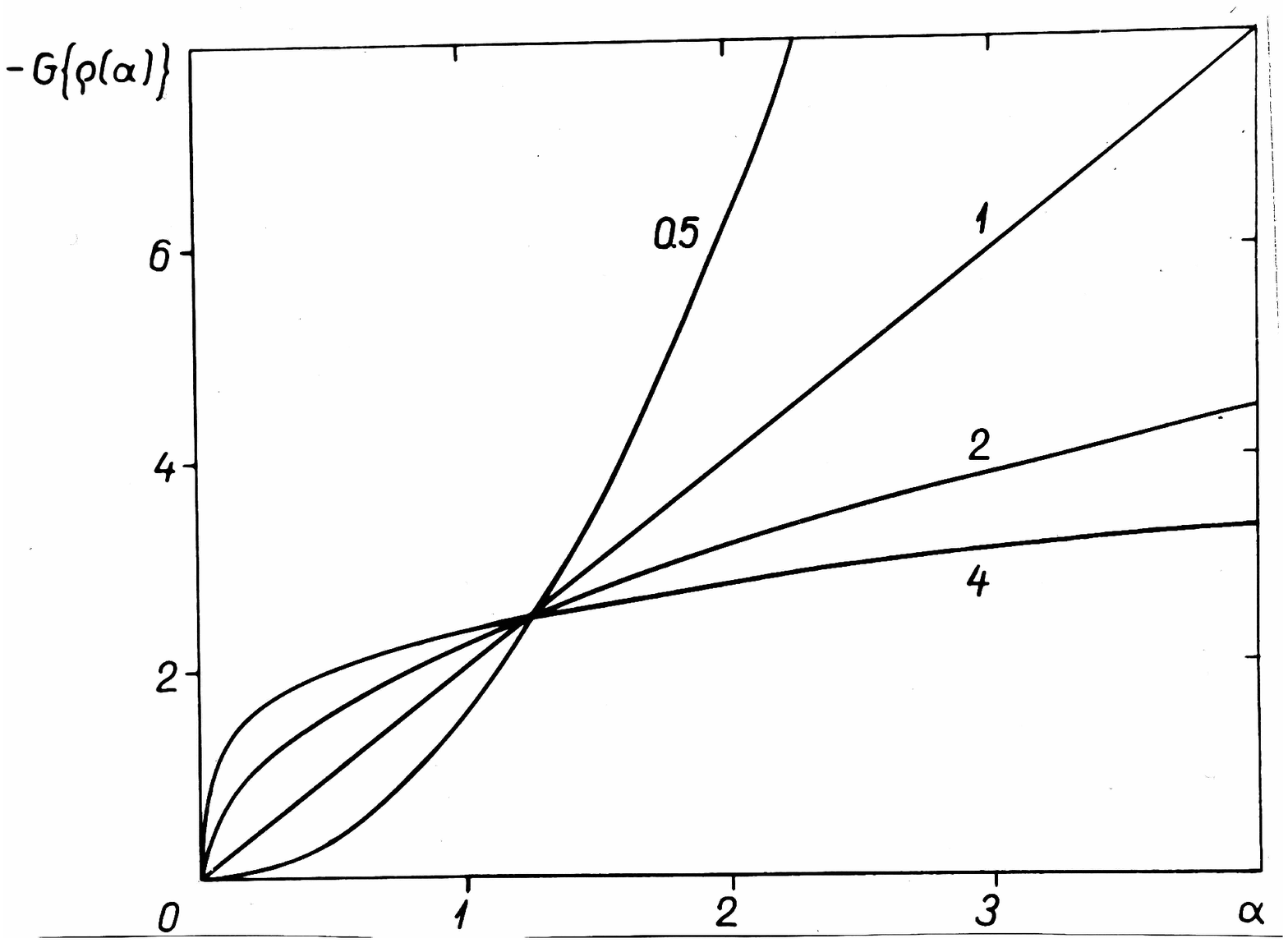}}
\caption{ {\em Left:} Mass function
  $\mu^\circ(t)$ of generalised exponential model. The shape parameter 
  $N$ is shown.  {\em Right:} The logarithmic density gradient $G\{\rho(\alpha)\}$ of the  generalised
  exponential model.} 
  \label{Fig15.2}
\end{figure*} 
}

Binomial models are defined by the description function
\be
g^\ast(\xi) = \xi^{\alpha}\,\left(1+1/\beta\,\xi^{\nu}
\right)^{-\beta},
\label{eq15.1}
\ee
where $\xi=a/a_0$ is the dimensionless distance, and $\beta$ and $\nu$
are structure parameters.  If $\beta \rightarrow \infty$, then the
binomial model reduces to the generalised exponential model
\be
g^\ast(\xi) =e^{-\xi^\nu}.
\label{15.2}
\ee
Often as the shape parameter, instead of $\nu$ its reciprocal value is used $N=1/\nu$.

{\begin{figure*}[ht] 
\centering 
\hspace{2mm}
\resizebox{0.44\textwidth}{!}{\includegraphics*{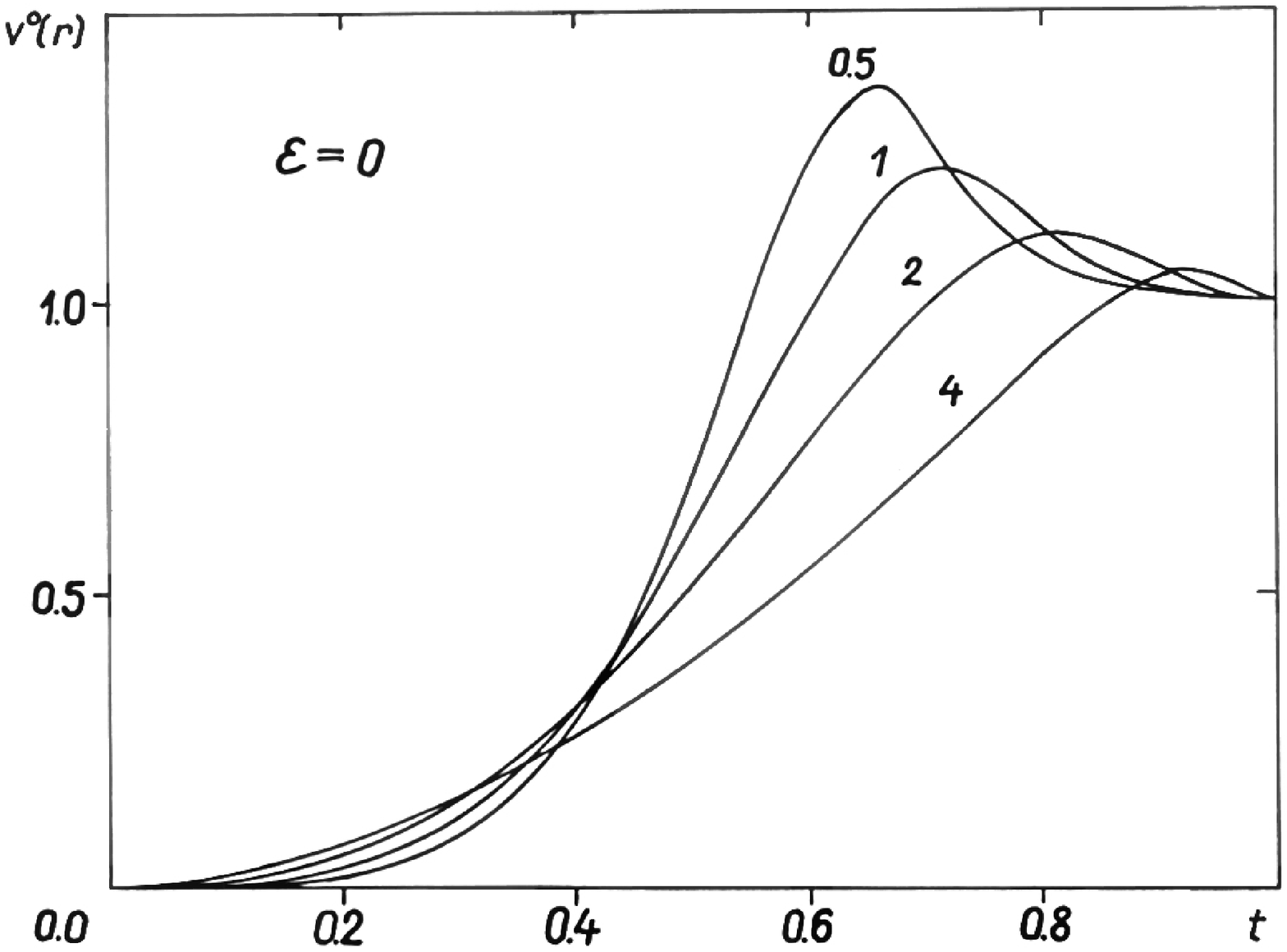}}
\hspace{2mm}
\resizebox{0.50\textwidth}{!}{\includegraphics*{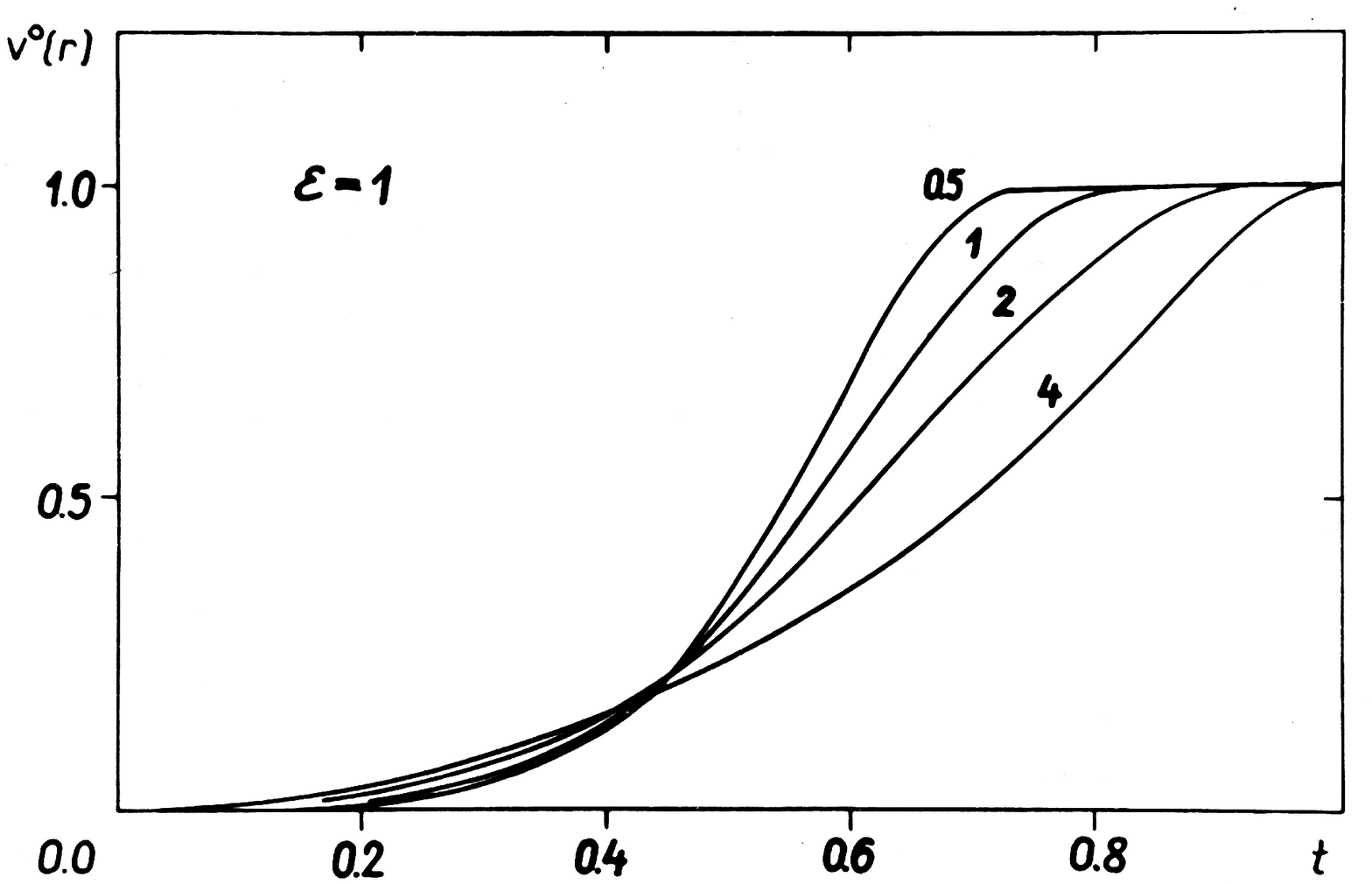}}
\caption{ Circular velocity function
  $v^\circ(t)$ of the generalised exponential model for two values of
  the axial ratio $\epsilon$. The shape parameter 
  $N$ is shown.  } 
  \label{Fig15.4}
\end{figure*} 
}

{\begin{figure*}[ht] 
\centering 
\hspace{2mm}
\resizebox{0.44\textwidth}{!}{\includegraphics*{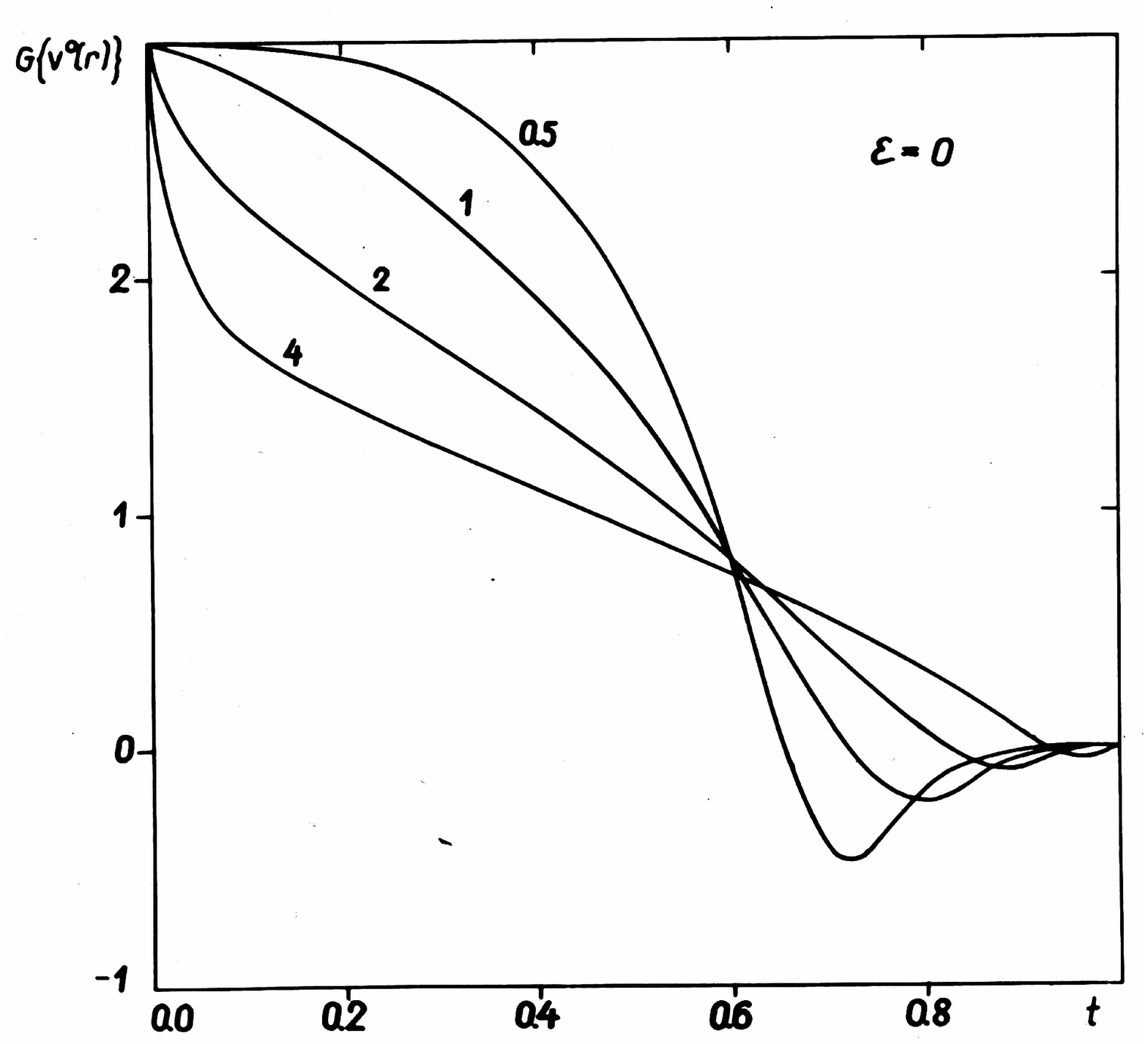}}
\hspace{2mm}
\resizebox{0.50\textwidth}{!}{\includegraphics*{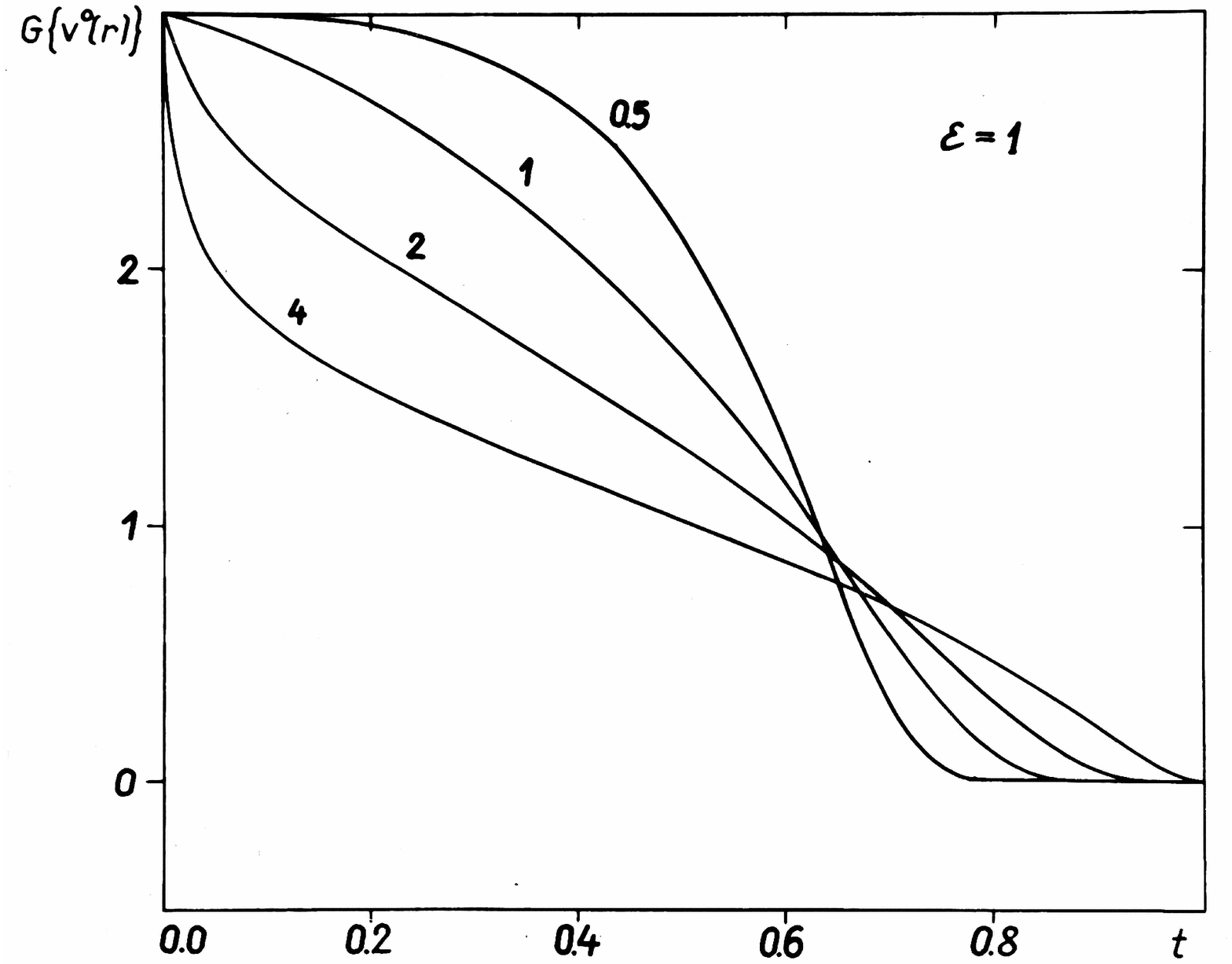}}
\caption{Logarithmic  gradient of the velocity  $G\{v^\circ(t)\}$ of the  generalised
  exponential model for two values of
  the axial ratio $\epsilon$. The shape parameter 
  $N$ is shown.  } 
  \label{Fig15.5}
\end{figure*} 
}

{\begin{figure*}[ht] 
\centering 
\hspace{2mm}
\resizebox{0.60\textwidth}{!}{\includegraphics*{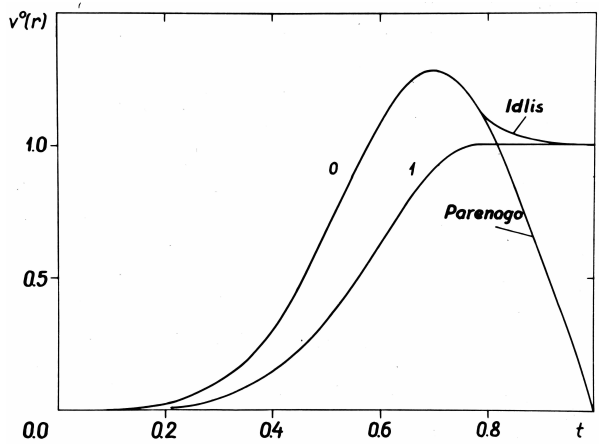}}
\caption{ Circular velocity function
  $v^\circ(\alpha)$ for Parenago and Idlis models. The axial ratio
  $\epsilon$ is shown. } 
  \label{Fig15.6}
\end{figure*} 
}

\vfill

Cases $\nu = 2$ and $\nu = 1$ are ordinary Gaussian and exponential
models. The case $\nu =1/4$, when applied to projected density, is the
\citet{de-Vaucouleurs:1948aa} profile for elliptical galaxies. It can
be used also for the spatial density, as done by
\citet{Einasto:1965aa, Einasto:1969aa}.

In the following Figures, several descriptive functions of the generalised
exponential function are given. As parameters we use $N=1/\nu$, 
$\alpha = a/a_0$ and $t=\alpha/(1+\alpha)$. Fig.~\ref{Fig15.1} shows the projected density,
Fig.~\ref{Fig15.2} left panel gives the relative mass function
$\mu^\circ(t)=\frac{1}{\mu_0}\mu(\alpha)\,(1+\alpha)^2$, where $\mu_0$
is a normalising constant. Fig.~\ref{Fig15.2} right panel shows the logarithmic
density gradient $G\{\rho(\alpha)\}$.  Fig.~\ref{Fig15.4} shows the
circular velocity function
\be
v^0(a)= \frac{a}{G\mm{M}}\,V^2(a),
\label{eq15.3}
\ee
where  $V$ is the circular velocity,  $G$ is the gravitation constant, and $\mm{M}$ is the mass of the
system. The circular velocity of a point mass $\mm{M}$ is $V^2(a)=
\frac{G\mm{M}}{a}$, thus the circular velocity function is the
relation of squares of circular velocities of the model and the 
point-mass of the same mass. This definition of the circular velocity
function was suggested by \citet{Perek:1962aa}. Next, in
Fig~\ref{Fig15.5} we give the logarithmic  gradient of the velocity
function, $G\{v^\circ(t)\}$. Finally, in Fig.~\ref{Fig15.6} we show circular velocity
functions of models by \citet{Parenago:1950aa} and \citet{Idlis:1956aa}.

\vskip 5mm
\hfill September 1967
\vskip 5mm
\hfill Revised September 1971

\chapter{Hydrodynamical models on the basis of the modified exponential
  function}\label{ch16}

In Chapters 5 and 7, I described methods to calculate spatial and
kinematical models of galaxies, using our own Galaxy as an example. In
this Chapter, I  shortly discuss some additional aspects of
constructing hydrodynamical models on the basis of the modified
exponential function.

The set of descriptive functions is fully determined if one dynamical
function is given. Theoretically, it is not important which function is
given initially. The remaining functions can be calculated by means of
the formulae given in Chapters 5 and 7. But from the practical point of view, it
is convenient to start from the luminosity distribution function and
from the data on the mass-to-light ratio.

The block diagram of the determination of a model of a stellar system
from observations is given in Fig. \ref{Fig16.10}, taken from 
\citet{Einasto:1969ab}. The observed functions and 
parameters are shown as circles, the calculated and preliminary adopted
quantities are presented by rectangular boxes. Three cases are considered:\\
a) simple (one component) model of an elliptical galaxy (classical method),\\
b) simple model of an oblate galaxy (classical method),\\
c) composed hydrodynamical model of a galaxy (new method).

{\begin{figure*}[h] 
\centering 
\resizebox{0.95\textwidth}{!}{\includegraphics*{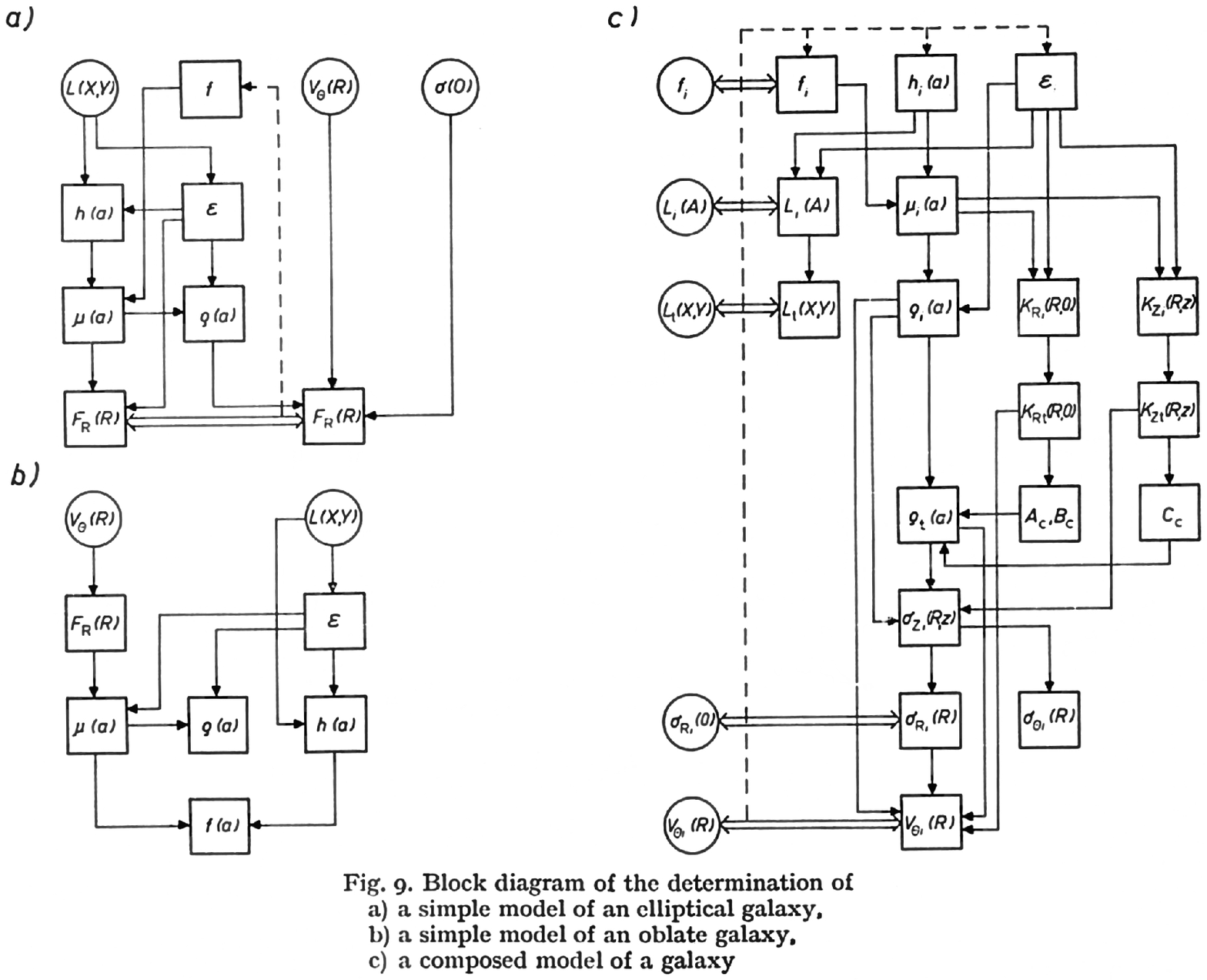}}
\caption{ Block diagram of the determination of a) a simple model of an
elliptical galaxy, b) a simple model of an oblate galaxy, c) a
composed model of a galaxy. }
 \label{Fig16.10}
\end{figure*} 
} 

In the first case, the flattening parameter $\epsilon$ and the
mass-to-light ratio $f$ is usually assumed to be constant. From
the observed projected distribution of luminosity $L(X,Y)$, the
corresponding spatial distribution $\mu(a)$ can be found by solving the
integral equation
\be
L(A)= {1 \over 2\pi\,E}\int_A^\infty {\mu(a)\dd{a} \over f\, a\sqrt{a^2-A^2}},
\label{eq16.1}
\ee
where $a$ and $A$ are major semiaxes of equidensity ellipsoids and
projected equidensity ellipsoids, respectively, 
$L(A)$ is the projected luminosity density, $f$ is the
mass-to-light ratio, $\mu(a) = f\,h(a)$ is the mass function, and
$h(a)$ the luminosity function.
If $f = const$, the mass distribution $\mu(a)$ is
similar to the luminosity distribution $h(a)$, and the radial
acceleration function
\be
F_R(R)= \frac{V_c^2R}{G}= \int_0^R{\mu(a)\dd{a} \over \sqrt{1-(\epsilon\,a/R)^2} }
\label{eq16.2}
\ee
can be found. Here $V_c$ is the circular velocity, and $G$ is the
gravitational constant. The $F_R(R)$ function can be
found independently from the kinematical data, using the
hydrodynamical equation
\be
V_\theta^2 - p\,\sigma_R^2 = V_c^2,
\label{eq16.3}
\ee
where $V_\theta$ is the rotation velocity of a galactic subsystem,
$\sigma_R$ is the velocity dispersion in $R$ direction of the
subsystem, and $p$ is a parameter, depending on the density gradient
and the shape of the velocity
ellipsoid, see Eqs.
(\ref{eq4.3}) and (\ref{eq11.1.4}). The comparison of the results for $F_R(R)$ gives the
possibility to determine $f$ and to check the initial assumptions $\epsilon =
const$, and $f= const$ (feedback coupling). 

In the second case, the usual assumption is $\sigma_R \ll V_\theta$;
from Eqs. (\ref{eq16.3}) and (\ref{eq4.3}) we see that then $V_\theta \approx V_c$. The
mass distribution function, $\mu(a)$, and the luminosity distribution
function, $h(a)$, can be determined independently from the integral
equations (\ref{eq16.2}) and (\ref{eq10.4}) respectively. The mass-to-light ratio
$f$ is to be considered as a function of the distance from the centre
of the system. 

Observations indicate that the axial ratio of isophotes $E$ of
elliptical and spiral galaxies is not constant. Therefore, the
assumption $\epsilon = const$ is justified only as the first rough
approximation. To obtain a better agreement with observations,
$\epsilon$ is to be considered as a variable, as it was taken 
 by \citet{Perek:1962aa}, \citet{Sizikov:1968wr} and
\citet{Kutuzov:1968ac}.  In this case, the block diagram of the model
determination will not change, but formulae should be replaced by more
complicated ones.

The variation of $E$ and $\epsilon$ is caused mainly by the presence of various
subsystems, having different flattening $\epsilon$, and different mass and
luminosity distribution. Therefore, the structure of a stellar system
can be more precisely described by a composed model. The determination
of a composed hydrodynamical model of a galaxy (case c in Fig.~\ref{Fig16.10}) differs from
the determination of a simple classical model in many respects.

As the distribution of luminosity of individual components $L_i(X,Y)$ of
the galaxy is only partially known, it is difficult to start the
model determination from these functions as in previous cases. It
would be better to calculate first the normalised descriptive
functions, $h^0(a)$, $L^0(A)$, $F^0_R(R)$ and $F^0_\theta(R)$,  for a set of concentration and
flattening parameters. The construction of the model reduces in this
case to the determination of parameters of the galaxy
components. For this purpose, both graphical and numerical methods can
be used, as done by \citet{Einasto:1968ad, Einasto:1969aa}.

The evaluation of $z$-velocity dispersion from the vertical acceleration
can be made by iterations, as stated by  \citet{Innanen:1968aa}. In
the first step, the term with $p$ in formula (\ref{eq16.3}) 
may be neglected. From the approximate run of velocity dispersions, the
correction terms can be calculated. The dispersions $\sigma_R,~\sigma_\theta$, and the
centroid velocity $V_\theta$ are to be found from $\sigma_z$ by using the formulae
(\ref{eq16.3}),
and
\be
\frac{1}{\sigma_z^2}=\frac{1}{\sigma_\theta^2} + \frac{1}{\sigma_R^2},
\label{eq16.4}
\ee
found from the theory of irregular forces for the case $z\ll R$
\citep{Kuzmin:1961aa}. Furthermore, the Lindblad formula
\be
k_\theta =\frac{-B}{A-B} = \frac{1}{2}[1+G\{V_\theta(R)\}]
\label{eq16.5}
\ee
can be used, where $G\{V_\theta(R)\}$ is the radial logarithmic gradient of the
rotation velocity $V_\theta(R)$.

A comparison of the calculated descriptive functions with the observed
ones (see case c in Fig. \ref{Fig16.10}) gives the possibility to improve the initial
descriptive parameters. So, the model will be determined by a
trial-and-error procedure.  To facilitate the determination of models of stellar systems, we are
computing the main dynamical descriptive functions for the exponential
model with $\nu$ and $\epsilon$ as parameters. This work is in good progress
now \citep{Einasto:1972ac}. Results are published for our Galaxy by
\citet{Einasto:1970ad} and are presented in Chapter 7, and for the Andromeda galaxy
M31 by \citet{Einasto:1969aa, Einasto:1970aa}, \citet{Einasto:1970ac,
  Einasto:1972ae} and are discussed in Chapters  18
and 20.

\vskip 5mm
\hfill May 1971
\vfill

\part{Spatial and kinematical structure of the Andromeda galaxy}

\chapter{Model of mass distribution of M31: Preliminary version}\label{ch17}

On the basis of the published photographic and photoelectric data on
the luminosity distribution along the major and minor axis of the
galaxy M31, the model of the latter was elaborated
\citep{Einasto:1969aa}, which is the topic of this Chapter. The model consists 
of four components: the nucleus, bulge, disc, and flat component. The
masses of components were derived from the velocity data,
collected from optical and radio sources. The velocity dispersion and
the mass-to-light ratio, spectroscopically obtained for the centre of
M31, were also used.

It was found that the circular velocity curve has a maximum $V_c=380$
km/sec at the distance of 4' from the centre. The rotational velocity
of the spheroidal component (the bulge) equals only 125 km/sec in
this region. The great difference between the circular and rotational
velocities can be explained by the great velocity dispersion and
radial density gradient of the spheroidal component. The dynamical
mass-to-light ratio 17.3 is in good agreement with the spectroscopical
one, 16.7.

For the mass of the galaxy M31, a value of $200\times 10^9$ solar masses is
found. Considerably greater values obtained by other authors (see
Table \ref{Tab17.3}) are biased by neglecting the fact that the galaxies are of
finite sizes.	

In the motion of interstellar hydrogen, local deviations from the circular motion occur.

\section{Introduction}

The study of the structure of the large Andromeda galaxy M31 is of
interest primarily because it is the closest outer spiral galaxy to
us. This allows us to find out  its structure in  details that are
not visible or difficult to study in other, more distant galaxies. In
addition, it is well known that the M31 galaxy is very similar in
structure to our Galaxy. Due to this circumstance, the study of M31
galaxy complements the study of our Galaxy and vice
versa.

Among the results obtained in the study of the general structure of
the M31 galaxy there are the following two contradictory
conclusions.

1. Estimates of the mass of the system, despite a very precisely
defined rotation curve, are very different, ranging from 200 to 600
billion solar masses.

2. According to the dynamical definition, the ratio of mass to
luminosity $f$ at the centre of the system is very small, while on the
periphery it approaches infinity (see Fig. \ref{Fig17.8}). On the other hand,
according to the spectral definition of the composition of the M31
core, the central value of $f$ is approximately equal to its mean value,
\ie the value of $f$ must be approximately constant. 

In the present series of articles, a model of the M31 galaxy will
be constructed, and an attempt will be made to clarify the reasons for
the above contradictions. It is assumed that the M31 galaxy consists
of four main components: the nucleus, bulge, disc, and 
flat component. In the course of the work, it turned out that 
the model can be constructed by successive
approximations. Therefore, this first article of the series describes a
preliminary model of the system. In the future, the model will be
refined and detailed.

\section{Description functions and the equations of the relationship
  between them}

From observations it is possible to determine the following
functions or their partial values: the projected luminosity density $L_S(X,Y)$
 in the photometric system $S$ ($X$ and $Y$ are rectangular visible
coordinates expressed in angular units, with the $X$ axis directed along
the visible major axis of the galaxy, and $Y$ along the minor axis); the
rotation speed of some subsystems $V_\theta$; the stellar velocities dispersion 
$\sigma$, and the stellar composition (for the galactic nucleus). It is also
possible to study the distribution and physical properties of
individual bright stars.

In order to model the galaxy, it is necessary to introduce simplifying
assumptions. In this series of works, it is assumed that the M31 galaxy
can be divided into a finite number of physically homogeneous
components, whose equidensity surfaces are similar
coaxial ellipsoids of rotation. The ratio of semiaxes of ellipsoids of
different components $\epsilon$ can be different, the density changes smoothly.

Since the main description functions are additive (except for rotation
velocity $V_\theta$ and velocity dispersion $\sigma$), 
for simplicity we will write their connection equations not for the
total values but for the individual components.

Let $\rho(x, y, z)$ be the spatial mass density of the component, and
$l_S(x, y, z)$ — the spatial luminosity density in the photometric system $S$
($x,~y,~z$ are rectangular galactocentric coordinates, the $z$ axis is
directed along the system axis).  With the above assumptions
\be
\rho(x,y,z)=\rho(a)=f_S\,l_S(a),
\label{eq17.2.1}
\ee
where
\be
a^2=x^2+y^2+\epsilon^{-2}z^2,
\label{eq17.2.2}
\ee
and $f_S$ is the mass-to-light ratio of the component. The mass
and luminosity densities are related as \citep{Einasto:1968aa}
\be
\mu(a)=4\pi\epsilon\,a^2\rho(a)
\label{eq17.2.3}
\ee
and
\be
h_S(a)=4\pi\epsilon\,a^2\,l_S(a).
\label{eq17.2.4}
\ee

The projected density of components is expressed \citep{Einasto:1968aa}
\be
L_S(A)=\frac{1}{2\pi\,E}\int_A^\infty\,{h_S(a)\dd{a} \over a\sqrt{a^2
    - A^2}  },
\label{eq17.2.5}
\ee
where
\be
A^2 = X^2 + E^{-2}Y^2,
\label{eq17.2.6}
\ee
\be
E^2 = \cos^2i +\epsilon^2\,\sin^2i,
\label{eq17.2.7}
\ee
where $i$ is the angle between the symmetry axis of the system and the line of
sight.

The rotation velocity of the component and the circular velocity of
the whole system, $V$, are related as \citep{Kuzmin:1962ac} (see also
Chapters 7 and 11): 
\be
V_\theta^2 -p\,\sigma_R^2 = V^2,
\label{eq17.2.8}
\ee
where $\sigma_R$ is the velocity dispersion of the component in the radial
direction ($R^2 =x^2 + y^2$), and the dimensionless parameter $p$ is expressed as
\be
p=\left(1 - \frac{\sigma_\theta^2}{\sigma_R^2}\right) + R\, \left(1 -
    \frac{\sigma_z^2}{\sigma_R^2}\right) \frac{\partial
      \alpha}{\partial z} + G\{\rho(R)\} +G\{\sigma_R^2(R)\},
    \label{eq17.2.9}
    \ee
 where $\alpha$ is the inclination angle of the vertex in respect to
  the plane of the system (outside the symmetry plane 
 $\alpha \neq 0$).
 
In the last equation we used for the logarithmic gradient $G$ the
expression
\be
G\{f(R)\} = \frac{\partial\,\ln f(R)}{\partial\,\ln R}.
\label{eq17.2.10}
\ee
The expression for $p$ in the form (\ref{eq17.2.9}) is inconvenient for practical
applications, since neither the ratio of velocity dispersion nor the
gradient of the angle $\alpha$ can be directly found from observations. When
transforming the expression $p$, we will use the relations found by
\citep{Kuzmin:1952ab, Kuzmin:1961aa}
\be
R\frac{\partial \alpha}{\partial z} =-\frac{1}{4}\,G\{\rho_t(R)\},
\label{eq17.2.11}
\ee
and
\be
\frac{1}{\sigma_z^2}=\frac{1}{\sigma_\theta^2}+\frac{1}{\sigma_R^2},
\label{eq17.2.12}
\ee
and the Lindblad equation
\be
\frac{\sigma_\theta^2}{\sigma_R^2} = \frac{-B}{A-B}.
\label{eq17.2.13}
\ee
Formulas (\ref{eq17.2.11}) - (\ref{eq17.2.13}) are derived for flat
subsystems. However, the calculations show that in the vicinity of the
Sun these formulas can be applied to less flattened subsystems as
well. Therefore, we can assume that the use of these formulas in
constructing the model of the M31 galaxy is not associated with large
errors.

In the Lindblad formula for planar subsystems, the Oort parameters $A$
and $B$ can be expressed through the circular velocity $\omega(R) =
V_\theta/R$ and the logarithmic gradient of the circular velocity function
$G\{v(R)\}$, with the circular velocity function defined by the formula
\citep{Einasto:1968aa} (see Eq.~\ref{eq15.3})
\be
v(R)= \frac{V^2\,R}{G\,\mm{M}},
\label{eq17.2.14}
\ee
where $G$ is the gravitation constant, and $\mm{M}$ the mass of the system. We get
\be
A(R)=\omega(R)=\frac{3-G\{v(R)\}}{4}
\label{eq17.2.15}
\ee
and
\be
B(R)= -\omega(R)\frac{1+G\{v(R)\}}{4},
\label{eq17.2.16}
\ee
and using Eq.~(\ref{eq17.2.12}) and (\ref{eq17.2.13})
\be
k_\theta = \frac{\sigma_\theta^2}{\sigma_R^2} = \frac{3-G\{v(R)\}}{4}
\label{eq17.2.17}
\ee
\be
k_z=\frac{\sigma_z^2}{\sigma_R^2} = \frac{1+G\{v(R)\}}{5+G\{v(R)\}}.
\label{eq17.2.18}
\ee

After these substitutions we get for $p$ the expression
\be
p= {3-G\{v\} \over 4} - {G\{\rho_t\} \over 5+G\{v\}} + G\{\rho\} + G\{\sigma_R^2\}.
  \label{eq17.2.19}
  \ee
The same expression is also valid for spherical subsystems if 
we suppose that $V_\theta \approx V$. In a more exact consideration,
in the expression for $p$
the circular velocity should be replaced by the rotational velocity of
the subsystem $V_\theta$.

The circular velocity $V(R)$ is related to the mass function by the
equation \citep{Einasto:1968aa}
\be
V^2(R)=\frac{G}{R} \int_0^R\,{\mu(a)\dd{a}  \over \sqrt{1-(ea/R)^2}},
\label{eq17.2.20}
\ee
where $G$ is the gravitation constant and  $e^2 - 1 - \epsilon^2$.

Finally, we also use the Poisson equation in terms of Oort-Kuzmin
parameters \citep{Kuzmin:1952ab}
\be
4\pi\,G\rho_t= C^2 -2\,(A^2 - B^2),
\label{eq17.2.21}
\ee
where $C$ is the Kuzmin parameter. For flat populations, it can be
found from the relation
\be
C=\sigma_z/\zeta,
\label{eq17.2.22}
\ee
where $\sigma_z$ and $\zeta$ are dispersions of $z$-velocities and
$z$-coordinates of stars, respectively.

\section{Choosing the form of the main description function}

A model of mass and luminosity distribution of the galaxy is
completely defined if the luminosity distribution of its subsystems is
known, as well as the mass-to-light ratio of 
subsystems. Using Eqs.  (\ref{eq17.2.1}), (\ref{eq17.2.4}) and (\ref{eq17.2.5}) the mass
distribution of the system can be found, and by Eqs. (\ref{eq17.2.3}) and
(\ref{eq17.2.20}) the circular velocity $V$ can be calculated. Due to the
proportionality of $\rho$ and $l_S$, we shall write formulas
only for one of them, $\rho$.

To build a hydrodynamical model, in addition to the mass distribution
function and related functions, we must also know the rotation
velocity and velocity dispersion of subsystems. Then by
Eqs. (\ref{eq17.2.8}), (\ref{eq17.2.12}) and (\ref{eq17.2.13}) we can
calculate other hydrodynamical functions of interest. The Poisson
equation (\ref{eq17.2.21}) allows us to check the obtained results.

To summarise, the hydrodynamical model is completely defined by specifying 
functions $\rho(a)$, $V_\theta(R)$, and parameters $\epsilon$ and $f$ of all components
of the galaxy.

The representativeness of the model depends essentially on the choice
of the type of these basic description functions. It is natural to
demand that description functions have no sharp jumps and kinks,
that $\rho(a) \ge 0$ and $\sigma_R^2(R)\ge 0$. Since real stellar systems have finite sizes
(due to the perturbing action of neighbouring systems), it is desirable
to choose for $\rho(a)$ an expression that decreases fast enough with
increasing $a$.  On the other hand, $\rho(a)$ 
should not decrease too fast, since in this case the circular motion is unstable. 

Taking into account all these considerations, we chose for $\rho(a)$ a
generalised exponential function \citep{Einasto:1965aa,
  Einasto:1968ac}
\be
\rho(a)= \rho_0\,\exp\left[-\left(\frac{a}{a_0k}\right)^\nu\right],
\label{eq17.3.1}
\ee
where $\rho_0$ (central density) and $a_0$ (effective radius) are
scale parameters,  
$k$  is a dimensionless normalising parameter (see
\citet{Einasto:1968ac},  and $\nu$  is a 
structural parameter of the model, determining the concentration of
mass to the centre. Structural parameters of the model 
also include $\epsilon$, which determines the thickness of the model.

In the case of the flat component, the simple ellipsoidal model
represents the density distribution poorly. It is known that in the central
regions of galaxies there are no representatives of flat subsystems
— emission nebulae and stellar associations  \citep{Arp:1964wi,
 van-den-Bergh:1964wl}. To take this 
circumstance into account in our model, we used an artificial method:
the density of the component was calculated as the difference of two 
ellipsoidal models:
\be
\rho(a)= \rho_+(a/a_0,\rho_0,\nu,\epsilon) -
\rho_{-}(a/(a_0\kappa),\rho_0,\nu,\kappa\epsilon),
\label{eq17.3.2}
\ee
where $\kappa >1$.  With such $\rho(a)$ automatically
$\rho_{R=0}(z)=0$. With a suitable choice of $\rho(a)$, the
conditions $\rho(a) \ge 0$ and $\partial \rho/\partial z^2 <0$  at $z
\neq 0$ are still satisfied.

In the framework of the preliminary model of the M31 galaxy, it is
sufficient to fix  hydrodynamical functions only for the spherical
component, the core. In this case, the rotation speed of the component
can be represented by the formula \citep{Brandt:1965ty}
\be
V_\theta = V_0 {R  \over [1+(R/R_0)^n]^{3/2n}  }.
\label{eq17.3.3}
\ee

\section{Observational data}

{\bf 1. The photometric data} on the brightness distribution along
the major and minor axes of the M31 galaxy were collected from all
available sources. Only the data that could be reduced to the UBV
system were used (see \citet{de-Vaucouleurs:1958uc} and 
\citet{Kinman:1965vv}).  In order to build a composite model, it would
be desirable to have photometric data in different colours. However, a
sufficiently wide brightness interval is covered only by the data in
blue rays, so we had to limit ourselves to a single photometric system
B.  We used from photographic observations results by
\citet{Redman:1937tg}, \citet{Fricke:1954tm}, \citet{Johnson:1961aa},
\citet{Richter:1963ud}, and from photoelectric observations those by
\citet{Thiessen:1955aa} and \citet{de-Vaucouleurs:1958uc}.

To form a summary curve of the brightness distribution of the M31
galaxy, the data for the NE and SW halves of the major axis, as well
as for the NW and SE halves of the minor axis were combined, and the
corresponding brightnesses were averaged. The agreement between the NE
and SW half-axes is good everywhere. The agreement between the NW and
SE hemispheres is less good, especially in the strong absorption
region of the NW hemisphere at 4.5 to 17' from the centre. In the
preliminary M31 model, this region was excluded. Similarly, the general uniform
absorption was not taken into account, neither in our Galaxy nor in the M31 galaxy.

The derived summary values of the projected luminosity along 
the major axis are indicated by dots in
Fig.~\ref{Fig17.1}. Fig.~\ref{Fig17.2} shows the change of the ratio
of the semi-axes of the isophotes $E$. As an argument we use
$R^{1/3}$, and respective angular distance from the centre of the
system along the major axis, expressed in arc minutes.

{\begin{figure*}[h] 
\centering 
\hspace{2mm}
\resizebox{0.70\textwidth}{!}{\includegraphics*{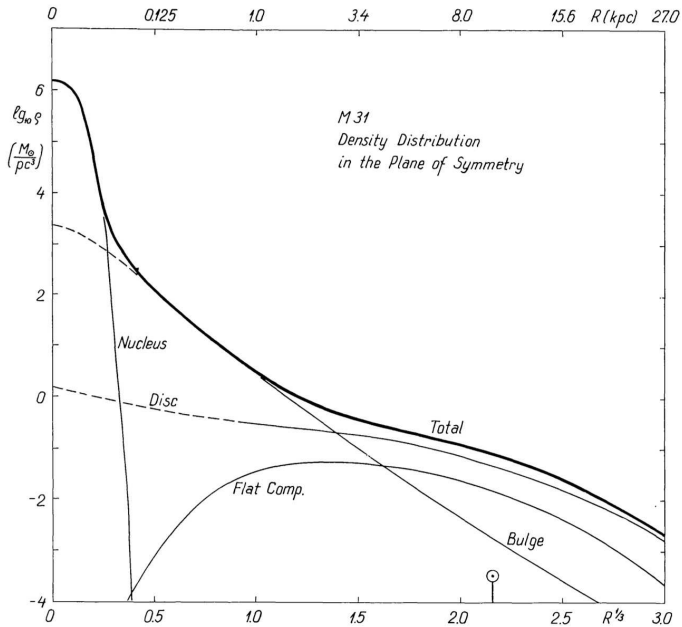}}
\caption{Luminosity distribution along the major axis of
  M31 \citep{Einasto:1970vz}.  } 
  \label{Fig17.1}
\end{figure*} 
}

{\begin{figure*}[h] 
\centering 
\hspace{2mm}
\resizebox{0.50\textwidth}{!}{\includegraphics*{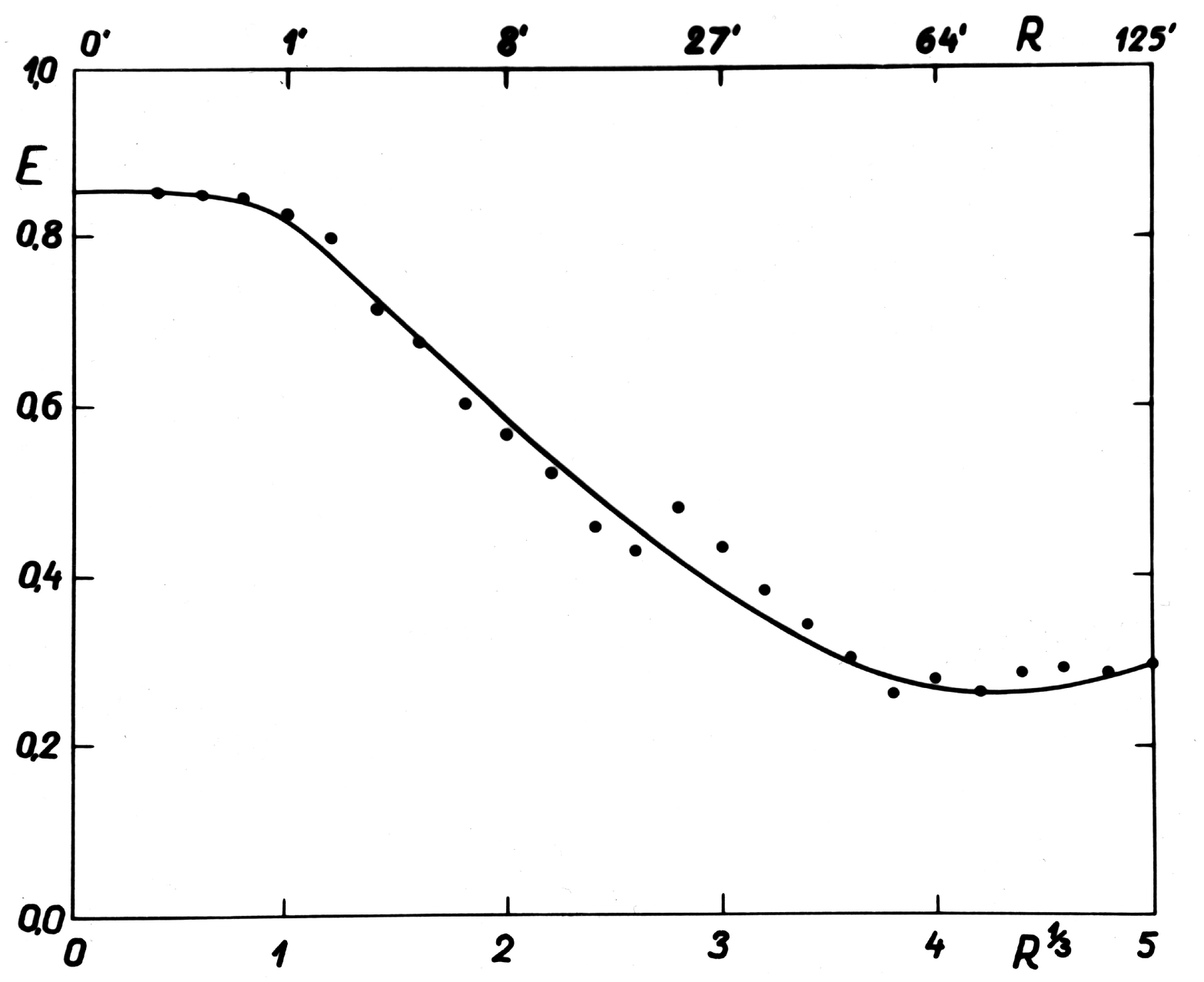}}
\caption{ Distribution of the axial ratio
  $\epsilon$. } 
  \label{Fig17.2}
\end{figure*} 
}

{\bf 2. The rotational velocity} was determined from optical 
(\citet{Babcock:1939wf}, \citet{Wyse:1942wd}, \citet{Mayall:1951wg}
and \citet{Lallemand:1960to})  and radio
data  (\citet{van-de-Hulst:1957vu}, \citet{Burke:1963vr},
\citet{Argyle:1965us}, \citet{Gottesman:1966vd} and \citet{Roberts:1966aa}).
In the central region ($R \le 10'$) only optical data were used,  
because the velocity changes rapidly and radio observations have too
low resolution. In the distance range $10' < R \le 50'$ both optical and radio
data were used, in the region $R > 50'$ only the radio data were used as
more accurate. The mutual consistency of the radio data obtained by
different authors is very good. For the velocity of the galaxy
centroid a value  $-300$ km/s was chosen. With this choice
the regions of maximum velocity on both sides of the centre agree
best. The observed rotational velocities are depicted in
Fig.~\ref{Fig17.3} by dots.

{\bf 3. The distance} of the M31 galaxy was assumed to be $d = 692$~kpc according
to the true distance modulus $(m - M)_0 = 24.2$ \citep{Baade:1963ud}.

{\bf 4. The inclination} of the galaxy was determined according to
the apparent ratio of semi-axes isophotes $E$, and from the apparent
distribution of emission nebulae \citep{Arp:1964}. The found value,
$i = 77.^\circ2$, agrees well with the \citet{Baade:1963ud} estimate,
$i = 77.^\circ3$, and with the result by \citep{Arp:1964},
$74^\circ<i<79^\circ$).

{\bf 5. Additional data} can be obtained for the central, brightest
regions of the system. The Minkowski velocity dispersion of the core
stars is $\sigma_R = 225$~ k/sec \citep{Lohmann:1964uw}, and the
\citet{Spinrad:1966wh} mass-to-light ratio  $f=16.7$.

{\begin{figure*}[h] 
\centering 
\resizebox{0.52\textwidth}{!}{\includegraphics*{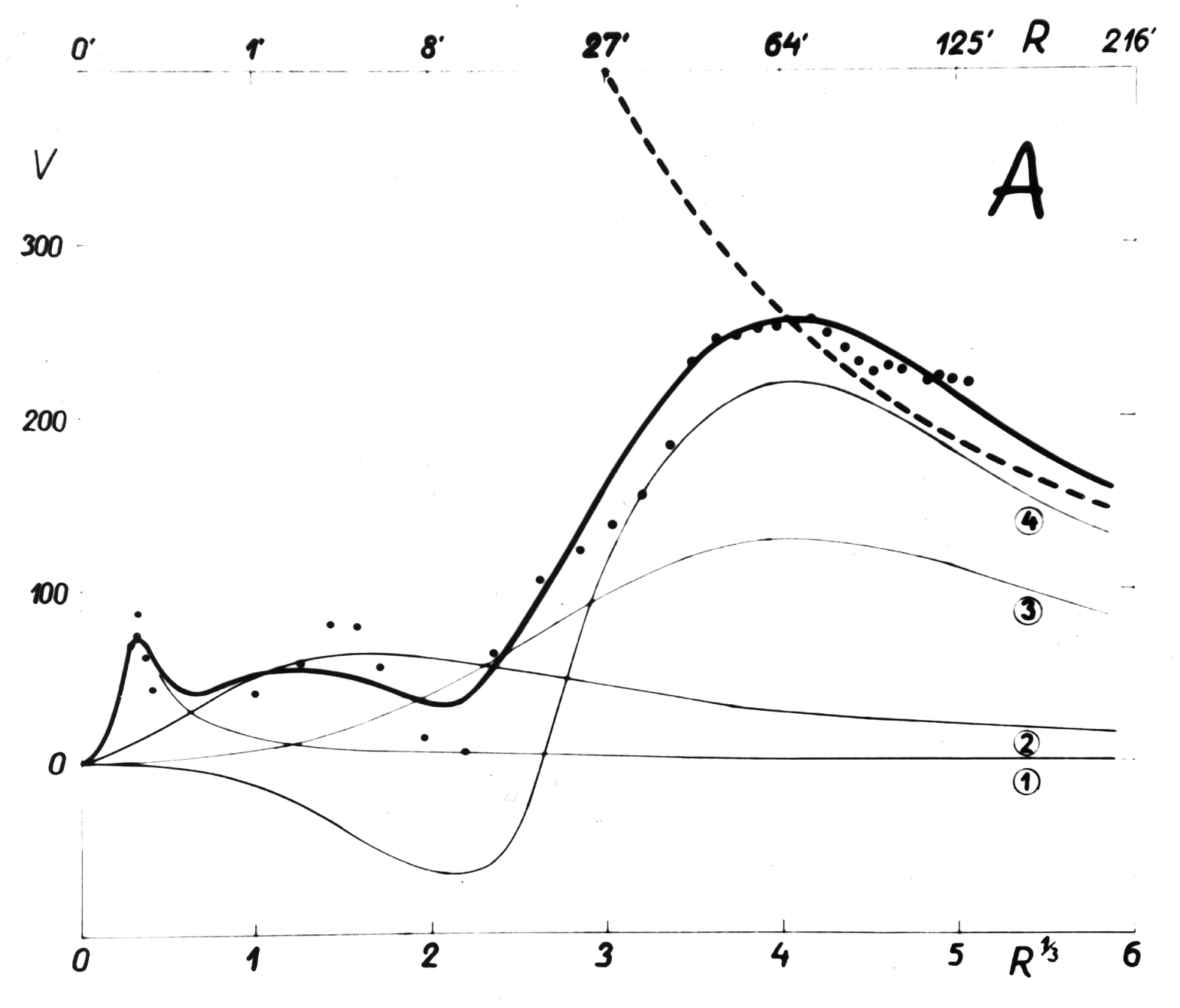}}\\
\resizebox{0.52\textwidth}{!}{\includegraphics*{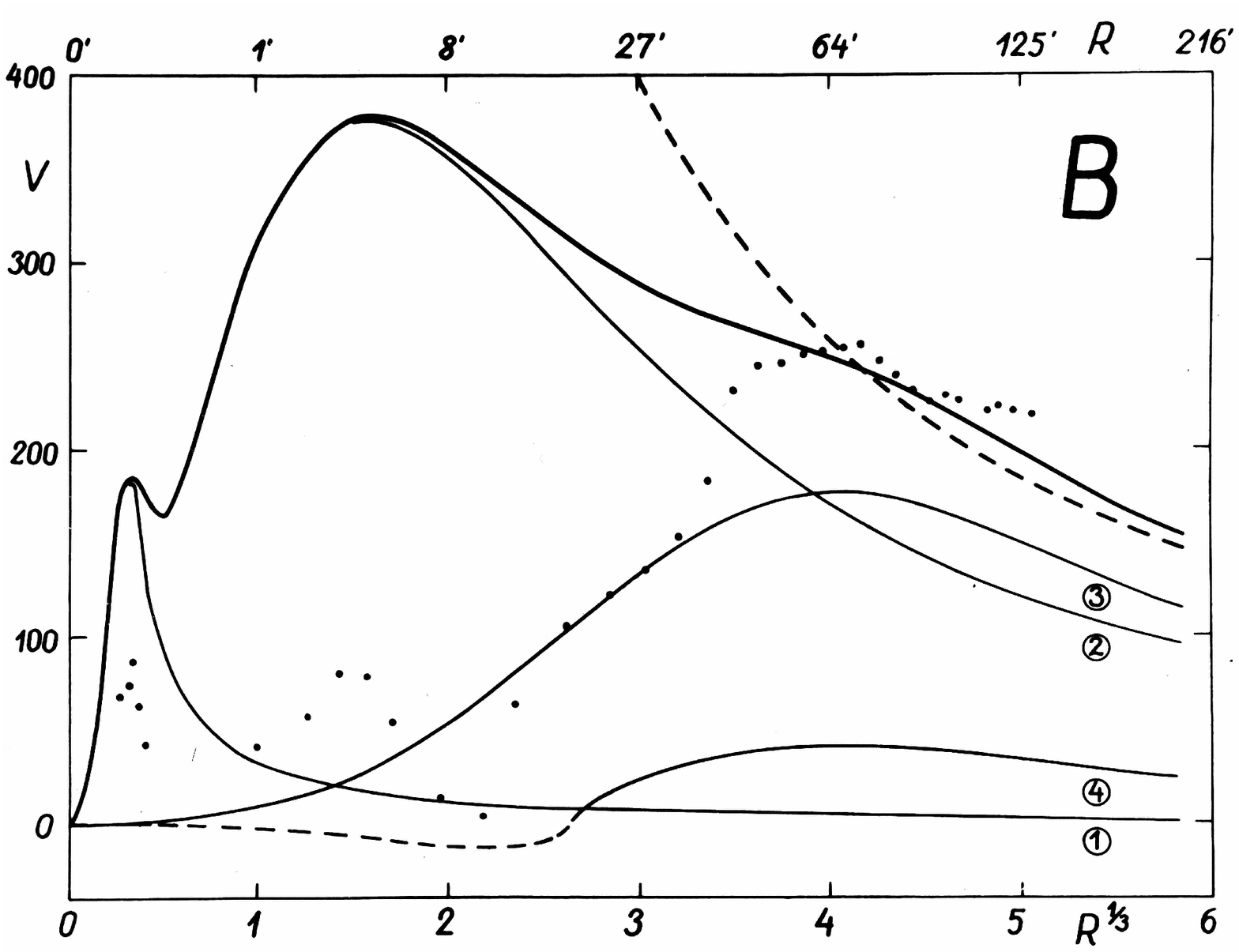}}
\caption{Distribution of the rotation velocity.
  Masses of components are taken for variants A and B, see Table
  \ref{Tab17.1}. 
Thin curves show contributions of components, the bold solid curve shows 
the rotation curve of the whole model. The dashed curve gives the Keplerian
rotation curve, corresponding to a point source with a mass, equal to
the mass of the model} 
  \label{Fig17.3}
\end{figure*} 
}

  \section{Model construction}

The construction of a model with a fixed analytical form of the main
description functions is reduced to the determination of parameters
of these functions. In this case we need to find the following 
parameters for all four components of the model:
\\
a) scale parameters $l_0$ and $a_0$;\\
b) structural parameters $\nu$ and $\epsilon$;\\
c) dynamical parameter $f$.\\

For the spherical component it is also necessary to know the
parameters of the rotation law (\ref{eq17.3.3}).

The analysis of observational data showed that both scale and
structural parameters of the components cannot be found with
sufficient accuracy from the rotation velocity. In addition to the
mass distribution, the relative motions of the stars also influence
the rotation velocity. \citet{de-Vaucouleurs:1958uc} showed that model
parameters can be found quite successfully from the photometric data.

Using these considerations we determined parameters $l_0,~a_0,~\nu_0$  and $\epsilon$
from the photometric material. The practical procedure was reduced to
the following.

As the first step, we calculated for a number of values of the parameter
$\nu$ the normalised projected density function $L^0(\alpha)$, and the
velocity function $v^0(\alpha)$, where $\alpha$ is the dimensionless
normalised distance (semi-major axis). The normalisation was performed
so that moments of order $-1$ and 0 of the mass function $\mu^0(a)$
were equal to one (see \citet{Einasto:1965aa}).

The calculated
functions $L^0(\alpha)$ were plotted in the logarithmic scale
$\log\,L^0(\log\,\alpha)$. Similar plots were made for the observed
projected luminosity M31 on the major and minor axes.
Since the scale transformations (offsets on the logarithmic scale) do
not change the shape of the $\log\,L^0(\log\,\alpha)$  curve, the
parameter $\nu$ can be found by the comparison of the model curve
with the observed one. The parameter of apparent flatness $E$ can be found
from the comparison of density distributions on minor and major
axes. The true axial ratio of semiaxis $\epsilon$ can be found from
Eq.~(\ref{eq17.2.7}). 

The practical difficulty in the calculation of the model is due to the
need to find parameters of all four components
simultaneously. However, density distributions of components are very
different, see Fig.~\ref{Fig17.1}. After several trials optimal
parameters of components were found.

This procedure cannot be applied to find the $\epsilon$ parameter for
the flat component. In this case $\epsilon$ was chosen in such a way that
the effective half-thickness of the components
\be
z_e= \frac{1}{2}\,\frac{P(R)}{\rho(R)},
\label{eq17.5.1}
\ee
has an acceptable value. Here $P(R)$ is the projected mass density of
the system.

\begin{table*}[h]
\centering    
\caption{Parameters of the components of M31} 
\begin{tabular}{lccccccc}
  \hline  \hline
  Quantity&Unit&Total&Nucleus&Bulge&Disc&Flat+&Flat$-$\\
  \hline
$\epsilon$&&            &0.84&0.57&0.09&0.01&0.02\\
  $\nu$&     &            & 1 & 1/4 &   1  &   1 &  1 \\
  $k$&         &            &0.5&$1.26\times10^{-4}$& 0.5&0.5&0.5\\
$a_0$       &kpc  &      &0.025&5&50&40&20\\
$L$&        $10^9\,L_\odot$                &13.13&0.003&4.95&6.46&2.29&$-0.57$\\
$\mm{M}_A$&$10^9\,\mm{M}_\odot$&201&0.009&2.4&58&188&$-47$\\
$\mm{M}_B$&$10^9\,\mm{M}_\odot$&201&0.05&85.5&111.5&5.73&$-1.43$\\
$f_A$&                                               &15.3&2.5&0.5&9&82&82\\         
$f_B$&                                                &15.3&17.3&17.3&17.3&2.5&2.5\\         
  \hline
\label{Tab17.1}   
\end{tabular}
\end{table*}

Parameter values for components are given in Table~\ref{Tab17.1} 
(mass-to-light ratios $f$ are in photometric system B). The
model is presented in a graphical way in Figs.~\ref{Fig17.1} and
\ref{Fig17.2}. In Fig.~\ref{Fig17.1} thin lines show the contributions
of components to the total luminosity (along the long axis), the bold line
shows the summed total luminosity.  We see that the model represents
observations (shown as points) rather well. Differences between the model
and observations are due to the fact that populations do not contain
information on individual spiral arms but only their mean structure.

Mass-to-light ratios of populations can be derived in two ways: A)
using the rotation speed; B) using the spectral information of stellar
components from independent sources.

In the first case, it is assumed that the rotation velocity is
equal to the circular velocity, $V_\theta =V$. In practical use, this
means that by a trial-and-error procedure values of $f$ are chosen
(at fixed parameter $a_0$ values), which yield for the circular
velocity values in accordance with the observed rotation velocity. Values
of $f$ found in this way are given in Table~\ref{Tab17.1} (variant A), respective
rotation curves are shown in Fig~\ref{Fig17.3}. As in
Fig.~\ref{Fig17.1}, thin lines show the contribution of individual
components, the bold line is the total calculated rotation curve, and
the dots are observations. The dashed line gives the Keplerian
rotation curve, corresponding to a point source with a mass, equal to
the mass of our model. We see that the theoretical curve represents
observations well.

It should be said that the flat component of the model has a toroidal
form. Inside the toroid, the attraction vector is directed not towards
the centre of the system but towards the nearest side of the
toroid. Therefore, in this region the velocity function is negative,
and circular motion is impossible (unless there are other components
that compensate for the negative region of the velocity function). In
Fig.~\ref{Fig17.3} this region of the component contribution to the
rotation curve is shown by a dashed line\footnote{Central holes in
  discs of spiral  galaxies were studied in more detail by
  \citet{Einasto:1980aa}}. 

In the second version B the parameters were found as follows. For the
nucleus of M31 $f$ was obtained by \citet{Spinrad:1966wh}  from spectral
observations, $f=16.7$. The flat component consists mainly of
hydrogen, whose mass according to \citet{van-de-Hulst:1957vu},
\citet{Argyle:1965us}, \citet{Gottesman:1966vd}, and \citet{Roberts:1966aa}
 can be taken as equal to $\mm{M}=3.7\times\,10^9~M_\odot$. 
The mass of the flat component of stars can be
estimated from the integral luminosity of the component, and the
initial luminosity and mass function (\citet{Salpeter:1955},
\citet{Sandage:1957aa}). As a result, we
obtained for the mass of the component $\mm{M} = 4.3\times\,10^9
~M_\odot$.  Since $L =
1.72\times10^9L_\odot$, we get $f = 2.5$. Knowing the luminosity of all components, and $f$
of the nucleus and the flat component, it is not difficult to
calculate $f$ of the disc. The result coincides almost exactly with the value of $f$
for the nucleus, found by Spinrad. Therefore, we assumed that all
components, except the flat component, have the identical mass-to-light ratio 
 (variant B in Table \ref{Tab17.1}). The corresponding rotation curves
are given in Fig.~\ref{Fig17.3}.

{\begin{figure*}[h] 
\centering 
\hspace{2mm}
\resizebox{0.50\textwidth}{!}{\includegraphics*{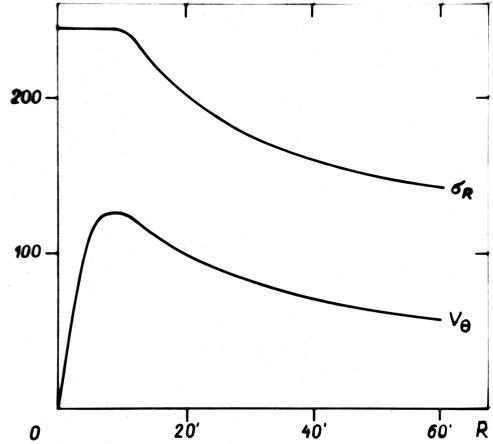}}
\caption{ Rotation velocity $V_\theta$ and velocity
 dispersion $\sigma_R$ for the spherical subsystem. } 
  \label{Fig17.5}
\end{figure*} 
}

In this variant, in the region $R < 50'$, the circular velocity is
noticeably greater than the rotation speed, and reaches  at $R =
4'$ a maximum value, $V = 380$~ km/sec. The observed rotational velocity
at $R = 4'$ is only $V_\theta = 80$ km/sec. Thus, a question arises, can the
velocity centroid shift reach such a large value, $\Delta\,V = V- V_\theta=  300$
km/sec?

Calculations using formulas (\ref{eq17.2.8}) and (\ref{eq17.2.19})
showed that, within the accuracy of the initial data, such a shift can
indeed be explained.  Fig.~\ref{Fig17.5} shows one possible variant of
the rotational velocity $V_\theta$, and the velocity dispersion
$\sigma$ of the spherical component of M31. For $V_\theta$, the
profile (\ref{eq17.3.3}) was taken; $\sigma_R$ was determined by
Eq. (\ref{eq17.2.8}). In our calculations, we took into account the
fact that the observed $V_\theta$ and $\sigma_R$ are lower than their
actual maximum values. This is caused by the fact that we observe some
mean value of the dispersion and velocity \citep{van-Houten:1961ug}
\be
V_\theta(R)= {\int_{-\infty}^\infty\,V_\theta(R,Z)\,l(R,Z)\dd{Z}
  \over \int_{-\infty}^\infty\,l(R,Z)\dd{Z} },
\label{eq17.5.2}
\ee
where $Z$ is the coordinate along the line of sight. In the region
$R<10'$, it was also necessary to slightly change the course of
the density gradient, which is also quite acceptable within the
accuracy of the available data.

{\begin{table*}[h] 
\caption{Description functions of the M31 model} 
\centering 
\hspace{2mm}
\resizebox{0.95\textwidth}{!}{\includegraphics*{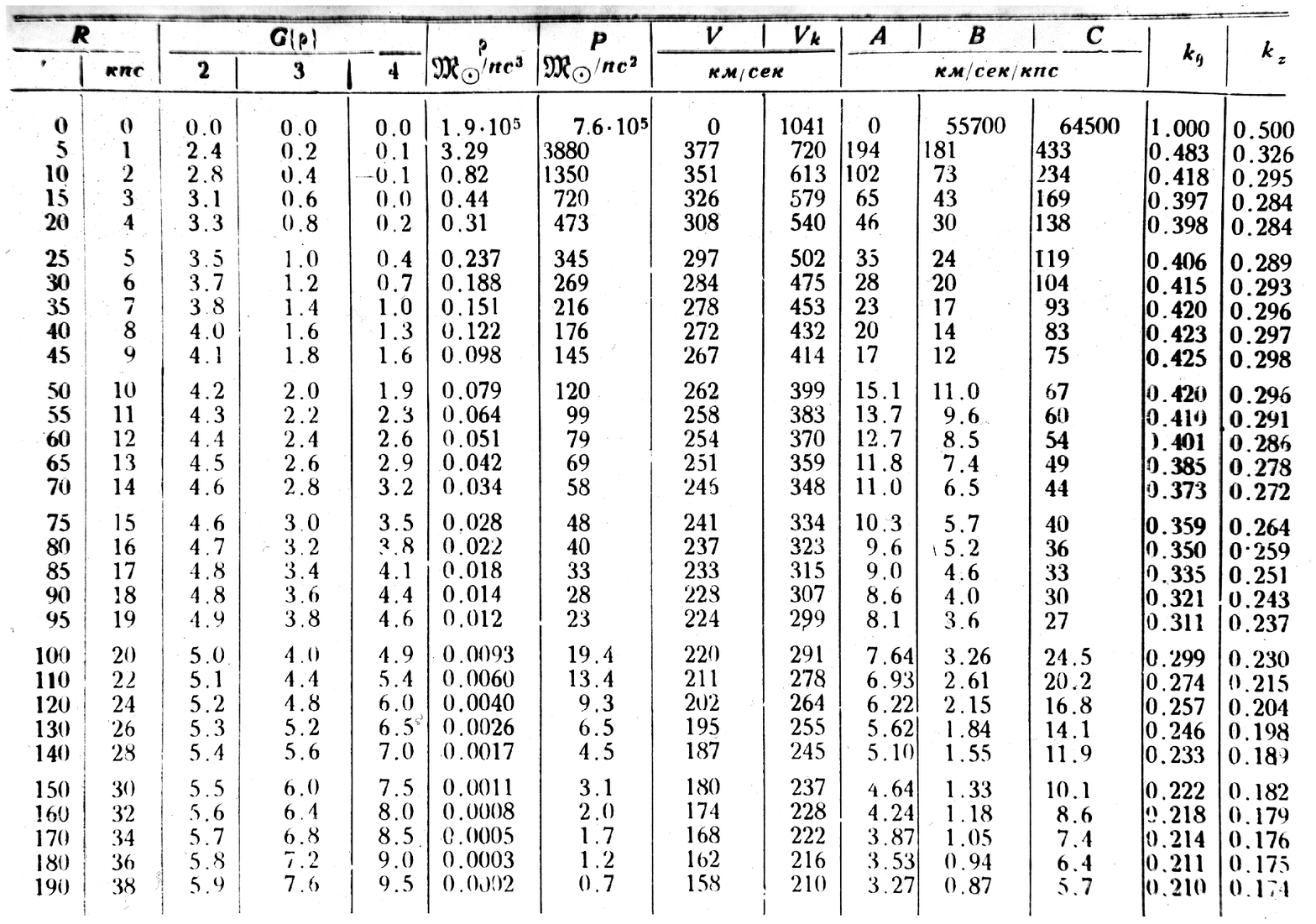}}
\label{Tab17.2}
\end{table*} 
}

\begin{table*}[h]
\centering    
\caption{Parameters of models of M31} 
\begin{tabular}{lcccc}
  \hline  \hline
  Author&$n$&$d$&$\mm{M}$&$\mm{M}^\ast$\\
              &             &kpc&$10^9\,M_\odot$&$10^9\,M_\odot$\\
  \hline
\citet{Lohmann:1964uw}     &3/2&  460  & 330& 500 \\
\citet{Schwarzschild:1954to}& &    460   &140&210 \\
\citet{Schmidt:1957uy}        &&    630   & 338 & 370 \\
\citet{Takase:1957vs}          &  &  540    &200 & 260 \\
\citet{Poveda:1958ui}          &  & 500 & 200 & 280 \\
\citet{Brandt:1960}             &3/2&600& 370 & 430 \\
\citet{Brandt:1965ty}          &3/2,~3,~10&630&580&640\\
\citet{Gottesman:1966vd}  & 3/2&630&480& 530\\
\citet{Roberts:1966aa}      & 3/2,\,3&690 &220&220\\
\citet{Einasto:1969aa}       &   & 690&200&200\\
  \hline
\label{Tab17.3}   
\end{tabular}
\\
Notes: $n=$ is the structural parameter of the generalised Bottlinger
model, Eq. (\ref{eq18.22}); $d$ is distance of M31; $\mm{M}^\ast$ is
mass of M31, reduced to  distance $d=690$~kpc.
\end{table*}

{\begin{figure*}[h] 
\centering 
\resizebox{0.48\textwidth}{!}{\includegraphics*{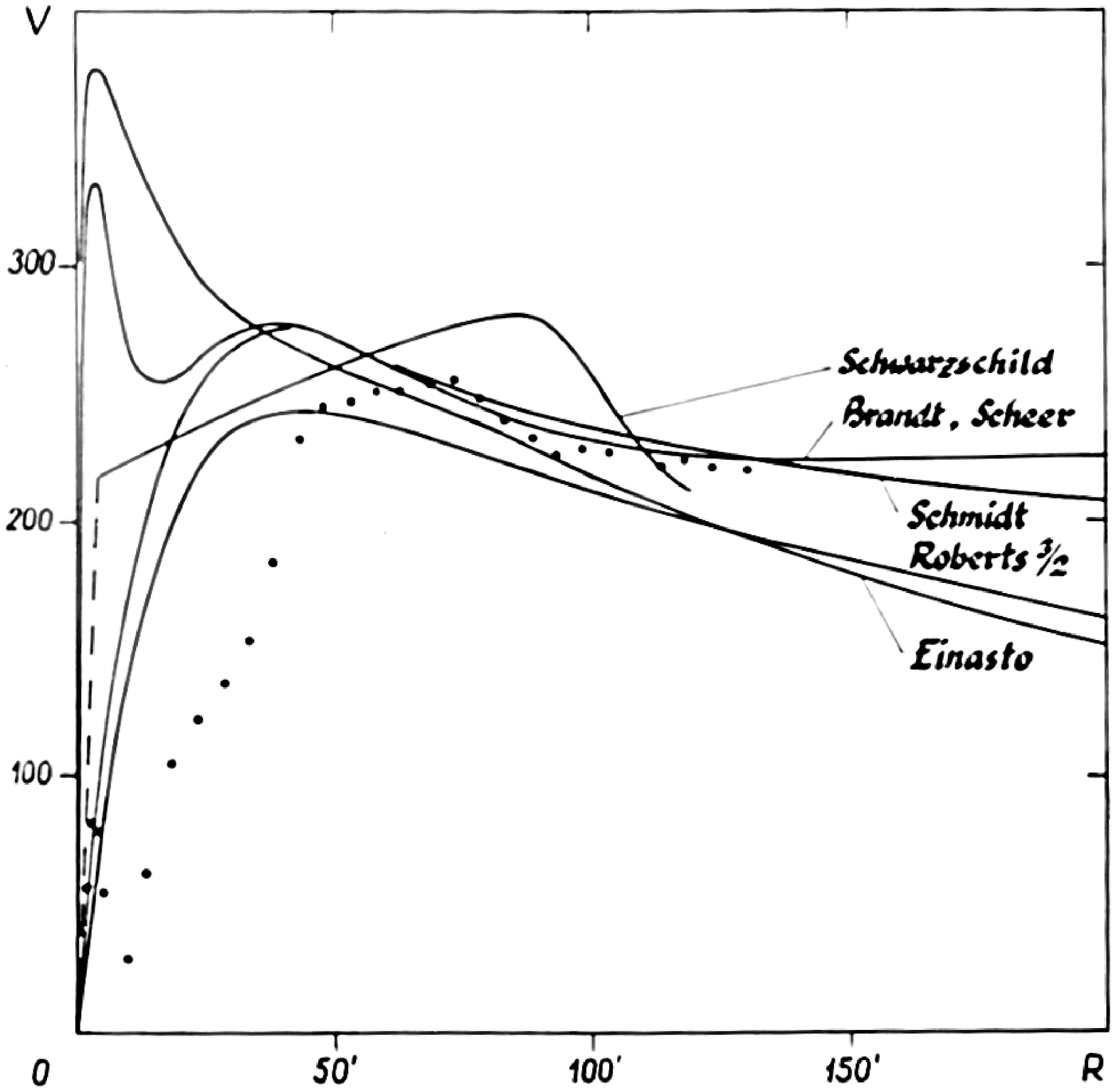}}
\resizebox{0.48\textwidth}{!}{\includegraphics*{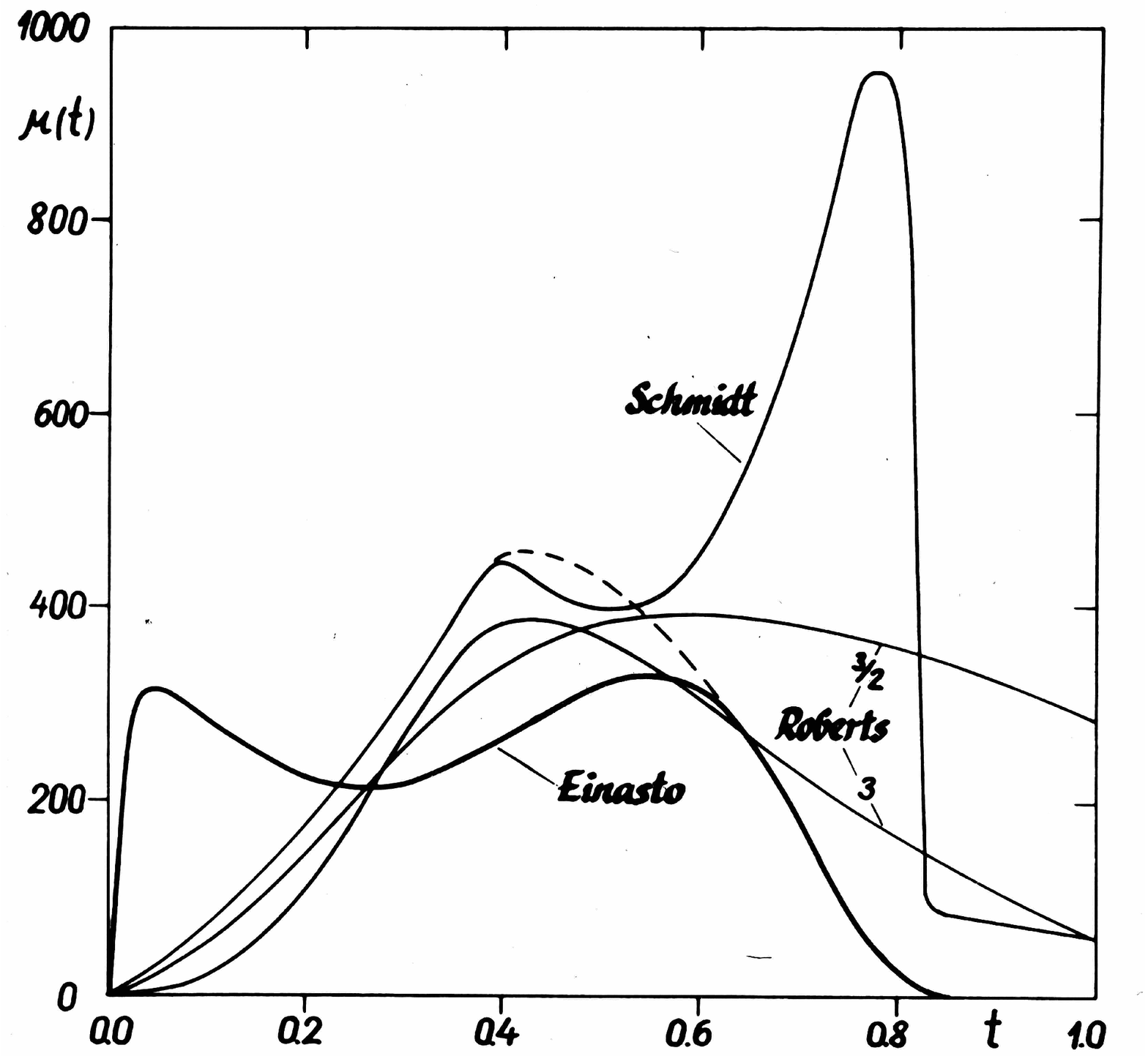}}
\caption{ {\em Left.} Circular velocity $V(R)$ of models by
  \citet{Schwarzschild:1954to}, \citet{Brandt:1965ty},
  \citet{Schmidt:1957uy}, \citet{Roberts:1966aa} ($n=3/2$) and
  \citet{Einasto:1969aa}  (variant B).  {\em Right:} Mass function
  $\mu(t)$ for models by  \citet{Schmidt:1957uy}, \citet{Roberts:1966aa} ($n=3/2$) and
  \citet{Einasto:1969aa}  (variant B). Mass is given in units
  $10^9\,\mm{M}_\odot$.  If the Schmidt model is extrapolated
  according to the dashed line, then its mass is equal to the mass of
  our model.
} 
  \label{Fig17.6}
\end{figure*} 
}

{\begin{figure*}[h] 
\centering 
\resizebox{0.48\textwidth}{!}{\includegraphics*{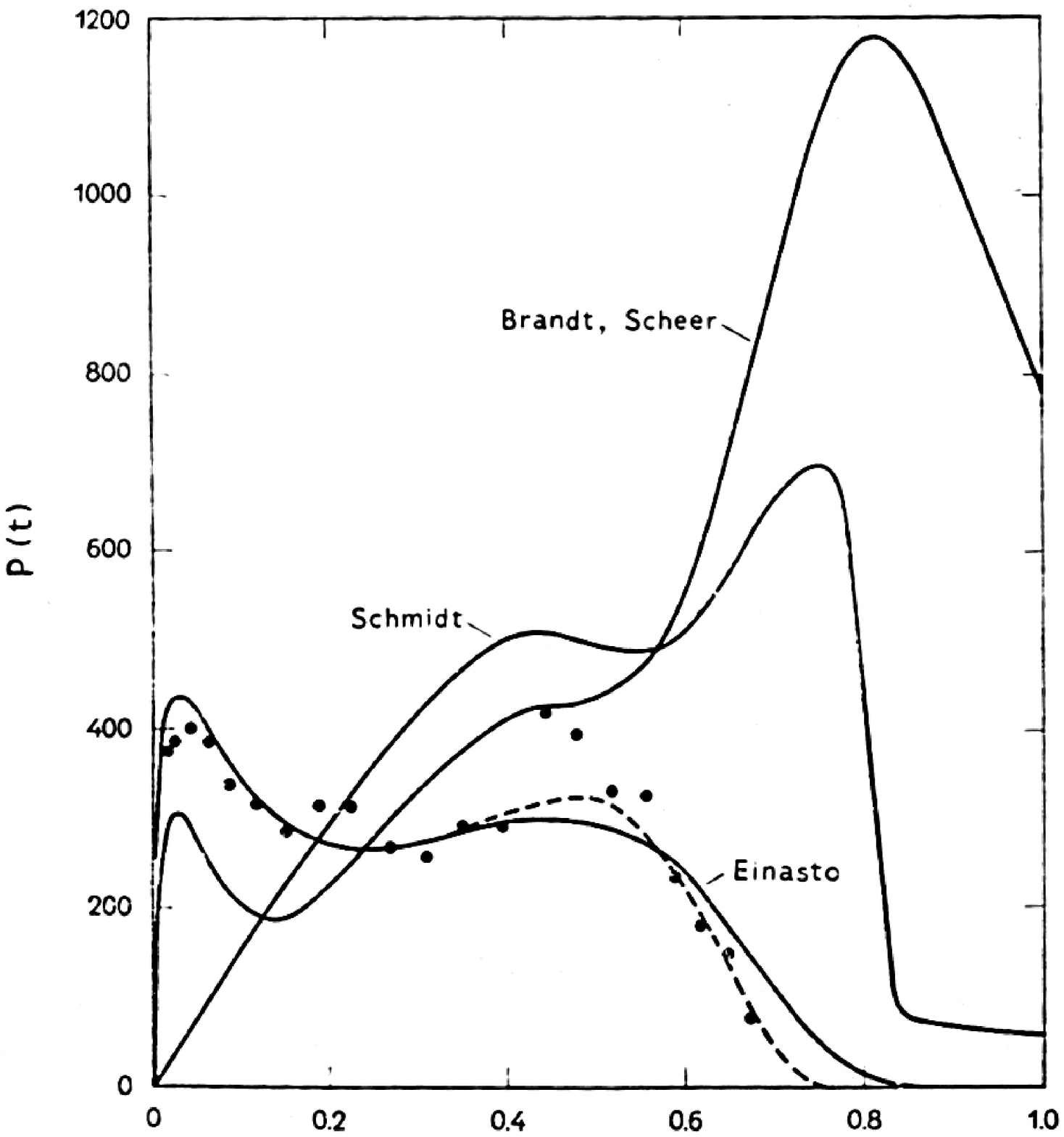}}
\resizebox{0.48\textwidth}{!}{\includegraphics*{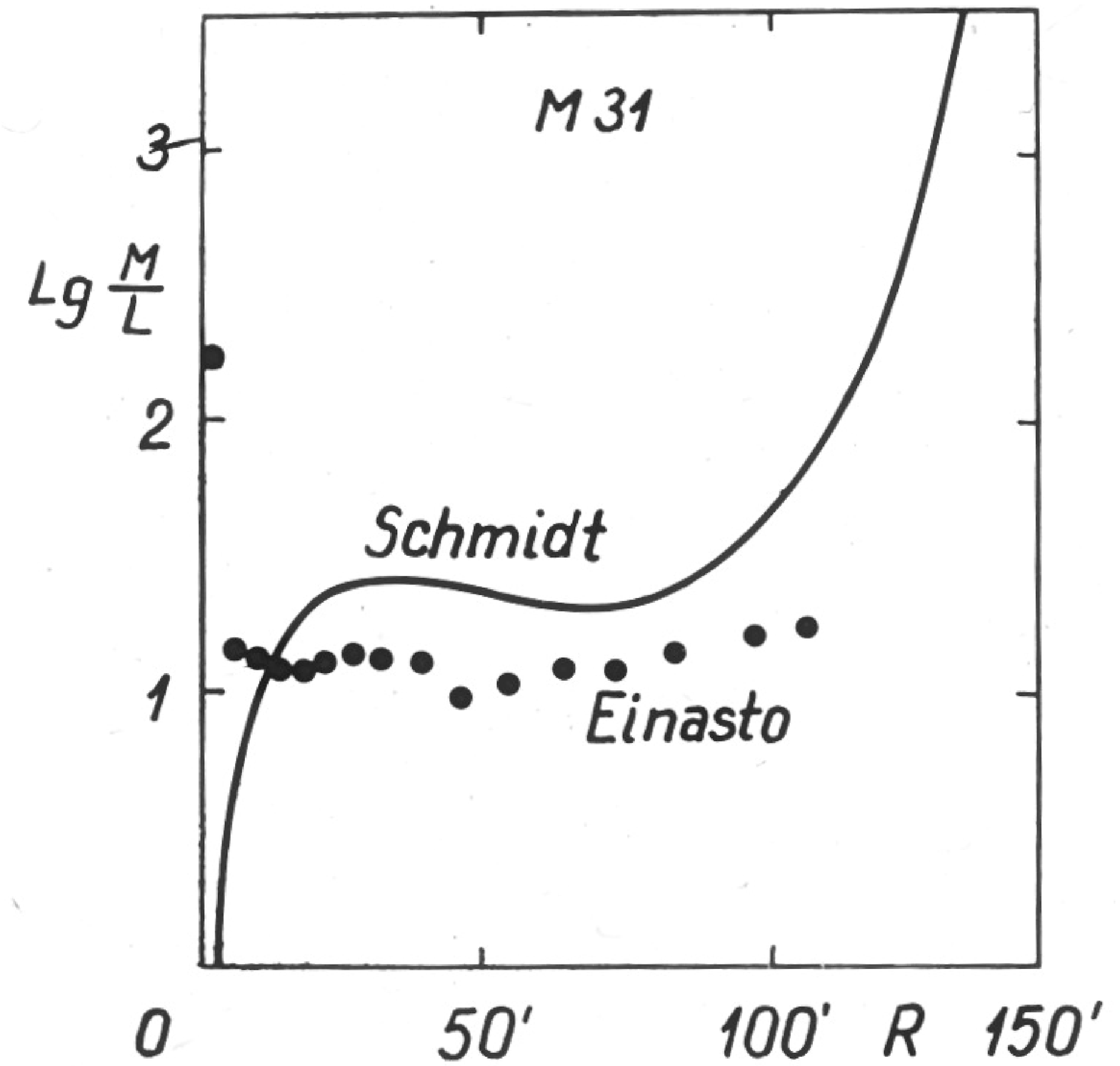}}
\caption{  {\em Left.} Distribution of the projected density $P(t)$ as
  function of $t$ for model by \citet{Brandt:1965ty},
  \citet{Schmidt:1957uy}, and \citet{Einasto:1969aa} (both variants)
  in units of $10^9\,M_\odot$. {\em Right:} Mass-to-light ratio
  $f$ of models by \citet{Schmidt:1957uy} (solid curve) and
  \citet{Einasto:1969aa} (dots). }
  \label{Fig17.8}
\end{figure*} 
}

We  conclude that the dynamical definition of $f$ agrees well with the
spectral definition in the variant B.
Calculated values of the
description function are shown in Table~\ref{Tab17.2}. Here, $V_k$ denotes the
critical velocity, calculated from the velocity function $v(R)$, using
the formula \citep{Einasto:1965aa}
\be
V_k^2(R)=2\,\int_R^\infty\,\frac{v(x)\dd{x}}{x^2}.
\label{eq17.5.3}
\ee
In the Table, logarithmic gradients of the density of 
components $G\{\rho(R)\}$ are also given. In studies of the 
Galaxy,  the gradients $m(R)=-\partial\,\ln\,\rho(R)/\partial\,R$ can
be easily calculated  by
the equation
\be
m(R)=-\frac{Mod}{R}\,G\{\rho(R)\}.
\label{eq17.5.4}
\ee
From the Table \ref{Tab17.2}  we see that the values of the description functions of
the Andromeda galaxy at a distance of 10 kpc from the centre agree
very well with the system of local circumsolar parameters in our
Galaxy  \citep{Einasto:1964aa,Einasto:1964wx}.

\section{Analysis of the model}

In Table~\ref{Tab17.3} we give main parameters of M31 models, as found
by various authors.  Here $d$ is the distance of M31, accepted by
authors, $\mm{M}$ is the mass of M31 according to authors, and
$\mm{M}^\ast$ is the mass, reduced to distance $d=690$~kpc.
In left panel of Fig.~\ref{Fig17.6} we show
circular velocity $V(R)$ of models by \citet{Schwarzschild:1954to},
\citet{Brandt:1965ty}, \citet{Schmidt:1957uy}, and
\citet{Roberts:1966aa} in comparison with our model. In right panel of
Fig.~\ref{Fig17.6} we show the corresponding mass functions
\be
\mu_t(t)=(1+a)^2\,\mu(a)=(1+a)^2\,\sum_{k=1}^n\,\mu_k(a)
\label{eq17.6.1}
\ee
and in Fig.~\ref{Fig17.8} 
the projected on the symmetry plane of the system mass density
\be
P_t(t)=2\pi\,a\,(1+a)^2\,P(a)=2\pi\,a\,(1+a)^2\,\sum_{k=1}^n\,P_k(a).
\label{eq17.6.2}
\ee
In these formulas $n$ is the number of components of the model, and we
use as argument 
\be
t=\frac{a}{1+a},
\label{eq17.6.3}
\ee
where  $a$ is expressed in degrees. This argument is used to represent 
better  distributions on  the periphery
of the model.

A comparison of the models shows that there are differences in the
mass distribution both inside the model and in the peripheral regions
of the model.  Differences of the first kind change the mass-to-light 
parameter $f$; total masses of systems, $\mm{M}$, do not depend on them
much. Differences of the second kind influence both the mass-to-light
ratio $f$ and
the mass of the system. Since these differences are caused by
different reasons, let us consider them separately.

{\bf A. The structure of the inner regions of the model} is determined to a
 large extent by the way of treatment — whether the centroid velocity
 asymmetric shift will be taken into account or not. In most of the works
 cited above, as well as in our version A, the presence of
  asymmetric shift was ignored. This way of treatment meets the following
 objections.
 
1) The inner regions of the galaxy are dominated by the bulge, 
whose stars have a large velocity dispersion. This
is also confirmed by direct determination of the dispersion. Therefore, in
equation (\ref{eq17.2.8}), the second term on the left is greater than
the first one 
and neglecting it is not justified.

2) The assumption of $V_\theta = V$ leads to the conclusion that in
central regions of the galaxy the mass-to-light ratio $f$ is small
(see \citet{Schmidt:1957uy} and Figs. \ref{Fig17.3} variant A, and
\ref{Fig17.8}). The study of elliptical galaxies, however, shows that
the mass-to-light ratio of stars with enhanced metal content of the
population II has a rather large value ($f>10$, see Chapter 22). This is confirmed by
direct spectral observations \citep{Spinrad:1966wh}. Therefore,
the assumption of $V_\theta= V$ leads in the central regions of
galaxies to unacceptable values of the mass-to-light ratio.

We conclude that option A cannot be accepted. Variant B, on the other
hand, leads, as we have seen above, to a mass distribution that is
acceptable from both the dynamical and physical points of
view. Hence, it follows that the concentration of mass to the centre
of M31 is much larger than previously thought.

{\bf B. The structure of the outer regions of the model} is determined
mainly by the extrapolation of the velocity function. The
extrapolation can be done in two ways — by the velocity function or by
the mass function (calculating the velocity function from the 
known mass distribution).

In previous studies of the structure of galaxies, the first of these
methods is usually used. In this case, the extrapolation is performed
by one or another circular velocity profile,  parameters
of which are chosen according to the range of $R$, covered by observations. In Figs. \ref{Fig17.3}
and \ref{Fig17.6} we see that the observed rotational velocity M31 at $R > 100'$
decreases very slowly with increasing distance from the
centre. Therefore, most authors have taken the circular velocity
also with a very small radial gradient.  This means that there
are significant masses at the periphery of the model, as shown in
Figs. \ref{Fig17.6} and \ref{Fig17.8}.

The mass distribution is particularly well seen in Fig. \ref{Fig17.8}. It follows from the
definition of the function $P_t(t)$ that
$\int_{t_1}^{t_2}\,P_t (t)\dd{t}$  is equal
to the mass enclosed between concentric cylinders of radii $R_1=t_1\,(1-t_1)$
and $R_2 = t_2\,(1 -t_2)$. In the Schmidt and Brandt and Scheer models, over half
the mass lies outside the observed region $R > 2^\circ$  ($t > 0.67$), significant
mass is located even at very large distances ($t \approx1$). Since all authors
have assumed ellipsoidal mass distributions, the presence of a massive
halo also affects the distribution of $P_t(t)$ at small $t$ ($P_t$ is obtained
by integrating the spatial density over $z$ from $-\infty$ to $+\infty$).

The question arises, can such a mass distribution correspond to
reality?

The presence of significant masses on the periphery meets the
following objection. It is known that the perturbing action of
neighbouring galaxies leads to the conclusion that the sizes of all galaxies
are finite. The radii (outer limits) of stellar systems obtained by
extrapolation of photometric data agree well with the dynamical
estimate of radii   \citep{King:1962aa}. In the case of M31, the photometric radius of
the system is of the order of $R = 150'$. It is unlikely that galaxies
have an ``invisible'' massive halo outside the photometric boundary of
the system. Otherwise we would get fantastically large values of the
mass-to-light ratio $f$ at the periphery of the galaxies, as seen in
Fig.~\ref{Fig17.8} in the case of the Schmidt model. The assumption of a
significant increase in $f$ at the periphery of the model requires for
its explanation the presence of an active mechanism of ``sorting'' of
stars by mass, which seems unlikely.

If we accept the mass distribution according to our model, then the
calculated radial gradient of the circular velocity at $R > 100'$ 
is greater than the observed rotation velocity gradient. It seems to us
that in this case there is a local deviation of the motion of the
objects of the planar component from the circular motion. Deviations
of the order of $5 - 10$ km/sec from circular motion take place in our
Galaxy as well. Such deviations can also explain the asymmetry of the
velocity curve noted by many authors
\citep{Gottesman:1966vd,Roberts:1966aa}.

The densities $P_t(t)$, calculated from the photometric material, assuming
a constant mass-to-light ratio $f=15.3$, are shown by dots in
Fig.~\ref{Fig17.8}. A comparison of our model with the photometric data shows that
even our model has a too large halo. This was expected since we
assumed an unbounded exponential law for the density. The assumed
actual course of the projected density is shown by the dashed line.

The differences between our model and the points in Fig.~\ref{Fig17.8} are
obviously caused by the fact that the parameter $f$ is not constant but
has local deviations, in particular, in the spiral branches. Aligning
the points with the calculated curve $P_t(t)$ we can obtain the ``observed''
values of $f$. They are shown in the logarithmic scale in Fig.~\ref{Fig17.8}
(dots). We see that the assumption of constancy of the mean value of $f$
is not badly fulfilled. The regions of minima of $f$ correspond to the
main spiral branches of the galaxy.

So, we conclude that the increase of $f$ at the periphery as well as
the large masses of the M31 galaxy, deduced by several authors, probably
do not correspond to reality\footnote{As described in the Epilogue, the reasons for
   contradictions in M31 models are the absence of dark matter population in
  earlier models of galaxies,  and erroneous high value of the observed velocity
  dispersion near the center.}.

\vskip 5mm
\hfill May 1968
\vfill 

\chapter{Hydrodynamical model of M31}\label{ch18}

This Chapter presents our first attempt to calculate a hydrodynamical 
model of a galaxy — M31. The method was described earlier by
\citet{Einasto:1968aa, Einasto:1968ad, Einasto:1969ab, 
Einasto:1970aa}. The method was applied to the Andromeda galaxy to
calculate a preliminary mass distribution model by
\citet{Einasto:1969aa}, and by \citet{Einasto:1970ac} to calculate the
hydrodynamical model, which formed the topic of the original Chapter
18.  The description of the method and the analysis of the
hydrodynamical model of M31 were improved by \citep{Einasto:1970vz}.
The present English version of Chapter 18 is based on the improved
version of the analysis by \citet{Einasto:1970vz}.

\section{Introduction}

The most convenient way to express the various observational data on galaxies 
in a condensed and mutually consistent form is the
construction of their models.  In the case of bright galaxies, there
are available the following data: the photometric data for the galaxy
as a whole, and for some subsystems (neutral and ionised hydrogen,
young bright stars, novae, cepheids), spectrophotometric data (mean
spectral type, stellar content) for the nucleus and the bulge, and
kinematical data (the systematic radial motion and the velocity
dispersion) for the gaseous component, the nucleus, and the bulge.

On the basis of these data, using the necessary dynamical and
geometric equations, it is possible to construct a composite
hydrodynamical model of the galaxy.

\section{Theory}

{ A. ASSUMPTIONS AND DESCRIPTIVE FUNCTIONS}

We assume that the galaxy has an axis and a plane of symmetry, common
for all subsystems, that the galaxy is in a steady state and consists
of a number of physically homogeneous subsystems. The equidensity
surfaces of the subsystems are similar concentric ellipsoids.

The hydrodynamical descriptive functions, determining the spatial density
of matter and the velocity dispersion tensor, are designated as
follows:\\
$\rho(a)$ — the spatial density of matter, $a$ being the major semiaxis
of the equidensity ellipsoid with the axial ratio $\epsilon=b/a$;\\
$\sigma_R,~\sigma_\theta,~\sigma_z$ — the velocity dispersions in a
galactocentric cylindrical coordinate system ($a^2 = R^2 +
\epsilon^{-2}z^2$);\\
$V_\theta$ — the rotation velocity;\\
$\alpha$ — the inclination angle of the major axis of the velocity
ellipsoid in respect to the plane of symmetry of the galaxy.
\\
\\
{ B. GEOMETRIC EQUATIONS}

The spatial density of matter can be found from the observed projected
luminosity density $L(A)$, where $A$ is the major semiaxis of the
projected equidensity ellipse with the apparent axial ratio $E$, $E^2
=\sin^2i+\epsilon^2\cos^2i$, where $i$ is the angle between the axis
of the system and the line of sight.  From geometric considerations,
neglecting the absorption of light, we have \citep{Einasto:1969ab}
\be
L(A)= \frac{2\epsilon}{Ef}\int_A^{A^0}{\rho(a) a\dd{a} \over \sqrt{a^2
-A^2}  },
\label{eq18.1}
\ee
where $f$ is the mass-to-light ratio of the subsystem, considered as a
constant, and $A^0$ the major semiaxis of the limiting ellipsoid of
the subsystem.  
\\
\\
{ C. HYDRODYNAMICAL EQUATIONS}

In a steady state galaxy, the gravitational attraction of the galaxy is
counterbalanced by the pressure (velocity dispersion) and the
rotation. In cylindrical coordinates, the hydrodynamical equilibrium
equations are:
\be
  \frac{1}{R} \left( \sr -\st \right) + \frac{1}{\rho} \pdif{}{R} \left( \rho \sr \right) + \frac{1}{\rho} \pdif{}{z} \left[ \rho\gamma \left( \sr - \sz \right)\right] - \frac{V^2_\theta}{R} = -K_R , 
  \label{eq18.2}
\ee
\be
  	\frac{1}{R} \gamma \left( \sr - \sz \right) + \frac{1}{\rho} \pdif{}{R} \left[ \rho\gamma \left( \sr - \sz \right)\right] + \frac{1}{\rho} \pdif{}{z}\left( \rho\sz\right) = -K_z .
  	\label{eq18.3}
\ee
where 
\be
\gamma =\frac{1}{2}\tan 2\alpha,
\label{eq18.4}
\ee
and $K_R$, $K_z$ are radial and vertical components of the
gravitational acceleration of the whole galaxy. The latter quantities
can be derived from the mass density distribution function
\citep{Einasto:1969ab}. In the steady state galaxy, the functions
$\sigma_R,~\sigma_\theta,~\sigma_z,~V_\theta,~\gamma$ fully determine
the velocity ellipsoid as two axes of the ellipsoid lie in the
meridional plane of the galaxy, and the radial and vertical components
of the centroid motion are equal to zero.
\\

D. ADDITIONAL EQUATIONS, CLOSING THE SYSTEM OF EQUATIONS

In order to obtain composite models of galaxies, the mass and light
distribution of subsystems is first to be determined from photometric
and spectroscopic data. Then the gravitational acceleration of the
whole galaxy can be found. Finally, the kinematical functions of
subsystems can be derived. Given the density and the acceleration, the
equations (\ref{eq18.2}) and (\ref{eq18.3}) involve five unknown
kinematical functions. As we have only two equations, the problem is
not closed: to solve the system of equations, three additional
equations are needed, see Chapter 11 for detailed discussion. 

It is convenient to give the additional equations for the functions,
which determine the orientation of the velocity ellipsoid, $\gamma$,
and its shape
\be
k_{\theta}(R,z) =
\sigma_\theta^2/\sigma_R^2,~~~~k_z(R,z)=\sigma_z^2/\sigma_R^2.
\label{eq18.5}
\ee

From the theory of the third integral of motion of stars
\citep{Kuzmin:1952ac}  follows
\be
\gamma = Rz/(R^2+z_0^2 -z^2),
\label{eq18.6}
\ee
where $z_0$ is a constant, depending on the gravitational potential of
the whole galaxy.

The equations for $k_\theta$ and $k_z$ are in the general case
complicated \citep{Einasto:1970ac}, see Chapter 11. In the present
paper, we have computed the kinematical functions for the plane and the
axis of the galaxy only. The theory of the steady state galaxy gives
\citep{Einasto:1969ab}
\be
k_\theta(R,0) = 1/2\left[1+ \frac{\partial \ln V_\theta}{\partial \ln
    R}\right].
\label{eq18.7}
\ee

We assume that in the first approximation, the centroid velocity
$V_\theta$ is proportional to the circular velocity $V_c$. In this case,
$k_\theta(R,0)$ are identical for all subsystems. From the symmetry
conditions on the axis of the galaxy we have
\be
k_\theta(0,z)=1.
\label{eq18.8}
\ee

For flat subsystems the ratio $k_z(R,0)$ can be found from the theory
of irregular gravitational forces. \citet{Kuzmin:1961aa} has derived
the following approximate relation
\be
[k_z(R,0]^{-1}=1+[k_\theta(R,0)]^{-1}.
\label{eq18.9}
\ee

On the other hand, from the theory of the third integral, we have for
the axis $R=0$, supposing the ellipsoidal distribution of velocities
\citep{Einasto:1970ac}
\be
k_z(0,z)=k_z(0,0)/k_z(\sqrt{z^2-z_0^2},0).
\label{eq18.10}
\ee

Formulae (\ref{eq18.9}) and (\ref{eq18.10}) can be used, if $R^2 \gg
z_0^2$, and $z^2 \ge z_0^2$ correspondingly. For small $R$ and $z$,
$k_z(R,z)$ is to be interpolated, using the value $k_z(0,0)$, derived
from the virial theorem.

\section{The model}

The theory outlined has been applied to a model of the Andromeda
galaxy, consisting of four components: the nucleus, the bulge, the
disc, and the flat component. Observational data used are published by
\citet{Einasto:1969aa}.  The distance 692 kpc of the galaxy is
accepted, corresponding to the true distance modulus $(m-M)_0 =
24.^m2$ \citep{Baade:1963ud}.

The inclination of the galaxy has been estimated by combining the data
on the axial ratio of isophotes in the outer region of the galaxy, and
the distribution of emission nebulae \citep{Baade:1964aa}. The value
$i=12.^\circ8$ has been found. It is in good agreement with an earlier
estimate by Baade $i=12.^\circ7$, quoted by
\citet{Schmidt:1957uy}. Somewhat larger values found by
\citet{Arp:1964wi} and by some other authors cannot be accepted, as in
this case the true axial ratio of the equidensity surfaces of the disc
population will be too small, of the order of 0.01. The disc component
of a galaxy consists of old population I stars. Their vertical
dispersion of velocities at the distance $R=10$~kpc from the centre is
of the order of 20~km/s. From these data, we can estimate the thickness
and the axial ratio of equidensity surfaces; the latter quantity
becomes of the order of 0.1.

The parameter $z_0$ was derived from the gravitational potential of
the system. An effective value $z_0=0.5$~kpc has been found.

The principal descriptive function, the spatial density of matter, has
been chosen in the form of a generalised exponential function
\be
\rho(a)= \rho_0\,\exp\left[-\left(\frac{a}{a_0k}\right)^\nu\right],
\label{eq18.20}
\ee
where $\rho_0$ is the central density of the component, $a_0$ — the
effective (harmonic mean) radius of the component, $\nu$ — the
structural parameter of the model, and $k$ — a dimensionless
parameter depending on $\nu$. The central density depends on the mass,
effective radius, and the axial ratio of the component:
\be
\rho_0 = \frac{h}{4\pi\epsilon}\,\frac{\mm{M}}{a_0^3},
\label{eq18.21}
\ee
where $h$ is a dimensionless parameter depending on $\nu$.

\begin{table*}[h]
\centering    
\caption{Parameters of the components of M31} 
\begin{tabular}{lccccccc}
  \hline  \hline
  Quantity&Unit&Total&Nucleus&Bulge&Disc&Flat+&Flat$-$\\
  \hline
$\epsilon$&&       &0.84&0.57&0.09&0.01&0.02\\
  $\nu$&     &       & 1 & 1/4 &   1  &   1 &  1 \\
  $k$&         &       &0.5&$1.26\times10^{-4}$& 0.5&0.5&0.5\\
$h$&           &       &4& 3112& 4&4&4\\
$a_0$       &kpc  &       &0.005&1&10&8&4\\
$L$&        $10^9\,L_\odot$&13.13&0.003&4.95&6.46&2.29&$-0.57$\\
$\mm{M}$&$10^9\,\mm{M}_\odot$&201.8&0.052&85.5&111.5&5.73&$-1.43$\\
$f$&                                 &15.4&17.3&17.3&17.3&2.5&2.5\\         
$\overline{\rho}$&$M_\odot/{pc^3}$  &&
                                      $1.2\times10^6$&35.8&0.296&0.267&$-0.267$\\
  \hline
\label{Tab18.1}   
\end{tabular}
\end{table*}

{\begin{figure*}[h] 
\centering 
\resizebox{0.49\textwidth}{!}{\includegraphics*{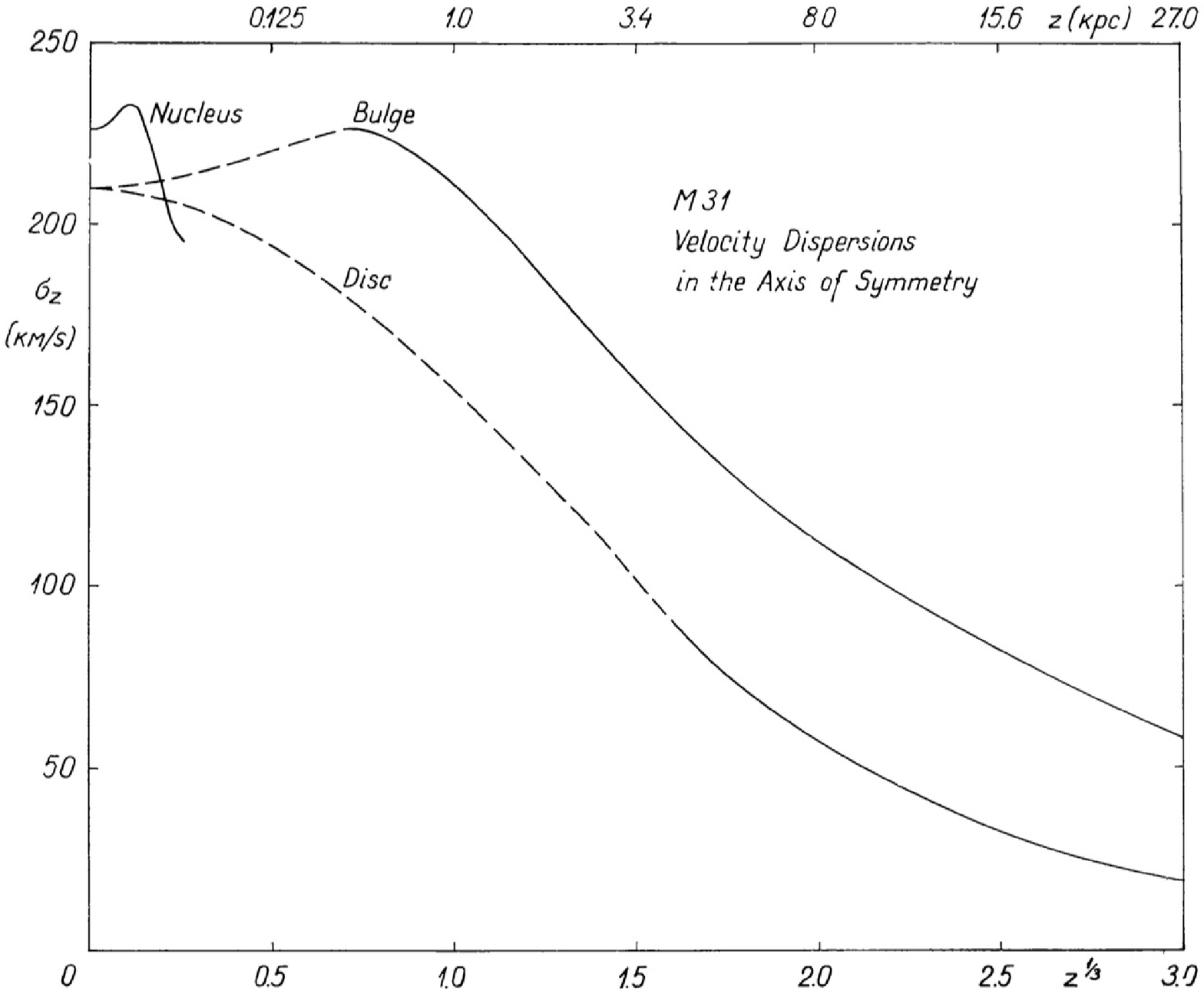}}
\resizebox{0.49\textwidth}{!}{\includegraphics*{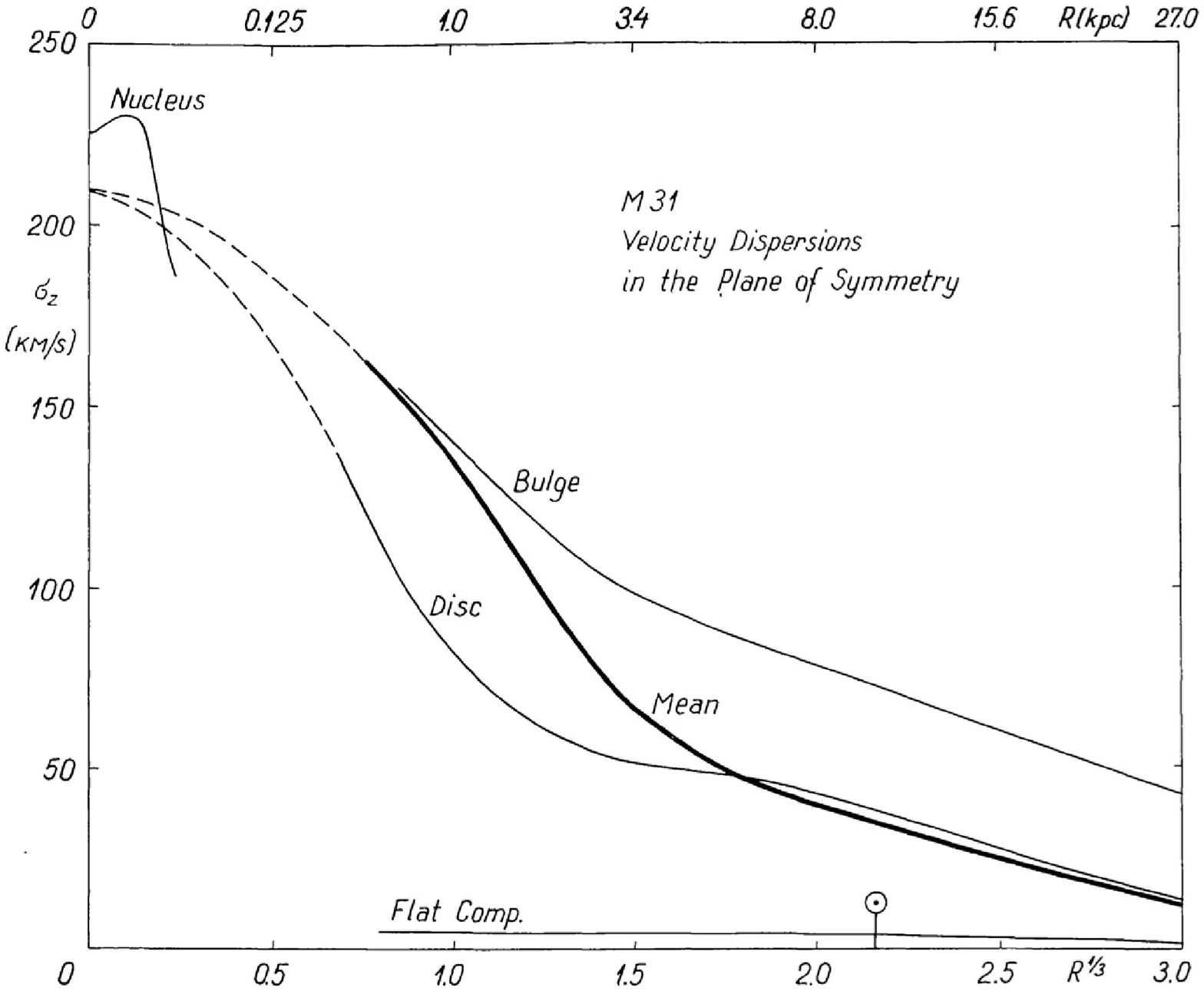}}
\caption{ {\em Left:} Vertical velocity dispersion $\sigma_z$  of the components of M31 in the 
  $z$-axis of symmetry. {\em Right:} Vertical velocity
  dispersion $\sigma_z$ of the components of M31 in the plane of
  symmetry.   }
  \label{Fig18.2}
\end{figure*} 
}

{\begin{figure*}[h] 
\centering 
\resizebox{0.49\textwidth}{!}{\includegraphics*{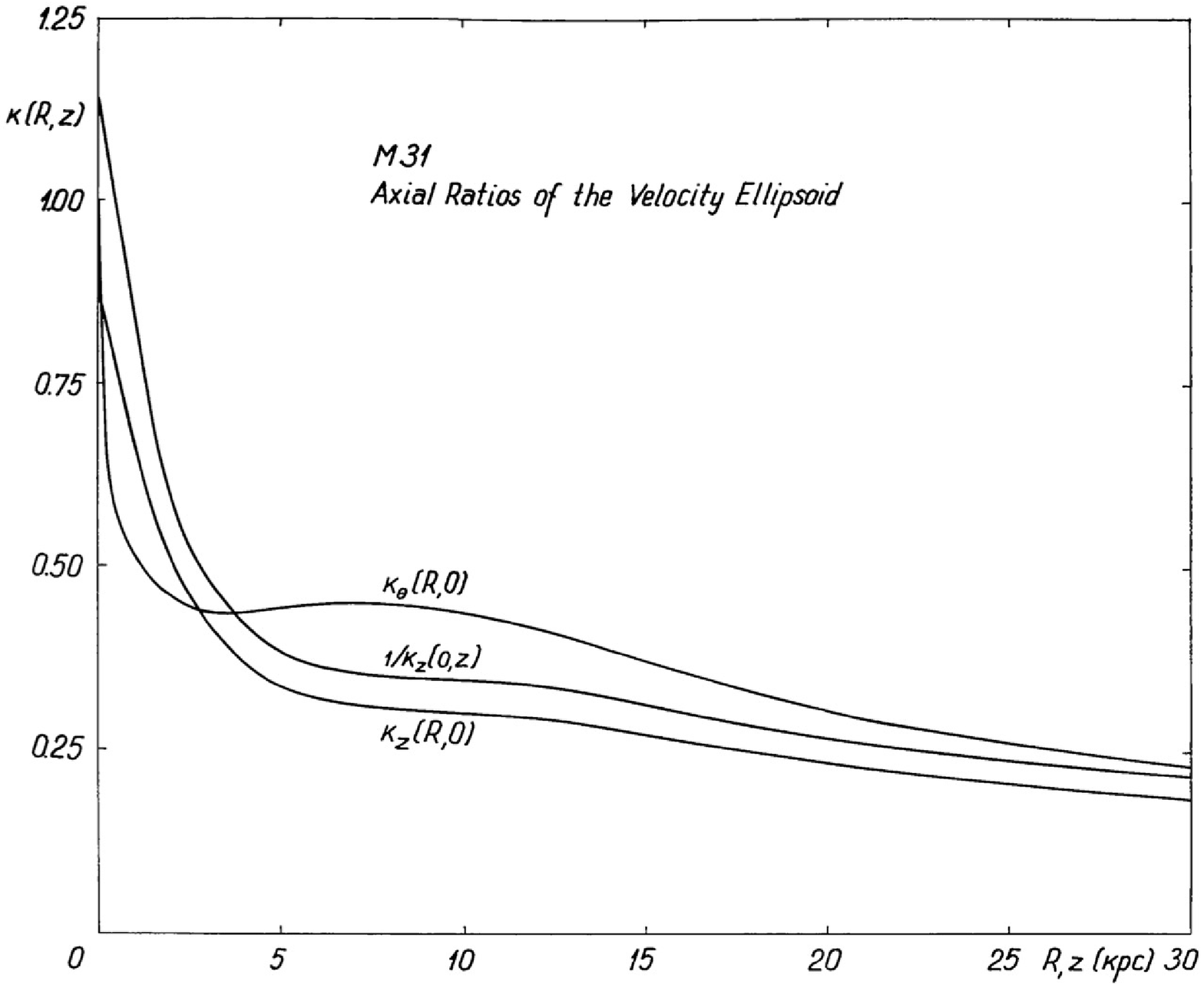}}
\resizebox{0.49\textwidth}{!}{\includegraphics*{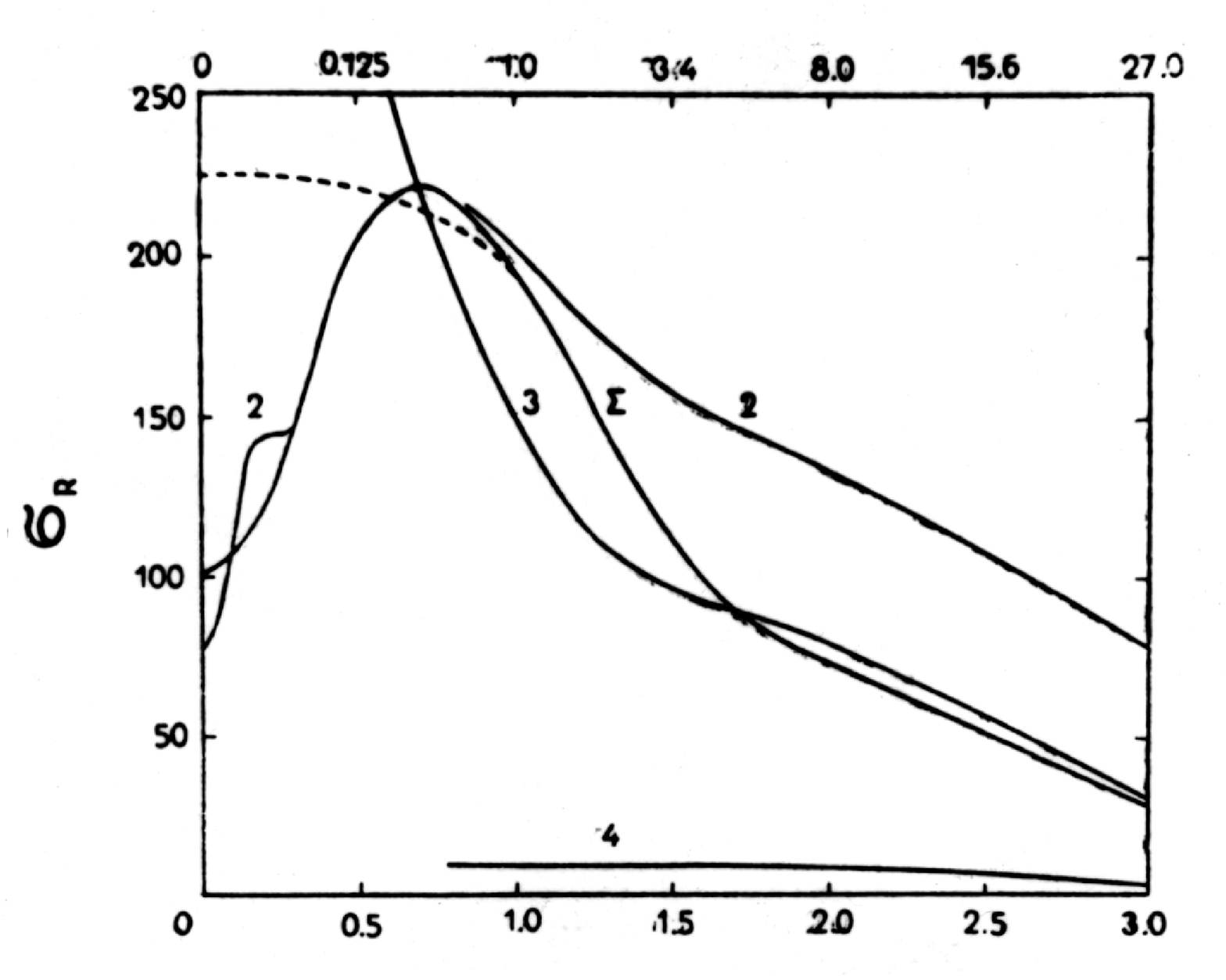}}
\caption{ {\em Left:} Axial ratios of the velocity ellipsoid in the
  plane of symmetry and in the axis of M31.
  {\em Right:} Radial velocity dispersion
  $\sigma_R$ as function of $R$ for components of M31: 2 — bulge and
  halo; 3 — disc; 4 — flat;  $\Sigma$ — galaxy as a whole. Horizontal
  scale is given in units $R^{1/3}$ and in $R$ on  
  bottom and top, respectively.  }
  \label{Fig18.4}
\end{figure*} 
}

{\begin{figure*}[h] 
\centering 
\hspace{2mm}
\resizebox{0.60\textwidth}{!}{\includegraphics*{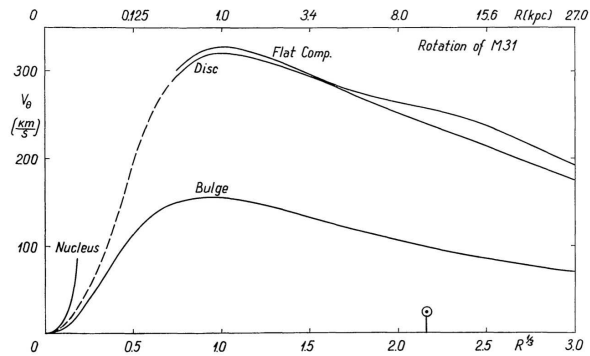}}
\caption{  Rotation velocity $V_\theta$ for components of M31
  in the plane of symmetry \citep{Einasto:1970vz}.
} 
  \label{Fig18.6}
\end{figure*} 
}

{\begin{figure*}[h] 
\centering 
\resizebox{0.51\textwidth}{!}{\includegraphics*{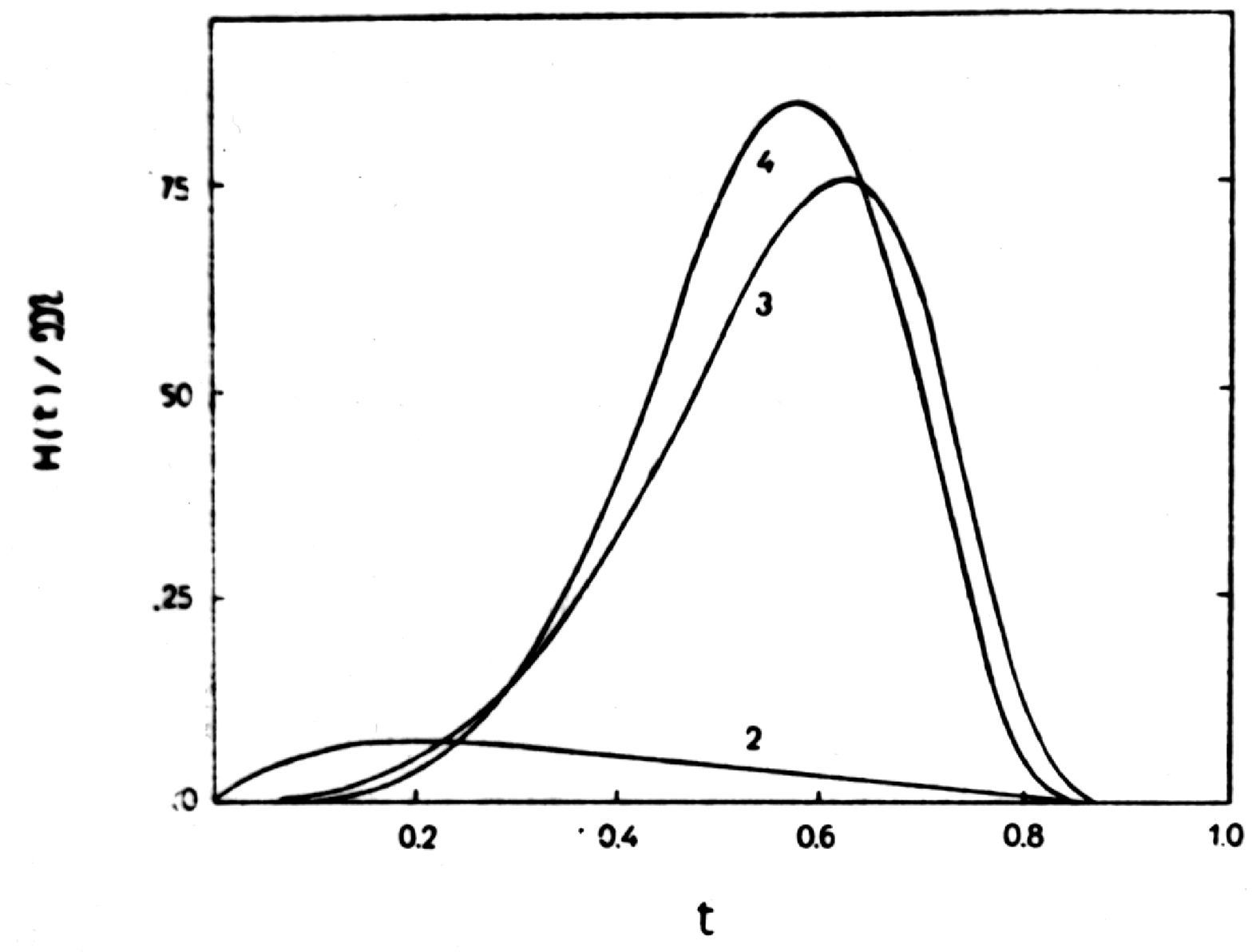}}
\resizebox{0.37\textwidth}{!}{\includegraphics*{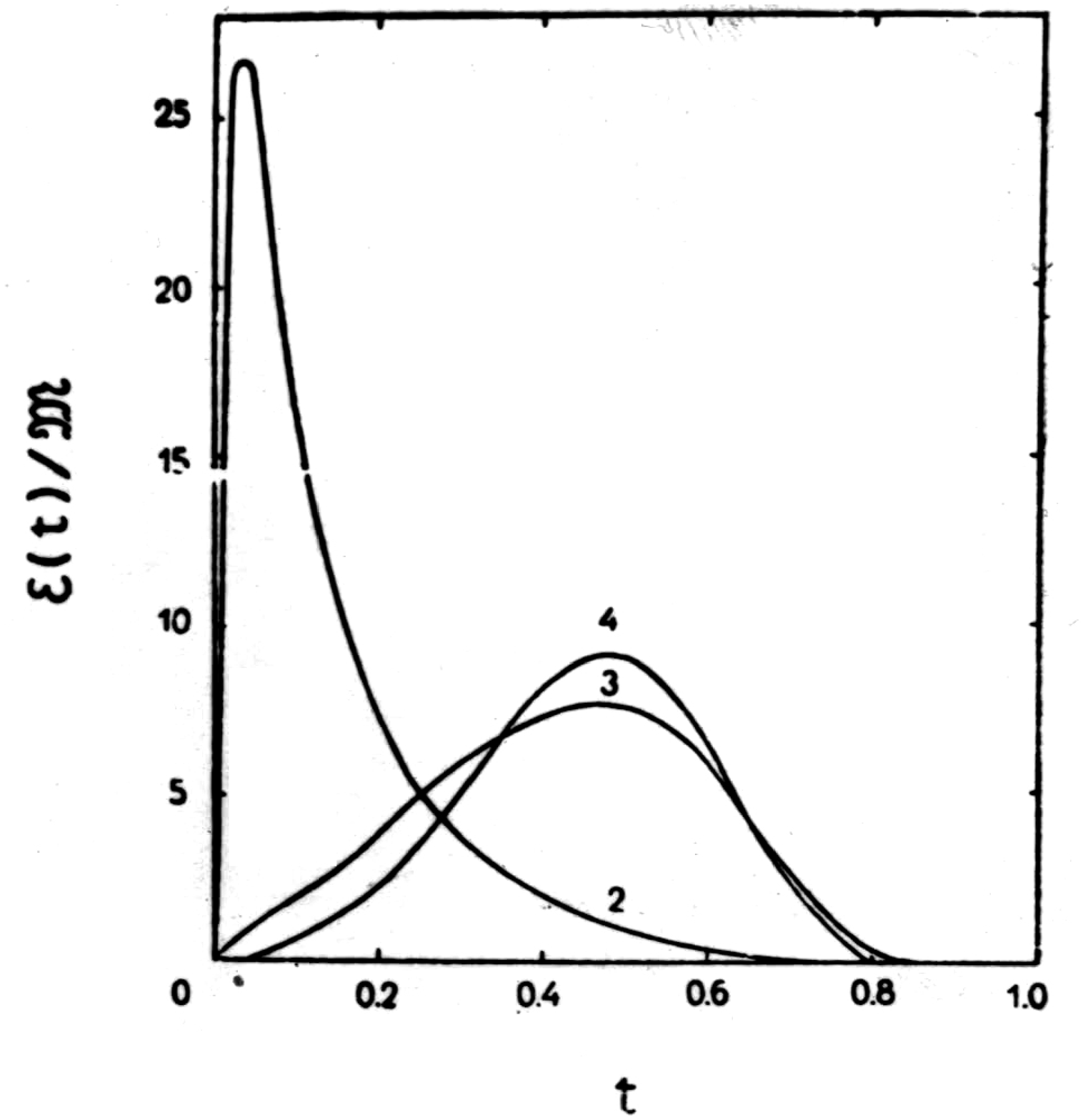}}
\caption{{\em Left:}  A specific  angular moment of components in units
  100~km/sec per kpc.   {\em Right:} Specific kinetic energy of M31
  components in units $10^4$~(km/sec)$^2$, both are expressed as
  functions of $t=a/(1+a)$.  }
  \label{Fig18.7}
\end{figure*} 
}

The derived parameters of the components are given in Table
\ref{Tab18.1}.

To obtain better agreement with observations, the flat component of
the galaxy has been represented by a sum of two functions
(\ref{eq18.20}), one of them being negative. This allows to describe
the ring-like shape of the flat population with zero density at the
centre. The parameters of this component are subject to the condition
that the ring-like mass distribution has everywhere non-negative
total density.

The mass of the nucleus has been determined by means of the virial
theorem. In an earlier paper \citep{Einasto:1969aa}, the mass has been
found from the luminosity of the nucleus and its accepted
mass-to-light ratio \citep{Spinrad:1966wh}. The mass
$\mm{M}=5.2\times 10^8~\mm{M}_\odot$, obtained from the virial theorem
for the nucleus, and the corresponding mass-to-light ratio
$f=170$, does not agree with the value $f=17$, derived
spectroscopically \citep{Spinrad:1966wh}. This discrepancy may be
removed, supposing that the nucleus contains besides stars an
invisible central body — a dead quasar \citep{Lynden-Bell:1969vs}. In
this case, the virial theorem must be modified, and we get for the
point mass $\mm{M}=1.4\times 10^8\,\mm{M}_\odot$, supposing
$\mm{M}=0.8\times 10^8~\mm{M}_\odot$ for the mass of the stellar
component of the nucleus.

\section{Discussion}

A. MASS DISTRIBUTION

Our model differs in two points from the models by
\citet{Schmidt:1957uy}, \citet{Brandt:1965ty}, \citet{Roberts:1966aa}
and \citet{Gottesman:1966vd}: the central concentration of mass is
much higher, and the total mass smaller \citep{Einasto:1969aa}, see
Table \ref{Tab18.1}. The differences can be explained by various
circular velocity curves adopted.

In the central region, the velocities found earlier from the 21-cm
radio line measurements are underestimated due to the insufficient
correction for the antenna smearing effect. The rotation velocities,
derived optically for the stellar component of the galaxy, cannot be
identified with the circular velocities, as the pressure term
(velocity dispersions) in hydrodynamical equations is predominating.

The great masses are found in most cases as a result of approximation
the observed rotation velocities with a generalised Bottlinger law
\be
V_\theta= {V_0\,R  \over [1+(R/R_0)^n]^{3/2n}  },
\label{eq18.22}
\ee
where $V_0$ and $R_0$ are constants, and rotation velocity is
identified with the circular velocity.

We have shown \citep{Einasto:1969ab} that the generalised Bottlinger
law cannot be applied to the circular velocity, as in this case great
masses at very large distances from the centre of the galaxy
occur. This is impossible due to the tidal effect of nearby galaxies
\citep{King:1962aa}.

The small radial gradient of the rotation velocity, observed in the
periphery of some galaxies, in particular, in the Andromeda galaxy, is
probably to be explained in another way, for instance, as the
appearance of systematic streaming motion in the galaxy.
\\
\\
B. MASS-TO-LIGHT RATIO

The mean mass-to-light ratio found, $f=15.4$, is normal for a Sb
galaxy. The flat population and the disc have also acceptable values,
$f=2.5$ and $f=17.3$, respectively.

The mass-to-light ratio for the nucleus, $f=17.3$, seems to be too large 
at first glance. To explain this value, we must suppose that the
nucleus: (a) consists of very old physically evolved stars, and (b) is
dynamically not evolved.

The mean relaxation time of the nucleus is of the same order
($10^{10}$ yr) as the age of the whole galaxy. Therefore, the nucleus
is dynamically indeed little evolved and has lost only a small
fraction of his low-mass stars. As the nucleus has had too little time
to form dynamically by star-star encounters, it must be formed in the
protogalaxy stage of the galaxy evolution.

The metal content of stars in the nucleus is normal
\citep{Spinrad:1966wh}. Therefore, if the high mass-to-light ratio and
the great age of the nucleus will be confirmed, we must conclude that
in the nucleus the metal enrichment has taken place in a very early
stage of the galaxy evolution.

\medskip
\hfill March 1969
\vskip 5mm
\hfill Revised January 1970

\chapter{The spiral structure of M31}\label{ch19}

The density distribution and the radial velocity field in the
Andromeda galaxy, M31, was studied on the basis of the 21-cm
radio-line data from Jodrell Bank and Green Bank by 
\citet{Einasto:1970tz}, which forms Chapter 19 of the Thesis. The true density 
has been obtained from the observed one by solving a two-dimensional
integral equation. As the resolving power of the radio telescopes is
too low to locate all spiral arms separately, optical data on the
distribution of ionised hydrogen clouds have also been used. The mean
radial velocities have been derived by solving a two-dimensional
non-linear integral equation with the help of hydrogen densities and
a model radial velocity field.

The inner concentrations of hydrogen form two patchy ring-like
structures with mean radii 30' and 50', the outer concentrations can be
represented as fragments of two {\em leading} spiral arms.

The rotational velocity, derived from the radial velocity field, in
the central region differs considerably from the velocity curves
obtained by earlier authors. The difference can be explained by the
fact that in this region, the correction for the antenna beam width is
much greater than adopted by previous investigators.

{\begin{figure*}[h] 
\centering 
\resizebox{0.47\textwidth}{!}{\includegraphics*{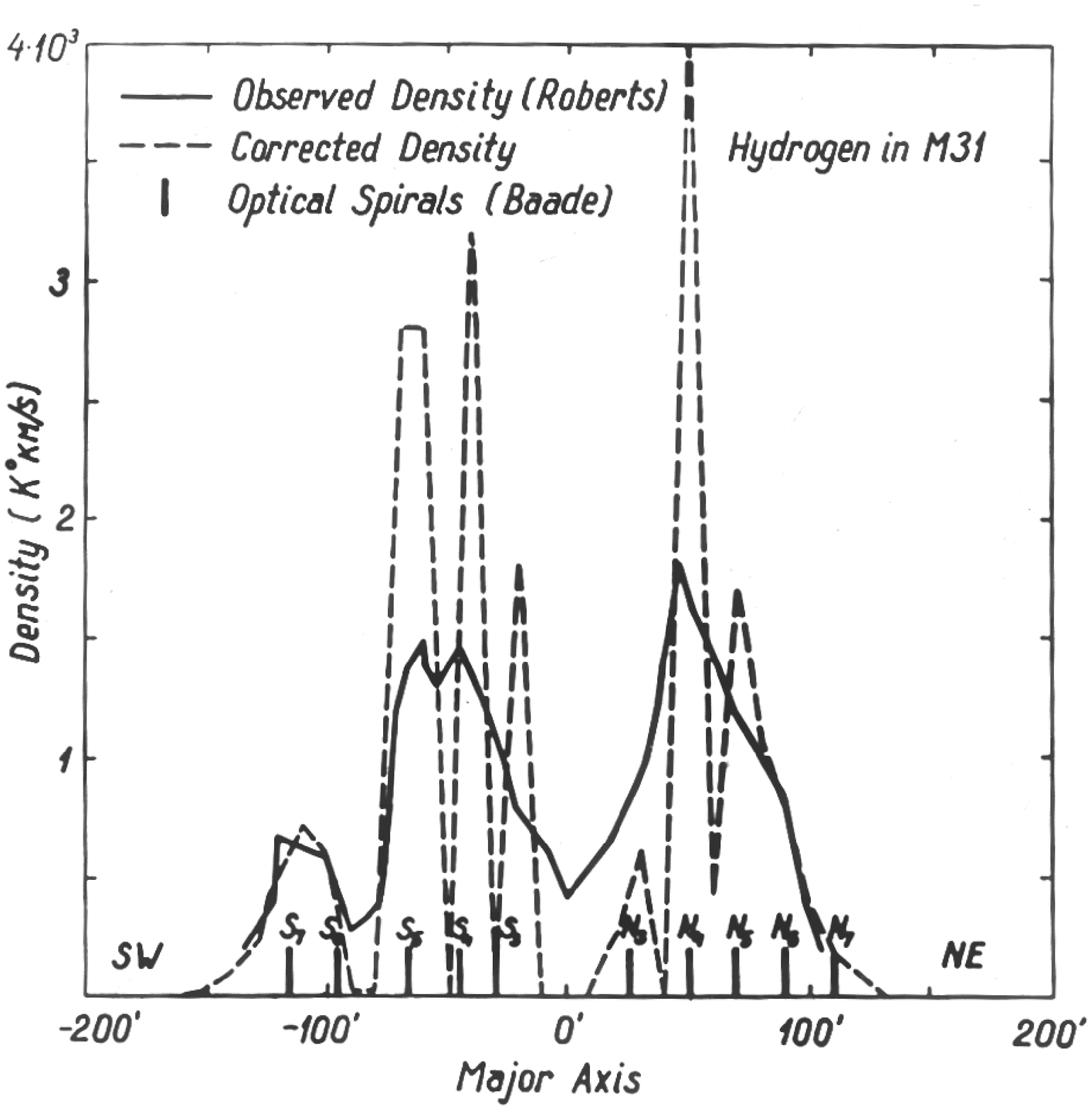}}
\resizebox{0.47\textwidth}{!}{\includegraphics*{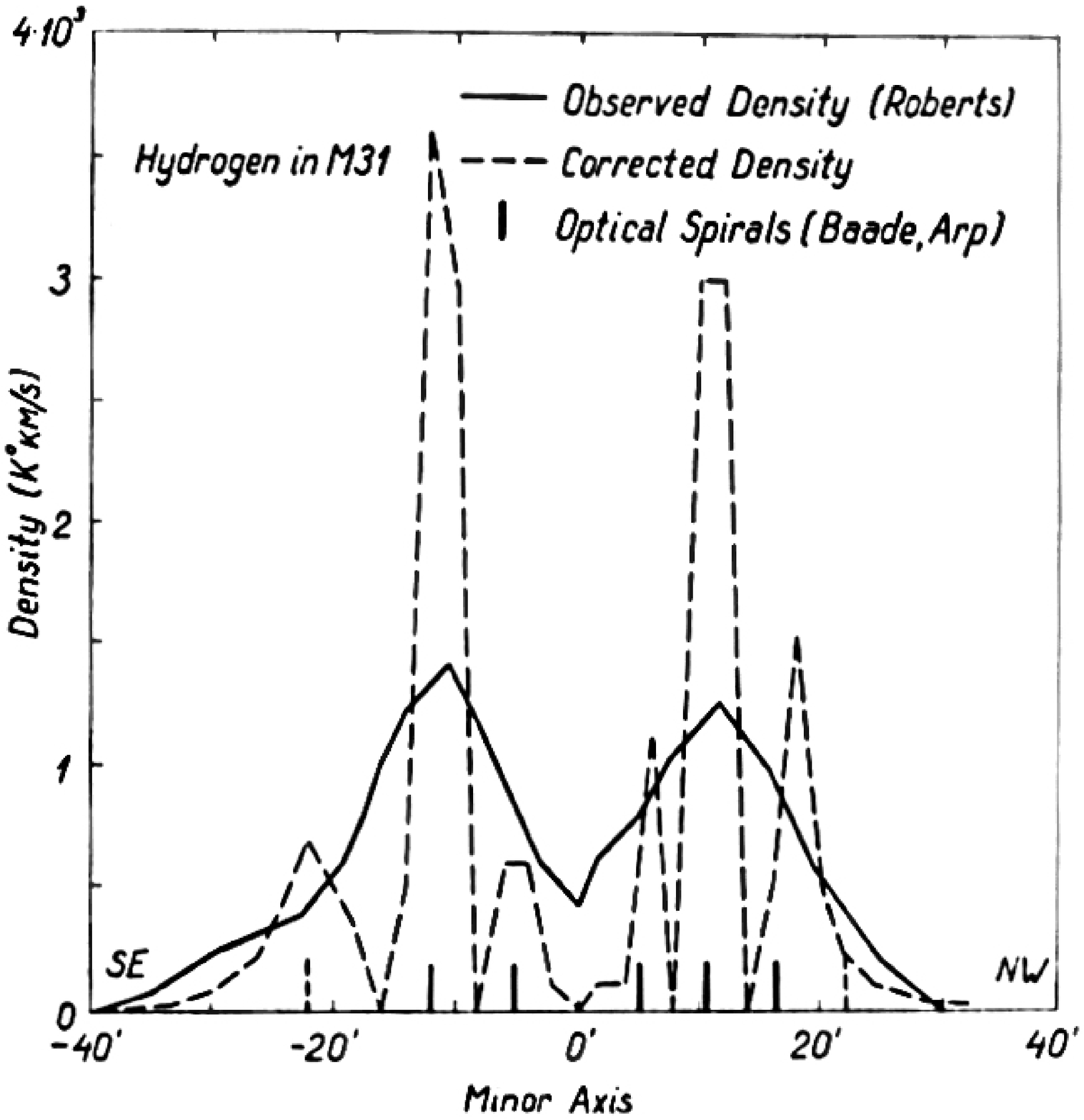}}
\caption{{\em Left;} Observed \citep{Roberts:1967aa}  and corrected distribution
  of projected density of neutral hydrogen along major axis of
  M31. Positions of optical spirals according to \citet{Baade:1964aa}
are also shown. {\em Right:} Observed \citep{Roberts:1967aa}  and corrected distribution
  of projected density of neutral hydrogen along minor axis of
  M31. Positions of optical spirals according to \citet{Baade:1964aa}
are also shown.} 
  \label{Fig19.1}
\end{figure*} 
}

{\begin{figure*}[h] 
\centering 
\resizebox{0.47\textwidth}{!}{\includegraphics*{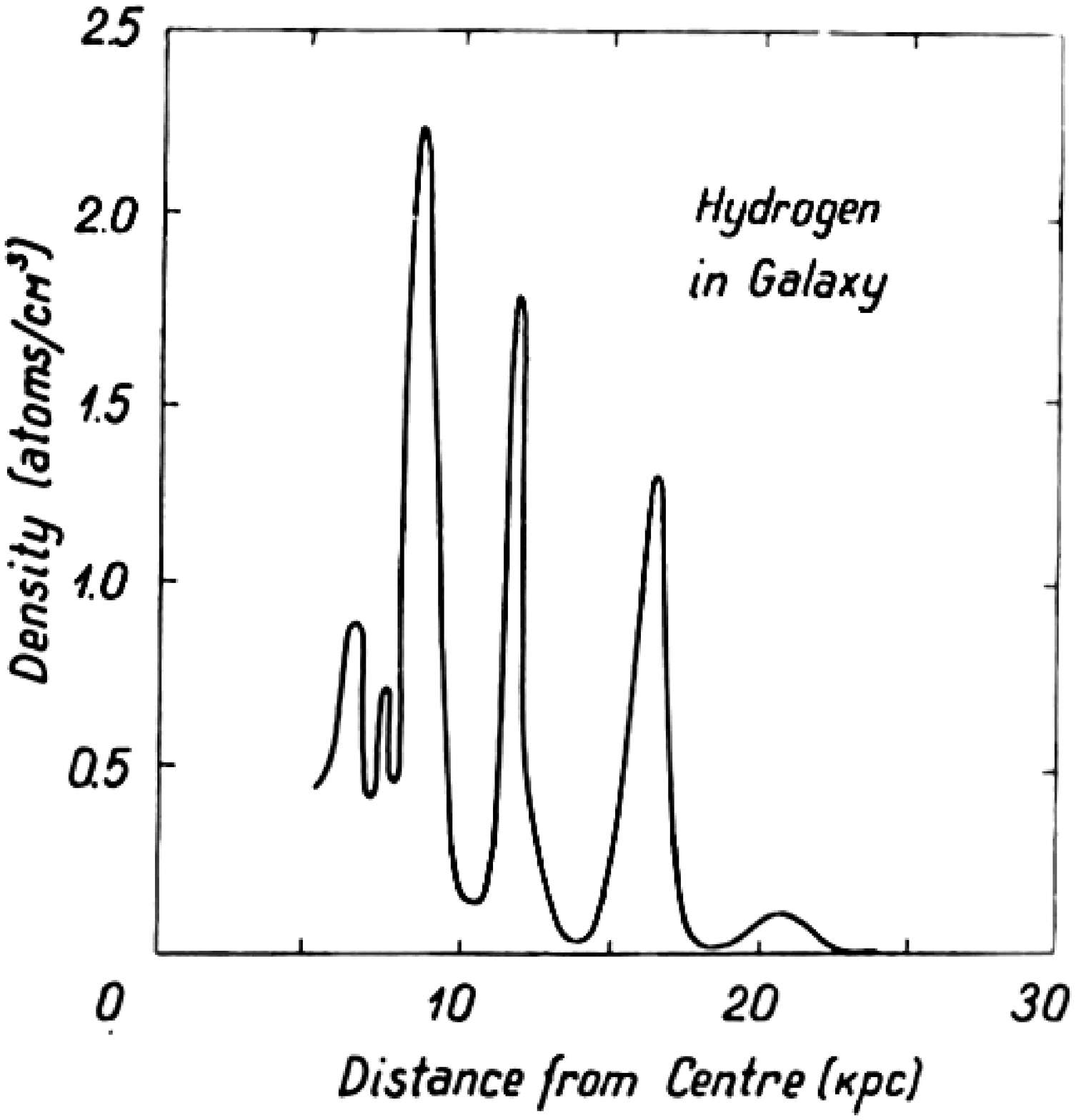}}
\resizebox{0.47\textwidth}{!}{\includegraphics*{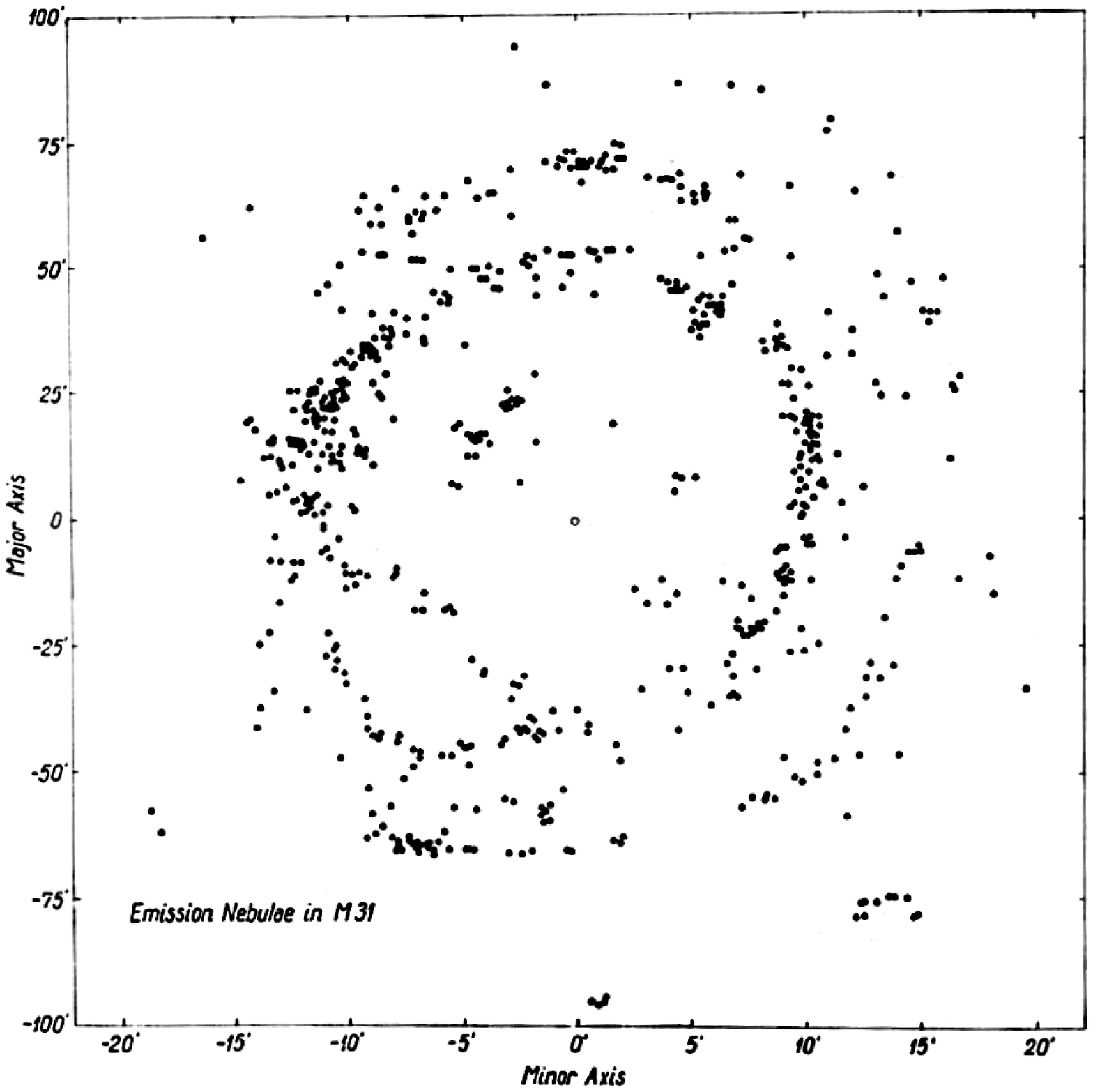}}
\caption{{\em Left;} Spatial density of neutral hydrogen in the Galaxy
  plane.  {\em Right:} Distribution of ionised hydrogen clouds in M31 according to
  \citet{Baade:1964aa}.  The $X$-scale is enlarged by a factor of 4.5
  to take into account the inclination of the plane. 
} 
  \label{Fig19.3}
\end{figure*} 
}

{\begin{figure*}[h] 
\centering 
\resizebox{0.47\textwidth}{!}{\includegraphics*{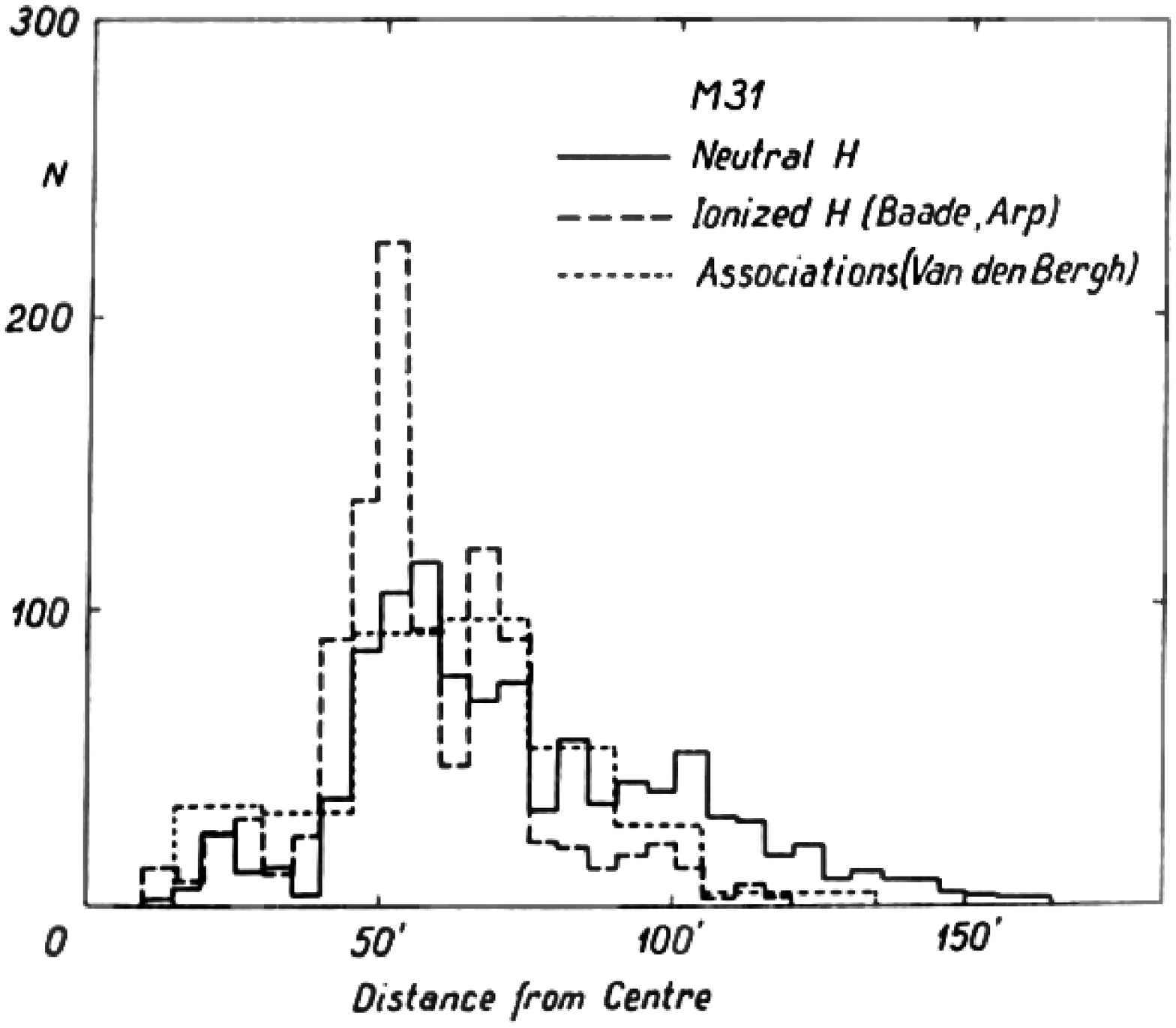}}
\resizebox{0.52\textwidth}{!}{\includegraphics*{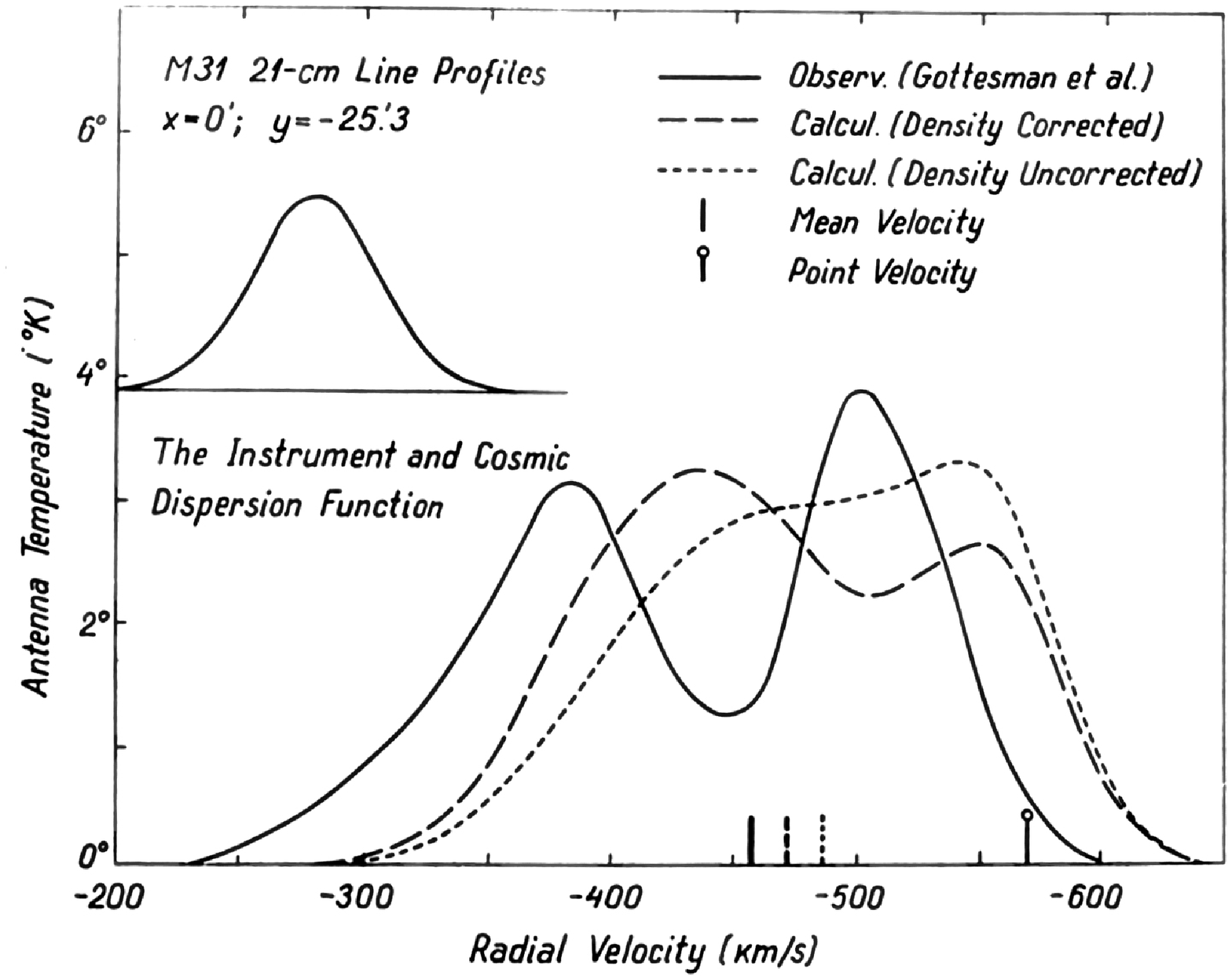}}
\caption{{\em Left;} Distribution of neutral and ionised hydrogen in
  M31 according to \citet{Baade:1964aa}.  Distribution of stellar
  associations according to \citet{van-den-Bergh:1964wl} is also
  shown.  {\em Right:} Profiles of 21-cm lines for one point at the
  major axis of M31. The profile due to observational and ``cosmic''
  errors is also shown.}
  \label{Fig19.6}
\end{figure*} 
}

{\begin{figure*}[h] 
\centering 
\resizebox{0.42\textwidth}{!}{\includegraphics*{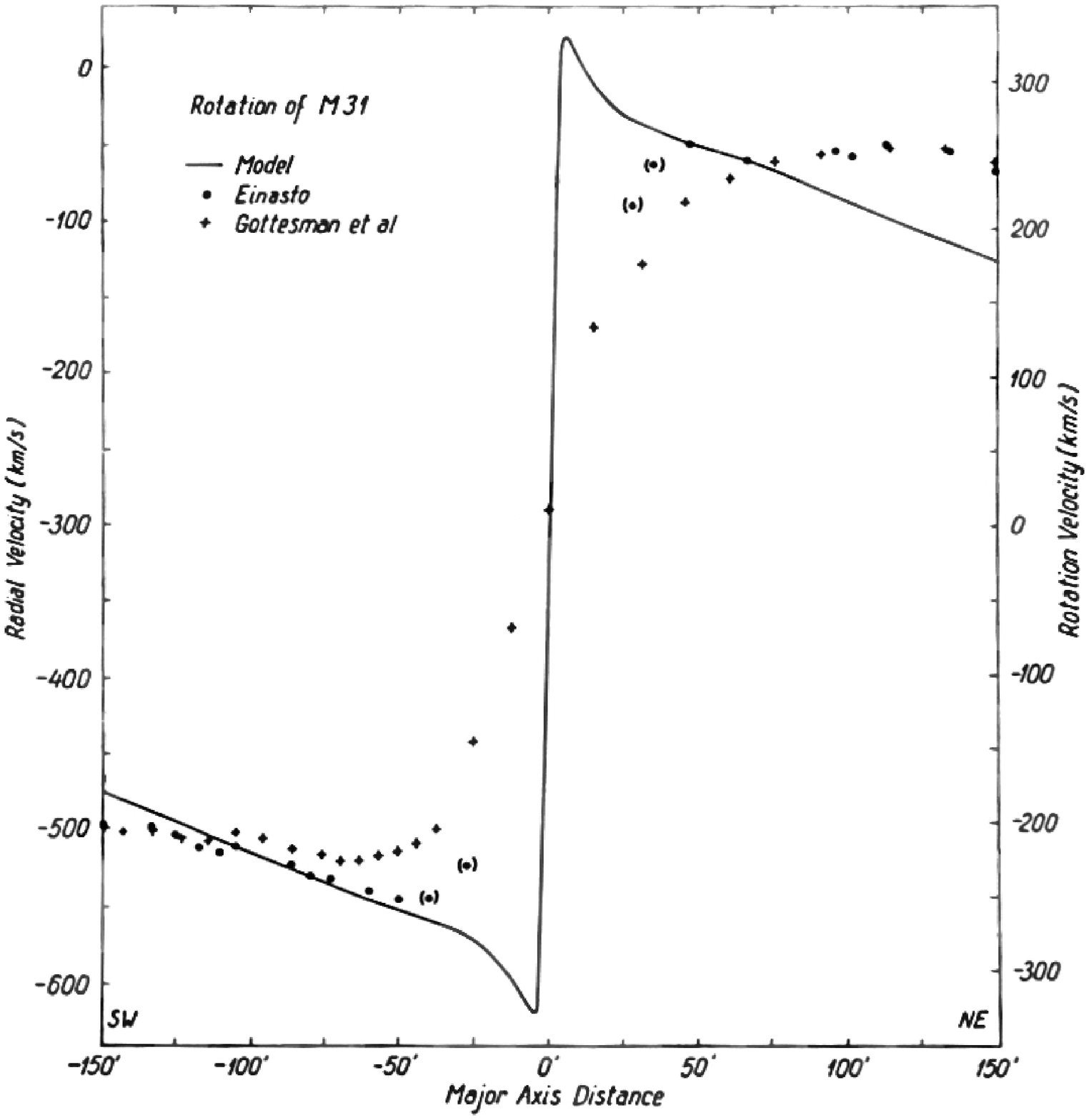}}
\resizebox{0.57\textwidth}{!}{\includegraphics*{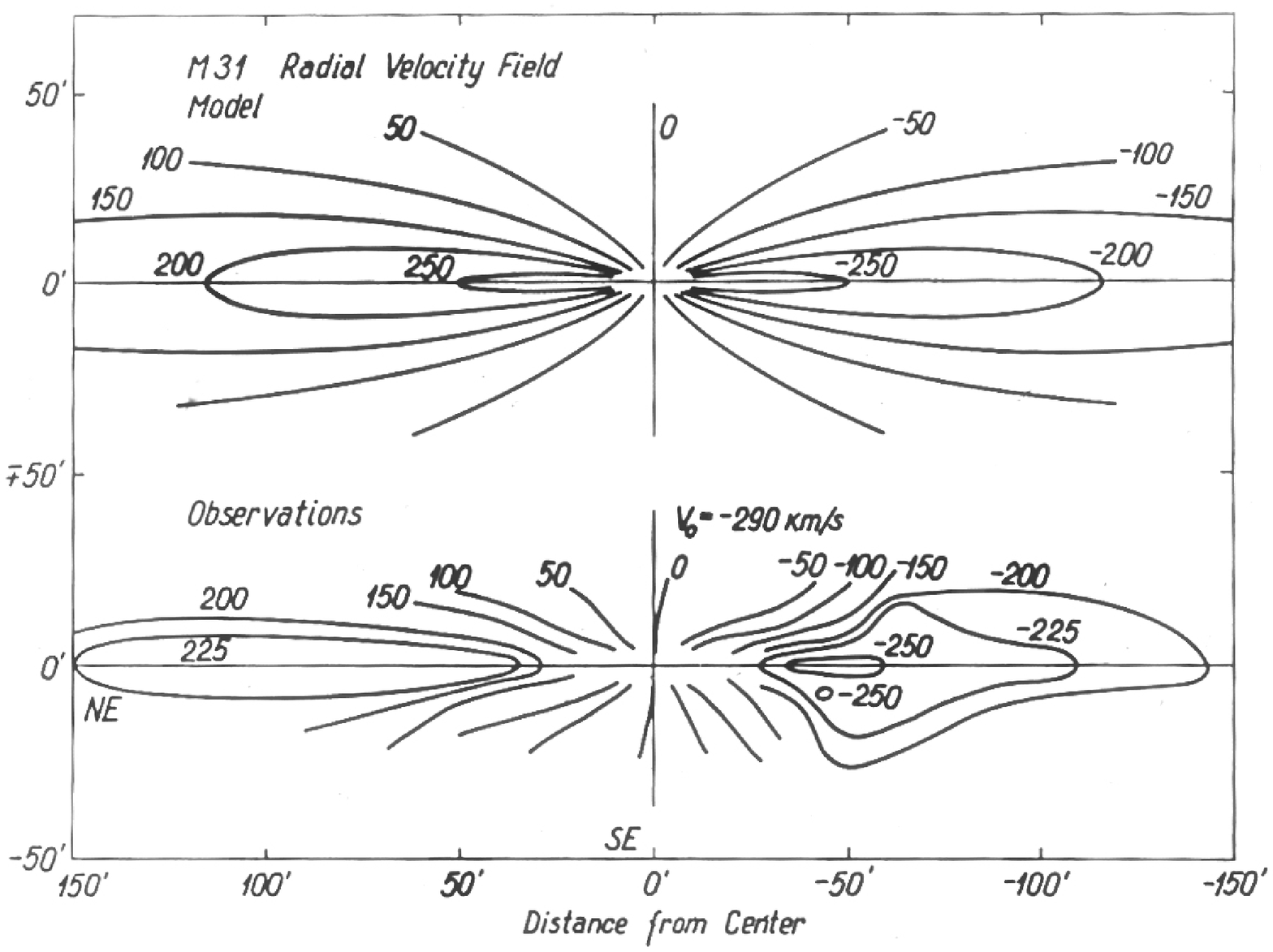}}
\caption{{\em Left;} Circular velocity of M31 according to models by
  \citet{Einasto:1970ac} and \citet{Gottesman:1966vd} and present work
  \citep{Einasto:1972ab}.
 {\em Right:} The radial velocity field of M31 according to model by
  \citep{Einasto:1972ab} and observations.} 
  \label{Fig19.8}
\end{figure*} 
}

{\begin{figure*}[h] 
\centering 
\resizebox{0.47\textwidth}{!}{\includegraphics*{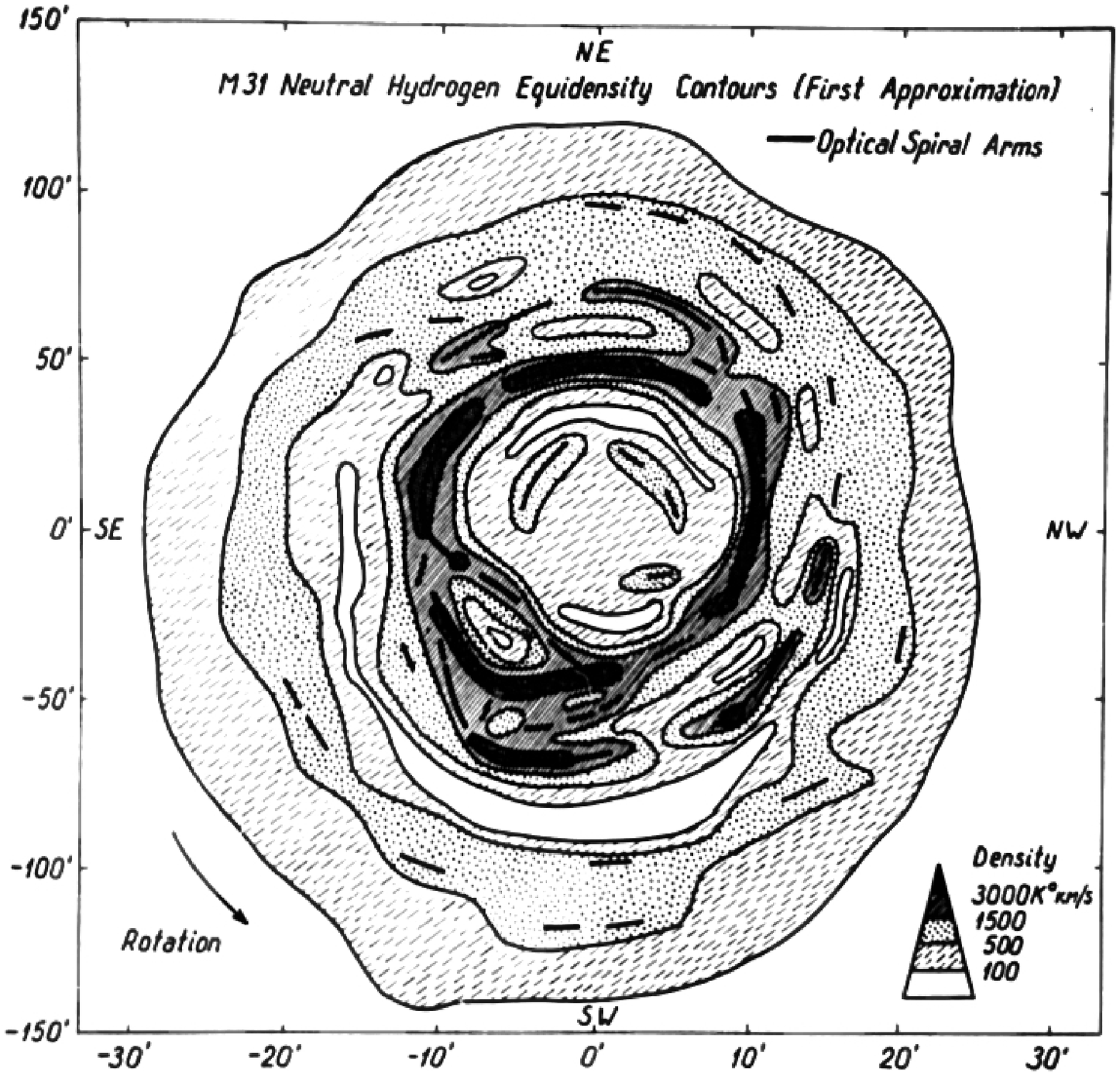}}
\caption{ Equidensity contours of neutral hydrogen in M31.
} 
  \label{Fig19.5}
\end{figure*} 
}

\section{Introduction}

In the present paper, the density distribution and the radial velocity
field of neutral hydrogen in the Andromeda galaxy, M31, have been
studied. The investigation is based on the 21-cm radio-line data from
Jodrell Bank and Green Bank Observatories, kindly sent to us by
Dr. R. G. Davies and Dr. M. S. Roberts. Optical data on the
distribution and motion of ionised hydrogen are also used.

When studying the distribution and motion of hydrogen in external
galaxies, it is necessary to take into consideration the angular
resolving power of radio telescopes. \citet{Gottesman:1966vd} found
that the correction for the beam width of the 250-feet Jodrell Bank
telescope both in the density and the velocity does not exceed
10\,\%.  Our calculations, however, have shown in some cases the
correction needed is much greater. This indicates that the Jodrell
Bank investigators have used a too simplified reduction method. The
Green Bank data have been reduced neglecting the antenna smearing
effect \citep{Roberts:1966aa}. For that reason, the available radio
data are to be reduced once again. At present the program is not
finished. In this paper the preliminary results are reported.

\section{The integral equations for the density and the mean radial 
velocity}

Let $X,Y$ be the rectangular galactocentric coordinates in minutes of
arc, the $Y-$axis being directed to the NE side of the major axis of
the galaxy; $V$ the true radial velocity; $D(X,Y)$ the true projected
density of neutral hydrogen; $E(V-\overline{V})$ the distribution
function of residual radial velocities in the direction $X$,~$Y$;
$\overline{V}= \overline{V}(X,Y)$ is the mean radial velocity in this
direction.

The radio telescope, directed to the point $X_p,~Y_p$ and disposed to
the frequency, corresponding to the radial velocity $V_k$, will record
the flux
\be
\ba{ll}
T(X_p,Y_p,V_k)&=\int\int\limits_{-\infty}^\infty\int\,D(X,Y)\,F(X-X_p,Y-Y_p)\\
&\times\,E[V-\overline{V}(X,Y)]\,G(V-V_k)\dd{X}\dd{Y}\dd{V},
\label{eq19.1}
\ea
\ee
where $F(X-X_p,Y-Y_p)$ is the angular sensitivity function of the
telescope, and $G(V-V_k)$ is the corresponding frequency sensitivity
function.

Integrating (\ref{eq19.1}) over all observed velocities $V_k$ we
obtain the observed projected density of hydrogen
$\overline{D}(X_p,Y_p)$, which is connected with the true density
$D(X,Y)$ by means of the equation
\be
\overline{D}(X_p,Y_p)=\int\int\limits_{-\infty}^\infty\,D(X,Y)\,F(X-X_p,Y-Y_p)\dd{X}\dd{Y}.
\label{eq19.2}
\ee

This is a two-dimensional homogeneous Fredhold integral equation of
the first kind for the determination of the true density $D(X,Y)$.  If
the density is known, the Eq,~(\ref{eq19.1}) can be considered as a
non-linear integral equation for the determination of the mean radial
velocity $\overline{V}(X,Y)$.

The observations of point radio sources indicate that the function $F$
can be fairly well approximated by a two-dimensional Gaussian with
half-intensity diameters 15' and 10' in the case of the Jodrell Bank
and Green Bank telescopes respectively
(\citet{Davies:1969ac}  and \citet{Roberts:1969ab}). The function $G$ has in the
case of the Jodrell Bank telescope also a Gaussian shape with
half-intensity width 200~kHz, which corresponds to a velocity
dispersion of 17~km/s.  The Green Bank telescope has a rectangular
shaped function $G$ of 95~kHz (20~km/s) wide.

\section{The density distribution}

From the analogy with our Galaxy, we may expect that the neutral
hydrogen in M31 is concentrated in spiral arms. The optical
observations of ionised hydrogen \citep{Baade:1963uz, Arp:1964wi}
indicate that the Andromeda galaxy has four to five spiral arms in both
sides of the galaxy. The mean distance between every two arms is
$20'=4$~kpc, in projection only $4' - 8'$, except the region around the
major axis. The ionised hydrogen arms coincide with the neutral
hydrogen arms within the actual distance of 5'; the neutral hydrogen
arms are situated closer to the centre of the galaxy
\citep{Roberts:1967aa}.

The resolving power of the radio telescopes used is not sufficient to
separate all spiral arms in the Andromeda galaxy; only the most dense
arms N4, S4 and S5 (designated after \citet{Baade:1963uz}) can be
``seen'' individually \citep{Roberts:1967aa}. To locate the outer
neutral hydrogen arms, the optical data on the distribution of the
ionised hydrogen clouds can be used \citep{Baade:1964aa}.

The true density distribution has been determined from the integral
equation (\ref{eq19.2}) by two methods. Near the minor axis, the
equidensity lines are almost parallel to the major axis, and the
two-dimensional equation can be reduced to the one-dimensional
one. Representing the observed density distribution by a sum of
Gaussian functions, we get the solution of the equation also in the form of
a sum of Gaussian functions.

For points far off from the minor axis, the solution of the
Eq.~(\ref{eq19.2}) has been found by successive approximations. The
arms have been located by combining optical and radio data, the
corrected densities have been derived from the observed radio
densities by a trial-and-error procedure. The densities have been
found for a network of points, placed in $X$ and $Y$ at intervals 2'
and 10' respectively.

The observed (Green Bank) and corrected density profiles (first
approximation) along the major and minor axes of the Andromeda galaxy
are shown in Figures \ref{Fig19.1}, left and right panels,  
respectively. The picture is quite similar to the neutral hydrogen
density profiles found for our Galaxy; an example of them, drawn on
the basis of the Dutch survey by \citet{Schmidt:1957aa} and
\citet{Westerhout:1957aa} is given in Fig.~\ref{Fig19.3}.

The $X,~Y$-distribution of ionised hydrogen clouds
\citep{Baade:1964aa} is given in the right panel of Fig.~\ref{Fig19.3}. The map of
equidensity contours of neutral hydrogen is presented in
Fig.~\ref{Fig19.5}. The $R$-distribution (integrated over all position
angles $\theta$) of neutral and ionised hydrogen as well as of the
stellar associations \citep{van-den-Bergh:1964wl} is plotted in the left
panel of 
Fig.~\ref{Fig19.6}. The original distributions are reduced to an equal
total number of objects, $N=1000$.

The inspection of the data obtained leads us to the following
conclusions:\\
(a) the spatial distribution of neutral hydrogen is similar to the
distribution of ionised hydrogen and stellar associations; at great
distances from the centre, the relative density of neutral hydrogen is
higher than that of the ionised hydrogen;\\
(b) the inner concentrations of hydrogen form two patchy ring-like
structures with the mean radii 30' (the arms N3, S3 after Baade) and
50' (the arms N4, S4);\\
(c) the outer hydrogen concentrations can be fairly well represented
as fragments of two {\em leading} spiral arms S5-N6, N5-S6.

\section{The radial velocity field}

The density distribution function $D$ and the angular sensitivity
function $F$ are independent of the velocity $V$, and in
Eq.~(\ref{eq19.1}) we can integrate first over the velocity
\be
\ba{ll}
T(X_p,Y_p,V_k)&=\int\limits_{-\infty}^\infty\int\,D(X,Y)\,F(X-X_p,Y-Y_p)\\
&\times H[V_k - \overline{V}(X,Y)]\dd{X}\dd{Y},
\label{eq19.3}\\
\ea
\ee
where
\be
H[V_k - \overline{V}(X,Y)]=\int_{-\infty}^\infty\,G(V-V_k)\,E[V-\overline{V}(X,Y)]\dd{V}.
\label{eq19.4}
\ee
If the velocity dispersion is independent of position $X,~Y$, the
Eq.~(\ref{eq19.3}) can be made more suitable for numerical
computations. Let us use instead of $X,~Y$ the variables
$S$,~$\overline{V}$, where $S$ is the length along the line
$\overline{V}(X,Y)=const$. We have
\be
\ba{ll}
T(X_p,Y_p,V_k)&=\int_{-\infty}^\infty\,H(V_k-\overline{V})\\
&\times\left[\int_S\,D(X,Y)\,F(X-X_p,Y-Y_p)\,J\left({X,Y \over
      S,\overline{V} }\right)\dd{S}\right]\dd{V},
\label{eq19.5}
\ea
\ee
Assuming the Gaussian form both for the functions $G$ and $E$ with
dispersions $\sigma_G$, $\sigma_E$, respectively, then the function
$H$ has also the Gaussian form with the dispersion
\be
\sigma_H^2=\sigma_G^2 + \sigma_E^2.
\label{eq19.6}
\ee
Interferometric observations show \citep{Deharveng:1969vl} that the
radial velocity dispersion has practically a constant value
$\sigma_E=17$~km/s (due to the projection effect, the dispersion
$\sigma_E$ is greater than the true radial velocity dispersion in a
small volume element of the galaxy).

The formula (\ref{eq19.5}) has been used to calculate the theoretical
21-cm line profiles. An effective radial velocity dispersion,
$\sigma_H=24$~km/s, the corrected hydrogen density field, and a model
radial velocity field have been used. The radial velocity field has
been calculated from a plane disc pure rotation model, using the
obvious formula
\be
\overline{V}(X,Y)=V_0+V(R)\frac{Y}{R}\cos i,
\label{eq19.7}
\ee
where $V(R)$ is the circular velocity at the distance $R$ from the
centre of the galaxy, $V_0$ — the mean radial velocity of the galaxy,
and $i$ — the tilt angle of the plane of symmetry of the galaxy to
the line of sight.  The velocity $V(R)$ was taken from our
four-component model of the Andromeda galaxy \citep{Einasto:1970ac},
the constants are chosen as follows: $i=12.^\circ8$, $V_0=-300$~km/s.

\citet{Gottesman:1966vd} derived for 231 points $X_p,~Y_p$ the line
profiles (spectra) $T(V_k|X_p,Y_p)$. For all these points, the
theoretical profiles have been calculated. These are quite similar to
the observed profiles but, in general, shifted in the velocity. The
comparison of the profiles enables us to correct the model radial
velocity field.

In this way, we have found a solution to the integral equation
(\ref{eq19.1}). From the corrected radial velocities near the major axis
points, a new improved rotation velocity curve has been derived.

The results are presented graphically. In the right panel of Fig.~\ref{Fig19.6} the 21-cm
line profiles for a major axis point are given. The theoretical
profiles are calculated by using both the corrected and uncorrected
(observed) hydrogen densities, the model velocity field being
identical. Mean radial velocities and the point velocity $V_p$ (the
model radial velocity at the point $X_p,Y_p$) are also indicated. In
Fig.~\ref{Fig19.8} the rotation curves are presented, and in the right
panel of 
Fig.~\ref{Fig19.8} the model and observed radial velocity field.

The analysis of the results can be summarised as follows:\\
(a) the change of the density causes both vertical and horizontal
shifts in the line profiles, therefore, an unbiased radial velocity
field can be derived only by using carefully corrected densities;\\
(b) when the radio telescope is directed to a point of low hydrogen
density or large density gradient, the mean radial velocity of the
profile does not coincide with the point velocity; in extreme cases
near the major axis the difference exceeds 100~km/s. This effect has
caused large systematic errors in the previous reductions of radio
data by \citet{Argyle:1965us}, \citet{Gottesman:1966vd} and
\citet{Roberts:1966aa};\\
(c) the corrected radial velocity field has great irregularities in
respect to the model field.

\medskip

\hfill July 1969
 

\chapter{Structural and kinematical properties of M31
 populations}\label{ch20}

This Chapter presents our second attempt to construct a model of the
Andromeda galaxy M31. It was developed step-by-step, adding more data
on structural and kinematical properties of populations. It was
presented in IAU Symposium on ``External galaxies and quasi-stellar
objects'', held in Uppsala August 10 - 14, 1970, and published by
\citet{Einasto:1972ab}.  Here the model is presented according to the
Thesis version of the Chapter, using also data and text of the
Symposium version.

\section{Introduction}

To construct a meaningful physical theory of the structure and
evolution of a galaxy, one needs reliable data on parameters,
describing the spatial and kinematical structure of the galaxy and its
subsystems of different ages. As such parameters one can adopt: the
mass of the subsystem $\mm{M}$, its mean radius $a_0$, the axial
ratio of equidensity ellipsoids $\epsilon$ (supposing equidensity
surfaces of subsystems to be ellipsoids of rotational symmetry and
constant axial ratio), and suitable structural parameters determining
the degree of concentration of the mass to the centre of the
system. As morphological parameters, we can consider colour and
mass-to-light ratio of subsystems. As descriptive functions, we
can use spatial density $\rho$, projected luminosity $L$, circular
velocity $V$, rotational velocity $V_\theta$, velocity dispersions
$\sigma_R,~\sigma_\theta,~\sigma_z$, and some others.

In a series of papers, \citet{Einasto:1969ab} and
\citet{Einasto:1970ac, Einasto:1970tz, Einasto:1970vz} studied the
structure of the Andromeda galaxy and its subsystems, and calculated
preliminary values for descriptive parameters. Recently, new
observational data have become available for the nucleus, the
subsystem of globular clusters, and interstellar hydrogen. It appears
reasonable to use these data to redetermine structural parameters of
the populations. Also the reconstruction of the physical
evolution of galaxies was made on the basis of the theory of evolution
of stars of different mass and composition, described in the following
Chapters. This allowed to calculate structural and kinematical parameters of
subsystems of the Andromeda galaxy M31, and to construct its new
model. Preliminary results of this work were published by
\citet{Einasto:1972ab}, here we describe the new model in more detail. 

\section{Reddening and luminosity dimming of M31}

\citet{Einasto:1969aa} did not correct data for the reddening effect,
\citet{Einasto:1972ab} used for correction data by \citet{Arp:1965aa}.
A collection of reddening determinations is given in Table~\ref{Tab20.1},
where we show the colour excess $E(B-V)$ of various objects in M31 and
its vicinity.

{\begin{table*}[h]
\centering    
\caption{Reddening determinations in M31} 
\begin{tabular}{lcl}
\hline  \hline
Object&  $E(B-V)$& References\\
  \hline
Halo clusters& $0.08 \pm 0.02$&\citet{van-den-Bergh:1969wi} \\
 Field stars&        $0.11 \pm 0.02$&   \citet{McClure:1968ti}     \\
Blue open clusters&$0.12 \pm 0.04$&\citet{Schmidt-Kaler:1967wa}\\
Nucleus&                0.13&\citet{Einasto:1972ab}\\
Nucleus&                0.20&\citet{Arp:1965aa}\\
Cepheids&              $0.16 \pm 0.03$&\citet{Baade:1963ud}\\
Cepheids&               0.15                   &\citet{van-den-Bergh:1968ab}\\
Mean M31& $0.17 - 0.20$                 &\citet{Einasto:1972ab}\\
          \hline
\label{Tab20.1}   
\end{tabular} 
\end{table*} 
}

The colour excess $E$ was determined as follows. Data by
\citet{van-den-Bergh:1969wi}, \citet{Kinman:1965vv},
\citet{Lallemand:1960to}, \citet{Sandage:1969aa} suggest that the
apparent colour excess of the nucleus of M31 is $B-V=1.04$ and
$U-B=0.78$, a mean colour excess of galaxies of Sb type is $B-V=0.97$,
$U-B=0.59$ \citep{de-Vaucouleurs:1961wr}, which gives $E(B-V)=0.07$
and $E(U-B)=0.19$. The colour excess $E(U-B)$ can be found from
$E(B-V)$, using the relation $E(U-B)/E(B-V)=0.72$, which yields
$E(B-V)=0.26$.  The mean value is $E(B-V)=0.13$, giving double weight
to the direct estimation of $E(B-V)$.
\citet{de-Vaucouleurs:1961wr} gives for the galaxy M31 as a whole
$B-V=0.91$ and $U-B=0.50$, and as a mean for Sb galaxies $B-V=0.81$
and $U-B=0.27$, which gives a value $E(B-V)=0.17$.

A slightly larger value was found in a different
way. \citet{Schmidt-Kaler:1967wa} used to find $E(B-V)$ the most blue
clusters of M31, which did not have reddening within M31, but only
within the Galaxy.  The reddening in the Galaxy can be found also
using halo clusters \citep{van-den-Bergh:1970aa} and field stars
\citep{McClure:1968ti}.  On the basis of these data, we find that the
reddening in the M31 direction within our Galaxy is $E(B-V)=0.10 \pm
0.02$.  On the other hand, according to \citet{Sharov:1968wy}, the mean
reddening in the line of sight within M31 is $A_V=1$. If we assume
that the reddening within M31 is due to a thin layer of dust near the
symmetry plane, then objects within M31 can be divided into two
classes, nearby ones with no reddening, and more distant ones with
reddening $A_V=1$. As the mean we get $A_V=0.39$ and $E(B-V)=0.13$.
However, some reddening occurs in spiral arms, thus outer regions of
M31 have smaller reddening.  Taking these considerations into account,
we accept the mean reddening within M31 $E(B-V)=0.10$, and together
with the reddening within our Galaxy $E(B-V)=0.20$.

This overview shows that the light of different M31 subsystems is
influenced by reddening differently. As a mean reddening, we accept
$E(B-V)=0.15$, which is twice less than accepted by
\citet{Einasto:1972ab}. Using the apparent distance modulus (after
cepheids) $(m-M)_0=24.2$ and $R=A_V/E(B-V)=3$ we get for the true
distance modulus $(m-M)_0=24.2\pm 0.05$.

{\begin{figure*}[h] 
\centering 
\resizebox{0.60\textwidth}{!}{\includegraphics*{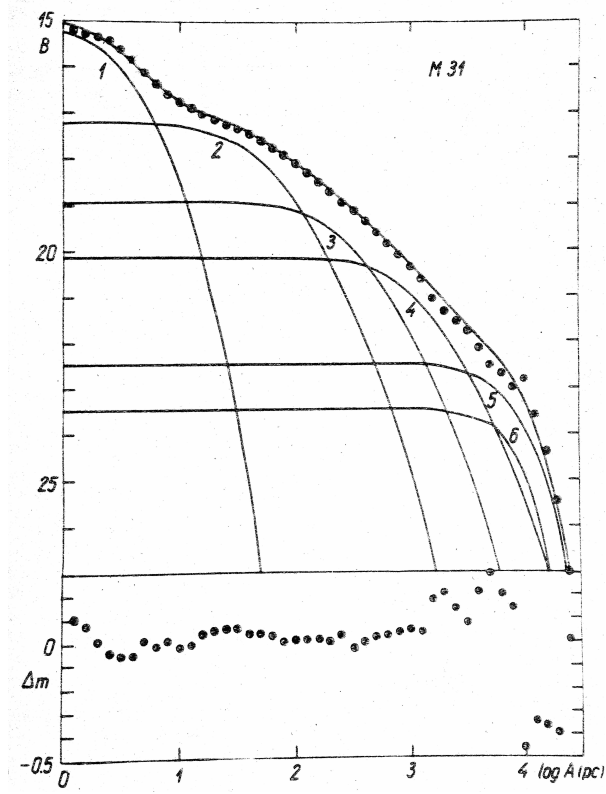}}
\caption{Luminosity profile of M31 in  photometric system B is shown by a bold
  curve, profiles of components by thin curves. The lower panel shows
  the difference $\Delta\,m = B_{Obs} - B_{Mod}$. Components are: 1 --
  nucleus, 2 -- core, 3 -- bulge, 4 -- halo, 5 -- disc, 6 -- flat.
} 
  \label{Fig20.1}
\end{figure*} 
}

\section{New model of M31}

In the new model, we used the photometrical profile in B system by
\citet{Einasto:1969aa}, using additional data to calculate improved
parameters of the M31 model. Parameters of the new model are given in
Table \ref{Tab20.2}, the new photometric profile in Fig.~\ref{Fig20.1},
and new axial ratio data in Table~\ref{Tab20.2}. We shall discuss
details on individual populations below. Here we consider principal
model parameters.

In addition to photometrical and kinematical data used earlier
\citep{Einasto:1969aa, Einasto:1970tz}, we now used the following new
data: spectro-photometrical and photometrical data on the stellar content
(\citet{van-den-Bergh:1970aa}, \citet{Sandage:1969aa},
\citet{Spinrad:1966wh,Spinrad:1969wa,Spinrad:1970aa}, 
\citet{McClure:1968ti} and \citet{de-Vaucouleurs:1969aa}), kinematical data
on globular clusters \citep{van-den-Bergh:1970aa} and interstellar
hydrogen (\citet{Deharveng:1969vl} and 
\citet{Rubin:1970tp}). Additionally, we used our results on the
physical evolution of galaxies, discussed in Chapter 22.

The main difficulty in the determination of parameters of our
preliminary models was related to calculations of mass-to-light
ratios of populations. New data allow to find more accurate values of
these parameters.

In Chapter 22,, we discussed three variants on the star formation
function with $S=0,~1$ and 2. To apply results of Chapter 22 to study
the structure of M31 we must have a choice between these parameter
values.

The mass of the neutral hydrogen in M31 can be taken equal to
$\mm{M}_{H_I}=3.7\times\,10^9~M_\odot$ (\citet{Argyle:1965us},
\citet{Gottesman:1966vd} and  \citet{Deharveng:1969vl}), using for the
distance $d=690$~kpc. The mass of ionised hydrogen is less by several
orders, and the total mass of interstellar matter is
$\mm{M}_G=5.3\times 10^9\,M_\odot$, if we accept normal chemical
abundance with $X= 0.70$. If we accept the total mass of the M31
$\mm{M}_t=218\times\,10^9\,M_\odot$ according to
\citet{Einasto:1972ab}, then we get for the mass of the gas
$\mm{M}_G=0.024\,\mm{M}_t$. If we neglect the gas, expelled from stars
during their evolution, and take the total age of M31 equal to
$10^{10}$ years, then with the exponential decrease of gas mass ($S=1$), this
gas fraction corresponds to star formation function parameter
$K=2.7\times\,10^9$~yr.  If we use $S=2$, we get for the
characteristic time $K=0.25\times\,10^9$~yr.

In our model, the flat component has $\epsilon=0.02$, and the disc has
$\epsilon=0.08$. We can attribute all populations with $\epsilon <0.4$
to the flat component, with  a maximal age $1.7\times10^9$~yr according
to \citet{Einasto:1970ad}. During last billions of years, the star
formation rate in galaxies like M31 was approximately constant, and we get
using data from Chapter 22 for the mass-to-light ratio
$f_B=0.43$. The total mass of stars formed during last
$1.7\times10^9$~yr is $\mm{M}_S=4.7\times10^9\,M_\odot$, and
luminosity $L_S=10.9\times\,10^9~L_\odot$, if we accept $S=1$ and
$K=2.7\times2.7\times10^9$~yr.  For $S=2$ and $K=0.25\times10^9$~yr we
get $\mm{M}_S=1.06\times10^9~M_\odot$ and
$L_S=2.46\times\,10^9~L_\odot$.  According to \citet{Einasto:1969aa}
the total luminosity of the flat component is
$L_B=3.0\times10^9~L_\odot$ (using $E(B-V)=0.15$). This comparison
suggests that the model with $S=2$ agrees much better with
observations.

Using data given in Chapter 22, we find that the mass-to-light ratio
of the whole galaxy is $f_B=9.5$, when we accept normal solar
composition $Z=0.02$,  $S=2$,  $T_G=10\times10^9$~yr, and star
formation parameter $K=0.25\times19^9$. From observations, using
$E(B-V)=0.15$, we get $f_B=11.08$, which corresponds to composition
with a slightly higher metal content. This result is expected, since
according to \citet{Spinrad:1969wa} the metal content in our Galaxy is
slightly higher than the solar content, and according to
\citet{van-den-Bergh:1969wi} the metal content in M31 is higher than
in the Galaxy.

Calculations by \citet{Cameron:1971aa} demonstrate that the
metal-enrichment of interstellar gas with heavy elements was very
rapid in the early phase of the evolution of the Galaxy. In later
stages of the evolution, the chemical content was almost constant with
time. For this reason, we can accept that the chemical composition of
flat and disc components did not change with time, and that this
content is representative for the whole galaxy. In this case we can
take that the sum of the flat and disc components, as well as the
galaxy as a whole, have constant ratio $f_B$. Accepting the values 
given above for the mass of the interstellar matter and the mass and
luminosity of young 
stars, we get for the mass of the flat component
$\mm{M}_F=6.36\times10^9~M_\odot$, and luminosity
$L_F=2.4\times10^9~L_\odot$.  For the disc, we have the luminosity from the
photometric profile, $L_D=7.57\times10^9~L_\odot$, and for the summed
luminosity of the disc and flat components:
$L_{FD}=10.03\times10^9~L_\odot$, which yields $f_B= 11.08$, and mass of
these components, $\mm{M}_{FD}=112.2\times10^9~M\odot$.

Now we consider data on spherical components. New data suggest that in
these components, there exist large local anomalies in the chemical
composition. For this reason, we cannot consider this component as a 
homogeneous one. Our data suggest that the spherical population can be
well described by three sub-systems. The inner-most subsystem has
chemical composition and mass-to-light ratio $f_B$, similar to the 
nucleus \citep{Spinrad:1971tv}. For the intermediate subsystem (bulge), we
accept a normal chemical composition and $f_B$, and for the most
extended subsystem (halo) we accept chemical composition and $f_B$, close to
that of halo globular star clusters.  After several trials, we accepted
mass and mass-to-light parameters, given in
Table~\ref{Tab20.2}\footnote{The Table is from the Thesis
  version.}. Following \citet{Einasto:1972af, Einasto:1974a} 
we designate the inner bulge as the core.  \citet{Einasto:1972ab}
calculated preliminary  versions of mass-to-light ratios and masses for
components, used to prepare  Fig.~\ref{Fig20.6}, and listed in Figure
caption.

\begin{table*}[h]
\centering    
\caption{Parameters of the components of M31} 
\begin{tabular}{lcccccccc}
  \hline  \hline
  Quantity&Unit&Nucleus&Core&Bulge&Halo&Disc&Flat&Total\\
  \hline
 $\epsilon$&      &0.80&0.80&0.80&0.30&0.08&0.02&\\
 $\nu$     &         & 1/2 & 1/4 & 1/4 &1/4 &  1&2& \\
 $a_0$       &kpc  &0.005&0.15&0.8&3&9.2&8&\\
$L_B$&$10^9\,L_\odot$&0.0057&0.501&2.94&6.18&7.57&2.46&19.65\\
$L_V$&$10^9\,L_\odot$&0.0073&0.642&3.63&6.34&9.08&1.88&21.57\\
$\mm{M}$&$10^9\,M_\odot$&0.306&27.0&58.4&19.8&105.8&6.36&217.7\\
$f_B$&                                 &53.8&53.8&19.88&3.20&13.98&2.59&11.08\\         
$f_V$&                                 &42.0&42.0&16.09&11.66&3.39&3.39&10.09\\         
$U-B$&                               &0.67&0.65&0.47&0.16&0.43&$-0.17$&  \\ 
$B-V$&                                &0.89&0.89&0.85&0.65&0.82&0.33&0.72\\  
  \hline
\label{Tab20.2}   
\end{tabular}
\end{table*}

{\begin{figure*}[h] 
 \centering 
\resizebox{0.50\textwidth}{!}{\includegraphics*{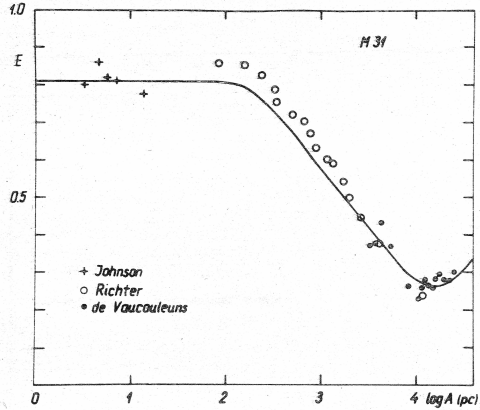}}
\caption{ The apparent axial ratio of isophotes $E$ as function of the
distance along major axis $A$. The curve is as found in our model,
symbols show observational determinations.} 
  \label{Fig20.2}
\end{figure*} 
}

{\begin{figure*}[h] 
\centering 
\resizebox{0.50\textwidth}{!}{\includegraphics*{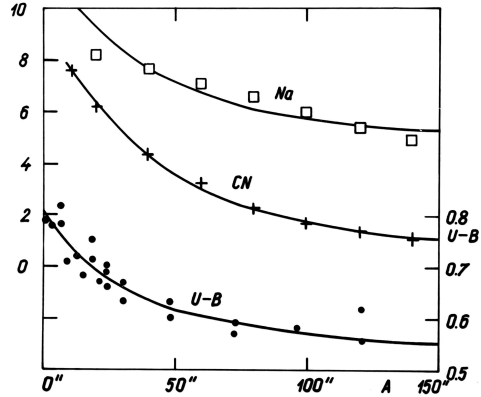}}
\caption{  Colour $U -
B$ and the intensity of spectral features $N\alpha$ and $CN$ in
central region of M31 as functions of the distance from the centre
$A$.  Model curves were found under the assumption that subcomponents
1 and 2 (nucleus and core) have a composition, rich in heavy
elements, and the  bulge has normal composition. } 
  \label{Fig20.3}
\end{figure*} 
}

{\begin{figure*}[h] 
\centering 
\resizebox{0.50\textwidth}{!}{\includegraphics*{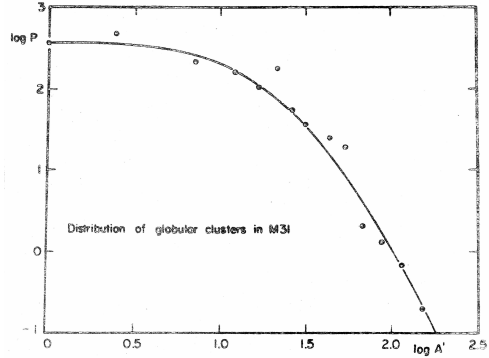}}
\caption{ The distribution of globular clusters in M31.
  } 
  \label{Fig20.4}
\end{figure*} 
}

{\begin{figure*}[h] 
\centering 
\resizebox{0.50\textwidth}{!}{\includegraphics*{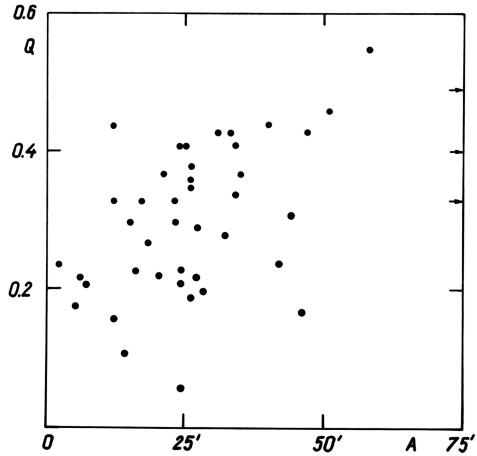}}
\caption{ The dependence of the metallicity parameter $Q$ of
  M31 globular clusters on the distance from the centre $A$. } 
  \label{Fig20.5}
\end{figure*} 
}

{\begin{figure*}[h] 
\centering 
\resizebox{0.70\textwidth}{!}{\includegraphics*{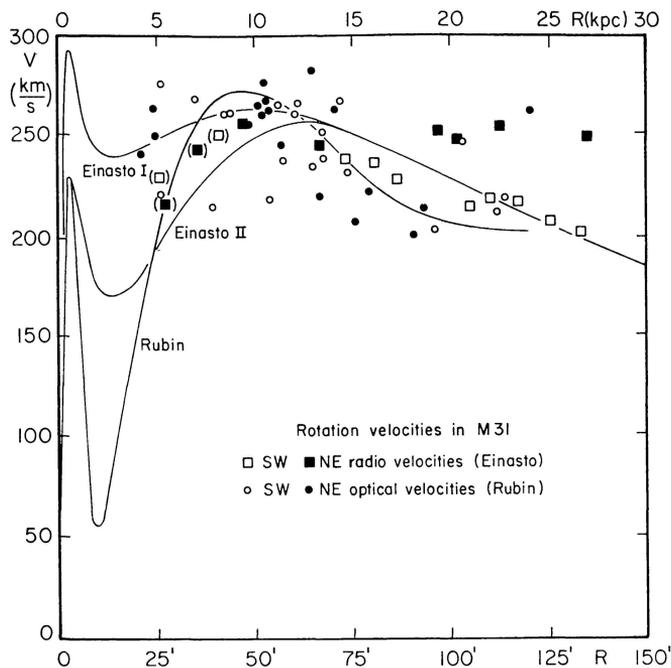}}
\caption{ The rotation curve of M31. Observed rotation data from
  optical and radio measurements according to a summary by
  \citet{Rubin:1970tp} are shown by various symbols. Solid curves show
rotation according to the model by \citet{Rubin:1970tp}, and two variants
of the model by \citet{Einasto:1972ab}. In this model   mass-to-light ratios $f_V$
of nucleus, bulge, halo, disc and flat
populations are in variant I --   42, ~4.2,~4.1,~10.1,~3.2, and in variant
II -- 42,~2.7,~0.07,~14.7,~6.5.
}
\label{Fig20.6}
\end{figure*} 
}

{\begin{figure*}[h] 
\centering 
\resizebox{0.60\textwidth}{!}{\includegraphics*{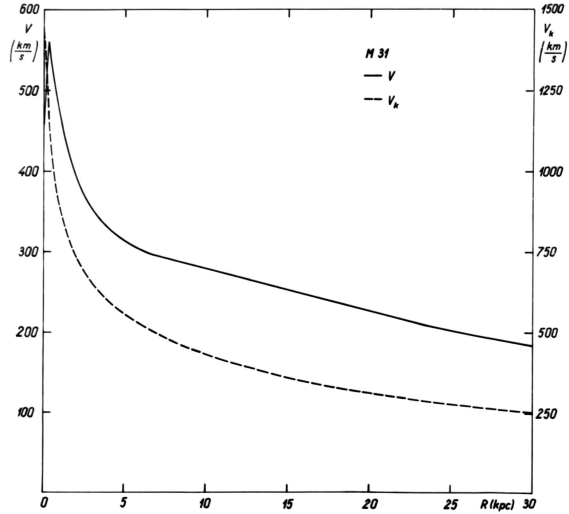}}
\caption{ Circular and escape velocities, $V$ and $V_k$
  of M31 as functions of distance from the centre $R$ according to
  the model by \citet{Einasto:1972ab}.}
  \label{Fig20.7}
\end{figure*} 
}

{\begin{figure*}[h] 
\centering 
\resizebox{0.70\textwidth}{!}{\includegraphics*{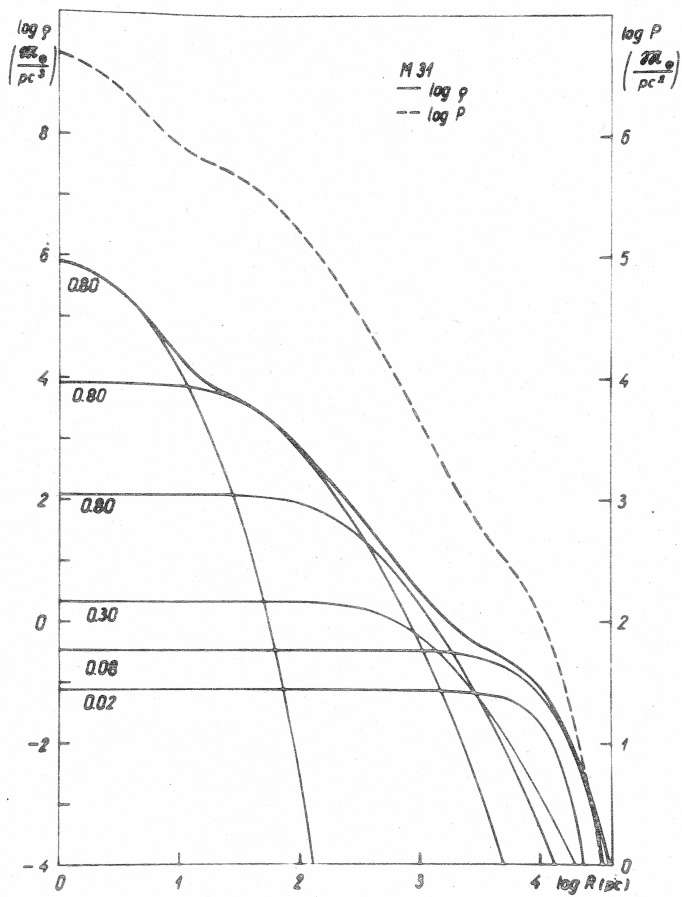}}
\caption{Distribution of the projected density of M31 and its components.  }
  \label{Fig20.8}
\end{figure*} 
}

{\begin{figure*}[h] 
\centering 
\resizebox{0.49\textwidth}{!}{\includegraphics*{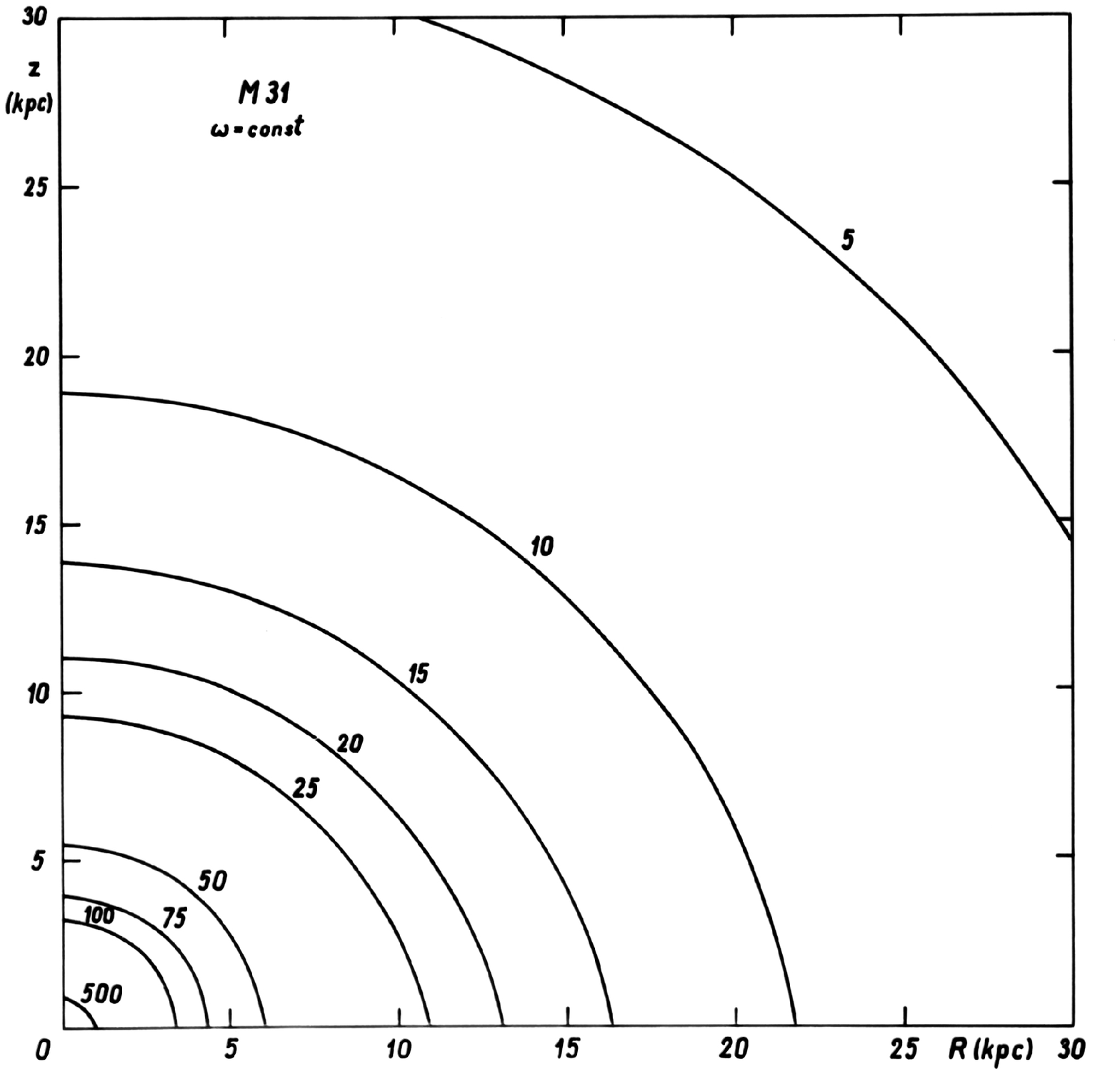}}
\resizebox{0.49\textwidth}{!}{\includegraphics*{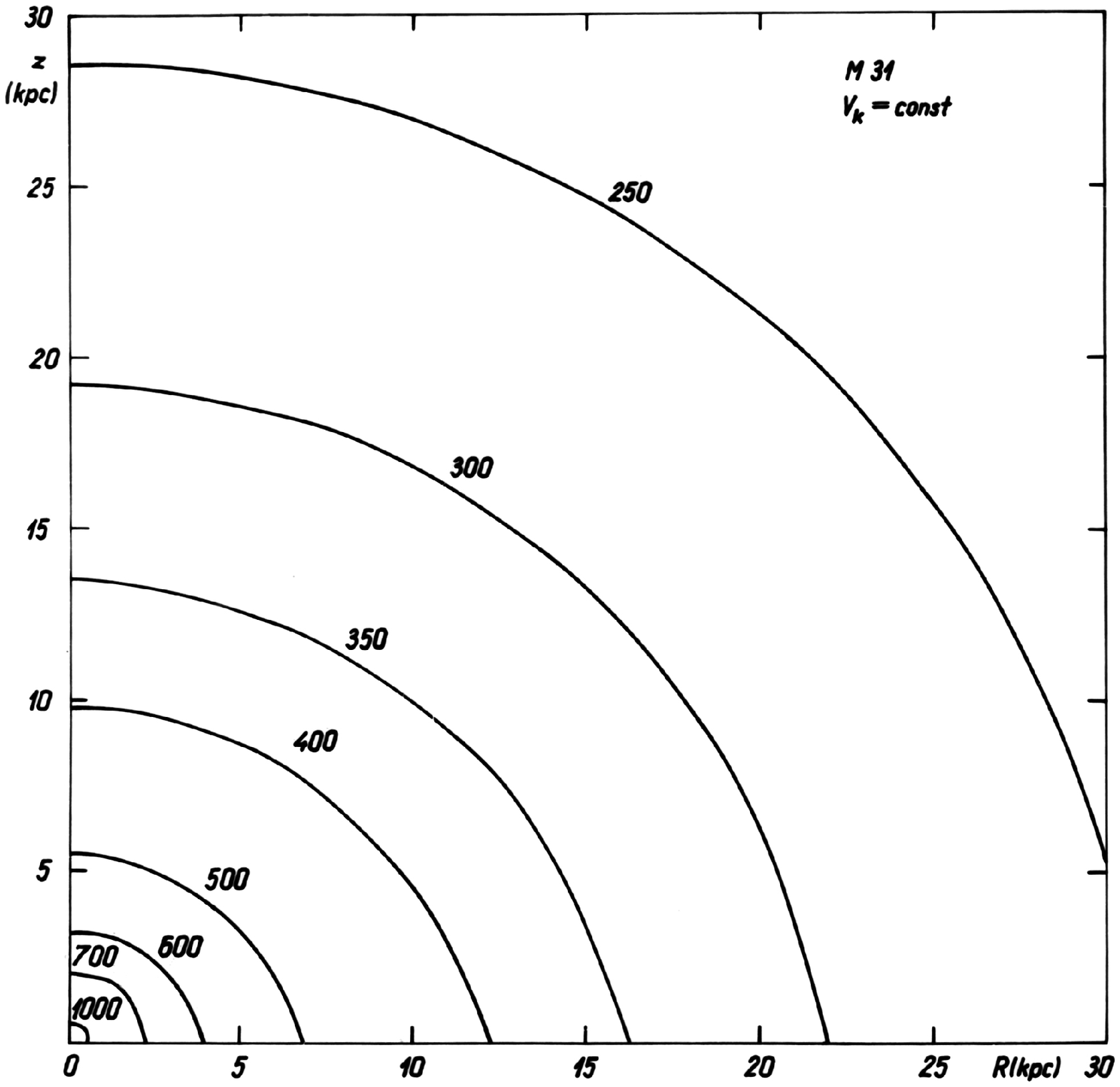}}
\caption{ Left and right panels show isolines of M31 angular velocity
  $\omega$ and escape velocity $V_k$, respectively.
 } 
  \label{Fig20.9}
\end{figure*} 
}

{\begin{figure*}[ht] 
\centering 
\resizebox{0.50\textwidth}{!}{\includegraphics*{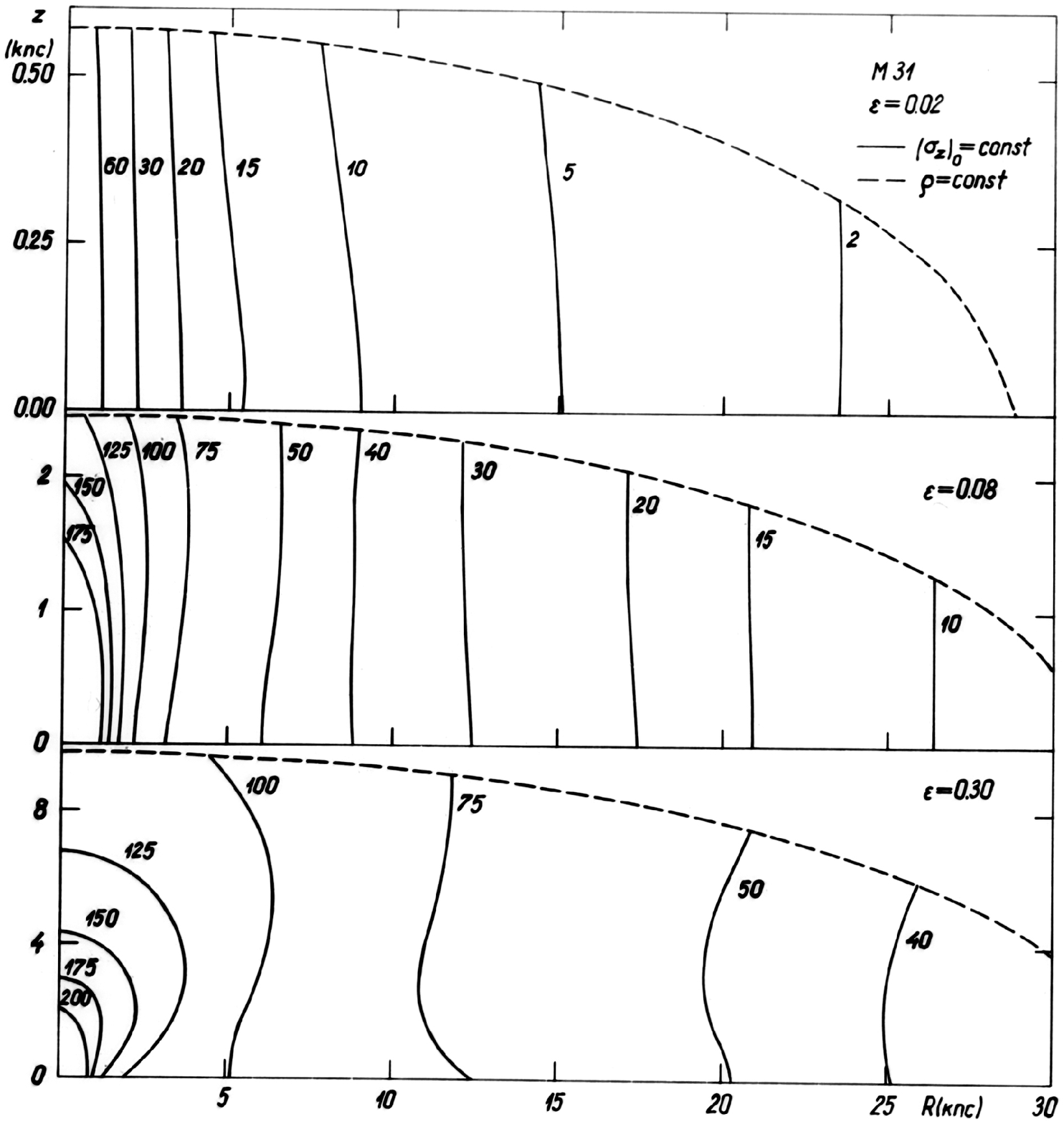}}\\
\resizebox{0.45\textwidth}{!}{\includegraphics*{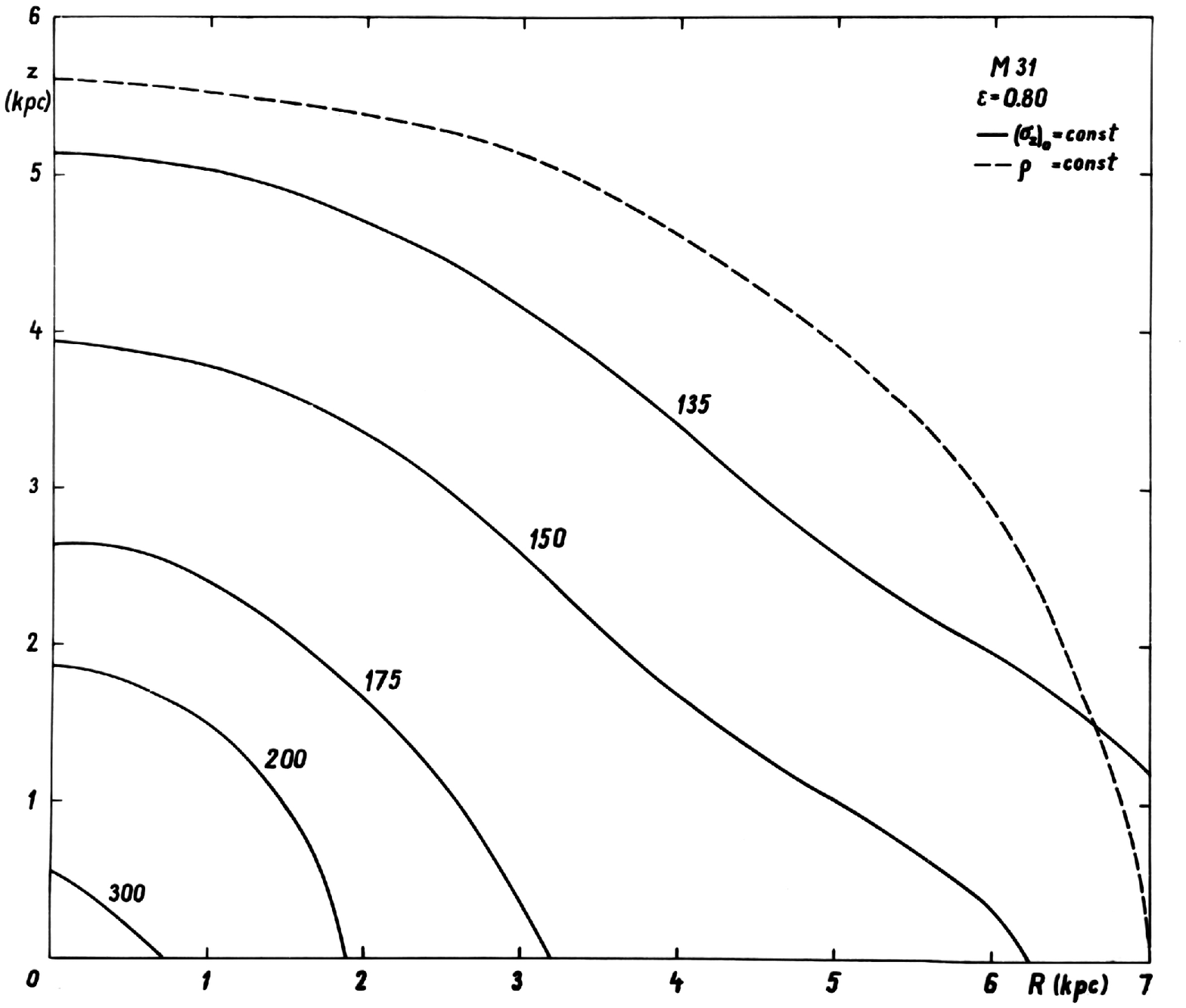}}
\caption{ {\em Top:} Isolines of velocity dispersion $\sigma_z(R,z)$
  and density $\rho(R,z)$ in the meridional plane $R,z$ of M31 for
  three test populations of flattening
  $\epsilon=0.02,~0.08,~0.30$. {\em Bottom:} Similar isolines for
  $\epsilon=0.80$. For explanations see Chapter 7.}
  \label{Fig20.11}
\end{figure*} 
}

The colour index of the nucleus and inner components of  spherical
components (core and bulge) were found from photometric observations by
\citet{Lallemand:1960to}, \citet{de-Vaucouleurs:1961wr},
\citet{Kinman:1965vv} and 
\citet{Sandage:1969aa}. For other components, there are no direct
observational data, and colours were calculated using the theory of the
evolution of integral colours of components, as described in Chapter
22.

Luminosities, effective radii and axial ratios of components were
found from photometric data. The radius of the metal-rich inner
spherical component (subpopulation 2 — core) was found from
spectrophotometric data by \citet{Spinrad:1971tv}. To get in the model
the minimum of the isophote axes ratio function, similar to the observed
minimum (see Fig.~\ref{Fig20.2}), the axial ratio of the halo was
decreased to $\epsilon=0.30$, and of the disc to $\epsilon=0.08$. Also,
it was needed to decrease the angle between the system plane and line
of sight from $i=12.^\circ7$ to $i=12.^\circ5$.

Using the virial theorem and hydrodynamical model for all subsystems,
mean velocity dispersions $\sigma_R,~\sigma_z,~\sigma_r$ were
calculated.  Velocity dispersions $\sigma_z(R,z)$ of four populations
in the meridional plane are shown in Fig.~\ref{Fig20.11}.

\section{Nucleus}

At the Basel IAU Symposium on Spiral Structure of the Galaxy, we argued
that different methods lead to different values for the mass of the
nucleus of M31 \citep{Einasto:1970vz}. On the basis of photometric
data by  \citet{Redman:1937tg}, \citet{Johnson:1961aa}, and 
\citet{Kinman:1965vv}, and spectrophotometric data on the stellar
content and mass-to-light ratio, $\mm{M}/L=f=16$ \citep{Spinrad:1966wh}
we obtained \citep{Einasto:1969aa} for the mass of the nucleus the
value $\mm{M}=5\times\,10^7~M_\odot $. On the other hand, the
known mean radial velocity dispersion of stars in the nucleus is 
$\sigma_r=225$~km/sec \citep{Minkowski:1962aa}, the mean radius is 
$a_0=5$~pc (obtained from photometric profile of the nucleus), and
axial ratio of equidensity ellipsoids is $\epsilon=0.8$. These data
enable us to apply the tensor virial theorem (see Chapter 12), which yield for the mass
$\mm{M}=5\times10^8~M_\odot $ \citep{Einasto:1970vz}.

The discrepancy in mass can be removed supposing, as
\citet{Lynden-Bell:1969vs} does, that a massive body exists in 
the centre of the galaxy. In this case, the tensor virial theorem is to be
modified. For the mean radial velocity dispersion of stars in the
nucleus we have
\be
\overline{\sigma}_r^2 = Ga_0^{-1}\beta_r(\mm{M}_C + H_0\,\mm{M}_0),
\label{20.4.1}
\ee
where $G$ is the gravitational constant, $a_0$ is the harmonic mean
radius of the stellar subsystem of the nucleus, $\mm{M}_C$  and
$\mm{M}_0$ are masses of the central body and the stellar population
of the nucleus, respectively, and $H_0$ and $\beta_r$ are
dimensionless parameters. From photometric observations, we get for the
nucleus $\epsilon=0.8$, and $i=12.^\circ8$, which gives
$\beta_r=0.375$. The spatial density of stellar population of the
nucleus can be well represented by the exponential model, for which 
$H_0=0.312$. Adopting for the mean radius and the mass of the stellar
component the values given above, we obtain for the central mass
$\mm{M}_C=1.4\times10^8~M_\odot$.

After the Basel Symposium, I asked Dr. H. Spinrad to determine the
maximum value of mass-to-light ratio, consistent with spectroscopic
observations. Recently new data became available
\citep{Spinrad:1971aa}. According to the new model of the stellar
content, the mass-to-light ratio increases to a value of $f_V=42$, due to
the necessity of adding faint M dwarfs. An upper limit to the number
of red dwarfs is given by $V-K,V-L$ colour observations
\citep{Sandage:1969aa}, which gives $f_V \le 65$.

A re-examination of the photometric data mentioned above gives for the
luminosity of the nucleus in visual light a value
$L_V=1.42\times10^7~L_\odot$.  This is four times higher than our
previous estimate. The difference is due to the absorption and colour
corrections: $A_V=0.6$ in our Galaxy, $A_V=0.3$ in M31
\citep{Arp:1965aa}, and $B-V=1.0$ \citep{Sandage:1969aa}. The
luminosity profile of M31 is shown in Fig.~\ref{Fig20.1}. Using these
data, we obtain for the mass of the stellar component of the nucleus
$\mm{M}_0=6.4\pm 2.1\times 10^8~M_\odot$, accepting
$E(B-V)=0.3$). On the other hand, on the basis of the virial theorem we get
$\mm{M}_0=5.7\pm1.9\times10^8~M_\odot$ (using $\mm{M}_C=0$).

Good agreement between these two independent estimates shows that it
is not necessary to suppose the existence in the centre of M31 of a
body of large mass. However, ultraviolet observations carried out from
the OAO indicate the presence of a UV-source in the nucleus of M31. If
this source has small dimensions compared with the mean radius of the
stellar component of the nucleus, Eq.~(\ref{20.4.1}) may be applied to
estimate an upper limit for the point mass, which may be attributed to
the UV-source. Assuming $\mm{M}_0 \ge 4.0\times10^8~M_\odot$, we
find $\mm{M}_C\le 0.5\times10^8~M_\odot$. Thus, presently
available data are not sufficient to confirm the presence of a
point-like mass in the centre of M31. However, indirect data (recent
explosion of the M82 galaxy) make the existence of such a body very
likely in galaxies of type Sb.

\citet{Sandage:1969aa} concluded from the $U-B$ colour variation along
the major axis that the stellar content of the nucleus differs from
that of the bulge. However, at present it is not clear whether the
variation is due to a difference in stellar content or to the
existence of a non-stellar UV-source in the centre of M31. A
difference in stellar content is not excluded because the nucleus is
dynamically almost isolated and there is no appreciable exchange of
stars between the nucleus and the bulge.

{\begin{table*}[h]
\centering    
\caption{Apogalactic distances from the nucleus centre} 
\begin{tabular}{cc}
\hline  \hline
  $V$ (km/sec)&$R_{apogal}$ (pc)\\
   \hline
  225& 1.8\\
  450&8\\
  1080&85\\
          \hline
\label{Tab20.4}   
\end{tabular} 
\end{table*} 
}

A rather great difference in chemical composition of stars on various
distances from the centre of galaxies raises the question: Can this
difference have a permanent character or will it change with time
due to a mixture of stellar orbits? If most stellar orbits in the
central region are radially extended, then stars in different regions
will be mixed. However, our calculations showed this does not occur
with stars of the nucleus as well with stars of inner region of the 
halo. In other words, stellar populations in different subsystems are
dynamically isolated. To demonstrate this, we 
calculated, with the aid of a model of the mass distribution and the
gravitational field of the nucleus, the apogalactic distances of stars
moving through the centre of the nucleus with various velocities, see
Table~\ref{Tab20.4}. The majority of the nucleus stars have velocities
of some hundred km/sec and do not move far off from the centre. Only
stars having large velocities exceeding the escape velocity with
respect to the nucleus, 1080~km/sec, go far away.

The mass density near the centre of M31 according to our model is
$2\times10^6~M_\odot$~pc$^{-3}$, the angular velocity is 12
km/sec/pc \citep{Lallemand:1960to}. Using the tensor virial theorem
and supposing a rigid-body rotation, we find that the rotation energy
is only 7.5~\% of the total kinetic energy of the nucleus. The binding
energy of the nucleus (total negative energy per unit mass) is
$7.5\times10^4$~(km/sec)$^2$.

\section{Bulge}

The bulge is the densest component of spheroidal populations of
spiral galaxies, the characteristic radius of the bulge of M31 is of
the order of 1~kpc. Photometrical data suggest that the bulge is
fairly homogeneous. However, spectrophotometric data indicate that
there exist differences in chemical composition
(\citet{Sandage:1969aa}, \citet{Spinrad:1971aa} and 
\citet{Spinrad:1971tv}). Chemical composition influences mass-to-light
ratios. Data by \citet{Spinrad:1971tv} suggest that in the inner part
of the bulge (at distance 1' from the centre of M31) $f_V=45$, but
data by \citet{Tinsley:1971} suggest $f_V=9.2$. If we attribute for
the whole bulge $f_V \approx 40 - 50$, we get for the mass of the
bulge an absurd value, larger than the mass of the whole galaxy
M31. On the other hand, there is no doubt that the inner part of
the bulge has high value of $f_V$. For this reason, we divide the bulge
into two subsystems.

In the choice of $f_V$ for the inner bulge, we shall use the value
$f_V=42$. If we accept the value $f_V=92$, suggested by
\citet{Tinsley:1971}, then we get for this subsystem velocity
dispersion $\sigma_r=450$~km/sec, which is too high.

Earlier we assumed (Chapter 17) that velocity dispersion can have a
minimum near the centre of the system. Later calculations (Chapter 18)
suggested that this is possible only in the case when the increase of the
density towards the centre is rather modest.  The projected luminosity
function has near the centre of M31 a high maximum, as in other spiral
and elliptical galaxies.  The minimum in the velocity dispersion near
the centre can be avoided only in case the mass-to-light ratio in
central regions is much lower than in the galaxy as a whole. Actually,
this is not the case, thus a minimum of the velocity dispersion near
the centre is unavoidable. This minimum is not very deep, as suggested
in our earlier model of M31, described in Chapters 17 and 18.

We determined the radius of the inner bulge using data on the chemical
composition of M31 (\citet{Sandage:1969aa} and 
\citet{Spinrad:1971tv}). In Fig.~\ref{Fig20.3} we show the colour
$U-B$, index $Na$ and $CN$ according to model (curve) and data
(points). 

There are no direct data on the chemical composition and mass-to-light
ratio for the outer component of the bulge. We accepted for this
component normal chemical composition, colour and parameters $f_B$ and
$f_V$, using data of calculations of the evolution in Chapter
22. Comparison with parameters of other components of the galaxy, and
the virial theorem suggest that parameters accepted for this
component cannot have large errors.

\section{Halo}

The halo is the outer subsystem of the spherical component of galaxies. We
found the effective radius, total luminosity and the axial ratio from
photometric data and colour. Also we used mass-to-light ratios of
globular clusters, which are essentially part of the halo component.

Halo population stars have the same physical characteristics as 
 globular clusters. Also spatial and kinematical properties of halo
 populations stars are close to similar properties of the globular
 cluster subsystem. The most characteristic property of halo stars
 is the very low metal content.  This follows from direct spectral
 observations as well as from the metal-index $Q=(U-B)-0.72\,(B-V)$
 \citep{van-den-Bergh:1969wi}. The mass-to-light ratio of globular
 clusters is approximately equal to unity
 \citep{Schwarzschild:1955aa}. Our calculations of the evolution of
 stellar systems suggest that this value is underestimated.  If we
 accept the heavy element content $Z=0.001$, exponential law of
 the star formation function with $S=1$, characteristic time
 $K=0.5\times10^9$~yr, and minimal mass of forming stars $M_0
 =0.1\, M_\odot$, then mass-to-light ratios are $f_B \approx f_V
 \approx 3$, see Chapter 22. This result seems to be acceptable, see
 Chapter 23.  If the metal content is higher, say $Z=0.005$, and
 parameter $K=0.5\times10^9$~yr, then with minimal mass of forming
 stars $M_0=0.3\, M_\odot$ we get $f_B\approx 3$, and with
 $M_0=0.1\, M_\odot$ we get  $f_B\approx 5$.  The true value of
 the mass-to-light ratio lies probably in-between.

In the Andromeda galaxy M31, we have the possibility of studying the
overall spatial distribution of an old population, using the sample of
globular clusters. Our sample of globular clusters has been collected
on the basis of \citet{Vetesnik:1962tf} catalogue.  Photometric data on
clusters were collected from various sources \citep{Sharov:1968aa}.
Probable open clusters are excluded from the general list; they lie in
the $V,(B-V)$ diagram to the left of the reddening line
$A_V/E_{B-V}=2.5$ \citep{van-den-Bergh:1969wi}, going through the
point $V=18.0$, $B-V=1.00$. $H_\alpha$-regions \citep{Haro:1950vk},
objects without any photometric data, and very faint clusters
($B>19.0$), were also excluded. The remaining sample was divided into
two groups: bright clusters ($B\le 17.5$), and faint clusters ($17.5 <
B \le 19.0$), which consist of 101 and 92 clusters respectively.

On the basis of the measured ($x,~y$) coordinates, published by
\citet{Vetesnik:1962tf}, the galactocentric coordinates $W,~U$ along
the major and minor axes of the galaxy, and the projected distance
from the centre $A=(W^2+E^{-2}U^2)^{1/2}$ were calculated. $E$ is the
apparent axial ratio of equidensity ellipses. We found $E=0.57$, which
corresponds to the true axial ratio $\epsilon=0.54$.

The distribution of clusters in $A$ is somewhat different in two
groups, which can perhaps be explained by a selection effect in
observations. In the central and outer halo region, the relative number
of faint clusters is smaller. After a slight correction of numbers of
faint clusters, both groups were united and general distribution was
found, see Fig.~\ref{Fig20.4}.  The distribution can be fairly well
represented by a modified exponential density law
\citep{Einasto:1970ae} with the mean radius $a_0=4.5$~kpc, $N=4$, and
$x_0=10.5$.

This result shows that the system of globular clusters has a much
greater mean radius than the spheroidal component of M31. As we have
no reason to believe that the spatial structure of the system of
globular clusters differs significantly from that of other old 
first-generation stars, we come to the conclusion that the subsystem of old
stars cannot be identified with the spheroidal component of the
galaxy. In the central part of the galaxy, star formation has probably
taken place much longer than in the halo, giving rise to the formation
of the bulge.

As to the kinematics of globular clusters,
\citet{van-den-Bergh:1969wi} has recently determined radial velocities
for 44 bright globular clusters in M31. From his data the mean
galactocentric radial velocity dispersion can be found. Adopting for
the systemic velocity of the galaxy $-300$~km/sec
(\citet{van-den-Bergh:1969wi}, \citet{Rubin:1970tp}), we get
$\sigma_r=122$~km/sec, which is only 0.54 of the mean radial velocity
dispersion in the nucleus \citet{Minkowski:1962aa}\footnote{More
  accurate measurements by \citet{Faber:1976} indicate that the
  velocity dispersion of the nucleus of M31 is $\sigma=180$ km/s.}.

Preliminary model calculations show that velocity dispersions of
the nucleus and the subsystem of globular clusters mainly depend on
the total mass of the galaxy, and are quite insensitive to the mass
concentration towards the centre.

The difference between the mean velocity dispersions of the nucleus
and the subsystem of globular clusters has one important
consequence. Velocity dispersions near the nucleus have been used to
determine the total mass of a galaxy. For example, by this method
\citet{Brandt:1969wv} obtained for M87
$\mm{M}=2.7\times10^{12}~M_\odot$, and $\mm{M}/L=85$. However, a
strong dependence of the dispersion on the distance $R$ shows that the
velocity dispersion near the centre characterises the mass and the
size of the nucleus only. If  relative velocity dispersion
in other galaxies is the same as in M31, the total mass of the galaxy
and the respective mass-to-light ratio obtained from the virial
theorem are approximately four times higher than the true ones.

\section{Disc and flat component}

We attribute to the disc all subsystems with true axial ratios $0.04 <
\epsilon \le 0.15$, and to the flat component subsystems with
$\epsilon \le 0.04$. Data on our Galaxy suggest that the flat
component includes interstellar matter and populations of very young
stars of ages up to $1.7\times10^9$~yr.  Oldest disc stars are only
slightly younger than the whole Galaxy. This suggests that star
formation in the disc started soon after the formation of the
Galaxy. In spite of the great spread of ages of disc stars, their mean
chemical composition is rather constant and close the Galaxy as a
whole. This allows to find colour and mass-to-light ratio of these
components.

Let us now discuss possible errors of parameters, in particular
mass-to-light ratios.

According to photometric data, the luminosity of the summed disc and
flat components is 51~\% of the total luminosity of the whole M31
galaxy. If these two components have the same mean value of $f$ as the
galaxy as a whole, then their summed mass is also 51~\% of the mass of
the whole galaxy.  This value seems to be too low, since usually it
is expected that spherical populations have the total mass of about 30~\%
of the whole galaxy. However, in this estimate it is assumed, that
all populations have identical $f_B=7$. Our data suggest that flat
and disc populations have $f_B \approx 15$. Thus, equal values of $f_B$ are
not likely. On the other hand, \citet{de-Vaucouleurs:1958uc} has found
that the mass of the spherical component has a mass about 70~\% of
the whole M31 galaxy. In this case for the mass-to-light ratio of the
bulge we get $f_B=30$, and for disc and flat components $f_B=6$. This
version of parameters means that outer parts of the bulge are similar
to inner parts, and that disc and flat components have a noticeable
deficit of metals in respect to solar abundance. These versions are
not confirmed by observations.

We come to the conclusion that accepted models of disc and flat
components cannot have large errors.  Our error estimate is of the
order of 10~\%, not including possible systematic errors in the
distance and absorption.

The total mass of M31 galaxy was determined using rotation data on the
flat component object, first of all the interstellar hydrogen.  In
addition to data used in our preliminary model, we used new radio
observations as well as optical measurements by \citet{Rubin:1970tp}
of radial velocities of ionised hydrogen, plotted in
Fig.~\ref{Fig20.6}.  If in one stellar association there were several
objects, we used the mean velocity and mean distance $R$ from the
centre of M31. Clouds, distant from the main axis of the galaxy, were
not used. We show in the Figure also the model by
\citet{Rubin:1970tp}, as well as two variants of our model
\citep{Einasto:1972ab}.  

New data confirm the measurements by \citet{Babcock:1939wf}, showing a
maximum of rotation velocity at the distance 0.5~kpc from the centre,
and a deep minimum of velocities at the distance about $R=2$~kpc. If we
identify the measured rotation velocity with the circular velocity, we
get for the spatial and projected densities negative values around
$R=2$ from the centre. Such a model cannot be accepted. First of all,
density cannot be negative. Thus, we cannot identify rotational and
circular velocities, since in central regions of the galaxy the
velocity dispersion is large, and the respective term in
hydrodynamical equations cannot be neglected.

The minimum in rotation velocity correspond to the maximum of the velocity
dispersion.  These anomalies in kinematical characteristics lead to
the conclusion that the density should also have at this distance some
anomaly, which is not confirmed by observations. On the other hand,
the anomaly in kinematics cannot be permanent — orbital motions of
stars swipe them out.  Thus, we come to the conclusion that the
anomaly is likely related to the motion of gas only, which is caused
not only by gravitational forces. One possibility to explain this
anomaly was suggested by Oort (see \citet{Rubin:1970tp}).  According
to this hypothesis, the gas was expelled from the nucleus of M31, and
the low rotational velocity of gas is due to low angular moment of the
expelled gas.

\section{Description functions}

Using the set of model parameters discussed above, we calculated all
principal descriptive functions of the model of the galaxy
M31. Several model descriptive functions are plotted in Figures
Fig.~\ref{Fig20.7} to \ref{Fig20.11}.

\medskip

\hfill August 1971
 

\part{Evolution of galaxies}

\chapter{Reconstruction of the dynamical evolution of the
  Galaxy\label{ch21}}

\section{Introduction}

The evolution of galaxies has been in the focus of astronomers' interest
for a long time. The first steps were the understanding of the nature of
galaxies as extragalactic objects, classification of galaxies and
determination of properties of galaxies. An early review of the
structure and evolution of galaxies was given by
\citet{Eigenson:1960aa}. However, very little was known of the
structure of galaxies, thus, these theories were rather speculative and
are of historical interest only.

In 1950s, the situation changed. The 5-m Hale telescope was
launched and the new field of radio astronomy gave the first important
results.  Considerable progress was made in understanding the
evolution of stars. This progress helped to investigate the galaxy
evolution as an observational problem.  Observational studies of the
evolution of stellar systems can be divided to five groups.

The first group of observational studies is devoted to the investigation
of the evolution of star clusters and associations.  The study of star
associations by \citet{Ambartsumian:1968aa} and
\citet{Parenago:1954ab}, and photometry of star clusters by
\citet{Eggen:1950aa,Eggen:1950ab,Eggen:1950ac} and
\citet{Sandage:1957aa,Sandage:1957ab} are examples of such
investigations.  Theoretical analyses by \citet{Opik:1938aa},
\citet{Schonberg:1942aa} and \citet{Hoyle:1955aa} helped to understand
the evolution of stars.  These studies helped to understand the main
character of the evolution of stars and star clusters.

The second approach to understand the evolution of galaxies is devoted
to the study of peculiar galaxies.  This group of studies includes
the work by \citet{Ambartsumian:1968aa} on nuclei of galaxies,
\citet{Lynds:1963aa} on exploding galaxies, and atlases of peculiar
galaxies by \citet{Vorontsov-Veljaminov:1959aa} and
\citet{Arp:1966aa}.  As a result of these studies, it was clear that in
some galaxies active processes are underway.

The third group of studies is concerned with the  statistics of galaxies, which helps
to find evolutionary sequences of galaxies of various type. An example
of such studies is the work by \citet{Holmberg:1964aa}.

There is a possibility to reconstruct the evolution of
galaxies by studying the spatial structure and kinematics of stars of
various ages in the Galaxy, as done by \citet{von-Hoerner:1960aa} and
\citet{Eggen:1962}.

Finally, to understand the evolution of galaxies, it is needed to investigate
evolutionary processes in galaxies.  As examples of such studies we
mention works by \citet{Spitzer:1953aa}, \citet{Gurevich:1964aa} and
\citet{Kuzmin:1961aa} on the influence of irregular gravitational
processes, and \citet{Lynden-Bell:1967aa} on statistical mechanics of
violent relaxation in globular clusters and elliptical galaxies.

The goal of this Chapter is to find the evolutionary conclusions that can
be made on the basis of kinematical and spatial data on galaxies,
using well-established theoretical results.

\section{Evolutionary conclusions from kinematics of flat population
  objects}  

Data presented in Chapter 4 suggest that there exist three population
groups in the Galaxy with different properties — flat populations,
disc and halo. Kinematical characteristics of flat and halo
populations considerably vary with age, whereas properties of disc
populations are relatively stable.

The relationship between the velocity dispersion and rotational speed
of populations was established by \citet{Stromberg:1924aa}.  The
greater the (negative) galactocentric rotational velocity $-V_\theta$, the larger 
the velocity dispersion of a population of stars. However,
there exist important deviations from this relationship: stars with
very small velocity dispersion and interstellar gas have rotational
speeds, smaller than stellar populations with velocity
dispersion of the order 15 km/s, see Fig.~\ref{Fig3.1}.  This problem was
known long ago, see \citet{Rootsmae:1961}.  \citet{Edmondson:1956aa}
suggested to explain this effect by non-accurate treatment of
differential galactic rotation.

\cite{Oort:1964aa} emphasised that this effect can be explained by the
hypothesis that interstellar gas rotates with a velocity  lower than
the circular velocity.  If this is the case, then young stars just
``fall'' in the direction of the Galactic centre after their
birth. \citet{Dixon:1965aa,Dixon:1966aa,Dixon:1967aa,Dixon:1967ab,Dixon:1968aa}
investigated this phenomenon and gave strong arguments in favour of the
Oort's hypothesis.  Using statistics of stars with known ages from
Str\"omberg photometry, he demonstrated that young stars are in their
orbits at apogalactic positions. Young stars populate in velocity space
an elliptical region around the point, which corresponds to circular
velocity.  From these data, Dixon concluded that gas rotates at the speed
by $\Delta V_\theta = 14$~km/s lower than the circular velocity. The
possible reason for this effect is electromagnetic forces which
support gas in addition to rotation.

A similar effect is observed in the vertical direction. Radio observation
suggests that the shell of interstellar gas does not coincide exactly with the
plane of the Galaxy and has a wave form \citep{Westerhout:1957aa}. In
the Solar neighbourhood, these waves have an amplitude of about 50 pc.
After formation, stars are free from electromagnetic forces and fall
towards the Galactic plane. Very young stars are still located close
to their places of origin \citep{Dixon:1967ab}. The frequency of such
oscillations, measured by Kuzmin parameter $C$, does not depend on
the amplitude (if it is small). Thus, young stars should oscillate
around the plane of the Galaxy. Such oscillations were actually found
by \citet{Joeveer:1968b, Joeveer:1972}.  The vertical speed of such oscillating stars
has a maximum, when it crosses the plane of the Galaxy.  This occurs
for the first time, when the age of stars $t_s$ is one quarter of the
full period. Taking $C=70$ km/s/kpc  \citet{Joeveer:1968b} found
$t_s=22\times 10^6$ years.  Using data on stellar evolution we
conclude that such mean age have stars of main sequence of spectral type B3.  Just B3 
subpopulation of main sequence stars has the maximal $\sigma_z/\zeta$ ratio 
\citep{Joeveer:1968b}.  

After formation, stars fall towards the centre and plane of the Galaxy
in a synchronous way.  After several circles around the axis, this
synchrony is lost, partly due to orbit mixing \citep{Kuzmin:1962ac},
partly due to difficulties to determine ages of stars accurately
enough.  Stars fill all regions in phase space, determined by initial
velocities and coordinates at birth.  Velocity dispersions $\sigma_R$
and $\sigma_\theta$ increase up to values, determined by epicyclic
orbits around the point of circular motion: $\sigma_\theta =
2/\pi\,\Delta V_\theta = 14$~km/s, $\sigma_R =
1/\sqrt{k_\theta}\,\sigma_\theta=9$ km/s. Dispersion $\sigma_z$
increases from the initial value, determined by the velocity dispersion of
gas clouds, $(\sigma_z)_0$, due to dispersion of positions in the vertical
direction: $\sigma_z^2 + C^2\,\zeta^2 = (\sigma_z)^2 +C^2\,z_0^2$.
The mixing process (real or mimicked, caused by the dispersion of ages
of stars in the subpopulation) and the approach of the subpopulation
to  a stationary stage takes  $2 - 4$ circles of stars around the centre
of the system, in our case $0.5 - 1$ billion years. During this time,
the velocity of the population around the centre of the Galaxy obtains
its stationary value, determined by hydrodynamical 
equations. Fig.~\ref{Fig4.3} shows that the actual increase of the
velocity dispersion is larger than expected from above considerations.
It is possible that this can be explained by irregular gravitational
forces.  These forces fill the hole in the velocity space around the
point with circular motion.  Such holes in velocity space are observed
only in very young populations.

It is well known that the rotational curve of the southern part of
the Galaxy is shifted in respect to the northern part by about 10 km/s
\citep{Westerhout:1957aa}. A similar picture is observed in other
galaxies \citep{Carranza:1968aa,Roberts:1969aa}. These data suggest
that the rotation velocity of interstellar hydrogen cannot be used as
an exact representative for circular velocity,  particularly  in models
of mass distribution of galaxies. In particular, very flat
rotation curves of galaxies, observed in some galaxies, including M31,
cannot probably be identified with circular velocities. If used as
circular velocities, corresponding to mass distribution, this would
raise mass-to-light ratios in the periphery of galaxies to very
high values of the order $f > 1000$, whereas physical data on the
distribution of mass of galactic populations show the opposite trend of
decreasing $f$ with increasing distance from the centre
\citep{Einasto:1969ab} (see Chapter 20).

\section{Evolution of the Galaxy from kinematics of disc and halo
  objects}

We mentioned earlier the work by \citet{von-Hoerner:1960aa} and
\citet{Eggen:1962}, who made an attempt to reconstruct the dynamical
evolution of the Galaxy by studying kinematical data on galactic
populations.  Their arguments are based on the following assumptions.

1. Stellar populations of forming stars get spatial and kinematic
characteristics of gas clouds from which they formed.

2. Spatial and kinematical characteristics of stellar populations do
not change with time or change very slowly. In particular, the angular 
momentum of stellar populations and eccentricity of their orbits do not
change, even if the gravitational field of the whole Galaxy changes.

3.  Mass and total angular momentum of the whole Galaxy is constant.

It must be said that these assumptions are valid not very accurately.
As we discussed in the previous section, gas is supported, in addition
to gravity, also by electromagnetic forces. Recently formed stars are
free from electromagnetic forces, and their populations obtain stable properties by
contracting both in radial and in the vertical direction. Due to
irregular gravitational forces, the spatial and kinematical properties of
stellar populations change slowly. However, these changes are small,
and stellar populations ``remember'' conditions during their
formation.

\citet{von-Hoerner:1960aa} and \citet{Eggen:1962} used in their study
ages of stars:  \citet{von-Hoerner:1960aa} used ages found from
models  of stellar evolution, \citet{Eggen:1962} applied qualitative
relative ages of stars, estimated from ultraviolet excess.  Authors
concluded that in the initial stage of its evolution the Galaxy collapsed
at least 10 times in radial direction and 25 times in vertical
direction. The collapse was very rapid, of the order of several
hundreds millions years.

Let us now consider what conclusions can be made, using data collected
in this work. In addition to assumptions, made by
\citet{von-Hoerner:1960aa} and \citet{Eggen:1962}, we assume that:

4. The ratio of the rotation velocity of a population, $V_{\theta\,i}(R)$, to the circular
velocity, $V_c(R)$, and respective ratios of angular momenta, do not depend on
distance $R$, \ie\ $V_{\theta\,i}(R) = \beta\,V_c(R)$, where $\beta$ is
a constant.

Using a mass distribution model, we can calculate the function $V_c(R)$,
and from kinematics of populations near the Sun find the parameter
$\beta$ for populations of various ages. In this way, we can calculate
the rotational velocity and relative angular momentum
\be
h_i(R) = R\,V_{\theta\,i}(R)
\label{eq21.3.1}
\ee
for all populations of interest. Results of these calculations are
shown in Fig.~\ref{Fig21.1} for seven test populations with
parameters, given in Table~\ref{Tab7.4}.

{\begin{figure*}[h] 
\centering 
\resizebox{0.75\textwidth}{!}{\includegraphics*{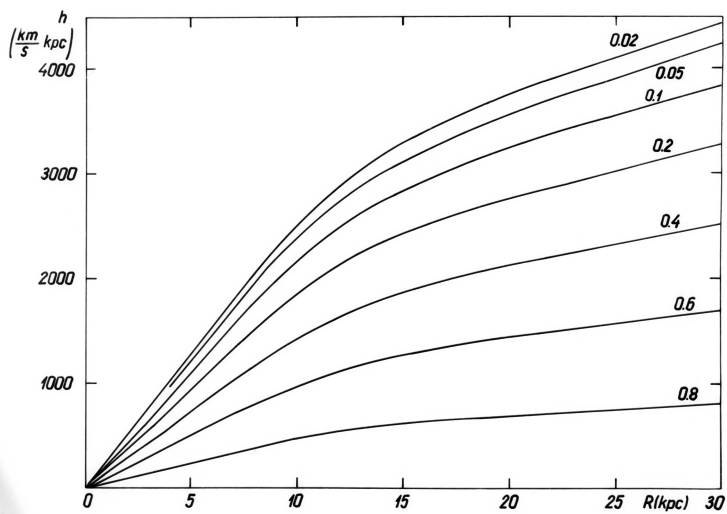}}
\caption{Angular momentum $h$ of populations of various flatness
  $\epsilon$ as function of the distance $R$. } 
  \label{Fig21.1}
\end{figure*} 
}

{\begin{figure*}[h] 
\centering 
\resizebox{0.75\textwidth}{!}{\includegraphics*{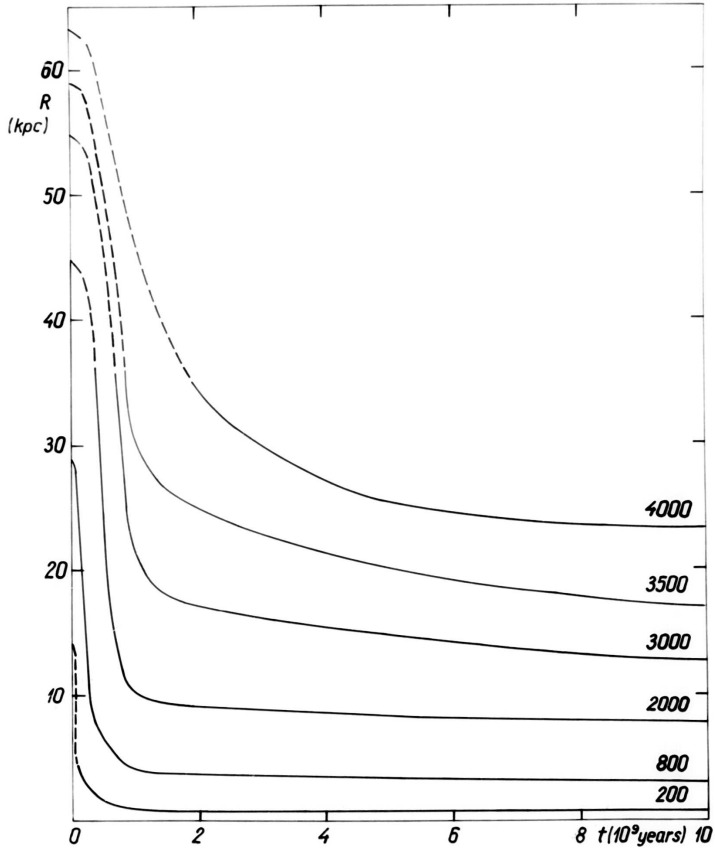}}
\caption{The evolution of the distance from galactic centre $R$ of gas clouds of various
angular momentum $h$. } 
  \label{Fig21.2}
\end{figure*} 
}

{\begin{figure*}[h] 
\centering 
\resizebox{0.85\textwidth}{!}{\includegraphics*{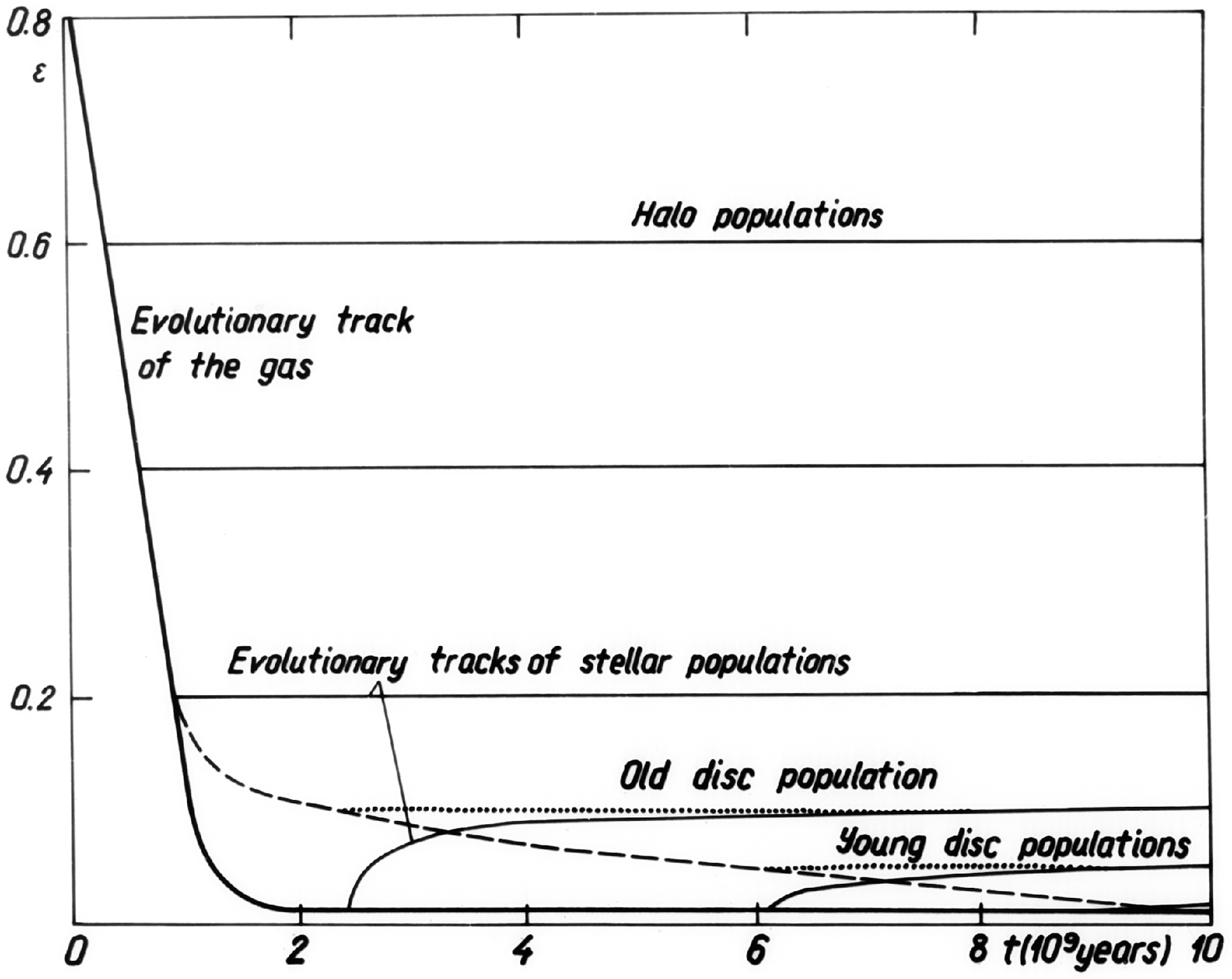}}
\caption{Possible evolution of the flatness $\epsilon$ of stellar and
  gas populations with time. The evolution of stellar populations is
  shown by thin lines, that of the gas by bold line.  Evolution is
  shown for two cases: the bold solid line is for the case when gas immediately arrives the final
  $\epsilon$, the bold dashed line is the case when for $\epsilon \le
  0.2$ gas  gets the flatness as given by respective present stellar
  population flatness. } 
  \label{Fig21.3}
\end{figure*} 
}

Every line in Fig.~\ref{Fig21.1} corresponds to a population of
certain age, see Table \ref{Tab7.4}.   According to our assumptions,
stellar populations are 
indicators of the gas at its formation time.  In this way, the
Fig.~\ref{Fig21.1} shows at which distance from the galactic centre, $R_g$,
gas clouds of certain angular momentum $h$ were at respective time of
the evolution $t$.  Using data shown in Fig.~\ref{Fig21.1}, we can
find the relationship between distance $R_g$ and time $t$ for gas
clouds with given angular momentum $h$. For six values of $h$ this
relationship is shown in Fig.~\ref{Fig21.2}.

In this way, we can reconstruct the evolution of gas population, 
starting from the formation of first generation of stars, using their 
respective specific 
angular momentum. Earlier evolution of gas clouds can be estimated by
the extrapolation using a certain model for the initial conditions of
gas clouds, and a certain regularity of the change of $R_g$ with
time.  We shall discuss this problem in the next section.  In
Fig.~\ref{Fig21.2} extrapolated parts of curves are plotted by dashed
lines.

Our results confirm and extend the evolution picture by
\citet{von-Hoerner:1960aa} and \citet{Eggen:1962}. In the initial
phase of the evolution, the collapse of the gas was very rapid.  In
later phases the speed of the collapse decreased; at the present epoch the
collapse has stopped completely.  The collapse rate, \ie the ratio of
initial radius $R_g$ to the present one $R$, is very large for the
nucleus of the Galaxy, and relatively modest for more distant
regions.

We can use our data to reconstruct the gas evolution in the
vertical direction too.

To characterise the vertical extent of populations, we shall use the
flatness (axial ratio $\epsilon$) of ellipsoids of various density.  In
Fig.~\ref{Fig21.3} we plot on vertical axis  $\epsilon$, and on
horizontal axis $t$ -- the age of the Galaxy, counting from the
formation epoch of oldest populations; the present age of the Galaxy
is taken equal to 10  billion years, see Chapter 23.  If we assume
that dynamical and spatial parameters of populations, including
flatness $\epsilon$,  do not change with time, then evolutionary
tracks of stellar populations in Fig.~\ref{Fig21.3} are horizontal lines, starting from the
point corresponding to the formation of the population. The length of
these lines shows the age of the population.  The solid bold line along
formation points shows the evolutionary track of the gas.  We see that
in vertical direction, gas also has collapsed and that the collapse 
was very rapid in the early phase of the evolution. From flatness
$\epsilon=0.2$ onwards the speed of vertical collapse decreased (dashed
bold line).

This picture corresponds to the assumption that dynamical parameters
of stellar populations do not change with time. In this case the
collapse of the gas after rapid initial phase proceeds very
slowly. Several authors, including \citet{Eggen:1962}, stress that the
collapse was very fast, and after the collapse the gas obtained its
equilibrium stage close to the present form. If we accept this
assumption, then the evolutionary track of the gas from point with
$\epsilon=0.20$ to point with $\epsilon=0.016$ is slowly decreasing;
this section of the curve in Fig.~\ref{Fig21.3} is drawn with a dashed
bold line.  The other possibility is that the gas obtained its present 
flat configuration immediately after the collapse, this variant is
plotted in Fig.~\ref{Fig21.3} by a solid bold line for
$\epsilon < 0.20$. Respective evolutionary tracks of stellar
populations are drawn with thin  lines.  We see that in the
second case the thickness of stellar populations increases with time.

Preliminary results of this study were presented by
\citet{Einasto:1970ad}, using a slightly different calibration of ages
of stellar populations.  The difference between extreme populations
was slightly larger than found now.

The actual evolutionary track of gas probably lies between extremal cases
considered.  Since some dynamical evolution of stellar populations is
probable, then the lower evolutionary track of the gas is probably
closer to the actual evolution. This problem was discussed by
\citet{Parker:1968aa}, see also Chapter 23.

Our conclusions on the radial and vertical collapse of the gas during
the evolution of the Galaxy were based on arguments, different from
arguments used by \citet{von-Hoerner:1960aa} and \citet{Eggen:1962}.
Our results on the fast initial collapse of the gas depend on the
accuracy of determinations of ages of old halo populations. Estimates
of absolute ages have a relative error about 10\,\%.  Errors of
determinations of chemical composition also increase errors. For this
reason we counted ages relative to oldest halo populations, for which
we used the age 10 billion years. In calculating relative ages we
take into account results of modelling the evolution of stars,
discussed in more detail in the next section.

\section{Model of the evolution of the protogalaxy}    

In order to have a choice to select a model for protogalaxy evolution, we have
to consider the initial conditions we can have from cosmological data.

The detection of relict radiation, chemical composition of the
primeval matter and a number of other data give strong support to the
primeval-fireball or ``Big Bang'' model of the formation of the
Universe. According to this model matter and radiation were in
thermodynamical equilibrium (\citet{Peebles:1965aa} and
\citet{Doroshkevich:1967aa}),  and formed an almost homogeneous medium.
There existed small density fluctuations, responsible for the
formation of galaxies \citep{Lifshitz:1946aa}.  The nature of
fluctuations is not clear, they can have a whirl character, as suggested
by \citet{von-Weizsacker:1951aa} and
\citet{Ozernoi:1967aa,Ozernoi:1968aa,Ozernoi:1970aa}, adiabatic, as
emphasised by 
\citet{Peebles:1965aa, Peebles:1970} and \citet{Silk:1968aa}, 
 or entropic (\citet{Doroshkevich:1967aa},
\citet{Peebles:1968aa}), for review see \citet{Ozernoi:1970cc}.  For
our goal, the nature of perturbations is of less interest, important is
their presence.

During the evolution, the Universe expands and its density and
temperature decrease. At a temperature of about 4000 K$^\circ$ the
recombination of gas starts. The matter is free from radiation, and
density fluctuation can start to amplify.  In this way,
the density initially decreases due to the expansion of the Universe,
and thereafter, the density increases in some regions due to the
self-gravity of gas clouds. 

To find the possible speed of the collapse, we consider a simple model,
starting from the moment of the largest expansion with its turnover
radius.  At this epoch fluctuations of the density were still small,
and we can use a model with homogeneous density, following
\citet{Mestel:1963aa}, \citet{Crampin:1964aa}, and
\cite{Innanen:1966ab}, rotating with constant angular velocity,
$\omega_0$. We assume that at the moment when the protogalaxy had its
largest extent at turnaround, it separated itself from the surrounding
gas, and its mass and total angular momentum obtained their present
constant values.

Present data are not sufficient to find for each protogalaxy its
initial radius $R^\circ$ and density $\rho_0$.  But radius $R^\circ$
can be found from observational data under certain assumptions. To
estimate $R^\circ$ we can use apogalactic distances of high-velocity
stars.  We can assume that these distances did not change essentially
during the galaxy evolution, since changes of the regular
gravitational field as well relatively weak irregular forces cannot
change orbits substantially. Calculations by \citet{Eggen:1962}
indicate that some stars have apogalactic distances of the order of $50
- 100$ kpc. It is clear that these values are not very accurate, since
observational errors and errors in the model of the Galaxy influence
results.

We can get independent estimates of $R^\circ$ in an indirect way, using
a model of the protogalaxy with $\omega_0=const$.  Since angular
moment keeps its value during the evolution, we have
\be
h=\omega\,R^2 = \omega_0\,R_0^2,
\label{eq21.4.1}
\ee
where $\omega$ and $R$ are angular velocity and distance of gas clouds
at the present epoch, and $\omega_0$ and $R_0$ at the initial
moment. $R_0$ values for various $h$ can be found from data,
presented in Fig.~\ref{Fig21.2}.  Results are given in Table~\ref{Tab21.1}.
  
{\begin{table*}[h]
\centering    
\caption{} 
\begin{tabular}{ccc}
\hline  \hline
  $h$& $R_0$ & $\omega_0$\\
  km/sec\,kpc & kpc & km/sec/kpc\\
  \hline
  200   & $>5$ & $<4$ \\  
  800   & 28 & 1.0\\
  2000 & 45 & 1.0\\
  3000 & $\ge 50$& $\ge 1.2$\\
  \hline
\label{Tab21.1}   
\end{tabular} 
\end{table*} 
} 
      
We see that $\omega_0 =1$ km/sec/kpc can be accepted.  The maximal
value of the angular momentum is slightly larger than 4000 km/sec~kpc,
which yields $R_0 \approx 65$ kpc.  This is in good agreement with the
estimate from apogalactic distances.

Using this $\omega_0$ value we calculated for all $h$ initial radii
$R_0$, and made extrapolation of evolutionary lines in
Fig.~\ref{Fig21.2} for early epochs.  Using Eq.~(\ref{eq21.4.1}) we
can calculate also the contraction degree of gas
\be
d={R_0 \over R}= \left({\omega \over \omega_0}\right)^{1/2}.
\label{eq21.4.2}
\ee
The initial distance $R_0$ and logarithm of the contraction degree $\log
d$ are shown in Fig.~\ref{Fig21.4} for four test populations of
flatness $\epsilon=0.02,~0.1,~0.55,~0.8$. As
argument we use the present distance $R$.

{\begin{figure*}[h] 
\centering 
\resizebox{0.75\textwidth}{!}{\includegraphics*{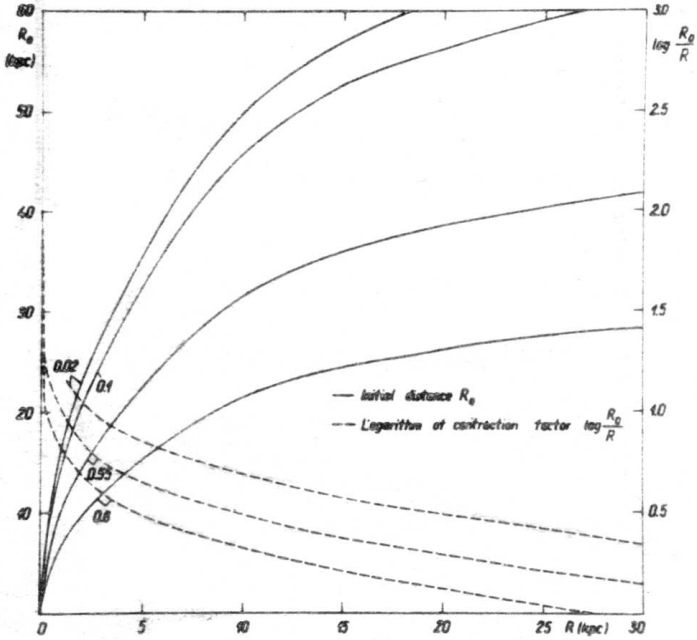}}
\caption{The initial distance, $R_0$ (solid lines), and contraction degree,  $\log\,d$,
  (dashed lines) of populations with various $\epsilon$ as function of the present
  distance $R$. } 
  \label{Fig21.4}
\end{figure*} 
}

We can use the model with constant angular momentum to estimate the
contraction time. The fastest contraction has the free-fall
model. This case is realised when the gas cooling time is much smaller
than the free-fall time. To simplify calculations, we assume that
during the contraction gas shells are not mixed, in other words, the
test particle falls together with the shell it belongs to. In this
case the internal mass $\mm{M}_{int}$ remains constant. Arguments for
this case were given by \citet{Crampin:1964aa}.

The contraction time from initial $R_0$ (apogalacticum) to distance
$r$ is as follows
\be
t_{contr}= \sqrt{{a_3 \over G\mm{M}_{int}}}(\arccos x+ e\sqrt{1-x^2}),
\label{eq21.4.3}
\ee
where $a$ and $e$ are major semiaxis and eccentricity of gas particle
orbits, and
\be
x = {r-a \over ae}.
\label{eq21.4.4}
\ee
Major semiaxis $a$ can be found from apogalactic distance $R_0$, and
we get
\be
t_{contr}=\sqrt{{R_0^3 \over G\mm{M}_{int}}}\,f(x),
\label{eq21.4.5}
\ee
where
\be
f(x)= {1 \over (1+e)^{3/2}}(\arccos x+ e\sqrt{1-x^2}).
\label{eq21.4.6}
\ee
The variable $x$ can be expressed through $e$ and contraction degree
$d$ as follows
\be
x={1+e-d \over e\,d},
\label{eq21.4.7}
\ee
and the eccentricity $e$
\be
1-e = {h^2 \over G\mm{M}_{int}\,R_0}.
\label{eq21.4.8}
\ee
For the internal mass we can take
\be
G\,\mm{M}_{int} = R^3\omega_c^2,
\label{eq21.4.9}
\ee
where $\omega_c$ is the angular velocity of circular motion. Using for
$h$ the expression (\ref{eq21.4.1}) through $R_0$ and $\omega_0$ we
get
\be
1-e= {d^3 \over d_0^4},
\label{eq21.4.10}
\ee
where $d_0$ is the contraction degree of objects of flat populations.

{\begin{figure*}[h] 
\centering 
\resizebox{0.75\textwidth}{!}{\includegraphics*{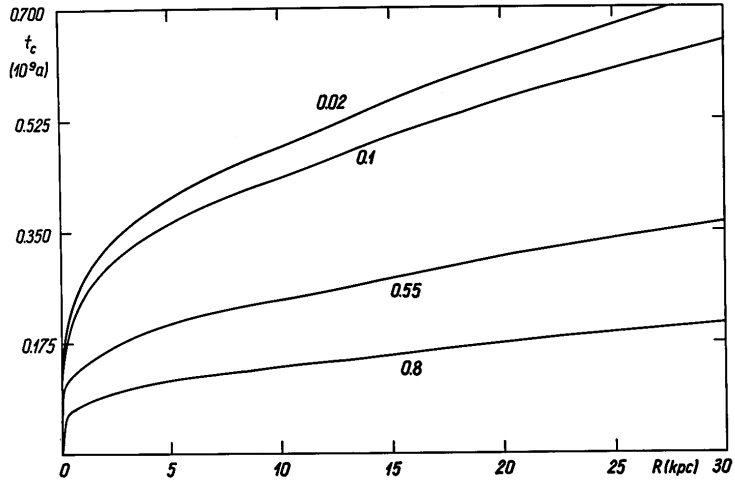}}
\caption{Contraction time of populations of various flatness
  $\epsilon$ as function of the distance $R$. Populations as in Fig.~\ref{Fig21.4}.} 
  \label{Fig21.5}
\end{figure*} 
}

Since $d_0 > d \gg 1$, then for all cases of practical interest
$e\approx 1$ and $x\approx -1$, and thus $f(x) \approx 1$.  This means
that we can ignore the factor $f(x)$ in the contraction
equation. Using Eq.~(\ref{eq21.4.9}) and expressing time in years and
angular velocity in km/sec/kpc, we can write Eq.~(\ref{eq21.4.5}) as
follows
\be
t_{contr}= {10^9 \over \omega_c}d^{3/2} = {10^9 \over
  \omega_c}(\omega/\omega_0)^{3/4}.
\label{eq21.4.11}
\ee

This equation was used to calculate contraction time, where $\omega(R)$ and
$\omega_c(R)$ were taken from our model of the Galaxy. For several
test populations results are given in Fig.~\ref{Fig21.5}.  Our
calculation suggest that the nucleus forms very quickly, in 10
millions years. Thereafter, new gas masses fall to the nucleus, 
shell after shell. In the Solar neighbourhood star formation  begins about
$10^8$ years after the condensation of the nucleus. The formation of
the halo takes about one billion years. The whole process of gas
contraction takes less than a billion years.

Similar calculations emphasise that the contraction of gas in vertical
direction proceeds approximately with the same speed.

The model is certainly rather crude. Actually, the contraction time can
be longer if the gas cools slowly. The time can be shorter if
initial volume of the gas was smaller that adopted in our model. In a
more accurate discussion of the problem, it must be taken into account
that the gas probably consisted of many clouds moving in respect to
each other. This aspect was recently studied by
\citet{Brosche:1970aa}. However, he discussed the contraction of the
whole galaxy, thus, his model cannot be used to understand the differential picture.

The contraction model was discussed by \citet{Sandage:1970aa} and
\citet{Rood:1968aa}. Sandage argued that the formation of globular
clusters and other halo objects was completed in about
$2\times10^8$~years, whereas Rood and Iben supported a much longer
formation period. When we compare data used by these authors, we see
that the controversy is not real. Inner objects of the halo, studied
by \citet{Eggen:1962} and  \citet{Sandage:1970aa}, can really be
condensed quickly, but the whole process of halo forming is much
longer.

Another aspect of the discussion on star formation is related with the
eccentricity $e$ of stellar orbits. As discussed above, during
the contractions of the protogalaxy, the eccentricity $e$ does not
change much. In this case halo stars should have had large eccentricity
$e$ already during their formation. In other words, gas clouds from
which halo stars formed should have moved in extended elliptical orbits, which
took place when the protogalaxy condensed quickly. If the gas was
subject to non-gravitational forces, then objects with large
eccentricity could form also in case the gas did not contract
\citep{Rood:1968aa}. However, the influence of non-gravitational
forces is strong enough only if the gas is hot, but star formation in
a hot gas is impossible \citep{Sandage:1970aa}.  We conclude that the
gas contraction was probably rather rapid. But here we do not have a
final answer, because of the lack of a quantitative theory of star
formation.

In connection with the contraction of the gas in the protogalaxy we
have two additional remarks.

High-velocity hydrogen clouds were recently detected in high galactic 
latitudes \citep{Dieter:1969aa}. The origin of these clouds is not
clear. But it is possible that they are remnants of protogalactic gas,
which now fall towards the plane of the Galaxy.  If this is the case,
then the total contraction time of the protogalaxy is longer than
accepted so far.

Some authors argue that spherical subsystems of the Galaxy were formed
by an outburst from the nucleus or disc \citep{Gurevich:1964aa}, or in
other words, that the star forming started only after the contraction
of the protogalaxy. However, as demonstrated by \citet{Rood:1971aa},
this hypothesis does not explain the differences in chemical composition
of stars.  Stars in the nucleus of the Galaxy are metal-rich, in
contrast to metal-poor halo stars. In the Solar neighbourhood all
stars with metal deficit have large $z$-velocities
\citep{Dixon:1965aa,Dixon:1966aa,Dixon:1967aa}. On the other hand,
minimal heavy element content of disc stars is about half of the Solar
content.  All these facts suggest that the halo cannot be formed by the
throw-out of stars from the nucleus or disc. Only runaway stars can be
formed in this way.

\section{Distribution of the angular momentum of M31}

\citet{Mestel:1963aa} and \citet{Crampin:1964aa}  found that the
distribution of the angular momentum in spiral galaxies is in good
agreement with the distribution of angular momentum in a homogeneous
ellipsoid, rotating with constant angular
velocity. \citet{Crampin:1964aa} interpreted this result as an
argument favouring the formation of galaxies from a homogeneous
ellipsoid of gas clouds, and that in the formation of galaxies the
turbulence did not play an important role, which could redistribute
the kinetic moment.

On the other hand, \citet{Lynden-Bell:1967cc}  noticed that all
elliptical galaxies and cores of spiral galaxies are very similar, if
we ignore differences in ellipticity. If we do not assume that
the formation conditions of elliptical galaxies and cores of spiral
galaxies were identical, we must conclude that the similarity is
caused by violent relaxation during the formation of
galaxies. \citet{Lynden-Bell:1967aa} studied this problem in detail,
and found that a fast change of the gravitation field during the
contraction of the protogalaxy leads to the observed effect. 
This theory was checked with numerical experiments by
\citet{Hohl:1968aa,Hohl:1968ab}, 
\citet{Cuperman:1969aa}, \citet{Henon:1969aa}, and 
\citet{Goldstein:1969aa}. These experiments confirmed that the
statistical method by Lynden-Ball describes well the observed
energy distribution of stellar systems.

Results by Lynden-Bell suggest that during the formation of elliptical
galaxies and cores of spiral galaxies the turbulence played an
important role.  For spiral galaxies this result is in contradiction
with the \citet{Crampin:1964aa} study. In order to have an independent check, we
can use our models of the Galaxy and M31 and calculate the
distribution of the angular momentum. It is better to use the model
of M31, since the distribution of mass in central regions of M31 is
known better.

Let us assume for simplicity that surfaces of $\omega =const$ are
concentric cylinders (\citet{Mestel:1963aa}, \citet{Crampin:1964aa}
and  \citet{Innanen:1966ab}).  An essential fraction of the kinetic moment
is concentrated in a thin disc, where the difference between the
actual surface of $\omega =const$ and respective surface in a cylinder
is small, see Chapters 7 and 20). A larger difference is expected only
in the central core, where it is better to identify surfaces $\omega
=const$ with surfaces $\rho=const$.

With these assumptions, the distribution of the mass as function of
the kinetic moment $h$ is the following (\citet{Mestel:1963aa} and 
\citet{Crampin:1964aa})
\be
m(h)=\mc{P}[R(h)]\,\frac{\dd{R}}{\dd{h}},
\label{eq21.5.1}
\ee
where
\be
\mc{P}(R)=2\pi\,R\,P(R)
\label{eq21.5.2}
\ee
is the mass of cylindric shell of unit thickness, and $P(R)$ is the
projected density. Further we have
\be
\frac{\dd{h}}{\dd{R}}= R\left(\frac{V_\theta}{R} +
  \frac{\dd{V_\theta}}{\dd{R}}\right).
  \label{eq21.5.3}
  \ee
  Here we can accept the approximation
  \be
  V_\theta=\beta\,V,
  \label{eq21.5.4}
  \ee
  where $V$ is the circular velocity. Applying Oort dynamical
  parameters $A,~B$ and using $k_\theta=-B/(A-B)$, we get
  \be
  \frac{\dd{R}}{\dd{h}}=\frac{1}{2k_\theta}\,\frac{R}{h}.
  \label{eq21.5.5}
  \ee

{\begin{figure*}[h] 
\centering 
\resizebox{0.70\textwidth}{!}{\includegraphics*{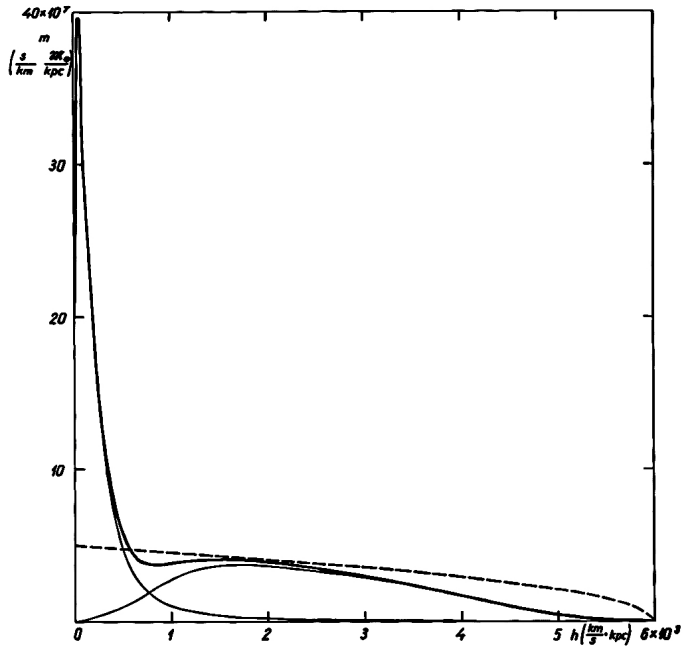}}
\caption{The mass distribution of M31 as function of the angular
  momentum $h$ (solid line). Dashed line shows a similar distribution
  of a body rotating with constant angular speed. } 
  \label{Fig21.6}
\end{figure*} 
}

The distribution of the mass as function of $h$ using
Eq.~(\ref{eq21.5.1}) and (\ref{eq21.5.5}) is shown in
Fig.~\ref{Fig21.6}. The bold curve shows the total distribution, thin
curves show the distribution for the first and second population. To
the first population we include the disc and flat components, for them
we used the value
$\beta=0$\footnote{Actually only the nucleus and inner core have
$\beta \approx 0$,  the rotation
  velocity of flat and disc components is almost equal to the circular
  velocity, their $\beta \approx 1$.   Thus in Fig.~\ref{Fig21.6} the curve
  for the first population should be replaced by two curves. The mass
  distribution of the whole galaxy M31 has a higher tail than the model
  presented, closer to the model  with constant angular speed
(correction made in October 2021).} on the basis of results from Chapters 4 and 7. The
second population includes the core (using data from Chapter 20 we
accepted $\beta=0.4$) and the halo. The halo was divided into two components
of different mass with values of the parameter $\beta=0.3$ and
$\beta=0.6$. We see that the distribution of the momentum of the first
and second population are very different.

\citet{Mestel:1963aa} and \citet{Crampin:1964aa} compared
distributions of the mass vs. kinetic moment of galaxies with similar
distributions for homogeneous spheroids, rotating with constant
angular velocity.  For this distribution, we have
\be
m(x)=3/2\,\mm{M}\sqrt{1-x},
\label{eq21.5.6}
\ee
where $\mm{M}$ is the mass of the spheroid, and
\be
x=h/h^0,
\label{eq21.5.7}
\ee
where $h^0$ is the maximal kinetic moment at the periphery of the
model.  The mass of the M31 model is known, and we get for the maximal
moment using data from Fig.~\ref{Fig21.6}: $h^0=6000$ km/sec per
kpc. This function is also plotted in Fig.~\ref{Fig21.6}. 

We see that the homogeneous model describes the function $m(h)$ only
in very general terms. In the periphery of the galaxy the function
$m(h)$ approaches zero more slowly than the homogeneous model. This
suggests that the protogalaxy did not have a sharp outer boundary, \ie
the density approached smoothly zero. Another difference between
models is observed near the centre of the system, where the model has
essential deficit of mass with low kinetic moment.  This difference
can be due to the presence of a dense nucleus in the protogalaxy. 
However, a more likely explanation of the mass excess with
small angular momentum is the redistribution of moment during the fast
contraction of the protogalaxy.  The distribution of the moment is in
agreement with the Lynden-Bell hypothesis on the violent
redistribution of matter during the formation of the protogalaxy.

However, we note that the violent relaxation happens only in the
central region of the Galaxy. This factor cannot strongly change our
picture on the general evolution of the protogalaxy.

\vskip 5mm
\hfill September 1971


\chapter{Physical evolution of stellar systems
  \label{ch22}}
 
\section{Introduction}

For the full description of the structure of galaxies and other
stellar systems it is necessary, in addition to kinematical and
spatial description functions, add also functions which describe the physical
structure of galaxies and their populations. Such functions include
the distribution of various spectral characteristics, chemical
content, age and other similar quantities. During the evolution of
galaxies these functions change with time.

The change of physical characteristics of galaxies with time is caused by
the dynamical evolution of galaxies (redistribution of mass), and by 
the change of physical characteristics of stars due to stellar
evolution. Advances in our understanding of stellar evolution permit
to follow the physical evolution of galaxies, and to build models of
the evolution of stars and gas in galaxies. Pioneering studies in this
direction were made by \citet{Limber:1960aa} and
\citet{Tinsley:1968}. In this Chapter, we describe our model of the
physical evolution of galaxies. To construct the model, we need to
know the rate of star formation, evolutionary tracks of stars of
various masses in the Herzsprung-Russell diagram, and bolometric data
and colours of stars as functions of the effective temperature and
bolometric luminosity.

\section{Data and method}

\subsection{Initial mass function of stars}

The initial distribution of stars according to their mass and luminosity
and the division of stars into giants and main-sequence stars was
discussed already by \citet{Opik:1938aa}. Using modern data, the
luminosity function and stellar evolution was discussed by
\citet{Salpeter:1955}. He found that the number of stars of mass $M$,
$F(M)$, formed in unit time interval per cubic parsec, can be
expressed by the following equation,
\be
F(M) = a\times M^{-n},
\label{eq22.1}
\ee
where $a$ and $n=2.35$ are constants. In the derivation of this
equation Salpeter assumed that the rate of star formation is constant
during the last $5 \times 10^9$ years and that stars move from main
sequence stars to giants when they have consumed 10~\% of their
hydrogen.

This result was confirmed by \citet{Sandage:1957aa} and
\citet{van-den-Bergh:1957aa} and other authors, using counts of stars
in young star clusters.  Later it was understood that the rate of
star formation is not constant but changes in time, see below.
However, this does not change the form of
Eq.~(\ref{eq22.1}). In an overview, \citet{Reddish:1966aa}  concluded
that the equation (\ref{eq22.1}) with exponent $n=2.5$ can be used in the
interval of stellar masses $M_0 \le M \le M_u$, where $M_0$ and $M_u$
are minimal and maximal masses of forming stars.  For medium mass
stars observational data are better represented using
Eq. ~(\ref{eq22.1}) with $n=2.33$. The total mass of forming stars is
equal to
\be
\int_{M_0}^{M_u} F(M)M\,\dd{M}.
\label{eq22.2}
\ee

We cannot use for the minimal mass a value $M_0=0$, since the integral
Eq.~(\ref{eq22.2}) is not converging in this case.  It is possible that the function
$F(M)M$ smoothly approaches zero when $M \Rightarrow 0$.  Our results
do not depend on the exact form of the function $F(M)$ for small $M$.
For this reason, we take $M_0$ as the effective lower limit of mass of
forming stars and take $F(M)=0$, if $M < M_0$. 
In solar neighbourhood of the Galaxy, we can use
$M_0=0.03$ in solar units \citep{Reddish:1966aa}.

According to \citet{Reddish:1966aa}, there exists an upper limit of
mass of forming stars, $M_u = 60 - 100~ M_\odot$.  A similar upper
limit is predicted by theory.  \citet{Stothers:1968ab} demonstrated
that blue supergiant stars are   unstable for pulsations when they
have masses in excess of 65\,$M_\odot$. However, as shown by
\citet{Talbot:1971aa}, stars cross the instability zone very fast and
cannot lose their mass during this period very much. For this reason, we take as
the upper limit of forming stars $100\,M_\odot$, not $65\,M_\odot$.
We chose the parameter $a$ in Eq. ~(\ref{eq22.1}) from the condition
that the integral Eq.~(\ref{eq22.2}) is equal to unity. In this case we
get
\be
a = (n-2)(M_0^{2-n} - M_u^{2-n})^{-1}.
\label{eq22.3}
\ee

The choice of parameters $M_0$, $M_u$ and $n$ is crucial in the
modeling  the physical evolution of galaxies. Earlier it was assumed
that these parameters are constants. However, already
\citet{Limber:1960aa} demonstrated that in this case it is impossible
to explain differences in mass-to-light ratios of globular
clusters ($f_V=\mm{M}/L_V \approx 1$),  dwarf galaxies
($f_V \approx 10$), and giant elliptical galaxies ($f_V \approx 100$),
all having approximately similar ages.

Differences in the parameters of star formation function can probably be 
explained by differences in the chemical composition of old stellar
populations of galaxies. \citet{van-den-Bergh:1961aa} hinted to the
fact that the fraction of heavy chemical elements is different in
globular clusters and in old open clusters, both having approximately
equal ages. Heavy elements are synthesised in stars.  Rapid enrichment
of interstellar gas by heavy elements is done by massive stars with
fast evolution. The large variability of chemical compositions of
stars of different old populations suggests that in the early period
of the evolution, various populations had different fractions of
massive stars, much higher than the present fraction of massive
stars. This conclusion was made by \citet{Schmidt:1963aa},
\citet{Truran:1970aa} and \citet{Cameron:1971aa}.

Such effects can be explained if we assume that the minimal mass of
forming stars, $M_0$, depends on time, or more accurately, on the
fraction of heavy elements in the interstellar gas during the
formation of stars of different populations.  \citet{Truran:1970aa}
suggested a mechanism to explain these differences.

\subsection{Star formation rate}

\citet{Salpeter:1955} assumed that the star formation rate in the Galaxy is
approximately constant. More accurate data showed that in the early
phase of the evolution of the Galaxy, the star formation rate was
considerably higher \citep{von-Hoerner:1960aa}.  In the early phase of
the evolution of the Galaxy, the density of interstellar matter was
much higher than in the present epoch.  Based on this argument,
\citet{Schmidt:1959aa} concluded that the star formation rate depends on the
density of interstellar matter.

We define the local star formation rate as the time derivation of the
density of stars. Following \citet{Schmidt:1959aa} we assume that the
star formation rate is proportional to the density of gas in power
$S$:
\be
R_l={\dd{\rho_s} \over \dd{t}}= \gamma \rho_g^S,
\label{eq22.4}
\ee
where $\rho_s$ and $\rho_g$ are star and gas densities, respectively,
and  $\gamma$ and $S$ are constants.  We assume that the full matter
density in a volume element, $\rho=\rho_s+\rho_g$, does not depend on
time.  In this case after integration of Eq.~(\ref{eq22.4}) we get
\be
\rho_g = \rho[1+(S-1)\tau]^{-1 \over S-1},
\label{eq22.5}
\ee
where
\be
\tau=t/K,
\label{eq22.6}
\ee
and for the characteristic time $K$ we have
\be
K=(\gamma \rho^{S-1})^{-1}.
\label{eq22.7}
\ee
In case $S=1$ we get
\be
\rho_g=\rho\,e^{-\tau}
\ee
and
\be
K=\gamma^{-1}.
\ee

Integrating Eq.~(\ref{eq22.4}) over the whole volume of the stellar
system, and assuming that the gas amount does not depend on time, we
get for the whole gas mass
\be
\mm{M}_g = \mm{M}[1+(S-1)\tau]^{-1 \over S-1},
\label{eq22.10}
\ee
where $\mm{M}=\mm{M}_s+\mm{M}_g$ is the full mass of
the galaxy, and $\mm{M}_s$ is its stellar mass. For the
characteristic time we get
\be
K=(\gamma \bar{\rho}^{S-1})^{-1},
\label{eq22.11}
\ee
where $\bar{\rho} = \mm{M}/V_\ast$.  Here $V_\ast$ is the mean
volume of the gas, which can be calculated as follows:
\be
V_\ast^{-(S-1)} = \mm{M}_g^{-s} \int  \rho_g^{S-1}
\dd{\mm{M}_g}.
\label{eq22.12}
\ee
In particular case $S=1$ we get
\be
\mm{M}_g = \mm{M}\,e^{-\tau}.
\ee

The characteristic time $K$ is the essential parameter which describes
the star formation rate in galaxies. The characteristic time depends
on the parameter $S$. If $S=0$, then the star formation rate is lower for
higher mass density; if $S=1$, then the characteristic time $K$ is
constant; if $S=2$, then the star formation rate is the higher, the higher
is the matter density.  Observational data on the density and mass of
stars and gas suggest that $S=2$.  This conclusion has been made by
\citet{Schmidt:1959aa} and \citet{Sanduleak:1969aa}. However, we shall
make calculations for all values, $S=0,~1,~2$. It is easy to show that
if $K \Rightarrow \infty$ then all variants lead to $S=0$.

\subsection{Evolutionary tracks of stars}

Presently several series of calculations of stellar evolutionary tracks
are available. The most used series of evolutionary tracks was calculated
by \citet{Iben:1965aa,Iben:1965ab,Iben:1966aa,Iben:1966ab,
  Iben:1966ac,Iben:1967aa,Iben:1967ab}.  For this series, the following
chemical abundance was used: $X=0.71$, $Y=0.27$ and $Z=0.02$; models
were calculated for stellar masses:
$0.5,~1.0,~1.25,~1.5,~2.25,~3,~5,~9,~15\,M_\odot$.  In another series,
Iben with collaborators calculated models of population II 
metal-poor stars: \citet{Faulkner:1966aa}, \citet{Iben:1968aa},
\citet{Rood:1968aa}, \citet{Iben:1968ab}, \citet{Iben:1970aa},
\citet{Rood:1970aa}, \citet{Iben:1970ab}, \citet{Simoda:1970aa} and
\citet{Iben:1971aa}. Chemical abundance parameters $X,~Y,~Z$ were
varied, most models were calculated for initial stellar masses around
the solar mass. Another similar series of models was calculated by
\citet{Demarque:1967aa}, \citet{Demarque:1968aa},
\citet{Demarque:1969aa}, \citet{Demarque:1969ab}, 
\citet{Demarque:1971aa}, and \citet{Demarque:1971ab,Demarque:1971ac}.  Recently a
series of evolution models was calculated by \citet{Paczynski:1968aa}
and \citet{Paczynski:1970aa, Paczynski:1970ab, Paczynski:1970ac,
  Paczynski:1971aa}, who accepted abundance parameters $X=0.70$,
$Y=0.27$ and $Z=0.03$, calculations were made for stellar masses
$0.8,~1.5,~3,~5,~7,~10,~15\,M_\odot$. 

{\begin{table*}[ht] 
\caption{Evolutionary tracks} 
\centering 
\resizebox{0.95\textwidth}{!}{\includegraphics*{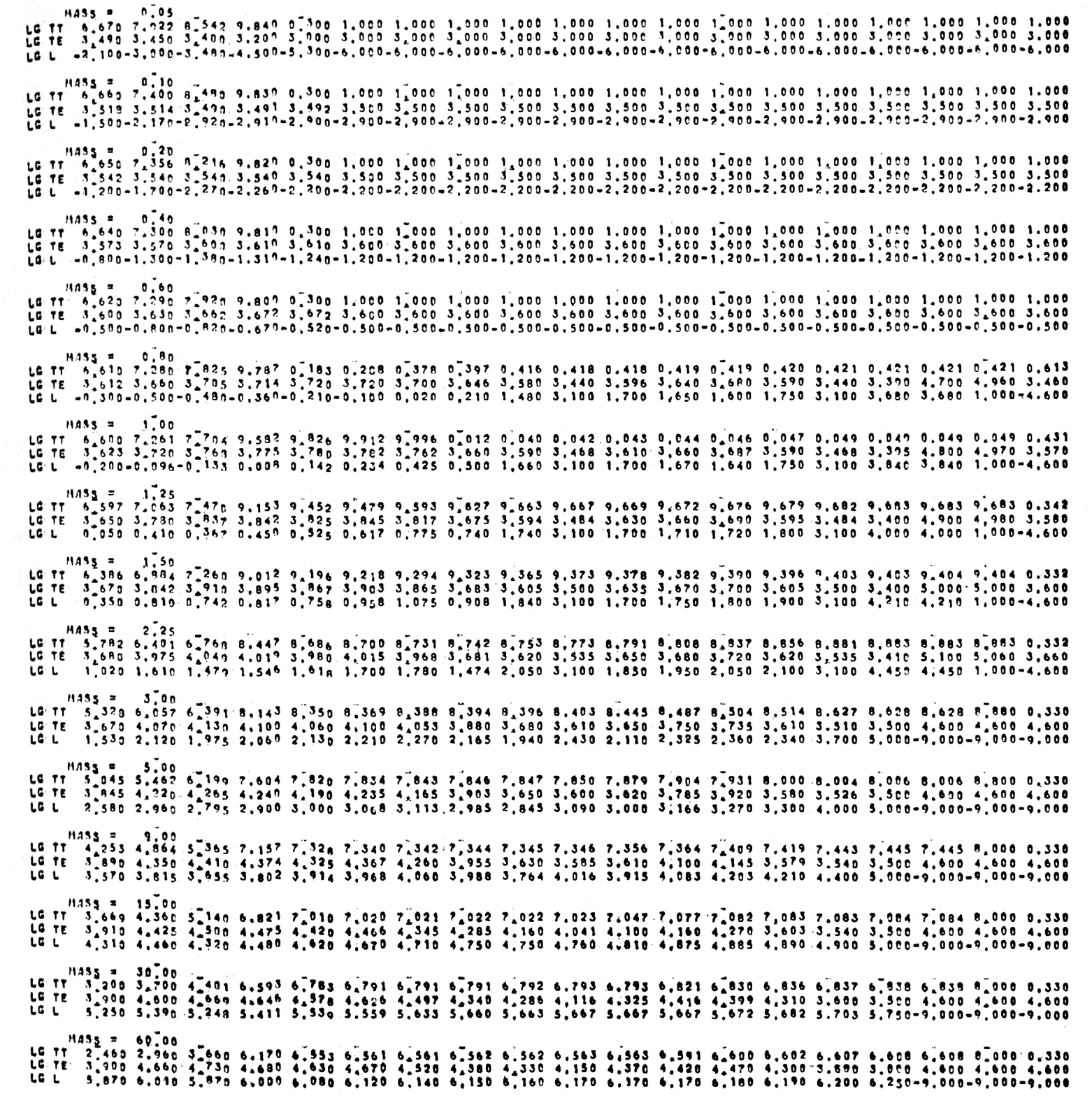}}
\label{Tab22.1}
\end{table*} 
}

In this work, we are mainly interested in the evolution of stars with
normal chemical abundance with $Z = 0.02 - 0.03$. Models by Iben and
Paczynski are slightly different. Paczynski calculated models until
the carbon ignition for massive stars, and until helium ignition for
less massive stars. As a rule, Iben's models were not calculated until
these late evolution phases. On the other hand, Iben has calculated
more models for stars of low mass. For the present work just the evolution
of low-mass stars is important. Taking all these arguments into
account we selected Iben's models as the basis for our use and used
Paczynski models for late evolutionary phases of stars. For very
massive stars ($M=30,~60\,M_\odot$), we used models by
\citet{Stothers:1963aa,Stothers:1964aa,Stothers:1966aa}  and
\citet{Stothers:1968aa,Stothers:1969aa}.  For stars of very small mass,
we used models calculated by various authors:
\citet{Kumar:1963aa,Kumar:1963ab}, \citet{Hayashi:1963aa},
\citet{Rose:1970aa},
\citet{Grossman:1970aa,Grossman:1970ab,Grossman:1971aa}.

Tracks used in this paper are given in Table~\ref{Tab22.1}.
We tabulated evolutionary tracks for stars of mass: 0.05,~0.1,~0.2, 
~0.4, ~0.6, ~0.8, ~1.0, ~1.25, ~1.5, ~2.25, ~3, ~5, ~9, ~15, ~30,
~$60\,M_\odot$.  In all tracks 19 points were given, each point
corresponds to a certain stage in the evolution of stars, for
comparison see Fig.~3 by \citet{Iben:1967ac}.  In our table point 3
corresponds to the arrival of star to the ``zero-age'' point on the
main sequence, point 13 corresponds to the maximum temperature in the
helium burning phase, point 16 to the tip of red giants (carbon
ignition for massive stars), point 18 to the beginning of the white dwarf
sequence, point 19 to the arrival of star to the cold red region. In
most cases, data were taken directly from respective papers, in some
cases interpolation and combination from various sources was needed.
Below we give some details on the interpolation and combination of
published evolutionary tracks.

{\bf A. Gravitational contraction phase.}  The time spent on the
gravitational contraction phase was calculated for stars of small mass
using data by \citet{Kumar:1963aa}. For massive stars, the contraction
time was estimated by extrapolation in Iben's models.

{\bf B. Core hydrogen burning.}  Most data on tracks were taken directly
from published sources. The track for mass $0.8~M_\odot$ was found by
interpolating tracks by Iben for mass 1 and $1.25~M_\odot$, and track
for $0.8~M_\odot$ by Paczynski. The time was estimated as follows:
\be
\log\,t_{0.02} = \log\,t_{0.03} + \Delta,
\ee
where $t_{0.03}$ is the  time according to the respective model by
Paczynski, and $\Delta=0.112$ is a correction, calculated as follows.
The time spent to burn $\Delta\,M$ solar mass of hydrogen by a star
of luminosity $L$ is equal to $\tau_H \propto \Delta\,M/L$. Most of
this time the star is located on the main sequence.  In the range of
stellar masses of interest $L \propto M^4$. If we suppose that in all
stars the same fraction of mass is burned, then $\tau_H = a\,M^{-3}$,
where $a$ is a certain constant. Relations given above have an
approximate character, and we get
\be
\tau_H\,M^3 = a(M),
\ee
where $a(M)$ is a slow function of mass and abundance. We found this
function using Paczynski models of $Z=0.03$ in the mass range $0.8 \le
M \le 3$, and using Iben models with $Z=0.02$ in the mass range $1 \le M
\le 3$. The correction $\Delta$ was estimated by the extrapolation of the
$Z=0.02$ curve toward stars of smaller mass.

{\bf C. Helium burning phase.} The initial phase of helium burning is
well studied.  The last phase of helium burning is less known. The best
data come from the Paczynski series. Iben models were calculated
only to early phases of helium burning. To find late phases of helium
burning of Iben models, an extrapolation
method is needed. This can be done using evolutionary tracks by Paczynski.

Calculations by Paczynski and others suggest that evolutionary tracks
of the last phases are just continuing tracks, found for earlier epochs
in the last phases of hydrogen burning in the giant
branch.   \citet{Uus:1970aa}   and \citet{Paczynski:1971aa} showed
that the speed of the growth of star luminosity in the giant branch is
almost independent of the mass of stars. Hydrogen and helium ignition
on this stage occurs at almost identical luminosity
\citep{Hayashi:1962aa, Paczynski:1970aa}. Authors found that for
$Z=0.03$ helium and hydrogen flashes occur at
$\log\,(L/L_\odot)=3.10$ and $\log\,(L/L_\odot)=5.0$,
respectively. Using these data and tracks found by Iben for early
stages on giant branch, it was possible to continue tracks for later
stages up to the tip of the red giant branch.

It was more difficult to estimate the time spent on helium burning
stage, since most Iben tracks were calculated only for the early stages of
helium burning. The most advanced track was found for the star of mass
15~$M_\odot$ \citep{Iben:1966ac}. To continue the track, only a short
last stage of evolution must be added. According to \citet{Uus:1970aa}
and \citet{Paczynski:1971aa} this stage is very short, thus an error
in the estimate of the time plays little role. The time spent in
helium burning, $\tau_{He}$, can be compared with the time spent in
hydrogen burning, $\tau_H$. For a star of mass $M=15\,M_\odot$ we get
\be
\left({\tau_H \over \tau_{He}}\right)_{002} = \left({\tau_H \over \tau_{He}}\right)_{003}.
\label{eq22.4.6}
\ee
\citet{Paczynski:1971aa} showed that for $Z=0.03$ and $M \ge
3\,M_\odot$ the following relation exists
\be
\log\,{\tau_H \over \tau_{He}} = -0.174+0.633\,\log\,M +
0.182\,(\log\,M)^2.
\label{eq22.4.7}
\ee
The time $\tau_{He}$ for $Z=0.02$ and $M \ge 3\,M_\odot$ was found
using Eq.~(\ref{eq22.4.6}) and (\ref{eq22.4.7}).

If we apply these equations for stars of smaller masses, then the time
$\tau_{He}$ is too large. On the other hand, \citet{Iben:1970ab}
showed that stars of approximate solar mass have $\tau_{He} \approx
12 \times 10^7$ years, almost independently of the chemical
composition. For $M=3\,M_\odot$ stars we found $\tau_{He} =1.71\times
10^7$ years.  Since $\tau_{He} \ll \tau_H$, then high accuracy of
$\tau_{He}$ plays little role, and we accepted for stars of mass $M
\le 3\,M_\odot$ $\tau_{He} = 17.1 \times 10^7$ years.

Evolutionary tracks of stars of mass $M < 3\,M_\odot$ were calculated
by Iben before the helium ignition. Thus, the whole red giant branch of
the evolution during the helium burning must be estimated from other
data. We used for stars of mass $0.8\,M_\odot$ the blue end of the
horizontal branch at point $\log\,L/L_\odot = 1.60$ and
$\log\,T_e =3.68$. For tracks of stars of other masses, we used
observational data of colour-magnitude diagrams of old star clusters
by \citet{Sandage:1962aa}, \citet{Eggen:1964ys}, and
\citet{Newell:1969aa}.

{\bf D. Last stages of stellar evolution.}  Available data suggest that
the final evolutionary stage of all stars is the degenerate white
dwarf. The path toward this stage can be different. Very low mass
stars of mass, $M \le 0.08\,M_\odot$, and  normal metal abundance come
to this stage of ``black'' dwarfs directly after the initial gravitational
contraction. The radius of such stars depends on its mass and chemical
abundance, the star in this stage uses its thermal energy. According
to \citet{Schwarzschild:1958kx} the time spent in this stage is
\be
\log\,\tau = b + 5/8\, \log\,(M/L),
\label{eq22.4.8}
\ee
where $b$ is a constant, depending on the chemical composition, and
$M$ and $L$ are expressed in solar units. Using data by
\citet{Schwarzschild:1958kx} we found that for red dwarfs of normal composition
$b=7.42$, if $\tau$ is expressed in years.

After the formation of carbon nucleus, stars of mass $M > 1.4\,M\odot$
explode as supernovae, and their nuclei become pulsars. Stars of lower
mass move after the exhausting of nuclear energy through the
planetary nebulae stage to white dwarfs. Due to mass loss during the
evolution, the limiting mass of stars at the stage where evolutionary
paths diverge, 
is higher than $1.4\,M_\odot$. According to \citet{Jones:1970aa}, the
initial mass of white dwarfs of Hyades is larger than $1.8\,M_\odot$.
Using novae statistics by \citet{Payne-Gaposhkin:1957}, 
\citet{Stothers:1963ab} concluded that the limiting mass of stars to
go through the supernova stage is $4\,M_\odot$. Now we shall find
the limiting mass using more recent data.

According to \citet{Tammann:1970aa}, the frequency of supernova
explosions in galaxies of type Sb is 0.09 for 100 years for mass unit
$10^{10}\,M_\odot$. This statistics is based on galaxy models by
\citet{Roberts:1969aa}, who used \citet{Brandt:1960} mass distribution
profile. As shown by \citet{Einasto:1969ab}, this mass distribution
model yields too high masses for galaxies.  For this reason, we use for
the Galaxy a mass $M=15 \times 10^{10}\,M_\odot$, and get the
frequency 1 supernova per 74 years.  This frequency is in good
agreement with the estimate by \citet{Caswell:1970aa} (40–80
years), based on radioastronomical data on the frequency of supernova
remnants in Galaxy.  Using star formation function Eqs.~(\ref{eq22.1})
and (\ref{eq22.3}) with parameters $n=7/3$, $M_0 =0.03\,M_\odot$ and
$M_u=100\,M_\odot$, and for the speed parameter of star formation in
Eq.~(\ref{eq22.11}) $K=0.25\times 10^9$ years, we find that in our Galaxy
in the present epoch 3.5 stars form per year.  The frequency 1
supernova per 74 years corresponds to the lower bound of mass of
supernova progenitor stars: $2.6\,M_\odot$.  We accept this value for
further calculations.  Optical luminosity of supernova remnants 
(pulsars) decreases very fast, and they are practically optically
invisible \citep{Pacini:1971aa}.  Their luminosity is not known. In
our calculation we used for their luminosity the value
$\log\,L/L_\odot = -9.$

Let us now discuss the last evolutionary stages of stars of intermediate
mass. According to \citet{Paczynski:1970aa} the luminosity of stars
during the path towards nuclei of planetary nebulae depends only on
the mass of the nucleus of the star $M_c$ (the future white dwarf):
\be
L/L_\odot = 59\,250\,M_c/M_\odot - 30\,950.
\ee
It follows from the same data that the mass of the nucleus depends on
the initial mass of stars. Interpolating \citet{Paczynski:1970aa}
data we get for stars of initial masses 2.25, 1.25 and 1~$M_\odot$
masses of nuclei 1.0, 0.7 and 0.65~$M_\odot$, respectively.  The time
spent on these evolutionary stages was taken from Table 2 of
\citet{Paczynski:1970aa}. 

Evolutionary tracks for white dwarfs were taken from
\citet{Schwarzschild:1958kx}.  We used data from Table 28 of
\citet{Schwarzschild:1958kx}, the time spent in this stage was
calculated using Eq,~({\ref{eq22.4.8}), where we used for the parameter $b$
a value $b=7.05$. This equation is evidently approximate. In the
last stage of their evolution, white dwarfs start to crystallise 
\citep{van-Horn:1968aa}, which initially slows the cooling but later
speeds it.  These details play practically no role in the
determination of integral characteristics of galaxies and can be
ignored. In our calculations we used as the last point of the
evolutionary track of white dwarfs $\log\,L/L_\odot = -4.6$. The total
time spent by stars to reach this point exceeds 20 billion years,
independent of the mass of stars. This was the maximal epoch used in
calculations of stellar evolutionary tracks.

{\begin{table*}[h] 
\caption{Intrinsic colours and bolometric corrections} 
\centering 
\hspace{2mm}
\resizebox{0.75\textwidth}{!}{\includegraphics*{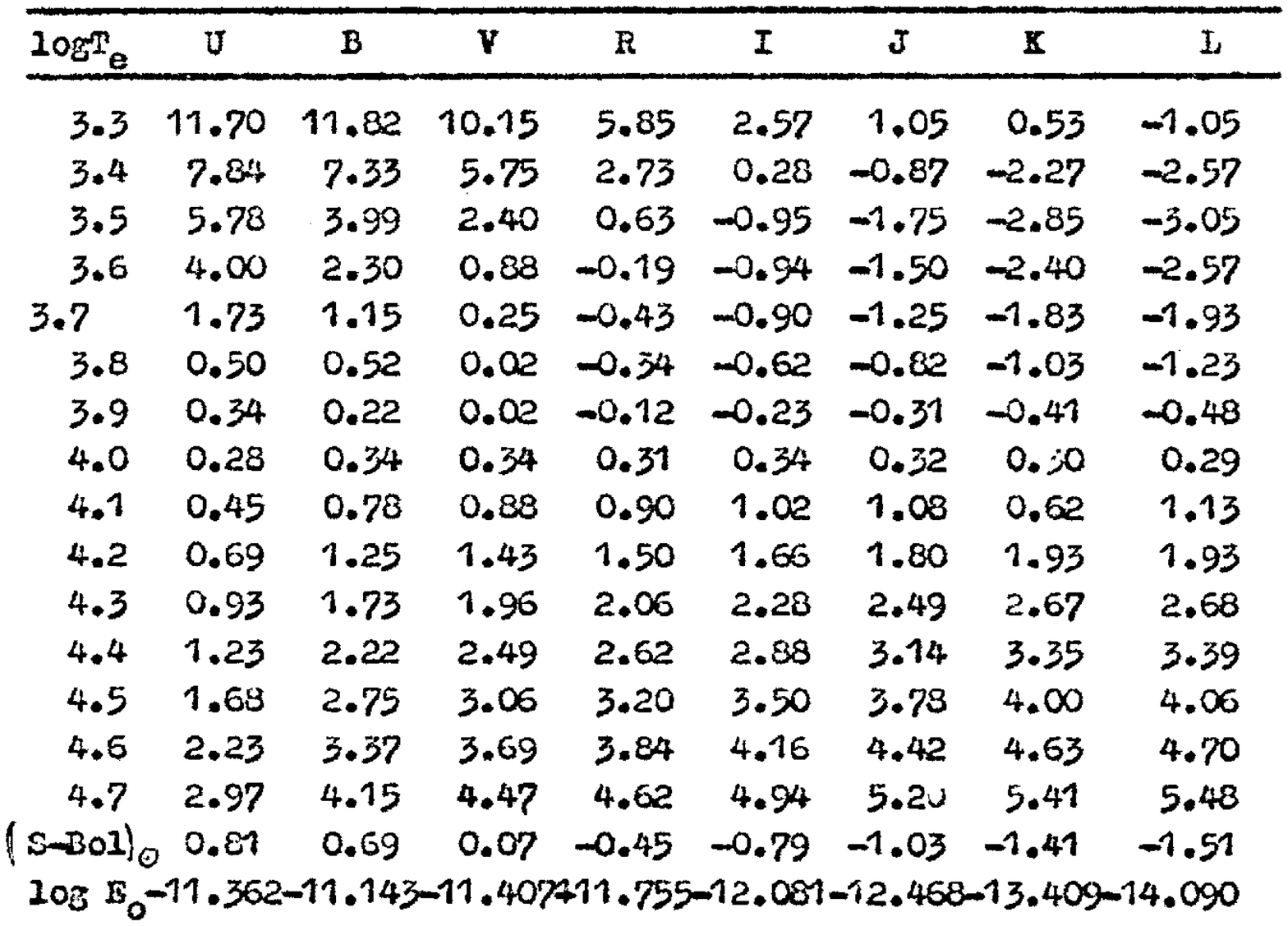}}
\label{Tab22.2}
\end{table*} 
}

\subsection{Bolometric corrections and colour indexes}

Theoretical evolutionary tracks are given in coordinates
$\log\,T_e,~\log\,L/L_\odot$, observational data are given as colours
and absolute magnitudes, $B-V,~M_V$.  To find the relationship between
these data as well as to calculate mass-to-light ratios of
populations and the energy distribution in spectra of model galaxies,
we need to know bolometric corrections, $BC = M_b -M_V$, and intrinsic 
colours, $(S-V)_0$, of stars of all types. In the last equation $S$
denotes the star magnitude in whatever colour in the magnitude system
UBVRIJKL by \citet{Johnson:1964aa}.

As the main source of bolometric corrections, intrinsic colours and effective
temperatures we used the compilation by \citet{Johnson:1966aa}.  Also,
we used some more recent sources which complement Johnson data in the
infrared and ultraviolet range of the spectrum. 

\citet{Johnson:1966aa} published $BC$, $(S-V)_0$ and $T_e$ as
functions of spectral type and luminosity class.  For our purpose we
need these data as functions of $\log\,T_e$ and $\log\,L/L_\odot$. We
found a respective relation by graphical interpolation. As the first step
we found the dependence of $\log\,T_e$ on spectral type, separately
for the main sequence (luminosity class V), giants (III) and supergiants
(I). Next we made graphs of $BC$ and $(S-V)_0$ as functions of
$\log\,T_e$.  Instead of intrinsic colours we used bolometric
corrections in the $S$ system
\be BC_S = M_b-M_S=BC_V - (S-V)_0.
\ee
Results of calculations are given in Table~\ref{Tab22.2}.
We give a logarithm of the luminosity in ergs for a star
of zero magnitude in the UBVRIJKL \citet{Johnson:1966aa} system,
and the intrinsic colour of the Sun according to \citet{Mendoza:1968}.
Data in this Table need some comments.

{\em Intrinsic colours.} Most intrinsic colours were taken from
\citet{Johnson:1966aa}.  Data for colours $(U-V)_0$ and $(B-V)_0$ were
taken from a more recent study by \citet{Fitzgerald:1970aa}. For red
stars we used the catalogue by \citet{Mendoza:1968} and data by
\citet{Greenstein:1970aa}. For A and F stars of the main sequence we
used data by \citet{Davis:1970aa}.

{\em Bolometric corrections and effective temperatures.}  The
effective temperature of a star is defined as the temperature of the
black body, which emits from the unit surface area the same amount of energy
as the star. To find effective temperatures of stars, their angular
sizes and bolometric luminosities are needed. Bolometric luminosities
can be calculated by integrating the spectral energy distribution over
wavelengths, expressed in absolute units. Angular diameters can be
found from interferometric observations. When Johnson made his summary,
diameters were available only for 14 stars, including
Sun. \citet{Mendoza:1968} used angular diameters of 27 stars.  These
data were not sufficient to find the scale of effective temperatures,
thus results from theoretical calculations of models of stellar
atmospheres were also used.

The scale of effective temperatures and bolometric corrections for
spectral classes O5–G2 was determined by \citet{Morton:1968aa},
using model atmosphere calculations.  We used these data for the main
sequence stars earlier 
than F2 type. For stars of type F2–G2, 
corrections from radiometric observations  were determined. The zero
point of theoretical BC was chosen in such a way that at spectral type
F2 theoretical model data coincide with observational data.

Recently \citet{van-Citters:1970aa} suggested that it is better to
connect BC directly with bolometric data on the Sun, BC$_\odot=
-0.07$. For this reason, we used for B0.5–B6 stars BC directly from
\citet{van-Citters:1970aa}, for O5–B0.5 stars from
\citet{Morton:1969aa}, and for B8–F stars from
\citet{Davis:1970aa}, who used the same method to determine BC.

For effective temperatures of B0.5–B7 stars, we used the scale by
\citet{Morton:1968aa}, as this scale coincides with the scale found
from angular diameters by \citet{Hanbury-Brown:1967aa}  and
\citet{van-Citters:1970aa}. For very bright blue stars of type O5–
B0.5 \citet{Morton:1969aa} determined a new temperature scale, which
is favoured over the theoretical scale. However,
\citet{Peterson:1971aa} determined effective temperatures of six stars
of type O5–09.5, and found that they are higher than temperatures
by \citet{Morton:1969aa}. We shall use the temperature scale by
\citet{Peterson:1971aa}, and to find BC we use the relationship
between $T_e$ and BC, found by  \citet{Morton:1969aa}.

For late M dwarfs we used $T_e$ and BC from the paper by
\citet{Greenstein:1970aa}.  For red giants of type M0–M6 we used
for $T_e$ and BC data by \citet{Lee:1970aa}. For M7–M8 giants
bolometric corrections were taken from \citet{Smak:1966aa}. For most
red giants we determined the temperature scale, applying procedures
given by \citet{Johnson:1966aa}.  The scale of effective temperatures
was based on angular diameters, given in Table V of
\citet{Johnson:1966aa} and in Table 2 of \citet{Mendoza:1968}.
Temperatures were calculated with the black-body fit method, using
data given in Tables VI and VII of \citet{Johnson:1966aa}.

{\begin{figure*}[h] 
\centering 
\resizebox{0.85\textwidth}{!}{\includegraphics*{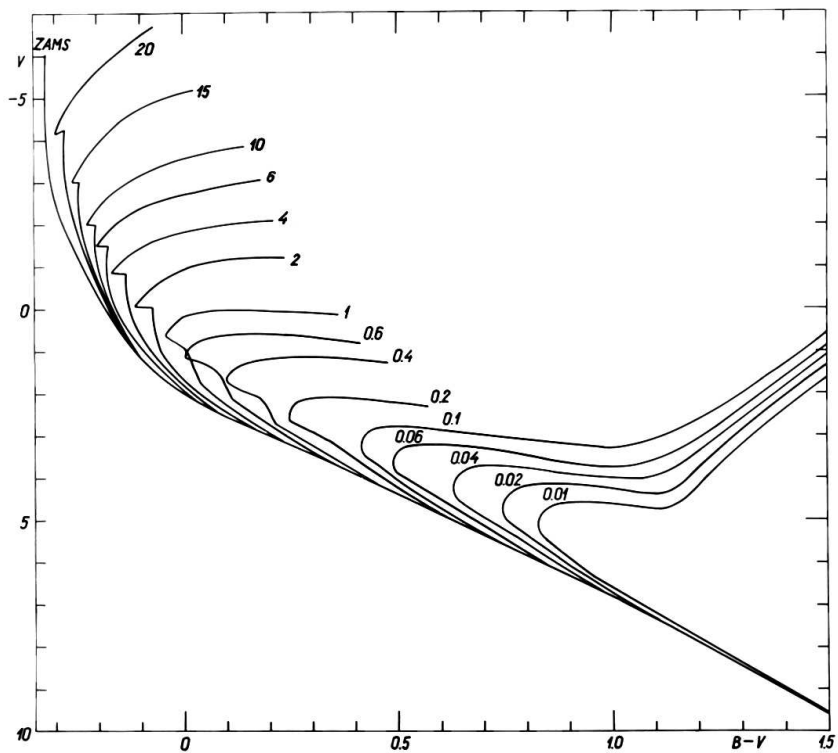}}
\caption{Isochrones of stars of normal chemical composition
  $Z=0.02$. Masses of stars are labeled 20, 15, 10 etc. }
  \label{Fig22.1}
\end{figure*} 
}

\subsection{Isochrones}

The collected information was used to derive isochrones of stars of
various masses in the model colour-luminosity $(B-V,M_V)$ diagram, and
to study the physical evolution of stellar systems. In this section we
discuss isochrones found, presented in Fig.~\ref{Fig22.1}. For
comparison we show in Fig.~\ref{Fig22.4} below colour-magnitude
diagrams for selected star clusters.

We show in the Figure isochrones for the main phase of hydrogen
burning of stellar evolution.  We also constructed isochrones for
late stages of stellar evolution, corresponding to helium burning,
white dwarfs, and final stages of red ``white'' stars. In these late
stages isochrones cross each other several times.  To avoid
overcrowding of the Figure these late stages are not shown.

Close to the hydrogen burning main sequence evolutionary tracks also
cross.  This effect was detected by \citet{Sandage:1969ac} from
theoretical models and later from empirical data. Around the turn-off
point of the main sequence isochrones have a zigzag form. This feature
was confirmed by observational data by \citet{van-den-Heuvel:1969aa},
see also \citet{Iben:1967ac}.

In the red giant region, our compiled data agree well with data by
\citet{Schlesinger:1969aa}.  However, in a later paper
\citet{Schlesinger:1969ab} showed that, in contrast to observations,
the helium sequence, found by \citet{Iben:1966aa}, extends too far
toward the blue region.  \citet{Schlesinger:1969ab} modified his
program and found for the blue tip of the helium main sequence for
$M=5\,M_\odot$ star a value $\log\,T_e =3.68$, whereas according to
\citet{Iben:1966aa} it is $\log\,T_e =3.92$.  The need to correct Iben's
data is confirmed by our calculations, which show that integral
colours of model galaxies become too blue. For this reason, we applied
for the blue end of the helium sequence of $M=5\,M_\odot$ stars the value
$\log\,T_e =3.68$ by \citet{Schlesinger:1969ab}.

Observational data by \citet{Cannon:1970aa} suggest that stars of
lower mass have a horizontal sequence, which is usually identified with
the helium burning phase of the giant branch. It is not easy to determine the blue
end of the horizontal sequence for more massive stars. In
this stage both hydrogen and helium burning phases are closely
located.  We accepted the more extreme case, and identified the 
yellow and red sequences of giant stars with the helium burning
phase.  According to \citet{Wildey:1964aa} the colour of these red
giants  is redder than $B-V=1.00$, which corresponds to
$\log\,T_e=3.68$.  We accepted this temperature value for the blue end
of the helium sequence of all stars of mass $M \ge 5\,M_\odot$.

{\begin{figure*}[h] 
\centering 
\resizebox{0.60\textwidth}{!}{\includegraphics*{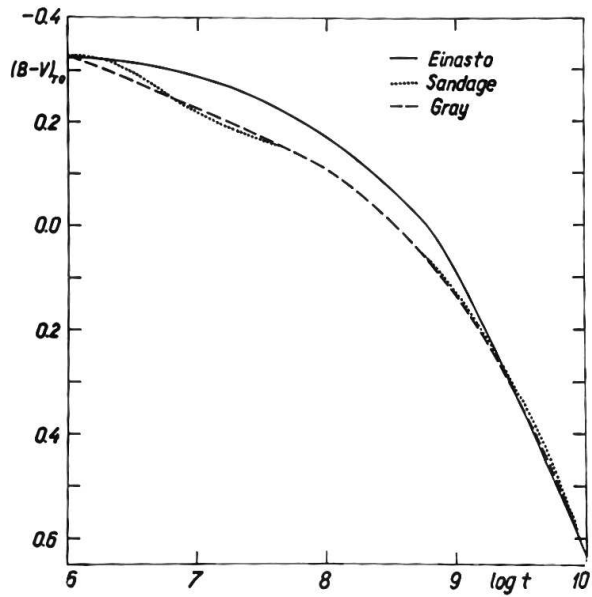}}
\caption{ Colours  of the main sequence turn-off
  point as functions of the star age according to the present paper,
  \citet{Sandage:1963aa} and \citet{Gray:1965aa}. } 
  \label{Fig22.2}
\end{figure*} 
}

The colour of the turn-off point of the main sequence is often used as
the criterion of the cluster age.  Using our model isochrones we made
a new calibration of this relationship, results are given in
Fig~\ref{Fig22.2}. For comparison we show here also calibrations
according to \citet{Sandage:1963aa} and \citet{Gray:1965aa}. 

\section{Time evolution of physical characteristics of model galaxies}

We made calculations to follow changes in physical characteristics of
model galaxies for a series of initial data. Below we describe initial
data used and results obtained.

\subsection{Initial data}

Most calculations were done using following initial data for star
formation rate: $n=7/3$, $M_0 =0.03\,M_\odot$, $M_u = 100\,M_\odot$,
$S=0$~ ($K=20$),~  $S=1$~ ($K=0.5,~3$),~  $S=2$~ ($K=0.3$);  here  the
characteristic time of galaxy formation $K$ is
expressed in billion years. We calculated model galaxies for ages 
0.01,~0.03,~0.3.~1,~2,~4,~6,~8,~9,~10,~15, and 20 billion years.

To find the dependence of physical properties of galaxies on the
values of star formation function parameters, we made calculations for
a range of $S$ and $K$, using $n=2.05$ instead of $n=7/3$. Some
calculations were made using a shifted blue end of the helium burning
sequence, as discussed above.

The upper limit of the star formation function, $M_u$, has little
influence on integral properties of galaxies. The change of the lower
mass limit, $M_0$, has little influence on colour and spectral energy
distribution, but has a large impact to the luminosity, $L$, and mass-to-light
relation, $f=M/L$, through the parameter $a$, see Eq.~(\ref{eq22.3}).
The luminosity of the model galaxy and its mass-to-light ratio
can be find as follows:
\be
L=L_0/\delta,~~~~f=\delta\,f_0,
\ee
where $L_0$ and $f_0$ are values of these parameters for lower mass limit
$M_0=0.03\,M_\odot$.  We give in Table~\ref{Tab22.3} coefficients
$\delta$ and $\Delta\,M = 2.5\,\log\,\delta$ for a series of minimal
masses of star formation function, $M_0$.

{\begin{table*}[h]
\centering    
\caption{} 
\begin{tabular}{lcr}
\hline  \hline
  $M_0$ & $\delta$ & $\Delta\,M$\\
  \hline\\
  0.50 & 0.348 & -1.145\\
  0.30 & 0.426 & -0.928\\
  0.10 & 0.644 & -0.478\\
  0.03 & 1.000 & 0.000\\
  0.01 & 1.473 & 0.420\\
  0.003& 2.234& 0.873\\
  0.001& 3.256 & 1.283\\
  \hline
\label{Tab22.3}   
\end{tabular} 
\end{table*} 
} 

Mass-to-light ratios of old stellar populations differ
considerably. According to \citet{Schwarzschild:1955aa} globular
clusters have $f=M/L \approx 1$, for giant elliptical galaxies
\citet{Page:1967} found values up to $f=M/L \approx 100$.
\citet{Tinsley:1968} explained the large variance of $f$ in old
populations with the large quantity of stars of low mass in giant
elliptical galaxies.  If we accept these high values of $f$, we have to
use for the lower end mass of star formation function values like $M_0
\approx 10^{-6}\,M_\odot$, which seem to be improbably  low.  If we
change $M_0$ within the range given in Table~\ref{Tab22.3}, it is
impossible to change mass-to-light in such large limits. The
change of the parameter $n$ in star formation rate function
cannot yield so large changes in $f$ either. \citet{Tinsley:1971}
demonstrated in the case of M31 that this parameter can be changed
only in a very limited range.

On the other hand, it is well known that evolutionary tracks of stars
of different chemical composition are different, which also changes
integral characteristics of stellar systems. In the present time, there
exists no complete series of evolutionary tracks for extremal chemical
compositions. Moreover, for such stars there are no reliable
bolometric corrections and intrinsic colours.  However, to get an idea
what possible changes of integral characteristics of galaxies and
stellar populations are expected for extreme chemical compositions, we
made calculations with the same program as before, but
with shifted evolutionary tracks.

To get tracks for population rich in heavy elements we added to tracks,
calculated for composition $Z=0.02$, the following corrections:
\be
  \Delta\,\log\,t = -0.12, \hspace{2mm}
\Delta\,\log\,T_e = -0.10, \hspace{2mm}
  \Delta\,\log\,L = -0.20.
\label{eq22.7.2}
\ee
These corrections were based on tracks found by \citet{Iben:1967ac},
\citet{Paczynski:1970aa} and \citet{Schlesinger:1969ab}.  We attribute
these corrections to stars of heavy element content $Z=0.08$. Since
there are presently no tracks for this composition, the shifted
model can correspond to some other value of $Z$.

{\begin{figure*}[h] 
\centering 
\resizebox{0.49\textwidth}{!}{\includegraphics*{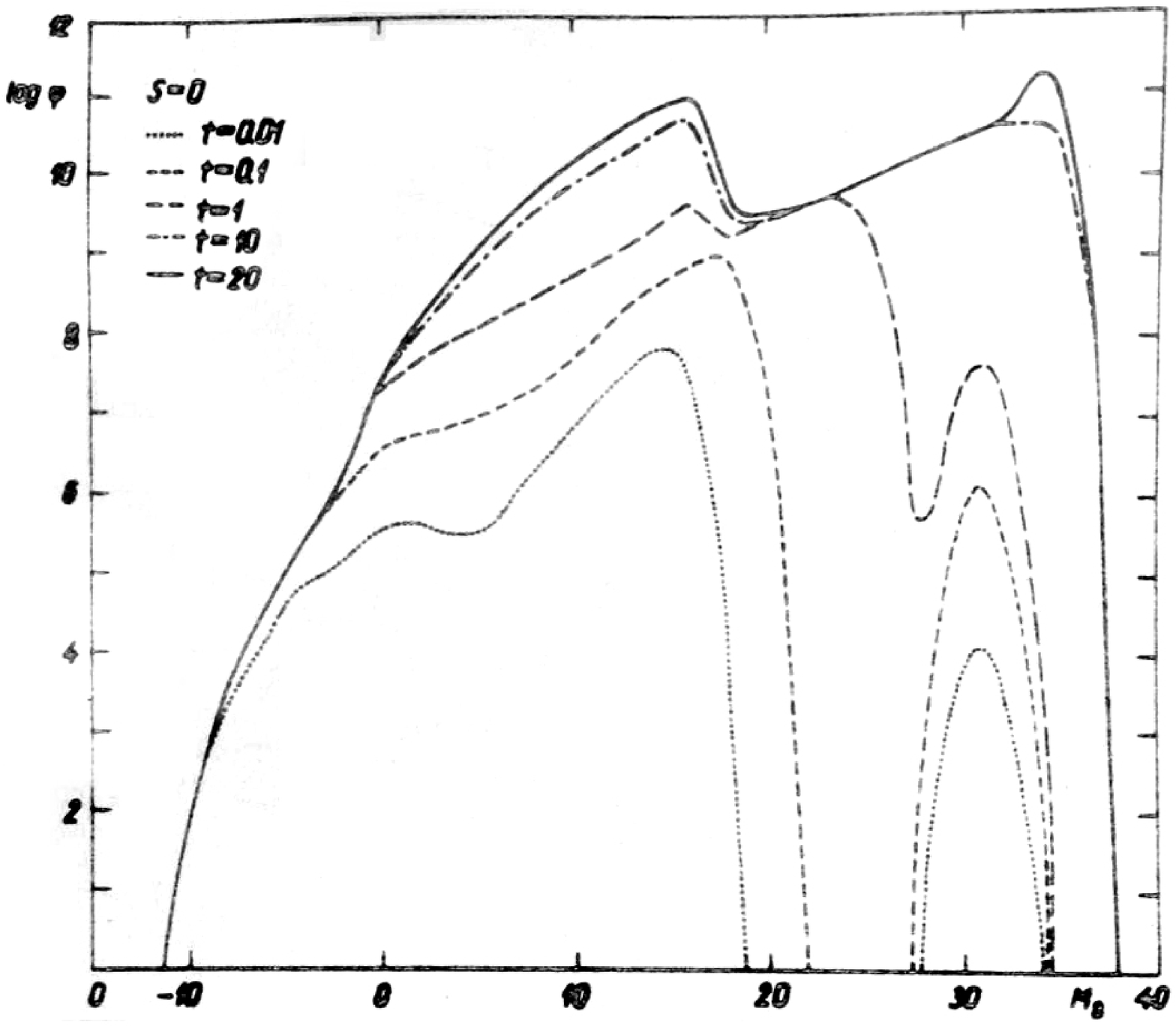}}
\resizebox{0.49\textwidth}{!}{\includegraphics*{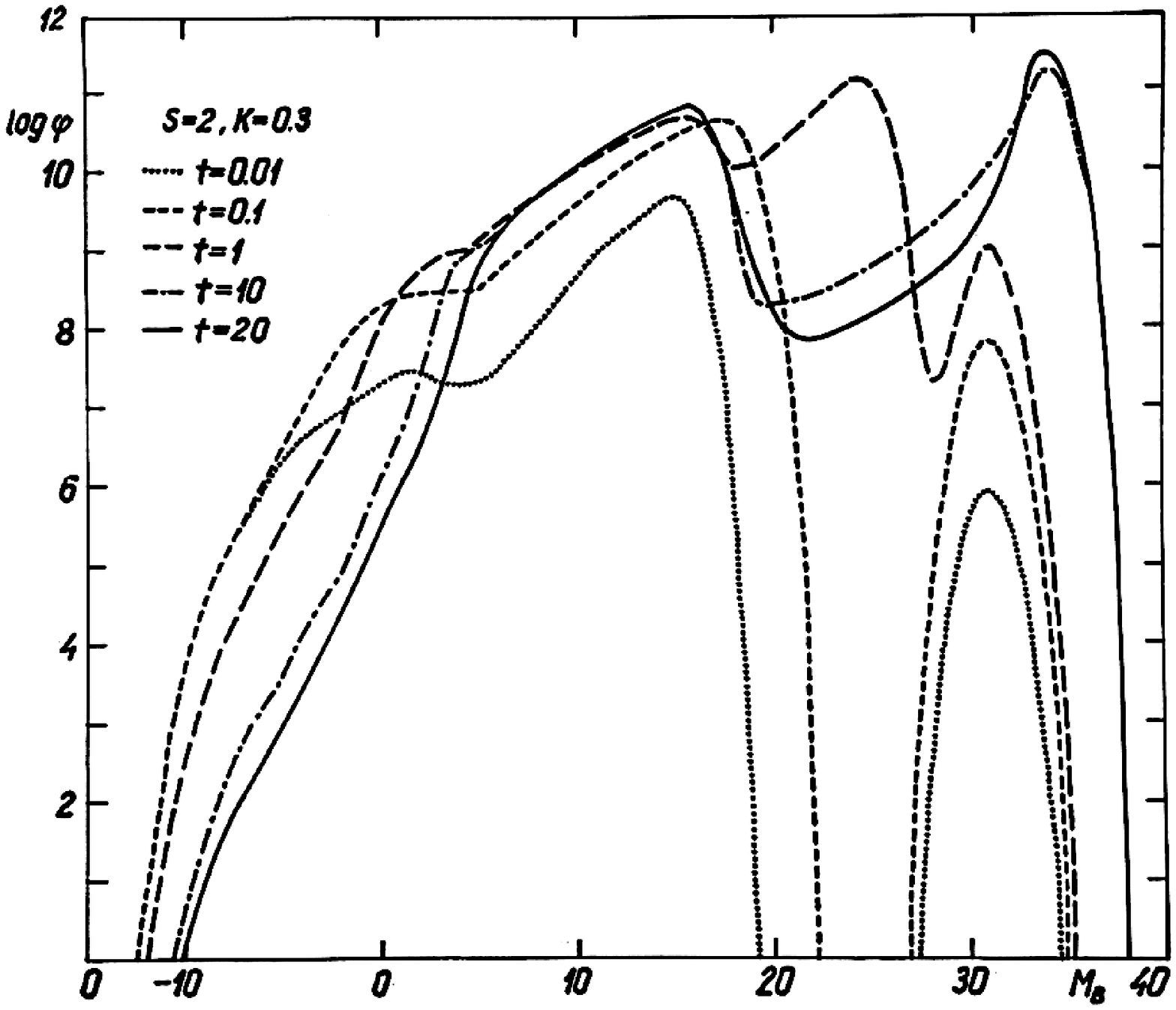}}
\caption{The dependence of the luminosity function of galaxies in B
  system on the age of the galaxy $t$ in billion years. As argument we
  use the absolute magnitude in B system. Star formation function
  parameters are taken as follows: {\em left }: $S=0$, $n=2.33$,
  $M_0=0.03\,M_\odot$, $R=5\,M_\odot$ per year; {\em right}
  $S=2$, $K=0.3\times 10^9$ years.  The mass of the galaxy is
  $M=10^{11}\,M_\odot$.}
  \label{Fig22.3}
\end{figure*} 
}

To get tracks for metal deficit stars we added to tracks for $Z=0.02$
corrections:
\be
 \Delta\,\log\,t = -0.22,\hspace{2mm}
\Delta\,\log\,L = 0.25.
\label{eq22.7.3}
\ee
Correction for $\log\,T_e$ was changed from zero-age main sequence
point 0.085 until the tip of the giant branch at 0.200. In addition,
the blue end of the horizontal giant branch was fixed at point
$\log\,T_e = 4.25$, and in the luminosity by $\Delta\,\log\,L =-0.15$
lower than for stars of normal composition. These corrections were found
using tracks by \citet{Demarque:1971aa},  and photometric observations
by \citet{Sandage:1970ab}.  We attributed these corrections to stars of
composition $Z=0.001$.

\subsection{Model results}

The following functions were calculated: luminosity function,
integrated luminosity in solar units and magnitudes, the contribution
of stars of different luminosity to the summed luminosity,
mass-to-light ratio. All functions were found in bolometric
units and in photometric system UBVRIJKL. Also the distribution of
energy in spectra of model galaxies using calibrations according to
Table IV by \citet{Johnson:1966aa}. Table~\ref{Tab22.2} gives a logarithm of the
luminosity in ergs for a star of zero magnitude in UBVRIJKL system,
and the intrinsic colour of the Sun according to
\citet{Mendoza:1968}.

The amount of calculations and output results
was rather large. We show the main results in a graphical form in
Figs.~\ref{Fig22.3} to \ref{Fig22.12}. The dependence of some quantities
(luminosity, mass-to-light ratio, colour) on time (age of the
model galaxy) is not very smooth.  The reason for this behaviour is
due to the use of a discrete distribution of masses of
stars in the program. For plotting we used in Figs.~\ref{Fig22.3} to \ref{Fig22.10}
smoothed functions.

The most interesting model calculation results are colour indexes
and energy distributions in spectra, as these quantities can be
directly compared with observations. This comparison is done in the
next section.

\section{Analysis of results}

In this section we discuss various functions and parameters of model
galaxies with observations.

{\bf A. Luminosity function}. 
An important property of luminosity functions is the continuation of
the function toward faint stars to very low luminosities. This
suggests that there exist stars of very low luminosity, contrary to
earlier estimates by van Rhijn and Luyten, who suggested that stars of
magnitude fainter than $M=16$ are very rare.

{\begin{figure*}[h] 
\centering 
\resizebox{0.44\textwidth}{!}{\includegraphics*{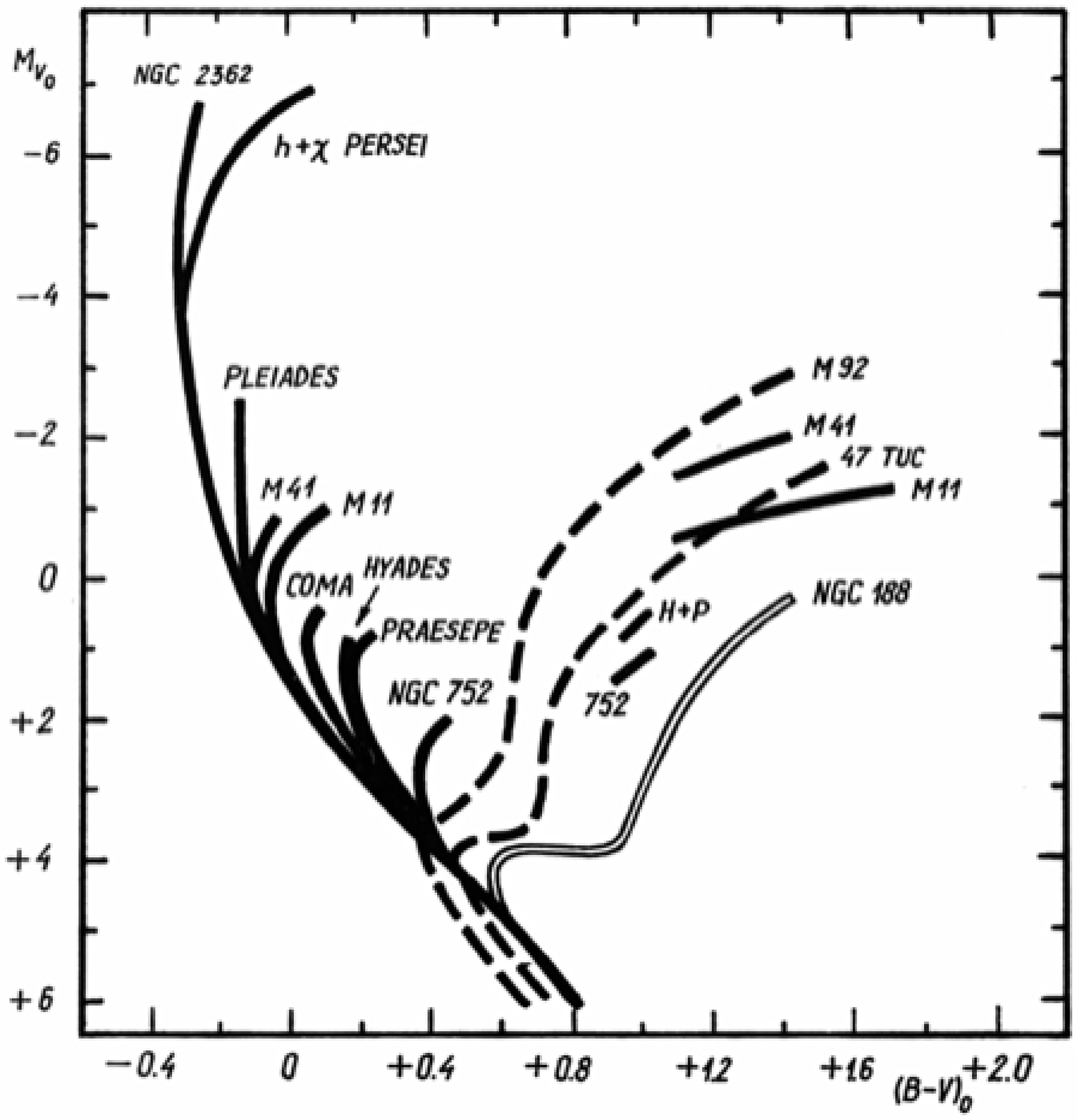}}
\resizebox{0.52\textwidth}{!}{\includegraphics*{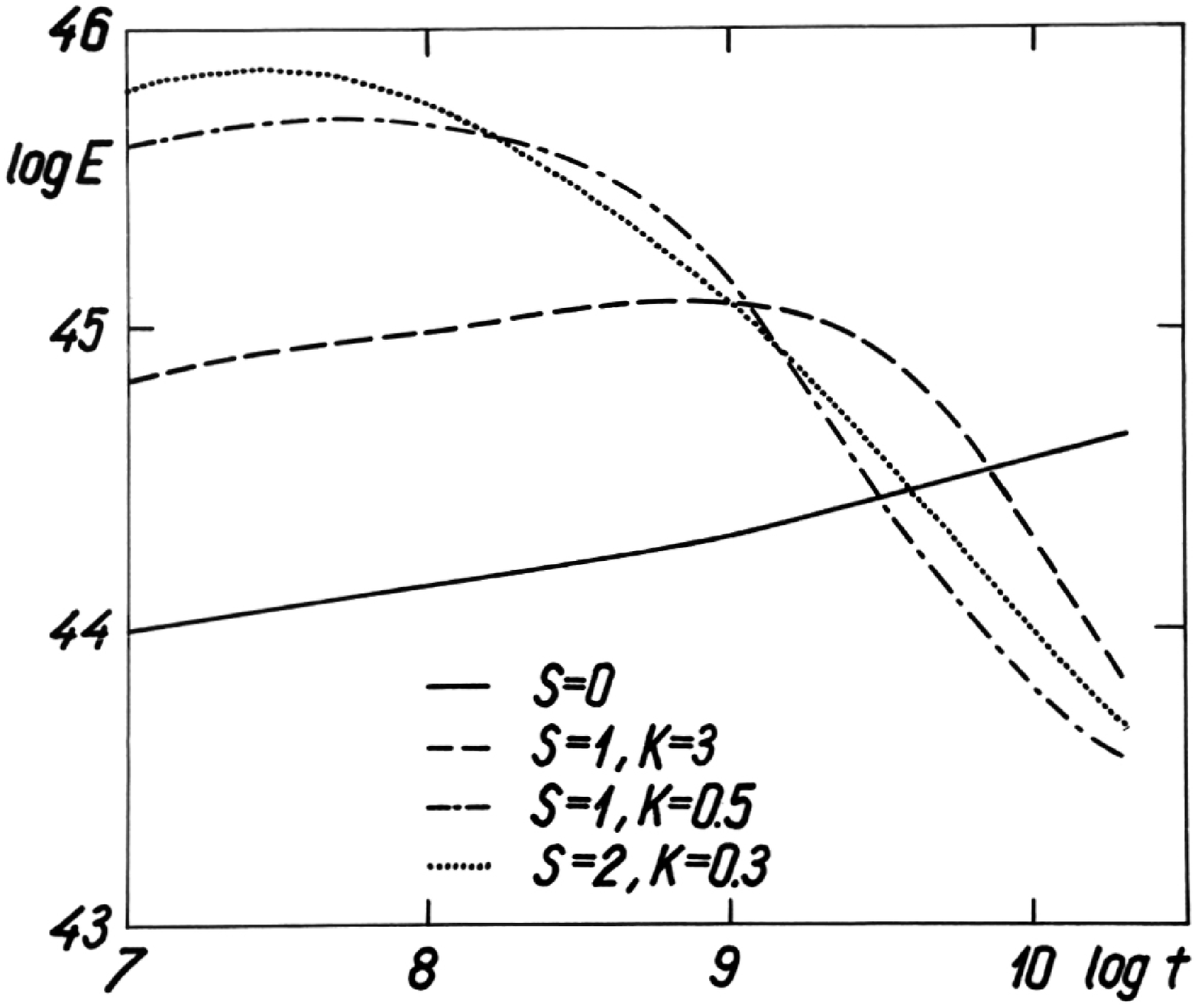}}
\caption{ {\em Left:} Colour-magnitude diagrams for selected star
  clusters.  {\em Right:} Integral bolometric luminosities of
  galaxies of mass $M=10^{11}\,M_\odot$ as function of the age $t$ in
   years for various parameters $S$ and $K$ of star formation
  function. }
  \label{Fig22.4}
\end{figure*} 
}

{\begin{figure*}[ht] 
\centering 
\resizebox{0.32\textwidth}{!}{\includegraphics*{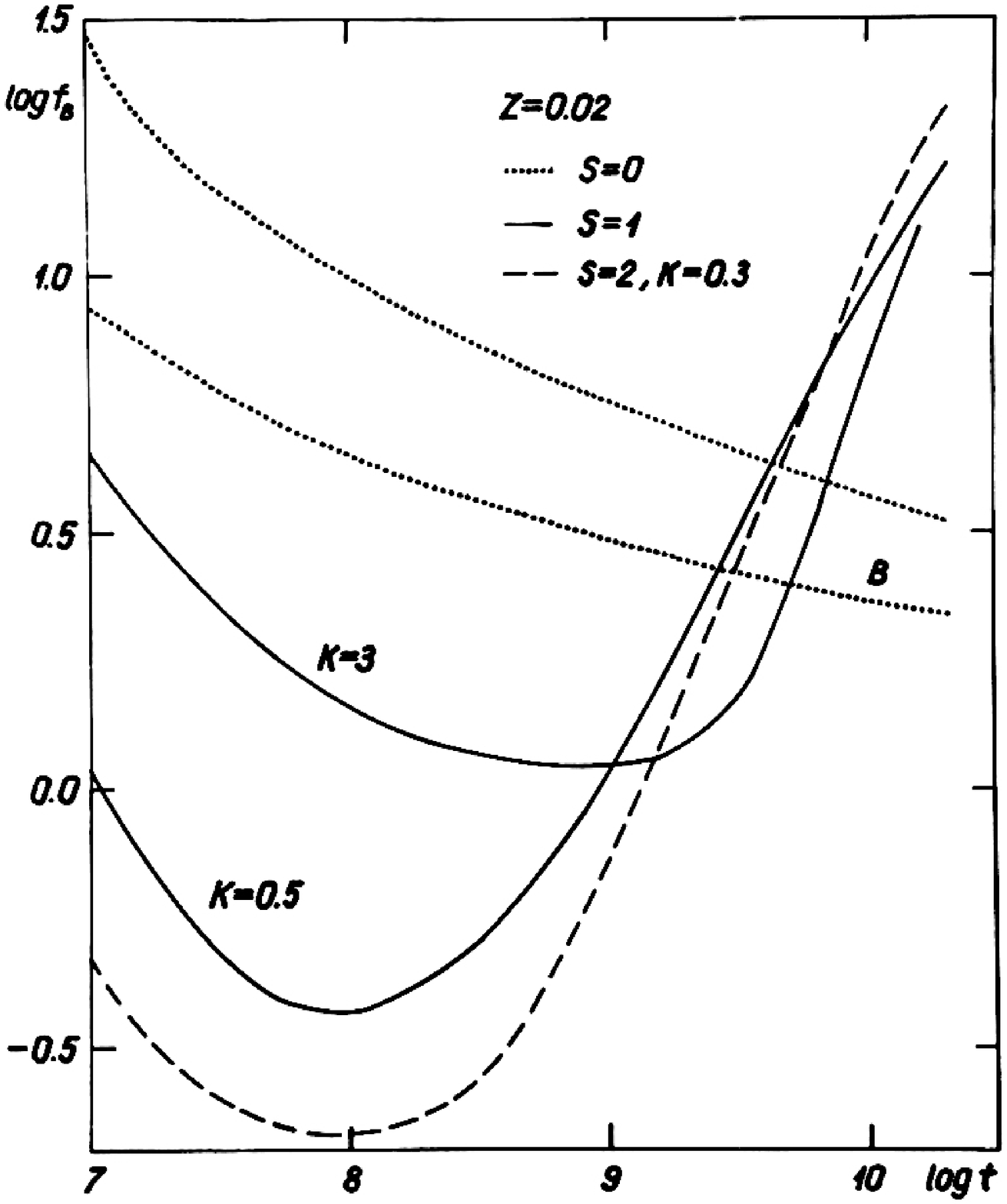}}
\resizebox{0.32\textwidth}{!}{\includegraphics*{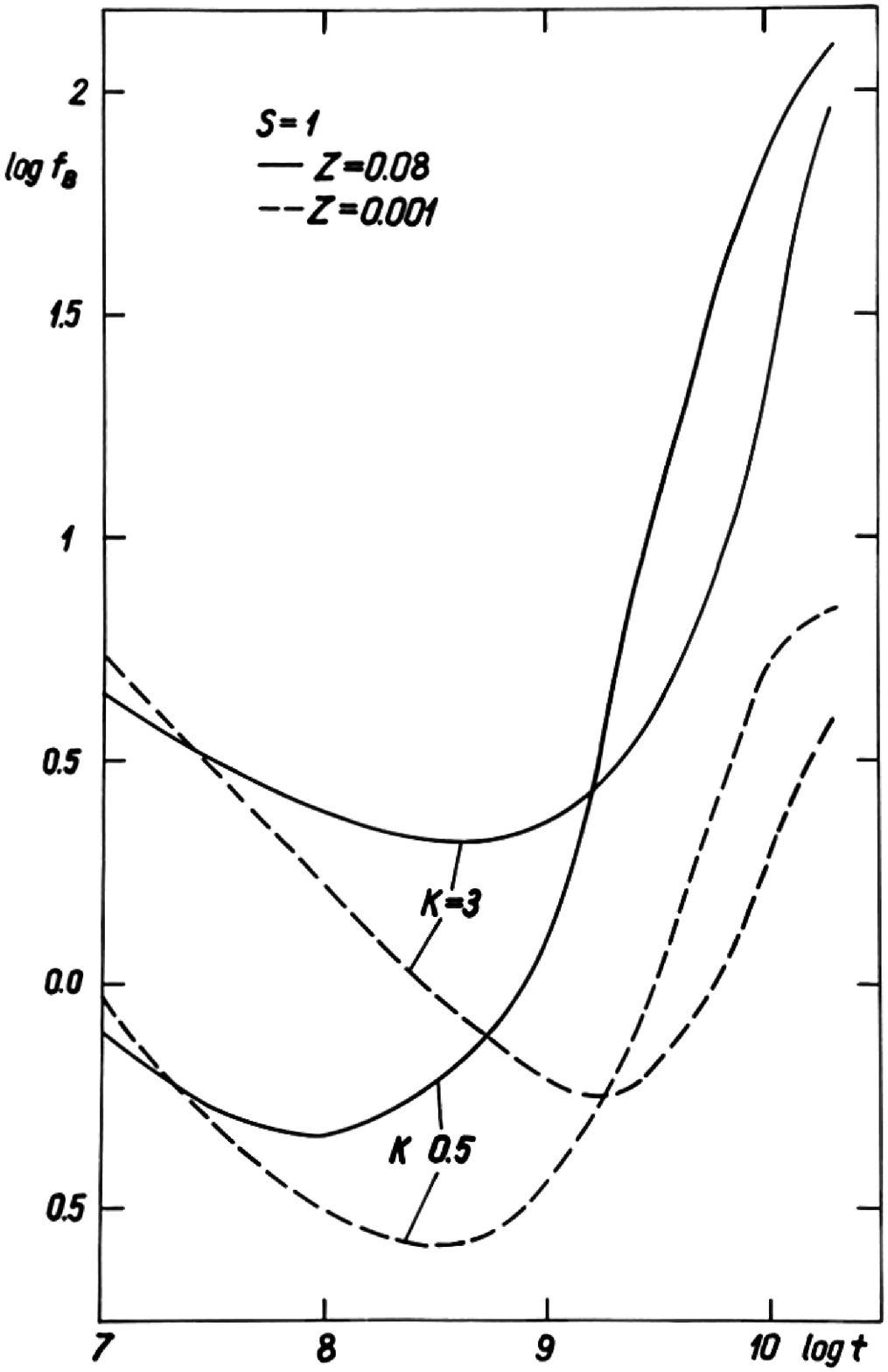}}
\resizebox{0.32\textwidth}{!}{\includegraphics*{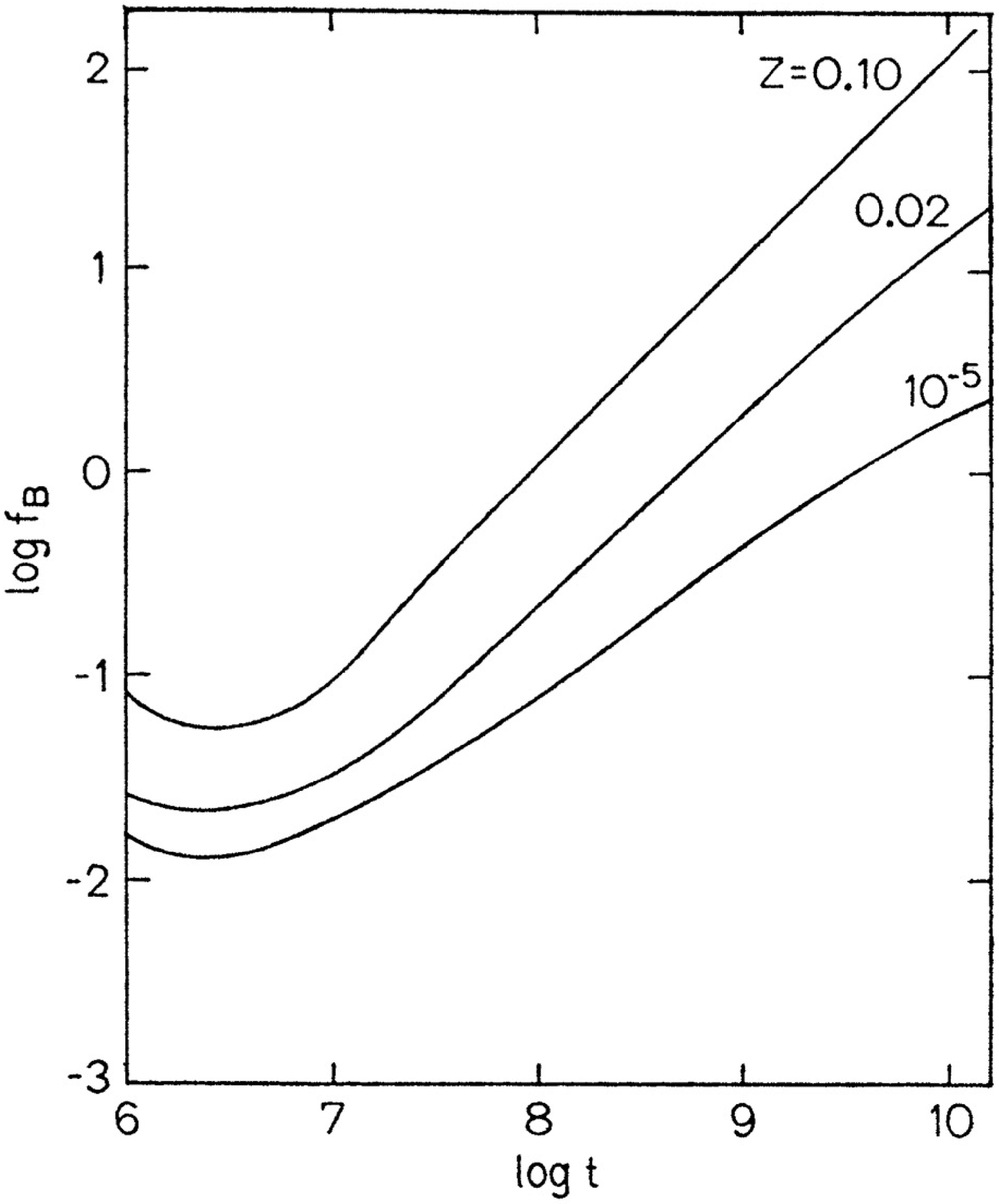}}
\caption{ {\em Left:}
Mass-to-light ratio of galaxies, $f_B=M/L_B$, as 
  function of the age $t$ for various parameters of the star formation
  function.   In {\em central and right panels} data are given for extremal values of the
  metal content $Z$. } 
  \label{Fig22.5}
\end{figure*} 
}

In very young star systems, stars of high luminosity are dominating,
there exist also pulsars – supernova remnants. In young systems low
mass stars are still in the stage of gravitational contraction, their
contribution to the luminosity of the system, is surprisingly
high. With increasing age of the system, the luminosity of red dwarfs
decreases, the number of pulsars increases, and first white dwarfs
appear.  All these developments increase the number of stars of low
luminosity.  During the whole existence of our Galaxy, the luminosity
of some types of stars has decreased by 38 magnitudes.

{\bf B. Bolometric luminosities}.
\citet{Tinsley:1968} calculated models of the evolution of galaxies in
the age interval 1 to 12 billion years with step 1 billion years.  Our
models cover a much larger age interval that allows to follow the
early phase of the evolution of galaxies.  In the early stage the
luminosity of galaxies increases very rapidly, but this stage is very
short.  In our models we used a constant density of matter during the
evolution.  In real galaxies in the early stage of the evolution the
density was smaller, and galaxies contracted during the evolution.
The contraction of the central region is fast, only a few millions of
years. For this reason, the overall change of the luminosity depends
only slightly on the variability of the density.

It is interesting to note that the maximal luminosity of the galaxy
was approximately a hundred times higher than in the present epoch (for a
galaxy with constant mass). 

{\bf C. Mass-to-light ratios}.
The dependence of the mass-to-light function on chemical
composition and star formation rate parameters is in good agreement
with expectations, see Fig.~\ref{Fig22.5}. The general diapason of the function $f=M/L$ of
old populations of different composition allows to
explain differences of this function for globular clusters and
elliptical galaxies of various mass.  Variations needed to explain
these very different populations can be obtained by changing
parameters of star formation function and respective evolutionary
tracks in reasonable limits.

{\bf D. Colours of model galaxies and star clusters}.  The dependence
of $U - B$ and $B - V$ colours of model galaxies on the age is shown
in Fig.~\ref{Fig22.7}, and in a two-colour diagram in
Fig.~\ref{Fig22.8}. To compare our results with observations, we
compiled similar data for real clusters. Stars of a given cluster have
identical chemical composition and age, and can be used to check
models of evolution of stars of different mass. Similarly, star
clusters can be used to check evolutions of whole stellar systems. For
this purpose, instead of individual stars we have to compare integrated
parameters of model clusters with integrated parameters with real
clusters.  We show in Fig.~\ref{Fig22.4} colour-magnitude diagrams for
selected star clusters of various ages and chemical composition.

{\begin{table*}[h] 
\caption{Data on clusters} 
{\centering 
\hspace{2mm}
\resizebox{0.65\textwidth}{!}{\includegraphics*{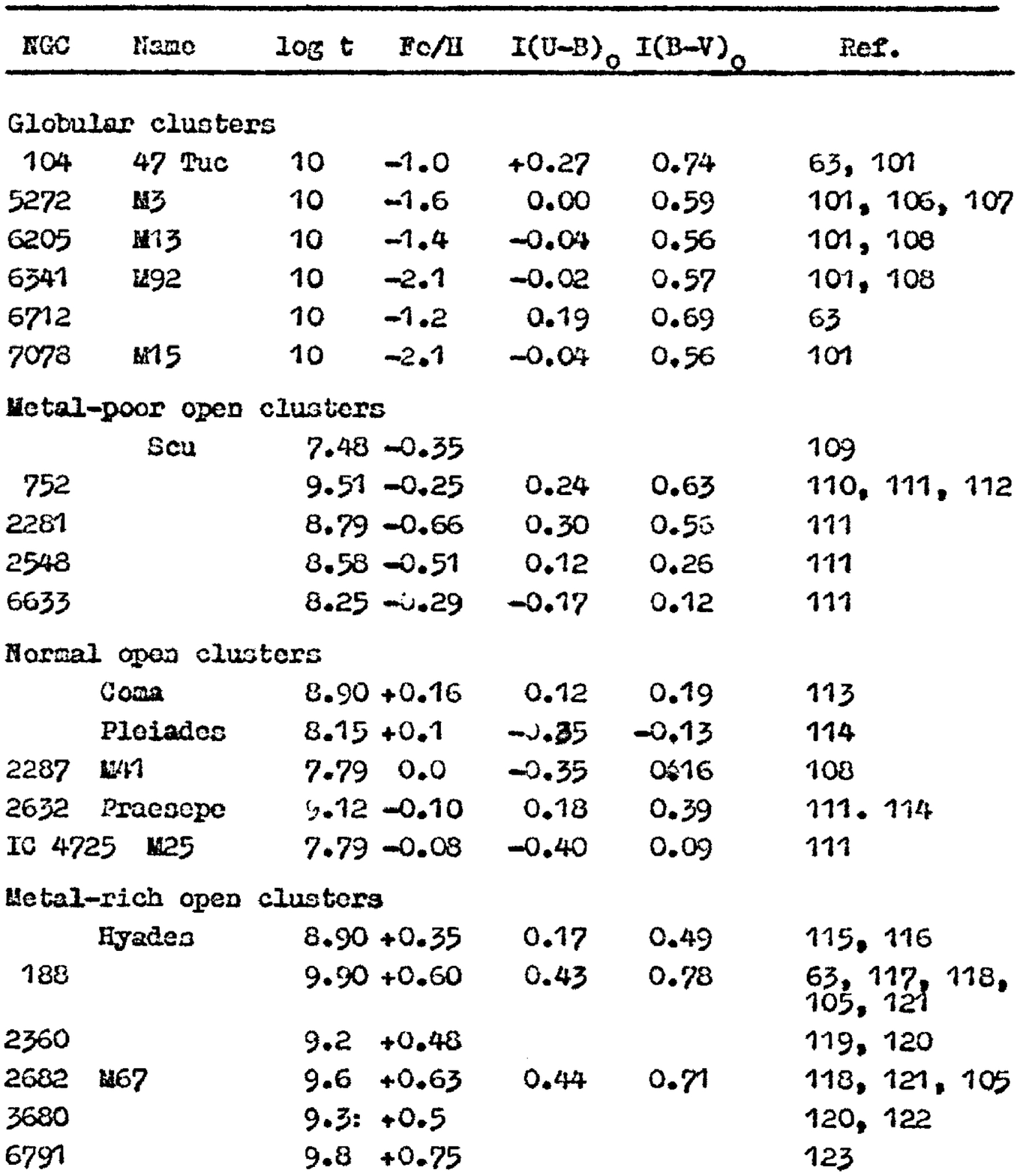}}
\label{Tab22.4}
\\
}
{\small
  References are: 63 \citet{Newell:1969aa}, 101
  \citet{Sandage:1970ab}, 105 \citet{Arp:1964}, 106
  \citet{Strom:1970aa}, 107 \citet{Sandage:1969ab}, 108
  \citet{Helfer:1959aa}, 109 \citet{Eggen:1970aa}, 110
  \citet{Gunn:1963aa}, 111 \citet{Wallerstein:1964aa}, 112
  \citet{Philip:1970aa}, 113 \citet{Nissen:1970ab}, 114
  \citet{Conti:1968aa}, 115 \citet{Nissen:1970aa}, 116
  \citet{Alexander:1967aa}, 117 \citet{Demarque:1969aa}, 118
  \citet{Aizenman:1969aa}, 119 \citet{Eggen:1968aa}, 120
  \citet{Demarque:1969ab}, 121 \citet{Spinrad:1970aa}, 122
  \citet{Eggen:1969aa}, 123 \citet{Spinrad:1971aa}.  
 } 
\end{table*} 
}

We compiled a list of star clusters with known integrated intrinsic
colours, $I(U - B)_0$ and $I(B-V)_0$. Our compilation is given in
Table~\ref{Tab22.4}. The age of clusters was found using the turn-off
point of main sequence, as seen in Fig.~\ref{Fig22.1}. For a number of
clusters, it was possible to find estimates of the metal content. We
give in the Table mean metal content as follows:
\be
[Fe/H] = (\log\,Fe/H)_{star} - (\log\,Fe/H)_\odot.
\ee
As usual, we accept the position that the iron content
characterises the content of all heavy elements. The metallicity 
parameter $Z$ is related with the $He/H$ parameter as follows:
\be
\log\,Z/Z_\odot = [Fe/H],
\ee
where we take $Z_\odot = 0.02$ according to \citet{Sandage:1969ac}.

{\begin{table*}[h] 
\caption{Mean metallicity and colour indexes } 
\centering 
\hspace{2mm}
\resizebox{0.50\textwidth}{!}{\includegraphics*{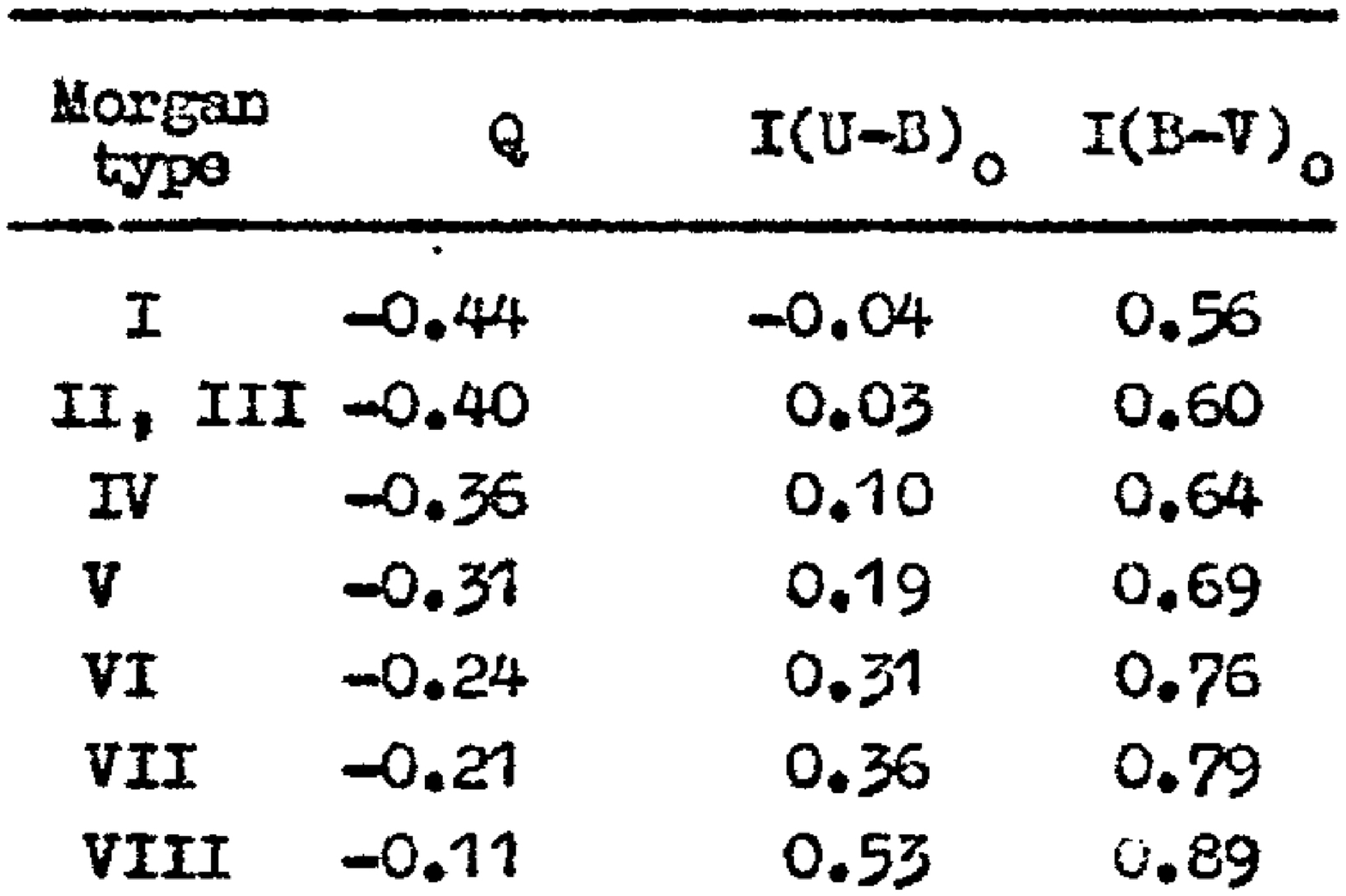}}
\label{Tab22.5}
\end{table*} 
}

{\begin{figure*}[h] 
\centering 
\resizebox{0.49\textwidth}{!}{\includegraphics*{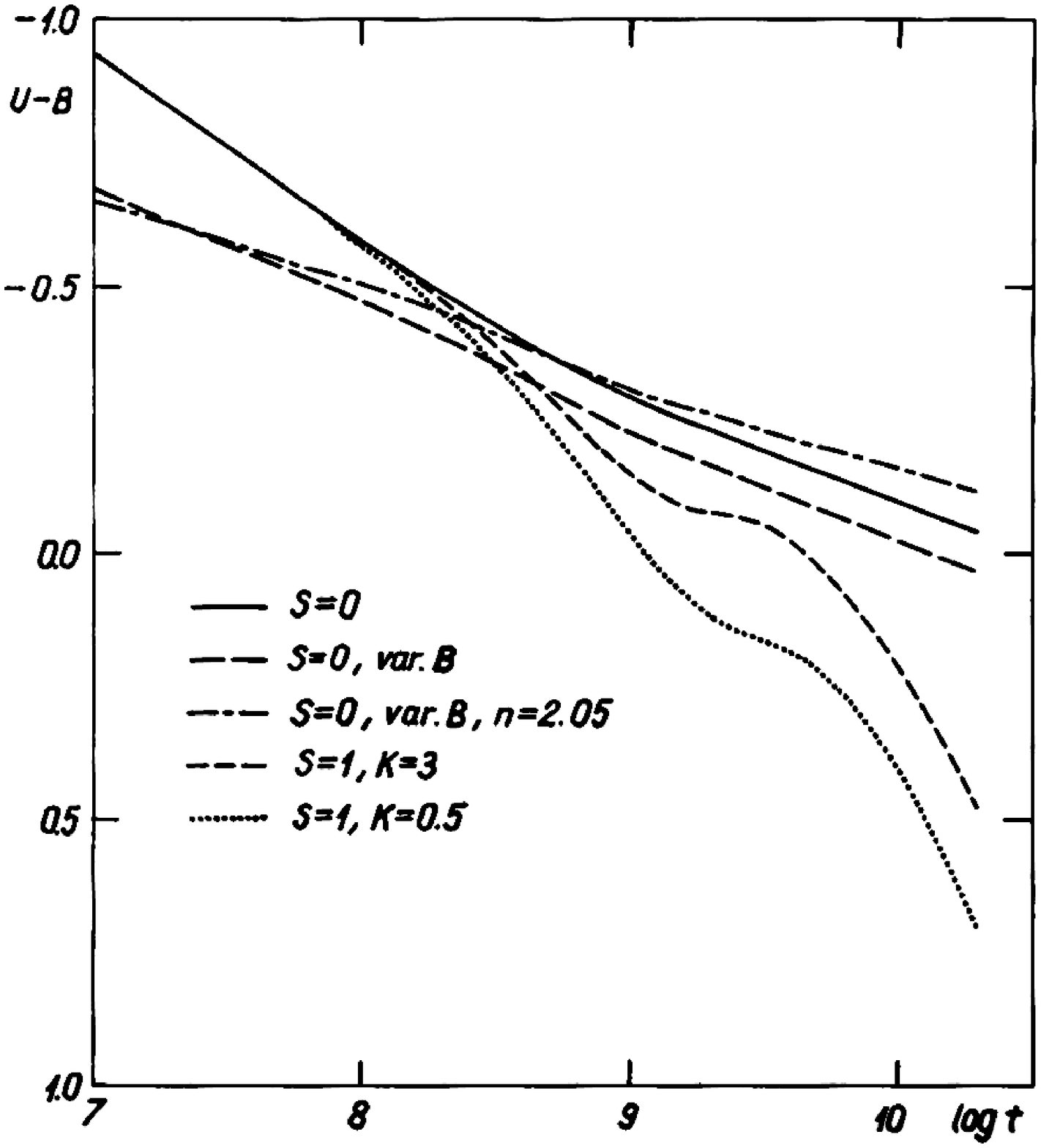}}
\resizebox{0.49\textwidth}{!}{\includegraphics*{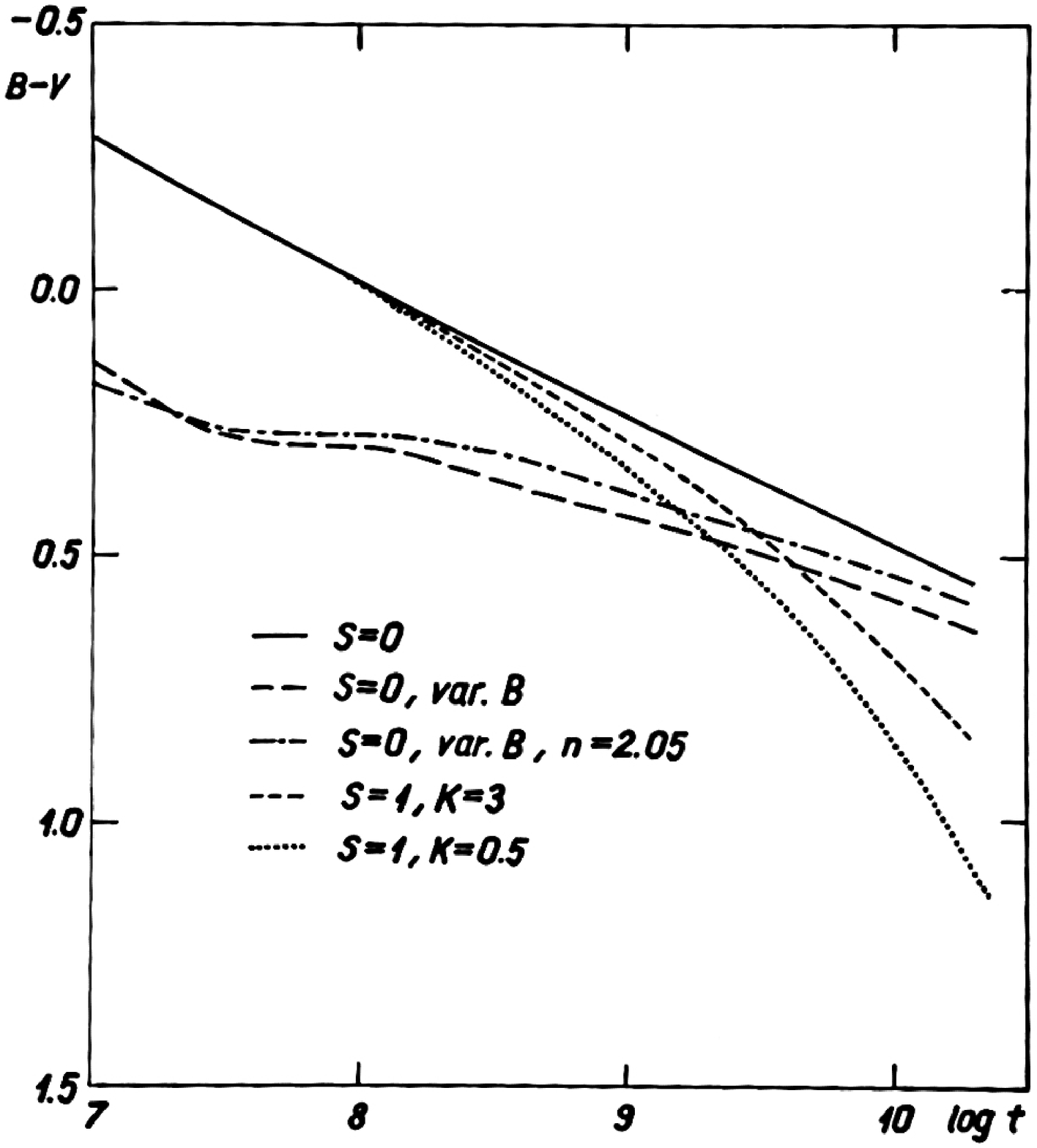}}
\caption{The dependence of colour  $U - B$ (left) and $B - V$ (right) of
model galaxies on the age for various star formation rate parameters. } 
  \label{Fig22.7}
\end{figure*} 
}

{\begin{figure*}[h] 
\centering 
\resizebox{0.40\textwidth}{!}{\includegraphics*{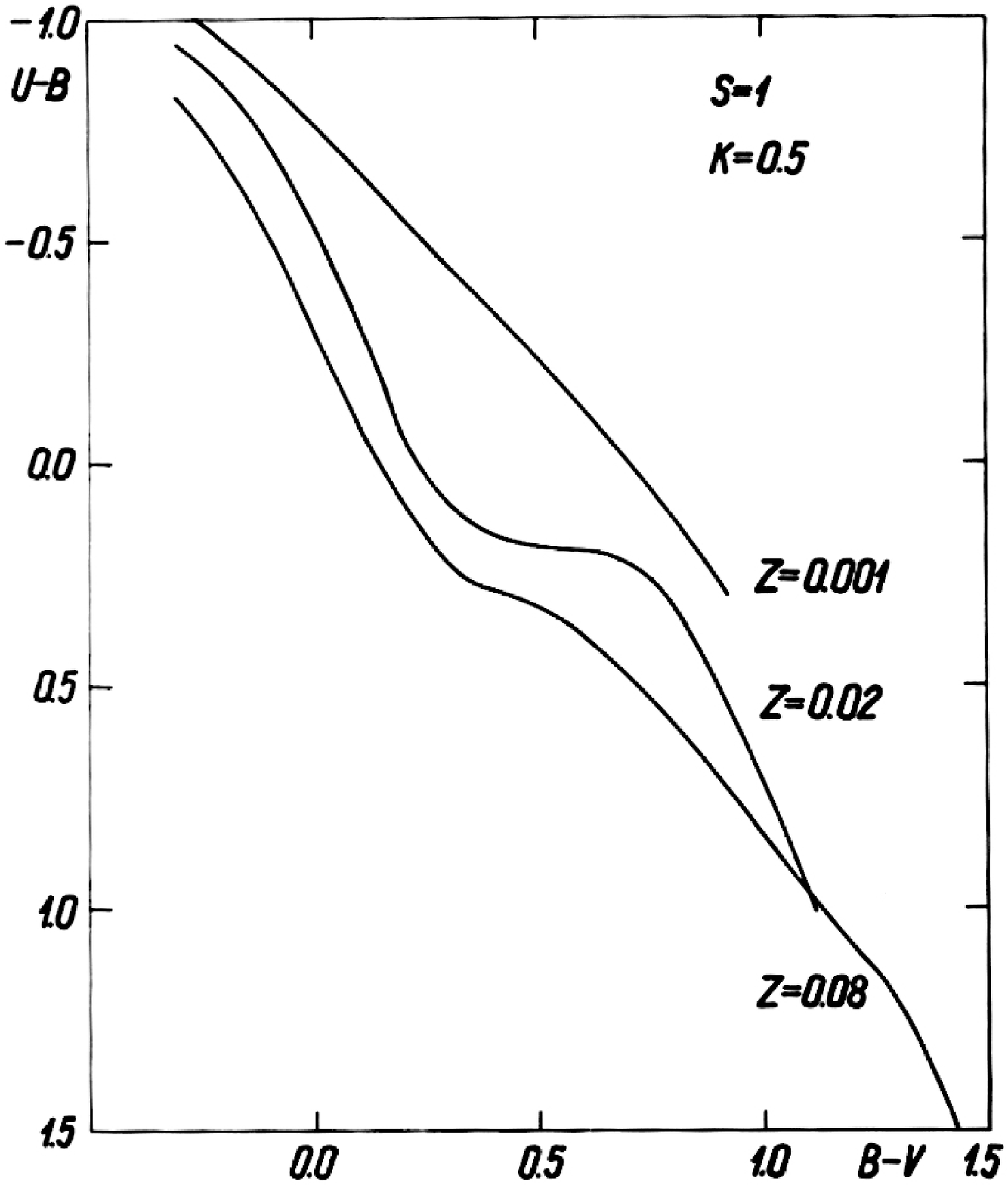}}
\resizebox{0.55\textwidth}{!}{\includegraphics*{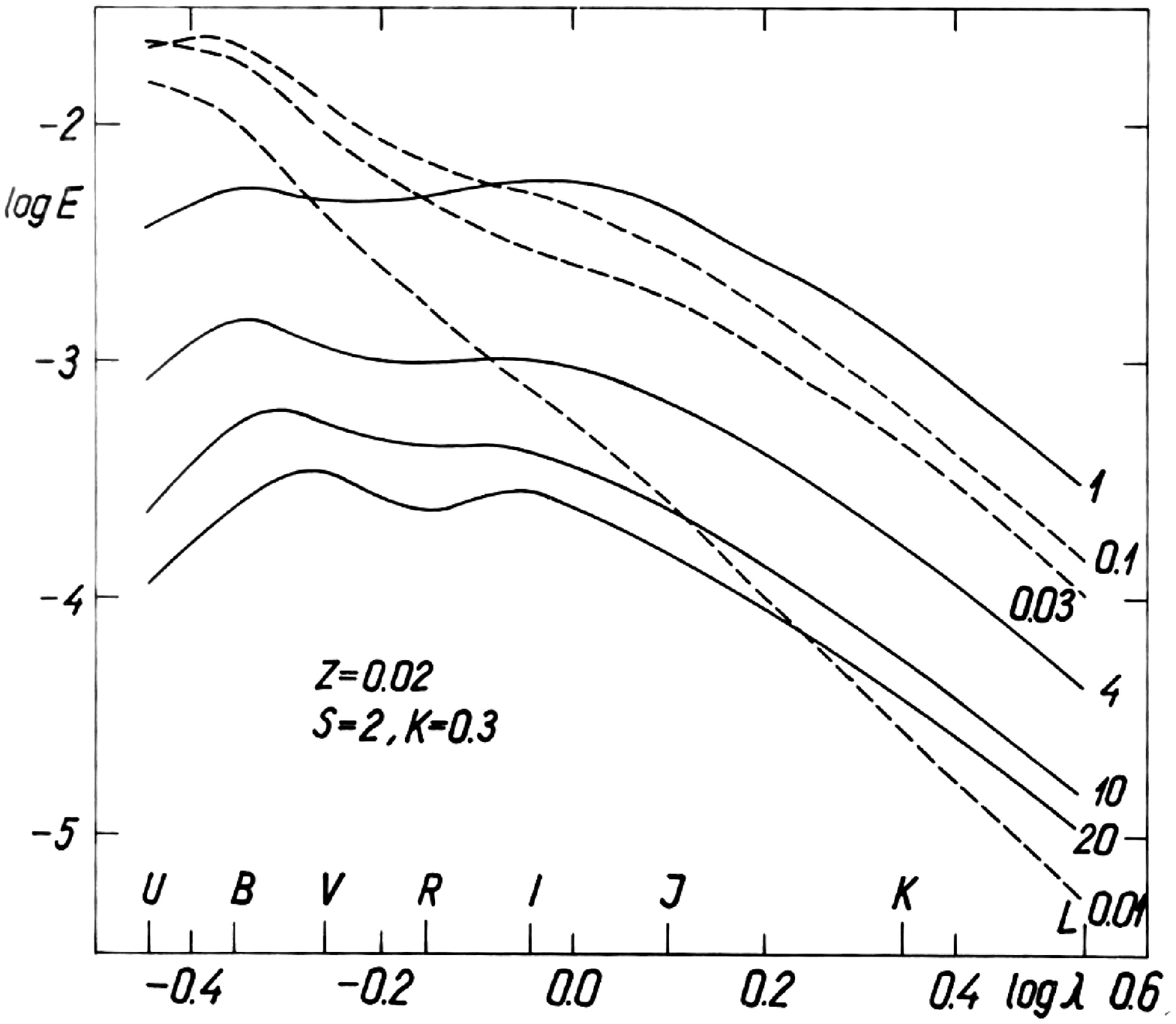}}
\caption{{\em Left:} Evolution tracks of model galaxies of different
  chemical composition in a two-colour diagram. {\em Right:} The energy
  distribution in spectra of a model galaxy of mass $10^{11}\,M_\odot$
  at different age, given in billion years. Wavelength is shown in
  microns and in watts per cm$^{-2}$~$\mu^{-1}$.   }
  \label{Fig22.8}
\end{figure*} 
}

{\begin{figure*}[h] 
\centering 
\resizebox{0.42\textwidth}{!}{\includegraphics*{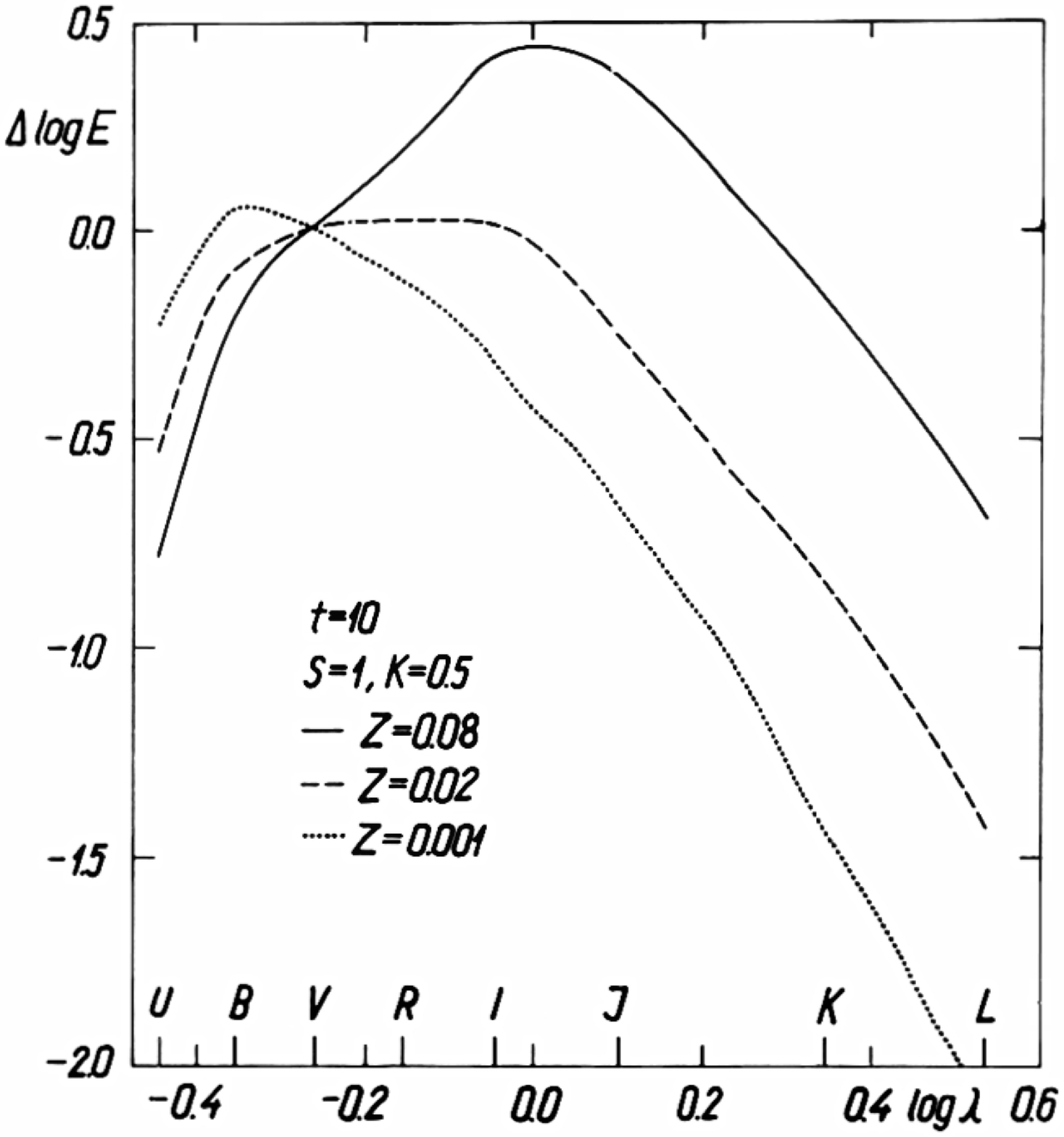}}
\resizebox{0.56\textwidth}{!}{\includegraphics*{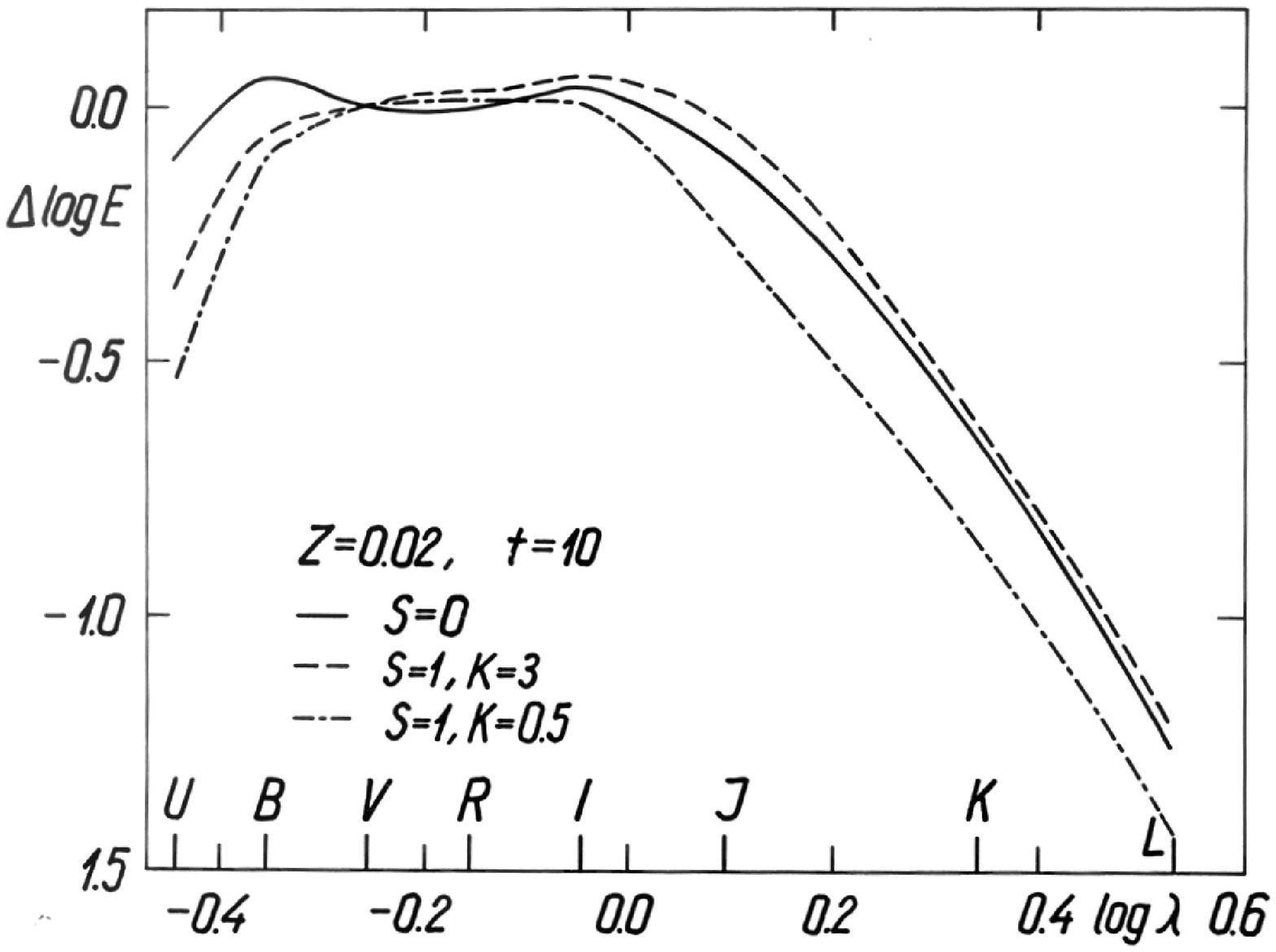}}
\caption{Energy distributions in model galaxies of identical age,
  $t=10\times 10^9$ years, but for  different
  composition  (left), and  for different
parameters of star formation function (right).} 
  \label{Fig22.10}
\end{figure*} 
}

{\begin{figure*}[ht] 
\centering 
\resizebox{0.42\textwidth}{!}{\includegraphics*{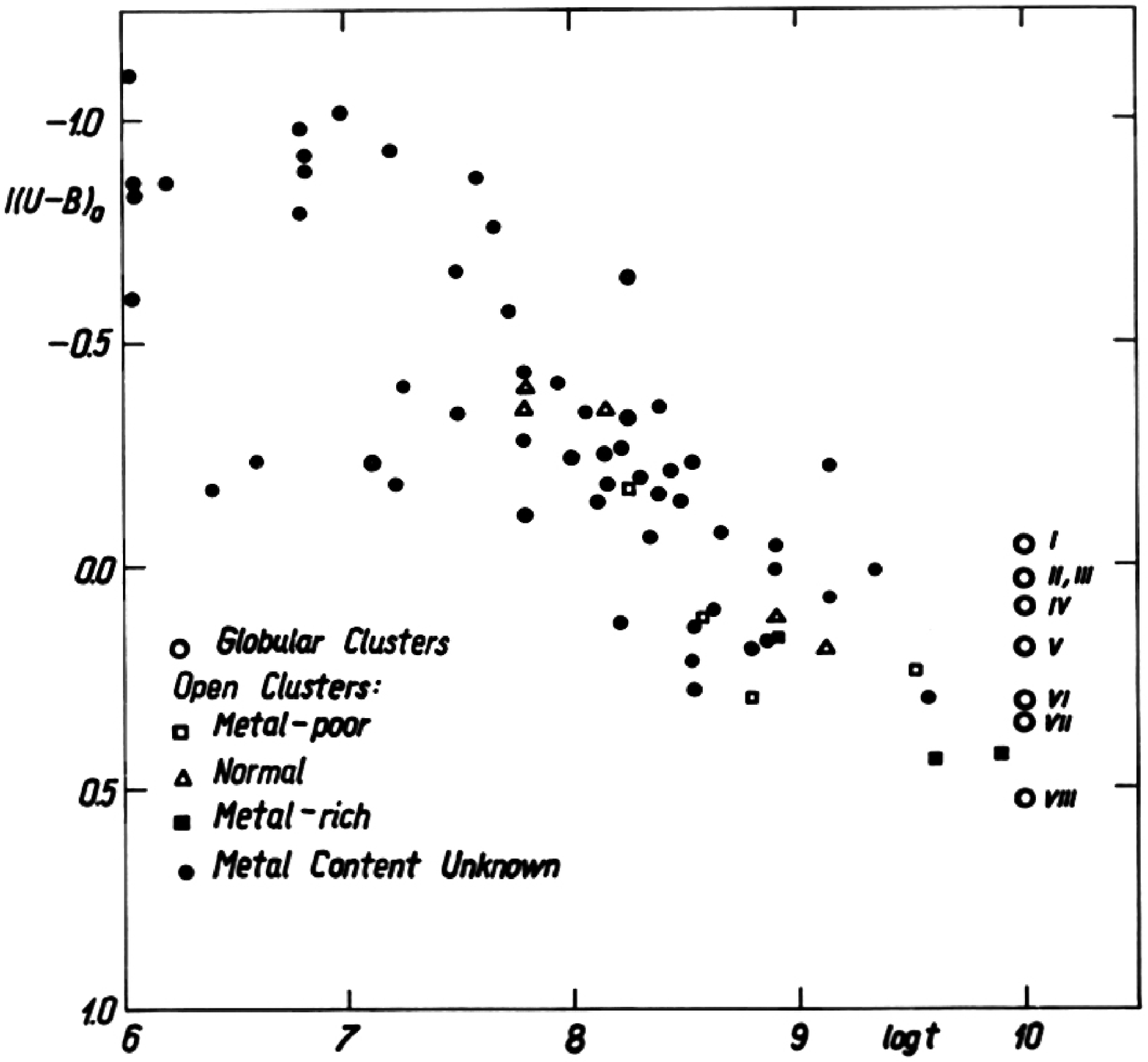}}
\resizebox{0.56\textwidth}{!}{\includegraphics*{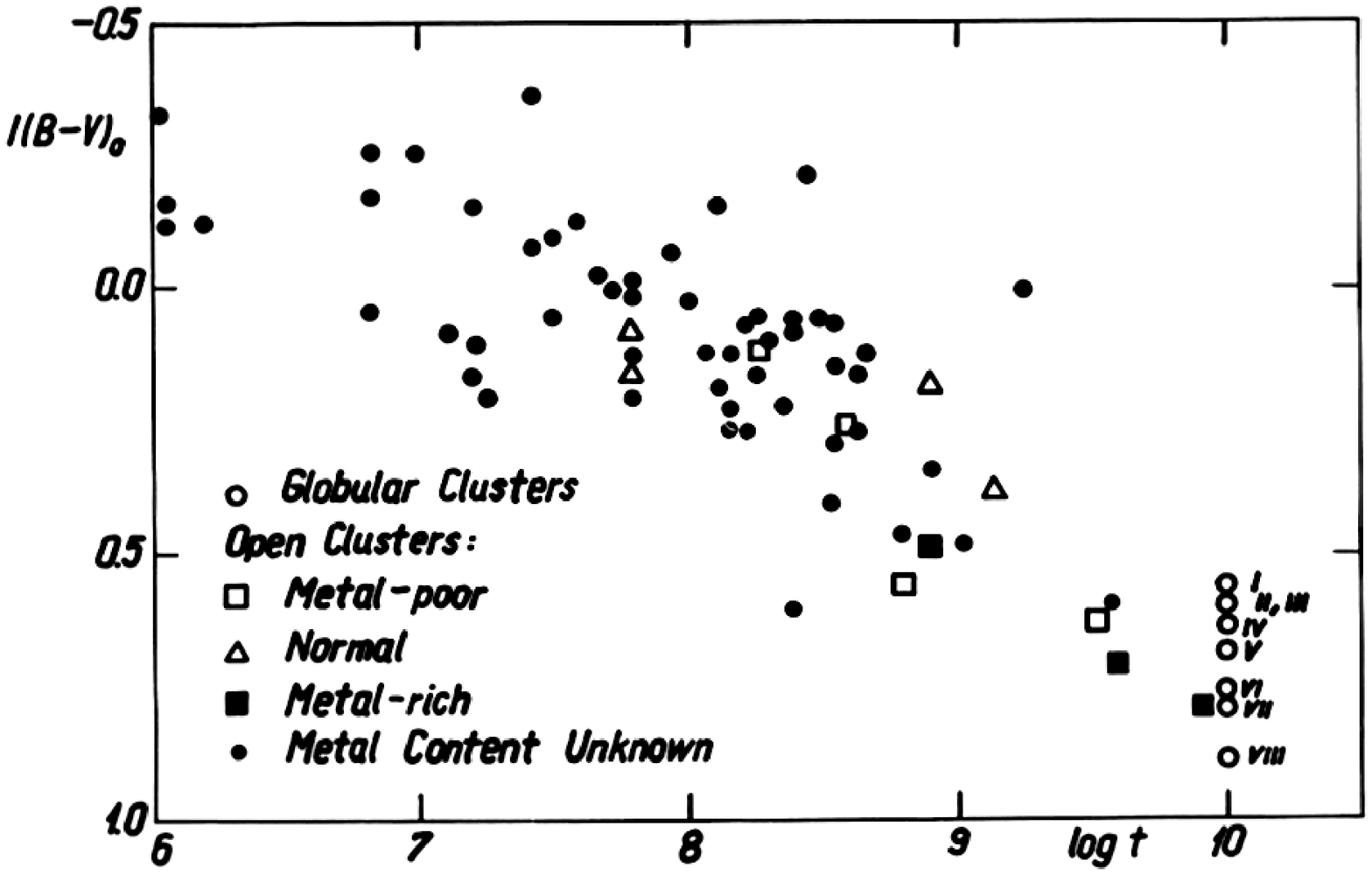}}
\caption{Age dependence of integral colours of star clusters of
  various ages: 
    $I(U-B)_0$ (left) and  $I(B-V)_o$ (right). Globular clusters are 
divided into groups according to Morgan metallicity index.} 
  \label{Fig22.12}
\end{figure*} 
}

The metal content in globular star clusters changes in rather large
limits. There are too few quantitative data on $[Fe/H]$  to find a
reliable relation between $[Fe/H]$ and the parameter
\be
Q=(U-B)-0.72\,(B-V),
\label{eq22.8.3}
\ee
which is often used as the metallicity index
\citep{van-den-Bergh:1967aa}. There are more quantitative data on the
\citet{Morgan:1959aa} metallicity type and the $Q$ parameter. Using available data
we found mean relations between Morgan type, metallicity index
$Q$, and integrated colours of clusters, the results are given in
Table~\ref{Tab22.5}. Here we give integrated colours, corrected for
interstellar reddening, $I(U-B)_0$ and $I(B-V)_0$, calculated using
Eq.~(\ref{eq22.8.3}).

Integrated colours are shown in Fig.~\ref{Fig22.12}. As
we see, theoretical relations between the colours are in good agreement with
observed ones.

{\bf E. Energy distribution} in spectra of model galaxies of age
$t=10\times 10^9$ years is shown in Fig.~\ref{Fig22.10} for different
compositions and parameters of the star formation function.  The
distribution is rather similar with the distribution found by
\citet{Tinsley:1968}. Model distributions can be compared with the
energy distribution at central regions of elliptical galaxies by
\citet{Sandage:1969aa} and \citet{Johnson:1966ab}.

\section{Conclusions}

We can summarise our results as follows.  By choosing in reasonable
limits parameters of star formation function, it is possible to calculate
model galaxies with integrated characteristics, similar to integrated
characteristics of star clusters. Integral parameters, including
colours and spectral energy distributions, depend on the age, chemical
compositions, and rate of star formation. Model galaxies of identical
colour can have ages which differ by an order of magnitude, see
Figs.~\ref{Fig22.5} and \ref{Fig22.10}. Our results support the result by
\citet{Sandage:1963aa} that blue colour of galaxies is not always an
argument to the small age of the system. Our data show that the view
by \citet{Tinsley:1968} that all galaxies have identical ages can be
distorted by the non-optimal choice of parameters of the star
formation function.

\vskip 5mm
\hfill August 1971
\chapter{Star formation function and galactic populations
  \label{ch23}}
 
\section{Introduction}

In earlier Chapters we discussed the reconstruction of the dynamical
and physical evolution of the Galaxy. The rate of star formation was
discussed in a general form. Calculations were done for a wide range
of parameters of the star formation function.

In the present Chapter we shall discuss the star formation function in
more detail. We shall find numerical values of parameters of the
function and estimate the dependence of this function on the chemical
composition of the interstellar gas. Also, we discuss on the basis
of star formation function how stellar populations of various age and
composition of the Galaxy could be formed.

An essential parameter of the Galaxy is its age. We start our
discussion with an overview of determinations of the age of the Galaxy.

\section{The age of the Galaxy}

We count the age of galaxies since the moment when star formation in
its protogalaxy started.  According to the present understanding on
the evolution of the Universe, most galaxies formed simultaneously when
fluctuations in the expanding primeval cooling matter density were
strong enough to start star formation in the densest regions, which
became nuclei of forming galaxies.

There exist today at least three independent methods to derive ages of
galaxies. The first is based on cosmological considerations: it is
clear that ages of galaxies are smaller than the age of the Universe. The
age of the Universe can be expressed as follows:
\be
T_U = \alpha(q_0)\, H^{-1},
\label{eq23.2.1}
\ee
where $H$ is the Hubble parameter (in the present epoch), and
$\alpha(q_0)$ is a dimensionless coefficient, its value depends
on the cosmic acceleration parameter $q_0$ (and of cosmological
density parameters). In Tabel~\ref{Tab23.1} we give some recent
determinations of the Hubble constant, and respective ages of the
Universe for three values of the acceleration parameter $q_0$.  If we
accept for the cosmological constant a value $\Lambda =0$, and for the
acceleration parameter values $q_0 =0.5$ and $q_o=1.5$
\citep{Sandage:1961aa, Peach:1970aa}, we get the age of the Universe
$T_U$ values given in Table~\ref{Tab23.1}. For comparison we give also
the age, corresponding to the acceleration parameter $q_0 =0$.

{\begin{table*}[h]
    {\small
      \centering    
\caption{} 
\begin{tabular}{ccccl}
\hline  \hline
  $H$~(km~s$^{-1}$~Mpc$^{-1}$)& & $T_U$ ($10^9$\,years)& & References\\
      & $q_0=0$& $q_0=0.5$& $q_0 = 1.5$& \\
  \hline
  95 & 10.5 &  6.9 & 5.3 & \citet{van-den-Bergh:1970aa}\\
  75 & 13.0 &  8.7 & 6.7 & \citet{Sandage:1968aa}\\
  47 & 20.8 & 13.9 &10.7 & \citet{Abell:1968aa}\\
  \hline
\label{Tab23.1}   
\end{tabular}
}
\end{table*} 
} 

We see that the accuracy of present determinations of parameters $H$
and $q_0$ is not sufficient to find the age of the Universe.  The
uncertainty is even larger when we take into account the possible
error in our assumption $\Lambda =0$. Most often it is assumed that
the Hubble constant is $H= 75$~km~s$^{-1}$~Mpc$^{-1}$, and the acceleration
parameter $q_0=0.5$, which yields $T_U=8.7 \pm 1.5\times\,10^9$
years. 

Another method to estimate the age of our Galaxy is to use
determinations of ages of its oldest halo populations.  This method
can be used for globular star clusters.

Earlier evolutionary tracks in Herzsprung-Russel diagram for globular
clusters were calculated using the assumption that the helium content
is very low, which yields to ages of the order $15 - 20 \times
10^9$~years \citep{Schwarzschild:1958kx}, exceeding the age of the
Universe $T_U$. Recent data suggest that the helium content in
globular clusters is the same as in stars of population I
\citep{Rood:1968aa,Sandage:1969aa}. This shows that the helium is of
primordial origin. New evolutionary tracks are in good harmony with
estimates of the age of the Universe from other sources. Table
\ref{Tab23.2} gives a summary of recent determinations of ages of
globular clusters. Different determinations vary in reasonable
limits. According to \citet{Sandage:1970ab} the probable age of
globular clusters is $T=10 \pm 0.8 \times 10^9$ years.

{\begin{table*}[h]
\centering    
\caption{} 
\begin{tabular}{cl}
\hline  \hline
  $T$ ($10^9$\,years) & References\\
  \hline
  8.5            & \citet{Rood:1968aa}\\
  $11 - 13$ &\citet{Iben:1970ab}\\
  9.6           &  \citet{Sandage:1970ab}\\
  11.6         &  \citet{Sandage:1970ab}\\
  \hline
\label{Tab23.2}   
\end{tabular} 
\end{table*} 
} 

The most accurate age estimates of the Galaxy come with the radiative
isotope method. The idea to use this method to determine the age of
the Solar system was expressed by \citet{Burbidge:1957aa}.  Initially
only the isotope $U^{235}$ was used, which gave for the age of the
Galaxy $T=6.6\times\,10^9$~years, if the uranium was formed rapidly in
the early phase of the history of the Galaxy.  If the uranium was formed
with constant speed, then the isotope age of the Galaxy is $T=11.5 -
18 \times\,10^9$ years.

A review of determinations of the age of the Galaxy with the isotope
method was given by \citet{Dicke:1969aa}.  We do not use older
determinations of the isotope age of the Galaxy, since the number of
isotopes used was too small, and errors in determinations of ages of
meteorites were large. 

The number of isotopes used in age determinations is increasing and
the accuracy in atomic parameters is improving, thus results are now more
accurate.  The theory of the method is given by
\citet{Schramm:1970aa}. The age estimates depend on the speed of the
formation of radioactive elements. If we suppose that most radioactive
elements were formed rapidly, then the age of the Galaxy is $9\times
10^9$ years. If isotopes were formed with constant speed, then the age
of the Galaxy increases by a factor of 1.5. This second possibility,
however, is excluded by other data \citep{Unsold:1969aa}. For this
reason we accept the first alternative.  Most recent data on isotope
ages are summarised in Table~\ref{Tab23.3}.  

{\begin{table*}[h]
\centering    
\caption{} 
\begin{tabular}{cl}
\hline  \hline
  $T$ ($10^9$\,years)& References\\
  \hline
  $8.7 \pm 0.7$  & \citet{Hohenberg:1969aa}\\
  $9.7 \pm 1.0$ & \citet{Wasserburg:1969aa}\\
  \hline
\label{Tab23.3}   
\end{tabular} 
\end{table*} 
} 

The mean value obtained with the isotope method is, $T=9.0 \pm 0.5
\times\,10^9$ years.  This age can be attributed to the disc of the
Galaxy, which according to our estimates is at the speed of the
contraction of the Galaxy by $0.5\times\,10^9$ years younger than the
Galaxy.  Thus we get for the whole Galaxy $T = 9.5 \pm 0.7\,\times
10^9$ years.

When we use results of all three methods, we get for the age of the
Galaxy $T = 9.5 \pm 0.75\,\times 10^9$ years.  Instead, we shall use in
further calculations a round value  $T = 10 \times 10^9$ years. If
$q_0 = 0.5$ and $\Lambda =0$, then this age corresponds to the Hubble
constant $H=65$~km~s$^{-1}$~Mpc$^{-1}$.

\section{Rate of star formation}

The local rate of star formation can be expressed as follows
\citep{Schmidt:1959aa,Schmidt:1963aa}: 
\be
R_l = {\dd{\rho_s} \over \dd{t}} = \gamma \rho_g^2,
\label{eq23.3.1}
\ee
where $\rho_s$ and $\rho_g$ are densities of stars and gas,
respectively, $\gamma$ is a coefficient with dimension (density x
time)$^{-1}$, and we accepted the  star formation parameter,
$S=2$. Since $\dd{\rho_g} = - \dd{\rho_s}$, then we can write
Eq.\,(\ref{eq23.3.1}) as follows:
\be
-{\dd{\rho_g} \over \rho_g^2} = \gamma\,\dd{t}.
\label{eq23.3.2}
\ee
Let us assume that the full density of matter in the given element of
space is constant and that at the initial time $t=0$ the whole matter
was in a gaseous form.  After integration we get:
\be
\rho_g = {\rho \over 1+ \tau},
\label{eq23.3.3}
\ee
where
\be
\tau = t/K
\label{eq23.3.4}
\ee
is the dimensionless time,  and the characteristic time is
\be
K={1 \over \gamma\,\rho}.
\label{eq23.3.5}
\ee
Using
(\ref{eq23.3.3}) we can write the equation (\ref{eq23.3.1}) as follows:
\be
{\dd{\rho_s} \over \dd{t}} = \gamma\,{\rho^2 \over
  (1+\gamma\,\rho\,t)^2}= {\rho \over K}\,{1 \over (1+\tau)^2}.
  \label{eq23.3.6}
\ee

Integrating Eq.\,(\ref{eq23.3.1}) along the line of sight we get in a
similar way
\be
P_g = {P \over 1+\tau}
\label{eq23.3.7}
\ee
and
\be
{\dd{P_s} \over \dd{t}} = \kappa\,{P^2 \over (1+\kappa\,P\,t)^2} = {P
  \over K}{1 \over (1+\tau)^2},
\label{eq23.3.8}
\ee
where $P_g$ and $P_s$ are projected densities of gas and stars, $P =
P_g +P_s$,
\be
\kappa ={\gamma \over 2\,\zeta^*},
\label{eq23.3.9}
\ee
and
\be
K= {1 \over \gamma \bar{\rho}}.
\label{eq23.3.10}
  \ee
  In the last equation
  \be
  \bar{\rho} = {P \over 2\zeta^*}
  \label{eq23.3.11}
  \ee
  and
  \be
  \zeta^* = {P_g \over 2\bar{\rho_g}},
  \label{eq23.3.12}
    \ee
    where
    \be
    \bar{\rho_g}= {\int\,\rho_g\,\dd{P_g} \over P_g},
    \label{eq23.3.13}
    \ee
    and integration is over the line of sight. If $\rho_g$ has normal
    distribution, then
    \be
    \zeta^*=\sqrt{\pi}\,\zeta,
    \label{eq23.3.14}
    \ee
    where $\zeta$ is the dispersion of positions of gas particles
    along the line of sight.

    We use now the modified exponential density profile
    \be
    \rho(a) = h \exp\left[x_0 - \left[x_0^{2N} + \left({a \over k\,a_0}\right)^2\right]^{1/(2N)}\right],
    \label{eq23.N}
    \ee
where $h$ and $k$ are normalising constants, $x_0$ and $N$ are
structural parameters, and $a_0$ is the major semiaxis of the
equidensity ellipsoid. The parameter $x_0$ regulates the density law
near the centre of the ellipsoid. If we take $x_0=0$, we get the
usual exponential density profile
\be
\rho(a)= h \exp\left[-\left({a \over k\,a_0}\right)^{1/N}\right].
\label{eq23.N2}
\ee

If the line of sight is directed along the rotation axis, we get for
large distance $R$
\be
\zeta = \epsilon \sqrt{N} R \left( {k\,a_0 \over R}\right)^{1/(2N)},
  \label{eq23.3.15}
  \ee
  and for the centre, $R=0$
  \be
  \zeta = \epsilon \sqrt{N}\,k\,a_0 x_0^{N-1 \over 2}.
  \label{eq23.3.15B}
  \ee
  
Integrating over the whole volume of the system we get
  \be
  \mm{M}_g = {\mm{M} \over 1 + \tau},
  \label{eq23.3.16}
  \ee
  and for the rate of star formation
  \be
  R={\dd{\mm{M}_s} \over \dd{t}} = \lambda {\mm{M}^2 \over
      (1 +\lambda\,\mm{M}\,t)^2} = {\mm{M} \over K} {1 \over
        (1+\tau)^2},
      \label{eq23.3.17}
      \ee
      where $\mm{M}_g$ and $\mm{M}_s$ are full masses of gas
      and star in the galaxy, and
      \be
      \lambda = \gamma/V^*,
      \label{eq23.3.18}
      \ee
      and $K$ is expressed by Eq.~\ref{eq23.3.10}. For the mean density
      we get now
      \be
      \bar{\rho} = {\mm{M} \over V^*},
      \label{eq23.3.19}
      \ee
      where
      \be
      V^*={\mm{M}_g \over \bar{\rho}_g},
      \label{eq23.3.20}
      \ee
      and
      \be
      \bar{\rho}_g = {\int\,\rho_g\,\dd{\mm{M}_g} \over
        \mm{M}_g}.
      \label{eq23.3.21}
      \ee

      Now we introduce standardised density and mass functions and get
      for the mean density
      \be
      \bar{\rho} = {1 \over 4\pi\,\epsilon} {\mm{M} \over
        a_0^3}\,\chi,
      \label{eq23.3.22}
      \ee
      where
      \be
      \chi=\int_0^\infty\left({\mu_0 \over
          \alpha}\right)^2\,\dd{\alpha},
      \label{eq23.3.23}
      \ee
      and $\epsilon$, $a_0$, and the standardised mass function $\mu_0$ describe
       the gas population, and only $\mm{M}$ is the whole mass
      of the galaxy.  Here we used the designation $\alpha = a/a_0$,
      where $a$ is the major semiaxis of the density ellipsoid, and
      $\mu(a)\,\dd{a}= 4\pi\,\epsilon\,a^2\,\rho(a)\,\dd{a}$ is the mass function --
      the mass of an ellipsoidal sheet of thickness $\dd{a}$ and ratio
      of vertical to horizontal axes $\epsilon$. 
      If the gas is distributed according to the
      exponential law, then we get
      \be
      \chi = {N\,h^2\,k^3 \over 2^{3N}}\,\Gamma(3N).
      \label{eq23.3.24}
    \ee
    For a series of $N$ values the function $\chi$ is given in
    Table~\ref{Tab23.4}.

{\begin{table*}[h]
\centering    
\caption{} 
\begin{tabular}{lc}
\hline  \hline
 $N$& $\chi$\\
  \hline
  0.5 & 0.5555\\
  1    & 0.5000\\
  2    & 0.7146\\
  3    & 0.6124\\
  4    & 0.7596\\
  5    & 0.9708\\
  6    & 1.2644\\
  \hline
\label{Tab23.4}   
\end{tabular} 
\end{table*} 
}

\section{Determination of the  parameter $\gamma$}

Direct observations allow to determine the full mass of the galaxy,
$\mm{M}$, the mass of the gas, $\mm{M}_g$, and the mean
density, $\bar{\rho}$.  In external galaxies, it is possible to find
also the projected density of gas and young stars. In our Galaxy it is
possible to find the spatial density of gas and young stars.  In all
equations, connecting these quantities, the parameter $\gamma$,
characterising the rate of star formation, plays an important role.
To determine the value of the parameter $\gamma$ there are several
integral and differential methods.  We shall apply the integral method
using data on M31 and the Small Magellan Cloud (SMC), and the
differential method using data on SMC.

Let us use first the integral method. Basic data used in calculations
are given in Table~\ref{Tab23.5}.  

{\begin{table*}[h]
\centering    
\caption{} 
\begin{tabular}{cccc}
\hline  \hline
 Quantity& Units&M31&SMC\\
  \hline
  $\mm{M}_H$ & $10^9\,M_\odot$ &  3.7 & 0.48\\
  $\mm{M}_g$ & $10^9\,M_\odot$ &  5.3 & 0.68\\
  $\mm{M}$    &  $10^9\,M_\odot$ &110  & 2.4\\
   $K$                    &   $10^9$ years      &       0.50 & 4 \\                                                                 %
$a_0$                    &   kpc                     &   9 & 1.5\\
$\epsilon$            &                              & 0.016 & 0.5\\
$\bar{\rho}$         &   $M_\odot$/pc$^3$ & 0.38 & 0.062\\
$\gamma$           & $\left({M_\odot \over pc^3} \times 10^9\,yr \right)^{-1}$ &  5.3 & 4.0 \\
  \hline
\label{Tab23.5}   
\end{tabular} 
\end{table*} 
}

The mass of hydrogen of M31 was taken from our recent model of M31
\citet{Einasto:1969aa}. The mass of hydrogen in SMC was taken from
\citet{Hindman:1967aa}. In the calculation of the full mass of gas,
the hydrogen parameter of chemical composition was taken $X=0.70$. The
total mass of M31 was taken equal to the sum of its disc and flat
populations.  The distribution of densities of these populations is
similar to the distribution of gas. The mass of SMC and its effective
radius $a_{0.5}$  (in de Vaucouleurs spirit) are taken from
\citet{deVaucouleurs:1962}. From $a_{0.5}$ we found the harmonic
radius $a_0$, accepting in the generalised exponential model the shape
parameter $N=0.5$.

The axial ratio of the density ellipsoid of the M31
gas was taken as equal to $\epsilon = 0.016$ in analogy to our Galaxy
\citep{Einasto:1970ad, Einasto:1972af}.  It was calculated using
Eq.~(\ref{eq23.3.15}), which for $N=0.5$ yields
\be
\epsilon = {\sqrt{2}\,\zeta \over k\,a_0}.
\ee
We calculated the dispersion of $z-$coordinates, $\zeta$, using
densities of hydrogen by \citet{Schmidt:1957aa}, with result
$\zeta=100$ pc. Effective radius of the gas population was taken
equal to $a_0 = 8$ kpc, parameter $k$ according to tables by
\citet{Einasto:1972ac,Einasto:1972ad}.  These calculations gave the
value $\epsilon = 0.0157$.

Table~\ref{Tab23.5} shows that values of the parameter $\gamma$
for M31 and SMC are rather similar in spite of the very different type of
these galaxies. 

Now we shall find the parameter $\gamma$ applying a differential method,
using data on young stars and projected density of gas in SMC.

From Eq.~(\ref{eq23.3.7}) to (\ref{eq23.3.9}) we get
\be
{\dd{P_s} \over \dd{t}} = {\gamma \over 2\,\zeta^*} P_g^2.
\label{eq23.4.2}
\ee
We found the projected density of the gas $P_g$ from the hydrogen
density $P_H$ \citep{Hindman:1967aa}, accepting the chemical abundance
parameter $X=0.70$. Parameter $\zeta^*$ characterises the
distribution of gas clouds along the line of sight, and was found as
follows. Fig.~2 by \citet{Hindman:1967aa} shows that hydrogen is
concentrated to clouds having approximately a round form.  This hints to
the spherical form of the hydrogen population (SMC as a total has a
flattened form). The hydrogen population has approximately normal
distribution (its shape parameter of the modified exponential function
is 
$N=0.5$). This allows to find effective radii $a_0$ of gas clouds, and
from these data the dispersion of $\zeta$ and parameter $\zeta^*$.  Values found
for individual clouds vary between $2\,\zeta^* = 1.0$ kpc to 2.0 kpc;
for the mean we accepted $2\,\zeta^* = 1.6$~kpc. Following
\citet{Hindman:1967aa} we accepted the distance to SMC $d=60$~kpc.

Next we have to find the projected density of stars formed in unit
time, $R_P = \dd{P_s}/\dd{t}$.

\citet{Sanduleak:1968aa} found from spectral observations the
distribution of bright stars of various spectral types in SMC.  His
list can be considered as complete up to magnitude $m_{pg} = 13.0$,
corresponding to $B=13.1$ \citep{Schmidt-Kaler:1965}.  For
distance modulus $(m-M)_{ph} = 19.0$ \citep{van-den-Bergh:1965} this
corresponds to completeness of data up to $M_B = -6.0$.

According to \citet{Sanduleak:1968aa}, the total number of stars
brighter than $M_B = -6.0$ is $P=126$.  We can compare this number
with the expected number from the theoretical star formation model. We
take in the model $S=0$, which corresponds to constant star formation
rate $R^\circ=5$ solar masses in a year.  Using the luminosity
function $\varphi(M_B)$ we found that the expected number of stars
brighter than $M_B = -6.0$ is $P^\circ=25\,500$. The actual star
formation rate in SMC is $P^\circ/P =200$ times lower than in our
model. This leads to SMC star formation rate
$R=0.025\,M_\odot$ in  year.

This star formation rate of SMC  can be underestimated, since the
\citet{Sanduleak:1968aa}  list did not include supergiant stars. To
estimate the possible selection effect, we calculated the star
formation rate $R$, applying  Eq.~(\ref{eq23.3.17}), and used as characteristic
time of star formation parameter $K$ our result from 
Table~\ref{Tab23.5}. We found $R=0.050\,M_\odot$ a year.  If
this estimate is correct, then the star formation rate, found from
\citet{Sanduleak:1968aa} list of stars,  is to be multiplied by a factor
of two. 

\citet{Sanduleak:1969aa} found the distribution of stars brighter than
$m_{ph} = 13$ for unit surface density, $P_l$. It is clear that
\be
{Q\,P_l \over P^\circ} = {R_P \over R^\circ},
\ee
where $Q \approx 2$ is the correction factor described above. We used this
equation together with Eq.~(\ref{eq23.4.2}) to estimate the parameter
$\gamma$. \citet{Sanduleak:1969aa} got for the power index of the star
formation rate $S=1.84 \pm 0.14$, which is almost equal to the value
we accepted, $S=2$. Using \citet{Sanduleak:1969aa} distributions of
$P_l$ and $P_g$,  and calibrating gas densities to surface densities, we
got $\gamma =2.9$ in units $(M_\odot\,pc^{-3}\,Gyr)^{-1}$. 

This estimate of the parameter $\gamma$ is in good agreement with data
given in Table~\ref{Tab23.5}.  We accept a mean value of all data,
\be
\gamma = 4\, (\mm{M}_\odot\,pc^{-3}\,Gyr)^{-1}.
\label{eq23.4.4}
\ee

Our result depends on the age of SMC (we accepted $t=10^{10}$ years),
and on the mean axial ratio of the equidensity ellipsoid $\epsilon$ of M31
gas. If we take for the age of SMC two times lower value, and for
$\epsilon$ of M31 gas two times larger value, then the parameter
$\gamma$ increases by a factor of two.  However, in this case the
selection factor $Q$ also increases by a factor of two, which is not
acceptable.  But we can check this result in another way. The total
luminosity of young stars in M31 is in good agreement with our model
using parameters $S=2$ and $K=0.25$, and using the total mass of M31.
If we use only the mass of the disc of M31, then a good agreement is
obtained using $K=0.5$.  The luminosity of young star population
depends on the star formation rate, which depends on the present mean
density $\bar{\rho}$.  Using data on gas density in M31, we found for
the parameter $\gamma$ a value, close to the value in
Eq.~(\ref{eq23.4.4}). In summary, we conclude that there are no reasons
to accept for SMC an age smaller than for other galaxies, and for
the mean axial ratio of the gas population a value, different from
the present one.

Finally, we note that recently \citet{Hartwick:1971aa} found for the
star formation parameter for M31 a value $S=3.5 \pm 0.12$, using a
method similar to the   \citet{Sanduleak:1969aa} method.  He used
\citet{Roberts:1966aa} data on the distribution of neutral hydrogen,
and \citet{Baade:1964aa} data on the distribution of ionised hydrogen
as an indicator for the presence of young stars.  The theoretical basis of
his analysis was criticised by  \citet{Talbot:1971aa}.   However, more
important is the effect of antenna smoothing, not used by
\citet{Roberts:1966aa} in the determination of the hydrogen
distribution. In this way, the hydrogen distribution was strongly
smoothed without any peaks of densities along spiral arms, see
\citet{Einasto:1972ab}. If the smoothing is properly taken into account, 
the value of the parameter $S$ becomes fully normal.  

\section{The dependence of the star formation rate on chemical
    composition of the gas}

The basis of our present understanding of the synthesis of chemical
elements in the Universe and chemical evolution of galaxies was
presented by \citet{Burbidge:1957aa}.  Authors demonstrated that all
chemical elements heavier than hydrogen are synthesised inside stars
as a result of various nuclear processes. Part of synthesis products
are expelled from stars, where in this way the interstellar gas is
enriched by heavier chemical elements. A bit later it was understood
that similarly to hydrogen, helium also has a primordial origin, and only
elements heavier than helium are produced in stars. Various nuclear
processes and  problems of the chemical evolution of galaxies are subjects of
intensive studies.

The synthesis of elements heavier than helium takes place in all stars
with nuclear activity.  The most important factor in the enrichment of
interstellar medium with heavy elements are massive stars, which after
their active life explode as supernovae. During the explosion,
elements of high atomic weight are produced. The study of stars of
various chemical compositions and ages gives us information on the
synthesis of heavy elements and the star formation function at various
stages of the evolution of galaxies.

The possibility of the use of compositions of stars of various age to
study the star formation history was explored by \citet{van-den-Bergh:1961aa}. He
noted that already in the early phase of the evolution of the Galaxy,
the chemical abundance of some open star clusters was close to the
normal composition in the present epoch. This hints to a high activity of
supernova explosions in the early phase of Galaxy evolution. The same
conclusion was made by \citet{Schmidt:1963aa}.  The chemical
composition of stars and gas in the early phase of the Galaxy
evolution was studied by \citet{Dixon:1965aa,
  Dixon:1966aa}. \citet{Truran:1970aa} and  \citet{Cameron:1971aa}
developed a model of the chemical evolution of the Galaxy.  According
to their model, in the early phase of the evolution, a large fraction
of almost pure hydrogen-helium stars evolved rapidly and enriched the
halo gas with heavy elements.

Until recently, the attention of astronomers was directed to understand
the chemical aspects of the problem. Relatively little attention was
given to possible changes of parameters of the star formation
function, which could cause the effects mentioned above.

Based on the arguments, given by \citet{Dixon:1965aa, Dixon:1966aa} and
\citet{Cameron:1971aa}, we can conclude that the heavy element content
after the initial rapid contraction of the protogalaxy was relatively
high, about three times less than the solar content.  To find possible
errors of this estimate, let us consider two variants: A) heavy element
content at the end of the contraction was $Z_h=0.010$, \ie two times
less than the present content, and B) $Z_h = 0.005$. The mass of the
halo of M31 is 10\,\% of the mass of the whole galaxy
\citep{Einasto:1970ae,Einasto:1972ab}, let us assume that the relative
halo mass of our Galaxy is the same (direct data are less accurate).
All heavy elements of the gas in the Galaxy in this epoch (90\,\% of
the total mass of Galaxy) were produced by first-generation stars
(10\,\% of the mass of the Galaxy).  If we assume that heavy elements
were not consumed by compact objects of the halo, then in variant A
the fraction $\psi_h=0.090$ of initial stellar mass was expelled from
stars as heavy elements.  In variant B the fraction is $\psi_h=0.045$.

The mean heavy element content in the disc of the Galaxy is $Z=0.02$
(in the present epoch the mean heavy element content of young stars is
slightly higher). The mass of the disc is about 53\,\% of the whole mass of
the Galaxy \citep{Einasto:1970ad, Einasto:1972ad, Einasto:1972af}.
Based on these data, we can calculate the total mass of heavy elements
in the disc. When we remove from this amount the mass of heavy
elements at the beginning of disc formation, we can find the amount
produced by disc stars. We find that in variant A, the fraction of mass
of disc stars, expelled as heavy elements, was $\psi_d = 0.009$, and
in variant B  $\psi_d = 0.014$.  We conclude that in comparison with
halo stars, the effectiveness of the synthesis of heavy elements has
decreased in variant A 10 times, and in variant B 3 times.

Can this decrease of the star formation rate be explained by changes
of parameters of the star formation function?

Applying the star formation function $F(M)$ of Chapter 22, we can calculate
the fraction of stars of large mass. Let us use for disc stars of
Galaxy the lower mass limit of forming stars, $M_0 = 0.03\,M_\odot$,
upper mass limit $M_u= 100\,M_\odot$, the exponent of the mass
function, $n=2.333$, and the minimal mass of stars as future
supernovae, $M_{SN} = 2.6\,M_\odot$. The total mass of stars with
masses $M \ge M_{SN}$, is for these parameters $E_d =0.17$ of the mass
of all stars – the disc supernova producing capability.  All heavy
elements are synthesised by these stars of masses $M \ge
M_{SN}$. Accepting for disc stars the heavy element synthesis fraction
$\psi_h$ as found above, we find that in variant A 5\,\% of star mass
is expelled as heavy elements, and in variant B 8\,\% of star mass.

Let us now consider possibilities to increase the fraction of heavy
elements. There are two possibilities for this: to increase the
fraction of massive stars, and to increase the heavy element synthesis
capacity of supernova explosions.

The fraction of stars with masses above $M_{SN}$ can be increased by the increase
of minimal star forming mass $M_0$, and by   decreases of index $n$ and
supernova mass limit $M_{SN}$.  Let us  take $M_0=0.3\,M_\odot$ (this is
the maximal possible value from globular cluster data), $n=2$ and
$M_{SN}=2\,M_\odot$  (only stars of this mass have time to evolve
during the time of the contraction of the halo – 1 billion years – to
reach the supernova explosion stage). Accepting these parameters we
find for the halo supernova producing capability $E_H = 0.68$, which
is four times higher than the disc capability.  This is more than needed
to explain the heavy element content, needed at the end of halo
contraction phase for the variant B, but not large enough for the
variant A. A further increase of $M_0$ or decrease of $M_{SN}$ are
not possible.  For this reason, if the variant A of the fraction of
mass expelled as heavy elements is to be favoured, then the only
possibility to explain the disc heavy element content in the framework
of the contracting halo scenario is to decrease $n$ or to increase the
effectiveness of the creation of heavy elements during supernova
explosions.

Requirements for changes of star formation function parameters are not
too restrictive. For this reason, we think that there are no need to
accept the hypothesis for the initial prompt formation of all heavy
elements, as suggested by \citet{Truran:1970aa} and
\citet{Cameron:1971aa}.

\section{Formation of galaxies and their populations}

The Schmidt Eq.~(\ref{eq23.3.1}) allows to explain in a quantitative way
the formation of galaxies of various morphological type as well as
the formation of galactic populations.

Let us discuss first the formation of galaxies of different
morphological type: ellipticals, spirals and irregulars.  Galaxies of
these types differ from each other by their mean density. Bulges of
elliptical and spiral galaxies have high mean density, for the bulge
of M31 we found $\bar{\rho} = 6\,M_\odot\,pc^{-3}$. The mean
density of the disc of spiral galaxies is lower by an order, and the
mean density of irregular galaxies is lower by another order. Using
these mean densities and Eq.~(\ref{eq23.3.5}) we calculated the
respective characteristic time $K$, and then using Eq.~(\ref{eq23.3.16})
the change of the relative amount of gas, and from Eq.\,(\ref{eq23.3.17})
the rate of star formation $R$.  Results of our calculations are given
in Figs.~\ref{Fig23.1} and \ref{Fig23.2}.

{\begin{figure*}[h] 
\centering 
\hspace{2mm}
\resizebox{0.55\textwidth}{!}{\includegraphics*{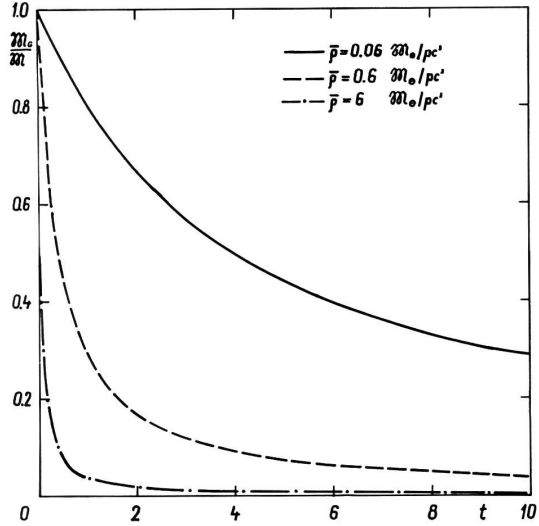}}
\caption{The fraction of gas mass in model galaxies of various mean density. } 
  \label{Fig23.1}
\end{figure*} 
}

{\begin{figure*}[h] 
\centering 
\hspace{2mm}
\resizebox{0.55\textwidth}{!}{\includegraphics*{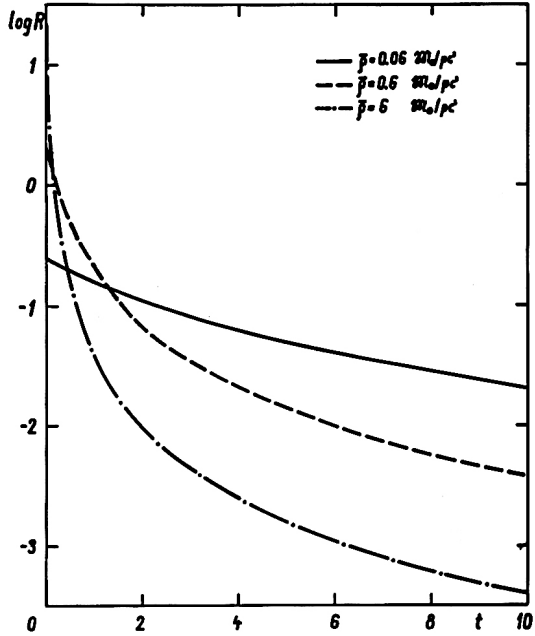}}
\caption{The rate of star formation $R(t)$ in galaxies of different
  mean densities. } 
  \label{Fig23.2}
\end{figure*} 
}

In Fig.~\ref{Fig23.3} we show the change of the total luminosity of
the galaxy with time. It was calculated as follows:
\be
\mathcal{L}_B(t) = \delta\,\int_{t_0}^t\,R\,\dd{t'}, \hspace{1cm} 
t_0=\left\{
  \ba{rc}
  0, & t\le \Delta\\
    t - \Delta, & t \ge \Delta.
    \ea
\right. 
\label{eq23.6.1}
\ee
Here the factor $\delta$ and time interval $\Delta$ were estimated
from the comparison of results of calculations with results in Chapter
23 for $K=0.3$.  Our calculations suggested that $\Delta$ must be
taken as a linear function of time
\be
\Delta = 0.1 + 0.14\,t,
\ee
where both $\Delta$ and $t$ are expressed in units $10^9$ years.

{\begin{figure*}[h] 
\centering 
\hspace{2mm}
\resizebox{0.55\textwidth}{!}{\includegraphics*{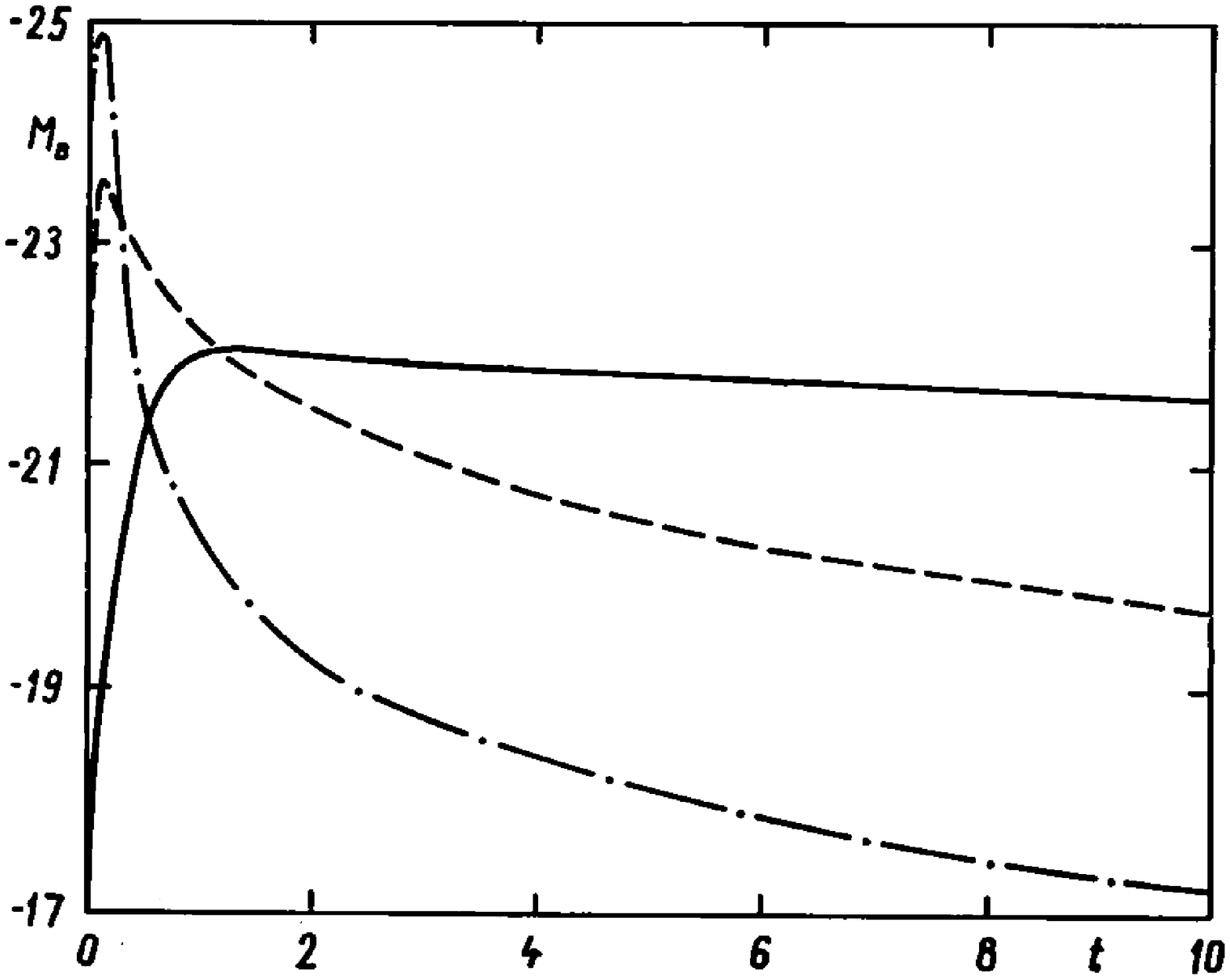}}
\caption{The evolution of galactic total luminosity for galaxies of
  different mean density. Lines as in Figs.~\ref{Fig23.1} and \ref{Fig23.2}. } 
  \label{Fig23.3}
\end{figure*} 
}

Properties of model galaxies for the age $t=10^{10}$ years represent
rather well the observed properties of elliptical, spiral and irregular
galaxies. This raises the question: How to explain differences in
density in these three types of galaxies?

It should be stressed that differences in mean densities can be
formed only in the gaseous phase of the evolution of galaxies.
Stellar populations are very conservative in this respect, as
kinematical and spatial characteristics of the structure of galaxies
change very slowly.

One of possible reasons for changes in mean densities could be
differences in primeval mass-angular moment distributions.
Protogalaxies with low primeval angular moment could contract
considerably and form elliptical galaxies. Protogalaxies with high
primeval angular moment could not contract in the radial direction. For
this reason, these galaxies formed after the contraction phase gas a
thin disc, which fragmented due to gravitational instability into
spiral arms. Theoretical aspects of the role of mass-angular moment
distribution were discussed by \citet{Lynden-Bell:1967aa}, observational
aspects were studied by \citet{Sandage:1970aa}.

Let us discuss now the formation of galactic populations. First we
consider the halo. Using data by \citet{Einasto:1970ae,
  Einasto:1972ab} we found the mean density of halo, $\bar{\rho} =
0.13\,M_\odot \,pc^3$, which yields $K=2\times 10^9$ years. At
the end of the halo forming the mass of gas was 90\,\% of the mass of
the galaxy, thus $\tau = 0.1$. From $\tau$ and $K$ we can find the 
time of halo formation $t=\tau\,K = 0.2\times10^9$ years.  This time is in
good agreement with our earlier estimate on the halo collapsing time.
Different regions of the halo formed at various times, first the
central  densest region, then more distant and less dense regions,
thus the contraction time is a certain mean value.

After the contraction phase, the remaining gas obtains the form which
is close to its present form. Thus, we do not make a considerable
error when we consider the projected density of gas, $P(A)$, as time
independent.  Based on these arguments, we consider the mean thickness
of gas, $\zeta^*$, also as time independent. Using Eq.~(\ref{eq23.3.8})
we find that the projected density of stars, formed in the time
interval $\Delta\,t$ at moment $t$ as follows:
\be
\Delta\,P_s =\gamma {P^2 \over 2\,\zeta^*}{\left[1+\gamma{P \over
      2\zeta^*}\,t\right]^{-2}}\,\Delta\,t.
\label{eq23.6.3}
\ee
Time in the equation is to be counted not from the formation of the
whole galaxy but from the beginning of the formation of the bulge and disc.

This equation allows to explain quantitatively the subsequent
formation of populations of increasing sizes\footnote{I am indebted to
  Grigori Kuzmin for the idea to use
   Eq.~(\ref{eq23.6.3}) for this purpose. }. If
$t$ is small, then $\tau = \gamma P\,t/(2\zeta^*) \ll 1$, and we have
\be
\Delta\,P_s = \gamma {P^2 \over 2\zeta^*}\,\Delta\,t.
\label{eq23.6.4}
\ee
Due to very rapid decrease of $P(A)$ with increasing $A$ star
formation occurs initially only in central regions –- in this way the
nucleus forms.  But in central regions of the galaxy, the density of
the gas decreases rapidly. For this reason, according to
Eq.~(\ref{eq23.6.3}), the central density of the following population also
decreases.  When $\tau \gg 1$, then we get for the projected density
using Eq.~(\ref{eq23.3.7})
\be P_g \approx  {2\,\zeta^* \over \gamma\,t}.
\label{eq23.6.5}
\ee

The density distribution of galactic populations can be described
by the modified exponential profile with $\nu =1/N \le 1$.  Based on
Eqs.~(\ref{eq23.3.14}) and (\ref{eq23.3.15}) we see that in this case the
parameter $\zeta^*$ decreases with increasing distance from the centre. In
this way during later phases of galactic evolution, the density of gas
in central regions of galaxies gets lower than in more distant
regions.  This situation is exactly observed in spiral galaxies like
M31 \citep{Einasto:1970ae,Einasto:1972ab}.   The active process of
star formation moves from central regions to more distant ones, and
the mean radius of recently formed galactic populations increases. In
peripheral regions of spiral galaxies, the gas density is very low, and
the star formation process has low intensity.  Thus, the fraction of gas in
the total projected density decreases with distance, see
Eq.~(\ref{eq23.3.7}).  For M31 at distance 25 kpc from the centre we
obtained: $2\,\zeta^* = 570$~pc, and $P=5\,\mm{M}_\odot/pc^2$, which
yields $P_g/P =0.75$. 

The star formation function gives us the possibility to determine the
distribution of galactic populations according to mass.

Let us discuss first the mass distribution of the disc populations of our
Galaxy.  We integrate Eq.~(\ref{eq23.3.17}) in time from $t=t_i - \Delta$
to $t=t_i$:
\be
\mm{M}_{Si}=\int_{t_i -\Delta}^{t_i}R\,\dd{t} = \mm{M}_S(t_i) -
\mm{M}_S(t_i-\Delta),
\label{eq23.6.6}
\ee
where $\mm{M}_{Si}$ is the mass of stellar population $i$, and
$\mm{M}_S(t_i)=\mm{M} - \mm{M}_g(t_i)$  is the total mass of stars of
the Galaxy at moment $t_i$, and the mass of gas $\mm{M}_g(t_i)$ is
calculated using Eq.~(\ref{eq23.3.16}).

In calculations, we made the following simplifying assumptions.
Effective radius of all disc subsystems was taken as equal to
$a_0=6.45$~kpc in accordance with our model described in Chapter
7. Structural parameters of all subsystems were also taken as equal:
$N=1$, $x_0=0$. The flattening parameter of the gas was taken equal to
$\epsilon=0.0157$, in accordance with the results discussed above.  The
characteristic time of disc star formation was found using
Eq.~(\ref{eq23.3.10}), where the mean density $\bar{\rho}$ was calculated
using Eq.~(\ref{eq23.3.22}).  We accepted the full mass of the disc
$\mm{M}=108\times 10^9\,\mm{M}_\odot$, and found
$\bar{\rho} = 1.0\,\mm{M}_\odot\,pc^{-3}$ and $K=0.25\times 10^9$
years. The total age of the disc of the Galaxy was taken equal to
$9\times 10^9$ years, and the time spent on the formation of disc
subpopulations was taken as equal to $\Delta=10^9$ years.  Results of
calculations are given in Table \ref{Tab23.6}.  We see that most
subpopulations of disc stars formed during the first billion years
after the start of disc formation. 

{\begin{table*}[h]
\centering    
\caption{} 
\begin{tabular}{cccc}
\hline  \hline
  $t_i$& $\epsilon_i$ & $\mm{M}_{Si}/\mm{M}$ & $\mm{M}_{Si}/\mm{M}$\\
  $10^9\,a$& & $K=0.25$ & $K=0.60$ \\
  \hline
  8.5 & 0.120 & 0.8000 & 0.6250 \\
  7.5 & 0.098 & 0.0889 & 0.1442 \\
  6.5 & 0.083 & 0.0342 & 0.0641 \\
  5.5 & 0.070 & 0.0181 & 0.0362\\
  4.5 & 0.057 & 0.0112 & 0.0233\\
  3.5 & 0.045 & 0.0076 & 0.0162\\
  2.5 & 0.035 & 0.0055 & 0.0120\\
  1.5 & 0.025 & 0.0042 & 0.0092\\
  0.5 & 0.018 & 0.0033 & 0.0073\\
  0.0 & 0.016 & 0.0270 & 0.0625\\
  \hline
\label{Tab23.6}   
\end{tabular} 
\end{table*} 
} 
      
In Chapter 7 we obtained the relationship between the age of
populations and the flattening parameter $\epsilon$  of iso-density
surfaces.  Using this relationship we calculated mean  $\epsilon$
values for disc subpopulations, results are given in
Table~\ref{Tab23.6}.  Using mass, radius, flattening and structural
parameters of subpopulations we can calculate the density, and by
summing over all subpopulations, find the total matter density.  We
made these calculations for the region near the Sun.  Adding to this
value the density of population II (halo), which according to
\citet{Oort:1958aa} is $\rho_{II} = 0.0015\,M_\odot\,pc^{-3}$, we
get for the mass density in the Solar vicinity
\be
\rho_\odot = 0.065\,M_\odot\, pc^{-3}.
\label{eq23.6.7}
\ee
Using Oort constants $A=15$~km/sec/kpc and $B=-10$~km/sec/kpc we find
for the Kuzmin parameter a value
\be
C = 61~km/sec/kpc.
\label{eq23.6.8}
\ee
We get  the total mass of interstellar matter
\be
\mm{M}_g = 2.9\times 10^9\,M_\odot,
\label{eq23.6.9}
\ee
and  the density of interstellar matter in the Solar vicinity,
$\rho_g=0.010\,M_\odot\, pc^{-3}$.

The comparison of these results with direct density estimates
discussed in Chapter 7 shows that all quantities are underestimated
except the total mass of gas, which according to
\citet{Westerhout:1957aa} is
$\mm{M}_g=1.4\times\,10^9\,M_\odot$.

To find possible reasons for this disagreement, we repeated
calculations using the gas density in Solar vicinity, $\rho_g =
0.023\, M_\odot\, pc^{-3}$.  For the parameter $K$ we got a value
$K= 0.6\times 10^9$ years. Results of calculations are given in
Table~\ref{Tab23.6}.  For the total mass density in Solar neighbourhood
we got now $\rho_g=0.082\,M_\odot\, pc^{-3}$, which yields Kuzmin
parameter value $C=68$~km/sec/kpc, and total gas mass in the Galaxy,
$\mm{M}_g =6.8\times 10^9\,M_\odot$. 

We see that values of the Kuzmin parameter and total density are fully
acceptable, but the total gas mass is too large. We recall that the
total gas mass of M31 is only
$\mm{M}_g = 5.3\times\,10^9\,M_\odot$, see Chapter 20. Andromeda
galaxy M31 is about 1.5 times more massive than our Galaxy. If
relative fractions of gas in both galaxies are equal, then the gas
mass of our Galaxy would be $\mm{M}_g =3.5\times\,10^9\,M_\odot$.
However, there exists arguments suggesting that the relative gas
content of our Galaxy is lower than in M31.  The linear size of M31 is
about 15\,\% larger than the size of our Galaxy, mutual distances of
neighbouring spiral arms of M31 are about two times larger than in our
Galaxy (see Chapter 19 and \citet{Westerhout:1957aa}). For this reason,
the estimated total mass of gas in Galaxy Eq.~(\ref{eq23.6.9}) is fully 
acceptable (observational estimate by \citet{Westerhout:1957aa} is
probably underestimated, as well as the total mass of Galaxy according
to \citet{Schmidt:1956}). 
The first variant of the distribution of masses of
disc subpopulations should be closer to reality. How can we explain
low values of gas density and total density in Solar neighbourhood
found for this variant?

Recently \citet{Woolley:1971aa} demonstrated on the basis of
statistics of nearby stars that the relative number of young stars
in Solar vicinity is larger than expected on the basis of the hypothesis
that the rate of star formation is constant. In other words, the Sun
is located in a region of enhanced star forming intensity. The
enhanced star forming activity in Solar neighbourhood is supported by the
presence of Gould Belt, as well as by the fact that the Sun is located inside
a spiral arm of Galaxy \citep{Becker:1970}. Apparently, this allows
to explain the disagreement between our theoretical density estimate
in a smooth Galaxy model, Eq.~(\ref{eq23.6.7}) and observations.

We conclude that the star formation function allows to explain
satisfactorily both integral properties of galaxies as well as
properties of galactic populations. The general picture of galaxy
evolution is similar to the view by \citet{Sandage:1970aa}.

\vskip 1mm

\hfill October 1971


\appendix

\setcounter{figure}{0}

\setcounter{table}{0}

\renewcommand{\thechapter}{\Alph{chapter}.\arabic{chapter}}
\setcounter{chapter}{0}

\appendix

\chapter{Epilogue}

The defence of the Thesis on March 17, 1972 was successful. However, two
related  problems remained — it was impossible to reproduce the
observed rotation curves of galaxies with known stellar populations,
and data on 
mass-to-light  ratios of populations were  uncertain. For
this reason, I started searches to find solutions to these open
questions immediately after the defence. The 
story of events after the defence of the Thesis is described in
detail in my book \citep{Einasto:2014} and in the review paper
\citep{Einasto:2018dz}. Here I give a short overview of the
development of ideas directly connected with the topic of the Thesis.

Both problems are connected with the possibility of the presence of
dark matter in galaxies.  
I had serious reasons to believe that there is only a limited quantity of dark matter in
galaxies like our own Galaxy. This problem had been studied by Tartu
astronomers long ago. \citet{Opik:1915} was one of the first to study
the dynamics of the Galaxy with the goal to find the density of matter
in Solar neighbourhood. He understood that due to the flat shape of the
Galaxy, the dynamical density can be determined from the comparison of
motions and spatial distributions of stars in the vertical direction. He
found that the vertical attraction of known stars is sufficient to
explain the observed distributions, and that there is no reason to add
invisible matter (the term ``dark matter'' had not yet been
suggested). \citet{Kuzmin:1955aa} and his student
\citet{Eelsalu:1958aa} repeated this study with new and better data and confirmed
\citet{Opik:1915} result. The problem was also discussed by
\citet{Oort:1960uy}, who found that the dynamical density near the Sun
is larger than found by \citet{Kuzmin:1952ab, Kuzmin:1955aa} and
\citet{Eelsalu:1958aa}.  In other words, there is a need for dark
invisible matter. Since the matter density and possible presence of
dark matter are of fundamental importance, my Tartu collaborator
\citet{Joeveer:1968b, Joeveer:1972} made a new analysis, using a
completely different method, see Chapter 21. Ages of young stars are
known, this allows to find parameters of vertical oscillations of
young B stars and cepheids, which led to parameter $C=70$~km/s/kpc and
dynamical density $\rho_{dyn}= 0.09\,M_\odot$/pc$^3$.  On the
basis of these studies, I supported the classical paradigm with no
large amounts of dark matter  in the Solar neighbourhood. More accurate
recent data support this conclusion \citep{Gilmore:1989}.

More data accumulated on rotation velocities of galaxies. New data
suggested the presence of almost flat rotation curves of galaxies,
thus, it was increasingly difficult to accept my previous solution of
the discrepancy with large non-circular motions.  I discussed the
problem with my colleague Enn Saar in spring 1972, who suggested abandoning my
earlier idea that galaxies have relatively sharp boundaries but may
have extended envelopes.

The possible presence of  dark matter  in the Galaxy in Solar vicinity at least
in some quantity was suggested by \citet{Oort:1932tb, Oort:1960uy}, and in clusters
of galaxies by \citet{Zwicky:1933}, \citet{Karachentsev:1966tv}, and
\citet{Rood:1972wr}.  A numerical study of the stability of flat
galaxies suggest the presence of massive halos of galaxies
\citep{Ostriker:1973}.  Rotation velocity measurements of galaxies by
\citet{Roberts:1966aa, Roberts:1967aa, Roberts:1969aa} and
\citet{Rubin:1970tp} suggest that galaxies have indeed large and
massive envelopes.  My detailed analysis of properties of
known stellar populations demonstrated that no known stellar
population can be responsible for flat rotation curves of galaxies.
As I discussed in \citet{Einasto:2018dz}, the tacit assumption in
earlier studies was that the stuff, responsible for this effect in
clusters, galaxies in general and near the plane of the Galaxy, is the
same everywhere.

After the discussion with Enn, I noticed that here lies a
controversy. Dynamical data suggest that eventual dark matter in Solar vicinity
is strongly concentrated toward the plane of the Galaxy, thus
dissipation is needed for its formation. By contrast, if the rotation
of galaxies in outer regions is influenced by a new hypothetical
population, then this population should form a large, massive, and an
almost spherical population. In particular, for its
formation, dissipation is not needed. Different size, shape,  mass and
dissipation properties  suggest a
different formation history and nature. Following these considerations,
I concluded that there must exist two types of dark matter: the
``local dark matter'' near the Sun close to the plane of the Galaxy,
and the ``global dark matter'', forming envelopes of galaxies and
clusters of galaxies \citep{Einasto:1972af, Einasto:1974a}.

To have a better reproduction of observed rotation curves, it would be
reasonable to look at which properties the population of global dark
matter should have using available data on known stellar populations
and galaxy rotation data.  To avoid confusion with the
known halo population, consisting of old metal-poor stars, I called the
new population ``corona'' \citep{Einasto:1972af, Einasto:1974a}. 
To check this possibility, I used my programs to calculate dynamical
models of galaxies.  It was easy to find a new set of models with one
addition component — dark corona. As the first approximation, I assumed
that the total mass of the M31 corona is equal to the mass of the sum
of known stellar populations \citep{Einasto:1972af, Einasto:1974a}. I made two
versions of models of galaxies in the Local Group and giant elliptical
galaxy M87, variant A without corona and variant B with corona, see
Fig.~\ref{Fig25.1}.  This calculation showed that the adding of
coronas improves model rotation curves, but not enough.

{\begin{figure*}[h] 
\centering 
\resizebox{0.65\textwidth}{!}{\includegraphics*{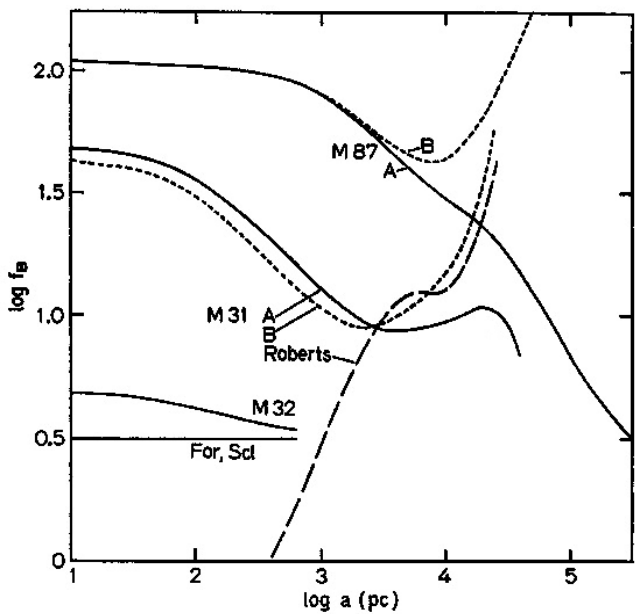}}
\caption{The distribution of mass-to-light ratio, $f_B = M/L_B$,
  in galaxies of the Local Group and M87: models without (A) and with
  (B) dark corona \citep{Einasto:1974a}. }
  \label{Fig25.1}
\end{figure*} 
}

I reported new results at the First European Astronomy Meeting in
Athens on September 8, 1972 \citep{Einasto:1974a}. It was clear that
coronas cannot be made of stars because outer stellar populations
consist of old halo-type stars with very low mass-to-light ratio, but
the mass-to-light ratio of the corona is very high. The coronal matter
cannot be in the form of neutral gas, since this gas would be
observable.  Initially I suspected that it could be ionised hot gas
\citep{Einasto:1972af, Einasto:1974a}.  However, the total mass of coronas
was not known yet, and the evidence for the presence of coronas was
not strong.

So far, I had concentrated my efforts on the study of the structure of
galaxies. It was now clear that the environment of galaxies was also
important. The dark matter problem was discussed a long time ago in
clusters of galaxies. Also masses of groups of
galaxies, measured from the velocity dispersion of galaxies, were
larger than summed masses of individual galaxies, see 
\citet{Holmberg:1937uq, Holmberg:1969ys} and
\citet{Karachentsev:1966tv}.  A similar discrepancy was 
found in the Local Group (the M31 - MW system) by
\citet{Kahn:1959}. Reading these papers on the mass discrepancy in
clusters, groups and galaxies, I realised that the problem of dark
matter in galaxies is the same as in clusters. This allows to find
masses and radii of dark coronas of galaxies. I noticed that if
galactic coronas are large enough, then companion galaxies should lie
inside coronas of the main galaxies. Thus, companion galaxies can be
considered as test particles to measure the gravitational attraction
of the main galaxy.

{\begin{figure*}[h] 
\centering 
\resizebox{0.55\textwidth}{!}{\includegraphics*{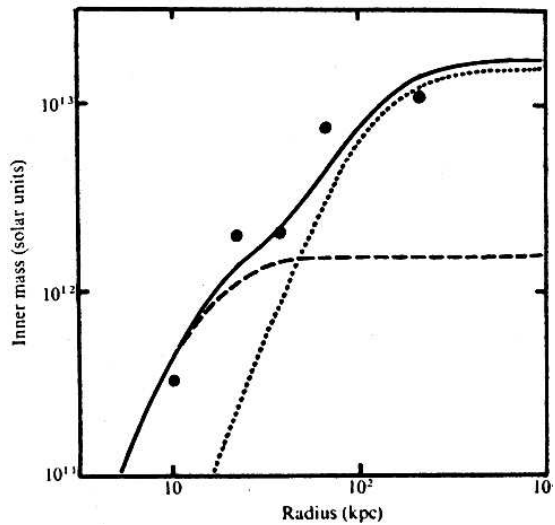}}
\caption{The mean internal mass $M(R)$ as a function of the radius $R$
  from the main galaxy in 105 
pairs of galaxies (dots). The dashed line shows the contribution of
visible populations, the dotted line the 
contribution of the dark corona, solid line the total distribution
\citep{Einasto:1974ad}. 
  }
  \label{Fig25.2}
\end{figure*} 
}

I collected data for pairs of galaxies.  The analysis was soon ready,
see Fig.~\ref{Fig25.2}.  Our analysis suggested that galactic coronas
have masses about ten times larger than masses of their visible
populations.  In those years, Soviet astrophysicists had the tradition
to organise Winter Schools.  In 1974, the School was held in the
Terskol winter resort. I presented my report on the masses of galaxies on
January 29, 1974. I stressed in my talk arguments, suggesting that
the presence of coronas around galaxies is a general phenomenon. Also,
I suggested that galactic coronas probably have the same origin as
dark matter in clusters and groups, and that coronas are probably not
of stellar origin.

Prominent Soviet astrophysicists like Yakov Zeldovich, Iosif
Shklovsky, and others participated in the Winter School. After my
talk, the 
atmosphere was as if a bomb had exploded. For Zeldovich and his group,
the presence of a completely new, massive non-stellar population was a
surprise.  Two questions dominated: What is the physical nature of the
dark matter?  and What is its role in the evolution of the Universe?

I had to hurry with the publication of our results, since massive
halos were already discussed by \cite{Ostriker:1973} to stabilise
orbits of flat population stars.  Following a suggestion by Yakov
Zeldovich we sent the paper to ``Nature'' \citep{Einasto:1974ad}.  Soon
it was clear that it was just in time. \cite{Ostriker:1974} got
similar results using similar arguments; their paper was published
several months after our ``Nature'' paper, and has a reference to our
preprint. Both papers suggest that the total cosmological density of
dark matter in galaxies is about 0.2 of the critical cosmological
density.

In the ``Nature'' paper, we noted that dark matter in clusters cannot be
explained by hot X-ray emitting gas, since its mass is insufficient to
stabilise clusters.  \citet{Ostriker:1974} did not notice that dark matter forms a
new population of unknown nature; authors write in the discussion that
the very great extent of spiral galaxies can perhaps be understood as due to a
giant halo of faint stars.

Soon the reaction to the results of both papers appeared:
\citet{Burbidge:1975} formulated difficulties of the dark corona/halo
concept.  The main problem is in the statistical character of
dynamical determinations of masses of multiple galaxies.  If companion
galaxies used in mass determination are not real physical companions
but random interlopers, as suggested by Burbidge, then the mean
velocity dispersion reflects random velocities of field galaxies, and
no conclusions on the mass distribution around giant galaxies can be
made.

{\begin{figure*}[h] 
\centering 
\resizebox{0.55\textwidth}{!}{\includegraphics*{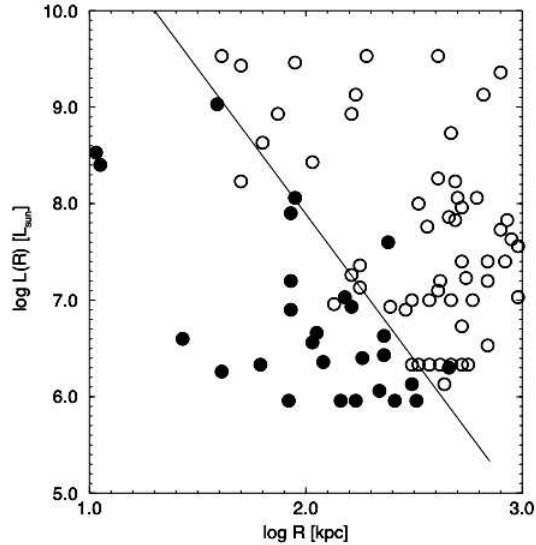}}
\caption{Distribution of luminosity of companion galaxies of
different morphology vs. distance from the central galaxy; spiral and
irregular companions are marked with open circles, elliptical
companions with filled circles \citep{Einasto:1974ae}.
  }
  \label{Fig25.3}
\end{figure*} 
}

Our ``Nature'' paper \citep{Einasto:1974ad} together with a similar paper
from the Princeton group by \citet{Ostriker:1974} and the response by
\citet{Burbidge:1975} started the ``dark matter'' boom.   As noted by
\citet{Kuhn:1970}, a scientific revolution begins when leading
scientists in the field start to discuss the problem and argue in
favour of the new over the old paradigm.

Difficulties connected with the statistical character of our arguments
were evident, thus we started a study of properties of companion
galaxies to find evidence for some other regularity in the satellite
system which surrounds giant galaxies.  Soon we discovered that
companion galaxies are segregated morphologically
\citep{Einasto:1974ae}. Elliptical (non-gaseous) companions lie close
to the primary galaxy whereas spiral and irregular (gaseous)
companions of the same luminosity have larger distances from the
primary galaxy. The distance of the segregation line from the primary
galaxy depends on the luminosity of the satellite galaxy, see
Fig.~\ref{Fig25.3}. This means that there exist physical interactions
between companions and the coronal gas of the main galaxy —
ram-pressure removal of gas from companion galaxies by the coronal gas
of the main galaxy.

It was also clear that coronas form an extended population of the
main central galaxy. But here lies a contradiction: inside a luminous
galaxy with its non-luminous corona there exist companion galaxies,
orbiting within the corona of the main galaxy.  In other words, a
dwarf galaxy inside the giant galaxy. To avoid confusion, we proposed
with Arthur Chernin to call giant galaxies together with their coronas
and satellites ``hypergalaxies'' \citep{Chernin:1976}. We found that
almost all dwarf galaxies are located near luminous galaxies.  This
led us to the conclusion that galaxies do not form in isolation, but
as systems, and that hypergalaxies are the main sites of galaxy
formation. However, the term ``hypergalaxies'' is not accepted by the
astronomical community, instead the term ``halo'' is used.

The dark matter problem was discussed in a special session of the Third
European Astronomy Meeting in Tbilisi, Georgia, in summer 1975. This
was the first international discussion between the supporters of the
classical paradigm with conventional mass estimates of galaxies, and
of the new one with dark matter. Arguments favouring the classical
paradigm were presented by \citet{Materne:1976wn}.  Their most serious
argument was: Big Bang nucleosynthesis suggests a low-density Universe
with the density parameter $\Omega \approx 0.05$; the smoothness of
the Hubble flow also favours a low-density Universe. If one excludes
inconvenient data by \citet{Zwicky:1933} on the Coma cluster,
\citet{Kahn:1959} data on the mass of the double system M31-Galaxy,
and recent data on flat rotation curves of galaxies by 
\citet{Roberts:1966aa} and \citet{Rubin:1970tp}, as written explicitly
by  \citet{Materne:1976wn},  then everything fits
well into this classical cosmological paradigm. It was clear that the
problem cannot be solved by dispute — new data were needed. 

Soon new radio measurements of neutral hydrogen for a large number of
galaxies were published by \citet{Bosma:1978}.  Another series of
extended rotation curves of spiral galaxies was made by
\citet{Roberts:1975uz} using radio data,  and by 
\citet{Rubin:1978, Rubin:1979mb, Rubin:1980} using optical
measurements. Observations confirmed the general trend that the mean
rotation curves remain flat over the whole observed range of distances
from the centre up to $\approx 40$ kpc for several galaxies. The
internal mass within the radius $R$ increases over the whole distance
interval. However, the nature of dark matter was still unknown.

The dark matter problem was discussed during the IAU General Assembly
in Grenoble, August 27, 1976, at the Commission 33 Meeting.  In my talk I presented
arguments for the non-stellar  
nature of dark corona \citep{Einasto:1976d}. After the lecture, Ivan
King came to me and asked  to repeat the main arguments against the
stellar origin of dark matter. The basic arguments are as follows.

{\begin{figure*}[ht] 
\centering 
\resizebox{0.80\textwidth}{!}{\includegraphics*{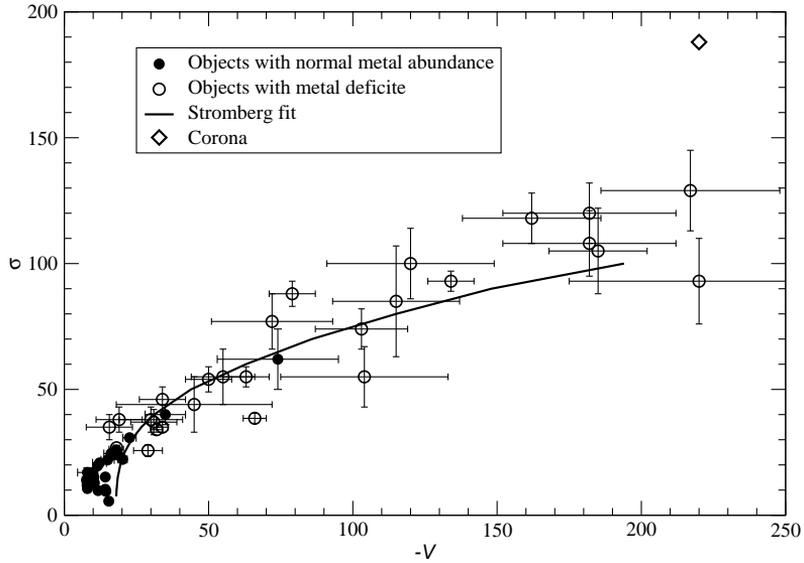}}
\caption{Str\"omberg diagram for galactic populations according to
  data presented in Chapter 4. Kinematical data for the corona are
  taken from the model by \citet{Einasto:1979ux}. The Str\"omberg fit was taken
  from original Russian version of Chapter 4, it does not take into
  account the non-stationary status of very young populations.
  }
  \label{FigAnew}
\end{figure*} 
}

Physical and kinematical properties of stellar populations depend
almost continuously on the age of the population, see
Fig.~\ref{Fig3.1}.  The continuity of stellar populations of various
age is reflected also in their kinematical characteristics, such as the
velocity dispersion and the heliocentric centroid velocity, expressed
in the Str\"omberg diagram.  The oldest halo populations have the
lowest metallicity and $M/L$-ratio, see Table~\ref{Tab20.2} and
Fig~\ref{Fig22.5}, the highest velocity dispersion, and the largest
(negative) heliocentric velocity, see Fig.~\ref{Fig4.1}  and \ref{FigAnew}.  There is no
place to put the new population into this sequence.  The dark
population is almost spherical and non-rotating.  It has a much larger
radius than all known stellar populations.  In order to be in
equilibrium in the Galactic gravitational potential, these coronal
stars must have a high velocity dispersion, about 
$\sigma \approx 200$ km/s, much more than all known stellar
populations, up to 125 km/s, see Figs.~\ref{Fig3.1}, 
\ref{Fig4.1} and \ref{FigAnew}. \citet{Jaaniste:1975} investigated the possible stellar
nature of the corona. Authors  found no fast moving stars, possible
candidates for coronal objects.

The $M/L$ value, and the spatial and kinematical distribution of the
dark population differ greatly from respective properties of all known
stellar populations, and there are no intermediate populations. Thus,
the corona must have been formed much earlier than all known
populations to form the gap in relations between various physical,
kinematical and spatial structure parameters. The total mass of the
new population exceeds the masses of known populations by an order of
magnitude, thus we have a problem: How to transform at an early stage
of the evolution of the Universe most of the primordial matter into
invisible stars? It is known that star formation is a very
inefficient process: in a star-forming gaseous nebula only about 1~\%
of matter transforms into stars.

As discussed above, neither neutral nor hot ionised gas is a suitable
candidate for dark matter. Thus, the nature of coronas remained
unclear. It was only much later that the non-baryonic nature of dark
matter became evident, as discussed in a conference in Tallinn, April
7 -- 10, 1981, and in Vatican Study Week, September 28 -- October 2,
1981.  Leading Soviet physicists and astronomers attended the Tallinn
conference.  Several talks were devoted to the formation of the
structure of the Universe with neutrinos as dark matter (Yakov
Zeldovich, Andrei Doroshkevich, Igor Novikov).
In the Vatican Study Week neutrinos as dark matter
candidates were discussed by Martin Rees,  Joe Silk, Jim Gunn and
Dennis Sciama.
%
Difficulties of the neutrino-dominated dark matter were evident, and
soon the Cold Dark Matter (CDM) was suggested by \citet{Bond:1982},
\citet{Pagels:1982}, \citet{Peebles:1982wa}, and
\citet{Blumenthal:1984}.

\begin{table*}[h] 
\caption{Galactic parameters} 
\label{Tab25.1}
{\small
  \centering 
\begin{tabular}{lccccl}
\hline  \hline
 Parameter&	Unit	&Observed&		Smoothed& Adopted & Reference\\
  \hline
$R_0$&	kpc             &	$8.8	\pm	0.7$&$8.5\pm  0.3$&  8.5 & 1,~2\\
$V$ &	km/sec      &	$220 \pm 10$	& $221\pm 5$&    225& 3\\
$W$ &	     ``          &	$120 \pm 15$&$133\pm 4$&  131.8& 4\\
$A$ &	km/sec/kpc&	$16	\pm	1$&$15.7\pm 0.4$& 15.5& 5 - 7\\
$C$ &	``                &	$70	\pm	5$&                       &74	& 14\\
$\omega$ &       ``      & $26 \pm 2$& $26.0\pm  0.7$&26.5& 8 -10\\
$k_z$ &	                 &	$0.282 \pm 0.020$&$0.285\pm 0.008$& 0.293&11\\
$\rho_0$ &  $M_\odot$/pc$^3$&$0.1\pm 0.02$&             &0.097 & 12, 13\\
\hline 
\end{tabular}\\
}
{References: 1. \citet{Oort:1975te}, 2. \citet{Harris:1976ti},
3. \citet{Einasto:1979ur}, 4. \citet{Haud:1984ve},
5. \citet{Crampton:1969tr}, 6. \citet{Balona:1974wq},
7. \citet{Crampton:1975vn}, 8. \citet{Asteriadis:1977wp},
9. \citet{Fricke:1977ts}, 10. \citet{Dieckvoss:1978uk},
11. \citet{Einasto:1972th}, 12. \citet{Joeveer:1974wa},
13. \citet{Woolley:1967tb}, 14. \citet{Joeveer:1974wa}.
}
\end{table*}

\begin{table*}[h]
\caption{Parameters of galactic components } 
\label{TabA.2}   
{\small
  \centering    
\begin{tabular}{lccccccc}
  \hline  \hline
  Quantity&Unit&Nucleus&Bulge&Halo&Disc&Flat&Corona\\
  \hline
 $\epsilon$&      &0.6&0.6&0.3&0.10&0.02& 1\\
 $N$     &         & 1 & 1 &4 &  1&0.5& 0.5\\
 $a_0$  &kpc  &0.005&0.21&1.9&4.62&6.4&75\\
$\mm{M}$&$10^{10}\,M_\odot$&0.009&0.442&1.2&7.68&1.0&110\\
  \hline
\end{tabular}\\
}
\end{table*} 

The presence of massive dark matter coronas influences galactic
models. Thus, I continued together with my collaborators Urmas Haud
and Ants Kaasik to develop new models which included dark coronas. To
develop the model of our Galaxy, a system of galactic parameters is
needed. One of important parameters is the circular velocity near the
Sun. Using the method described in Chapter 7 and shown in
Fig.~\ref{Fig7.1}, we found for the circular
velocity $V_0 = 220 \pm 7$~km/sec \citep{Einasto:1979ur}.  This value
is lower than our previous estimate, discussed in Chapter 7, due to
the addition of dark corona in the new model.  The model of
the Galaxy was described in the preliminary form by \citet{Einasto:1976d}
and in a more polished form by \citet{Einasto:1979ux}. In this model, we
found an improved system of galactic parameters with $R_0=8.5$~kpc,
presented in Table \ref{Tab25.1}.  Parameters of galactic populations
according to this model are given in
Table~\ref{TabA.2}.  For disc and flat populations parameters are
given for positive mass components.

{\begin{figure*}[ht] 
\centering 
\resizebox{0.45\textwidth}{!}{\includegraphics*{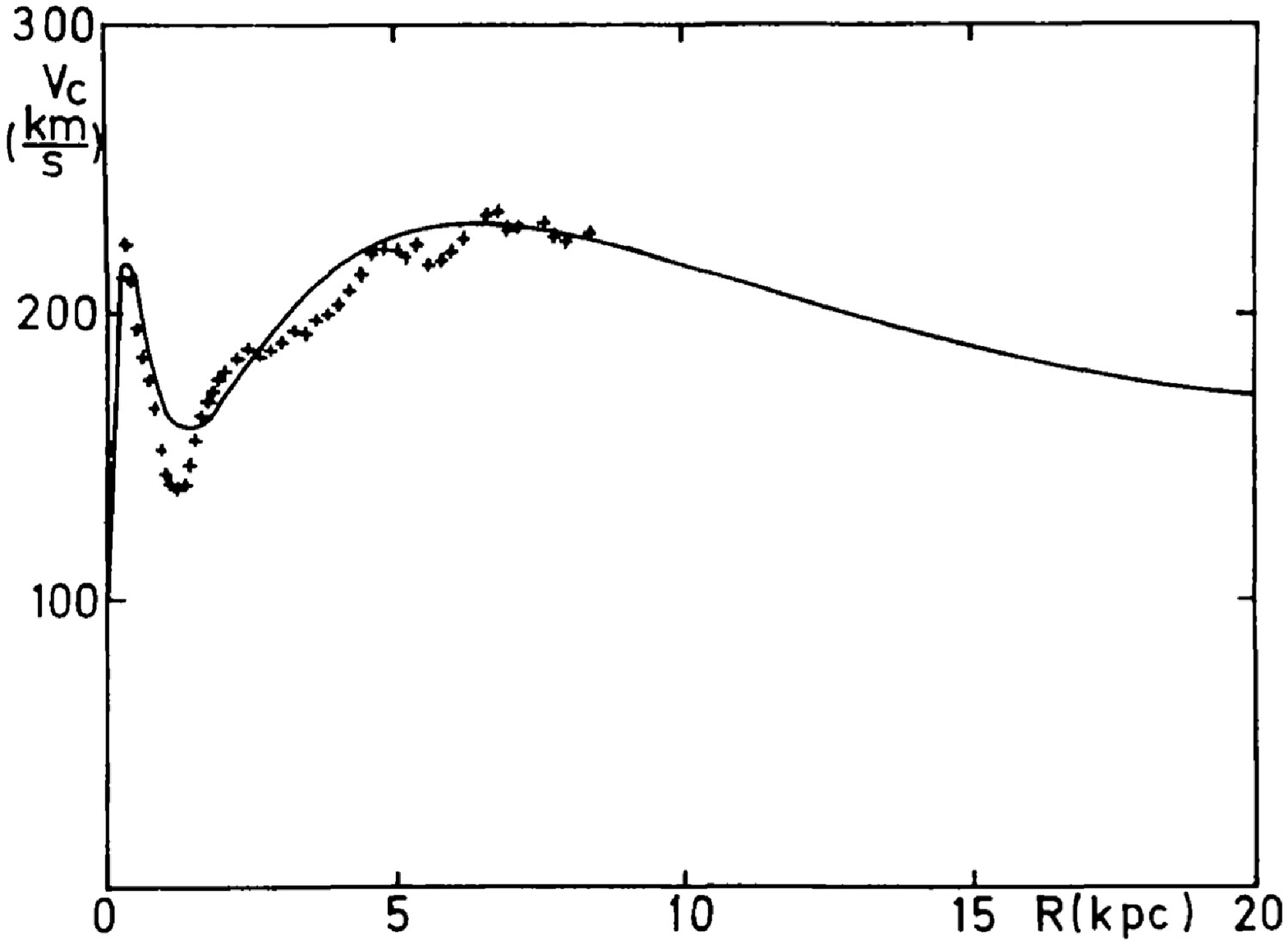}}
\resizebox{0.47\textwidth}{!}{\includegraphics*{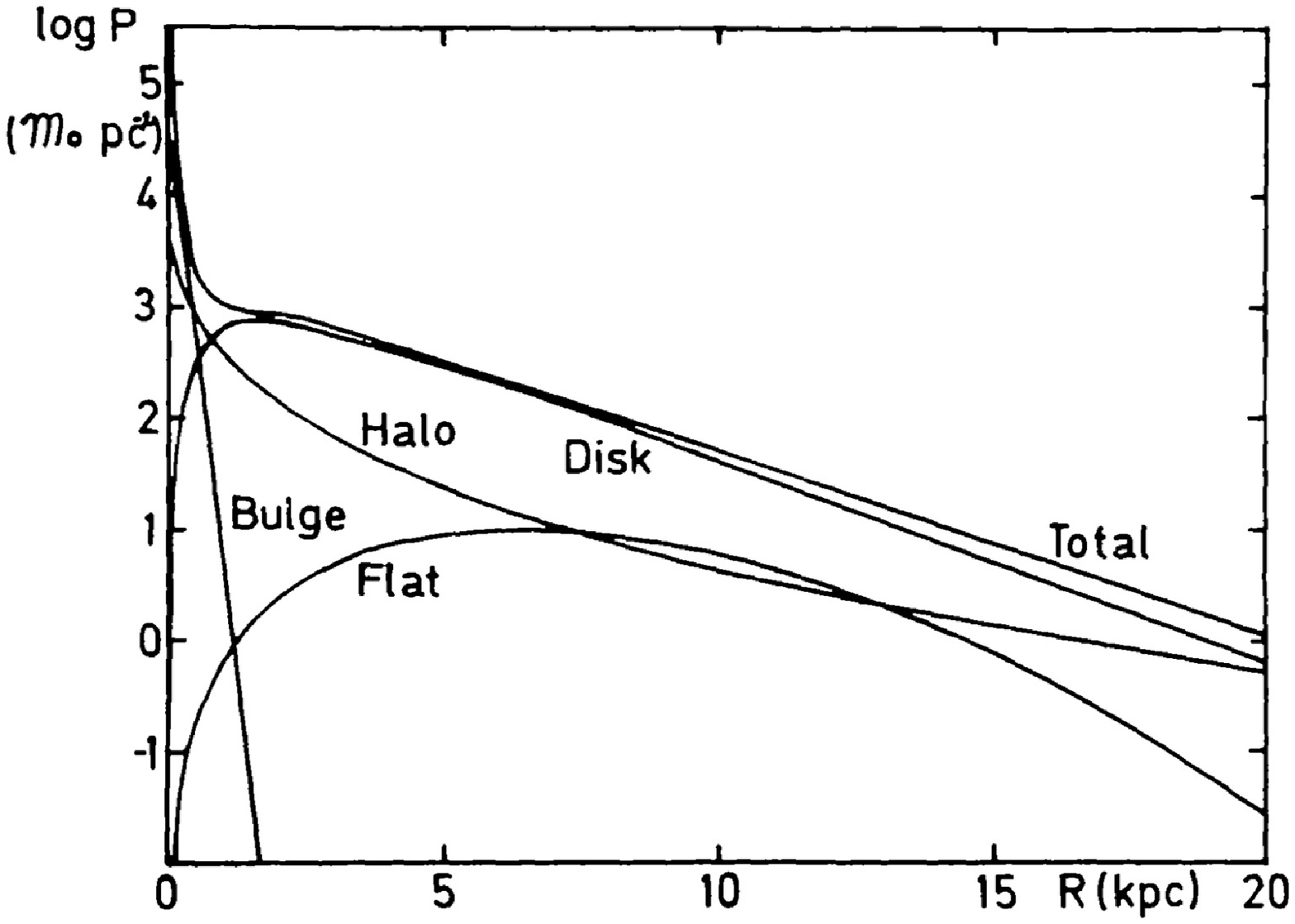}}\\
\resizebox{0.50\textwidth}{!}{\includegraphics*{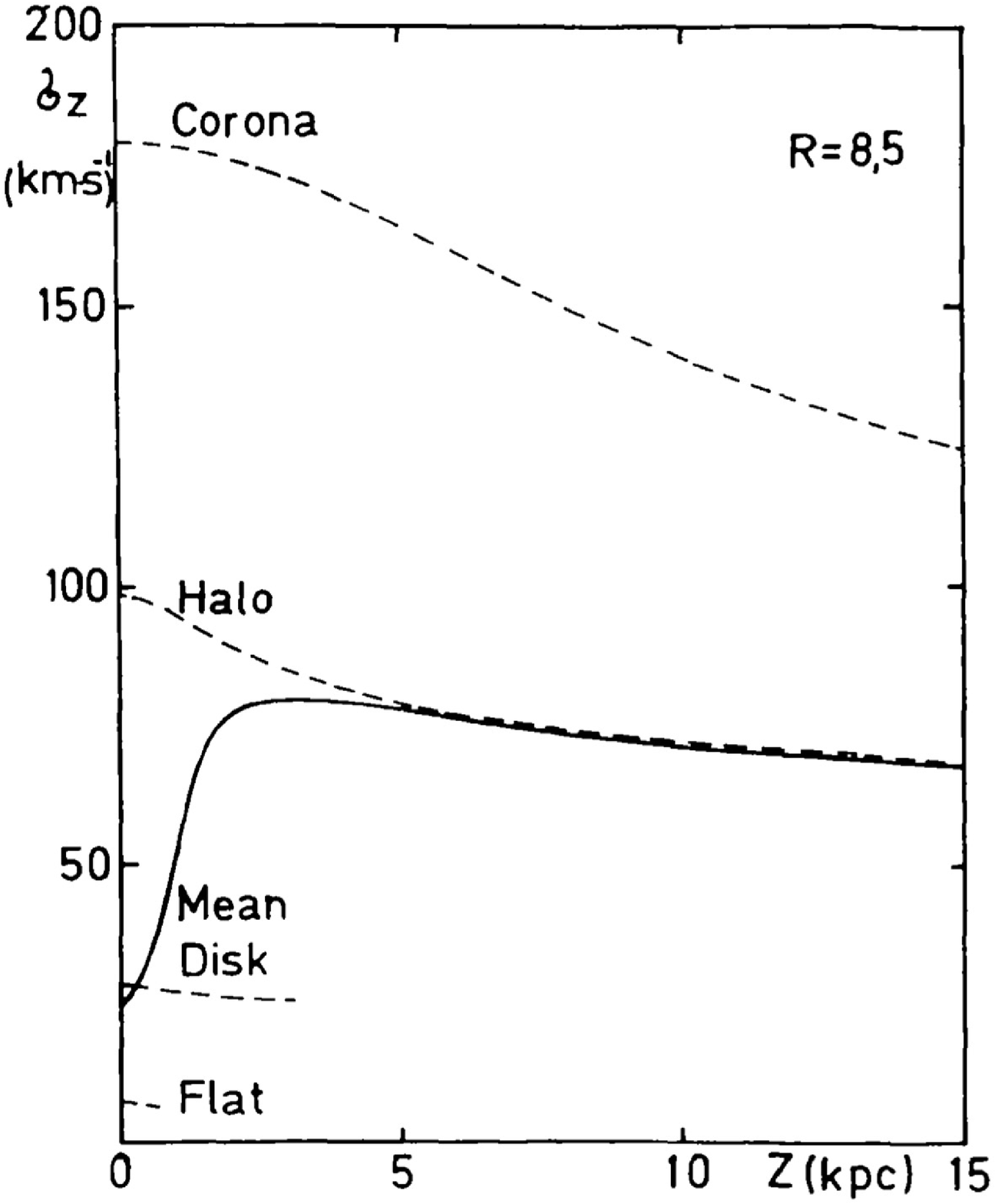}}
\resizebox{0.45\textwidth}{!}{\includegraphics*{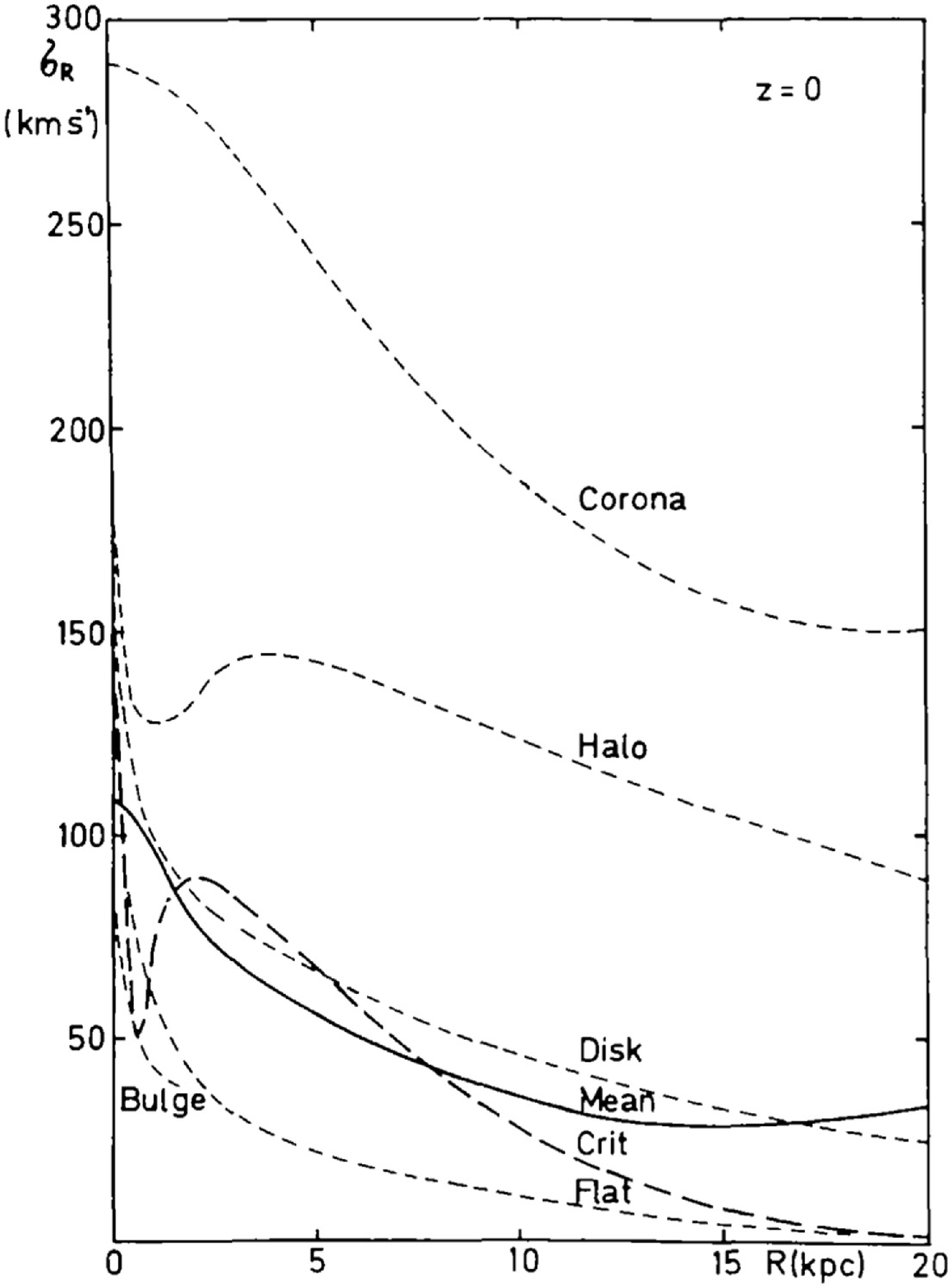}}
\caption{ {\em Top panels:} Circular velocity surface density of the Galaxy
  and its components. {\em Bottom panels:} Velocity dispersion, $\sigma_z$
  and  $\sigma_R$, of galactic populations \citep{Einasto:1979ux}.
  }
  \label{FigA02}
\end{figure*} 
}

In the top  left panel of Fig.~\ref{FigA02} we show the circular velocity
(solid line) and observed rotation velocity (symbols) of the model by
\citet{Einasto:1979ux}.  In the right panel of the Figure we give the
surface density of the Galaxy and its components of the same model.
In the bottom left panel of Fig.~\ref{FigA02} we show the velocity dispersion
$\sigma_z$ as the function of the distance from the galactic plane,
and in the right panel the velocity dispersion $\sigma_R$ as the function
of the distance from galactic centre \citep{Einasto:1979ux}. Data are
given for the main populations: flat, bulge, disc, halo and
corona. Also we show the mean dispersion, and the critical dispersion
by \citet{Toomre:1964qy}. We see that the mean velocity dispersion is
larger than the critical Toomre dispersion, thus the model is stable
against small radial perturbations.

The critical point in model construction is the determination of
mass-to-light ratios for individual populations. For the nucleus and
core this ratio can be determined from observations by two methods,
from spectrophotometric data and from virial theorem. For the halo we
can use the value for globular clusters, determined from velocity
dispersions.  For the disc and bulge we can use the value for open clusters, as
found from velocity dispersion,  from the rotation velocity at
distance from the center where the disc or bulge dominate, and from
calculations of the physical evolution of populations. We assume that
the bulge and disc have the same chemical composition and
mass-to-light ratio as open clusters with similar colour and spectral
properties. The dependence of $f_B$ of individual galactic populations
on the total mass of galaxies is shown in Fig.~\ref{FigA9} and on B-V
and U-B colours in Fig.~\ref{FigA10} \citep{Einasto:1974a}.

{\begin{figure*}[h] 
\centering 
\resizebox{0.50\textwidth}{!}{\includegraphics*{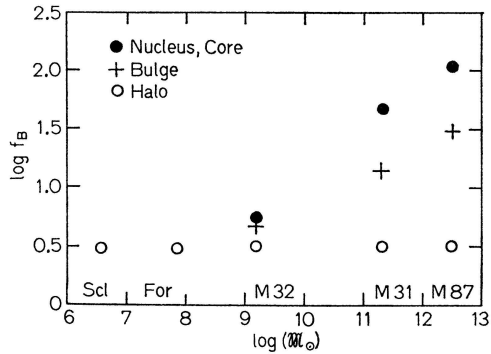}}
\caption{Dependence of mass-to-light ratio $f_B$ of old galactic
  populations on the total mass of galaxies \citep{Einasto:1974a}. }
  \label{FigA9}
\end{figure*} 
}

{\begin{figure*}[h] 
\centering 
\resizebox{0.40\textwidth}{!}{\includegraphics*{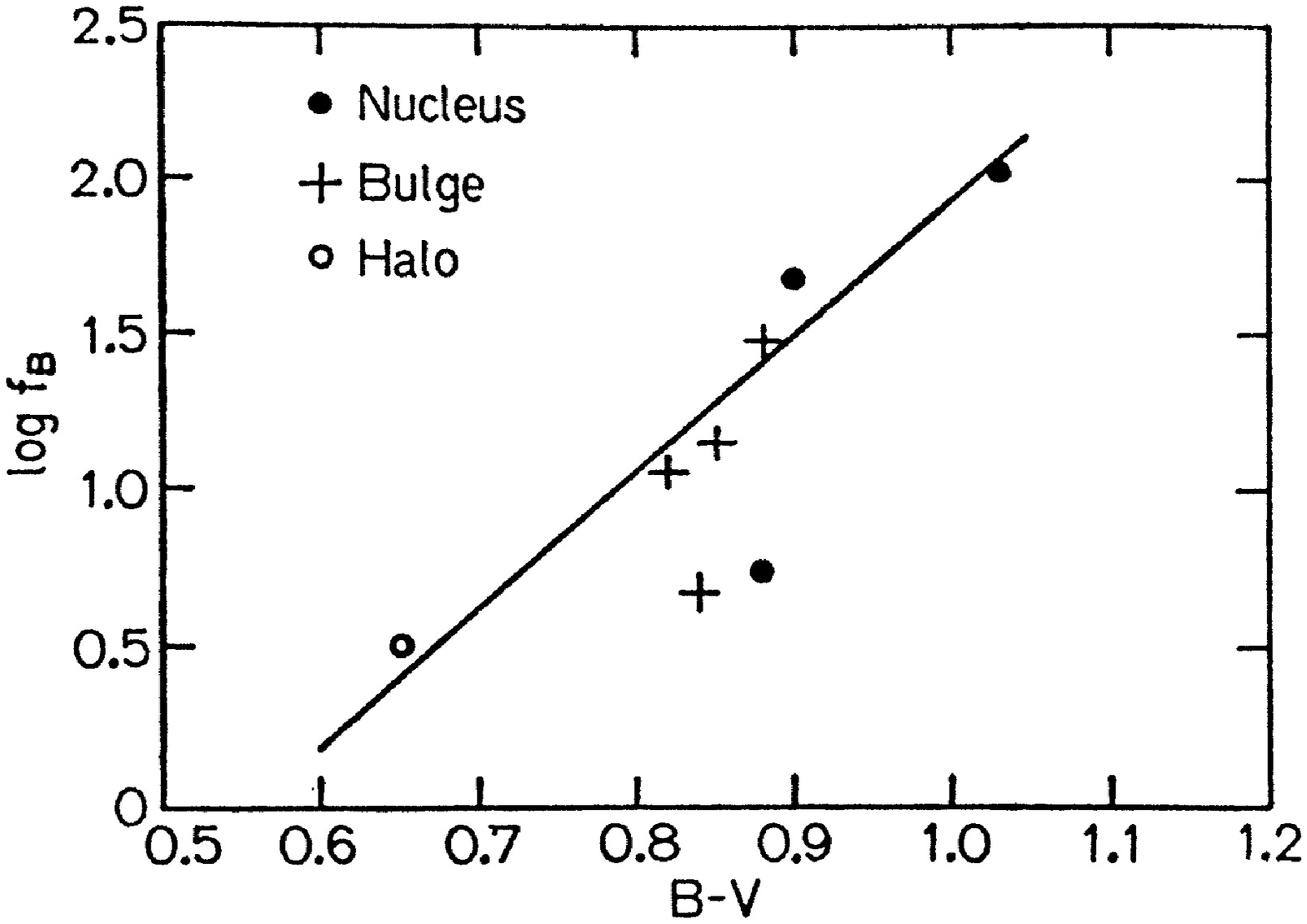}}
\resizebox{0.40\textwidth}{!}{\includegraphics*{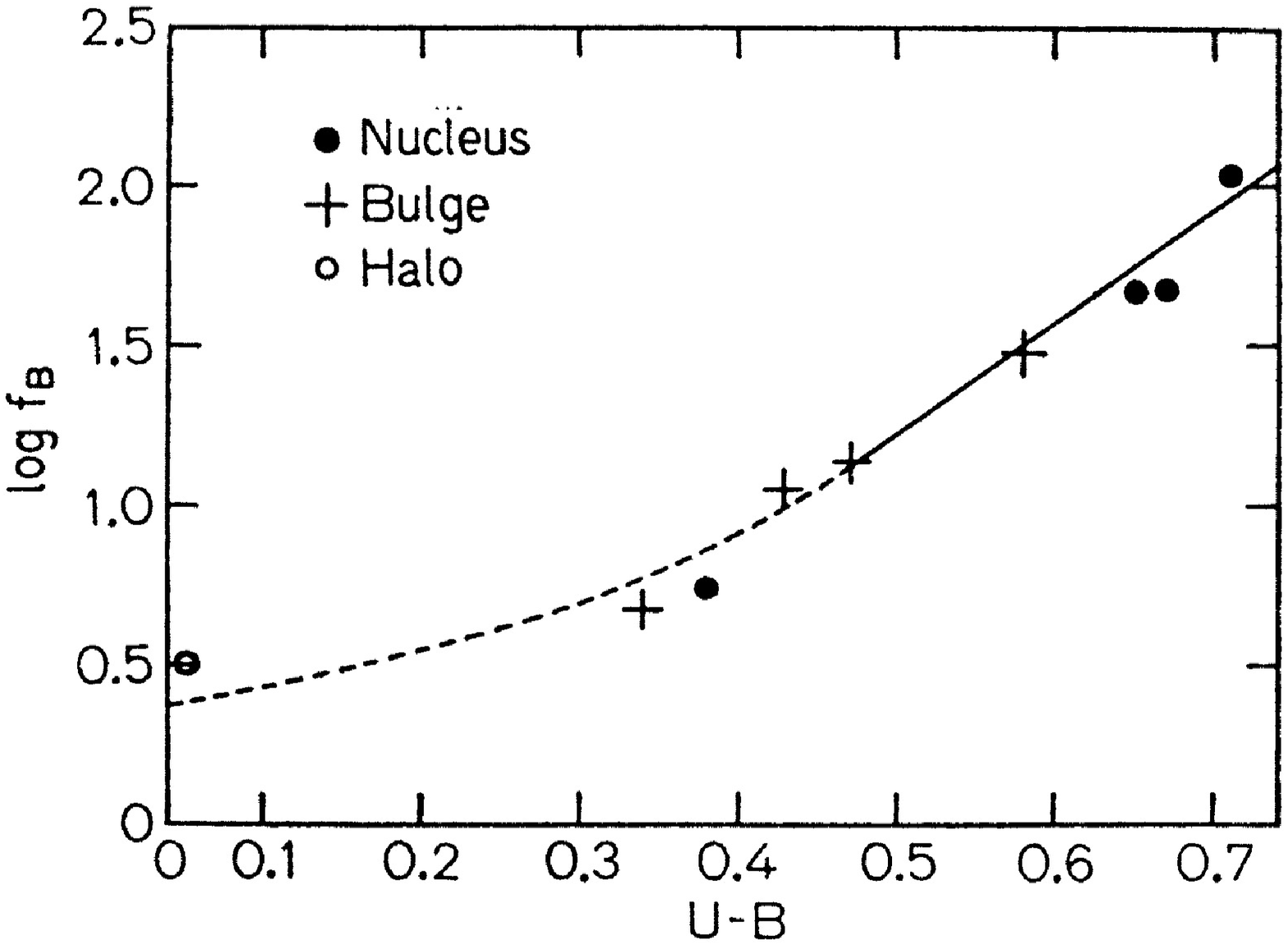}}
\caption{Dependence of mass-to-light ratio $f_B$ of old galactic
  populations on their B-V and U-B colours \citep{Einasto:1974a}. }
  \label{FigA10}
\end{figure*} 
}

Mass-to-light ratios 
$f_B=M/L_B$ of galactic populations are formed during
the evolution of stars, and are incorporated in dynamical models of
galaxies. $M/L_B$-ratios depend on the age and the chemical content of
populations, and are fixed by the minimal mass of stars in the
star-formation function, $M_0$, see Eqs.~(\ref{eq22.2}) and  (\ref{eq22.3}).  I accepted
for stellar populations with a normal metal content a lower star formation
limit $M_0 = 0.03~M_\odot$, for metal-rich populations a limit
$M_0 = 0.001~M_\odot$, and for metal-poor populations a limit
$M_0 = 0.1~M_\odot $. Most limits are lower than the lowest masses
needed to start hydrogen burning in stars,
$M^\ast = 0.08~M_\odot $. Using these lower mass limits, I got
for old metal-poor populations $M/L_B \le 3$, for old intermediate
populations $M/L_B \le 10$, and for old extremely metal-rich
populations $M/L_B \le 100$, see Figs.~\ref{Fig22.5} and \ref{FigA9}. The spatial
distribution of mass in populations is well determined, and
$M/L_B$-ratios can be checked by independent velocity dispersion data
in small systems of different age and chemical  content 
(open and globular clusters, nuclei of galaxies). As our model
calculations showed, it is impossible to reproduce with known
populations the observed flat rotation curves of spiral galaxies. In
contrast, models based on rotation velocities (\citet{Schmidt:1957uy},
\citet{Brandt:1965ty}, \citet{Roberts:1966aa}, \citet{Rubin:1970tp})
have a very rapid increase of $M/L_B$-ratios on the periphery of M31,
see Fig.~\ref{Fig17.8}. But these models contain no hint to understand
how to explain this increase.

During one of 1976 IAU General Assembly meetings, Sandra Faber
discussed her recent measurements of spectra of elliptical galaxies
\citep{Faber:1976}.  New data suggested that velocity dispersions of
the nuclei of elliptical galaxies are much lower than accepted so
far, which  leads to a considerable decrease of mass-to-light
ratios of elliptical galaxies. This suggests that corrections are
needed to my previous galaxy evolution models. This can be done by
changing the lower mass limit of forming stars, and using for all
populations identical lower mass limits,
$M_0 \approx 0.1~M_\odot$, which yields  lower $M/L$ values for all
populations. A very detailed review of masses and mass-to-light ratios
of galaxies is given by \citet{Faber:1979}.  Their Table 1 gives
$M/L_B$ values within Holmberg radius of galaxies with extended
rotation curves. These measured mass-to-light values lie in the
interval $0.6 \le M/L_B \le 12$, with a mean value about 4, which
corresponds to the disc of galaxies.

In galactic models, the main task is the determination of parameters
of populations.  First, a crude preliminary model is calculated, model
functions are compared with observed functions, and differences are
found.  In earlier models a simple trial-and-error procedure was
applied to find proper values of population parameters. In late 1970s,
Urmas Haud suggested applying an automatic procedure for model
parameters search. As in previous model calculations, first
preliminary values of model parameters are selected, model functions
are calculated and compared with observed functions.  To estimate the
degree of consistence of the model with observational data, the sum of
squares of relative deviations is calculated. Next, each model
parameter was changed by a small correction, the degree of consistency
was found, and a new model was calculated. This procedure was made for
all model parameters, one at a time. In this way, optimal values of
all model parameters were found. The iterations were completed when
the change of an arbitrary parameter by 1 percent did not reduce the
sum of the squares of relative deviations. The development of the
iteration program demanded much effort and time, thus the method was
published only in late 1980s by \citet{Einasto:1989qu}.  With this
method, first a new model of the Galaxy was found by
\citet{Haud:1989fo}, models of other galaxies were published by
\citet{Tenjes:1991zu,Tenjes:1994do,Tenjes:1998ij}.

We show in Table~\ref{TabA.3} parameters of components of the model of
M31 by \citet{Tenjes:1994do}, and in Fig~\ref{FigA6} rotation and
mass-to-light curves according to the model.  Parameters for the disc
and flat subsystems are for positive mass components.  In new models
the most essential change in comparison with earlier models is the
decrease of masses and mass-to-light ratios of the nucleus, core and
bulge, and the addition of a massive corona.

\begin{table*}[h]
\caption{Parameters of  components of M31} 
\label{TabA.3}   
{\small
  \centering    
\begin{tabular}{lcccccccc}
  \hline  \hline
  Quantity&Unit&Nucleus&Core&Bulge&Halo&Disc&Flat&Corona\\
  \hline
 $\epsilon$&      &0.69&0.82&0.67&0.47&0.10&0.02& 1\\
 $N$     &         & 1.2 & 1.5 & 2.4 &4.9 &  1.3&0.3& \\
 $a_0$  &kpc  &0.0039&0.10&0.75&4.8&4.1&11.1&60\\
$\mm{M}$&$10^{10}\,M_\odot$&0.031&0.20&1.0&0.8&8.4&0.75&320\\
$f_B$&  $M_\odot/L_\odot$&32&13&2.6&2.0&15&1.1&\\         
$U-B$&                               &0.88&0.80&0.54&0.21&0.90&$-0.38$&  \\ 
$B-V$&                                &       &1.03&0.97&0.79&1.01&0.45&\\  
  \hline
\end{tabular}\\
}
\end{table*}

{\begin{figure*}[h] 
\centering 
\resizebox{0.53\textwidth}{!}{\includegraphics*{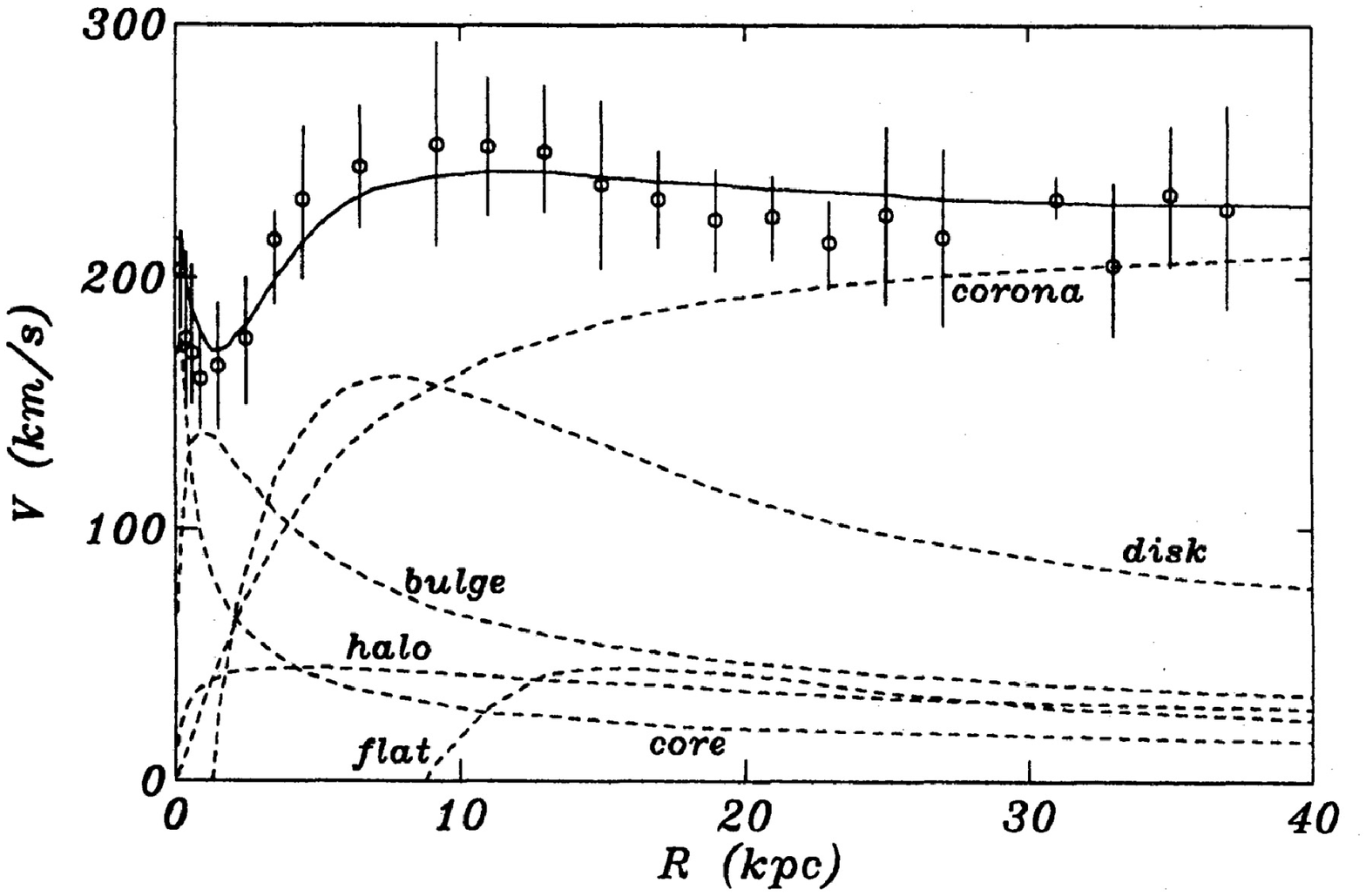}}
\resizebox{0.43\textwidth}{!}{\includegraphics*{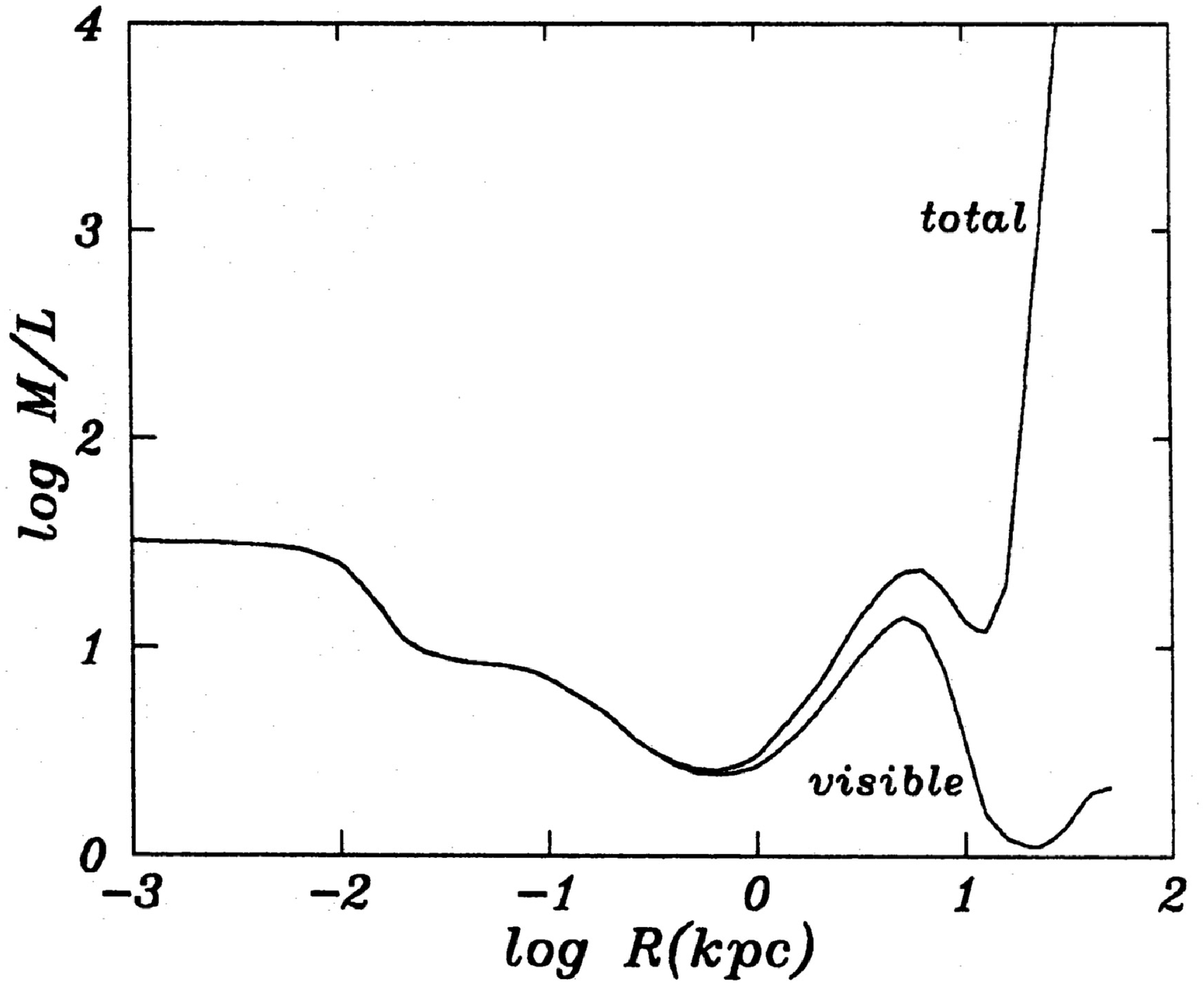}}
\caption{{\em Left:}. The rotation curve of M31 according to the model by
  \citet{Tenjes:1994do}. Open circles -- observations, thick line --
   model, dashed lines -- model curves of components.
  {\em Right:} Local mass-to-light ratios for visible
  populations and for the model with dark corona. 
  }
  \label{FigA6}
\end{figure*} 
}

In the case of the M31
model the decrease of masses and mass-to-light ratios reduces the
height of the peak of the model circular velocity at small distances
from the center, and the addition of the corona improves the model
circular velocity on large distance from the center.  These changes
are well seen when we compare Fig.~\ref{Fig17.3} variant B,
Fig.~\ref{Fig17.6} and Fig.~\ref{Fig20.7}, and Tables \ref{Tab17.1}
and \ref{Tab20.2} of previous models with respective data in the new
model, presented in Fig.~\ref{FigA6} and Table \ref{TabA.3}.

In galactic models, we used the modified exponential model,
Eq.~(\ref{eq7.3.1}), with a parameter $x_0$ to improve the shape of
the density profile near the centre.  This automatic model parameter
search showed that for all stellar populations the optimal value of
the parameter is $x_0=0$. In other words, there is no need for this
modification of the exponential profile. Presently this profile is
called ``Einasto profile'', and is used mainly to describe the spatial
density distribution of dark matter halos \citep{Merritt:2005fy}.

The study of the morphology of satellite galaxies was our first step
in the investigation of the environment of galaxies.  Following a
suggestion by Iosif Shklovsky we studied the dynamics of the
Magellanic Stream, discovered by \citet{Mathewson:1974ug}.  The
Magellanic Stream is a huge strip of gas through Magellanic Clouds.
We noticed that most companions of the Galaxy, including Magellanic
Clouds, the Magellanic Stream, and an another stream of high-velocity
hydrogen clouds lie close to a plane that is almost perpendicular to
the Galactic plane \citep{Einasto:1976vm}.  In this paper we used
velocities of satellite galaxies and Magellanic Stream gas clouds to
determine the mass of the Galaxy together with its satellites -- our
Local Hypergalaxy: $M_{tot}= 1.2\pm 0.5 \times 10^{12}\,M_\odot$.
Inspired by this pioneering work Urmas Haud continued the study of the
dynamics of high-velocity hydrogen clouds surrounding the Galaxy.

{\begin{figure*}[h] 
\centering 
\resizebox{0.40\textwidth}{!}{\includegraphics*{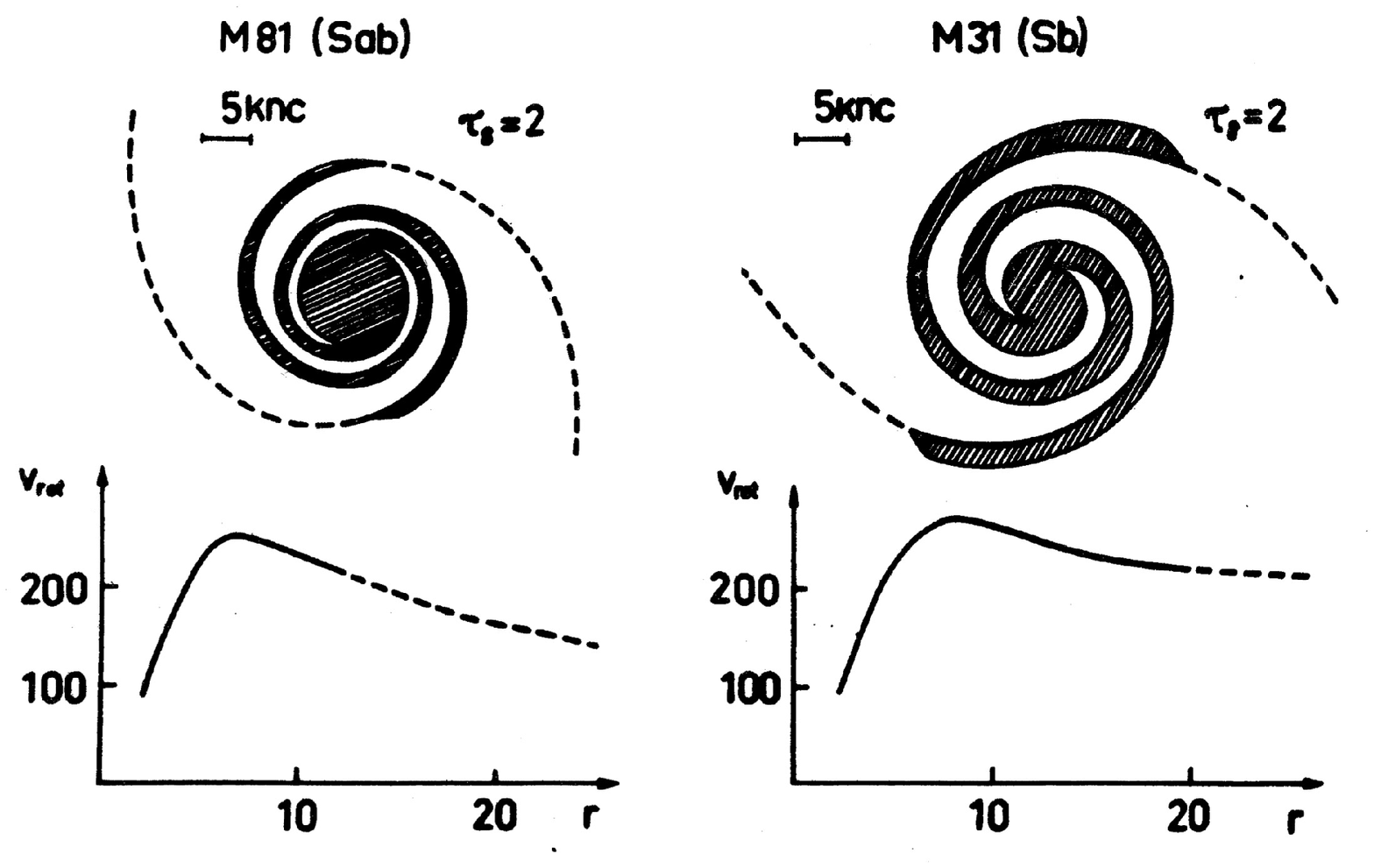}}
\resizebox{0.40\textwidth}{!}{\includegraphics*{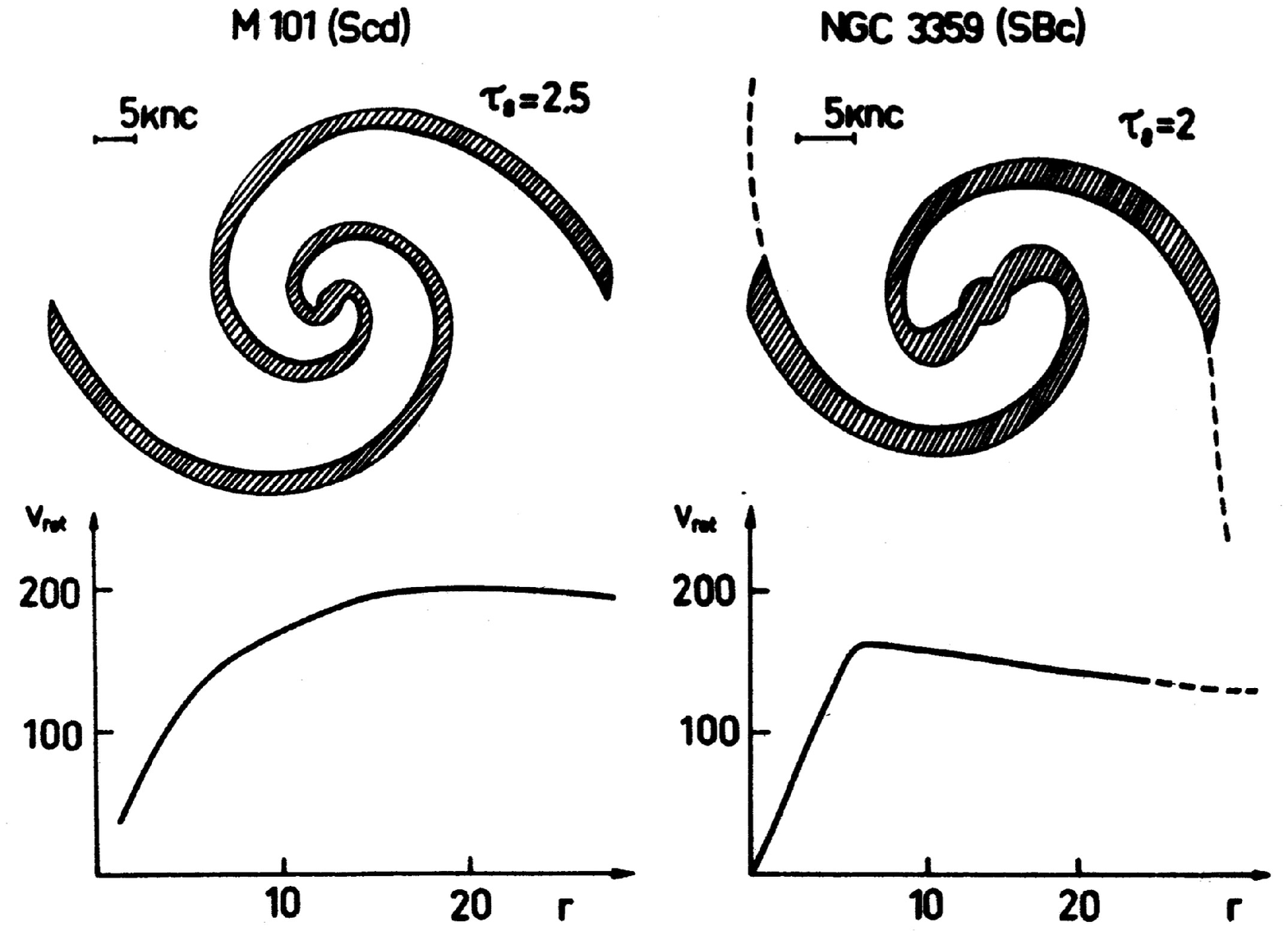}}
\caption{The spiral pattern of four galaxies according to
  \citet{Jaaniste:1976va}. Bold lines indicate the loci of new-born
  stars. The thickness of spiral arms (shaded areas) is determined by
  the lifetime of massive stars. Dotted curves are hydrogen spiral arms.
  }
  \label{FigA07}
\end{figure*} 
}

Another inspiration suggested by the Magellanic Stream phenomenon
concerns the formation of spiral structure of galaxies.
\citet{Jaaniste:1976va} and \citet{Einasto:1976d} noticed that in many giant galaxies dwarf
satellite companions are located close to planes perpendicular to the
main plane of the central giant galaxy.  It is natural to assume that
similar to the Magellanic Stream also other giant galaxies have
gaseous streams surrounding the main galaxy near the plane of
satellites.  Gas in these streams falls to the central galaxy along
the intersection of planes of the main galaxy and its satellites. These
streams initiate perturbations in the gas of the central galaxy and
give rise to star formation.  Due to the rotation of central galaxies
 star forming regions form a spiral
pattern. \citet{Jaaniste:1976va}  suggested that the accretion of gas can
be the main physical mechanism in the formation of spiral structure of
galaxies.   Using observed rotation curves of several galaxies  authors calculated
the expected form of spiral pattern.  Results are close to actually
observed spiral pattern of galaxies, see Fig.~\ref{FigA07}.

The study of the environment of galaxies was actually only an 
introduction to a much wider research area — the distribution of
galaxies on large scales.  Our involvement in these studies was
emphasised by Yakov Zeldovich.  After my report on dark matter in
galaxies in the Terskol winter school he turned to me and asked to
collaborate with him in the study of the Universe.  He was developing
a theory for the formation of galaxies \citep{Zeldovich:1970},
alternative theories were suggested by \citet{Peebles:1970} and
\citet{Ozernoi:1974qy}, and he was interested to find some
observational evidence that can be used to discriminate between these
theories.

Initially, we did not know how we can contribute to the problem of
galaxy formation. The expected consequences of the Zeldovich model were
discussed by \citet{Doroshkevich:1974fk} in the IAU Cosmology
Symposium in Krakow 1973. According to this scenario, the first forming
objects are superclusters of galaxies which fragment into galaxies.
The \citet{Peebles:1970} scenario suggests that the first forming objects
are small systems (galaxies or even star clusters), which by
gravitational clustering form superclusters of galaxies.
\citet{Ozernoi:1974qy} model did not predict any spatial distribution
of galaxies. When discussing the problem with my Tartu collaborators, I
remembered my previous experience in the study of galactic
populations: kinematical and structural properties of galactic
populations evolve only slowly, and thus remember their previous
state.  Large aggregates of galaxies remember their history better,
since the crossing time in these systems is larger.
Thus we had a leading idea for the search — we have to search
for regularities in the large-scale distribution of galaxies.

In this way, we started to collect data on spatial distribution of
galaxies in the nearby Universe. This resulted in the discovery of the
cosmic web by \citet{Joeveer:1977py}, \citet{Joeveer:1978a} and
\citet{Joeveer:1978b}.  The observed pattern of the distribution of
galaxies has some similarity with the expected distribution, as found
with numerical experiments by \citet{Doroshkevich:1977uc}. Following
this similarity, we called the observed distribution as
``cellular''. Subsequently we used the  term ``supercluster-void
network'' \citep{Einasto:1980jg}.  Presently the structure is called
``cosmic web'', following a suggestion by \citet{Bond:1996}.

The development of our understanding of the structure and evolution of
the Universe is described in detail by
\citet{Einasto:2014, Einasto:2018dz}. This forms a natural extension to
my earlier studies on the structure and evolution of galaxies.

  \vskip 5mm
\hfill November 2021

\backmatter

\chapter{Conclusions}\label{ch24}

In this Thesis, I have combined three previously independent areas of research in astronomy into one frame: kinematics and spatial
properties of Galaxy populations, development of dynamical models of
galaxies, and the study of the physical evolution of galaxies. The main
results of the study can be divided into methodical and astronomical.

\vskip 0.5cm
{\bf A. Methods of practical stellar dynamics}

\begin{enumerate}

\item{} A method has been developed to use tangential velocities to
  determine kinematical parameters of star samples (Chapter 1 \citep{Einasto:1954ts}).
  
\item{} A method has been developed to determine the mean velocity
  dispersion of samples of star using radial, tangential or spatial
  velocities, taking into account observational errors (Chapter 2
  \citep{Einasto:1955ty}). 

\item{} The concept of the system of galactic parameters has been
  elaborated, and a method to find the system  developed
  (Chapters 3,~5,~6
  \citep{Einasto:1961aa, Einasto:1964aa, Einasto:1964wc, Einasto:1964wx}).

\item{} A method has been developed to extrapolate the mass
  distribution function beyond the Sun's distance, and to determine the circular velocity
  at the Sun's distance from the Galactic centre (Chapter 7 \citep{Einasto:1965aa}).

\item{} The method to construct mass distribution models of galaxies
  is refined (Chapter 7 \citep{Einasto:1965aa}).

\item{} A classification of  models of stellar systems,  
  and conditions of physical correctness of models are developed
  (Chapters 8 and 13 \citep{Einasto:1969ab, Kutuzov:1968aa}). 

\item{} A method has been developed to construct spatial and
  hydrodynamical models of stellar systems (Chapters 10,~11
  \citep{Einasto:1968ad, Einasto:1970aa}). 

\item{} The virial theorem has been modified to apply it to components
  of galaxies (Chapter 12). 

\item{} It is shown that most models of stellar systems are particular
  cases of two model families: polynomial and binomial
  models (Chapters 13—15 \citep{Einasto:1968ab, Einasto:1968ac,
    Einasto:1968ad}). 
  
\item{} The generalised exponential model is suggested, and its
  descriptive functions are determined (Chapters 7, 16
  \citep{Einasto:1965aa, Einasto:1972ac}). 

\item{} Methods to analyse radio observations are refined 
  to determine density and
  velocity fields of galaxies (Chapter 19
  \citep{Einasto:1970tz,Einasto:1970vz}). 

\item{} The method to reconstruct the dynamical evolution of galaxies
  on the basis of the structure and kinematics of star populations of
  different ages is refined (Chapter 21). 

\item{} The method to investigate the physical evolution of galaxies
  is refined (Chapter 22).

\item{} A method has been developed to determine parameters of star
  formation function (Chapter 23 \citep{Einasto:1972aa}).

  \end{enumerate}
\vskip 1cm

  {\bf B. The study of the structure and evolution of regular galaxies}

  \begin{enumerate}
    
\item{} It is demonstrated that samples of stars of the main sequence
  later than F spectral class are kinematically heterogeneous (Chapter
  1 \citep{Einasto:1954ts}).

\item{} The relationship between kinematical characteristics and ages
  of stellar populations is found (Chapters 3,~4
  \citep{Einasto:1954ts,Einasto:1955tz}). 

\item{} A new system of galactic parameters is found (Chapters 5,~7
  \citep{Einasto:1964aa, Einasto:1964wc, Einasto:1964wx, Einasto:1965aa}). 

\item{} Models of the Galaxy are critically analysed and  new models
  are suggested in two approximations (Chapters 5,~7
  \citep{Einasto:1965aa, Einasto:1969ab, Einasto:1970ad}). 

\item{} The kinematical and spatial structure of the Andromeda galaxy
  M31 is studied and its spatial and hydrodynamical models developed in
  two approximations (Chapters 17 -- 20 \citep{Einasto:1969aa,
    Einasto:1970ac, Einasto:1972ab, Einasto:1972ae}).

\item{} The dynamical evolution of the Galaxy is reconstructed using
  kinematical characteristics of stellar populations of different ages
  (Chapter 21). 

\item{} On the basis of stellar evolutionary tracks and star formation
  function, a theory of the evolution of galaxies is elaborated
  (Chapter 22).

\item{} Parameters of star formation function are refined (Chapter 23
  \citep{Einasto:1972aa}).

  \end{enumerate}

  Results obtained in this series of studies and incorporated in the
  Thesis were discussed in astronomical seminars in Tartu Observatory,
  Leningrad State University, Sternberg Astronomical Institute, and in
  conferences and symposia in Alma-Ata, in IAU General Assembly in
  Hamburg 1964 \citep{Einasto:1964aa, Einasto:1964wc, Einasto:1964wx}
  and in Brighton 1970 \citep{Einasto:1970ad, Einasto:1970ae}, in IAU
  Symposium on Spiral Structure of Our Galaxy in Basel 1969
  \citep{Einasto:1970tz,Einasto:1970vz}, and in IAU Symposium on
  External Galaxies and Quasi Stellar Objects in Uppsala 1972
  \citep{Einasto:1972ab}.

  \vskip 5mm
\hfill November 1971

\chapter{Acknowledgements}

First of all I would like to thank my mentors Professors Taavet
Rootsm\"ae, Ernst \"Opik, Aksel Kipper and Grigori Kuzmin. Professors
Rootsm\"ae and Kipper created a very high moral and ethical atmosphere
in Tartu Observatory. Prof. Rootsm\"ae's moto was ``Science is carried
by the search for truth that is as sincere and honest as Nature
itself'', and Prof. Kipper emphasised, ``Let a hundred flowers bloom,
for we cannot foresee, which blossom will come to bear the best
fruit.''  Grigori Kuzmin was a student of Ernst \"Opik and advised me
on how to solve a new problem. First, the problem must be simplified
so that only the main factors are taken into account. On the basis of
this, one can find a preliminary answer. Thereafter, other factors
influencing the process under study can be taken into account, step by
step. This allows one to select all important factors and eliminate
less important ones. Fritz Zwicky called this approach to solve
scientific problems as ``morphological''. I tried to follow this
experience in my studies. When starting a new program, I first tried 
to find the answer myself and only later looked at what others have
done in the respective area. This helped me to have a fresh outlook on
a problem and avoid errors made by earlier investigators.

I am also very grateful to Grigori Kuzmin for his advice on
improving my Thesis manuscript. During the work on these problems, we
often had long discussions. He always gave me freedom to think for
myself and independently. His role was to help me to find errors in my
work and 
give hints on how to do things better.

My sincere gratitude goes to my colleagues at Tartu Observatory Maret
Einasto,  Mirt Gramann, Jaak
Jaaniste, Mihkel J\~oeveer, Urmas Haud, Gert H\"utsi, Ants Kaasik, Lev Kofman,
Sergei Kutuzov, Lauri-Juhan Liivam\"agi, Dmitri Pogosyan, Enn Saar,  Ivan Suhhonenko, 
Erik Tago, Antti Tamm, Elmo Tempel, Peeter Tenjes,  Peeter Traat and Jaan Vennik
for the fruitful collaboration and their contribution to the
studies, to Peeter Kalamees, Aivo Kivila, Urve R\"ummel,  Hilju
Silvet and Margus Sisask for their help in computations
and forming manuscripts, to Piret Zettur for scanning the
manuscript and producing with OCR software readable text,  to Kerli
Linnat for proofreading the English version of the Thesis, and to
Marja-Liisa Plats for help in preparing the text for printing.

During the study I had contacts with astronomers in other centres
within Soviet Union — Viktor Ambartsumian, Arthur Chernin, Igor
Karachentsev, Evgeny Kharadze, Andrei Linde, Kirill Ogorodnikov, Pavel
Parenago, Josif Shklovsky, Alexei Starobinsky, and many others.  In
later years I collaborated closely with Yakov Zeldovich and his team
of young physicists Andrei Doroshkevich, Anatoly Klypin, Sergei
Shandarin, Rashid Sunyaev and Alexei Starobinsky.  Yakov Zeldovich
insisted that in solving problems, attention is to be given to
essential factors, and when results are available, they must be
promptly published. He also emphasised that major results must be
published in major journals. He was interested in the nature of
phenomena and was quick to change his opinion, if new data suggested a
revision.  In a conference in Tallinn in April 1981, the non-baryonic nature of
dark matter was suggested, and at the conference banquet Zeldovich gave an
enthusiastic speech:  {\em Observers work hard in sleepless nights to
collect data; theorists interpret observations, are often in error,
correct their errors and try again; and there are only very rare
moments of clarification. Today it is one of such rare moments when we
have a holy feeling of understanding secrets of Nature.}

I benefitted from contacts with astronomers from other
countries. These contacts started already during the Soviet period of
my life by visiting conferences and exchanging letters, and continued
after the independence of Estonia was restored. Most contacts 
started after the Thesis was finished and our
Tartu team was involved in the study of dark matter and cosmic web. 
I had discussions with George Abell, Heinz Andernach, Neta and John
Bahcall, Ed Bertschinger, Peter Brosche, Margaret and Geoffrey
Burbidge, George Contopoulos, Gerard de Vaucouleurs, Tim de Zeeuw,
Sandy Faber, Margaret Geller,
Wilhelm Gliese, John Huchra, Bernard Jones, Rocky Kolb, Dave Latham, Donald
Lynden-Bell, Vicent Martinez, Dick Miller, Volker M\"uller, Jan Henrik
Oort, Jerry Ostriker, Changbom Park, Jim Peebles, Lubo\^s Perek, Joel
Primack, Martin Rees, Mort Roberts, Vera Rubin, Remo Ruffini,  Hyron Spinrad, Alex
Szalay, Gustav Andreas Tammann, Beatrice Tinsley, Alar Toomre,
Virginia Trimble, Sidney van den Bergh, Rien van de Weygaert, Simon
White, and many others. My sincere thanks to all my friends and
colleagues — interactions with them helped to develop the present
concept of the structure of the universe.

I thank Dr. Tiiu Kaasik and Prof. Toomas Asser who helped me to stay
fit and in good health. 

My special thanks goes to my wife Liia, both for her participation in
the early stages of the research and for her support and understanding
during our common life for more than 50 years.  I am very grateful to
my daughter Maret and grandchildren Peeter, Triin and Stiina for their
help in many ways.

I acknowledge the financial support by the Estonian Science Foundation
and the Estonian Ministry of Education and the European Regional
Development Fund (TK133).


\addcontentsline{toc}{chapter}{Bibliography}
\bibliographystyle{aa2}

\end{document}